\documentclass[reprint,showpacs,superscriptaddress,english,twocolumn,notitlepage,longbibliography,nofootinbib]{revtex4-1}
\usepackage{times}
\usepackage{graphicx}
\usepackage{float}
\usepackage{latexsym,amsmath,amssymb,bm,euscript}
\usepackage{color}
\usepackage{subfigure}
\usepackage{epstopdf}
\usepackage[colorlinks=true,linkcolor=blue,citecolor=blue,urlcolor=blue]{hyperref}
\usepackage{hyperref}
\usepackage{ulem}
\usepackage[dvipsnames]{xcolor}
\usepackage{dcolumn}     
\usepackage{bbm}

\usepackage{tikz}
\usetikzlibrary{decorations.pathreplacing}
\newcommand{\tikznode}[3][inner sep=0pt]{\tikz[remember
picture,baseline=(#2.base)]{\node(#2)[#1]{$#3$};}}
\usepackage{ytableau}

\newcommand{\be}[0]{\begin{equation}}
\newcommand{\ee}[0]{\end{equation}}
\def\ba#1\ea{\begin{align*}#1\end{align*}}  
\def\baa#1\eaa{\begin{align}#1\end{align}}  
\newcommand{\up}[0]{\uparrow}
\newcommand{\dn}[0]{\downarrow}
\newcommand{\bmat}[0]{\begin{bmatrix}}
\newcommand{\emat}[0]{\end{bmatrix}}

\newcommand{\beq}{\begin{equation}}
\newcommand{\eneq}{\end{equation}}

\def\qq{\mathbf{q}}
\def\kk{\mathbf{k}}

\def\RR{\mathbf{R}}

\def\rr{\mathbf{r}}

\def\QQ{\mathbf{Q}}

\def\kk{\mathbf{k}}
\def\qq{\mathbf{q}}

\def\QQ{\mathbf{Q}}
\def\RR{\mathbf{R}}

\def\ee{\epsilon}

\def\spin{{\varsigma}}

\def\up{\uparrow}

\def\ee{\epsilon}

\def\qq{\mathbf{q}}
\def\kk{\mathbf{k}}

\def\RR{\mathbf{R}}

\def\rr{\mathbf{r}}

\def\QQ{\mathbf{Q}}

\def\spin{{\varsigma}}

\def\hH{{ \hat{H} }}

\def\UC{{\hat{\Theta}}}
\def\UF{{\hat{\Sigma}}}

\newcounter{SMmark}
\makeatletter

\@addtoreset{equation}{SMmark}
\@addtoreset{figure}{SMmark}
\@addtoreset{table}{SMmark}
\@addtoreset{page}{SMmark}
\@addtoreset{section}{SMmark}

\makeatother

\begin{document}

\title{Kondo Lattice Model of Magic-Angle Twisted-Bilayer Graphene: Hund's Rule, Local-Moment Fluctuations, and Low-Energy Effective Theory}

\author{Haoyu Hu}
\affiliation{Donostia International Physics Center, P. Manuel de Lardizabal 4, 20018 Donostia-San Sebastian, Spain}

\author{B. Andrei Bernevig}
\email{bernevig@princeton.edu} 
\affiliation{Department of Physics, Princeton University, Princeton, New Jersey 08544, USA}
\affiliation{Donostia International Physics Center, P. Manuel de Lardizabal 4, 20018 Donostia-San Sebastian, Spain}
\affiliation{IKERBASQUE, Basque Foundation for Science, Bilbao, Spain}

\author{ Alexei M. Tsvelik}
\affiliation{Division of Condensed Matter Physics and Materials Science, Brookhaven National Laboratory, Upton, NY 11973-5000, USA}

\begin{abstract}
We apply a generalized Schrieffer–Wolff transformation to the extended Anderson-like topological heavy fermion (THF) model for the magic-angle $(\theta=1.05^{\circ})$ twisted bilayer graphene (MATBLG) (Phys. Rev. Lett. {\bf 129}, 047601 (2022)), 
to obtain its Kondo Lattice limit. In this limit localized $f$-electrons on a triangular lattice interact with topological conduction $c$-electrons. By solving the exact limit of the THF model, we show that the integer fillings $\nu=0, \pm 1, \pm 2$ are controlled by the heavy $f$-electrons, while $\nu = \pm 3$ is at the border of a phase transition between two $f$-electron fillings. For $\nu=0, \pm 1, \pm 2$, we then calculate the RKKY interactions between the $f$-moments in the full model and analytically prove the SU(4) Hund's rule for the ground state which maintains that two $f$-electrons fill the same valley-spin flavor. Our (ferromagnetic interactions in the) spin model dramatically differ from the usual Heisenberg antiferromagnetic interactions expected at strong coupling. We show 
the ground state in some limits can be found exactly by employing a positive semidefinite "bond-operators" method.  We then compute the excitation spectrum of the $f$-moments in the ordered ground state, prove the stability of the ground state favored by RKKY interactions, and discuss the properties of the Goldstone modes, the (reason for the accidental) degeneracy of (some of) the excitation modes, and the physics of their phase stiffness. We develop a low-energy effective theory for the $f$-moments 
and obtain analytic expressions for the dispersion of the collective modes. We discuss the relevance of our results to the spin-entropy experiments in twisted bilayer graphene. 
\end{abstract}


\maketitle

{\it Introduction---}
The discovery of the correlated insulating phase~\cite{Cao2018} and superconductivity~\cite{Cao2018_sc} in the MATBLG~\cite{bistritzer2011moire} has driven considerable theoretical~\cite{tbgiii,HEJ19,PAD20,DOD18,SOE20,EUG20,HUA20a,ZHA21,KAN21,HOF21,ZHA19,WU19,FER20,WIL20,WAN21a,BER21a,KAN21,MATBLG_hf_DS} 
and experimental efforts~\cite{CAO20,CAO21,LU19,YAN19,STE20,XIE21a,JIA19,WON20,ZON20,DAS21,PAR21c,LU21,Otteneder2020,Lisi2021,PhysRevResearch.3.013153,Seifert2020,Otteneder2020,Hesp2021,diez2021magnetic,PhysRevLett.127.197701,PhysRevMaterials.6.024003,Jaoui2022,PhysRevLett.128.217701,Grover2022,DiBattista2022}
to understand 
its topology~\cite{NUC20,CHO21a,SAI21,WU21a, ZOU18,EFI18,KAN18,SON19,PO19,PAD18,HEJ21,KAN20a,SON21,LIA20a,HEJ19a,Balents2020,Andrei2021}
and correlation physics~\cite{POL19,LIU21e, XIE19, CHO19, NUC20, KER19,PO18a,KAN19,XIE21,VEN18,YOU19,BUL20a,THO18,DA19,DA21,YUA18,Balents2020,Andrei2021}.
Theoretically, correlated insulators~\cite{XIE20a,OCH18,LIU19,SEO19,CHR20,LIU21,CEA20,ZHA20,tbgiv,tbgv,GUO18,PhysRevLett.129.076401}, ferromagnetic order~\cite{PIX19,WU20b,REP20,tbgiv,LIU19a}, superconductivity~\cite{LIA19,WU18,XU18,GUI18,ISO18,LIU18,GON19,KHA21,CHR20,LEW21,LIU21a,KEN18,HUA19,JUL20,XIE20,KON20,CHI20a}, and other exotic quantum phases 
~\cite{ABO20,KWA21,REP20a,LED20,BAL20,CHA21,xie2022phase,SHE21}
 have been identified and systematically studied which all point to rich physics~\cite{Balents2020} of the MATBLG.
The recent experiments~\cite{Saito2021,ROZ21,XIE19,WON20} have provided evidence for fluctuating local moments and Hubbard-like physics. 
Meanwhile the theoretical understanding is challenging since the stable topology~\cite{SON19,SON21} of the flat bands obstructs 
the symmetric real-space description. A real-space extended Hubbard model~\cite{VAF21,KAN19,KOS18,XU18b,VEN18} can still be constructed, but a certain symmetry ($C_{2z}T$ or $P$) becomes non-local.
To address this problem 
the authors of Ref.~\cite{HF_MATBLG} have introduced an exact mapping of MATBG to a THF model. THF is a version  of the extended Anderson lattice model describing localized $f$-electrons interacting with topological conduction $c$-electrons. The $f$-electrons have zero kinetic energy and strong Hubbard interactions; they admit a description in terms of localized Wannier orbitals centered at the AA-stacking region.  
The topological flat bands can be recovered from the hybridization between the $f$- and the $c$-electrons~\cite{HF_MATBLG}.

In this letter, we map the THF model to a Kondo lattice model using a generalized Schrieffer-Wolff (SW) transformation which takes into account the density-density interaction term between $f$- and $c$-electrons. In this limit, the dynamics of the localized orbitals becomes the one  of the  $f$-moments.  By solving exactly a particular limit of the THF model, we show that the integer fillings $\nu=0,\pm 1,\pm2$ are controlled by the $f$-electrons, while the situation is drastically different for $\nu=\pm 3$ which sits at the phase transition between two $f$-electrons fillings. 
In the Kondo lattice model of MATBLG, the local moments formed by the fully-localized $f$-electrons interact with topological conduction electron bands via Kondo superexchange  and direct ferromagnetic exchange interaction. These two types of exchange interactions induce an RKKY interaction. At $\nu=0,-1,-2$, the RKKY interaction dominates the physics and stabilizes the ferromagnetic ground states that obey the  Hund's rule.
This result provides an analytic derivation of the Hund's rule found  numerically in the Hartree-Fock calculations of THF model \cite{HF_MATBLG}. 
We then proceed to investigate fluctuations of the $f$-moments 
 in the symmetry broken ground states
by developing the low-energy effective theory  and calculating the excitation spectrum.

{\it Schrieffer–Wolff transformation and Kondo lattice model---}
The single-particle Hamiltonian of the THF model contains the kinetic term $\hH_c$ describing the  topological conduction $c$-electron bands (SM~\cite{SM}, Sec.~I), and the hybridization between the $f$- and the $c$-electrons $\hH_{fc}$~\cite{HF_MATBLG,SM}: 
\baa 
&\hH_{fc} 
=  \sum_{\substack{|\kk|<\Lambda_c, \RR \\{i,\xi,\xi'}} }\bigg( \frac{e^{i\kk \cdot \RR -\frac{|\kk|^2 \lambda^2}{2}}\tilde{H}_{\xi\xi'}^{(fc)}(\kk)}{\sqrt{N_M}}  
 \psi_{\RR,i}^{f,\xi,\dag} \psi_{\kk, i}^{c',\xi'}  +\text{h.c.}\bigg) \nonumber \\ 
 &\tilde{H}^{(fc)}(\kk) = 
\begin{bmatrix}
    \gamma & v_\star^\prime (k_x-ik_y)  \\ 
    v_\star^\prime (k_x+ik_y) & \gamma 
\end{bmatrix}
\, .
\label{eq:hyb_mat} 
\eaa 
where $\psi_{\kk,i}^{c',\xi,\dag}$ creates $\Gamma_3$ ``conduction" $c$-electron with momentum $\kk$, valley-spin flavor $i\in\{1,2,3,4\}$ (with $(1,2,3,4)$ corresponding to $(+\up, -\up, +\dn,-\dn)$), and "orbital" index $\xi=(-1)^{a+1}\eta$ (with $a=1,2$ are original orbital indices and $\eta=\pm$ are valley indices defined in Ref.~\cite{HF_MATBLG}). $\psi_{\RR,i}^{f,\xi,\dag}$ creates $f$-electron at moir\'e unit cell $\RR$ with valley-spin flavor $i$, and orbital index $\xi=(-1)^{a+1}\eta$, where $a=1,2,\eta =\pm$~\cite{HF_MATBLG}. 
$\Lambda_c$ is the momentum cutoff, $N_M$ is the total number of moir\'e unit cells and $\lambda$ is the damping factor~\cite{HF_MATBLG}.
In the hybridization matrix $\tilde{H}^{(fc)}(\kk)$, we only keep the first two terms ($\gamma$ and $v_\star^\prime$) in the expansion in powers of $\kk$.
The flat band limit is realized by setting $M=0$, where $M$ is taken as a parameter of $\hH_c$.

The interaction Hamiltonian of the THF model is
$
\hH_I = \hH_U + \hH_J +\hH_V +\hH_W 
$,
where $\hH_U, \hH_J$ describe respectively  the on-site Hubbard interaction of the $f$-electrons
($U=57.95$meV), and  the ferromagnetic exchange  between the $f$- and the $c$-electrons
($J=16.38$meV), $\hH_V, \hH_W$ describe respectively  the repulsion between the $c$-electrons 
($\sim 48$meV)
and  the repulsion between the $f$- and the $c$-electrons 
($W=47$meV)~\cite{HF_MATBLG,dumitru_2022}.
The full Hamiltonian is  $\hH_c+\hH_{fc}+\hH_I$. The model possesses the  $U(4)\times U(4)$ symmetry in the chiral-flat limit ($M=0$, $v_\star^\prime=0$), a flat 
$U(4)$ symmetry in the nonchiral-flat limit ($M=0,v_\star^\prime \ne 0$), a chiral $U(4)$ symmetry in the chiral-nonflat limit $(M\ne 0,v_\star^\prime =0$) and a $U(2)\times U(2)$ symmetry in the nonchiral-nonflat limit  $(M\ne 0,v_\star^\prime \ne 0$)~\cite{HF_MATBLG, tbgiii, tbgiv,BUL20,VAF20,VAF21,SM}.

 At sufficiently strong on-site Coulomb interaction $U$, the $f$-electrons are fully localized and give rise to local $f$-moments which are defined
 as
\baa
\UF_{\mu\nu}^{(f,\xi\xi')}(\RR) =\sum_{i,j}\frac{1}{2}T_{ij}^{\mu\nu} \psi_{\RR,i}^{f,\xi,\dag} \psi_{\RR,j}^{f,\xi'} 
\label{eq:f_moment_def_main}\,.
\eaa 
$\{T^{\mu\nu}_{ij}\}$ with $\mu,\nu \in\{0,x,y,z\}$ are given in SM~\cite{SM}, Sec.~I.
Eq.~\ref{eq:f_moment_def_main} are the generators of the $U(8)$ group ($8=2 \text{(orbital)} \times 2 \text{(valley)} \times 2\text{(spin)}$) where the $U(1)$ charge component can be gauged away for the fully-localized $f$-electrons.

\begin{figure}
    \centering
    \includegraphics[width=0.5\textwidth]{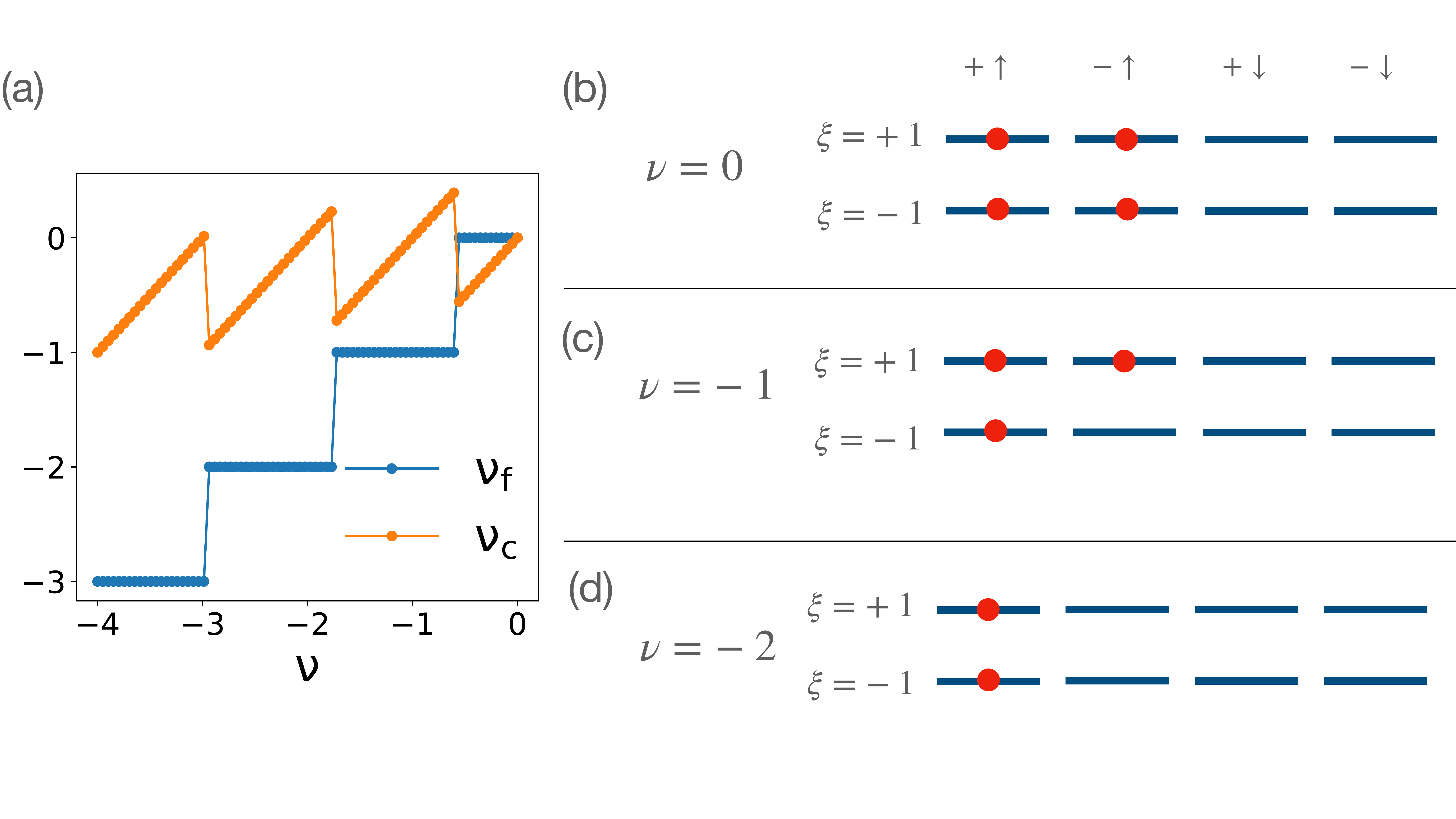}
    \caption{(a) Filling of $f$ electrons($\nu_f$) and $c$-electrons($\nu_c$) as a function of total filling($\nu$) in the zero hybridization model. (b), (c), (d) Illustrations of ground states at $\nu=0,-1,-2$. The red dot means the filling of one $f$-electron.}
    \label{fig:gnd_state}
\end{figure}

We first analyze the zero-hybridization limit of the model 
($\gamma=0,v_\star^\prime=0$). $v_\star^\prime$ is small~\cite{HF_MATBLG} while $\gamma$ changes rapidly and goes through zero at the value of $w_0/w_1=0.9$ close to the actual MATBLG value $w_0/w_1=0.8$~\cite{dumitru_2022}. Hence this limit can be thought as a meaningful approximation close to the MATBLG.
The zero-hybridization model is exactly solvable at zero Coulomb repulsion between $c$-electrons.
Here, we treat $\hH_V$ in the mean-field approximation ($\hH_V^{MF}$) and drop the $\hH_J$ which is relatively weak. 
We then solve the model under the assumption that each site is filled with $\nu_f+4$ $f$-electrons with an integer $\nu_f$
(SM.~\cite{SM}, Sec.~II). We use $\nu_c$ to denote the filling of the $c$-electrons and use $\nu=\nu_f+\nu_c$ to denote the total fillings.
In Fig.~\ref{fig:gnd_state} (a), we plot $\nu_f$ and $\nu_c$ of the ground state as a function of $\nu$. At $\nu=0,-1,-2,-3$, the ground state has $\nu_f=\nu$ and $\nu_c=0$. 
$\nu =-3$ is close to the transition point between $\nu_f=-3$ state and $\nu_f=-2$ states. 
Thus, at $\nu=-3$, our assumption of uniform charge distribution may be violated~\cite{xie2022phase}, and a Kondo model description fails. 
This is consistent with the special place that $\nu=-3$ has in the TBG physics~\cite{xie2022phase}. 


At $\nu=0,-1,-2$, we fix the filling of the $f$-electrons (according to Fig.~\ref{fig:gnd_state} (a)) and perform a generalized SW transformation (SM.~\cite{SM}, Sec. IV), which leads to
the following Kondo lattice Hamiltonian:
\baa 
\hH_{Kondo} = \hH_{c} +\hH_{cc} +\hH_{K}  +\hH_J +\hH^{MF}_V +\hH_W 
\label{eq:kondo_ham_main}
\eaa 
where $\hH_K$ and $\hH_{cc}$ are the Kondo interaction and one-body scattering term generated by the SW transformation
respectively~\cite{SW_transf,SM}. The Kondo interaction $\hH_K$ takes the form of
\baa 
&\hH_K
= \sum_{\RR}\sum_{|\kk|<\Lambda_c, |\kk+\qq|<\Lambda_c}\sum_{\mu\nu \xi\xi'}
\frac{e^{-i\qq \cdot \RR} e^{-\frac{|\kk|^2 +|\kk+\qq|^2}{2}\lambda^2 }}{N_M}
\nonumber  \\
&
\bigg[ 
\frac{\gamma^2}{D_{\nu_c,\nu_f}} :\UF_{\mu\nu}^{(f,\xi\xi')}(\RR) : :\UF_{\mu\nu} ^{(c',\xi'\xi)}(\kk,\qq): + \label{eq:Kondo_main}\\
&\bigg( \frac{v_\star^\prime \gamma (k_x-i\xi k_y)}{D_{\nu_c,\nu_f}} :\UF_{\mu\nu}^{(f,\xi\xi')}(\RR): :\UF_{\mu\nu} ^{(c',\xi'-\xi)}(\kk,\qq):
+\text{h.c.}
\bigg) 
\bigg] \nonumber 
\eaa 
where the colon symbols represent the normal ordering and $\Sigma_{\mu\nu}^{(c',\xi'\xi)}(\kk,\qq)$ (SM~\cite{SM}, Sec.~I) is the $U(8)$ moment of $c$-electrons defined in the  manner similar to  $\Sigma_{\mu\nu}^{(f,\xi'\xi)}(\RR)$ in Eq.~\ref{eq:f_moment_def_main}. 
The parameter $D_{\nu_c,\nu_f}$ at $\nu = \nu_f=0,-1,-2$ is defined as
\baa 
\frac{1}{D_{\nu_c=0,\nu_f}}
=-\frac{1}{ (U-W)\nu_f-\frac{U}{2}} +\frac{1}{ (U-W)\nu_f+\frac{U}{2}}
\label{eq:defD_main} \, . 
\eaa
The distinct feature of this expression is the presence $W$ absent in the standard Kondo Hamiltonians. 
In addition, we perform a $\kk$-expansion in the square bracket of Eq.~\ref{eq:Kondo_main}, and keep only the zeroth and the linear order terms in $\kk$; as was done in the expression for  the hybridization matrix $\tilde{H}^{(fc,\eta)}(\kk)$ (Eq.~\ref{eq:hyb_mat}).
The zeroth order Kondo coupling 
has strength $\gamma^2/D_{\nu_c=0,\nu_f} =42.3\mathrm{meV}, 49.3\mathrm{meV}, 98.6\mathrm{meV}$ at $\nu=0,-1,-2$ respectively.

\begin{figure*}
    \centering
    \includegraphics[width=1.0\textwidth]{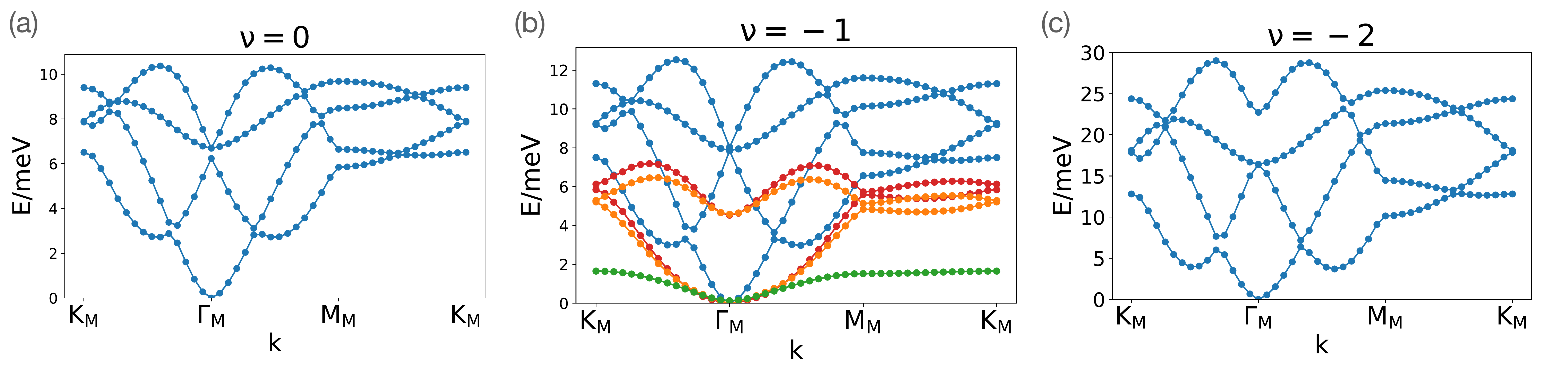}
    \caption{Excitation spectrum at $\nu=0,-1,-2$. Blue, orange, red and green denote the fluctuations in the full-empty sector, half-empty sector, full-half sector and half-half sector respectively. } 
    \label{fig:spin_spec}
\end{figure*}

{\it The RKKY interactions and the Hund's rule---} 
By integrating out the c-electrons in the Kondo Hamiltonian(Eq.~\ref{eq:kondo_ham_main}),
one induces an RKKY interaction between the $f$-moments~\cite{RKKY_1,RKKY_2,RKKY_3} . We restrict ourselves to deriving  the RKKY interactions~\cite{SM} in the leading order in $\hH_K,\hH_J$ at integer fillings $\nu=0,-1,-2$. 
In the chiral-flat limit ($v_\star^\prime =0, M=0$), the RKKY interactions can be described by the following $U(4)\times U(4)$ symmetric Hamiltonian 
\baa 
\hH_{RKKY}^{v_\star^\prime=0,M=0}= &\sum_{\substack{\RR,\RR' \\ \mu\nu,\xi\xi'}}[J^{RKKY}_{0}(\RR')+J^{RKKY}_{1}(\RR')\delta_{\xi,\xi'}] \nonumber \\
&:\Sigma_{\mu\nu}^{(f,\xi\xi')}(\RR)::\Sigma_{\mu\nu}^{(f,\xi'\xi)}(\RR+\RR') : \, ,
\eaa 
where both $J^{RKKY}_0(\RR') (\le 0)$ and $J^{RKKY}_1(\RR') (\le 0)$ can be analytically obtained and are ferromagnetic (SM~\cite{SM}, Sec.V). Using bond operators $A_{\RR,\RR'}^{\xi,\xi'} = \sum_i \psi_{\RR,i}^{f,\xi,\dag}\psi_{\RR',i}^{f,\xi'}$, we show that $\hH_{RKKY}^{v_\star^\prime=0,M=0}$ is a Positive Semidefinite Hamiltonian~\cite{tbgiii,KAN19}, where the exact ground state can be obtained~\cite{SM,tbgiii,KAN19,tbgiv}.
The grounds states
are ferromagnetic with the form of (SM~\cite{SM}, Sec. VI)
\baa 
\prod_\RR \bigg[ \prod_{n=1}^{\nu_f^++2} \psi_{\RR,i_n}^{f,+,\dag} \prod_{m=\nu_f^++3}^{\nu_f+4} \psi_{\RR,i_m}^{f,-,\dag} \bigg] |0\rangle 
\label{eq:gnd_nu_0} \, .
\eaa 
where $\nu_f^+$ denotes the filling of the $f$-electrons with index $\xi =1$, and $\nu_f^+=0$ at $\nu=0$, $\nu_f^+=-1,0$ at $\nu=-1$, and $\nu_f^+=-1$ at $\nu=-2$. $\{i_n\}$ are chosen arbitrarily and $|0\rangle$ is the vacuum with $\psi_{\RR,i}^{f,\xi}|0\rangle =0$. We note that our ground states form a subset of the ground states in Ref.~\cite{tbgiv}, due to the additional non-zero kinetic energy generated by $f$-$c$ hybridizations in our model. 

We then consider the nonchiral-flat limit ($v_\star^\prime \ne 0$, $M=0$), where the RKKY interactions are flat-$U(4)$ symmetric. They lift the ground-state degeneracy in 
Eq.~\ref{eq:gnd_nu_0}. We obtain the ground states by treating $v_\star^\prime$-induced RKKY interactions as perturbations (SM~\cite{SM}, Sec. VI) 
, and the true ground state is selected by the following RKKY interactions 
\baa 
\sum_{\substack{\RR,\RR'\\\mu\nu,\xi}}  J^{RKKY}_2(\RR')
:\UF_{\mu\nu}^{(f,\xi\xi)}(\RR):
:\UF_{\mu\nu}^{(f,-\xi-\xi)}(\RR+\RR'): 
\label{eq:hund_rule_int}
\eaa 
where $J_2^{RKKY}(\RR) (\le 0)$ is analytically obtained and ferromagnetic (SM~\cite{SM}, Sec.~V). $J^{RKKY}_2(\RR_2-\RR)$ tends to align two $f$-moments $:\UF_{\mu\nu}^{(f,++)}(\RR):$ and $:\UF_{\mu\nu}^{(f,--)}(\RR_2):$
and stabilize the ground states obeying the Hund's rule (see Fig.~\ref{fig:gnd_state}). 
The corresponding ground states are consistent with Ref.~\cite{tbgiv, KAN19} (nonchiral-flat limit). 

Moreover, we also derive the RKKY interactions in the zero-hybridization limit ($\gamma=0,v_\star^\prime=0$) with non-zero $J(\ne 0)$ (SM~\cite{SM}, Sec.~III), where the corresponding ground states are consistent with Ref.~\cite{tbgiv} (chiral-nonflat limit). 


{\it Fluctuations of the $f$-moments---}
We check the stability of the ferromagnetic ground state derived from RKKY Hamiltonian by studying small fluctuations. 
We restrict ourselves to the flat-$U(4)$ nonchiral-flat limit $M=0,v_\star^\prime \ne 0$ at $\nu=0,-1,-2$. 
To describe the fluctuations we introduce for each site a $ 8\times 8$ traceless Hermitian matrix $u_{i\xi,j\xi'}(\RR)$, where 
$i,j\in \{1,2,3,4\}$ are valley-spin flavors 
and $\xi,\xi' \in \{+,-\}$. The $f$-moments in Eq.~\ref{eq:f_moment_def_main} can then be written as (SM~\cite{SM}, Sec. VIII).
\baa 
\UF_{\mu\nu}^{(f,\xi\xi')}(\RR) = \sum_{ij}T^{\mu\nu}_{ij} [e^{iu(\RR)}\Lambda e^{-iu(\RR) }]_{i\xi,j\xi'} 
\label{eq:def_uf_u}
\eaa 
where $\Lambda$ is an $8\times 8$ matrix defined as $\Lambda_{i\xi,j\xi'} = \langle \psi_0 |:\psi_{\RR,i}^{f,\xi,\dag} \psi_{\RR,j}^{f,\xi'}:
|\psi_0\rangle/2 $
and $|\psi_0\rangle$ is the ground state.
A non-zero $u_{i\xi,j\xi'}(\RR)$ will generate fluctuations by rotating the $f$-moments from their ground-state expectation values. 

The ferromagnetic order in the ground state
opens a gap in the single-particle spectrum of the $c$-electrons which allows us to safely integrate them out~\cite{SM} and to  develop an effective theory for small fluctuations by expanding the action to the second order in $u_{i\xi,j\xi'}(\RR,\tau)$, where $\tau$ is the imaginary time (SM~\cite{SM}, Sec. VIII). 
The Lagrangian of the effective theory is provided in SM~\cite{SM}, Sec.~VIII, 
where we find the diagonal components $u_{i\xi,i\xi}(\RR,\tau)$ only contribute a total derivative and we will focus on the off-diagonal components: $u_{i\xi,j\xi'}(\RR,\tau)$ with $i\xi \ne j\xi'$. 

We then introduce two sets $S_{fill}$ and $S_{emp}$ to 
characterize the ground state, where $S_{fill}$ and $S_{emp}$ denote the sets of $i\xi$ indices that are filled with one and zero number of the $f$-electrons at each site, respectively. 
A fluctuation generated by $u_{i\xi,j\xi'}(\RR)$ ($i\xi \ne j\xi'$) is described by the procedure of moving one $f$-electron from $j\xi'$ flavor at site $\RR$ to $i\xi$ flavor at the same site (SM~\cite{SM}, Sec.~VIII).
This procedure can only be valid when $i\xi \in S_{emp},j\xi' \in S_{fill}$. 
Consequently, only $u_{i\xi,j\xi'}(\RR,\tau)$ with $j\xi' \in S_{fill},i\xi \in S_{emp}$ and also its complex conjugate $u_{j\xi',i\xi}(\RR,\tau)$ that describes the reverse procedure appear in our effective Lagrangian. 
We diagonalize the Lagrangian and plot the excitation spectrum in Fig.~\ref{fig:spin_spec}. 

We first analyze the spectrum at $\nu=0,-2$. The Lagrangian density in the long-wavelength limit and in momentum $(\kk)$ and frequency $(\omega)$ spaces has the form of $L = L_{Goldstone} + L_{gapped}$:
\begin{eqnarray}
&& L_{Goldstone} = \sum_{\substack{j \xi \in {S}_{fill}\\ i \xi\in {S}_{emp}}}u_{j,i}^\dag({\bf k},\omega)\Big(\omega  - \frac{k^2}{2m_0}\Big)u_{j,i}({\bf k},\omega), \nonumber\\
&&L_{gapped} = \sum_{\substack{j \xi \in {S}_{fill}\\ i \xi\in {S}_{emp}}}U_{j,i}^\dag({\bf k},\omega)[\omega \hat I - \hat H(\kk)]U_{j,i}({\bf k},\omega), \nonumber \\
&& H(\kk)
= \left(
\begin{array}{ccc}
\frac{k_+k_- }{2m_1}+\Delta_1 & Vk_+& - Vk_- \\
 V k_- &\frac{k_+k_-}{2m_2} +\Delta_2 & \frac{k_-^2}{2m_3}\\
- V k_+ & \frac{k_+^2}{2m_3} & \frac{k_+k_-}{2m_2} +\Delta_2 
\end{array}
\right) 
\label{eq:lag_nu_even}
\end{eqnarray}
where $u_{j,i} = (u_{(j+,i+)}+u_{(j-,i-)})/\sqrt 2$, and  $U_{j,i}^T =((u_{(j+,i+)}-u_{(j-,i-)})/\sqrt 2, u_{(j+,i-)},u_{(j-,i+)})$ and $k_{\pm} = k_x \pm ik_y$, $k=|\kk|$. 
$m_0,m_1,m_2,V,\Delta_1,\Delta_2,V$
are analytically defined constants that depend on the parameters of the original THF Hamiltonian (SM~\cite{SM}, Sec.~VIII). The Goldstone modes  with quadratic dispersion 
decouple from the rest (gapped modes). Their stiffness is $1/m_0=5.9\mathrm{meV}\cdot a_M^2$ (at $\nu =0$), $13.2\mathrm{meV}\cdot a_M^2$ (at $\nu = -2$), where $a_M$ is the moir\'e lattice constant.   

The gaps at $\kk =0$ are $\Delta_1$ and $\Delta_2$, with $\Delta_2$ corresponding to two-fold degenerate modes. This feature is reproduced in Fig.~\ref{fig:spin_spec} (c). The exceptional case when all three modes are degenerate ($\Delta_1 =\Delta_2$) is depicted in Fig.~\ref{fig:spin_spec} (a) and the  condition for the degeneracy is $\alpha =[ -0.37  + \sqrt{ 0.14 + 0.23  (v_\star^\prime) ^2/(\gamma \lambda)^2 } ]\gamma^2/(JD_{\nu_c\nu_f}) =1 $. Using the realistic values of parameters, we find $\alpha =1.07\approx 1$. 
By directly evaluating the Lagrangian, we also observe "roton" minima in the gapped modes (most obviously at $\nu=-2$) at $|\kk|\sim 0.3|\bm{b}_{M1}|$ with $\bm{b}_{M1}$ the moir\'e reciprocal lattice vector.
The roton mode has small anisotropy with the minimum located along the $\Gamma_M$-$M_M$ line. 

We next discuss the number of the Goldstone modes. Each combination of $(i,j)$ that satisfies $i+,i-\in S_{emp}$, $j+,j- \in S_{fill}$ produce a Goldstone mode~\cite{SM}. This leads to
four Goldstone modes at $\nu=0$ and three Goldstone  modes at $\nu=-2$~\cite{tbgv}.
Furthermore, all excitation modes depicted in Fig.~\ref{fig:spin_spec} are four-fold degenerate at $\nu=0$ and three-fold degenerate at $\nu=-2$, due to the remaining $U(2)\times U(2)$ symmetries at $\nu=0$ and the remaining $U(1)\times U(3)$ symmetries at $\nu=-2$. 

We now analyze the spectrum at $\nu=-1$. Unlike $\nu=0,-2$ where each valley-spin flavor is filled with either two or zero $f$- electrons, at $\nu=-1$, there is one valley-spin flavor (denoted by $i=1$) filled with two $f$-electrons and one valley-spin flavor (denoted by $i=2$) filled with one $f$-electron and two empty valley-spin flavors (denoted by $i=3,4$) as shown in Fig.~\ref{fig:gnd_state} (c). 
This allows us to classify 
$u_{i\xi,j\xi'}(\RR,\tau)$ at $\nu=-1$ into four sectors: (1) full-empty sector with $i = 3,4, j=1$ or $i= 1, j=3,4$; (2) half-empty sector with $i = 2, j=3,4$ or $i = 3,4, j=2$ ; (3) full-half sector with $i=1,j=2$ or $j=2,i=1$; (4) half-half sector with $i=2,j=2$. 
In Fig.~\ref{fig:spin_spec}, we label the excitation in different sectors with different colors. Due to the remaining $U(1)\times U(1)\times U(2)$ symmetry of the $\nu=-1$ ground state
, each mode is two-fold degenerate in the full-empty and half-empty sectors and is non-degenerate in the full-half and half-half sectors. We find $2$ degenerate Goldstone modes with stiffness $6.7\mathrm{meV}\cdot{a_M^2}$ in the full-empty sector, $2$ degenerate Goldstone modes with stiffness $1.3\mathrm{meV}\cdot a_M^2$ in the half-empty sector, and $1$ Goldstone mode with stiffness $1.8\mathrm{meV}\cdot {a_M^2}$ in the full-half sector~\cite{tbgv}. 

Several remarks are in order. Firstly, some of the gapped modes at $\nu=0,-1,-2$ are relatively flat as shown in Fig.~\ref{fig:spin_spec}. 
 Secondly, at $\nu=-1$, one of the flat modes (green curve in Fig.~\ref{fig:spin_spec} (b)) with eigenfunction $u_{2-,2+}(\kk)$ has a tiny gap $0.12\text{meV}$ at $\Gamma_M$ point and a very narrow bandwidth $1.5$meV. 
 Thirdly, the dispersion along $M_M$ to $K_M$ is also relatively flat which is related to the approximate $C_{\infty}$ symmetry of the excitation modes. 
 

{\it Summary and discussions---}
We have constructed and studied a Kondo lattice model for  MATBLG. Its distinct feature is the Dirac character of the $c$-electron spectrum: at integer fillings, there is no Fermi surface.  
Hence the Kondo screening is irrelevant and the  low energy physics is dominated by the  RKKY interactions, which is also responsible for the Hund's rule and ferromagnetism of the ground states. We have developed an effective theory describing small fluctuations of the local moments and found their excitation spectrum. We have also discussed the properties of the Goldstone and gapped modes and their  degeneracies. 
We believe that our work provides insight into the correlated ground states and the local-moment fluctuations of the MATBLG.

We also comment on the connection with previous works~\cite{tbgv,KHA20}.  Some features, including the soft modes and accidental degeneracy of gapped modes at $\Gamma_M$, also appear in the projected Coulomb model~\cite{tbgv}, but the roton modes do not. We note that the projected Coulomb model~\cite{tbgv} has ignored the effect of remote bands.
Roton modes have also been seen in Ref.~\cite{KHA20}, where the remote bands have been included. However, our model gives a larger bandwidth of the gapped modes than Ref.~\cite{KHA20}. We point out that, we take a different set of parameter values (including dielectric constant, and gate distance). 
Besides, in our model, all Goldstone modes have quadratic dispersions, in contrast to Ref.~\cite{KHA20} which found a linear one.  The difference in the results has its origin in the different symmetry properties. We consider the flat limit of the model, and the quadratic dispersion comes from the broken flat $U(4)$ symmetry. Ref.~\cite{KHA20} takes non-flat bands, and the linear Goldstone modes come from the broken $U(1)$ valley symmetry. 

We conclude the paper with a discussion of the relevance of our results to the recent entropy experiments~\cite{Saito2021,ROZ21}. Experimentally, the entropy in TBG near $\nu=-1$ has been found to be of the order of Boltzmann’s constant and is suppressed by the applications of magnetic fields. This large entropy can be explained by the presence of soft mode at $\nu =-1$, which will be suppressed by the magnetic field.
Finally, we point out that the fluctuations of the $f$-moments could potentially generate attractive interactions between $c$-electrons and drive the system to the superconducting phase. 
Thus, our work also establishes a platform for understanding the
superconductivity.

{\it Note added---}
After finishing this work, we have learned that related, but not identical, results had  recently been obtained by the S. Das Sarma's~\cite{das_sarma_kondo}, P. Coleman's~\cite{coleman_kondo} and Z. Song's groups~\cite{song_kondo}.

{\it Acknowledgements---}
We thank Z. Song for discussions. We also thank P. Coleman and S. Das Sarma for discussions. 
B. A. B.’s work was primarily supported by the DOE Grant No. DE-SC0016239. H. H. was supported by the European Research Council (ERC) under the European Union’s
Horizon 2020 research and innovation program (Grant Agreement No. 101020833). 
 Further support was provided by the Gordon and Betty Moore Foundation through the EPiQS Initiative, Grant GBMF11070 and the European Research Council (ERC) under the European Union’s
Horizon 2020 research and innovation program (Grant Agreement No. 101020833)
A. M. T. was supported by the Office of Basic Energy Sciences, Material Sciences and Engineering Division, U.S. Department of Energy (DOE) under Contract No. DE-SC0012704.

\bibliography{ref}

\clearpage 
\onecolumngrid
\begin{center}
\textbf{Supplementary Materials}
\end{center}

\stepcounter{SMmark}
\stepcounter{page}

\newcommand{\hh}{\textcolor{black}}
\newenvironment{hhc}{\par\color{black}}{\par}
\newenvironment{hbb}{\par\color{black}}{\par}

\newcommand{\hhb}{\textcolor{black}}

\newcommand{\hb}{\textcolor{black}}

\newcommand{\bh}{\textcolor{black}}
\renewcommand{\thefigure}{S\arabic{figure}}

\renewcommand{\thetable}{S\arabic{table}}
\renewcommand{\tablename}{Supplementary Table}

\renewcommand{\thesection}{S\arabic{section}}

\renewcommand{\theequation}{S\arabic{equation}}

\tableofcontents

\clearpage

\section{Model and notation}

The topological heavy-fermion model introduced in Ref.~\cite{HF_MATBLG} takes the following Hamiltonian 
\baa  
\hH = \hH_c  +\hH_{fc} + \hH_U +\hH_J + \hH_W + \hH_V
\eaa  
The single-particle Hamiltonian of conduction $c$-electrons has the form of 
\baa 
 &\hat{H}_c = \sum_{\eta,s,a,a',|\kk |<\Lambda_c} 
 H_{a,a'}^{(c,\eta)}(\kk )c_{\kk a\eta s}^\dag c_{\kk a'\eta s} 
 \quad ,\quad 
 H^{(c,\eta)}(\kk ) = \begin{bmatrix}
 0_{2\times 2 } & v_\star(\eta k_x\sigma_0+ik_y\sigma_z)\\
 v_\star(\eta k_x\sigma_0-ik_y\sigma_z) & M\sigma_x .
 \end{bmatrix}  
 \label{eq:hc_def}
\eaa  
where $\sigma_{0,x,y,z}$ are identity and Pauli matrices.
$c_{\kk a\eta s}$ represents the annihilation operator of the $a(=1,2,3,4)$-th conduction band basis of the valley $\eta (=\pm)$ and spin $s(=\up,\dn)$ at the moir\'e momentum $\kk$. At $\Gamma_M$ point ($\kk=0$) of the moir\'e Brillouin zone, $c_{\kk 1 \eta s},c_{\kk 2 \eta s}$ form a $\Gamma_3$ irreducible representation (of $P6'2'2$ group),
$c_{\kk 3 \eta s},c_{\kk 4 \eta s}$ form a $\Gamma_1\oplus \Gamma_2$ reducible (into $\Gamma_1$ and $ \Gamma_2$ - as they are written, the  $c_{\kk 3 \eta s},c_{\kk 4 \eta s}$ are just the $\sigma_x$ linear combinations of $\Gamma_1 \pm  \Gamma_2$ ) representation (of $P6'2'2$ group). $\Lambda_c$ is the momentum cutoff for the $c$-electrons. $N$ is the total number of moir\'e unit cells. 
Parameter values are 
$v_\star = -4.303 \mathrm{eV}\cdot\mathrm{\mathring{A}}$, $M=3.697$meV.  

The hybridization between $f$ and $c$ electrons has the form of 
\baa  
\hat{H}_{fc} =\frac1{\sqrt{N_M}} \sum_{\substack{|\kk|<\Lambda_c\\ \RR}} \sum_{\alpha a \eta s} \bigg( e^{i\kk\cdot\RR -\frac{|\kk|^2\lambda^2}2} H^{(fc,\eta)}_{\alpha a} (\kk)  f_{\RR \alpha\eta s}^\dag c_{\kk a\eta s} + h.c. \bigg) \ ,
\label{eq:def_hfc}
\eaa  
where $f_{\RR \alpha\eta s}$ represents the annihilation operators of the $f$ electrons with orbital index $\alpha(=1,2)$, valley index $\eta(=\pm)$ and spin $s(=\up,\dn)$ at the moir\'e unit cell $\RR$. $N_M$ is the number of moir\'e unit cells and $\lambda=0.3376a_M$ is the damping factor, where $a_M$ is the moir\'e lattice constant.
The hybridization matrix $H^{(fc,\eta)}$ has the form of 
\begin{equation} 
   H^{(fc,\eta)}(\kk) = \begin{pmatrix}
     \gamma \sigma_0 + v_\star'(\eta k_x\sigma_x + k_y\sigma_y), & 0_{2\times 2} 
    \end{pmatrix} 
    \label{eq:def_hyb_mat}
\end{equation}
which describe the hybridization between $f$ electrons and $\Gamma_3$ $c$ electrons $(a=1,2)$. Parameter values are $\gamma=-24.75$meV, $v_\star'=1.622 \mathrm{eV}\cdot\mathrm{\mathring{A}}$. 

$\hH_{U}$ ($U = 57.89\mathrm{meV}$) describes the on-site interactions of $f$-electrons.
    \begin{equation}
    \hH_{U} =  \frac{U}2 \sum_{\RR} :n^f_{\RR}: :n^f_{\RR}:  ,
    \end{equation} 
     where $n^f_{\RR} = \sum_{\alpha\eta s} f_{\RR \alpha\eta s}^\dagger f_{\RR \alpha\eta s}$ is the $f$-electrons density and the colon symbols represent the normal ordered operator with respect to the normal state: $:f^\dagger_{\RR\alpha_1 \eta_1 s_1} f_{\RR \alpha_2 \eta_2 s_2}: = f^\dagger_{\RR\alpha_1 \eta_1 s_1} f_{\RR \alpha_2 \eta_2 s_2} - \frac{1}{2} \delta_{\alpha_1 \eta_1 s_1; \alpha_2 \eta_2 s_2}$. 

The ferromagnetic exchange interaction between $f$ and $c$ electrons $\hH_J$ is defined as 
\baa  
 H_J = - \frac{J}{2N_M} \sum_{\RR s_1 s_2} \sum_{\alpha\alpha'\eta\eta'} \sum_{|\kk_1|,|\kk_2|<\Lambda_c }   
     e^{i( \kk_1- \kk_2 )\cdot\RR } 
     ( \eta\eta' + (-1)^{\alpha+\alpha'} )
     :f_{\RR \alpha \eta s_1}^\dagger f_{\RR \alpha' \eta' s_2}:  :c_{\kk_2, \alpha'+2, \eta' s_2}^\dagger  c_{ \kk_1, \alpha+2, \eta s_1}: 
     \label{eq:hj_def}
\eaa  
where $J=16.38$meV and $ :c_{\kk_2, \alpha'+2, \eta' s_2}^\dagger  c_{ \kk_1, \alpha+2, \eta s_1}: =  c_{\kk_2, \alpha'+2, \eta' s_2}^\dagger  c_{ \kk_1, \alpha+2, \eta s_1}-\frac{1}{2}\delta_{\kk_1,\kk_2}\delta_{\alpha,\alpha'}\delta_{\eta,\eta'} \delta_{s_1,s_2} $

The repulsion between $f$ and $c$ electrons $\hH_W$ has the form of 
\baa  
&\hat{H}_W = \sum_{\eta,s,\eta',s',a,\alpha}\sum_{|\kk |<\Lambda_c, |\bm{\kk+\qq}|<\Lambda_c } 
W_ae^{-i\qq \cdot \RR} :f_{\RR, a\eta s}^\dag f_{\RR ,a\eta s}:
:c_{\kk+\qq, a\eta's'}^\dag c_{\kk,a\eta's'} :
\label{eq:hw_def}
\eaa 
\bh{where $W_1=W_2,W_3=W_4$~\cite{HF_MATBLG}. We further require $W_1=W_2=W_3=W_4= W=47$meV (the difference is about 10-15\% )}.

The Coulomb interaction between $c$ electrons has the form of
\baa  
&     \hH_{V} = \frac{1}{2\Omega_0 N_M} \sum_{\eta_1 s_1 a_1} \sum_{\eta_2 s_2 a_2} \sum_{|\kk_1|, |\kk_2|<\Lambda_c} \sum_{\substack{\qq \\ |\kk_1+\qq|,|\kk_2+\qq|<\Lambda_c}} V(\qq) 
        :c_{\kk_1 a_1\eta_1 s_1}^\dagger c_{\kk_1+\qq a_1 \eta_1 s_1}:
        :c_{\kk_2+\qq a_2 \eta_2 s_2}^\dagger c_{\kk_2 a_2 \eta_2 s_2}:
        \label{eq:hv_def}
\eaa 
where $\Omega_0$ is the area of the moir\'e unit cell and $V(\qq=0)/\Omega_0= 48.33\mathrm{meV}$.\hb{
We will always take a mean-field treatment of $\hH_V$, which gives 
\baa  
\hH_V^{MF} =- \frac{V(0)}{2\Omega_0}N_M\nu_c^2 +\frac{V(0)}{\Omega_0}\nu_c\sum_{|\kk|<\Lambda_c,a\eta s} (c_{\kk,a \eta s}^\dag c_{\kk, a \eta s}-1/2)
\label{eq:hv_mf_def}
\eaa 
where $\nu_c $ is the filling of $c$-electrons and $|\psi_0\rangle$ 
\baa 
\nu_c = \langle \psi_0| \hat{\nu_c} |\psi_0\rangle \,. 
\eaa 
$|\psi_0\rangle$ is the ground state and the density operator $\hat{\nu_c}$ of $c$-electrons is defined
\baa  
\hat{\nu}_c = \frac{1}{N_M}\sum_{|\kk|<\Lambda_c,a\eta s}c_{\kk,a\eta s}^\dag c_{\kk, a \eta s}-1/2 
\label{eq:def_hat_c}
\eaa  
and $|\psi_0\rangle$ is the ground state.
Clearly, at the mean-field level $\hH_V^{MF}$ is equivalent to a chemical potential shifting. In addition, the energy loss from $\hH_V^{MF}$ is
\baa  
\langle \psi_0| \hH_{V}^{MF} |\psi_0\rangle =\frac{V(0)}{2\Omega_0}N_M \nu_c^2 
\label{eq:hv_mf_exp}
\eaa 
}

\hb{ The model has a $U(4)\times U(4)$ symmetry at chiral-flat limit $M=0,v_\star^\prime =0$, a flat $U(4)$ symmetry at flat limit $M=0$ and a chiral $U(4)$ symmetry at $v_\star^\prime=0$. In addition, the particle-hole conjugation transformation will map the ground state of the model at $\nu$ (total filling of $f$ and $c$ electrons) to the ground state at $-\nu$. Thus we only consider $\nu\le 0$ in this work. 
}

\subsection{$U(4)$ moments}

To observe the symmetry of the system, we introduce the following flat $U(4)$ moments \hb{(flat $U(4)$ symmetry at $M=0$)}
\begin{equation}  \label{eq:U4op-maintext}
\begin{aligned}
\UF_{\mu\nu}^{(f,\xi)}(\RR) =& \frac{\delta_{\xi, (-1)^{\alpha-1}\eta}}2 A^{\mu\nu}_{\alpha\eta s, \alpha'\eta's'} f_{\RR \alpha \eta s}^\dagger f_{\RR \alpha' \eta' s'} \\ 
\UF_{\mu\nu}^{(c\prime,\xi)}(\qq) =& \frac{\delta_{\xi, (-1)^{a-1}\eta}}{2N_M} \sum_{|\kk|<\Lambda_c}A^{\mu\nu}_{a\eta s, a'\eta's'} c_{\kk+\qq a \eta s}^\dagger c_{\kk a' \eta' s'},\; (a,a'=1,2) \\
\UF_{\mu\nu}^{(c\prime\prime,\xi)}(\qq) =& \frac{\delta_{\xi, (-1)^{a-1}\eta}}{2N_M} \sum_{|\kk|<\Lambda_c}B^{\mu\nu}_{a\eta s, a'\eta's'} c_{\kk+\qq a \eta s}^\dagger c_{\kk a' \eta' s'},\; (a,a'=3,4)
\end{aligned}
\end{equation}
where repeated indices should be summed over and $A^{\mu\nu},B^{\mu\nu}$ ($\mu,\nu=0,x,y,z$) are eight-by-eight matrices 
\begin{equation} \label{eq:flat-U4-maintext}
\begin{aligned}
A^{\mu\nu} =& \{ \sigma_0 \tau_0 \spin_\nu, \sigma_y \tau_x \spin_\nu, \sigma_y \tau_y \spin_\nu, \sigma_0 \tau_z \spin_\nu \} \\
B^{\mu\nu} =& \{ \sigma_0 \tau_0 \spin_\nu, - \sigma_y \tau_x \spin_\nu, - \sigma_y \tau_y \spin_\nu, \sigma_0 \tau_z \spin_\nu \}\ ,
\end{aligned}
\end{equation}
with $\sigma_{0,x,y,z}$, $\tau_{0,x,y,z}$, $\spin_{0,x,y,z}$ being the Pauli or identity matrices for the orbital, valley, and spin degrees of freedom, respectively.  
The $\pm1$ valued index $\xi$, equal to $(-1)^{\alpha-1}\eta$ or $(-1)^{a-1}\eta$ in the \hh{generators the $f$ and $c$ electrons respectively}, labels different fundamental representations of the flat-U(4) group. 
The global flat-U(4) rotations are generated by
\baa
\UF_{\mu\nu} = \sum_{\xi=\pm1} \bigg( \UF_{\mu\nu}^{(f,\xi)} + \UF_{\mu\nu}^{(c\prime,\xi)} + \UF_{\mu\nu}^{(c\prime\prime,\xi)}\bigg).
\label{eq:full_flat_u4_gen}
\eaa 

The chiral $U(4)$ moments \hb{(chiral limit with $v_\star^\prime=0$ and different other parameters~\cite{HF_MATBLG})} can be defined in a similar manner 
\begin{equation}  \label{eq:U4op-maintext}
\begin{aligned}
\UC_{\mu\nu}^{(f,\xi)}(\RR) =& \frac{\delta_{\xi, (-1)^{\alpha-1}\eta}}2 \Theta^{\mu\nu,f}_{\alpha\eta s, \alpha'\eta's'} f_{\RR \alpha \eta s}^\dagger f_{\RR \alpha' \eta' s'} \\ 
\UC_{\mu\nu}^{(c\prime,\xi)}(\qq) =& \frac{\delta_{\xi, (-1)^{a-1}\eta}}{2N_M}\sum_{|\kk|<\Lambda_c} \Theta^{\mu\nu,c'}_{a\eta s, a'\eta's'} c_{\kk+\qq a \eta s}^\dagger c_{\kk a' \eta' s'},\; (a,a'=1,2) \\
\UC_{\mu\nu}^{(c\prime\prime,\xi)}(\qq) =& \frac{\delta_{\xi, (-1)^{a-1}\eta}}{2N_M} \sum_{|\kk|<\Lambda_c} \Theta^{\mu\nu,c''}_{a\eta s, a'\eta's'} c_{\kk+\qq a \eta s}^\dagger c_{\kk a' \eta' s'},\; (a,a'=3,4)
\end{aligned}
\end{equation}
where $\mu,\nu=0,x,y,z$, and the repeated indices should be summed over. In addition,
\begin{equation*}
\Theta^{(0\nu,f)} = \sigma_0 \tau_0 \spin_{\nu}, \qquad
\Theta^{(0\nu,c')} =  \sigma_0 \tau_0 \spin_{\nu} \qquad \Theta^{(0\nu,c'')}= \sigma_0 \tau_0 \spin_{\nu}  \ , 
\end{equation*}
\begin{equation*}
\Theta^{(x\nu,f)} = \sigma_x \tau_x \spin_{\nu}, \qquad
\Theta^{(x\nu,c')} = \sigma_x \tau_x \spin_{\nu}  \qquad \Theta^{(x\nu,c'')}=  - \sigma_x \tau_x \spin_{\nu}  \ , 
\end{equation*}
\begin{equation*}
\Theta^{(y\nu,f)} = \sigma_x \tau_y \spin_{\nu}, \qquad
\Theta^{(y\nu,c')} = \sigma_x \tau_y \spin_{\nu}  \qquad \Theta^{(y\nu,c'')}=  - \sigma_x \tau_y \spin_{\nu}  \ , 
\end{equation*}
\begin{equation*}
\Theta^{(z\nu,f)} = \sigma_0 \tau_z \spin_{\nu}, \qquad
\Theta^{(z\nu,c')} = \sigma_0 \tau_z \spin_{\nu}  \qquad \Theta^{(z\nu,c'')}= \sigma_0 \tau_z \spin_{\nu}  \ . 
\end{equation*}
The global chiral-$U(4)$ rotations are generated by $\UC_{\mu\nu} = \sum_{\xi=\pm1} \bigg( \UC_{\mu\nu}^{(f,\xi)} + \UC_{\mu\nu}^{(c\prime,\xi)} + \UC_{\mu\nu}^{(c\prime\prime,\xi)}\bigg)$. 

The relations between chiral and flat $U(4)$ moments are
\baa 
&\UF_{0\nu}^{f,\xi}(\RR) =\UC_{0\nu}^{f,\xi}(\RR) \quad \quad \UF_{z\nu}^{f,\xi}(\RR) =\UC_{z\nu}^{f,\xi}(\RR)  \quad \quad \UF_{x\nu}^{f,\xi}(\RR) = \xi \UC_{y\nu}^{(f,\xi)}(\RR) \quad \quad \UF_{y\nu}^{f,\xi}(\RR) = -\xi \UC_{x\nu}^{(f,\xi)}(\RR) \nonumber \\
&\UF_{0\nu}^{c',\xi}(\RR) =\UC_{0\nu}^{c',\xi}(\qq)  \quad \quad \UF_{z\nu}^{c',\xi}(\qq) =\UC_{z\nu}^{c',\xi}(\qq)  \quad \quad \UF_{x\nu}^{c',\xi}(\qq) = \xi \UC_{y\nu}^{(c',\xi)}(\qq) \quad \quad \UF_{y\nu}^{c',\xi}(\qq) = -\xi \UC_{x\nu}^{(c',\xi)}(\qq) \nonumber \\
&\UF_{0\nu}^{c'',\xi}(\qq) =\UC_{0\nu}^{c'',\xi}(\qq)  \quad \quad \UF_{z\nu}^{c'',\xi}(\qq) =\UC_{z\nu}^{c'',\xi}(\qq)  \quad \quad \UF_{x\nu}^{c'',\xi}(\qq) = \xi \UC_{y\nu}^{(c'',\xi)}(\qq) \quad \quad \UF_{y\nu}^{c'',\xi}(\qq) = -\xi \UC_{x\nu}^{(c'',\xi)}(\qq) 
\label{eq:rel_u4}
\eaa 
\hhb{We note that even though two $U(4)$ moments are related via Eq.~\ref{eq:rel_u4}, they actually characterize two different $U(4)$ symmetries. }


\subsection{{$\psi$ basis and $U(8)$ moments}}
\label{sec:u4mom}
\bh{ 
In this section, we introduce more general $U(8)$ moments, where 8=2(orbital)$\times$ 2(valley) $\times $2(spin). 
}
We first consider the following convenient basis of electron operators \bh{ $\psi^{f,\xi}_{\RR,n},\psi^{c',\xi}_{\kk,n},\psi^{c'',\xi}_{\kk,n}$ } ($n=1,2,3,4$, $\xi = \pm $):
\baa  
& (\psi^{f,+}_{\RR, 1},\psi^{f,+}_{\RR, 2},\psi^{f,+}_{\RR, 3},\psi^{f,+}_{\RR, 4}) = (f_{\RR,1+\up}, f_{\RR,2-\up} , f_{\RR,1+\dn}, f_{\RR,2-\dn} ) \quad \nonumber \\
& (\psi^{f,-}_{\RR, 1},\psi^{f,-}_{\RR, 2},\psi^{f,-}_{\RR, 3},\psi^{f,-}_{\RR, 4}) = (f_{\RR,2+\up}, -f_{\RR,1-\up} f_{\RR,2+\dn}, -f_{\RR,1-\dn} ) \nonumber  \\ 
& (\psi^{c',+}_{\kk, 1},\psi^{c',+}_{\kk, 2},\psi^{c',+}_{\kk, 3},\psi^{c',+}_{\kk, 4}) = (c_{\kk,1+\up}, c_{\kk,2-\up} , c_{\kk,1+\dn}, c_{\kk,2-\dn} ) \nonumber \\
&  (\psi^{c',-}_{\kk, 1},\psi^{c',-}_{\kk, 2},\psi^{c',-}_{\kk, 3},\psi^{c',-}_{\kk, 4}) = (c_{\kk,2+\up}, -c_{\kk,1-\up} , c_{\kk,2+\dn}, -c_{\kk,1-\dn} ) \nonumber \\
& (\psi^{c'',+}_{\kk, 1},\psi^{c'',+}_{\kk, 2},\psi^{c'',+}_{\kk, 3},\psi^{c'',+}_{\kk, 4}) = (c_{\kk,3+\up}, -c_{\kk,4-\up} , c_{\kk,3+\dn}, -c_{\kk,4-\dn} ) \quad
\nonumber \\
&
(\psi^{c'',-}_{\kk, 1},\psi^{c'',-}_{\kk, 2},\psi^{c'',-}_{\kk, 3},\psi^{c'',-}_{\kk, 4}) = (c_{\kk,4+\up}, c_{\kk,3-\up} , c_{\kk,4+\dn}, c_{\kk,3-\dn} )  
\label{eq:psi_basis_def}
\eaa  
\bh{Here, the index $\xi=\eta (-1)^{\alpha+1}$ 
 can be understood as the index of Chern basis~\cite{tbgiii}. 
}

\bh{ 
The single-particle Hamiltonian $\hH_c$ (Eq.~\ref{eq:hc_def}) and $\hH_{fc}$ (Eq.~\ref{eq:hc_def}) in the $\psi$ basis takes the form of 
\baa  
\hH_c =& \sum_{|\kk|<\Lambda_c,n,\xi}  (\Psi_{\kk}^{c,\xi})^{\dag}v_\star  
\begin{bmatrix}
    0_{4\times 4} &( k_x+i\xi k_y )\mathbb{I}_{4\times 4}\\
    (k_x-i\xi k_y)\mathbb{I}_{4\times 4} &0_{4\times 4} 
\end{bmatrix} \Psi_{\kk}^{c,\xi} 
+ \sum_{|\kk|<\Lambda_c,n,\xi}(-1)^{n+1}M \psi_{\kk,n}^{c'',\xi,\dag} \psi_{\kk,n}^{c'',-\xi } \nonumber \\ 
\hH_{fc} =&\frac{1}{\sqrt{N_M}} \sum_{\substack{|\kk|<\Lambda_c, \RR, n } }\bigg( e^{i\kk \cdot \RR -\frac{|\kk|^2 \lambda^2}{2}} \tilde{H}_{\xi\xi'}^{(fc)}(\kk)  \psi_{\RR,n}^{f,\xi,\dag} \psi_{\kk, n}^{c',\xi'}  +\text{h.c.}\bigg),\quad \tilde{H}^{(fc)}(\kk) = 
\begin{bmatrix}
    \gamma & v_\star^\prime k_-  \\
    v_\star^\prime k_+ & \gamma 
\end{bmatrix}
\eaa 
where $\Psi^{c,\xi}_\kk =
\begin{bmatrix}
    \psi_{\kk,1,...,4}^{c',\xi} & \psi_{\kk,1,...,4}^{c'',\xi}
\end{bmatrix}^T
$ and $k_{\pm} = k_x \pm ik_y $. 
}




\hb{ 
We now define the $U(8)$ moments with $\psi$ basis ($\mu,\nu = 0,x,y,z$):
\baa  
&\UF_{\mu\nu}^{(f,\xi\xi')}(\RR) = \frac{1}{2} \sum_{mn}\psi^{f,\xi,\dag}_{\RR,n} [T^{\mu\nu} ]_{nm} \psi^{f,\xi'}_{\RR,m} \nonumber  \\
&
\UF_{\mu\nu}^{(c',\xi\xi')}(\kk,\qq) = \frac{1}{2} \sum_{mn}\psi^{c',\xi,\dag}_{\kk+\qq,n} [T^{\mu\nu} ]_{nm} \psi^{c',\xi'}_{\kk,m}
\quad,\quad 
\UF_{\mu\nu}^{(c'',\xi\xi')}(\kk,\qq) = \frac{1}{2} \sum_{mn}\psi^{c'',\xi,\dag}_{\kk+\qq,n} [T^{\mu\nu} ]_{nm} \psi^{c'',\xi'}_{\kk,m} \nonumber \\
&\{T^{\mu\nu}\} = \{ \varsigma'_\nu\rho_0,  \varsigma'_\nu \rho_y, -\varsigma'_\nu \rho_x,  \varsigma'_\nu \rho_z\}
\label{eq:flat_u4_general}
\eaa  
}and we let \bm{$\rho_{x,y,z,0}$} be the Pauli matrices and the identity matrix defined in the subspace of $(1+,2-)$ for $\xi=+1$ and $(2+,1-)$ for $\xi=-1$. \bh{$\varsigma'_{x,y,z,0}$ are the Pauli matrices and identity matrix acting in the  spin subspace.}

The previous flat $U(4)$ moments in Eq.~\ref{eq:U4op-maintext} can be written with new electron basis as
\baa  
&\UF_{\mu\nu}^{(f,\xi)} (\RR) =\frac{1}{2} \sum_{mn}\psi^{f,\xi,\dag}_{\RR,n} [T^{\mu\nu} ]_{nm} \psi^{f,\xi}_{\RR,m} = \UF_{\mu\nu}^{(f,\xi\xi)}(\RR)
\nonumber 
\\
&
\UF_{\mu\nu}^{(c',\xi)} (\qq) =\frac{1}{2N_M}\sum_{mn,\kk}\psi^{c',\xi,\dag}_{\kk+\qq,n} [T^{\mu\nu} ]_{nm} \psi^{c',\xi}_{\kk,m} = \frac{1}{N_M}\sum_{\kk} \UF_{\mu\nu}^{(c',\xi\xi)}(\kk,\qq) \nonumber \\ 
&
\UF_{\mu\nu}^{(c'',\xi)} (\qq) =\frac{1}{2N_M}  \sum_{mn}\psi^{c'',\xi,\dag}_{\kk+\qq,n} [T^{\mu\nu} ]_{nm} \psi^{c'',\xi}_{\kk,m} = \frac{1}{N_M}\sum_{\kk} \UF_{\mu\nu}^{(c'',\xi\xi)} (\kk,\qq)
\label{eq:flat_u4_new_basis}
\, .
\eaa  
\hh{The advantage of using the $\psi$ basis is that the flat $U(4)$ moments of $f,c',c''$ electrons have the same matrix structure $T^{\mu\nu}$.}

\hh{ 
It is also useful to consider the following Fourier transformation of electron operators and $f$-moments.
}
\baa  
&\psi_{\rr,n}^{c',\xi} = \frac{1}{\sqrt{N_M}}\sum_\kk e^{i\kk \cdot \rr } \psi_{\kk,n}^{c',\xi} \nonumber \\ &\psi_{\rr,n}^{c'',\xi} = \frac{1}{\sqrt{N_M}}\sum_\kk e^{i\kk \cdot \rr } \psi_{\kk,n}^{c'',\xi} \nonumber \\
&\UF_{\mu\nu}^{(c',\xi\xi')}(\rr,\rr') = \frac{1}{N_M}\sum_{\kk,\qq} \UF_{\mu\nu}^{(c',\xi\xi')}(\kk,\qq) e^{i\kk \cdot \rr' - i (\kk +\qq) \rr } = \frac{1}{2}\sum_{mn}\psi_{\rr,n}^{c',\xi,\dag} [T^{\mu\nu}]_{nm} \psi_{\rr',m}^{c',\xi'} \nonumber \\
&
\UF_{\mu\nu}^{(c'',\xi\xi')}(\rr,\rr') = \frac{1}{N_M}\sum_{\kk,\qq} \UF_{\mu\nu}^{(c'',\xi\xi')}(\kk,\qq) e^{i\kk \cdot \rr' - i (\kk +\qq) \rr }= \frac{1}{2}\sum_{mn}\psi_{\rr,n}^{c'',\xi,\dag} [T^{\mu\nu}]_{nm} \psi_{\rr',m}^{c'',\xi'}\, . 
\label{eq:u4_ft}
\eaa 
\hb{Since the momentum of $c$-electron has a finite cutoff, this definition is for convenience. However, if we only consider the $c$-electrons in the first moir\'e Brillouin zone, the completeness of $c$-electron Fourier transformation is guaranteed and a well-defined inverse Fourier transformation also exists for $c$ electrons. For what follows, unless specifically mentioned, we only consider $c$-electrons in the first moir\'e Brillouin zone. }

Now, we utilize the following relation
\baa  
\sum_{\mu\nu  } [T^{\mu\nu}]_{ab}[T^{\mu\nu}]_{cd} = 4\delta_{a,d}\delta_{b,c}
\eaa  
and find
\baa 
&\sum_{\mu\nu} \UF_{\mu\nu}^{(f,\xi\xi')}(\RR)  \UF_{\mu\nu}^{(f,\xi_2' \xi_2)}(\RR_2) = \sum_{a,b} \psi_{\RR,a}^{f,\xi, \dag} \psi_{\RR, b}^{f,\xi'} \psi_{\RR_2,b}^{f,\xi'_2,\dag} \psi^{f,\xi_2}_{\RR_2,a}  \nonumber \\
&\sum_{\mu\nu} \UF_{\mu\nu}^{(c',\xi\xi')}(\rr_1,\rr_1')  \UF_{\mu\nu}^{(c',\xi_2' \xi_2)}(\rr_2',\rr_2) = \sum_{a,b} \psi_{\rr_1,a}^{c',\xi, \dag} \psi_{\rr_1', b}^{c',\xi'} \psi_{\rr_2',b}^{c',\xi'_2,\dag} \psi^{c',\xi_2}_{\rr_2,a}  \nonumber \\
&\sum_{\mu\nu} \UF_{\mu\nu}^{(f,\xi\xi')}(\RR)  \UF_{\mu\nu}^{(c'',\xi_2' \xi_2)}(\rr,\rr') = \sum_{a,b,\RR} \psi_{\RR,a}^{f,\xi, \dag} \psi_{\RR, b}^{f,\xi'} \psi_{\rr,b}^{c'',\xi'_2,\dag} \psi^{c'',\xi_2}_{\rr',a}  
\label{eq:u4_mom_summ}
\eaa 

We change from $\psi$ basis to $f$ and $c$ basis, \hh{then Eq.~\ref{eq:u4_mom_summ} becomes}\bh{ (which will be used in Sec.~\ref{sec:sw_transf})}
\baa  
&\sum_{\mu\nu} \sum_{\xi,\xi'} \UF_{\mu\nu}^{(f,\xi\xi')}(\RR)\UF_{\mu\nu}^{(f,\xi'\xi)}(\RR_2)=\sum_{\alpha\alpha'\eta \eta' s s'} f_{\RR,\alpha \eta s}^\dag f_{\RR,\alpha'\eta' s'}. f_{\RR_2,\alpha'\eta' s'}^\dag f_{\RR_2,\alpha \eta s} \nonumber \\
&\sum_{\mu\nu} \sum_{\xi,\xi'} \UF_{\mu\nu}^{(f,\xi\xi')}(\RR)\UF_{\mu\nu}^{(c',\xi'\xi)}(\rr,\rr')=\sum_{\alpha\alpha'\eta \eta' s s'} f_{\RR,\alpha \eta s}^\dag f_{\RR,\alpha'\eta' s'}. c_{\rr,\alpha'\eta' s'}^\dag c_{\rr',\alpha \eta s} 
\nonumber 
\\
&\sum_{\mu\nu} \sum_{\xi,\xi'} \UF_{\mu\nu}^{(f,\xi\xi')}(\RR)\UF_{\mu\nu}^{(f,-\xi'\xi)}(\RR_2)
=\sum_{\alpha\eta s, \alpha'\eta's'} f_{\RR,\alpha\eta s}^\dag f_{\RR,\alpha'\eta's'}f_{\RR_2,\alpha_2\eta' s'}^\dag  f_{\RR_2,\alpha\eta s} \eta' [\sigma_x]_{\alpha',\alpha_2}
\nonumber 
\\
&\sum_{\mu\nu} \sum_{\xi,\xi'} \UF_{\mu\nu}^{(f,\xi\xi')}(\RR)\UF_{\mu\nu}^{(c',-\xi'\xi)}(\rr,\rr')=\sum_{\alpha\eta s, \alpha'\eta's'} f_{\RR,\alpha\eta s}^\dag f_{\RR,\alpha'\eta's'}c_{\rr,\alpha_2\eta' s'}^\dag  c_{\rr',\alpha\eta s} \eta' [\sigma_x]_{\alpha',\alpha_2} 
\nonumber 
\\
&\sum_{\mu\nu} \sum_{\xi,\xi'} \UF_{\mu\nu}^{(f,\xi\xi')}(\RR)\UF_{\mu\nu}^{(c',-\xi'-\xi)}(\rr,\rr' )
= \sum_{\alpha\eta s, \alpha'\eta's'} f_{\RR,\alpha\eta s}^\dag f_{\RR,\alpha'\eta's'}c_{\rr,\alpha_2\eta' s'}^\dag  c_{\rr',\alpha_2'\eta s} \eta'\eta [\sigma_x]_{\alpha',\alpha_2} [\sigma_x]_{\alpha_2',\alpha}  
\nonumber 
\\
&\sum_{\mu\nu} \sum_{\xi,\xi'} \UF_{\mu\nu}^{(f,\xi\xi')}(\RR)\xi'\UF_{\mu\nu}^{(c',-\xi'\xi)}(\rr,\rr')=\sum_{\alpha\eta s, \alpha'\eta's'} f_{\RR,\alpha\eta s}^\dag f_{\RR,\alpha'\eta's'}c_{\rr,\alpha_2\eta' s'}^\dag  c_{\rr',\alpha\eta s}[i\sigma_y]_{\alpha',\alpha_2} 
\nonumber 
\\
&\sum_{\mu\nu} \sum_{\xi,\xi'} (-\xi)  \UF_{\mu\nu}^{(f,\xi\xi')}(\RR)\xi'\UF_{\mu\nu}^{(c',-\xi'\xi)}(\rr,\rr')=\sum_{\alpha\eta s, \alpha'\eta's'} f_{\RR,\alpha\eta s}^\dag f_{\RR,\alpha'\eta's'}c_{\rr',\alpha_2\eta' s'}^\dag  c_{\rr,\alpha_{2}'\eta s}[i\sigma_y]_{\alpha',\alpha_2}[i\sigma_y]_{\alpha_2',\alpha} 
\nonumber 
\\
&\sum_{\mu\nu} \sum_{\xi,\xi'}  \UF_{\mu\nu}^{(f,\xi\xi')}(\RR)\xi'\UF_{\mu\nu}^{(c',-\xi'\xi)}(\rr,\rr')=\sum_{\alpha\eta s, \alpha'\eta's'} f_{\RR,\alpha\eta s}^\dag f_{\RR,\alpha'\eta's'}c_{\rr,\alpha_2\eta' s'}^\dag  c_{\rr',\alpha_{2}'\eta s}[i\sigma_y]_{\alpha',\alpha_2}[\eta\sigma_x]_{\alpha_2',\alpha} 
\label{eq:u4_mom_sum_2}
\eaa

\section{Zero-hybridziation limit}

The \hb{$f$-$c$ hybridization} parameter $\gamma$ vanishes at $w_0/w_1 =0.9$ which is  close to the actual value $w_0/w_1= 0.8$. Vanishing $\gamma$ corresponds to the gap closing between the remote and the flat bands in the continuum single particle model. We solve the model exactly at $\gamma =0$ and then treat it as a perturbation. 
In this limit, the $f$ and $c$ electrons are coupled only through the interacting terms.
We now neglect  $\hH_{V} $ except for a mean-field treatment- for the same reason as in Ref.~\cite{HF_MATBLG}, that it will cause only a velocity renormalization of the linear conduction fermion, and keep only the \hb{remaining} terms \hhb{$\hH_U,\hH_W,\hH_J,\hH_V^{MF}$ where $\hH_V^{MF}$ denotes the $\hH_V$ with mean-field approximation.}


This now becomes a polynomially solvable Hamiltonian. Some remarks: \textbf{1.} This is more complicated than the zero hybridization limit of the Anderson \hhb{lattice} model in which the Hilbert space factorizes. This is because in both  $\hH_{W}$ \hb{(we have work in the limit $W_1=W_2=W_3=W_4$)}, and $\hH_{J}$ terms, the $f$ occupation number will influence the electron Hamiltonian, but in a one-body fashion, since the $c$ operators are quadratic. 
The strategy is then to solve the Hamiltonian in this limit and then add the $f$-$c$ hybridization perturbatively \hb{via Schrieffer–Wolff transformation (SW) transformation~\cite{SW_transf}}. 

To solve the Hamiltonian we first need to find the on-site configurations of the $f$ fermions (which are good quantum numbers) which then, after adding $\hH_{U}+ \hH_{W}+ \hH_{J}$ and the $c$-electron kinetic term 
    \begin{eqnarray}
& H_c = \sum_{\eta s} \sum_{aa'} \sum_{|\kk|<\Lambda_c} (H^{(c,\eta)}_{a,a'}(\kk) - \mu \delta_{aa'}) c_{\kk a\eta s}^\dagger c_{\kk a'\eta s}  
- \mu  N_f 
\end{eqnarray}
give the lowest energy. This is a minimization problem similar in spirit to the Lieb theorems~\cite{Lieb} with flux $\pi$ (used in the Kitaev model~\cite{kitaev2006anyons}) where it is shown that the flux $\pi$ configuration of a noninteracting fermion model in a background plaquette flux with each plaquette having possibly different flux is minimal at flux $\pi$ per all plaquettes. Our problem, at least in the first try, might be amenable to Monte Carlo sampling.

\subsection{Symmetries of the zero-hybridization Limit}

We can rewrite the  $\hH_{U}+ \hH_{W}+ \hH_{J}$

\begin{eqnarray}
& \hH_{U}  = \frac{U}{2} \sum_{\RR} :n_{f\RR}: :n_{f\RR},\;\;\; \hH_{W} = \frac1{N} \sum_{\RR } :n_{f\RR}: \sum_{|\kk|, |\kk'|<\Lambda_c}  \sum_{\eta_2s_2 a} 
    W_a  e^{-i(\kk-\kk')\cdot\RR} 
    :c_{\kk a\eta_2 s_2}^\dagger  c_{\kk' a \eta_2 s_2}:\ , \nonumber \\ & H_J = - \frac{J}{2N} \sum_{\RR s_1 s_2} \sum_{\alpha\alpha'\eta\eta'} \sum_{|\kk_1|,|\kk_2|<\Lambda_c }   
     e^{i( \kk_1- \kk_2 )\cdot\RR } 
     ( \eta\eta' + (-1)^{\alpha+\alpha'} )
     :f_{\RR \alpha \eta s_1}^\dagger f_{\RR \alpha' \eta' s_2}:  :c_{\kk_2, \alpha'+2, \eta' s_2}^\dagger  c_{ \kk_1, \alpha+2, \eta s_1}:  
\end{eqnarray}
Where we have defined the occupation number of the electron on-site $\RR$
\begin{equation}
n_{f\RR}= \sum_{\alpha\eta s} 
    f_{\RR \alpha \eta s}^\dagger f_{\RR \alpha\eta s} \in 0,1\ldots 8
\end{equation} which is a symmetry of the system \hb{even when $W_1 \ne W_3$}:
\begin{equation}
  [\hH_{U}, n_{f\RR}]=0,\;\;  [\hH_{W}, n_{f\RR}]=0,\;\;  [\hH_{J}, n_{f\RR}]=0,\;\;  [H_c,n_{f\RR} ]=0
\end{equation} 
The system has a large $U(1)^{N_M}$ symmetry where $N_M$ is the number of moir\'e unit cells. An eigenstate of the system is first indexed by $n_{f\RR}$. Then this couples through the $\hH_{W}$ and (with a smaller coefficient) through $\hH_{J}$. 

There are now several questions
\begin{itemize}
    \item If first we neglect $\hH_{J}$ we have a large symmetry, $U(8)^{N_M}$, whose generators are  \bh{$\UF_{\mu\nu}^{(\xi\xi')}(\RR) = \UF_{\mu\nu}^{(f,\xi\xi')}(\RR)+\UF_{\mu\nu}^{(c',\xi\xi')} (\RR)+\UF_{\mu\nu}^{(c'',\xi\xi')} (\RR)$ (Eq.~\ref{eq:flat_u4_general}).
    }
    \begin{equation}
        [\hH_{U},\UF_{\mu\nu}^{(\xi\xi')}(\RR)]=0\quad,\quad  [\hH_{W}, \UF_{\mu\nu}^{(\xi\xi')}(\RR)]=0\quad,\quad  [H_c,\UF_{\mu\nu}^{(f,\xi\xi')}(\RR) ]=0 
    \end{equation}
which gives \hb{huge number} of degenerate states, all with the same occupation number $n_{f\RR}$ on-site distributed around the different $\alpha, \eta, s.$ 

\item If we take $W_{1}=W_3=W$ then we have that 
\begin{eqnarray}
    & \hH_{W} =W  \frac1{N} \sum_{\RR } :n_{f\RR}: \sum_{|\kk|, |\kk'|<\Lambda_c}  \sum_{\eta_2s_2 a} 
     e^{-i(\kk-\kk')\cdot\RR} 
   \sum_{\eta s a}  :c_{\kk a\eta s}^\dagger  c_{\kk' a \eta s}:\nonumber \\ &= W  \frac1{N} \sum_{\RR } :n_{f\RR}: \sum_{|\kk|, |\kk'|<\Lambda_c}  \sum_{\eta_2s_2 } 
     e^{-i(\kk-\kk')\cdot\RR} 
   \sum_{n_1,n_2} (\sum_{a} U^{ \eta ,*}_{\kk, a n_1}  U^{ \eta ,*}_{\kk' ,a n_2})  \gamma_{\kk, n_1 \eta s}^\dagger \gamma_{\kk', n_2 \eta s}
\end{eqnarray} 
\hb{
where we have introduced the eigenvectors $U^{\eta}_{\kk,an}$ and eigenvalues $\epsilon_{\kk,n}^\eta $ of $\hH^{(c,\eta)}(\kk)$ 
\ba 
\sum_{a'}H^c_{aa'}(\kk)U^\eta_{\kk,a'n} = \epsilon^\eta_{k,n}U_{\kk,an}\, ,
\ea 
and the operator in the band basis 
\ba 
\gamma_{\kk,n \eta s} = \sum_{a}U_{\kk,an }^*c_{\kk,a\eta s} 
\,. 
\ea 
}
$\hH_W$ is now a one-body term for $c$ or $\gamma$, which depends on the distribution of $n_{\RR}^f$. Through Monte Carlo sampling, we can find the ground-state exactly and check whether the ground state involve uniform $n_\RR^f$ or not. 

\item If furthermore, we assume uniform $n^f_{\RR} = n_f$ then the summation over $\RR$ gives us a $\delta_{\kk,\kk'} $ and we obtain an $n^f$-dependent chemical potential of the electrons:

\begin{eqnarray}
    & \hH_{W} = W n_f  \sum_{|\kk|, |\kk'|<\Lambda_c}  \sum_{n \eta s } 
  \gamma_{\kk n \eta s}^\dagger \gamma_{\kk n \eta s}
\end{eqnarray}
We can now find easily the admixture of $f, c$ fermions at any filling $N= N_f+ N_c$ with $N_f= N_M n_f$. This allows us to see, as a function of the total filling, analytically, the distribution between the $f,c$ electrons \hb{in the ground state}. We can then add the $\hH_{J}$ under the stronger assumption that 
\begin{eqnarray}
    f_{\RR \alpha \eta s_1}^\dagger f_{\RR \alpha' \eta' s_2} = A_{\alpha \eta s_1, \alpha' \eta' s_2}
\, .
\end{eqnarray} 
We can then find the $A_{\alpha \eta s_1, \alpha' \eta' s_2}$ which will minimize the state energy and break the $U(8)$ on-site symmetry to \hh{$U(1)$ symmetry}.

\item We can assume $n_{\RR}^f$ not constant, and \hb{adopt} charge density wave (CDW) or other types of translational symmetry breaking. \hb{We leave it for future study} 

\item We can then add the $f^\dag c$ term through Schrieffer-Wollf \hb{(SW) transformation} , \hb{which will be discussed in the Sec.~\ref{sec:sw_transf}}. 

\end{itemize}

\begin{figure}[!htbp]
    \centering
    \includegraphics[width=\linewidth]{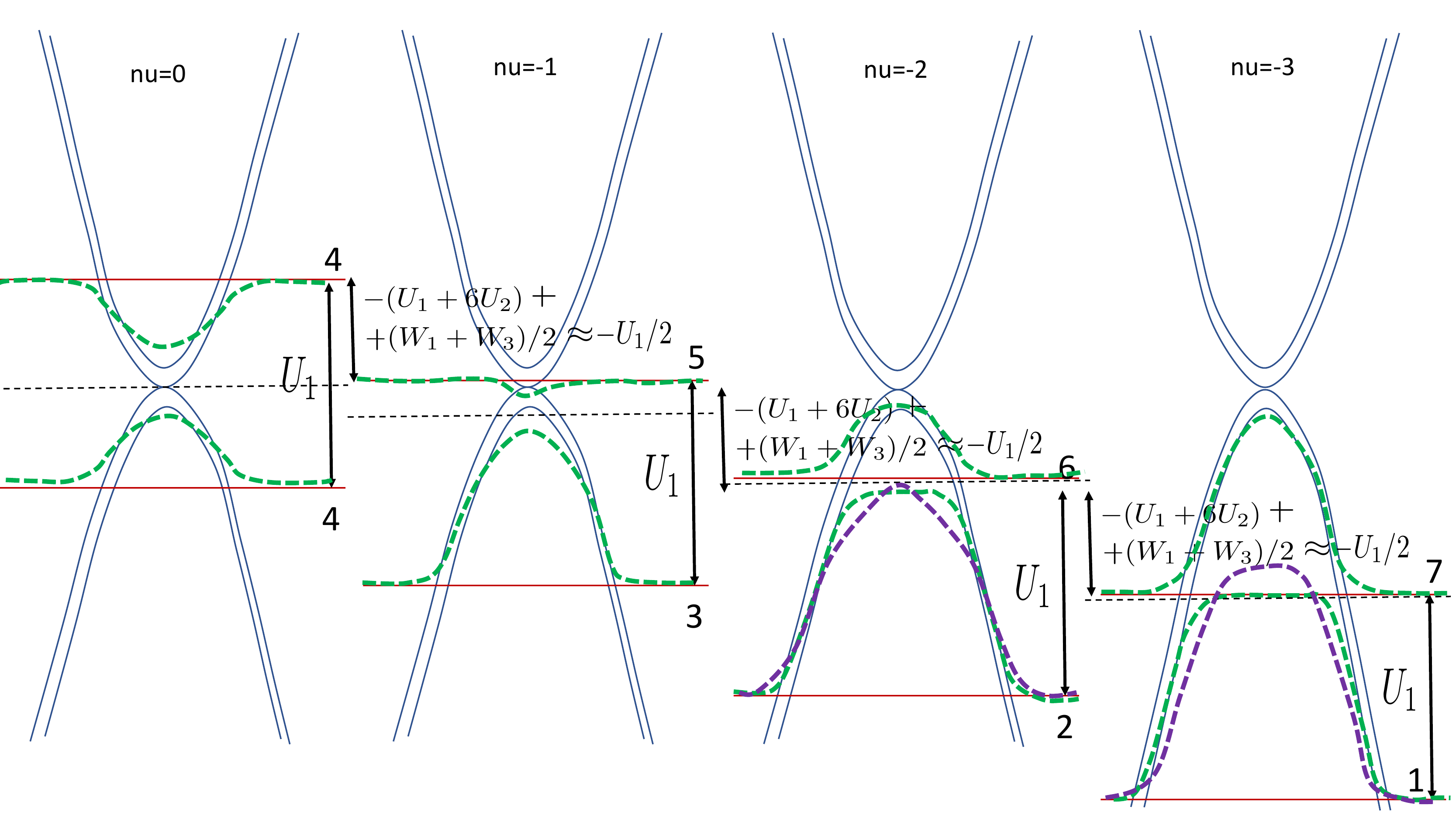}
    \caption{Zero hybridization $\gamma=0$ model at different fillings of the Heavy fermions (the state assumed as the parent state carries all the filling in the heavy fermions, unlike the exact state at $\gamma=0$ which contains both $c$ and $f$ fermions). The Fermi level is a horizontal dashed black line. The solid lines are zero hybridization scenarios while the green dashed lines are the level splittings occurring after small increase of $\gamma$ from nonzero, but still smaller than the realistic value of $\gamma= 24.6 $\bh{meV}. At $\nu=0$ both minima of the bands are made by $c$ electrons. At $\nu=- 1+ \epsilon$ the band is very flat while at $\nu=-1- \epsilon$ the minimum of the band is formed by the $c$ electrons. This last situation changes at $\nu=-2,-3$, at least in the absence of hybridization: both band edges are made up of heavy fermions. At small hybridization (green dashed) the band edges at $\nu=-2 \pm \epsilon$ and $\nu=-3 \pm \epsilon$ are still made up by heavy fermions. Once $\gamma$ reaches its large $24.6$\bh{meV} value, we find that at $\nu=-2- \epsilon$ the band has changed character from heavy fermion to light fermion due to large heavy-light mixing (see violet dashed line, large curvature). At $\nu=-3- \epsilon$ however, the portion of the heavy band as a ratio of the Brillouin zone is larger and hence upon hybridization, the portion of the mixing is smaller than at $\nu=-2- \epsilon$, and the band has more $f$-character. The CDW~\cite{xie2022phase} obtained at $\nu=-3$ most likely has $f$-$f$ CDW correlations between the $\nu=-3+ \epsilon$ band edge (which is around $K,M$ \bh{points} and hence made up of heavy fermions) and the $\nu=-3- \epsilon$ band edge (around $\Gamma$ but still made up of heavy fermions)} 
    \label{heavyfermionzerohybrid1}
\end{figure}

\subsection{Zero hybridization limit without $\hH_J$} 
We first solve the zero hybridization model at $J=0$.
The total Hamiltonian of the zero-hybridization model reads 
\baa 
&\hat{H}_c +\hat{H}_U + \hat{H}_W +\hat{H}_V^{MF} 
\label{eq:H0}
 \eaa 
We consider the solution with uniform charge distribution \hb{of $f$-electrons $n_\RR^f=\nu_f+4$, with integer $\nu_f \in [-4,...,4]$. We note that, in the zero-hybridization model, the filling of $f$-electron of each site is a good quantum number and takes integer values.  }. 
The trial wavefunction of the ground state we proposed is 
\baa 
|\nu_f,\nu\rangle = [\prod_\RR |\nu_f\rangle_{\RR}] |\Psi[\nu-\nu_f,\nu_f]\rangle_c. 
\label{eq:uniform_charge_trial}
\eaa  
where $\nu_f$ denotes the filling of $f$ electrons, $\nu$ denotes the total filling of $f$ and $c$ electrons, and $\nu_c =\nu-\nu_f$ is the filling of $c$ electrons. 
\begin{itemize}
    \item $|\nu_f\rangle_\RR$ describes a $f$ state at site $\RR$ with filling $\nu_f$:
    \ba 
    \sum_{\alpha \eta s}:f_{\RR, \alpha\eta s}^\dag f_{\RR, \alpha \eta s}: |\nu_f\rangle _\RR= \nu_f |\nu_f\rangle_\RR. 
    \ea 
    and we take a uniform charge distribution, so for each site, the fillings of $f$-electrons are the same. 
    \item $|\Psi[\nu_c,\nu_f] \rangle_s$ denotes a Slater determinant state corresponding to the ground state of the following one-body Hamiltonian of $c$ electrons at filling $\nu_c=\nu-\nu_f$: 
    \ba 
    H_{c}' = \hat{H}_c + \sum_{\eta,s,a}\sum_{|\kk |<\Lambda_c } 
    W \nu_f 
    :c_{\kk a\eta s}^\dag c_{\kk a\eta s} :. 
    \ea  
        
    Above one-body Hamiltonian can be diagonalized.
    (Note that $W_a$ is taken to be orbital independent):
    \ba 
    H_c' = \sum_{|\kk |<\Lambda_c,\eta ,s,n} (E_{\kk  n\eta}+W \nu_f) \gamma_{\kk n \eta s}^\dag \gamma_{\kk  n\eta }
    \ea  
    where $n$ is the band index and \hb{$\gamma$ is the operator in the band basis} 
    \begin{equation}
        \gamma_{\kk  n \eta s} = \sum_a U^{ \eta,*}_{\kk  a n}c_{\kk  a \eta s },\;\;\;  c_{\kk  a \eta s }  = \sum_n U^{ \eta}_{\kk  a n} \gamma_{\kk  n \eta s} ,\;\;\;  \sum_{a' } H^{(c,\eta)}_{a,a'}(\kk )U^{\eta}_{\kk  a' n} = E_{\kk  n \eta} U^{\eta}_{\kk  a n}  
    \end{equation}
\end{itemize}

The energy of $|\nu_f,\nu\rangle$ state is then 
\ba 
&E_{\nu_f,\nu}/N_M =\langle \nu_f,\nu|\hH_{U} |\nu_f,\nu\rangle/N_M
+ \langle \nu_f,\nu|\hH_{V} |\nu_f,\nu\rangle/N_M
+\langle \nu_f,\nu|\hH_{W} |\nu_f,\nu\rangle/N_M
+\langle \nu_f,\nu|\hH_{c} |\nu_f,\nu\rangle/N_M
\\
&\langle \nu_f,\nu|\hH_{U} |\nu_f,\nu\rangle/N_M = \frac{U}{2}\nu_f^2 \quad,\quad 
\langle \nu_f,\nu|\hH^{MF}_{V} |\nu_f,\nu\rangle/N_M = \frac{V_0}{2\Omega_0}\nu_c^2 = \frac{V_0}{2\Omega_0}(\nu-\nu_f)^2 \\
&\langle \nu_f,\nu|\hH_{W} |\nu_f,\nu\rangle /N_M=  W\nu_f\nu_c = W\nu_f (\nu-\nu_f)\\ 
&\langle \nu_f,\nu|\hH_{c} |\nu_f,\nu\rangle/N_M = \frac{1}{N_M} \sum_{\kk , n\eta s}(E_{\kk  n\eta} )\langle\Psi[\nu_c,\nu_f| \gamma_{\kk n\eta s}^\dag \gamma_{\kk  n \eta s} |\Psi[\nu_c,\nu_f]\rangle \\
&E_{\nu_f,\nu}/N_M = \frac{U_1}{2}\nu_f^2  + \frac{V_0}{2\Omega_0} (\nu-\nu_f)^2 +W\nu_f (\nu-\nu_f)+ \frac{1}{N_M} \sum_{\kk , n\eta s}(E_{\kk  n\eta} )\langle\Psi[\nu_c,\nu_f| \gamma_{\kk n\eta s}^\dag \gamma_{\kk  n \eta s} |\Psi[\nu_c,\nu_f]\rangle 
\ea 
For a given total filling $\nu$, we compare the energies of different $\nu_f$ ($\nu_f$ can only be an integer) and take the one with the lowest energy as our ground state. 

\begin{figure}
    \centering
    \includegraphics[width=0.7\textwidth]{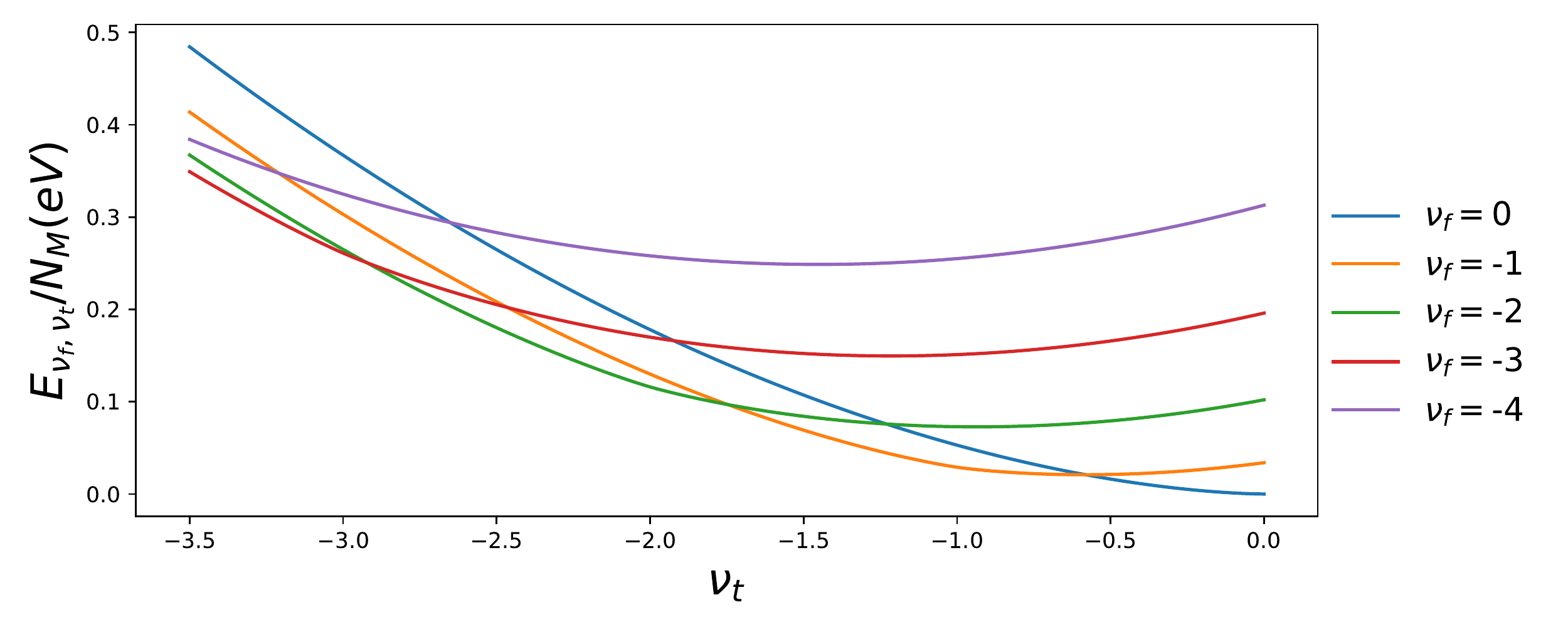}
    \caption{
    Evolution of energy as a function of $\nu$ at different $\nu_f$ and $M= 0$. For fixed $\nu$, the ground state is determined by $\nu_f$ that gives the lowest energy.
    }
     \label{fig:fill_1}
\end{figure}

In the $M=0$ limit, the analytical expression of $E_{\nu_f,\nu}/N_M$ can be easily given  At $M=0$, $c$ electron dispersion becomes $\pm v_{\star} |k|$. Filling $\nu_c=\nu-\nu_f$ conduction electrons is equivalent to fill the 8-fold ($8=2\times2\times 2$, 2 for spin, 2 for valley, 2 for orbital) linear-dispersive bands up to certain momentum $k_0$. \hh{Depending} on the sign of $\nu_c$, we either \hh{fill} electrons ($\nu_c>0$) or holes $(\nu_c<0)$ to the Dirac sea.
This gives the following equations to determine $k_0$
\baa  
&\frac{8}{A_{MBZ}} \int_{|\kk|<k_0} dk_x dk_y  =|\nu_c| 
\Rightarrow k_0=\sqrt{\frac{|\nu_c|A_{MBZ} }{8\pi}}
\label{eq:k0_def}
\eaa  
where $A_{MBZ}$ is the size of moir\'e Brillouin zone.  The energy \hb{loss} from conduction bands are then (factor 8 for 8-fold degenerate bands)
\ba 
 \frac{1}{N_M} \sum_{\kk , n\eta s}(E_{\kk  n\eta} )\langle\Psi[\nu_c,\nu_f| \gamma_{\kk n\eta s}^\dag \gamma_{\kk  n \eta s} |\Psi[\nu_c,\nu_f]\rangle =&\frac{8}{A_{MBZ}} \int_{|\kk| < k_0} |v_{\star}| |\kk| dk_x dk_y 
 = \frac{16}{A_{MBZ}} \pi |v_{\star}|\frac{k_0^3}{3}
\ea 
Finally, we have 
\baa 
E_{\nu_f,\nu}/N_M &=  \frac{U_1}{2}\nu_f^2  + \frac{V_0}{2\Omega_0} (\nu-\nu_f)^2 +W\nu_f (\nu-\nu_f) + \frac{16 \pi |v_{\star}| k_0^3 }{3A_{MBZ} } \nonumber  \\
&=\frac{U_1}{2}\nu_f^2  + \frac{V_0}{2\Omega_0} (\nu-\nu_f)^2 +W\nu_f (\nu-\nu_f) + \frac{16 \pi |v_{\star}| }{3A_{MBZ} }\bigg[
\frac{|\nu-\nu_f|A_{MBZ}}{8\pi} 
\bigg]^{3/2} 
\eaa 
The behaviors of $E_{\nu_f,\nu}/N_M$ as a function of $\nu$ at various $\nu_f$ is shown in Fig.~\ref{fig:fill_1}.  

\begin{figure}
    \includegraphics[width=0.4\textwidth]{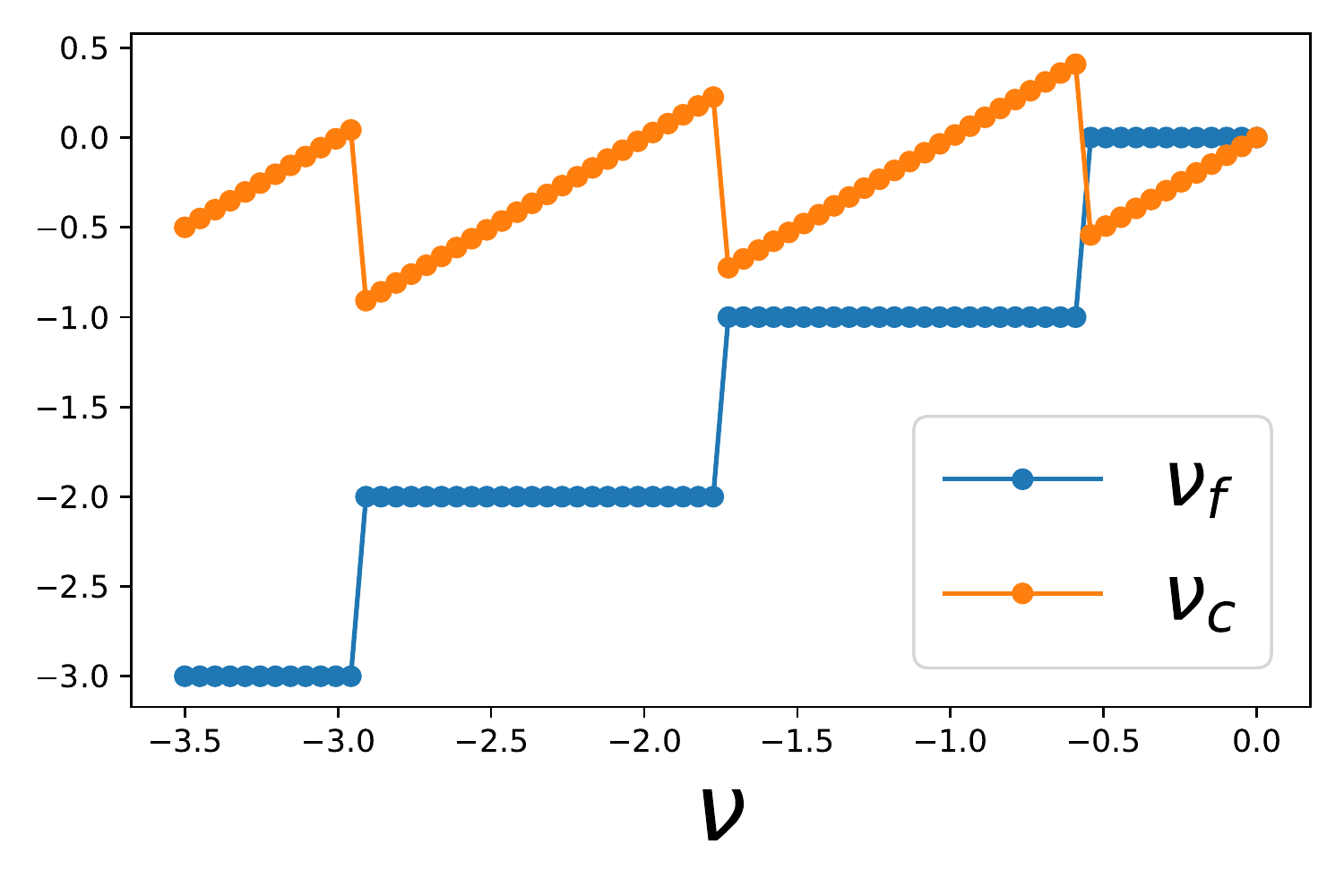}
    \caption{ 
     Evolution of the $f$ and $c$ filling as a function of $\nu$. We observe that $\nu=-3$ is near the transition point between $\nu_f=-2,-3$ states.
    }
    \label{fig:fill_2}
\end{figure}

\bh{ 
We next consider non-zero $M$. Now the dispersions of conduction electrons become $\frac{\pm M \pm \sqrt{M^2+4|v_\star|^2 |\kk|^2} }{2}$. Without loss of generality, we consider the hole doping and fill the bands with energy $E < \mu$ with $\mu<0$. Then the relation between $\nu_c$ and $\mu$ is
\baa  
\nu_c =& \frac{-4}{A_{MBZ}} \int_{|\kk|<\Lambda}
\theta( |\mu|- \frac{-M +\sqrt{M^2+4|v_\star|^2|\kk|^2} }{2} )
dk_xdk_y  \nonumber \\
&+ \frac{-4}{A_{MBZ}} \int_{|\kk|<\Lambda}
\theta( |\mu|- \frac{M +\sqrt{M^2+4|v_\star|^2|\kk|^2} }{2} )
dk_xdk_y
\eaa  
(where $4$ comes from 2 valleys and 2 spins, and we fill the hole bands with energy $E>-\mu$). For $|\mu| <M$, we find 
\baa  
\nu_c = - \frac{ 4\pi}{A_{MBZ}}\frac{|\mu|(|\mu|+M)}{|v_\star|^2}
\label{eq:zero_hyb_nuc_1}
\eaa  
For $|\mu|>M$, we find 
\baa  
\nu_c = - \frac{ 4\pi}{A_{MBZ}}\frac{|\mu|(|\mu|+M)}{|v_\star|^2} - \frac{ 4\pi}{A_{MBZ}}\frac{|\mu|(|\mu|-M)}{|v_\star|^2}
\label{eq:zero_hyb_nuc_2}
\eaa  
Combining Eq.~\ref{eq:zero_hyb_nuc_1} and Eq.~\ref{eq:zero_hyb_nuc_2}, we find 
\baa  
\nu_c = - \frac{ 4\pi}{A_{MBZ}}\frac{|\mu|(|\mu|+M)}{|v_\star|^2} - \frac{ 4\pi}{A_{MBZ}}\frac{|\mu|(|\mu|-M)}{|v_\star|^2}\theta (|\mu|-M)
\label{eq:zero_hyb_mu}
\eaa 
Then the energy loss from conduction $c$ electron is
\baa  
E_c=&\frac{1}{N_M} \sum_{\kk , n\eta s}(E_{\kk  n\eta} )\langle\Psi[\nu_c,\nu_f| \gamma_{\kk n\eta s}^\dag \gamma_{\kk  n \eta s} |\Psi[\nu_c,\nu_f]\rangle \nonumber \\
=&
\frac{4}{A_{MBZ}}\int_{|\kk|<\Lambda_c} \theta( |\mu|- \frac{-M +\sqrt{M^2+4|v_\star|^2|\kk|^2} }{2} )\frac{-M +\sqrt{M^2+4|v_\star|^2|\kk|^2} }{2}
dk_xdk_y \nonumber \\
&+ \frac{4}{A_{MBZ}} \int_{|\kk|<\Lambda}
\theta( |\mu|- \frac{M +\sqrt{M^2+4|v_\star|^2|\kk|^2} }{2} )\frac{M +\sqrt{M^2+4|v_\star|^2|\kk|^2} }{2}
dk_xdk_y 
\eaa  
For $|\mu|<M$, we find 
\baa  
E_c = \frac{4\pi}{A_{MBZ}} \frac{|\mu|(|\mu| +M) }{|v_\star|^2}
\label{eq:ec_1}
\eaa  
For $|\mu|>M$, we find 
\baa  
E_c = \frac{4\pi}{A_{MBZ}} \frac{\mu^2 +M|\mu| }{|v_\star|^2}
+\frac{4\pi}{A_{MBZ}}\frac{|\mu|(|\mu|-M)}{|v_\star^2|}
\label{eq:ec_2}
\eaa  
Combing Eq.~\ref{eq:ec_1} and Eq.~\ref{eq:ec_2}, we have 
\baa  
E_c = \frac{4\pi}{A_{MBZ}} \frac{\mu^2 +M|\mu| }{|v_\star|^2}
+\frac{4\pi}{A_{MBZ}}\frac{|\mu|(|\mu|-M)}{|v_\star^2|}\theta( |\mu|-M) \label{eq:ec_final}
\eaa  
Then the energy of the system is 
\baa 
E_{\nu_f,\nu}/N_M &=  \frac{U_1}{2}\nu_f^2  + \frac{V_0}{2\Omega_0} (\nu-\nu_f)^2 +W\nu_f (\nu-\nu_f) + E_c
\eaa 
where $E_c$ is given in Eq.~\ref{eq:ec_final} and depends on $\mu$, with $\mu$ solved from Eq.~\ref{eq:zero_hyb_mu}.
Solutions at nonzero $M$ are shown in Fig.~\ref{fig:fill_2}. 
}

We notice two rules:
\begin{itemize}
    \item At integer $\nu$, the states with $\nu_f = \nu$ would minimize the energy.  
    \item $\nu=-3$ is very close to the transition point (see Fig.~\ref{fig:fill_2}) \hh{between $\nu_f=-3$ and $\nu_f=-2$}.
\end{itemize}
We will consider $\nu = 0,-1,-2$ in this work. The $\nu=-3$ needs separate treatment due to it being near the transition point between $\nu_f=-2,-3$. We point out that the $\nu=-3$ case is also known to exhibit a competition of orders~\cite{xie2022phase}.


\subsection{Solution of the one $f$-electron problem above (or below) uniform integer filling}
\label{sec:sol_one_f}

We now consider one more electron than integer filling. The state reads
\begin{equation}
    |\RR, \alpha \rangle = f^\dagger_{\RR,\alpha} |\nu_f\rangle |c \rangle
    \label{eq:one_fermi_st}
\end{equation} where $|\nu_f\rangle $ is the state with $\nu_f+4$ fermions per site and \hh{$|c\rangle$ is the conduction electron ground state of $\hH_c$ (Eq.~\ref{eq:hc_def}) with filling $\nu_c=0$}. We then have:
\begin{equation}
  \sum_{\alpha }:f_{\RR_1,\alpha }^\dag f_{\RR_1,\alpha }: |\RR, \alpha\rangle =( \delta_{\RR_1, \RR} + \nu_f) |\RR, \alpha\rangle
\end{equation} (here $\alpha = 1\ldots 8$ over all valleys and spins and flavors)

\begin{eqnarray}
\hH_{U} |\RR, \alpha\rangle  &=& \frac{U}{2} \sum_{\RR} :n_{f\RR}: :n_{f\RR}: |\RR, \alpha\rangle = ( N_M \nu_f^2 \frac{U}{2}  + \frac{U}{2} (1+ 2 \nu_f)  )  |\RR, \alpha\rangle
\nonumber \\ 
\hH_{W} |R, \alpha\rangle & =& \frac1{N} \sum_{\RR } :n_{f\RR}: \sum_{|\kk|, |\kk'|<\Lambda_c}  \sum_{\eta_2s_2 a} 
    W_a  e^{-i(\kk-\kk')\cdot\RR} 
    :c_{\kk a\eta_2 s_2}^\dagger  c_{\kk' a \eta_2 s_2}:\  |\RR, \alpha\rangle = \nonumber \\  &=& \frac1{N} \sum_{\RR } (\delta_{R_1, R} + \nu_f)  \sum_{|\kk|, |\kk'|<\Lambda_c}  \sum_{\eta_2s_2 a} 
    W_a  e^{-i(\kk-\kk')\cdot\RR} 
    :c_{\kk a\eta_2 s_2}^\dagger  c_{\kk' a \eta_2 s_2}:\  |\RR, \alpha\rangle  = \nonumber \\ & =&  \nu_f  \sum_{|\kk|<\Lambda_c}  \sum_{\eta_2s_2 a} 
    W_a 
    :c_{\kk a\eta_2 s_2}^\dagger  c_{\kk a \eta_2 s_2}:\  |\RR, \alpha\rangle  +  \frac1{N}  \sum_{|\kk|, |\kk'|<\Lambda_c}  \sum_{\eta_2s_2 a} 
    W_a  e^{-i(\kk-\kk')\cdot R} 
    :c_{\kk a\eta_2 s_2}^\dagger  c_{\kk' a \eta_2 s_2}:\  |\RR, \alpha\rangle  \nonumber  \\
\end{eqnarray} This gives the one electron Hamiltonian for the $c$-electrons without the \hb{ferromagnetic exchange interaction term $\hH_J$ (Eq.~\ref{eq:hj_def}). The Coulomb repulsion $\hH_{V}$ has been treated at the mean-field level (Eq.~\ref{eq:hv_mf_def}) and is absorbed by the chemical potential of conduction $c$-electrons}.

\begin{eqnarray}
&&H|\RR, \alpha\rangle \nonumber \\ &=& ( N_M \nu_f^2 \frac{U}{2}  + \frac{U}{2} (1+ 2 \nu_f)  )  |\RR, \alpha\rangle +  (\sum_{\eta s} \sum_{aa'} \sum_{|\kk|<\Lambda_c} (H^{(c,\eta)}_{a,a'}(\kk) - \mu \delta_{aa'} + \nu_f W_a \delta_{a a'}) c_{\kk a\eta s}^\dagger c_{\kk a'\eta s}  +  \nonumber \\ && +  \frac1{N}  \sum_{|\kk|, |\kk'|<\Lambda_c}  \sum_{\eta_2s_2 a} 
    W_a  e^{-i(\kk-\kk')\cdot \RR} 
    :c_{\kk a\eta_2 s_2}^\dagger  c_{\kk' a \eta_2 s_2}:\  ) |\RR, \alpha\rangle   \nonumber \\ & =& ( N_M \nu_f^2 \frac{U}{2}  + \frac{U}{2} (1+ 2 \nu_f)  )  |\RR, \alpha\rangle + \sum_{\eta s, \eta' s'} \sum_{a, a'} \sum_{|\kk|, |\kk'|<\Lambda_c  }  (A_{\kk a\eta s, \kk' a'\eta' s'} + B_{\kk a\eta s, \kk' a'\eta' s'} (\RR) ) c_{\kk a\eta s}^\dagger c_{\kk' a'\eta' s'} 
\end{eqnarray} We see that the Hamiltonian action on the one $f$ particle above the integer state contains two $c$-electron terms. First:
\begin{equation}\label{firsttermsinglepartham1}
 A_{\kk a\eta s, \kk' a'\eta' s'} =  (H^{(c,\eta)}_{a,a'}(\kk) - \mu \delta_{aa'} + \nu_f W_a \delta_{a a'})\delta_{k, k'}\delta_{\eta \eta'} \delta_{s,s'}
\end{equation}
has a continuous spectrum, given by $    H^{(c,\eta)}_{a,a'}(\kk)$ and shifted by an $a$-dependent chemical potential $W_a$. When $W_1= W_2$, the term is actually transformed into the density, and it is a bona-fide chemical potential. Let us call the eigenvalues of this term in Eq.~\ref{firsttermsinglepartham1} $\nu_1\ge \nu_2\ge \ldots \nu_{16 N _M}$ as the Hamiltonian has $16 N _M$ eigenvalues, \bh{16 for each momentum}.

The second term is 
\begin{equation}\label{firsttermsinglepartham2}
 B_{\kk a\eta s, \kk' a'\eta' s'} ({\RR})=     \frac1{N_M}  \sum_{|\kk|, |\kk'|<\Lambda_c} 
    W_a  e^{-i(\kk-\kk')\cdot \RR} \delta_{a, a'} \delta_{\eta, \eta'} \delta_{s,s'}
\end{equation} Unfortunately Eq.~\ref{firsttermsinglepartham1} and Eq.~\ref{firsttermsinglepartham2} do not commute, hence obtaining the spectrum analytically is hard. However, we can still \hb{derive clear theorems}. First we rewrite 

\begin{equation}\label{firsttermsinglepartham3}
 B_{\kk a\eta s, \kk' a'\eta' s'}(\bh{\RR})=      \sum_{\eta'', s'' a'' } P_{\kk a \eta s; \eta'', s'' a''   } (\bh{\RR}) P^\dagger_{ \eta'', s'' a'' ; \kk' a'; \eta; s'  }(\bh{\RR}) ,\;\;\; P_{\kk a \eta s; \eta'', s'' a''   }(\bh{\RR})  = \sqrt{\frac{W_a}{N_M} } e^{-i \kk \cdot \RR} \delta_{a a''}\delta_{\eta \eta''}\delta_{s s''} 
\end{equation} 
Since $B= P(\bh{\RR}) P(\bh{\RR}) ^\dagger$, where $P\bh{\RR}$ is a Rank $16$ ($a=1\ldots 4, \eta = \pm, a= \pm $) matrix, 
it is highly rank deficient and has a large number of zero eigenvalues. The nonzero eigenvalues are the same as those of the matrix $P^{\dagger}(\bh{\RR}) P(\bh{\RR})$:
\begin{equation}
    (P^{\dagger}(\bh{\RR}) P(\bh{\RR}))_{\eta' s' a'; \eta s a} = W_a  \delta_{a a'}\delta_{\eta \eta'}\delta_{s s'}
\end{equation} where we have summed over all the momentum values $\kk$ in the first moir\'e Brillouin zone (i.e we took $\Lambda_c$ to be entire first moir\'e Brillouin zone). Since $B$ is a positive semidefinite matrix, we confirm all eigenvalues are non-negative.  We now know that there are $16$ nonzero eigenvalues, $8$ of them being $W_3$, larger than the other $8$ of them which are $W_1$. We now order them as such: $\rho_1= \ldots =\rho_8= W_3>\rho_9= \ldots = \rho_{16}= W_1 > \rho_{17}= \ldots \ \rho_{16 N_M}=0$. 

The existence of only a finite ($N_{orbitals}= N_{orb}=16$) number of nonzero eigenvalues of the second matrix $B$ with eigenvalues $\rho_i$ and a continuum spectrum of eigenvalues of the first matrix $A$ with eigenvalues $\nu_i$ allows for strong theorems of the spectrum. We employ Weyl's inequalities that say that the eigenvalues $\mu_1 \ge \mu_2 \ge \ldots \ge \mu_{N_{orb} N_M}$ of the sum of two matrices $A,B$ satisfy:

\begin{equation}
    \nu_j + \rho_k \le \mu_i \le \nu_r + \rho_{s}, \;\;\;\; \text{if} \;\;\; j+ k- N_{orb} N_M \ge i \ge r+s -1 
    \label{eq:eigen_relation}
\end{equation}
\hb{For the $A,B$ matric in Eq.~\ref{firsttermsinglepartham1} and Eq.~\ref{firsttermsinglepartham2}}, we pick $k= N_{orb} N_M \implies \rho_k = 0$ and $s= N_{orb}+1 \implies \rho_s= 0$. We then have:
\begin{equation}
    \nu_j \ge \mu_i \ge \nu_r, \;\;\; \text{if} \;\;\; j \ge i \ge r+ N_{orb}
\end{equation} 
In Eq.~\ref{eq:eigen_relation}, we pick $j=i, r= i - N_{orb}$ and we have
\begin{equation}
    \nu_i \le \mu_i \le \nu_{i - N_{orb}} , \;\;\; i> N_{orb}
\end{equation}
Now we see that most of the $\mu$ eigenvalues form a continuum themselves, as they are bounded between the eigenvalues of the matrix $A$, which form a continuum. In fact, $\mu_{N_{orb}+1} \ge\mu_{N_{orb}+2} \ge \ldots \ge \mu_{N_{orb} N_M}$ are bounded by the eigenvalues of $A$, $\nu_i, \nu_{i- N_{orb}}$; since these latter eigenvalues form a continuum, the most (most of the times they are degenerate as they represent equal energies in $k$-space, i.e. Fermi surfaces of the continuum Hamiltonian) they can have in energy separation is $N_{orb} v_\star 2 \pi/\sqrt{N_M}$ where we have used the linear spectrum of $A$ and the momentum quantum $2 \pi/\sqrt{N_M}$. When $N_M \rightarrow \infty$ the two eigenvalues are the same and hence to a good approximation $\mu_i \approx \nu_i, \;\; i> N_{orb} $. (A similar proof is an observed reason why the particle-particle continuum is formed by sums of the particle-particle energies, and the bound/anti bound states form a finite number). The number of eigenvalues that can be away from a $\nu_i$ is small, $\mu_i , i= 1,2\ldots N_{orb}$: $\nu_i \le \mu_i \le \nu_i + W$. These are anti-bound states ($i =1,2\ldots N_{orb}$ and hence they are at the top of the spectrum, as they should be since $B$ \bh{(Eq.~\ref{firsttermsinglepartham2})} is a positive semidefinite matrix.

Since the $c$-spectrum of the one-$f$ fermion state in Eq.~\ref{eq:one_fermi_st} is unchanged except for the upper parts of the spectrum, adding one electron to the integer-filled heavy fermion ground-state at integer total fillings costs $ \nu_f^2 \frac{U}{2}  + \frac{U}{2} (1+ 2 \nu_f)  $ per unit cell. As such it is beneficial to add $c$ electrons, at least at $\nu_f=0$. 

\hb{We next comment on the effect of $\hH_V$. At the mean-field level ($\hH_V\approx \hH_V^{MF}$, Eq.~\ref{eq:hv_mf_def}), $\hH_V^{MF}$ corresponds to a chemical potential term. The $c$-spectrum remains gapless and forms a continuum even after including $\hH_V^{MF}$. Therefore, our statement remains valid.
}

\subsection{ Charge density waves?}

Consider charge neutrality $\nu_f = \nu_c=0 $. As we showed, it costs $\frac{U}{2} $ to add one $f$ electron, hence upon changing the filling, it is easy to add $c$ electrons. 
\bh{ Suppose we add $c$ electron up to $k_0$ defined in Eq.~\ref{eq:k0_def}, adding one more $c$ electron will cost 
\baa  
E_{c}= v_\star k_0 + \frac{V(0)}{\Omega_0}\nu_c = v_\star k_0 + \frac{V(0)}{\Omega_0}\nu_c = v_\star \sqrt{\frac{|\nu_c|A_{MBZ} }{8\pi}} + \frac{V(0)}{\Omega_0}\nu_c 
\eaa  
where the second term comes from the $\hH_V^{MF}$ (Eq.~\ref{eq:hv_mf_def}). For sufficient large $\nu_c$, we have $E_c>U_1/2$. Then} it becomes advantageous to add $f$-electrons, and they can be added (if we neglect $W$) from $:\nu_f:=0$ all the way to $:\nu_f:=1$ at $W=0$ limit. At non-zero $W$, the process of adding electrons is not so straightforward. 
We use a trial state 
\begin{equation}
    |\psi\rangle = \prod_{ (\RR, \alpha \eta s) \in S_{fill} } f^{\dagger}_{\RR,\alpha \eta s }|:\nu_f:=0\rangle |k_0, \text{  c-filled}\rangle 
\end{equation} where \bh{$S_{fill}$ is a set with $N_M\nu_f$ elements, that characterize the site and flavor (orbital, valley, and spin) } where the $f$-electron is filled. 
We have that:
\begin{eqnarray}
    \hH_{W} |\psi \rangle &=& \frac1{N} \sum_{\RR }  :n^f_{\RR}: \sum_{|\kk|, |\kk'|<\Lambda_c}  \sum_{\eta_2s_2 a} 
    W_a  e^{-i(\kk-\kk')\cdot\RR} 
    :c_{\kk a\eta_2 s_2}^\dagger  c_{\kk' a \eta_2 s_2}:\  |\psi \rangle \nonumber \\ 
    & =& \frac1{N} \sum_{\RR,\alpha \eta s } \sum_{(\RR_1 ,\alpha_1\eta_1s_1) \in S_{fill}}\delta_{\RR_1, \RR}  \delta_{\alpha,\alpha_1}\delta_{s,s_1}\delta_{\eta ,\eta_1} \sum_{|\kk|, |\kk'|<\Lambda_c}  \sum_{\eta_2s_2 a} 
    W_a  e^{-i(\kk-\kk')\cdot\RR} 
    :c_{\kk a\eta_2 s_2}^\dagger  c_{\kk' a \eta_2 s_2}:\  |\psi \rangle  = \nonumber \\ 
    & =&  \frac1{N}  \sum_{|\kk|, |\kk'|<\Lambda_c}  \sum_{(\RR,\alpha\eta s)\in S_{fill}} \sum_{\eta_2s_2 a} 
    W_a  e^{-i(\kk-\kk')\cdot \RR} 
    :c_{\kk a\eta_2 s_2}^\dagger  c_{\kk' a \eta_2 s_2}:\  |\psi \rangle
\end{eqnarray} 
The spectrum of this matrix is again very simple, only $W_1, \bm{W_3}$ and zero eigenvalues, by the same proof as in the previous section \bh{(Sec.~\ref
{sec:sol_one_f})}. However, there is now a thermodynamic number of them, $N_{orb} :\nu_f: N_M$ and the Weyl theorems do not give nontrivial results, as they only bound the new eigenvalues in between $\nu_i$ and $\nu_{i - N_{orb}:\nu_f: N_M}$ which can now be a large interval. \bh{At large fillings (say $\nu_f=-3$), it could become convenient~\cite{xie2022phase} to fill unbounded eigenvalues rather than just fill the particle continuum and hence a CDW or other translational breaking states could be competitive. }

\section{Zero-hybridization Model with $\hH_J$} 
\label{sec:sec_zero_hyb_j} 
We next study the zero hybridization model with $\hH_J$. The Hamiltonian is given below 
\baa  
\hH_c +\hH_U +\hH_W +\hH_V +\hH_J
\eaa  
where $\hH_J$ describes a ferromagnetic exchange coupling between flat U(4) moments (Eq.~\ref{eq:U4op-maintext} of $f$-electrons and $\Gamma_1\oplus\Gamma_2$ $c$-electrons ($a=3,4$)~\cite{HF_MATBLG} 
\begin{equation} \label{eq:HJ}
\hat H_J = - J\sum_{\RR \qq }\sum_{\mu\nu} \sum_{\xi=\pm} e^{-i\qq \cdot\RR }:\hat{\Sigma}_{\mu\nu}^{(f,\xi)}(\RR ): :\hat{\Sigma}_{\mu\nu}^{(c\prime\prime,\xi)}(\qq ):
\end{equation}

\subsection{Coherent states of the $f$-moments.} 
Since $J\sim 16.38meV$ is relatively small compared to $U$ and $W$, we assume that $\hH_J$ term will not destroy the uniform charge distribution of the $f$-electrons in the plateau region of Fig.~\ref{fig:fill_2} \hb{which includes $\nu=0,-1,-2$}. To find the ground state at small but non-zero $J$, we do the following 
\begin{itemize}
    \item Find $\nu_f$ at given $\nu$ at $J=0$ limit \bh{(which is given in Fig.~\ref{fig:fill_2})}. 
    \item Turn on a small but nonzero $J$ and assume the filling of $f$ electron at each site is fixed to be $\nu_f$. Find the optimal solution.
\end{itemize}

We now describe the second step in detail.
We start by \hh{introducing} the coherent state in a similar manner as spin coherent state of $SU(2)$ spin. \hh{We assume that, for each site $\RR$, we have a fixed integer number of $f$ electrons ($\nu_f+4$), which allows us to label each site via its expectation value of chiral $U(4)$ moments.} In other words, for a given a state \hh{$|\psi_\RR \rangle$} of the $f$-electron at site $\RR$, we label it as $|\{\theta_{\mu\nu}^{(f,\xi)}(\RR)\}_{\mu,\nu,\xi}\rangle$, where $\theta_{\mu\nu}^{(f,\xi)}(\RR)$ is given by
\hh{
\baa  
\langle \psi_{\RR} |:\hat{\Theta}_{\mu\nu}^{(f,\xi)} (\RR): |\psi_{\RR} \rangle = \theta_{\mu\nu}^{(f,\xi)} (\RR)
\label{eq:coh_def}
\eaa  
}
\hh{We will} call the states with such labeling coherent states.

\subsection{Trial wavefunction}
Using the coherent state defined in the previous section, we now propose the trial wavefunction of the ground state. Again, we fix the filling of $f$ to be $\nu_f$ and the total filling to be $\nu$, and assume uniform charge distribution of $f$. 

We first consider the following state of $f$ electrons
\baa  
|\vartheta \rangle  = \prod_{\RR} |\{\theta_{\mu\nu}^{f,\xi}(\RR)\}_{\mu\nu\xi}\rangle_{\RR} 
\label{eq:trial_state_f}
\eaa  
where we simply tensor product the coherent state of each site and we use $\vartheta $ ($= \{\{\theta_{\mu\nu}^{f,\xi}(\RR)\}_{\mu\nu\xi}\}_\RR$) to denote the set of $\theta_{\mu\nu}^{f,\xi}(\RR)$. 
We now act the Hamiltonian on it and calculate:
\baa  
\langle \vartheta |\hH |\vartheta \rangle .
\eaa 
$|\vartheta\rangle$ only describes $f$ states, so $\langle \vartheta | \hH |\vartheta\rangle$ is an operator acting on the Hilbert space of $c$ electrons. 
Each term in $\langle \vartheta | \hH |\vartheta\rangle$ becomes 
\baa  
&\langle \vartheta |\hH_U |\vartheta \rangle = N_M \frac{U_1}{2}\nu_f^2 \nonumber \\
&\langle \vartheta |\hH_V |\vartheta \rangle = N_M \frac{V_0}{2\Omega_0}\hat{\nu_c}^2 \text{ \quad (mean-field level)} \nonumber \\
&\langle \vartheta |\hH_W |\vartheta \rangle = N_M W \nu_f \hat{\nu_c} \nonumber \\
&\langle \vartheta |\hat{H_c}|\vartheta \rangle= \hat{H_c}\quad \quad  \text{ ($H_c$ only contains conduction electron operators)} \nonumber \\
&\langle \vartheta |\hH_J |\vartheta \rangle = - J\sum_{\RR \qq }\sum_{\mu\nu} \sum_{\xi=\pm} e^{-i\qq \cdot\RR }\theta^{f,\xi}_{\mu\nu}(\RR)  :\hat{\Theta}_{\mu\nu}^{(c\prime\prime,\xi)}(\qq ):  
\eaa 
Here, we rewrite $\hH_J$ in terms of chiral $U(4)$ moment.  $\theta_{\mu\nu}^{f,\xi}(\RR)$ is a complex number 
\hh{but $:\hat{\Theta}_{\mu\nu}^{(c\prime\prime,\xi)}(\qq ):$ is an operator}. 
Since we fix the total filling to be $\nu$
\hh{ and the Hamiltonian without $f$-$c$ hybridization preserves $c$-electron particle numbers}
, we can replace operator $\hat{\nu}_c$ with a number $\nu_c = \nu-\nu_f$. Now, $\langle \vartheta |\hH |\vartheta \rangle $ can be separated into two parts: a constant term $E_0[\nu_f,\nu]$ (which is a function of $\nu_f$, $\nu$) \bh{and a one-body} Hamiltonian $\hat{H}''_c[\vartheta]$ of $c$-electron: 
\baa
&\langle \vartheta |\hH |\vartheta \rangle 
=E_0[\nu_f,\nu] +\hat{H}_c''[\vartheta] \nonumber  \\
&E_0[\nu_f,\nu] =  N_M \frac{U_1}{2}\nu_f^2
+N_M W \nu_f {\nu_c} 
+
N_M \frac{V_0}{2\Omega_0}{\nu_c}  \nonumber \\
&\hat{H}_c''[\vartheta] = \hat{H}_c
- J\sum_{\RR \qq }\sum_{\mu\nu} \sum_{\xi=\pm} e^{-i\qq \cdot\RR }\theta^{f,\xi}_{\mu\nu}(\RR)  :\hat{\Theta}_{\mu\nu}^{(c\prime\prime,\xi)}(\qq ): 
\eaa
We then write $\hat{H}''_c[\vartheta] $ in a more compact form

\baa  
 \hat{H}_c''[\vartheta]= \sum_{\kk,\kk', aa',\eta\eta',ss'}c_{\kk a\eta s}^\dag \bigg([h_0]_{\kk a \eta s,\kk' a' \eta' s'} +[h_1]_{\kk a \eta s,\kk' a' \eta' s'}\bigg) c_{\kk' \eta' s'}  +\frac{J}{N_M}\sum_{\RR \xi =\pm.|\kk|<\Lambda_c} \theta_{00}^{(f,\xi)}(\RR) 
 \label{eq:hcpp_theta}
\eaa  
$h_0, h_1$ are understood as a matrix with row index $\kk a \eta s$ and column index $\kk' a' \eta' s'$. The explicit form of $h_0,h_1$ are 
\baa  
&[h_0]_{\kk a \eta s,\kk' a' \eta' s'} = \delta_{\kk,\kk'} \delta_{\eta\eta'}\delta_{ss'}H^{(c,\eta)}_{aa'}(\kk) \nonumber \\
&[h_1]_{\kk a \eta s , \kk' a'\eta' s'} = -J\sum_{\xi=\pm,\RR} e^{-i\qq \cdot\RR }\theta^{f,\xi}_{\mu\nu}(\RR) [\Theta_{\mu\nu}^{(c\prime\prime,\xi)}]_{a\eta s, a'\eta's'}\frac{\delta_{\xi,(-1)^{a-1}\eta} \delta_{\xi,(-1)^{a'-1}\eta'} }{2N_M} \delta_{\kk, \kk'+\qq} 
\eaa 
$h_0$ describes the noninteracting part of conduction electrons. $h_1$ describes the contribution from $\hH_J$. The last term of Eq.~\ref{eq:hcpp_theta}, comes from normal ordering. 
\hb{To observe the origin of the last term, we note 
\baa 
&-J\sum_{\RR \qq ,\mu\nu,\xi}e^{-i\qq \cdot \RR}\theta_{\mu\nu}^{(f,\xi)}(\RR):\UC_{\mu\nu}^{(c'',\xi)}(\qq): \nonumber\\
= &\frac{-J}{2 N_M}\sum_{\RR \qq ,\mu\nu,\xi}e^{-i\qq \cdot \RR}\theta_{\mu\nu}^{(f,\xi)}(\RR)\sum_{|\kk|<\Lambda_c}\sum_{a\eta s,a'\eta's'}\delta_{\xi,(-1)^{a-1}\eta}\frac{\Theta^{\mu\nu,c''}_{a\eta s, a'\eta's'}}{2}[c_{\kk+\qq,\alpha \eta s} c_{\kk,\alpha'\eta's'}-\frac{1}{2}\delta_{\qq,0} \delta_{\alpha,\alpha'}\delta_{\eta,\eta'}\delta_{s,s'}]  \nonumber \\
= &-J\sum_{\RR \qq ,\mu\nu,\xi}e^{-i\qq \cdot \RR}\theta_{\mu\nu}^{(f,\xi)}(\RR)\UC_{\mu\nu}^{(c'',\xi)}(\qq) - \frac{1 }{4N_M}\sum_{\kk,\RR}\sum_{a\eta s,a'\eta's'} \theta^{(f,\xi)}_{\mu\nu}(\RR)\Theta^{\mu\nu,c''}_{a\eta s, a'\eta's'} \nonumber \\
=& 
\UC_{\mu\nu}^{(c'',\xi)}(\qq) - \frac{\delta_{\qq,0} }{N_M}\delta_{\mu\nu,00}\sum_{\kk<\Lambda_c} \theta^{(f,\xi)}_{\mu\nu}(\RR)
\eaa  
}
$h_0+h_1$ \bh{in Eq.~\ref{eq:hcpp_theta}} can be diagonalized. Let $E_n$ and $[v_n]_{\kk a\eta s}$ be $n$-th eigenvalue and $n$-th eigenvector
\baa  
\sum_{\kk'a'\eta's'}[h_0+h_1]_{\kk a\eta s,\kk'a'\eta' s'} [v_n]_{\kk'a'\eta's'} = E_n [v_n]_{\kk a \eta s} 
\eaa 
where we assume $E_1\le E_2\le ...$. Then we have 
\baa  
 &\hat{H}_c''[\vartheta]=\sum_{n} E_n \gamma_n^\dag \gamma_n +\frac{J}{N_M}\sum_{\RR \xi =\pm.|\kk|<\Lambda_c} \theta_{00}^{f,\xi}(\RR) \nonumber  \\
 &\gamma_n = \sum_{\kk a \eta s} c_{\kk a \eta s }\bigg([v_n]_{\kk a \eta s}\bigg)^*
\label{eq:def_hc_pp_J}
\eaa  
For fixed $\nu_c$, the Slater determinant states that minimize the energy of $\hat{H}_c''$ are 
\baa  
|\Psi[\vartheta, \nu_c]\rangle := \prod_{n \le N_c} \gamma_n^\dag |0\rangle 
\label{eq:slater_det_state_j}
\eaa  
where  $\gamma_n|0\rangle =0 $ for all $n$ and $N_c = (\nu_c+8)N_M$ denotes the total particle numbers of $c$ (without normal ordering). Based on the states we constructed
\bh{in Eq.~\ref{eq:trial_state_f} and Eq.~\ref{eq:slater_det_state_j} }, we define the following trial wavefunction of the ground state:
\baa 
|\vartheta\rangle |\Psi[\vartheta, \nu_c]\rangle .
\label{eq:trial_state}
\eaa 
Its energy is 
\baa  
E[\vartheta,\nu_f,\nu]& = 
\langle \Psi[\vartheta,\nu_c] |\langle \vartheta | \hH
|\vartheta \rangle |\Psi[\vartheta,\nu_c]\rangle \nonumber \\
&
=  N_M \frac{U_1}{2}\nu_f^2
+N_M W \nu_f {\nu_c} 
+
N_M \frac{V_0}{2\Omega_0}{\nu_c} 
+\langle \Psi[\vartheta,\nu_c]|\hat{H}_c''[\vartheta]|\Psi[\vartheta,\nu_c]\rangle .  \nonumber \\
&
=  N_M \frac{U_1}{2}\nu_f^2
+N_M W \nu_f {\nu_c} 
+
N_M \frac{V_0}{2\Omega_0}{\nu_c} 
+\frac{J}{N_M}\sum_{\RR |\kk|<\Lambda_c,\xi =\pm} \theta_{00}^{f,\xi}(\RR)+ \sum_{n\le (\nu_c+8)N_M} E_n \gamma_n^\dag \gamma_n 
\eaa  
where $\nu_c=\nu-\nu_f$. For given $\vartheta,\nu_f,\nu$, \hh{the ground state} will minimize the energy. Remember, the energy contains two terms: $\langle \vartheta |\hH|\vartheta\rangle = E_0[\nu_f,\nu] + \hat{H_c}''[\vartheta]$. The first term is fixed for given $\nu_f,\nu$, and the Slater determinant state we choose will always minimize the energy from $\hat{H_c}''[\vartheta]$ at fixed filling. 

In order to find the optimal states at fixed $\nu_f,\nu$, we need to vary $\vartheta$ and find the configuration of $f$ states that minimize $E[\vartheta, \nu_f,\nu]$. In principle, this could be efficiently done numerically via simulated annealing and kernel polynomial methods~\cite{KMP}. Here, instead, we tackle this problem via perturbation theory.

\subsection{Perturbation theory} 
Instead of numerically solving the problem, we could expand the energy function $E[\vartheta,\nu_f,\nu]$ in powers of $J$. We note that
\ba 
E[\vartheta,\nu_f,\nu]&  
=  N_M \frac{U_1}{2}\nu_f^2
+N_M W \nu_f {\nu_c} 
+
N_M \frac{V_0}{2\Omega_0}{\nu_c} 
+\langle \Psi[\vartheta,\nu_c]|\hat{H}_c''[\vartheta]|\Psi[\vartheta,\nu_c]\rangle \, .
\ea 
For fixed $\nu_f, \nu$, the ground state minimize $\langle \Psi[\vartheta,\nu_c]|\hat{H}_c''[\vartheta]|\Psi[\vartheta,\nu_c]\rangle$. Then we find the ground state by calculating the ground state energy of $\hat{H}_c''[\vartheta]$ at fixed filling $\nu_c$. To do so, it is more convenient to first calculate the free energy \bh{at the fixed filing of $c$ electrons $\nu_c$} and at a finite temperature. We then set the temperature to zero, and find the ground state energy. We take the following partition function 
\baa  
Z = \text{Tr}[e^{-\beta \hat{H}''_c[\vartheta]}P ]
\eaa  
where $P$ is a projection operator \hb{that fix the total filling of $c$ electrons to be $\nu_c = \nu - \nu_f$ being an integer. $\beta$ is the inverse temperature. For given $\nu_c$, $P$ is defined as 
\baa  
P =\frac{1}{A_{\nu_c}} \prod_{\nu_c' \in S_{\nu_c} } (\hat{\nu_c} - \nu_c' )\quad,\quad A_{\nu_c} = \prod_{\nu_c' \in S_{\nu_c} } ({\nu_c} - \nu_c' )
\eaa  
where $\hat{\nu}_c$ is the density operator of $c$ electrons (defined in Eq.~\ref{eq:def_hat_c}) and $S_{\nu_c}$ denotes the set of all possible fillings of $c$ electrons that not equals to $\nu_c$.} 
The free energy is
\ba 
F = -\frac{1}{\beta} \log(Z)
\ea 
and the energy is
\ba 
\langle \Psi[\vartheta,\nu_c]|\hat{H}_c''[\vartheta]|\Psi[\vartheta,\nu_c]\rangle=E_{H''_c} = \lim_{\beta \rightarrow \infty }F
\ea 
\hh{We calculate the free energy by performing expansion in $J$.}
To do an expansion in powers of $J$,
we separate $\hat{H}''_c[\vartheta]$ into two parts, $J$ independent and $J$ dependent part. 
\baa 
&\hat{H}''_c[\vartheta] =\hH_0 +\hH_1 \nonumber  \\
&\hH_0 = \hat{H}_c \nonumber \\
&\hH_1 = 
- J\sum_{\RR \qq }\sum_{\mu\nu} \sum_{\xi=\pm} e^{-i\qq \cdot\RR }\theta^{f,\xi}_{\mu\nu}(\RR)  :\hat{\Theta}_{\mu\nu}^{(c\prime\prime,\xi)}(\qq ): 
\label{eq:j_h0_h1}
\eaa  

Then we expand the partition functions as
\baa  
&Z = \sum_{n=0}^\infty Z_n \nonumber \\
&Z_n = (-1)^n\int_0^{\beta}d\tau_1 \int_{\tau_1}^{\beta}d\tau_{2}...\int_{\tau_{n}-1}^{\beta}d\tau_n \text{Tr}\bigg[e^{-(\beta-\tau_n)\hH_0}\hH_1e^{-(\tau_n-\tau_{n-1})\hH_0}...e^{-(\tau_2-\tau_1)\hH_0}\hH_1e^{-\tau_1\hH_0} P\bigg] \label{eq:z_exp}
\eaa 

\bh{ 
We now prove Eq.~\ref{eq:z_exp}. For convenience, we introduce an artificial parameter $g=1$. We first perform a Suzuki–Trotter expansion. 
\baa  
Z= \lim_{N\rightarrow \infty}
\text{Tr}[ \prod_{n=i}^{N}e^{-\Delta \tau (\hH_0+g\hH_1)}P] 
\eaa  
where $\Delta \tau = \beta /N$. We next expand partition functions in powers of $g$ via a Taylor expansion ($Z=\sum_n Z_n$). The $n$-th order term is 
\ba 
Z_n = &\lim_{N\rightarrow \infty}\frac{1}{n!}\frac{\partial^n}{\partial g^n} Z \bigg|_{g=0}= \frac{ (-\Delta \tau)^n}{n!} n!\sum_{i_1
\le i_2...\le i_n}\text{Tr}[\bigg(\prod_{n={i_{n}+1}}^{N} e^{-\Delta  \tau\hH_0 }\bigg) \hH_1 \bigg(\prod_{n=i_{n-1}+1}^{i_n} e^{-\Delta  \tau \hH_0}\bigg)\hH_1...\hH_1\bigg(\prod_{n=1}^{i_1} e^{-\Delta  \tau \hH_0}\bigg) P] \nonumber \\
=&\lim_{N\rightarrow \infty}
(-\Delta \tau)^n\sum_{i_1
\le i_2...\le i_n}\text{Tr}[ e^{-\Delta  \tau (N-i_{n+1})\hH_0 } \hH_1  e^{-\Delta  \tau (i_n-i_{n-1})\hH_0}\hH_1...\hH_1\bigg( e^{-\Delta  \tau i_1\hH_0 }\bigg) P] \nonumber \\
\ea 
where $i_1,...,i_n$ denotes the time-slices where the derivative has acted on. Since we have assumed a specific order of $i_1\le i_2...\le i_n $, permutation of $i_1,...,i_n$ leads to an additional $n!$ factor. In the limit of $N\rightarrow \infty$, we reach Eq.~\ref{eq:z_exp}. }

Since $\hH_1$ has a coefficient $J$, $Z_n \propto J^n$. We truncate the expansion to the second order: $Z=Z_0 +Z_1+Z_2 +o(J^2)$.  The free energy becomes
\baa
F = -\frac{1}{\beta}\log[Z_0+Z_1+Z_2+o(J^2)]=-\frac{1}{\beta}\log[Z_0] -\frac{1}{\beta} \frac{Z_1}{Z_0} + \frac{1}{2}\frac{1}{\beta}[\frac{Z_1}{Z_0}]^2 - \frac{1}{\beta}\frac{Z_2}{Z_0} +o(J^2)
\label{eq:free}
\eaa 

Now we want to compute $Z_0,Z_1/Z_0,Z_2/Z_0$. Before moving on, we first introduce the following notation. 
For an operator $\hat{O}$, its expectation value with respect to the Hamiltonian $\hH_0 (=\hH_c)$ at inverse temperature $\beta$ and filling $\nu_c =\nu-\nu_f$ is 
\ba 
\langle \hat{O} \rangle_0 :=\frac{1}{Z_0}\text{Tr}[e^{-\beta H_0}OP]. 
\ea 
where $Z_0= \text{Tr}\log[e^{-\beta H_0}P]$. \hh{The ground state information can be obtained by setting $\beta\rightarrow \infty$.}
The operator in the interaction picture $\hat{O}(\tau)$ is defined as
\ba 
\hat{O}(\tau) := e^{\tau H_0}\hat{O}e^{-\tau H_0}
\ea 
Now, we calculate $Z_0,Z_1/Z_0,Z_2/Z_0$

\subsubsection{$Z_0$}
$Z_0 = \text{Tr}[e^{-\beta H_0}P]$.
This is the partition of the conduction electrons at filling $\nu_c$ with $J=0$. 

\subsubsection{$Z_1/Z_0$}
    \ba 
    \frac{Z_1}{Z_0} & = - \frac{1}{Z_0}\int_0^\beta d\tau  \text{Tr}\bigg[e^{-\beta \hH_0}[e^{\tau \hH_0}\hH_1e^{-\tau \hH_0}]P \bigg]  \\
    &=- \frac{1}{Z_0}\int_0^\beta d\tau  \text{Tr}\bigg[e^{-\beta \hH_0}\hH_1(\tau)P \bigg] 
    =
    -\int_0^{\beta} \langle \hH_1(\tau) \rangle_0  d\tau 
    \ea 
    Using the explicit expression of $\hH_1$, we find
    \ba 
    \frac{Z_1}{Z_0} = J\sum_{\RR \qq \kk}\sum_{\mu\nu}\sum_{\xi=\pm}\sum_{a,a'=3,4}\sum_{\eta\eta'ss'}e^{-i\qq \cdot \RR}\delta_{\xi,(-1)^{a-1}\eta} \frac{\theta_{\mu\nu}^{(f,\xi)}(\RR)}{2N_M}\int_0^\beta d\tau \langle :
    c_{\kk+\qq a\eta s}^\dag (\tau)\Theta^{\mu\nu, c''}_{a\eta s, a'\eta' s'}c_{\kk a'\eta' s'}(\tau): 
    \rangle_0 d\tau 
    \ea 
    where $c_{\kk a\eta s}^\dag(\tau)$ is defined as $c_{\kk a\eta s}(\tau) = e^{\tau H_0}c_{\kk a \eta s}e^{-\tau H_0}$. Now we want to evaluate 
    \ba 
    \frac{1}{2N_M}\sum_{\kk,\qq}\sum_{a,a'=3,4}\sum_{\eta\eta'ss'}\delta_{\xi,(-1)^{a-1}\eta}\int_0^\beta d\tau \langle : 
    c_{\kk+\qq a\eta s}^\dag (\tau)\Theta^{\mu\nu, c''}_{a\eta s, a'\eta' s'}c_{\kk a'\eta' s'}(\tau) :
    \rangle_0
    \ea 
    Due to the space translation symmetry of $H_0$, only $\qq=0$ gives a non-zero contribution. \hh{In addition, for any operator $\hat{O}$, $\langle \hat{O}(\tau)\rangle_0$ is $\tau$-independent due to the imaginary-time translational symmetry. This is because
    $
    \langle \hat{O}(\tau) \rangle = \langle e^{\tau  \hH_0 } \hat{O} e^{-\tau \hH_0}\rangle_0 =  \text{Tr}[ e^{-(\beta -\tau)  \hH_0 } \hat{O} e^{-\tau \hH_0}]/Z_0 =  \text{Tr}[ e^{-\beta  \hH_0 } \hat{O} ]/Z_0 = \langle \hat{O}\rangle_0 
    $.} 
   Therefore, we have
    \ba 
    &\frac{1}{2N_M}\sum_{\kk}\sum_{a,a'=3,4}\sum_{\eta\eta'ss'}\delta_{\xi,(-1)^{a-1}\eta}\int_0^\beta d\tau \langle : 
    c_{\kk+\qq a\eta s}^\dag (\tau)\Theta^{\mu\nu, c''}_{a\eta s, a'\eta' s'}c_{\kk a'\eta' s'}(\tau) :
    \rangle_0 \\
    =&\delta_{\qq,0}\beta \langle : 
    \frac{1}{2N_M}\sum_{\kk}\sum_{a,a'=3,4}\sum_{\eta\eta'ss'}\delta_{\xi,(-1)^{a-1}\eta} c_{\kk a\eta s}^\dag \Theta^{\mu\nu, c''}_{a\eta s, a'\eta' s'}c_{\kk a'\eta' s'} :
    \rangle_0\\
    = &\beta \langle :\hat{\Theta}^{(c'',\xi)}_{\mu\nu}(\qq=0):\rangle 
    \ea 
   This term gives the expectation value of chiral $U(4)$ moment of $a=3,4$ orbitals. 
   
   Since the fermion-bilinear Hamiltonian $\hH_0$ has chiral U(4) symmetry, not all the operators have a non-zero expectation value.  We utilize the following statement: Given operator $\hat{O}$, if there exists a chiral U(4) transformation $g$, such that $g\hat{O}g^{-1}=-\hat{O}$, then $\langle \hat{O}\rangle_0 =\langle g\hat{O}g^{-1}\rangle_0 =-\langle \hat{O}\rangle_0= 0 $. 
    We prove that, for \hh{$\UC_{\mu\nu}^{(c'',\xi)}(\qq) = \sum_{\kk,a\eta s,a'\eta's'}c_{\kk+\qq,a\eta s}\Sigma_{a\eta s,a'\eta's'}^{(\mu\nu,c'')}c_{\kk,a'\eta's'}\delta_{\xi,(-1)^{a-1}\eta} /(2N_M)$ with $\mu\nu\ne 00$}, we can always find such chiral U(4) transformation. 
\baa 
&e^{-i\pi \UC_{z0}} \quad:\quad    \UC_{x\mu}^{(c'',\xi)}(\qq),\UC_{y\mu}^{(c'',\xi)}(\qq) \rightarrow 
        -\UC_{x\mu}^{(c'',\xi)}(\qq),-\UC_{y\mu}^{(c'',\xi)}(\qq) \nonumber \\
&e^{-i\pi \UC_{x0}}   \quad:\quad  \UC_{z\mu}^{(c'',\xi)}(\qq)\rightarrow -\UC_{z\mu}^{(c'',\xi)}(\qq)\nonumber \\
&e^{-i\pi \UC_{0z}}  \quad:\quad
        \UC_{0x}^{(c'',\xi)}(\qq),\UC_{0y}^{(c'',\xi)}(\qq) \rightarrow -\UC_{0x}^{(c'',\xi)}(\qq),-\UC_{0y}^{(c'',\xi)}(\qq)\nonumber \\
&e^{-i\pi \UC_{0x}}  \quad:\quad \UC_{0z}^{(c'',\xi)}(\qq)\rightarrow -\UC_{0z}^{(c'',\xi)}(\qq)
\label{eq:u4_flip_sign}
\eaa 
    Therefore, 
    \baa 
     \langle :\hat{\Theta}^{(c'',\xi)}_{\mu\nu}(\qq):\rangle_0 =0 \quad,\quad \mu\nu \ne 00
    \label{eq:zero_linear_uc}
    \eaa 
    \bh{Here, we comment that, in order to prove $\langle O\rangle_0=0$, we require the chiral $U(4)$ transformation to flip the sign of the operator. Suppose we consider an chiral $U(4)$ transformation that changes $\hat{O}$(one components of $U(4)$ momentum) to $\hat{O}'$ (another $U(4)$ momentum). Symmetry only indicates $\langle \hat{O}\rangle_0 = \langle \hat{O}'\rangle_0 $, but does not indicate the expectation value goes to zero. }
    
    We now conclude the only non-zero component is 
    \baa 
    \delta_{\qq,0}\beta \langle :\hat{\Theta}^{(c'',\xi)}_{00}(\qq=0):\rangle =  \delta_{\qq,0}\beta \sum_{a = 3,4}\frac{1}{2N_M}\sum_{\xi, s}\delta_{\xi,(-1)^{a-1}\eta} 
    \langle 
    : c_{\kk a\eta s}^\dag c_{\kk a \eta s}:
    \rangle_0 
    \eaa 
    By only including the non-vanishing contributions. 
    \baa 
    \frac{Z_1}{Z_0} =J\beta \sum_{\RR}\sum_{\xi=\pm} \frac{\theta_{00}^{f,\xi}(\RR)}{2N_M}\sum_{\kk,\eta s}\sum_{a=3,4}\delta_{\xi,(-1)^{a-1}\eta}\langle : 
    c_{\kk a\eta s}^\dag c_{\kk a\eta s}:
    \rangle_0 
    \label{eq:z1z0} 
    \eaa

\subsubsection{$Z_2/Z_0$}
    \baa  
    Z_2/Z_0 &= \frac{1}{Z_0}\int_0^{\beta} d\tau_2 \int_0^{\tau_2} d\tau_1 \text{Tr}\bigg[e^{-\beta \hH_0} (e^{\tau_2\hH_0}\hH_1e^{-\tau_2\hH_0})(e^{\tau_1\hH_0}\hH_1 e^{-\tau_1\hH_0} )P\bigg] \nonumber  \\
    &=\frac{1}{Z_0}\int_0^{\beta} d\tau_2 \int_0^{\tau_2} d\tau_1 \text{Tr}\bigg[e^{-\beta \hH_0}\hH_1(\tau_2)\hH_1(\tau_1)P\bigg] \nonumber \\
    &=\int_0^{\beta} d\tau_2 \int_0^{\tau_2} d\tau_1
    \langle \hH_1(\tau_2)\hH_1(\tau_1)\rangle_0  
    \eaa  
    Using the explicit formula of $\hH_1$\hh{ (Eq.~\ref{eq:j_h0_h1})}, we have 
    \baa  
     Z_2/Z_0 
     =& J^2 \sum_{\RR,\RR_2,\qq,\qq_2,\mu\nu,\mu_2\nu_2}\sum_{\xi=\pm,\xi_2=\pm} \theta_{\mu\nu}^{(f,\xi)}(\RR) \theta_{\mu_2\nu_2}^{(f,\xi_2)}(\RR_2) e^{-i\qq \cdot \RR -i \qq_2 \cdot \RR_2} 
     \int_0^\beta d\tau_2 \int_0^{\tau_2} d\tau_1
     \langle 
     :\UC_{\mu\nu}^{c'',\xi}(\qq,\tau_2): :\UC_{\mu_2\nu_2}^{c'',\xi_2}(\qq_2,\tau_1) :
     \rangle_0 
     \label{eq:hj_z2_z0}
     \eaa   
     We note that 
     \baa 
     \langle:\UC_{\mu\nu}^{c'',\xi}(\qq):: \UC_{\mu_2\nu_2}^{c'',\xi_2}(\qq_2):\rangle_0\propto \delta_{\mu,\mu_2}\delta_{\nu,\nu_2}
     \label{eq:u4_moment_munu_diag}
     \eaa
    \hb{ To prove Eq.~\ref{eq:u4_moment_munu_diag}, we first define
     \baa 
     \hat{N}_{\mu\nu,\mu_2\nu_2} = :\UC_{\mu\nu}^{c'',\xi}(\qq):: \UC_{\mu_2\nu_2}^{c'',\xi_2}(\qq_2):\, .
     \eaa 
     We now show, for $\mu\nu \ne \mu_2\nu_2$, there exists an chiral-$U(4)$ transformation $g$, such that $g\hat{N}_{\mu\nu,\mu_2\nu_2}g^{-1} = -\hat{N}_{\mu\nu,\mu_2\nu_2}$, Then $\langle\hat{N}_{\mu\nu,\mu_2\nu_2}\rangle_0 = - \langle\hat{N}_{\mu\nu,\mu_2\nu_2}\rangle_0=0$. We consider different cases of $\mu\nu,\mu_2\nu_2$, and show that in the following cases $\langle \hat{N}_{\mu\nu,\mu_2\nu_2}\rangle_0=0$ if $\mu\nu \ne \mu_2\nu_2$.}
     \begin{itemize}
         \item $\mu \nu = 00, \mu_2\nu_2 \ne 00$ or  $\mu\nu\ne 00, \mu_2 \nu_2 =00$.  We take the case of $\mu\nu=00,\mu_2\nu_2\ne 00$ as an example. $:\hat{O}_{\mu\nu=(00)}^{c'' ,\xi}(\qq_2):$ is invariant under all the chiral U(4) transformation. However, $:\UC_{\mu_2\nu_2}^{c'',\xi_2}(\qq_2,\tau_1) :$ would change sign under certain chiral U(4) transformation (as shown in \hh{Eq.\ref{eq:u4_flip_sign}}). Therefore, the expectation value of $\hat{N}_{\mu\nu,\mu_2\nu_2}=:\UC_{\mu\nu}^{c'',\xi}(\qq):: \UC_{\mu_2\nu_2}^{c'',\xi_2}(\qq_2):$ goes to zero.
         
         \item $\mu \ne \mu_2$ and $\mu\ne 0, \mu_2\ne 0 $. \bh{For $g=e^{-i\pi \UC_{\mu0}}$, we have $\langle \hat{N}_{\mu\nu,\mu_2\nu_2}\rangle_0 
         = \langle g\hat{N}_{\mu\nu,\mu_2\nu_2}g^{-1}\rangle_0 
         = - \langle\hat{N}_{\mu\nu,\mu_2\nu_2}\rangle_0=0$.}

         \item $\mu \ne \mu_2$ and $\mu= 0, \mu_2\ne 0 $. 
         F\bh{or $g=e^{-i\pi \UC_{\mu_30}}$ with $\mu_3 \ne 0,\mu_3\ne \mu_2$, we find $\langle \hat{N}_{\mu\nu,\mu_2\nu_2}\rangle_0 
         = \langle g\hat{N}_{\mu\nu,\mu_2\nu_2}g^{-1}\rangle_0 
         = - \langle\hat{N}_{\mu\nu,\mu_2\nu_2}\rangle_0=0$.}

        \item $\mu \ne \mu_2$ and $\mu\ne 0, \mu_2= 0 $. 
        \bh{For $g=e^{-i\pi \UC_{\mu_30}}$ with $\mu_3 \ne 0,\mu_3\ne \mu$, we find $\langle \hat{N}_{\mu\nu,\mu_2\nu_2}\rangle_0 
         = \langle g\hat{N}_{\mu\nu,\mu_2\nu_2}g^{-1}\rangle_0 
         = - \langle\hat{N}_{\mu\nu,\mu_2\nu_2}\rangle_0=0$.}

          \item $\nu\ne \nu_2$ and $\nu \ne0$, $\nu_2 \ne 0$.
          \bh{For $g=e^{-i\pi \UC_{0\nu}}$, we $\langle \hat{N}_{\mu\nu,\mu_2\nu_2}\rangle_0 
         = \langle g\hat{N}_{\mu\nu,\mu_2\nu_2}g^{-1}\rangle_0 
         = - \langle\hat{N}_{\mu\nu,\mu_2\nu_2}\rangle_0=0$.}

         \item $\nu\ne \nu_2$ and $\nu =0$, $\nu_2 \ne 0$.
         \bh{ For $g=e^{-i\pi \UC_{0\nu_3 }}$ with $\nu_3 \ne 0,\nu_3\ne \nu_2$, we find $\langle \hat{N}_{\mu\nu,\mu_2\nu_2}\rangle_0 
         = \langle g\hat{N}_{\mu\nu,\mu_2\nu_2}g^{-1}\rangle_0 
         = - \langle\hat{N}_{\mu\nu,\mu_2\nu_2}\rangle_0=0$.}

         \item $\nu\ne \nu_2$ and $\nu \ne 0$, $\nu_2 = 0$.
          \bh{For $g=e^{-i\pi \UC_{0\nu_3 }}$ with $\nu_3 \ne 0,\nu_3\ne \nu$, we find $\langle \hat{N}_{\mu\nu,\mu_2\nu_2}\rangle_0 
         = \langle g\hat{N}_{\mu\nu,\mu_2\nu_2}g^{-1}\rangle_0 
         = - \langle\hat{N}_{\mu\nu,\mu_2\nu_2}\rangle_0=0$.}
     \end{itemize}
     
     Therefore, the only nonvanishing term comes from the case with $\mu\nu = \mu_2\nu_2$. Then we can rewrite \bh{Eq.~\ref{eq:hj_z2_z0}} as 
     \ba 
     Z_2/Z_0 =& J^2 \sum_{\RR,\RR_2,\mu\nu}\sum_{\xi=\pm,\xi_2=\pm} \theta_{\mu\nu}^{(f,\xi)}(\RR) \theta_{\mu\nu}^{(f,\xi_2)}(\RR_2) e^{-i\qq \cdot \RR -i \qq_2 \cdot \RR_2} 
     \int_0^\beta d\tau_1 \int_0^{\tau_1} d\tau_2
     \langle 
     :\UC_{\mu\nu}^{c'',\xi}(\qq,\tau_1): :\UC_{\mu\nu}^{c'',\xi_2}(\qq_2,\tau_2) :
     \rangle 
     \ea  
     One can further show the expectation value for all the $\mu\nu$ with $\mu\nu\ne 00$ are the same. This is because, for a chiral U(4) transformation $g$ and operator $\hat{O}$, we have 
     \ba 
     \langle g^{-1}\hat{O} g\rangle_0 = \langle \hat{O} \rangle_0
     \ea
    (note that $\hH_0$ has chiral U(4) symmetry). We let $\hat{O}^{\mu\nu}_{\xi,\xi_2,\qq,\qq_2,\tau_1,\tau_2}:=:\UC_{\mu\nu}^{c'',\xi}(\qq,\tau_1): :\UC_{\mu\nu}^{c'',\xi_2}(\qq_2,\tau_2) :$. We then show $\hat{O}^{\mu\nu}_{\xi,\xi_2,\qq,\qq_2,\tau_1,\tau_2}$ with different $\mu\nu$ index are connected by chiral U(4) symmetry (except for $\mu\nu=00$). \hh{This is because $\{\hat{\Theta}^{c'',\xi}_{\mu\nu}(\qq,\tau)\}_{\mu\nu} \ne 00$ form a irreducible representation of chiral $SU(4)$ group. We can find a chiral $SU(4)$ transformation $g$, such that $g:\hat{\Theta}^{c'',\xi}_{\mu\nu}(\qq,\tau)\rightarrow  \hat{\Theta}^{c'',\xi}_{0z}(\qq,\tau)$ and then 
    \baa  
    g:\hat{O}^{\mu\nu}_{\xi,\xi_2,\qq,\qq_2} \rightarrow \hat{O}^{0z}_{\xi,\xi_2,\qq,\qq_2}
    \label{eq:g_o_0z}
    \eaa , where we have pick $\mu\nu=0z$ as a representative.}
    To show it we first introduce the following chiral transformations:
    \baa  
    & e^{i\pi/2 \UC_{y0} }:\hat{O}^{x\mu }_{\xi,\xi_2,\qq,\qq_2} \rightarrow \hat{O}^{z\mu}_{\xi,\xi_2,\qq,\qq_2} \quad,\quad
    e^{-i\pi/2 \UC_{x0}} :\hat{O}^{y\mu }_{\xi,\xi_2,\qq,\qq_2} \rightarrow \hat{O}^{z\mu }_{\xi,\xi_2,\qq,\qq_2} 
    \label{eq:chiral_1}
    \\
    & e^{i\pi/2 \UC_{0y} }:\hat{O}^{\mu x }_{\xi,\xi_2,\qq,\qq_2} \rightarrow \hat{O}^{\mu z}_{\xi,\xi_2,\qq,\qq_2} \quad,\quad
    e^{-i\pi/2 \UC_{0x}} :\hat{O}^{\mu y }_{\xi,\xi_2,\qq,\qq_2} \rightarrow \hat{O}^{\mu z }_{\xi,\xi_2,\qq,\qq_2} 
    \label{eq:chiral_2}
    \\
    & e^{-i\pi\UC_{xx}/2 + i\pi \UC_{yy}/2 }: \hat{O}^{0z}_{\xi,\xi_2,\qq,\qq_2} \rightarrow \hat{O}^{z0}_{\xi,\xi_2,\qq,\qq_2} 
    \quad,\quad e^{-i\pi /2\UC_{x0} + i\pi /2\UC_{xz} }: \hat{O}^{zz}_{\xi,\xi_2,\qq,\qq_2} \rightarrow \hat{O}^{z0}_{\xi,\xi_2,\qq,\qq_2}
    \label{eq:chiral_3}
    \eaa 
    Then 
    \hb{
    \baa  
   &\hat{O}^{x\mu }_{\xi,\xi_2,\qq,\qq_2} ,\hat{O}^{y\mu }_{\xi,\xi_2,\qq,\qq_2}  \xrightarrow{\text{Eq.~\ref{eq:chiral_1}}}\hat{O}^{z\mu }_{\xi,\xi_2,\qq,\qq_2}  \xrightarrow{\text{Eq.~\ref{eq:chiral_2}}}\hat{O}^{zz}_{\xi,\xi_2,\qq,\qq_2}\text{ or }\hat{O}^{z0}_{\xi,\xi_2,\qq,\qq_2} \xrightarrow{\text{Eq.~\ref{eq:chiral_3}}} \hat{O}^{z0}_{\xi,\xi_2,\qq,\qq_2} \nonumber \\ 
   &\hat{O}^{\mu x }_{\xi,\xi_2,\qq,\qq_2} ,\hat{O}^{\mu  y}_{\xi,\xi_2,\qq,\qq_2}  \xrightarrow{\text{Eq.~\ref{eq:chiral_2}}}\hat{O}^{\mu z }_{\xi,\xi_2,\qq,\qq_2}  \xrightarrow{\text{Eq.~\ref{eq:chiral_1}}}\hat{O}^{zz}_{\xi,\xi_2,\qq,\qq_2}\text{ or }\hat{O}^{0z}_{\xi,\xi_2,\qq,\qq_2} \xrightarrow{\text{Eq.~\ref{eq:chiral_3}}} \hat{O}^{z0}_{\xi,\xi_2,\qq,\qq_2} \nonumber \\ 
   & \hat{O}^{0z}_{\xi,\xi_2,\qq,\qq_2} ,\hat{O}^{zz}_{\xi,\xi_2,\qq,\qq_2} \xrightarrow{\text{Eq.~\ref{eq:chiral_3}}} \hat{O}^{z0}_{\xi,\xi_2,\qq,\qq_2} 
    \eaa  
    We thus prove Eq.~\ref{eq:g_o_0z}}.
     Therefore 
     \baa 
     \langle  :\UC_{\mu\nu}^{c'',\xi}(\qq,\tau_1): :\UC_{\mu\nu}^{c'',\xi_2}(\qq_2,\tau_2) : \rangle_0  = \langle \ :\UC_{0z}^{c'',\xi}(\qq,\tau_1): :\UC_{0z}^{c'',\xi_2}(\qq_2,\tau_2) : \rangle_0 ,\quad \text{whenever } \mu\nu \ne 00
     \label{eq:uc_correlator_0z}
     \eaa 
     (Here we pick $\mu\nu = 0z$ as a representative.)

     Now the expression becomes 
     \baa  
       Z_2/Z_0 =& J^2 \sum_{\RR,\RR_2}\sum_{\qq,\qq_2}\sum_{\mu\nu \ne 00}\sum_{\xi=\pm,\xi_2=\pm} \theta_{\mu\nu}^{(f,\xi)}(\RR) \theta_{\mu\nu}^{(f,\xi_2)}(\RR_2) e^{-i\qq \cdot \RR -i \qq_2 \cdot \RR_2} 
     \int_0^\beta d\tau_1 \int_0^{\tau_1} d\tau_2
     \langle 
     :\UC_{0z}^{c'',\xi}(\qq,\tau_1): :\UC_{0z}^{c'',\xi_2}(\qq_2,\tau_2) :\rangle_0 \nonumber 
     \\
     &+ J^2 \sum_{\RR,\RR_2}\sum_{\qq,\qq_2}\sum_{\xi=\pm,\xi_2=\pm} \theta_{00}^{(f,\xi)}(\RR) \theta_{00}^{(f,\xi_2)}(\RR_2) e^{-i\qq \cdot \RR -i \qq_2 \cdot \RR_2} 
     \int_0^\beta d\tau_1 \int_0^{\tau_1} d\tau_2
     \langle 
     :\UC_{00}^{c'',\xi}(\qq,\tau_1): :\UC_{00}^{c'',\xi_2}(\qq_2,\tau_2) :
     \rangle_0 
     \eaa  
     
     Since $\hat{H}_0$ is also translational invariant, the moment conservation requires $\qq = -\qq_2$. This leads to
      \baa  
       Z_2/Z_0 =& J^2 \sum_{\RR,\RR_2}\sum_{\qq}\sum_{\mu\nu \ne 00}\sum_{\xi=\pm,\xi_2=\pm} \theta_{\mu\nu}^{(f,\xi)}(\RR) \theta_{\mu\nu}^{(f,\xi_2)}(\RR_2) e^{-i\qq \cdot \RR+i \qq \cdot \RR_2} 
     \int_0^\beta d\tau_1 \int_0^{\tau_1} d\tau_2
     \langle 
     :\UC_{0z}^{c'',\xi}(\qq,\tau_1): :\UC_{0z}^{c'',\xi_2}(-\qq,\tau_2) :\rangle_0 \nonumber 
     \\
     &+ J^2 \sum_{\RR,\RR_2}\sum_{\qq}\sum_{\xi=\pm,\xi_2=\pm} \theta_{00}^{(f,\xi)}(\RR) \theta_{00}^{(f,\xi_2)}(\RR_2) e^{-i\qq \cdot \RR +i \qq \cdot \RR_2} 
     \int_0^\beta d\tau_1 \int_0^{\tau_1} d\tau_2
     \langle 
     :\UC_{00}^{c'',\xi}(\qq,\tau_1): :\UC_{00}^{c'',\xi_2}(-\qq,\tau_2) :
     \rangle_0 
     \label{eq:z2_z0_j_zero_hyb}
     \eaa  

     Then we only need to evaluate the following four-fermion correlation functions:
     \ba 
     \langle 
     :\UC_{00}^{c'',\xi}(\qq,\tau_1): :\UC_{00}^{c'',\xi_2}(-\qq,\tau_2) :
     \rangle_0  \\
      \langle 
     :\UC_{0z}^{c'',\xi}(\qq,\tau_1): :\UC_{0z}^{c'',\xi_2}(-\qq,\tau_2) :\rangle_0
     \ea 
     where 
     \ba 
       :\UC_{00}^{c'',\xi}(\qq,\tau_1):
        =\sum_{\kk,}\sum_{a_1 =3,4}\sum_{\eta s } \frac{1}{2N_M} \delta_{\xi, (-1)^{a-1}\eta}  : c^\dag_{a\eta s,\kk+\qq}(\tau_1) c_{a\eta s,\kk}(\tau_1):
        \\
           :\UC_{0z}^{c'',\xi}(\qq,\tau_1):
        =\sum_{\kk,}\sum_{a_1 =3,4}\sum_{\eta s }s \frac{1}{2N_M} \delta_{\xi, (-1)^{a-1}\eta} : c^\dag_{a\eta s,\kk+\qq}(\tau_1) c_{a\eta s,\kk}(\tau_1):
     \ea 
      one is the U(1) charge, the other one is the spin moment along $z$ direction of $a=3,4$ orbital with index $\xi$, momentum $\qq$.
     
     We will evaluate the four-fermion correlation functions using Wick's theorem. To do so we first introduce single-particle Green's function 
       \ba 
        G_{ aa',\eta\eta',ss'}(\kk,\tau_1-\tau_2) = -\langle T_\tau c_{\kk a\eta s}(\tau_1) c^\dag_{\kk a'\eta' s'}(\tau_2)\rangle_0
        \ea 
        where $T_\tau$ denotes time-ordering. We now explore the structure of the single-particle Green's function.
         \hh{The single-particle Green's functions are calculated with respect to the ground state of Hamiltonian $\hH_0 = \hH_c$ \bh{(We also calculate the single-particle Green's function with respect to the symmetry broken state in Sec.~\ref{sec:eff_thy} and Sec.~\ref{sec:green_order})}. The ground state of non-interacting conduction-electron Hamiltonian $\hH_c$ (Eq.~\ref{eq:hc_def}) always has chiral-$U(4)$ symmetry, and thus has $SU(2)$ spin and $U(1)$ valley symmetries. The corresponding single-particle Green's function will also have $SU(2)$ spin and $U(1)$ valley symmetries.}
        Due to the $SU(2)$ spin symmetry, $G_{\kk,aa',\eta\eta',ss'}(\tau_1-\tau_2) \propto \delta_{ss'}$ and $G_{\kk,aa',\eta\eta',\up\up}(\tau_1-\tau_2) = G_{\kk,aa',\eta\eta',\dn\dn}(\tau_1-\tau_2)$ 
        Due to the $U(1)$ valley symmetry, $G_{\kk,aa',\eta\eta',ss'} \propto \delta_{\eta\eta'}$.

        This leads to a simplified notation of \bh{the} Green's function:
        \ba 
       &  G_{ aa',\eta\eta',ss'}(\kk,\tau_1-\tau_2)  = \delta_{\eta\eta'}\delta_{ss'} G_{aa',\eta}(\kk,\tau_1-\tau_2) \\
      &  G_{aa',\eta}(\kk,\tau_1-\tau_2) = -\langle T_\tau c_{\kk a\eta s}(\tau_1) c^\dag_{\kk a'\eta s}(\tau_2)\rangle_0.
        \ea 
        The analytical formula of \bh{the} Green's function at zero temperature and infinity momentum cutoff is \hh{given in Sec. \ref{sec:green}}.
    
    Now we are in the position to calculate two correlation functions. 
    The first correlation function
     \baa  
     &
     \langle 
     :\UC_{00}^{c'',\xi}(\qq,\tau_1): :\UC_{00}^{c'',\xi_2}(-\qq,\tau_2) :
     \rangle_0   \nonumber \\
     =& \sum_{\kk,\kk'}\sum_{a_1,a_2 =3,4}\sum_{\eta\eta_2 s s_2 } \frac{1}{2N_M}\frac{1}{2N_M} \delta_{\xi, (-1)^{a-1}\eta} \delta_{\xi_2,(-1)^{a_2-1}\eta_2} \langle : c^\dag_{a\eta s,\kk+\qq}(\tau_1) c_{a\eta s,\kk}(\tau_1): :c^\dag_{a_2\eta_2 s_2 \kk'-\qq }(\tau_2) c_{a_2\eta_2s_2 \kk'}(\tau_2):\rangle_0  \nonumber \\
     &\text{[Use \bh{Wick's} theorem]}  \nonumber \\
     =&
     \sum_{\kk,\kk'}\sum_{a_1,a_2 =3,4}\sum_{\eta\eta_2 s s_2 } \frac{1}{2N_M}\frac{1}{2N_M} \delta_{\xi, (-1)^{a-1}\eta} \delta_{\xi_2,(-1)^{a_2-1}\eta_2} \langle : c^\dag_{a\eta s,\kk+\qq}(\tau_1) c_{a\eta s,\kk}(\tau_1):\rangle_0\langle  :c^\dag_{a_2\eta_2 s_2 \kk'-\qq }(\tau_2) c_{a_2\eta_2s_2 \kk'}(\tau_2):\rangle_0  \nonumber  \\
     -&\sum_{\kk,\kk'}\sum_{a_1,a_2 =3,4}\sum_{\eta\eta_2 s s_2 } \frac{1}{2N_M}\frac{1}{2N_M} \delta_{\xi, (-1)^{a-1}\eta} \delta_{\xi_2,(-1)^{a_2-1}\eta_2}
     \langle c_{a_2\eta_2s_2\kk'}(\tau_2)c_{a\eta s,\kk+\qq}^\dag(\tau_1) \rangle_0 \langle c_{a\eta s \kk}(\tau_1) c^\dag_{a_2\eta_2s_2\kk'-\qq}(\tau_2) \rangle_0  \nonumber \\
     &\text{[Rewrite 1st term]}  \nonumber \\
     =&
    \langle 
     :\UC_{00}^{c'',\xi}(\qq,\tau_1): \rangle_0 \langle  :\UC_{00}^{c'',\xi_2}(-\qq,\tau_2) :
     \rangle_0 
      \nonumber \\ 
     -&\sum_{\kk,\kk'}\sum_{a_1,a_2 =3,4}\sum_{\eta\eta_2 s s_2 } \frac{1}{2N_M}\frac{1}{2N_M} \delta_{\xi, (-1)^{a-1}\eta} \delta_{\xi_2,(-1)^{a_2-1}\eta_2}
     \langle c_{a_2\eta_2s_2\kk'}(\tau_2)c_{a\eta s,\kk+\qq}^\dag(\tau_1) \rangle_0 \langle c_{a\eta s \kk}(\tau_1) c^\dag_{a_2\eta_2s_2\kk'-\qq}(\tau_2) \rangle_0 \nonumber\\
      &\text{[Rewrite second term via single-particle Green's function]}\nonumber\\
     =&\langle 
     :\UC_{00}^{c'',\xi}(\qq,\tau_1): \rangle_0 \langle  :\UC_{00}^{c'',\xi_2}(-\qq,\tau_2) :
     \rangle_0 
      \nonumber \\ 
     -&\sum_{\kk,\kk'}\sum_{a_1,a_2 =3,4}\sum_{\eta\eta_2 s s_2 } \frac{1}{2N_M}\frac{1}{2N_M} \delta_{\xi, (-1)^{a-1}\eta} \delta_{\xi_2,(-1)^{a_2-1}\eta}
     G_{a_2a, \eta}(\kk+\qq,\tau_2-\tau_1)G_{aa_2,\eta}(\kk, \tau_1-\tau_2) \delta_{\eta_2,\eta} \delta_{s,s_2}\delta_{\kk',\kk+\qq} \nonumber  \\
     &\text{[Further simplification]}  \nonumber \\
     =&\langle 
     :\UC_{00}^{c'',\xi}(\qq,\tau_1): \rangle_0 \langle  :\UC_{00}^{c'',\xi_2}(-\qq,\tau_2) :
     \rangle_0 
     \nonumber  \\ 
     -&\sum_{\kk}\sum_{a_1,a_2 =3,4}\sum_{\eta s  } \frac{1}{2N_M}\frac{1}{2N_M} \delta_{\xi, (-1)^{a-1}\eta} \delta_{\xi_2,(-1)^{a_2-1}\eta}
     G_{a_2a,\eta}(\kk+\qq,\tau_2-\tau_1)G_{aa_2,\eta }(\kk, \tau_1-\tau_2)  
     \label{eq:chiral_u4_corre_1}
     \eaa

     The second correlation functions
     \baa 
     &
     \langle 
     :\UC_{0z}^{c'',\xi}(\qq,\tau_1): :\UC_{0z}^{c'',\xi_2}(-\qq,\tau_2) :
     \rangle_0 \nonumber   \\
     =& \sum_{\kk,\kk'}\sum_{a_1,a_2 =3,4}\sum_{\eta\eta_2 s s_2 } \frac{1}{2N_M}\frac{1}{2N_M} \delta_{\xi, (-1)^{a-1}\eta} \delta_{\xi_2,(-1)^{a_2-1}\eta_2}ss_2 \langle : c^\dag_{a\eta s,\kk+\qq}(\tau_1) c_{a\eta s,\kk}(\tau_1): :c^\dag_{a_2\eta_2 s_2 \kk'-\qq }(\tau_2) c_{a_2\eta_2s_2 \kk'}(\tau_2):\rangle_0  \nonumber \\
     &\text{[Use \bh{Wick's} theorem]}\nonumber  \\
     =&
     \sum_{\kk,\kk'}\sum_{a_1,a_2 =3,4}\sum_{\eta\eta_2 s s_2 } \frac{1}{2N_M}\frac{1}{2N_M}ss_2 \delta_{\xi, (-1)^{a-1}\eta} \delta_{\xi_2,(-1)^{a_2-1}\eta_2} \langle : c^\dag_{a\eta s,\kk+\qq}(\tau_1) c_{a\eta s,\kk}(\tau_1):\rangle_0\langle  :c^\dag_{a_2\eta_2 s_2 \kk'-\qq }(\tau_2) c_{a_2\eta_2s_2 \kk'}(\tau_2):\rangle_0 \nonumber  \\
     -&\sum_{\kk,\kk'}\sum_{a_1,a_2 =3,4}\sum_{\eta\eta_2 s s_2 } \frac{1}{2N_M}\frac{1}{2N_M}ss_2 \delta_{\xi, (-1)^{a-1}\eta} \delta_{\xi_2,(-1)^{a_2-1}\eta_2}
     \langle c_{a_2\eta_2s_2\kk'}(\tau_2)c_{a\eta s,\kk+\qq}^\dag(\tau_1) \rangle_0 \langle c_{a\eta s \kk}(\tau_1) c^\dag_{a_2\eta_2s_2\kk'-\qq}(\tau_2) \rangle_0  \nonumber \\
     &\text{[Rewrite 1st term]} \nonumber \\
     =&
    \langle 
     :\UC_{0z}^{c'',\xi}(\qq,\tau_1): \rangle_0 \langle  :\UC_{0z}^{c'',\xi_2}(-\qq,\tau_2) :
     \rangle_0 
     \nonumber  \nonumber \\ 
     -&\sum_{\kk,\kk'}\sum_{a_1,a_2 =3,4}\sum_{\eta\eta_2 s s_2 } \frac{1}{2N_M}\frac{1}{2N_M}ss_2 \delta_{\xi, (-1)^{a-1}\eta} \delta_{\xi_2,(-1)^{a_2-1}\eta_2}
     \langle c_{a_2\eta_2s_2\kk'}(\tau_2)c_{a\eta s,\kk+\qq}^\dag(\tau_1) \rangle_0 \langle c_{a\eta s \kk}(\tau_1) c_{a_2\eta_2s_2\kk'-\qq}^\dag(\tau_2) \rangle_0  \nonumber \\
      &\text{[1st term goes to zero \hh{(as shown in Eq.~\ref{eq:zero_linear_uc})}. Rewrite the second term via single-particle Green's function]} \nonumber \\
     =&-\sum_{\kk,\kk'}\sum_{a_1,a_2 =3,4}\sum_{\eta s  } \frac{1}{2N_M}\frac{1}{2N_M} \delta_{\xi, (-1)^{a-1}\eta} \delta_{\xi_2,(-1)^{a_2-1}\eta}ss_2
     G_{a_2a, \eta}(\kk+\qq,\tau_2-\tau_1)G_{aa_2,\eta }(\kk, \tau_1-\tau_2) \delta_{\eta_2,\eta} \delta_{s,s_2}\delta_{\kk',\kk+\qq}  \nonumber \\
     &\text{[Further simplification]} \nonumber \\
     =&
     -\sum_{\kk}\sum_{a_1,a_2 =3,4}\sum_{\eta s } \frac{1}{2N_M}\frac{1}{2N_M} \delta_{\xi, (-1)^{a-1}\eta} \delta_{\xi_2,(-1)^{a_2-1}\eta}
     G_{a_2a,\eta}(\kk+\qq,\tau_2-\tau_1)G_{aa_2,\eta }(\kk, \tau_1-\tau_2)  
      \label{eq:chiral_u4_corre_2}
     \eaa  
     
     Comparing two correlation functions (\hb{Eq.\ref{eq:chiral_u4_corre_1} and Eq.\ref{eq:chiral_u4_corre_2}}), we find 
     \baa  
     & \langle 
     :\UC_{00}^{c'',\xi}(\qq,\tau_1): :\UC_{00}^{c'',\xi_2}(-\qq,\tau_2) :
     \rangle_0 
     - \langle 
     :\UC_{00}^{c'',\xi}(\qq,\tau_1):\rangle_0 \langle  :\UC_{00}^{c'',\xi_2}(-\qq,\tau_2) :
     \rangle_0 = \langle 
     :\UC_{0z}^{c'',\xi}(\qq,\tau_1): :\UC_{0z}^{c'',\xi_2}(-\qq,\tau_2) :
     \rangle_0 \nonumber  \\
     =-&\sum_{\kk}\sum_{a_1,a_2 =3,4}\sum_{\eta\eta_2 s s_2 } \frac{1}{2N_M}\frac{1}{2N_M} \delta_{\xi, (-1)^{a-1}\eta} \delta_{\xi_2,(-1)^{a_2-1}\eta}
     G_{a_2a,\eta}(\kk+\qq,\tau_2-\tau_1)G_{aa_2,\eta }(\kk, \tau_1-\tau_2)  
     \eaa 
    We \hh{then define} $\chi_c(\qq ,\tau_1-\tau_2)$
    \baa 
    &\chi_c(\qq,\tau_1-\tau_2,\xi,\xi_2) \nonumber  \\
    =& \langle  :\UC_{00}^{c'',\xi}(\qq,\tau_1): :\UC_{00}^{c'',\xi_2}(-\qq,\tau_2) :
     \rangle_0 
     - \langle 
     :\UC_{00}^{c'',\xi}(\qq,\tau_1):\rangle_0 \langle  :\UC_{00}^{c'',\xi_2}(-\qq,\tau_2) :
     \rangle_0 = \langle 
     :\UC_{0z}^{c'',\xi}(\qq,\tau_1): :\UC_{0z}^{c'',\xi_2}(-\qq,\tau_2) :
     \rangle_0 \nonumber  \\
     =&-\sum_{\kk}\sum_{a_1,a_2 =3,4}\sum_{\eta s   } \frac{1}{2}\frac{1}{2N_M} \delta_{\xi, (-1)^{a-1}\eta} \delta_{\xi_2,(-1)^{a_2-1}\eta}
     G_{a_2a,\eta}(\kk+\qq,\tau_2-\tau_1)G_{aa_2,\eta }(\kk, \tau_1-\tau_2)  \label{eq:chic}
    \eaa 
    \bh{such that}
    \bh{ 
    \baa  
    & \langle  :\UC_{00}^{c'',\xi}(\qq,\tau_1): :\UC_{00}^{c'',\xi_2}(-\qq,\tau_2) :
     \rangle_0 = \chi_c(\qq,\tau_1-\tau_2,\xi,\xi_2)+\langle 
     :\UC_{00}^{c'',\xi}(\qq,\tau_1):\rangle_0 \langle  :\UC_{00}^{c'',\xi_2}(-\qq,\tau_2) :
     \rangle_0 \nonumber \\ 
    & \langle 
     :\UC_{0z}^{c'',\xi}(\qq,\tau_1): :\UC_{0z}^{c'',\xi_2}(-\qq,\tau_2) :
     \rangle_0 =  \chi_c(\qq,\tau_1-\tau_2,\xi,\xi_2) 
     \label{eq:theta_in_chi}
    \eaa 
    }
    The Fourier transformation of $\chi_c(\qq,\tau_1-\tau_2)$ is defined as
    \baa 
    \chi_c(\RR,\tau_1-\tau_2,\xi,\xi_2) =\frac{1}{N_M} \sum_{\kk} \chi_c(\qq,\tau_1-\tau_2,\xi,\xi_2)e^{i\qq\cdot \RR } 
    \label{eq:chic_real}
    \eaa 
    \hh{The analytical formulas of $\chi_c(\RR,\tau,\xi,\xi_2)$ \bh{at zero temperature} are derived in Sec.~\ref{sec:chiral_u4_corre}, Eqs.~\ref{eq:chic_ana},~\ref{eq:chic_ana_2}
    }
    \bh{and are given below 
    \baa  
 \chi_c(\RR,\tau,\xi,\xi) \approx & \frac{\pi^2}{A_{MBZ}^2}\frac{|v_\star \tau|^2}{(|v_\star \tau|^2 +r^2)^{3} }  \nonumber \\ 
\chi_c(\RR,\tau,\xi,-\xi) \approx &
-\frac{\pi^2M^2}{4A_{MBZ}^2 |v_\star|^2} \frac{r^4}{\bigg( |v_\star \tau|^2 + r^2\bigg)^3}
\label{eq:chic_app}
    \eaa  
    where we perform an expansion in powers of $M$ and truncate to second orders.
    }

     Combining Eq.~\ref{eq:z2_z0_j_zero_hyb} and Eq.~\ref{eq:theta_in_chi},
     \ba 
     Z_2/Z_0 =& J^2 \sum_{\RR,\RR_2}\sum_{\mu\nu \ne 00 }\sum_{\xi=\pm,\xi_2=\pm} \theta_{\mu\nu}^{(f,\xi)}(\RR) \theta_{\mu\nu}^{(f,\xi_2)}(\RR_2)
     \int_0^\beta d\tau_1 \int_0^{\tau_1} d\tau_2 \chi_c(\RR_2-\RR,\tau_1-\tau_2,\xi,\xi_2) \\
     &+J^2 \sum_{\RR,\RR_2}\sum_{\mu\nu =00  }\sum_{\xi=\pm,\xi_2=\pm} \theta_{\mu\nu}^{(f,\xi)}(\RR) \theta_{\mu\nu}^{(f,\xi_2)}(\RR_2)
     \int_0^\beta d\tau_1 \int_0^{\tau_1} d\tau_2 \chi_c(\RR_2-\RR,\tau_1-\tau_2,\xi,\xi_2) \\
  &  + J^2  \sum_{\RR,\RR_2,\QQ}\sum_{\xi=\pm,\xi_2=\pm} \theta_{00}^{(f,\xi)}(\RR) \theta_{00}^{(f,\xi_2)}(\RR_2) e^{-i\qq \cdot \RR+i \qq \cdot \RR_2}  \int_0^\beta d\tau_1 \int_0^{\tau_1} d\tau_2  \langle 
     :\UC_{00}^{c'',\xi}(\qq,\tau_1):\rangle_0 \langle  :\UC_{00}^{c'',\xi_2}(-\qq,\tau_2) :
     \rangle_0 \\
   =& J^2 \sum_{\RR,\RR_2}\sum_{\mu\nu  }\sum_{\xi=\pm,\xi_2=\pm} \theta_{\mu\nu}^{(f,\xi)}(\RR) \theta_{\mu\nu}^{(f,\xi_2)}(\RR_2) 
     \int_0^\beta d\tau_1 \int_0^{\tau_1} d\tau_2 \chi_c(\RR_2-\RR,\tau_1-\tau_2,\xi,\xi_2) \\
  &  + J^2  \sum_{\RR,\RR_2}\sum_{\xi=\pm,\xi_2=\pm} \theta_{00}^{(f,\xi)}(\RR) \theta_{00}^{(f,\xi_2)}(\RR_2) e^{-i\qq \cdot \RR+i \qq \cdot \RR_2}  \int_0^\beta d\tau_1 \int_0^{\tau_1} d\tau_2  \langle 
     :\UC_{00}^{c'',\xi}(\qq,\tau_1):\rangle_0 \langle  :\UC_{00}^{c'',\xi_2}(-\qq,\tau_2) :
     \rangle_0 \\
 \ea 
 The last term can be written as \hb{
\ba 
& J^2  \sum_{\qq,\RR,\RR_2}\sum_{\xi=\pm,\xi_2=\pm} \theta_{00}^{(f,\xi)}(\RR) \theta_{00}^{(f,\xi_2)}(\RR_2) e^{-i\qq \cdot \RR +i \qq \cdot \RR_2}   \int_0^\beta d\tau_1 \int_0^{\tau_1} d\tau_2 \langle 
     :\UC_{00}^{c'',\xi}(\qq,\tau_1):\rangle_0 \langle  :\UC_{00}^{c'',\xi_2}(-\qq,\tau_2) :
     \rangle_0 \\ 
=&J^2  \frac{1}{2}\sum_{\qq,\RR,\RR_2}\sum_{\xi=\pm,\xi_2=\pm} \theta_{00}^{(f,\xi)}(\RR) \theta_{00}^{(f,\xi_2)}(\RR_2) e^{-i\qq \cdot \RR +i \qq  \cdot \RR_2}   \int_0^\beta d\tau_1 \int_0^{\tau_1} d\tau_2 \langle 
     :\UC_{00}^{c'',\xi}(\qq,\tau_1):\rangle_0 \langle  :\UC_{00}^{c'',\xi_2}(-\qq,\tau_2) :
     \rangle_0 \\ 
&+J^2  \frac{1}{2}\sum_{\qq,\RR,\RR_2}\sum_{\xi=\pm,\xi_2=\pm} \theta_{00}^{(f,\xi)}(\RR) \theta_{00}^{(f,\xi_2)}(\RR_2) e^{-i\qq \cdot \RR +i \qq  \cdot \RR_2}   \int_0^\beta d\tau_1 \int_0^{\tau_1} d\tau_2 \langle 
     :\UC_{00}^{c'',\xi}(\qq,\tau_1):\rangle_0 \langle  :\UC_{00}^{c'',\xi_2}(-\qq,\tau_2) :
     \rangle_0 \\  
=&J^2  \frac{1}{2}\sum_{\qq,\RR,\RR_2}\sum_{\xi=\pm,\xi_2=\pm} \theta_{00}^{(f,\xi)}(\RR) \theta_{00}^{(f,\xi_2)}(\RR_2) e^{-i\qq \cdot \RR +i \qq  \cdot \RR_2}   \int_0^\beta d\tau_1 \int_0^{\tau_1} d\tau_2 \langle 
     :\UC_{00}^{c'',\xi}(\qq,\tau_1):\rangle_0 \langle  :\UC_{00}^{c'',\xi_2}(-\qq,\tau_2) :
     \rangle_0 \\ 
&+J^2  \frac{1}{2}\sum_{\qq,\RR,\RR_2}\sum_{\xi=\pm,\xi_2=\pm} \theta_{00}^{(f,\xi)}(\RR) \theta_{00}^{(f,\xi_2)}(\RR_2) e^{-i\qq \cdot \RR +i \qq  \cdot \RR_2}   \int_{\tau_2}^\beta d\tau_1 \int_0^{\beta} d\tau_2 \langle 
     :\UC_{00}^{c'',\xi}(\qq,\tau_1):\rangle_0 \langle  :\UC_{00}^{c'',\xi_2}(-\qq,\tau_2) :
     \rangle_0 \\  
=&J^2  \frac{1}{2}\sum_{\qq,\RR,\RR_2}\sum_{\xi=\pm,\xi_2=\pm} \theta_{00}^{(f,\xi)}(\RR) \theta_{00}^{(f,\xi_2)}(\RR_2) e^{-i\qq \cdot \RR +i \qq  \cdot \RR_2}   \int_0^\beta d\tau_1 \int_0^{\tau_1} d\tau_2 \langle 
     :\UC_{00}^{c'',\xi}(\qq,\tau_1):\rangle_0 \langle  :\UC_{00}^{c'',\xi_2}(-\qq,\tau_2) :
     \rangle_0 \\ 
&+J^2  \frac{1}{2}\sum_{\qq,\RR,\RR_2}\sum_{\xi=\pm,\xi_2=\pm} \theta_{00}^{(f,\xi)}(\RR) \theta_{00}^{(f,\xi_2)}(\RR_2) e^{-i\qq \cdot \RR +i \qq  \cdot \RR_2}   \int_{\tau_1}^\beta d\tau_2 \int_0^{\beta} d\tau_1 \langle 
     :\UC_{00}^{c'',\xi}(\qq,\tau_2):\rangle_0 \langle  :\UC_{00}^{c'',\xi_2}(-\qq,\tau_1) :
     \rangle_0 \\  
=&J^2  \frac{1}{2}\sum_{\qq,\RR,\RR_2}\sum_{\xi=\pm,\xi_2=\pm} \theta_{00}^{(f,\xi)}(\RR) \theta_{00}^{(f,\xi_2)}(\RR_2) e^{-i\qq \cdot \RR +i \qq  \cdot \RR_2}   \int_0^\beta d\tau_1 \int_0^{\tau_1} d\tau_2 \langle 
     :\UC_{00}^{c'',\xi}(\qq,\tau_1):\rangle_0 \langle  :\UC_{00}^{c'',\xi_2}(-\qq,\tau_2) :
     \rangle_0 \\ 
&+J^2  \frac{1}{2}\sum_{\qq,\RR,\RR_2}\sum_{\xi=\pm,\xi_2=\pm} \theta_{00}^{(f,\xi)}(\RR) \theta_{00}^{(f,\xi_2)}(\RR_2) e^{+i\qq \cdot \RR -i \qq  \cdot \RR_2}   \int_{\tau_1}^\beta d\tau_2 \int_0^{\beta} d\tau_1 \langle 
     :\UC_{00}^{c'',\xi}(-\qq,\tau_2):\rangle_0 \langle  :\UC_{00}^{c'',\xi_2}(\qq,\tau_1) :
     \rangle_0 \\
     =&J^2  \frac{1}{2}\sum_{\qq,\RR,\RR_2}\sum_{\xi=\pm,\xi_2=\pm} \theta_{00}^{(f,\xi)}(\RR) \theta_{00}^{(f,\xi_2)}(\RR_2) e^{-i\qq \cdot \RR +i \qq  \cdot \RR_2}   \int_0^\beta d\tau_1 \int_0^{\tau_1} d\tau_2 \langle 
     :\UC_{00}^{c'',\xi}(\qq,\tau_1):\rangle_0 \langle  :\UC_{00}^{c'',\xi_2}(-\qq,\tau_2) :
     \rangle_0 \\ 
&+J^2  \frac{1}{2}\sum_{\qq,\RR,\RR_2}\sum_{\xi=\pm,\xi_2=\pm} \theta_{00}^{(f,\xi)}(\RR) \theta_{00}^{(f,\xi_2)}(\RR_2) e^{-i\qq \cdot \RR +i \qq  \cdot \RR_2}   \int_{\tau_1}^\beta d\tau_2 \int_0^{\beta} d\tau_1 \langle 
     :\UC_{00}^{c'',\xi}(-\qq,\tau_2):\rangle_0 \langle  :\UC_{00}^{c'',\xi_2}(\qq,\tau_1) :
     \rangle_0 \\
=&J^2  \sum_{\qq,\RR,\RR_2}\sum_{\xi=\pm,\xi_2=\pm} \theta_{00}^{(f,\xi)}(\RR) \theta_{00}^{(f,\xi_2)}(\RR_2) e^{-i\qq \cdot \RR +i \qq  \cdot \RR_2} \frac{1}{2}  \int_0^\beta d\tau_1 \int_0^{\beta} d\tau_2 \langle 
     :\UC_{00}^{c'',\xi}(\qq,\tau_1):\rangle_0 \langle  :\UC_{00}^{c'',\xi_2}(-\qq,\tau_2) :
     \rangle_0 \\ 
\\
 =&\frac{1}{2}\Bigg[ \int_0^\beta d\tau 
   \sum_{\qq,\RR}\sum_{\xi=\pm} \theta_{00}^{(f,\xi)}(\RR)e^{-i\qq \cdot \RR }  \langle 
     :\UC_{00}^{c'',\xi}(\qq,\tau_1):\rangle_0 
 \Bigg] ^2 =\frac{1}{2} \bigg(\frac{Z_1}{Z_0}\bigg)^2 
\ea 
}
Therefore
\baa 
Z_2/Z_0 = J^2 \sum_{\RR,\RR_2}\sum_{\mu\nu  }\sum_{\xi=\pm,\xi_2=\pm} \theta_{\mu\nu}^{(f,\xi)}(\RR) \theta_{\mu\nu}^{(f,\xi_2)}(\RR_2) 
     \int_0^\beta d\tau_1 \int_0^{\tau_1} d\tau_2 \chi_c(\RR_2-\RR,\tau_1-\tau_2,\xi,\xi_2) +\frac{1}{2}(\frac{Z_1}{Z_0})^2
\label{eq:z2z0}
\eaa

\subsubsection{Free energy: $F$}
In summary 
\ba 
&Z_0 = \text{Tr}[e^{-\beta H_0}P] \\
&Z_1/Z_0 =J\beta \sum_{\RR}\sum_{\xi=\pm} \frac{\theta_{00}^{f,\xi}(\RR)}{2N_M}\sum_{\kk,\eta s}\sum_{a=3,4}\delta_{\xi,(-1)^{a-1}\eta}\langle : 
    c_{\kk a\eta s}^\dag c_{\kk a\eta s}:
    \rangle_0 \\
& Z_2/Z_0 = J^2 \sum_{\qq,\RR,\RR_2}\sum_{\mu\nu  }\sum_{\xi=\pm,\xi_2=\pm} \theta_{\mu\nu}^{(f,\xi)}(\RR) \theta_{\mu\nu}^{(f,\xi_2)}(\RR_2) e^{-i\qq \cdot \RR +i \qq \cdot \RR_2} 
     \int_0^\beta d\tau_1 \int_0^{\tau_1} d\tau_2 \chi_c(\qq,\tau_1-\tau_2,\xi,\xi_2) +\frac{1}{2}(\frac{Z_1}{Z_0})^2
\ea 

\hh{ Combining Eqs.~\ref{eq:free}, ~\ref{eq:z1z0} and~\ref{eq:z2z0}, we have}
\ba 
F = & -\frac{1}{\beta}\log[Z_0] -\frac{1}{\beta} \frac{Z_1}{Z_0} + \frac{1}{2}\frac{1}{\beta}[\frac{Z_1}{Z_0}]^2 - \frac{1}{\beta}\frac{Z_2}{Z_0} +o(J^2)\\
F =&F_0 - J\sum_{\RR}\sum_{\xi=\pm} \frac{\theta_{00}^{f,\xi}(\RR)}{2N_M}\sum_{\kk,\eta s}\sum_{a=3,4}\delta_{\xi,(-1)^{a-1}\eta}\langle : 
    c_{\kk a\eta s}^\dag c_{\kk a\eta s}:
    \rangle_0 +\frac{1}{2\beta}(\frac{Z_1}{Z_0})^2 \\
    &
    -\frac{1}{\beta }  J^2 \sum_{\qq,\RR,\RR_2}\sum_{\mu\nu  }\sum_{\xi=\pm,\xi_2=\pm} \theta_{\mu\nu}^{(f,\xi)}(\RR) \theta_{\mu\nu}^{(f,\xi_2)}(\RR_2) e^{-i\qq \cdot \RR +i \qq \cdot \RR_2}
     \int_0^\beta d\tau_1 \int_0^{\tau_1} d\tau_2 \chi_c(\qq,\tau_1-\tau_2,\xi,\xi_2) - \frac{1}{2\beta}(\frac{Z_1}{Z_0})^2 \\ 
    =&F_0 - J\sum_{\RR}\sum_{\xi=\pm} \frac{\theta_{00}^{f,\xi}(\RR)}{2N_M}\sum_{\kk,\eta s}\sum_{a=3,4}\delta_{\xi,(-1)^{a-1}\eta}\langle : 
    c_{\kk a\eta s}^\dag c_{\kk a\eta s}:
    \rangle_0 \\
    &
    -\frac{1}{\beta }  J^2 \sum_{\qq,\RR,\RR_2}\sum_{\mu\nu  }\sum_{\xi=\pm,\xi_2=\pm} \theta_{\mu\nu}^{(f,\xi)}(\RR) \theta_{\mu\nu}^{(f,\xi_2)}(\RR_2) e^{-i\qq \cdot \RR +i \qq \cdot \RR_2}
     \int_0^\beta d\tau_1 \int_0^{\tau_1} d\tau_2 \chi_c(\qq,\tau_1-\tau_2,\xi,\xi_2)
\ea 
 where $F_0= -\frac{1}{\beta}\log[Z_0]$ denotes the free energy of the conduction electron system at $J=0$ (with Hamiltonian $\hH_0=\hH_c$). Finally we set $\beta = \infty$. The energy is 
 \ba 
 &\langle \Psi[\vartheta,\nu_c]|\hat{H}_c''[\vartheta]|\Psi[\vartheta,\nu_c]\rangle\\
 =&\lim_{\beta\rightarrow \infty}F\\
 =& E_0 - J\sum_{\RR}\sum_{\xi=\pm} \frac{\theta_{00}^{f,\xi}(\RR)}{2N_M}\sum_{\kk,\eta s}\sum_{a=3,4}\delta_{\xi,(-1)^{a-1}\eta}\langle : 
    c_{\kk a\eta s}^\dag c_{\kk a\eta s}:
    \rangle_0
    + \sum_{\RR\RR',\xi\xi_2}J_{RKKY}(\RR-\RR_2,\xi,\xi_2)\theta_{\mu\nu}^{(f,\xi)}(\RR) \theta_{\mu\nu}^{(f,\xi_2)}(\RR_2)
    \\
    &+o(J^2)
 \ea 
 where $E_0=\lim_{\beta\rightarrow \infty} F_0$ is the energy of conduction electron system with Hamiltonian $\hH_0=\hat{H}_c$ and filling $\nu_c$. The RKKY type of interactions are defined as
 \hhb{ 
 \baa 
 J_{RKKY}(\RR-\RR_2,\xi,\xi_2)=&\lim_{\beta\rightarrow \infty} \sum_{\qq}-\frac{1}{\beta}J^2 \int_0^\beta d\tau_1 \int_0^{\tau_1}d\tau_2 \chi_c(\qq,\tau_1-\tau_2,\xi,\xi_2) e^{-i\qq \cdot \RR +i \qq \cdot \RR_2}\nonumber  \\
 =&\lim_{\beta\rightarrow \infty} \sum_{\qq}-\frac{1}{\beta}J^2 \frac{1}{2}\int_0^\beta d\tau_1 \int_0^{\beta}d\tau_2 \chi_c(\qq,\tau_1-\tau_2,\xi,\xi_2) e^{-i\qq \cdot \RR +i \qq \cdot \RR_2} \nonumber \\
 &[\text{Using$\Delta \tau = \tau_2-\tau_2$ and $\chi(\qq,\tau, +\beta ,\xi,\xi_2) = \chi(\qq,\tau, \xi,\xi_2) $} ]\nonumber \\
 =&\lim_{\beta\rightarrow \infty} \sum_{\qq}-\frac{1}{\beta}J^2 \frac{1}{2}\int_0^\beta d\tau_1 \int_{-\beta/2}^{\beta/2}d\Delta \tau \chi_c(\qq,\Delta \tau ,\xi,\xi_2) e^{-i\qq \cdot \RR +i \qq \cdot \RR_2} \nonumber \\
 &[\text{Replace $\Delta \tau$ with $\tau$ ]} \nonumber \\
 =&\lim_{\beta\rightarrow \infty} \sum_{\qq}-\frac{J^2}{2}\int_{-\beta/2}^{\beta/2} d\tau \chi_c(\qq,\tau,\xi,\xi_2) e^{-i\qq \cdot (\RR-\RR_2)} 
 \label{eq:rkky_def}
 \eaa  
 }
 \hh{where $\chi_c(\qq,\tau,\xi,\xi_2)$ is defined in Eq.~\ref{eq:chic}. 
 }
 
 In summary, for a given $\vartheta, \nu_f,\nu$, the energy of the proposed trial wavefunction can be written as
 \ba 
 E[\vartheta,\nu_f,\nu]& = 
\langle \Psi[\vartheta,\nu_c] |\langle \vartheta | \hH
|\vartheta\rangle |\Psi[\vartheta,\nu_c]\rangle\\
&
=  N_M \frac{U_1}{2}\nu_f^2
+N_M W \nu_f {\nu_c} 
+
N_M \frac{V_0}{2\Omega_0}{\nu_c} 
+\langle \Psi[\vartheta,\nu_c]|\hat{H}_c''[\vartheta]|\Psi[\vartheta,\nu_c]\rangle  \\
&
\approx  N_M \frac{U_1}{2}\nu_f^2
+N_M W \nu_f {\nu_c} 
+
N_M \frac{V_0}{2\Omega_0}{\nu_c} 
+ E_0 \\
&- J\sum_{\RR}\sum_{\xi=\pm} \frac{\theta_{00}^{f,\xi}(\RR)}{2N_M}\sum_{\kk,\eta s}\sum_{a=3,4}\delta_{\xi,(-1)^{a-1}\eta}\langle : 
    c_{\kk a\eta s}^\dag c_{\kk a\eta s}:
    \rangle_0\\
    &
    + \sum_{\RR\RR',\xi\xi_2,\mu\nu}J_{RKKY}(\RR-\RR_2,\xi,\xi_2)\theta_{\mu\nu}^{(f,\xi)}(\RR) \theta_{\mu\nu}^{(f,\xi_2)}(\RR_2) 
 \ea 
where $E_0$ is the ground state energy of one-particle Hamiltonian $
\hat{H}_0 =\hat{H}_c$
with filling $\nu_c$. The expectation value $\langle \rangle_0$ is taken for the ground state of $\hat{H}_0$ at filling $\nu_c$.

\hh{We note that, for both $M=0$ or $M\ne0$}, the electrons equally fill two types of fermion($\xi=\pm$). When $\nu_c=0$ the filling of each type of fermion is zero. Therefore.
\ba 
\frac{1}{N_M}\sum_{\kk a\eta s}\delta_{\xi, (-1)^{a-1}\eta} \langle :c_{\kk a \eta s}^\dag c_{\kk a \eta s} :\rangle_0 = 0 ,\quad \text{\bh{when $\nu_c=0$}}
\ea 
In other words the linear $J$ term in $E[\vartheta,\nu_f,\nu]$ vanishes. Therefore, for fixed \hb{integer fillings} $\nu_f =\nu= 0,-1,-2$, the ground state energy is determined by RKKY interactions. We define the energy coming from RKKY interactions as
\baa  
E_{RKKY}[\vartheta] :=  \sum_{\RR\RR',\xi\xi_2}J_{RKKY}(\RR-\RR_2,\xi,\xi_2)\theta_{\mu\nu}^{(f,\xi)}(\RR) \theta_{\mu\nu}^{(f,\xi_2)}(\RR_2)  \label{eq:rkky_energy}
\eaa 
\hh{ 
Using the analytical expressions of $\chi_c(\RR,\tau,\xi,\xi_2)$ derived in section~\ref{sec:chiral_u4_corre}, Eqs.~\ref{eq:chic_ana},~\ref{eq:chic_ana_2} and the definition of RKKY interactions (Eq.~\ref{eq:rkky_def}, we find the analytical expression of $J_{RKKY}(\RR,\xi,\xi_2)$. $J_{RKKY}(\RR-\RR_2,\xi,\xi_2)$ at $\nu_c=0$, zero temperature $\beta=\infty$ and infinity momentum cutoff $\Lambda_c=\infty$:}
 \baa 
 &J_{RKKY}(\RR,\xi,\xi)  = -\frac{\pi^3}{16A_{MBZ}^2 |v_\star|}\frac{1}{|\RR|^3}+o(M^2)  \nonumber \\
  &J_{RKKY}(\RR,\xi,-\xi) = \frac{3\pi^3M^2}{32A_{MBZ}^2|v_\star|^3|\RR|}+o(M^2) 
  \label{eq:rkky_1}
 \eaa 
 where $A_{MBZ}$ is the area of the first moir\'e Brillouin zone \hh{and we expand the results in powers of $M$ and keep the leading-order contributions}.
 
\hh{ 
Here, we comment that a non-zero $M$ hybridizes two conduction electrons with opposite $\xi$ indices. Therefore, it induces an RKKY interaction between two chiral U(4) moments with opposite $\xi$ indices. However, such RKKY interaction still preserves the chiral $U(4)$ symmetry but breaks flat $U(4)$ symmetry,
}\hb{as $M\ne 0, v_\star^\prime=0$ preserves the chiral $U(4)$ but not flat $U(4)$ symmetry.}

At $\nu_f=\nu = 0,-1,-2$, we observe 
\begin{itemize}
    \item $J_{RKKY}(\RR,\xi,\xi)$ decays as $1/|\RR|^3$ and is always $\le 0$, in other words, is ferromagnetic.
    \item $J_{RKKY}(\RR,\xi,-\xi)$ decays as $1/|\RR|$ and is always $\ge 0$ (antiferromagnetic). In addition, it goes to zero at $M=0$.
    \item In the derivation of the above analytical formula, we take the momentum cutoff to be infinity (therefore, it diverges at $r=|\RR|=0$. This divergence will be regularized to a finite value once we take a finite cutoff in momentum space $\Lambda_c$). 
\end{itemize} 

\hh{ In addition, the energy function in Eq.~\ref{eq:rkky_energy} can be described by the following effective Hamiltonian
\begin{equation}
    \hH_{RKKY} = \sum_{\RR\RR',\xi\xi_2} J_{RKKY}(\RR-\RR_2,\xi,\xi_2) :\UC_{\mu\nu}^{f,\xi}(\RR): :\UC_{\mu\nu}^{f,\xi_2}(\RR_2):
\label{eq:ham_rkky}
\, .
\end{equation}
where we simply replace $\theta_{\mu\nu}^{(f,\xi)}(\RR)$ with the corresponding operator $\UC_{\mu\nu}^{f,\xi}(\RR)$.
}

\subsection{Path integral formula} 
The effective Hamiltonian \hh{in Eq.~\ref{eq:ham_rkky}} can also be derived by directly integrating out conduction $c$-electrons. The action of the system is 
\baa  
S& = S_f +S_c+ S_J \nonumber \\
S_f& = \sum_{\RR, \alpha \eta s } \int_0^\beta d\tau f_{\RR,\alpha\eta s}^\dag(\tau) (\partial_\tau-\mu)  f_{\RR,\alpha \eta s}(\tau) + \int_0^\beta H_U(\tau) d\tau  
\nonumber \\
S_c &= \sum_{\kk,a\eta s} \int_0^{\beta} d\tau c_{\kk, a\eta s}^\dag(\tau)  (\partial_\tau-\mu) c_{\kk , a \eta s}(\tau) + \int_0^\beta (H_c(\tau)+H_V(\tau) +H_W(\tau)) d\tau \nonumber  \\
S_J&= \int_0^{\beta} \hH_J(\tau) d\tau \, .
\label{eq:action_v0}
\eaa 
\hh{Here, 
$f_{\RR,\alpha \eta s}(\tau), c_{\kk,a\eta s}(\tau)$ are $\tau$ dependent Grassmann numbers.
$H_U(\tau),H_c(\tau),H_V(\tau),H_W(\tau)$ are defined by replacing $f_{\RR,\alpha \eta s}, c_{\kk,a\eta s}$ with $f_{\RR,\alpha \eta s}(\tau), c_{\kk,a\eta s}(\tau)$ in the expressions of $\hH_U,\hH_c,\hH_V,\hH_W$, $\hH_J$ shown in Eq.~\ref{eq:H0} and Eq.~\ref{eq:HJ}. $\mu$ is the chemical potential. 
\hb{ 
The partition function then reads
\baa  
Z = &\int D[f_{\RR,\alpha \eta s}^\dag,f_{\RR,\alpha \eta s},c_{\kk,a,\eta, s}^\dag ,c_{\kk,a,\eta, s} ]e^{-S_f+S_c+S_J} \delta( \sum_{\alpha \eta s} f_{\RR,\alpha \eta s}^\dag(\tau)f_{\RR,\alpha \eta s}(\tau)-4 -\nu_f) ) \nonumber \\
=&\int D[f_{\RR,\alpha \eta s}^\dag,f_{\RR,\alpha \eta s},c_{\kk,a,\eta, s}^\dag ,c_{\kk,a,\eta, s} ]e^{-S_f+S_c+S_J} \int D[\lambda_\RR(\tau)] e^{-i \int_0^\beta 
\lambda_{\RR}(\tau) [ \sum_{\alpha \eta s} f_{\RR,\alpha \eta s}^\dag(\tau)f_{\RR,\alpha \eta s}(\tau)-4 -\nu_f) d\tau }
d\tau \nonumber \\
=&\int D[f_{\RR,\alpha \eta s}^\dag,f_{\RR,\alpha \eta s},c_{\kk,a,\eta, s}^\dag ,c_{\kk,a,\eta, s} ,\lambda_{\RR}(\tau)]e^{-S_f+S_c+S_J-i \int_0^\beta \lambda_{\RR}(\tau)
 [ \sum_{\alpha \eta s} f_{\RR,\alpha \eta s}^\dag(\tau)f_{\RR,\alpha \eta s}(\tau)-4 -\nu_f) } 
\eaa 
where we use the Dirac Delta function $ \delta( x )$ to fix the filling of $f$-electron and then replace the Dirac-Delta function with a Lagrangian multiplier in the second line.
}
}
\hb{ 
We combine the Lagrangian multiplier term and $S_f$, and then define the following new $S_f$}
\hh{
\baa 
S_f = \sum_{\RR, \alpha \eta s } \int_0^\beta d\tau f_{\RR,\alpha\eta s}^\dag(\tau)  \partial_\tau  f_{\RR,\alpha \eta s}(\tau)  d\tau 
+i\sum_{\RR}\int_0^\beta \lambda_\RR(\tau)
[ \sum_{ \alpha \eta s } f_{\RR,\alpha \eta s}^\dag(\tau)f_{\RR,\alpha \eta s}(\tau)-4 -\nu_f)] d\tau 
 \label{eq:action_sf_lag}
\eaa 
where \hb{we drop the constant contribution from $\hH_U$ and $\mu$ (Note that the filling of $f$ is fixed)}. \hb{We also comment that the Lagrangian multiplier $\lambda_{\RR}(\tau)$ is only introduced to fix the filling of $f$ electrons in the path integral formula, and will not be explicitly used in the calculations of this section. }
}

\hh{ 
 We can also rewrite $H_W$ as 
\ba 
H_W =  \int_0^{\beta} W\nu_f \sum_{\kk,a,\eta s}:c_{\kk a \eta s}^\dag c_{\kk a \eta s} : d \tau 
\ea 
where we have take $W_{a=1,2,3,4}=W$. As for $H_V$, we treat this term at the mean-field level:
\ba 
H_V\rightarrow  \frac{ V_0}{\Omega_0} \nu_c \sum_{\eta ,s ,a,\kk } c_{\kk a\eta s}^\dag c_{\kk a'\eta s} - \frac{ V_0N_M}{2\Omega_0} \nu_c^2 -  \frac{V_0}{\Omega_0}\nu_c \sum_{\kk} 8
\ea 
At the mean-field level, the action of conduction electron only contains fermion bilinear term 
\baa 
S_c = &\sum_{\kk,a\eta s} \int_0^{\beta} d\tau c_{\kk, a\eta s}^\dag(\tau)  \partial_\tau c_{\kk , a \eta s}(\tau)d\tau   + \int_0^\beta \sum_{\eta,s,a,a',\kk}\bigg(H^{(c,\eta)}_{a,a'}
+(-\mu + W \nu_f + \frac{V_0}{\Omega_0}\nu_c)\delta_{a,a'}
\bigg) 
c_{\kk,a\eta s}^\dag (\tau)
c_{\kk,a'\eta s}(\tau) d\tau \nonumber \\
 &- \frac{ V_0N_M}{2\Omega_0} \nu_c^2 - (W\nu_f +\frac{V_0}{\Omega_0}\nu_c)\sum_{\kk} 8
 \label{eq:action_sc}
\eaa 
where the constant term comes from the normal ordering.
} 

\hh{ 
Using $S_f$ and $S_c$ defined in Eq.~\ref{eq:action_sf_lag} and Eq.~\ref{eq:action_sc}, the partition function is written as 
\ba 
Z=&\int D[c_{\kk,a\eta s}^\dag (\tau) ,c_{\kk,a\eta s}(\tau)]
D[f_{\RR,\alpha\eta s}^\dag (\tau),f_{\RR,\alpha\eta s}(\tau),\lambda_\RR(\tau)]e^{-S_f-S_c -S_J } 
\ea 
}
\hh{We next integrate out conduction $c$-electrons:
\ba 
Z=&\int D[c_{\kk,a\eta s}^\dag (\tau) ,c_{\kk,a\eta s}(\tau)]
D[f_{\RR,\alpha\eta s}^\dag (\tau),f_{\RR,\alpha\eta s}(\tau),\lambda_\RR(\tau)]e^{-S_f-S_c -S_J } \\
=&Z_0 \int
D[f_{\RR,\alpha\eta s}^\dag(\tau) ,f_{\RR,\alpha\eta s}(\tau),\lambda_{\RR}(\tau)]e^{-S_f}\bigg(
\frac{1}{Z_0} \int D[c_{\kk,a\eta s}^\dag(\tau) ,c_{\kk,a\eta s}(\tau)]e^{-S_J} e^{-S_c}\bigg) 
\\
=&Z_0 \int
D[f_{\RR,\alpha\eta s}^\dag(\tau) ,f_{\RR,\alpha\eta s}(\tau),\lambda_{\RR}(\tau)]e^{-S_f}\langle  e^{-S_J}\rangle_0 \\
=& Z_0 \int
D[f_{\RR,\alpha\eta s}^\dag(\tau) ,f_{\RR,\alpha\eta s}(\tau),\lambda_{\RR}(\tau)]e^{-S_{eff} } 
\ea 
where the expectation value of a given operator $\hat{O}$ is defined as
\ba 
\langle \hat{O}\rangle_0 = \frac{1}{Z_0} \int D[c_{\kk,a\eta s}^\dag(\tau) ,c_{\kk,a\eta s}(\tau)]
Oe^{-S_c} \, ,
\ea  }
\hhb{and in the path integral formula, $\hat{O}$ has been replaced by the corresponding $c$-numers or Grassmann numbers $O$.}
\hh{$Z_0 =\int D[c_{\kk,a\eta s}^\dag ,c_{\kk,a\eta s}]e^{-S_c} $ and the effective action is defined as 
\ba 
S_{eff} = S_f - \log \langle e^{-S_J}\rangle_0.
\ea 
} 
\hh{ 
Since $S_J \propto J$, we expand $S_{eff}$ in powers of $J$ and truncate to the second order of $J$
\ba 
S_{eff} &= S_f - \log \langle 
e^{-S_J}\rangle_0 = 
S_f - \log\bigg( 1 - \langle S_J\rangle_0 + \frac{1}{2} \langle S_J^2\rangle_0 +o(J^2) \bigg) \\
&=S_f +\langle S_J\rangle_0 - \frac{1}{2} 
\bigg[ 
\langle S_J^2\rangle_0 - ( \langle S_J\rangle_0)^2
\bigg] +o(J^2)
\ea 
}

The first-order term is
\ba 
\langle S_J\rangle_0 = -J\int_{-\beta/2}^{\beta/2} d\tau  \sum_{\RR \qq}\sum_{\mu\nu} \sum_{\xi=\pm} e^{-i\qq \RR} :\UC_{\mu\nu}^{(f,\xi)}(\RR, \tau):
\langle :\UC_{\mu\nu}^{(c'',\xi)} (\qq,\tau) : \rangle_0 
\ea 
\hh{ 
where
\ba 
&:\UC^{(f,\xi)}_{\mu\nu}(\RR,\tau): =\sum_{\alpha, \alpha',\eta, \eta', s ,s'} \frac{\delta_{\xi,(-1)^{\alpha-1}\eta}}{2} \Theta^{(\mu\nu,f)}_{\alpha \eta s,\alpha'\eta's'}
\bigg[ 
f_{\RR,\alpha \eta s}^\dag (\tau)f_{\RR,\alpha'\eta ' s'}(\tau) -\frac{1}{2}\delta_{\alpha,\alpha'}\delta_{\eta,\eta'}\delta_{s,s'}\bigg] \\
&:\UC^{(c'',\xi)}_{\mu\nu}(\qq,\tau): =\sum_{a, a'\in \{3,4\}}\sum_{\eta, \eta', s ,s'} \frac{\delta_{\xi,(-1)^{a-1}\eta}}{2N_M} \Theta^{(\mu\nu,c'')}_{a \eta s,a'\eta's'}
\bigg[ 
c_{\kk+\qq,a \eta s}^\dag (\tau)c_{\kk,a'\eta ' s'}(\tau) -\frac{1}{2}\delta_{\qq,0}\delta_{a,a'}\delta_{\eta,\eta'}\delta_{s,s'} 
\bigg] \, .
\ea 
}
Due to the time translational symmetry, $\langle :\UC_{\mu\nu}^{(c'',\xi)} (\tau, \qq) : \rangle_0 
$ is time independent. Due to momentum conservation only $\qq =\bm{0}$ components are finite. Then we have
\ba 
\langle S_J\rangle_0 = -J\int_{-\beta/2}^{\beta/2} d\tau  \sum_{\RR }\sum_{\mu\nu} \sum_{\xi=\pm} :\UC_{\mu\nu}^{(f,\xi)}(\RR, \tau):
\langle :\UC_{\mu\nu}^{(c'',\xi)} (\bm{0},0) : \rangle_0 
\ea 
As proved 
\hh{in Eq.~\ref{eq:zero_linear_uc}}, only $\mu\nu = 00 $ component are finite, we have (\hb{at $\nu_c=0$})
\ba 
\langle S_J\rangle_0 = -J\int_{-\beta/2}^{\beta/2}d\tau  \sum_{\RR } \sum_{\xi=\pm}  :\UC_{00}^{(f,\xi)}(\RR, \tau):
\langle :\UC_{00}^{(c'',\xi)} (\bm{0},{0}) : \rangle_0 
\ea

For the next order 
\ba 
\langle S_J^2\rangle_0 &= J^2\int_{-\beta/2}^{\beta/2} d\tau 
\int_{-\beta/2}^{\beta/2}
d\tau_2  \sum_{\RR \qq,\RR_2\qq_2}\sum_{\mu\nu,\mu_2\nu_2} \sum_{\xi=\pm,\xi_2=\pm} e^{-i\qq \RR-i\qq_2\RR_2} 
\langle :\UC_{\mu\nu}^{(c'',\xi)} (\qq,\tau) :  :\UC_{\mu_2\nu_2}^{(c'',\xi_2)} (\qq_2,\tau_2) : \rangle_0 \\
&:\UC_{\mu\nu}^{(f,\xi)}(\RR,\tau): :\UC_{\mu_2\nu_2}^{(f,\xi_2)}(\RR_2,\tau_2):
\ea 
As described \hh{around Eq.~\ref{eq:u4_moment_munu_diag}}, only terms with $\mu\nu = \mu_2\nu_2$ remain finite. Due to momentum conservation, only $\qq=-\qq_2$ component produces a non-zero contribution. Then we have
\ba 
\langle S_J^2\rangle_0 &= J^2\int_{-\beta/2}^{\beta/2} d\tau \int_{-\beta/2}^{\beta/2}d\tau_2  \sum_{\RR \qq,\RR_2}\sum_{\mu\nu,\mu_2\nu_2} \sum_{\xi=\pm,\xi_2=\pm} e^{-i\qq (\RR-\RR_2)} 
\langle :\UC_{\mu\nu}^{(c'',\xi)} ( \qq,\tau) :  :\UC_{\mu\nu}^{(c'',\xi_2)} ( -\qq,\tau_2) : \rangle_0 \\
&:\UC_{\mu\nu}^{(f,\xi)}(\RR,\tau): :\UC_{\mu\nu}^{(f,\xi_2)}(\RR_2,\tau_2):
\ea

\hh{We calculate $\langle S_J^2\rangle_0 $ and $\langle S_J^2\rangle_0$ at integer $\nu_f$ and $\nu_c=0$.} The first-order term becomes 
\baa  
\langle S_J\rangle_0 =& -J\int_{-\beta/2}^{\beta/2} d\tau  \sum_{\RR } \sum_{\xi=\pm}  :\UC_{00}^{(f,\xi)}(\RR, \tau):
\langle :\UC_{00}^{(c'',\xi)} (\bm{0},{0}) : \rangle_0  \nonumber \\
=&-J\int_{-\beta/2}^{\beta/2}a d\tau \sum_{\RR}\sum_{\xi= \pm} 
:\UC_{00}^{(f,\xi)}(\RR, \tau): \sum_{a = 3,4,\eta ,s} \frac{1}{2N_M}\langle  c_{a\eta s}^\dag c_{\kk,a\eta s} - \frac{1}{2}\rangle_0
 \nonumber \\
 =&-J\int_{-\beta/2}^{\beta/2} d\tau \sum_{\RR}\sum_{\xi= \pm} 
:\UC_{00}^{(f,\xi)}(\RR, \tau): 0 = 0
\label{eq:seff_s1}
\eaa 
where we use the fact that the filling of $c$-electrons in orbital $3,4$ are $0$ at $\nu_c=0$.

The second order term is 
\ba 
\langle S_J^2\rangle_0 &= J^2 \int_{-\beta/2}^{\beta/2} d\tau 
\int_{-\beta/2}^{\beta/2}
d\tau_2  \sum_{\RR \qq,\RR_2}\sum_{\mu\nu} \sum_{\xi=\pm,\xi_2=\pm} e^{-i\qq (\RR-\RR_2)} 
\langle :\UC_{\mu\nu}^{(c'',\xi)} (\tau, \qq) :  :\UC_{\mu\nu}^{(c'',\xi_2)} (\tau_2, -\qq) : \rangle_0 \\
&:\UC_{\mu\nu}^{(f,\xi)}(\RR,\tau): :\UC_{\mu\nu}^{(f,\xi_2)}(\RR_2,\tau_2): 
\ea 
For the same reason as we give in \hh{Eq.~\ref{eq:uc_correlator_0z}}, all the correlators with $\mu\nu\ne 00$ are the same \hhb{as the one with $\mu\nu=0z$
\baa 
\langle :\UC_{\mu\nu}^{(c'',\xi)} (\tau, \qq) :  :\UC_{\mu\nu}^{(c'',\xi_2)} (\tau_2, -\qq) : \rangle_0  =\langle :\UC_{0z}^{(c'',\xi)} (\tau, \qq) :  :\UC_{0z}^{(c'',\xi_2)} (\tau_2, -\qq) : \rangle_0,\quad \mu\nu \ne 00
\eaa 
}
Then we have
\ba 
\langle S_J^2\rangle_0 =& J^2\int_{-\beta/2}^{\beta/2}d\tau \int_{-\beta/2}^{\beta/2}d\tau_2  \sum_{\RR \qq,\RR_2}\sum_{\mu\nu \ne 00 } \sum_{\xi=\pm,\xi_2=\pm} e^{-i\qq (\RR-\RR_2)} 
\langle :\UC_{0z}^{(c'',\xi)} (\tau, \qq) :  :\UC_{0z}^{(c'',\xi_2)} (\tau_2, -\qq) : \rangle_0 \\
&:\UC_{\mu\nu}^{(f,\xi)}(\RR,\tau): :\UC_{\mu\nu}^{(f,\xi_2)}(\RR_2,\tau_2):  \\
&+J^2\int_0^\beta d\tau \int_0^{\beta}d\tau_2  \sum_{\RR \qq,\RR_2} \sum_{\xi=\pm,\xi_2=\pm} e^{-i\qq (\RR-\RR_2)} 
\langle :\UC_{00}^{(c'',\xi)} (\tau, \qq) :  :\UC_{00}^{(c'',\xi_2)} (\tau_2, -\qq) : \rangle_0 \\
&:\UC_{00}^{(f,\xi)}(\RR,\tau): :\UC_{00}^{(f,\xi_2)}(\RR_2,\tau_2):  \\
\ea  
Using the expression of $\chi_c(\qq,\tau,\xi,\xi_2)$ defined in \hh{Eq.~\ref{eq:chic_real}},
we find 
\baa 
\langle S_J^2\rangle_0 =& 
\frac{J^2}{N_M}\int_{-\beta/2}^{\beta/2}d\tau \int_{-\beta/2}^{\beta/2}d\tau_2  \sum_{\RR \qq,\RR_2}\sum_{\mu\nu } \sum_{\xi=\pm,\xi_2=\pm} e^{-i\qq (\RR-\RR_2)} 
\chi_c(\qq, \tau-\tau_2,\xi,\xi_2):\UC_{\mu\nu}^{(f,\xi)}(\RR,\tau): :\UC_{\mu\nu}^{(f,\xi_2)}(\RR_2,\tau_2):  \nonumber\\
=&J^2\int_{-\beta/2}^{\beta/2}d\tau \int_{-\beta/2}^{\beta/2}d\tau_2  \sum_{\RR,\RR_2}\sum_{\mu\nu } \sum_{\xi=\pm,\xi_2=\pm}
\chi_c(\RR_2-\RR, \tau-\tau_2,\xi,\xi_2):\UC_{\mu\nu}^{(f,\xi)}(\RR,\tau): :\UC_{\mu\nu}^{(f,\xi_2)}(\RR_2,\tau_2):
\label{eq:seff_s2}
\eaa 

\hh{Combining Eq.~\ref{eq:seff_s1} and Eq.~\ref{eq:seff_s2}}, we find the following effective action
\baa  
S_{eff} =& \langle S_J\rangle_0 -  \frac{1}{2}\bigg( \langle S_J^2\rangle_0 -(\langle S_J\rangle_0)^2\bigg) \nonumber \\
=&S_f - \frac{1}{2}J^2\int_{-\beta/2}^{\beta/2}d\tau \int_{-\beta/2}^{\beta/2}d\tau_2  \sum_{\RR ,\RR_2}\sum_{\mu\nu } \sum_{\xi=\pm,\xi_2=\pm}
\chi_c(\RR_2-\RR, \tau-\tau_2,\xi,\xi_2):\UC_{\mu\nu}^{(f,\xi)}(\RR,\tau): :\UC_{\mu\nu}^{(f,\xi_2)}(\RR_2,\tau_2): 
\label{eq:seff_j_zero_hyb}
\eaa  
\bh{where $\chi_c$ is the same correlation function given before in Eq.~\ref{eq:chic_app}.}

\hb{
We next take the low-frequency limit of $\chi_c$. We first perform Fourier transformation 
\baa 
\chi_c(\RR_2-\RR,i\omega_n,\xi,\xi_2) = \int_{-\beta/2}^{\beta/2} \chi_c(\RR_2-\RR,\tau,\xi,\xi_2) e^{i\omega_n\tau}d\tau  
\eaa  
with $\omega_n = 2n\pi/\beta$. The inverse Fourier transformation gives 
\baa  
\chi_c(\RR_2-\RR,\tau,\xi,\xi_2) = \frac{1}{\beta}\sum_n \chi_c(\RR_2-\RR,i\omega_n,\xi,\xi_2)e^{-i\omega_n \tau}
\eaa 
We only keep the $i\omega_n=0$ contributions and then have 
\baa  
\chi_c(\RR_2-\RR,\tau,\xi,\xi_2) \approx  \frac{1}{\beta} \chi_c(\RR_2-\RR,i\omega_n=0,\xi,\xi_2) = \frac{1}{\beta}  \int_{-\beta/2}^{\beta/2}\chi_c(\RR_2-\RR,\tau',\xi,\xi_2) d\tau'
\label{eq:long_wave_chic}
\eaa 
}
Using Eq.~\ref{eq:long_wave_chic}, the second term of the effective action (Eq.~\ref{eq:seff_j_zero_hyb}) now becomes
\hhb{ 
\ba 
&-\frac{1}{2} J^2 \int_0^\beta d\tau \int_0^{\beta}d\tau_2  \sum_{\RR \qq,\RR_2}\sum_{\mu\nu,\mu_2\nu_2} \sum_{\xi=\pm,\xi_2=\pm} e^{-i\qq (\RR-\RR_2)} \chi_{c}(\qq, \tau_1-\tau_2,\xi,\xi_2) :\UC_{\mu\nu}^{(f,\xi)}(\RR,\tau): :\UC_{\mu\nu}^{(f,\xi_2)}(\RR_2,\tau_2): \nonumber \\ 
\approx 
&-\frac{1}{2} J^2 \frac{1}{\beta}\int_{-\beta/2}^{\beta/2}d\tau' \int_0^\beta d\tau \int_0^{\beta}d\tau_2 \sum_{\RR \qq,\RR_2}\sum_{\mu\nu,\mu_2\nu_2} \sum_{\xi=\pm,\xi_2=\pm} e^{-i\qq (\RR-\RR_2)} \chi_{c}(\qq, \tau',\xi,\xi_2) :\UC_{\mu\nu}^{(f,\xi)}(\RR,\tau): :\UC_{\mu\nu}^{(f,\xi_2)}(\RR_2,\tau_2):  \nonumber \\ 
\approx &-\frac{J^2\beta }{2}  \sum_{\xi,\xi_2,\mu\nu.\RR,\RR_2}\bigg( \frac{1}{\beta}\int_0^\beta 
:\UC_{\mu\nu}^{(f,\xi)}(\RR,\tau): d\tau \bigg) 
\bigg( \frac{1}{\beta}\int_{\beta}^{0}:\UC_{\mu\nu}^{(f,\xi_2)}(\RR_2,\tau_2): d\tau_2 \bigg)\\
&\sum_{\qq}\int_{-\beta/2}^{\beta/2}\chi_c(\qq, \tau',\xi,\xi_2)e^{-i\qq (\RR-\RR_2)}d\tau' 
\\
=&\beta \sum_{\xi,\xi_2,\mu\nu,\RR,\RR_2}\bigg( \frac{1}{\beta}\int_{-\beta/2}^{\beta/2}
:\UC_{\mu\nu}^{(f,\xi)}(\RR,\tau): d\tau \bigg) 
\bigg( \frac{1}{\beta}\int_{-\beta/2}^{\beta/2}:\UC_{\mu\nu}^{(f,\xi_2)}(\RR_2,\tau_2): d\tau_2 \bigg) J_{RKKY}(\RR-\RR_2,\xi,\xi_2)\\
\ea 
}
where we have defined $J_{RKKY}(\RR-\RR_2,\xi,\xi_2)$ as
\ba 
J_{RKKY}(\RR-\RR_2,\xi,\xi_2) =& -\int_{-\beta/2}^{\beta/2}\frac{J^2}{2} \frac{1}{\beta}\sum_{\qq}\chi_c(\qq,\tau',\xi,\xi_2) e^{-i\qq(\RR-\RR_2)}d\tau'
\ea 

Transforming the action to the Hamiltonian, we find the following RKKY interaction terms \hh{which is the same Hamiltonian as we derived in Eq.~\ref{eq:ham_rkky} \hh{for both $M=0,M\ne 0$, at integer filling $\nu=\nu_f=0,-1,-2$} }
\baa  
\hH_{RKKY} = \sum_{\RR,\RR_2,\mu\nu,\xi,\xi_2} J_{RKKY}(\RR-\RR_2,\xi,\xi_2) :\UC_{\mu\nu}^{(f,\xi)}(\RR): :\UC_{\mu\nu}^{(f,\xi_2)}(\RR_2):
\label{eq:ham_rkky_}
\eaa  

We can also rewrite the RKKY interaction with flat $U(4)$ moment, using the relation we derived in Eq.~\ref{eq:rel_u4}
\baa  
\hH_{RKKY} = & \sum_{\RR,\RR_2,\xi}\sum_{\mu\nu} J_{RKKY}(\RR-\RR_2,\xi,\xi_2) :\UF_{\mu\nu}^{(f,\xi)}(\RR): :\UF_{\mu\nu}^{(f,\xi)}(\RR_2): \nonumber \\
&+ \sum_{\RR,\RR_2,\xi}
\sum_{\mu\nu \in \{00,0x,0y,0z,z0,zx,zy,zz\}}
J_{RKKY}(\RR-\RR_2,\xi,\hhb{-\xi})  :\UF_{\mu\nu}^{(f,\xi)}(\RR): :\UF_{\mu\nu}^{(f,-\xi)}(\RR_2): \nonumber \\
&+ \sum_{\RR,\RR_2,\xi}
\sum_{\mu\nu \in \{x0,xx,xy,xz,y0,yx,yy,yz\}}\hhb{-}
J_{RKKY}(\RR-\RR_2,\xi,\hhb{-\xi}) :\UF_{\mu\nu}^{(f,\xi)}(\RR): :\UF_{\mu\nu}^{(f,-\xi)}(\RR_2):
\label{eq:spin_ham_flat}
\eaa  
We observe that
\begin{itemize}
    \item The interactions between two chiral $U(4)$ moment with different $\xi$ indices are antiferromagnetic.
    \item The interactions between two chiral (or two flat) $U(4)$ momentum with the same $\xi$ indices are ferromagnetic. 
   \bh{ \item The interactions between two chiral $U(4)$ moments with the opposite $\xi$ indices are antiferromagnetic. However, the interactions between two flat $U(4)$ moments are antiferromagnetic for the $00,0x,0y,0z,z0,zx,zy,zz$ components and are ferromagnetic for the $x0,xx,xy,xz,y0,yx,yy,yz$ components. 
    \item The interactions between two chiral (or flat) $U(4)$ moments with opposite $\xi$ indices are induced by $M$ term and vanish in the flat limit $M=0$. 
    }
    \item We also notice that 
    \hh{, in both the zero-hybridization model ($\gamma=v_\star^\prime=0)$ and the effective spin model in Eq.~\ref{eq:ham_rkky_}, the density operator of $f$ electrons with index $\xi$ at each site $\RR$,  $\hat{\nu}_f^{\xi}(\RR) =\sum_{\alpha \eta s}\delta_{\xi,(-1)^{\alpha-1}\eta} :f_{\RR \alpha \eta s}^\dag f_{\RR \alpha \eta s}:$, commutes with all the terms in the Hamiltonian and is good quantum number. The average filling of each $\xi$ is also a good quantum number: $\hat{\nu}_f^\xi = \sum_{\RR} \hat{\nu}_f^\xi(\RR)/N_M$. Therefore, we can replace $\hat{\nu}^f_\xi(\RR)$, $\hat{\nu}_f^\xi$ with real numbers $\nu^f_\xi(\RR), \nu_f^\xi$ respectively. } 
\end{itemize}

\subsection{Ground state of the zero hybridization model at $M=0$, $\nu_f=0,-1,-2$} 
We now discuss the ground state of RKKY Hamiltonian in Eq.~\ref{eq:ham_rkky} \hh{at $M=0$} and $\nu_c=0,\nu_f=0,-1,-2$.
At $M=0$, $J_{RKKY}(\RR-\RR_2,\xi,-\xi)=0$. 
\hh{In this case we only have a ferromagnetic interaction between $U(4)$ moments with same $\xi$. The Hamiltonian becomes}
\baa 
\hH_{RKKY} \bigg|_{M=0}= \sum_{\RR_1,\RR_2,\mu\nu}J_{RKKY}(\RR_1-\RR_2,\xi,\xi) :\UC_{\mu\nu}^{f,\xi}(\RR): :\UC_{\mu\nu}^{f,\xi}(\RR_2):
\label{eq:ham_spin_m0}
\eaa  
with $J_{RKKY}(\RR-\RR_2,\xi,\xi)\le 0$. 
\hh{ 
We next introduce the bond operators 
\baa 
B_{\RR,\RR_2}^{\xi,\xi} = \sum_{\alpha, \eta ,s } f_{\RR,\alpha \eta s}^\dag f_{\RR_2,\alpha \eta s}\delta_{\xi, (-1)^{\alpha+1}\eta } \label{eq:bond_1}
\eaa 
where $
(B_{\RR,\RR}^{\xi,\xi})^\dag B_{\RR,\RR}^{\xi,\xi} = (\hat{\nu}_f^\xi(\RR) + 2)^2 
$. \hb{We note that the $\hat{\nu}_f^\xi(\RR)$ is measured with respect to the charge neutrality point, and at the charge neutrality point, each $\xi$ sector has $2$ $f$-electrons, so we have a $+2$ coming with $\hat{\nu}_f^\xi(\RR) $. }
Via bond operators, we find 
\baa  
&\sum_{\mu\nu}:\UC_{\mu\nu}^{f,\xi}(\RR) : : \UC_{\mu\nu}^{f,\xi}(\RR_2) : 
\nonumber \\
=& \sum_{\mu\nu,\alpha \eta s,\alpha'\eta' s', \alpha_2\eta_2s_2,\alpha_2'\eta_2's_2'} \frac{1}{4} \delta_{\xi, (-1)^{\alpha -1 } \eta} \delta_{\xi, (-1)^{\alpha_2-1}\eta_2}\nonumber \\
&(f^\dag_{\RR,\alpha \eta s} f_{\RR,\alpha'\eta' s'}-\frac{1}{2}\delta_{\alpha,\alpha'}\delta_{\eta,\eta'}\delta_{s,s'}) (f^\dag_{\RR_2,\alpha_2\eta_2 s_2} f_{\RR_2,\alpha_2'\eta_2' s_2'} -\frac{1}{2}\delta_{\alpha_2,\alpha'_2}\delta_{\eta_2,\eta'_2}\delta_{s_2,s'_2})\Theta^{\mu\nu,f}_{\alpha \eta s, \alpha'\eta' s'} \Theta^{\mu\nu,f}_{\alpha_2\eta_2 s_2,\alpha_2'\eta_2's_2'}\nonumber \\
=&1 - \frac{2}{4}  \times (\hat{\nu}_f^\xi(\RR) +\hat{\nu}_f^\xi(\RR_2)+4) + \frac{1}{4} \sum_{\alpha \eta s, \alpha'\eta' s'}\delta_{\xi,(-1)^{\alpha-1}\eta} \delta_{\xi,(-1)^{\alpha'-1}\eta'} 4 f_{\RR,\alpha \eta s}^\dag f_{\RR,\alpha'\eta' s'} f_{\RR_2,\alpha'\eta' s'}^\dag f_{\RR_2,\alpha \eta s}\nonumber  \\
=&-1 -\frac{1}{2}(\hat{\nu}_f^\xi(\RR) +\hat{\nu}_f^\xi(\RR_2) ) 
+  \sum_{\alpha \eta s, \alpha'\eta' s'}\delta_{\xi,(-1)^{\alpha-1}\eta} \delta_{\xi,(-1)^{\alpha'-1}\eta'} f_{\RR,\alpha \eta s}^\dag f_{\RR,\alpha'\eta' s'} (\delta_{\alpha'\eta's',\alpha\eta s} - f_{\RR_2,\alpha \eta s} f_{\RR_2,\alpha'\eta' s'}^\dag ) \nonumber \\
=& -1 -\frac{1}{2}(\hat{\nu}_f^\xi(\RR) +\hat{\nu}_f^\xi(\RR_2) ) 
+ (\hat{\nu}_f^\xi(\RR) + 2)
+ \nonumber \\ 
&\sum_{\alpha \eta s, \alpha'\eta' s'}\delta_{\xi,(-1)^{\alpha-1}\eta} \delta_{\xi,(-1)^{\alpha'-1}\eta'} f_{\RR,\alpha \eta s}^\dag f_{\RR_2,\alpha \eta s} 
(\delta_{\RR,\RR_2} -f_{\RR_2,\alpha'\eta' s'}^\dag
f_{\RR,\alpha'\eta' s'}  )  \nonumber \\ 
=& -1 -\frac{1}{2}(\hat{\nu}_f^\xi(\RR) +\hat{\nu}_f^\xi(\RR_2) ) 
+ (\hat{\nu}_f^\xi(\RR) + 2)
+ 4\delta_{\RR,\RR_2}(\hat{\nu}_f^\xi(\RR)+2)\nonumber \\
&-\sum_{\alpha \eta s, \alpha'\eta' s'}\delta_{\xi,(-1)^{\alpha-1}\eta} \delta_{\xi,(-1)^{\alpha'-1}\eta'} f_{\RR,\alpha \eta s}^\dag  
f_{\RR_2,\alpha \eta s} f_{\RR_2,\alpha'\eta' s'}^\dag
f_{\RR,\alpha'\eta' s'}   \nonumber \\
=& -1 -\frac{1}{2}(\hat{\nu}_f^\xi(\RR) +\hat{\nu}_f^\xi(\RR_2) ) 
+ (\hat{\nu}_f^\xi(\RR) + 2)
+ 4\delta_{\RR,\RR_2}(\hat{\nu}_f^\xi(\RR)+2)-(B_{\RR_2,\RR}^{\xi,\xi})^\dag  B_{\RR_2,\RR}^{\xi,\xi} 
\eaa  
Therefore, we can rewrite the Hamiltonian as 
\baa  
\hH_{RKKY} \bigg|_{M=0} = &\sum_{\RR_1,\RR_2} J_{RKKY}(\RR-\RR_2,+1,+1) 
\bigg( 2+ \delta_{\RR,\RR_2} (\nu_f + 4)  -\sum_{\xi} (B_{\RR_2,\RR}^{\xi,\xi})^\dag B_{\RR_2,\RR}^{\xi,\xi}  \bigg) \nonumber  \\
=& \sum_{\RR} J_{RKKY}(\textbf{0},+1,+1) 
\bigg(2+ \nu_f +4 -\sum_{\xi} (\nu_f^\xi +2)^2  \bigg) 
+ \sum_{\RR_1 \ne \RR_2} J_{RKKY}(\RR-\RR_2,+1,+1) 
\nonumber \\
&\bigg( 2 -\sum_{\xi} (B_{\RR_2,\RR}^{\xi,\xi})^\dag B_{\RR_2,\RR}^{\xi,\xi}  \bigg) \nonumber  \\
=& const - \sum_{\RR} J_{RKKY}(0,+1,+1) \sum_{\xi} (\nu_f^\xi +2)^2
- \sum_{\RR\ne \RR_2} J_{RKKY}(\RR-\RR_2,+1,+1) \sum_{\xi} (B_{\RR_2,\RR}^{\xi,\xi})^\dag B_{\RR_2,\RR}^{\xi,\xi} \nonumber \\
\label{eq:ham_rkky_bond_m0_j}
\eaa 
where we use the fact that $J_{RKKY}(\RR-\RR_2,+1,+1) = J_{RKKY}(\RR-\RR_2,-1,-1)$ and replace $\hat{\nu}_f^\xi(\RR)$ with $\nu_f^\xi$. $const$ denotes the constant term that only depends on $\nu_f$. We next prove that the following state is the ground state of $\hH_{RKKY}$ at $M=0$ \bh{(with chiral $U(4)$ and flat $U(4)$ symmetry)}: 
\baa 
|\psi_0\rangle = \prod_\RR \bigg(\prod_{i=1}^{\nu_f^{+1}+2} f^\dag_{\RR,\alpha_i\eta_is_i} \prod_{i=\nu_f^{+1}+3 }^{\nu_f+4} f^\dag_{\RR,\alpha_i\eta_is_i}\bigg)|0\rangle 
\label{eq:gnd_state_m0}
\eaa 
with the filling being
\baa  
&\nu_f=-2: (\nu_f^{+1},\nu_f^{-1}) =(-1,-1) \nonumber \\
&\nu_f=-1:  (\nu_f^{+1},\nu_f^{-1}) =(-1,0)\text{ or } (0,-1)\nonumber \\ 
&\nu_f=0:  (\nu_f^{+1},\nu_f^{-1}) =(0,0)
\label{eq:fill_req_m0}
\eaa  
and the orbital, valley indices satisfying
\baa  
&1=\eta_i(-1)^{\alpha_i+1} \quad\text{ for }\quad i = 1,...,\nu_f^+ \nonumber \\
&-1 = \eta_i(-1)^{\alpha_i+1} \quad\text{ for }\quad i=\nu_f^+ +1,...,\nu_f+4 
\label{eq:ind_req_m0}
\eaa 
We note that the filling requirements in Eq.~\ref{eq:fill_req_m0} minimize the second term in Eq.~\ref{eq:ham_rkky_bond_m0_j}: 
\baa  
E_{\nu_f^+,\nu_f^-} =-\sum_\RR J_{RKKY}(0,+1,+1)\sum_\xi (\nu_f^\xi+2)^2
 \eaa  
 \bh{
 We show the values of $E_{\nu_f^+,\nu_f^-} $ at different fillings 
 \baa  
 \nu_f=0 :& E_{-2,2} = E_{2,-2} = -\sum_\RR 16 J_{RKKY}(0,+1,+1) ,\quad  E_{-1,1} = E_{1,-1}= -\sum_\RR 10J_{RKKY}(0,+1,+1), \nonumber \\
 &\quad E_{0,0} = -\sum_\RR 8J_{RKKY}(0,+1,+1)\nonumber \\
  \nu_f=-1 :& E_{-2,1} = E_{1,-2} = -\sum_\RR 9 J_{RKKY}(0,+1,+1) ,\quad  E_{0,-1} = E_{-1,0}= -\sum_\RR 5J_{RKKY}(0,+1,+1), \nonumber \\
  \nu_f=-2 :& E_{-2,0} = E_{0,-2} = -\sum_\RR 4 J_{RKKY}(0,+1,+1) ,\quad  E_{-1,-1} = -\sum_\RR 2J_{RKKY}(0,+1,+1) \nonumber  \\
  nu_f=-3 :& E_{-2,-1} = E_{-1,-2} = -\sum_\RR J_{RKKY}(0,+1,+1) 
  \label{eq:fill_zero_hhyb_rkky}
 \eaa  
 From Eq.~\ref{eq:fill_zero_hhyb_rkky}, we proved that the filling requirements in Eq.~\ref{eq:fill_req_m0} minimize $E_{\nu_f^+,\nu_f^-}$.
 }
} 

\hb{
We next prove that, for $\RR\ne \RR_2$, $B_{\RR_2,\RR}^{\xi,\xi}|\psi_0\rangle=0$. We first consider
\baa  
f_{\RR,\alpha \eta s}^\dag f_{\RR_2,\alpha \eta s} |\psi_0\rangle 
\eaa 
with $\RR \ne \RR_2$. Clearly, $f_{\RR,\alpha \eta s}^\dag f_{\RR_2,\alpha \eta s}$ will move one $f$ electron at $\RR_2$ in orbital $\alpha $ valley $\eta$ spin $s$ to site $\RR$ and same orbital, valley, spin. There are two possibilities, $|\psi_0\rangle$ has zero electron at $\RR_2,\alpha \eta s$, then $f_{\RR,\alpha \eta s}^\dag f_{\RR_2,\alpha \eta s} |\psi_0\rangle =0$. If $|\psi_0\rangle$ has one electron at $\RR_2,\alpha \eta s$, then there must also be one electron at $\RR,\alpha \eta s$ (from Eq.~\ref{eq:gnd_state_m0}). Consequently, $f_{\RR,\alpha \eta s}^\dag f_{\RR_2,\alpha \eta s} |\psi_0\rangle =0$. Thus we conclude $f_{\RR,\alpha \eta s}^\dag f_{\RR_2,\alpha \eta s} |\psi_0\rangle =0$ for $\RR\ne \RR_2$. Then we have 
\baa 
B_{\RR_2,\RR}^{\xi,\xi} |\psi_0\rangle = \sum_{\alpha \eta s}f_{\RR,\alpha \eta s}^\dag f_{\RR_2,\alpha \eta s} \delta_{\xi,(-1)^{\alpha+1}\eta}|\psi_0\rangle =0
\eaa 
}
and also $\langle \psi_0|B_{\RR_2,\RR}^{\xi,\xi} |\psi_0\rangle $.
\hh{Since 
\ba 
\langle \psi_0| (B_{\RR_2,\RR}^{\xi,\xi}) ^\dag B_{\RR_2,\RR}^{\xi,\xi} |\psi_0\rangle = 0
\ea 
and $J_{RKKY}(\RR-\RR_2,+1,+1) \le 0 $. $|\psi_0\rangle $ minimize energy of the third term in Eq.~\ref{eq:ham_rkky_bond}: $- \sum_{\RR\ne \RR_2} J_{RKKY}(\RR-\RR_2,+1,+1) \sum_{\xi} (B_{\RR_2,\RR}^{\xi,\xi})^\dag B_{\RR_2,\RR}^{\xi,\xi} $. In summary, $|\psi_0\rangle $ (Eq.~\ref{eq:gnd_state_m0}) is the ground state that minimizes the energy. \bh{We mention that the ground states we derived here form a subset (that of zero Chern number) of the ground states of the projected Coulomb model in the chiral-flat limit~\cite{tbgiv}. We mention that in the topological heavy-fermion model, the effect of remote bands is included, whose effect is absent in the projected Coulomb model. }
}

\subsection{Ground state of the zero hybridization model at $M\ne 0, \nu_f=0,-1,-2$}
We now solve the ground state at $M\ne 0$. The Hamiltonian is
\begin{equation}
    \hH_{RKKY} = \sum_{\RR\RR_2,\xi\xi_2} J_{RKKY}(\RR-\RR_2,\xi,\xi_2) \UC_{\mu\nu}^{f,\xi}(\RR) \UC_{\mu\nu}^{f,\xi_2}(\RR_2)
\label{eq:ham_rkky_2}
\, .
\end{equation}
where $J_{RKKY}(\RR-\RR_2,\xi,\xi) \le 0 $ and $J_{RKKY}(\RR-\RR_2,\xi,\xi) \ge 0 $. 
However, since $J_{RKKY}(\RR-\RR_2,+1,-1)$ are induced by the non-zero $M$, and $J_{RKKY}(\RR-\RR_2,\xi,-\xi)$ is relatively weak comparing to $J_{RKKY}(\RR-\RR_2,\xi,\xi)$. Therefore, we can treat $J_{RKKY}(\RR-\RR_2,\xi,\xi)$ perturbatively.
\hh{ 
We separate the Hamiltonian into two parts 
\baa  
&\hH_{RKKY} = \hH_{M=0} + \hH_{+-} \nonumber  \\
&\hH_{M=0} =  \sum_{\RR\RR_2,\xi} J_{RKKY}(\RR-\RR_2,\xi,\xi) \UC_{\mu\nu}^{f,\xi}(\RR) \UC_{\mu\nu}^{f,\xi}(\RR_2) \nonumber \\
&\hH_{+-} =  \sum_{\RR\RR_2,\xi} J_{RKKY}(\RR-\RR_2,\xi,-\xi) \UC_{\mu\nu}^{f,\xi}(\RR) \UC_{\mu\nu}^{f,-\xi}(\RR_2)
\label{eq:jm_pert_ham}
\eaa  
}
\hb{Becaue the Hamiltonian $\hH_{RKKY}$ breaks flat $U(4)$ symmetry but keeps the chiral $U(4)$ symmetry, so we work with chiral $U(4)$ moment $\UC_{\mu\nu}^{f,\xi}$. Flat $U(4)$ breaking can be observed from Eq.~\ref{eq:spin_ham_flat}, where we find $J_{RKKY}(\RR-\RR_2,\xi,-\xi)$ leads to anisotropic interactions (different $\mu\nu$ components have different interaction strength) between flat $U(4)$ moments. 
} 

\hh{
$\hH_{M=0}$ has degenerate ground states as defined in Eq.~\ref{eq:gnd_state_m0}. We let $|\psi_{0,i}\rangle$ be the ground states of $\hH_{M=0}$. Based on the perturbation theory of degenerate levels, we define the following matrix
\baa 
[H_{+-}]_{ij} = \langle \psi_{0,i} | \hH_{+-} |\psi_{0,j}\rangle \, .
\label{eq:pert_ham_def}
\eaa 
and determine the ground state by finding the lowest eigenstate of $\hH_{+-}$.
}

\begin{hbb} 
We first note that $|\psi_{0,i}\rangle$ (given in Eq.~\ref{eq:gnd_state_m0}) can be written as a product state 
\baa  
|\psi_{0,i}(\RR)\rangle = \prod_{\RR}|\psi_{0,i}(\RR)\rangle 
\eaa  
where $|\psi_{0,i}(\RR)\rangle $ is the corresponding state at $\RR$ and 
\baa  
\langle \psi_{0,j}|\psi_{0,i}(\RR)\rangle =0,\quad \text{when } i\ne j \,.
\eaa  
Then, for $(\RR_1\ne\RR_2)$ $i \ne j$, we find
\baa  
&\langle \psi_{0,j} | \UC_{\mu\nu}^{f,\xi}(\RR_1) \UC_{\mu\nu}^{f,-\xi}(\RR_2)|\psi_{0,i}\rangle  \nonumber \\
=& \prod_{\RR\ne \RR_1,\RR \ne \RR_2} 
\langle \psi_{0,j}(\RR) |\psi_{0,i} (\RR)\rangle 
\langle \psi_{0,j}(\RR_1)| \UC_{\mu\nu}^{f,\xi}(\RR_1) |\psi_{0,i} (\RR_1)\rangle \langle \psi_{0,j}(\RR_2)| \UC_{\mu\nu}^{f,-\xi}(\RR_2) |\psi_{0,i} (\RR_2)\rangle 
=0 \nonumber \\ 
&\langle \psi_{0,j} | \UC_{\mu\nu}^{f,\xi}(\RR_1) |\psi_{0,i}\rangle  \nonumber \\
=& \prod_{\RR\ne \RR_1,\RR \ne \RR_2} 
\langle \psi_{0,j}(\RR) |\psi_{0,i} (\RR)\rangle 
\langle \psi_{0,j}(\RR_1)| \UC_{\mu\nu}^{f,\xi}(\RR_1) \UC_{\mu\nu}^{f,-\xi}(\RR_1) |\psi_{0,i} (\RR_1)\rangle
=0 
\label{eq:pert_exp_vanish}
\eaa 
Combining Eq.~\ref{eq:pert_exp_vanish} and Eq.~\ref{eq:jm_pert_ham}, we have $\langle \psi_{0,i}|\hH_{+-}|\psi_{0,j}\rangle =0$ for $i\ne j$. 

For $i=j$, we find
\baa  
&\langle \psi_{0,i} | \UC_{\mu\nu}^{f,\xi}(\RR_1) \UC_{\mu\nu}^{f,-\xi}(\RR_2)|\psi_{0,i}\rangle  \nonumber \\
=& \prod_{\RR\ne \RR_1,\RR \ne \RR_2} 
\langle \psi_{0,i}(\RR) |\psi_{0,i} (\RR)\rangle 
\langle \psi_{0,i}(\RR_1)| \UC_{\mu\nu}^{f,\xi}(\RR_1) |\psi_{0,i} (\RR_1)\rangle \langle psi_{0,j}(\RR_2)| \UC_{\mu\nu}^{f,-\xi}(\RR_2) |\psi_{0,i} (\RR_2)\rangle 
=0 \nonumber \\ 
&\langle \psi_{0,i} | \UC_{\mu\nu}^{f,\xi}(\RR_1) |\psi_{0,i}\rangle  \nonumber \\
=& \prod_{\RR\ne \RR_1,\RR \ne \RR_2} 
\langle \psi_{0,i}(\RR) |\psi_{0,i} (\RR)\rangle 
\langle \psi_{0,i}(\RR_1)| \UC_{\mu\nu}^{f,\xi}(\RR_1) \UC_{\mu\nu}^{f,-\xi}(\RR_1) |\psi_{0,i} (\RR_1)\rangle
\label{eq:pert_exp_vanish_2}
\eaa  
 According to Eq.~\ref{eq:U4op-maintext}, $\UC_{\mu\nu}^{f,-\xi}(\RR_1)$ with $\mu\nu \ne 00,0z,zz$ will move $f$ electron from one $\alpha \eta s$ flavor with $(-1)^{\alpha+1}\eta=-\xi$ to another $\alpha' \eta' s'$ flavor $(-1)^{\alpha'+1}\eta'=-\xi$ at the same site. Therefore $\langle \psi_{0,i}(\RR_1)|\UC_{\mu\nu}^{f,-\xi}(\RR_1) |\psi_{0,i} (\RR_1)\rangle =0$ because $\UC_{\mu\nu}^{f,-\xi}(\RR_1)$ will change the wavefunction of $|\psi_{0,i} (\RR_1)\rangle$ or annihilate it. Similarly, $ \UC_{\mu\nu}^{f,\xi}(\RR_1) \UC_{\mu\nu}^{f,-\xi}(\RR_1)$ with $\mu\nu \ne 00,0z,zz$ will either change the configuration of $|\psi_{0,i}\rangle$ (and leads to a state orthogonal to $|\psi_{0,i}\rangle$) or annihilate $|\psi_{0,i}\rangle$. Thus 
 \baa 
 &\langle \psi_{0,i} | \UC_{\mu\nu}^{f,\xi}(\RR_1) \UC_{\mu\nu}^{f,-\xi}(\RR_2)|\psi_{0,i}\rangle  =0 ,\quad \mu\nu \notin \{00,0z,zz\}\nonumber \\
&\langle \psi_{0,i} | \UC_{\mu\nu}^{f,\xi}(\RR_1) |\psi_{0,i}\rangle  =0,\quad \mu\nu \notin \{00,0z,zz
 \}
 \label{eq:pert_exp_vanish_2}
 \eaa 
 Combining Eq.~\ref{eq:pert_exp_vanish} and Eq.~\ref{eq:pert_exp_vanish_2}, the only non-vanishing components of $\hH_{+-}$ in Eq.~\ref{eq:pert_ham_def} are 
 \baa  
 &[\hH_{+-}]_{ij} = \langle \psi_{0,i}|\hH_{+-}|\psi_{0,i}\rangle \nonumber \\
 =&\delta_{i,j}\sum_{\RR \RR_2,\xi}\sum_{\mu\nu \in \{00,0z,z0,zz\}} J_{RKKY}(\RR-\RR_2,\xi,-\xi)\langle \psi_{0,i}|\UC_{\mu\nu}^{f,\xi}(\RR) \UC_{\mu\nu}^{f,-\xi}(\RR_2) |\psi_{0,i}\rangle 
 \eaa 
 Therefore $[\hH_{+-}]_{ij}$ is a diagonal matrix.
 
We next take the state in Eq.~\ref{eq:gnd_state_m0} and calculate 
\baa 
E_{+-}=[\hH_{+-}]_{ii} \, .
\eaa 
The ground states are the states that minimize $E_{+-}$. 

We provide the values of $E_{+-}$ of the states shown in Eq.~\ref{eq:gnd_state_m0}.

At $\nu_f=0$, we find the following states have $E_{+-} =\sum_{\RR\RR_2}2J_{RKKY}(\RR-\RR_2,+,-) $ 
\baa  
&\{1+\up,2-\up,2+\up,1-\up \}, \{1+\up,1+\dn,2+\up,2+\dn \}, \{1+\up,2-\dn,2+\up,1-\dn\},\nonumber \\
&\{2-\up,2-\dn,1-\up,1-\dn\},\{1+\dn,2-\dn,1-\dn,2+\dn \}
\eaa  
where we have characterized the states with $\{\alpha_i\eta_is_i\}$ (Eq.~\ref{eq:gnd_state_m0}). 
At $\nu=0$, the following states have $E_{+-}= -\sum_{\RR\RR_2}2J_{RKKY}(\RR-\RR_2,+,-)$
\baa  
&\{ 1+\dn,2-\dn,2+\up,1-\up \}, \{2-\up,2-\dn,2+\up,2+\dn\}, \{1+\dn,2-\up,2+\up,1-\dn\},\{1+\up,2-\dn,1-\up,2+\dn \} ,\nonumber \\
&\{1+\up,1+\dn,1-\up,1-\dn\}
,\{1+\up,2-\up,1-\dn,2+\dn\}
\label{eq:gnd_nu0_jh_m}
\eaa  
All other states in Eq.~\ref{eq:gnd_state_m0} at $\nu=0$ have $E_{+-}=0$. Since $J_{RKKY}(\RR-\RR_2,+,-)\ge 0$, Eq.~\ref{eq:gnd_nu0_jh_m} gives the ground states at $\nu_f=0$.

At $\nu_f=-1$, we find the following states in Eq.~\ref{eq:gnd_state_m0} have energy $E_{+-}=\sum_{\RR\RR_2}J_{RKKY}(\RR-\RR_2,+,-) $
\baa  
&\{ 1+\up,1+\dn, 1-\up\}, \{ 1+\up,1+\dn, 1-\dn\}
\nonumber \\ 
&\{ 1+\up,2-\up, 2+\dn\},\{ 1+\up,2-\up, 1-\dn\}
\nonumber \\ 
&\{ 1+\up,2-\dn, 1-\up\},\{ 1+\up,2-\dn, 2+\dn\}
\nonumber \\ 
&\{ 1+\dn,2-\up, 2+\up\},\{  1+\dn,2-\up, 1-\dn\}
\nonumber \\ 
&\{ 1+\dn,2-\dn, 2+\up\},\{  1+\dn,2-\dn, 1-\up\}
\nonumber \\ 
&\{ 2-\up,2-\dn, 2+\up\},\{ 2-\up,2-\dn, 2+\dn \}
\label{eq:gnd_nu1_jh_m}
\eaa 
The remaining states in Eq.~\ref{eq:gnd_state_m0} at $\nu_f=-1$ have $E_{+-}=-\sum_{\RR\RR_2}J_{RKKY}(\RR-\RR_2,+,-)$. Thus the ground states at $\nu=-1$ are the states in Eq.~\ref{eq:gnd_nu1_jh_m}.

At $\nu_f=-2$, we find the following states have $E_{+-}=0 $
\baa  
&\{ 1+\up, 2+\dn\} ,\{ 1+\up, 1-\up\} ,\{ 1+\up, 1-\dn\} ,\{ 1+\dn, 2+\up\} ,\{ 1+\dn, 2+\dn\} ,\{ 1+\dn, 1-\dn\},
\nonumber \\ 
&\{ 2-\up, 2+\dn\} ,\{ 2-\up, 2+\up\} ,\{ 2-\up, 1-\dn\} ,\{ 2-\dn, 2+\up\} ,\{ 2-\dn\dn, 2+\dn\} ,\{ 2-\dn, 1-\up\}
\label{eq:gnd_nu2_jh_m}. 
\eaa 
The remaining states in Eq.~\ref{eq:gnd_state_m0} at $\nu_f=-2$ have $E_{+-}=\sum_{\RR\RR_2}2J_{RKKY}(\RR-\RR_2,+,-)$. Thus the ground states at $\nu_f=-2$ are the states in Eq.~\ref{eq:gnd_nu2_jh_m}.  

At $\nu_f=-3$, all states Eq.~\ref{eq:gnd_state_m0} in have the same $E_{+-}$

In summary, we find $f$ electrons at the same site tend to fill different spin-valley flavors, in order to minimize the energy from $J_{RKKY}(\RR-\RR_2,\xi,-\xi)$. \hh{In a more compact form, the following states (from Eq.~\ref{eq:gnd_nu0_jh_m},Eq.~\ref{eq:gnd_nu1_jh_m},Eq.~\ref{eq:gnd_nu2_jh_m}) are the ground state at $M\ne 0$}
\end{hbb}
\hh{ 
\baa 
\prod_\RR \bigg(\prod_{i=1}^{\nu_f^++2 }f^\dag_{\RR,1\eta_is_i} \prod_{i=\nu_f^++3 }^{\nu_f+4} f^\dag_{\RR,2\eta_is_i}\bigg)|0\rangle
\label{eq:gnd_state_mn0}
\eaa 
where $\nu_f^+$ and $\{\alpha_i\eta_is_i\}_{i=1,...,\nu_f+4}$ need to satisfy the Eq.~\ref{eq:ind_req_m0}, Eq.~\ref{eq:fill_req_m0} and
\baa  
(\eta_i, s_i) \ne (\eta_j, s_j) \quad \text{ for }\quad  i\ne j. 
\label{eq:ind_req_mn0}
\eaa 
}
\bh{We point out that the ground states of $f$-electrons here give rise to the same ground states of the projected Coulomb model in the chiral-nonflat limit~\cite{tbgiv}. }

\section{Schrieffer-Wolff transformation}
\label{sec:sw_transf}
\hh{ 
In this section, we treat the model of non-zero hybridization $\gamma \ne 0, v_\star^\prime \ne 0$ with the Schrieffer-Wolff transformation~\cite{SW_transf}. We focus on the $\nu_f=0,-1,-2$. At $\nu_f = -3$ and in the zero-hybridization limit, $\nu_f=-3$ is close to the transition point as shown in Fig.~\ref{fig:fill_2}. Therefore, a uniform charge distribution of $f$ electrons may not be energetically favorable at $\nu_f=-3$ and the SW transformation could fail at this \hb{filling}.}

\subsection{Hamiltonian}
We first separate the Hamiltonian \hh{with non-zero hybridization} into two parts
\baa  
\hH = \hat{H}^{complete}_0 +\hat{H}_1 
\eaa 
In the heavy-fermion model with non-zero hybridization, we let 
\baa 
&\hh{\hat{H}^{complete}_0} = \hH_c + \hH_U + \hH_V + \hH_W  - \mu \sum_{\RR,\alpha,\eta ,s}f_{\RR,\alpha \eta s}^\dag f_{\RR,\alpha \eta s} -\mu \sum_{\RR,a,\eta ,s}c_{\kk,a\eta s}^\dag c_{\kk, a \eta s} 
\hh{+P_H\hH_{fc}P_H + \hH_J }
\nonumber \\
&\hat{H}_1 = P_L \hH_{fc} + \hH_{fc} P_L \nonumber \\
&\hat{H}_{fc} = \frac{1}{\sqrt{N_M}}\sum_{\kk,\RR, \alpha \eta s}e^{i\kk\RR}H_{\alpha a}^{(fc,\eta)}(\kk) f^\dag_{\RR \alpha  \eta s}c_{\kk a \eta s} +h.c \quad,\quad H^{(fc,\eta)} (\kk) = \begin{bmatrix}
\gamma \sigma_0 +v_\star^\prime (\eta k_x \sigma_x +k_y\sigma_y)
& 0_{2 \times 2}
\end{bmatrix}.
\label{eq:ham_sw}
\eaa 
where $\mu$ is the chemical potential. The projection operator for a given filling of $f$-electrons $\nu_f$ is defined as
\baa 
\bh{P_L = \prod_{\RR} \bigg[\alpha_{\nu_f}\prod_{ n \ne \nu_f+4 }
\bigg( 
\sum_{\alpha \eta s}f_{\RR,\alpha \eta s}^\dag f_{\RR,\alpha \eta s} -n
\bigg) \bigg]
\quad, \quad P_H = \mathbb{I} -P_L  \, .
\label{eq:proj}}
\eaa 
\hh{The \bh{projection operator} $P_L$ will only keep the states with filling of $f$ electron being $\nu_f$ for each site and $\alpha_{\nu_f}^{-1} = \prod_{n\ne \nu_f+4} (\nu_f+4-n)] = (\nu_f+4)!(4-\nu_f)!(-1)^{\nu_f} $ is the normalization factor ensuring $P_L^2=P_L$.} \bh{Here we also introduce the notation of low-energy space and high-energy scape. Low-energy space is defined as
\ba 
\mathcal{H}_L = \bigg\{ |\psi\rangle \bigg|  P_L|\psi\rangle =|\psi\rangle \bigg\} 
\ea  
For all states in $\mathcal{H}_L$, the filling of $f$-electrons is $\nu_f$ at each site. We define the high-energy space as 
\baa  
\mathcal{H}_H = \bigg\{ |\psi\rangle \bigg| P_L|\psi\rangle =0\bigg\} 
\eaa  
which is formed by all states that are not in $\mathcal{H}_L $. We use low-energy (high-energy) states to denote the state in low-energy (high-energy) space.} 
Then the Hamiltonian $\hH_0^{complete}$ maps a low-energy state to a low-energy state\hb{, or maps} a high-energy state to a high-energy state; $\hH_1$ maps a low-energy state to a high-energy state or a high-energy state to a low-energy state~\cite{coleman2015introduction}. Then after SW transformation, we \bh{are} able to derive an effective Hamiltonian that maps a low-energy state to a low-energy state or a high-energy state to a high-energy state. In other words, we eliminate the off-diagonal term, $\hH_1$, during the SW transformation~\cite{coleman2015introduction}.

However, we comment that the original hybridization term $\hH_{fc}$ can map a high-energy state to either a high-energy state or a low-energy state. To observe this, we act $\hH_{fc}$ on a state with filling $\nu_f +1$. The resulting state can has filling $\nu_f -1$ or $\nu_f+2$ (if $\nu_f+2 \le 4$). Therefore, it can map a high-energy state with filling $\nu_f+1$ to another high-energy state with filling $\nu_f+2$. The same argument also applies to the state with $\nu_f-1$. It violates the requirement of $\hH_1$ that we explained earlier. However
\bh{, we could separate $\hH_{fc}$ into two parts: $\hH_{fc} = \bigg(P_L\hH_cP_H +P_L\hH_cP_H\bigg) + P_H\hH_{fc}P_H$ (Note that $\hH_{fc}$ will change the filling of $f$ electrons by $1$, so $P_L\hH_{fc}P_L=0$). We treat the first part $P_L\hH_cP_H +P_L\hH_cP_H$ as $\hH_1$, and put the second part $P_H\hH_{fc}P_H$ into $\hH_0^{complelte}$. Then we can perform SW transformation by treating $\hH_1$ as a perturbation. }  

One may wonder what happens if we take $\hH_{fc}$ as $\hH_1$ and perform SW transformation. In principle, we can find the operator $S$ that satisfies $[\hH_0,S]=\hH_1$. However, $[\hH_1,S]$, which appears in the effective Hamiltonian, would contain terms with the form of $f^\dag f^\dag cc $ or $f f c^\dag c^\dag$. $f^\dag f^\dag cc $ or $f f c^\dag c^\dag$ which maps a low-energy state to a high-energy stat. And the effective Hamiltonian after transformation can still map a low-energy state to a high-energy state or a high-energy state to a low-energy state. Therefore, we can not treat $f$ as a local moment in the effective Hamiltonian. The such issue will not emerge if we let $\hH_1 =P_L \hH_{fc} + \hH_{fc} P_L $. We remark that, in the standard one-orbital Anderson model, we always have $\hH_{fc}=P_L \hH_{fc} + \hH_{fc} P_L$ and it is not necessary to explicitly introduce the projection operator.

In addition, instead of working with $\hH_0^{complete}$, we first drop both $P_H\hH_{fc}P_H$ and $\hH_J$ terms in the procedure of SW transformation, and define 
\baa  
\hat{H}_0 = \hH_c + \hH_U + \hH_V + \hH_W  - \mu \sum_{\RR,\alpha,\eta ,s}f_{\RR,\alpha \eta s}^\dag f_{\RR,\alpha \eta s} -\mu \sum_{\RR,a,\eta ,s}c_{\kk,a\eta s}^\dag c_{\kk, a \eta s}  \, .
\eaa  
If we keep both terms, it would be complicated to derive the effective Hamiltonian, since it contains $f$-$c$ hybridization and $f$-$c$ interactions.  

\subsubsection{$\hat{H}_0$ }
We provide the explicit form of $\hH_0$ in this section. We simplify the notation by using single index $m$ to label orbital $\alpha$, valley $\eta$ and spin $\sigma$ and let 
$
f_{\RR,m} := f_{\RR,\alpha\eta s},
c_{\kk,m} := c_{\kk, a \eta s}
$. The corresponding density operator is defined as 
\baa  
&n^{f}_{\RR,m} = f_{\RR,m}^\dag f_{\RR,m} \quad ,\quad n_{\RR}^{f} = \sum_m n^f_{\RR m }\quad,\quad N_{f} = \sum_{\RR,m} f_{\RR,m}^\dag f_{\RR,m} \nonumber \\ 
&N_{c} = \sum_{\kk,a} \gamma_{\kk,a }^\dag \gamma_{\kk, a} 
=\sum_{\kk,a} c_{\kk,a }^\dag c_{\kk, a} \, . 
\eaa

The kinetic term of conduction electrons now becomes
\baa  
\hH_c = \sum_{\kk,  m,m'} H^c_{mm'}(\kk)c_{\kk,m}^\dag c_{\kk,m'}.
\eaa  
where $H^c_{mm'}(\kk)$ is the hopping matrix of conduction electrons with new label $m,m'$. We next introduce eigenvalues and eigenvectors of $H^c_{mm'}(\kk)$,
\baa  
\sum_{\alpha}H^c_{mn}(\kk)U_{\kk,n a} = \epsilon_{k,a}U_{\kk,ma}\, ,
\eaa 
and the operator in the band basis 
\baa 
\gamma_{\kk,a} = \sum_{n}U_{\kk,na }^*c_{\kk,n} 
\,. 
\eaa 
Then we have $\hH_c = \sum_{\kk, a} \epsilon_{\kk,a} \gamma_{\kk,a}^\dag \gamma_{\kk,a}$.

The Hubbard interactions between $f$ electrons now become
\baa 
\hat{H}_U =& \frac{U}{2} \sum_{\RR} \bigg(:\sum_{m} f_{\RR,m}^\dag f_{\RR,m}:\bigg)^2=\frac{U}{2}\sum_{\RR}(\sum_m (f_{\RR,m}^\dag f_{\RR,m} - 1/2))^2  \nonumber \\
=&
\frac{U}{2}\sum_{\RR}\sum_{m',m} n_{\RR,m}^f n^f_{\RR,m'} 
+\frac{U}{2}\sum_{\RR} 4^2 -\frac{U}{2}\sum_{\RR,m} 2f_{\RR,m}^\dag f_{\RR,m}4 \nonumber \\
=
&\frac{U}{2}\sum_{\RR}\sum_{m',m,m'\ne m} n_{\RR,m}^f n^f_{\RR,m'} 
+\frac{U}{2}\sum_{\RR,m} n_{\RR,m}^f +8UN_M -\frac{U}{2}\sum_{\RR,m} 8n^f_{\RR,m} \nonumber \\
=&\frac{U}{2}\sum_{\RR,m',m,m\ne m'} n^f_{\RR,m}n^f_{\RR,m'}  -\frac{7U}{2}N_f +8UN_M \, .
\eaa 
where $N_M$ is the total number of moir\'e unit cells.

For the Coulomb interactions between $c$ electrons, we only include the $\qq=0$ contribution: $V(\qq) =V_0\delta_{\qq,0}$. This leads to
\baa  
\hat{\hH}_V =& \frac{V_0}{2\Omega_0}\frac{1}{N_M}(\sum_{\kk,m}:c_{\kk,m}^\dag c_{\kk,m}:)^2
= \frac{V_0}{2\Omega_0}\frac{1}{N_M}\bigg( 
N_c - \sum_{\kk} 8 
\bigg)^2
=\frac{V_0}{2\Omega_0N_M}N_c^2 -\bigg(\frac{8V_0}{\Omega_0N_M}\sum_{\kk} 1\bigg) N_c + \frac{V_0}{2\Omega_0N_M}(\sum_{\kk} 8)^2 
\eaa  
where $\Omega_0$ is the area of the first moir\'e Brillouin zone.


The density-density interactions between $f,c$ are
\ba 
\hH_W =& \frac{1}{N_M}W\sum_{a,\RR,\qq,\kk} :n_{\RR}^f: :c_{\kk+\qq, a}^\dag c_{\kk,a}: e^{-i\qq \RR}
=W \frac{1}{N_M}\sum_{a,\RR,\qq, \kk, a } (n_{\RR}^f-4) (c_{\kk+\qq,a}^\dag c_{\kk,a}-\frac{\delta_{\qq,0}}{2})e^{-i\qq \RR} \\
&= \frac{1}{N_M}W\sum_{a,\RR,\qq, \kk } n_{\RR}^f c_{\kk+\qq,a}^\dag c_{\kk,a}e^{-i\qq \RR} 
+ \frac{W}{N_M}\sum_{a,\RR,\kk }n_{\RR}^f(-\frac{1}{2}) -  \frac{4W}{N_M}\sum_{a,\RR,\qq,\kk}c_{\kk+\qq,a}^\dag c_{\kk,a} e^{-i\qq\RR} \nonumber \\ 
&+ \frac{W}{N_M}\sum_{a,\RR,\qq,\kk}(-4)(-\frac{\delta_{\qq,0}}{2})e^{-i\qq \RR} \\
&= \frac{1}{N_M}W\sum_{a,\RR,\qq, \kk } n_{\RR}^f c_{\kk+\qq,a}^\dag c_{\kk,a}e^{-i\qq \RR}
-8W \frac{1}{N_M}\sum_{\RR,\kk}n_{\RR}^f-4WN_c +32W\sum_\kk 1 
\ea 
Here is the derivation, \bh{we only consider the $c$-electrons in the first moir\'e Brillouin by taking a specific momentum cutoff $\Lambda_c$.} 
Due to the first term in $\hH_W$ (which contains $c_{\kk+\qq,a}^\dag c_{\kk,a}$), \bh{it is complicated} to find an analytical expression of the SW transformation. However, if we stick to the case, where the fillings of $f$ fermion are the same at each site, we can replace $n_{\RR}^f$ by its average value $\frac{1}{N_M}\sum_{\RR'}n_{\RR'}^f$. This leads to  
\ba 
 \frac{W}{N_M}\sum_{a,\RR,\qq, \kk } n_{\RR}^f c_{\kk+\qq,a}^\dag c_{\kk,a}e^{-i\qq \RR} 
\rightarrow & \frac{W}{N_M}\sum_{a,\RR,\qq, \kk } \bigg(\frac{\sum_{\RR'}n_{\RR'}^f}{N_M}\bigg) c_{\kk+\qq,a}^\dag c_{\kk,a}e^{-i\qq \RR} 
= \frac{W}{N_M}\sum_{\RR'}n_{\RR'}^f\sum_{a,\qq, \kk }  c_{\kk+\qq,a}^\dag c_{\kk,a}\delta_{\qq,0} \\
&=\frac{W}{N_M}\sum_{\RR}n_{\RR}^f \sum_{\kk,a} c_{\kk,a}^\dag c_{\kk,a} =\frac{W}{N_M}\sum_{\RR} n_{\RR}^f \sum_{\kk, a}\gamma_{\kk,a}^\dag \gamma_{\kk,a}
\ea 
Now, we define our new $\hH_W$ as
\ba 
\hH_W =\frac{W}{N_M}\sum_{\RR,\kk,a} n_{\RR}^f\gamma_{\kk,a}^\dag \gamma_{\kk,a}-8W \frac{1}{N_M}\sum_{\RR,\kk}n_{\RR}^f-4WN_c +32W\sum_\kk 1  \\
\ea 

Combining all terms and rearranging them, we have 
\baa  
&\hH_0 = \hat{h}_U + \hat{h}_{E_f} +\hat{h}_{V_c} +\hat{h}_c +\hat{h}_W + \hat{h}_{const} \nonumber \\
&\hat{h}_U = \frac{U}{2}\sum_{\RR} \sum_{m,m',m\ne m'} n_{\RR m}^f n_{\RR m'}^f 
\quad ,\quad \hat{h}_{E_f} = E_{f}N_f  \nonumber  \\
&\hat{h}_{V_c} =E_c N_c +V_cN_c^2
\quad,\quad 
\hat{h}_c = \sum_{\kk, a} \epsilon_{k,a} \gamma_{\kk,a}^\dag \gamma_{\kk,a}  \nonumber \\
&\hat{h}_{W} =\frac{1}{N_M}W\sum_{a,\RR,\qq, \kk } n_{\RR}^f c_{\kk+\qq,a}^\dag c_{\kk,a}e^{-i\qq \RR} \quad,\quad \hat{h}_{const} = 8UN_M+ \frac{V_0}{2\Omega_0N_M}(\sum_{\kk}8)^2 +32W\sum_{\kk} 1 
\label{eq:sw_ham_0}
\eaa  
where 
\baa 
&E_f = -\frac{7U}{2}- \frac{8W}{N_M}\sum_{\kk} 1 -\mu
\quad,\quad 
E_c = -4W -\frac{8V_0}{\Omega_0N_M}\sum_{\kk}1 -\mu\quad,\quad V_c = \frac{V_0}{2\Omega_0N_M} 
\label{eq:sw_ham_const}
\eaa

\subsubsection{$\hH_1$}
$\hH_1$ can be written as
\baa  
\hat{H}_1& = \frac{1}{\sqrt{N_M}} \sum_{\kk,\RR}\sum_{m,m'} e^{i\kk \RR} H_{mm'}^{fc}(\kk)(P_Lf_{\RR,m}^\dag c_{\kk,m'}+f_{\RR,m}^\dag c_{\kk,m'}P_L) +\text{h.c.} \nonumber \\
=&
\frac{1}{\sqrt{N_M}} \sum_{\kk,\RR}\sum_{m,n} e^{i\kk \RR}(\sum_{m'} H_{mm'}^{fc}(\kk)U_{\kk,m'n})(P_Lf_{\RR,m}^\dag \gamma_{\kk,n} +f_{\RR,m}^\dag \gamma_{\kk,n}P_L)+\text{h.c.} \nonumber \\ 
&=
 \sum_{\kk,\RR}\sum_{m,n} V_{\RR\kk,mn}(P_Lf_{\RR,m}^\dag \gamma_{\kk,n} +f_{\RR,m}^\dag \gamma_{\kk,n}P_L) +\text{h.c.} 
 \label{eq:sw_ham_1}
\eaa  
where $H^{fc}(\kk)$ is the hybridization matrix (with damping factor included, Eq.~\ref{eq:def_hfc}, Eq.~\ref{eq:def_hyb_mat}) with new label $m,m'$. $V_{\kk,mn}$ is defined as $V_{\RR\kk,mn} =\frac{1}{\sqrt{N_M}} \sum_{m'}e^{i\kk \RR}H^{fc}_{mm'}(\kk)U_{\kk,m'n}$.

\subsection{Procedure of Schrieffer–Wolff transformation} 
We now perform SW transformation. We aim to find operator $S$ such that 
\baa   
S = -S^\dag \quad,\quad [\hH_0,S] = \hH_1  \, .
 \label{eq:sw_s_req}
\eaa 
This allows us to introduce the following unitary transformations 
\baa  
 e^S \hH e^{-S} \approx \hH_0 + (\hH_1 - [\hH_0,S] )  + [S,\hH_1] + \frac{1}{2}[[\hH_0,S],S] = \hH_0 + \frac{1}{2}[S,\hH_1]
 \label{eq:sw_eff_ham}
 \, ,
\eaa  
and obtain the effective Hamiltonian 
\baa 
\hH_{eff} = \hH_0 + \frac{1}{2} [S,\hH_1]
\label{eq:sw_def_eff_ham}
\eaa

\subsection{Expression of $S$}
In this section, we provide the analytical expression of $S$. We first decompose $S$ based on its anti-hermitian property 
\baa  
S = S_1-S_1^\dag \, . 
\label{eq:sw_def_s}
\eaa  
We then assume 
\baa  
&S_1^\dag = P_Ls^\dag_1 + s^\dag_1P_L \nonumber \\
&s_1^\dag =\sum_{\RR,\kk,m,n}V_{\RR\kk,mn}R_{\RR\kk,mn}f_{\RR,m}^\dag \gamma_{\kk,n}
\label{eq:sw_def_s1}
\eaa  
where $V_{\RR\kk,ma}$ is the hybridization matrix between $f$-fermion and $\gamma$-fermion. $R_{\RR\kk,mn}$ is an {\it operator} with assumed commutation relations $[R_{\RR\kk,mn},\gamma^\dag_{\kk+\qq,n}\gamma_{\kk',n'}]=0$, $[R_{\RR\kk,mn}, f_{\RR',n'}^\dag f_{\RR',n'}]=0$
and
$[R_{\RR\kk,mn}, P_L]=0$. $R_{\RR\kk,mn}$ are determined by requiring $[\hH_0,S]=\hH_1$.

To find the expression of $R_{\RR\kk,mn}$, we directly calculate $[\hat{H}_0,s_1]$. Before moving to the calculations, we first list several useful commutation relations:
\baa  
&[n_{\RR,m}^f , f_{\RR',m'}^\dag ]=\delta_{\RR,\RR'}\delta_{m,m'} f_{\RR',m'}^\dag\nonumber  \\
&[\gamma_{\kk+\qq,a}^\dag \gamma_{\kk, a'}, \gamma_{\kk',b} ] = -\delta_{\kk+\qq, \kk'}\delta_{a,b} \gamma_{\kk'-\qq,a'} \nonumber \\
&\text{Given operators } A,B,C,D \text{ with } [A,D]=[B,C]=0,\text{ we have } [AB,CD] =[A,C]DB+CA[B,D] +[A,C][B,D] 
\label{eq:commute_rel}
\eaa 

For each term in $\hat{H}_0$ \hh{(Eq.~\ref{eq:sw_ham_0}), we use Eq.~\ref{eq:commute_rel} and find }
\ba 
[\hat{h}_U,s_1^\dag ] &= 
\sum_{\RR',m,m',m\ne m'}\sum_{\RR,\kk,a}UV_{\RR\kk,ma}R_{\RR\kk,ma}n_{\RR,m'}^f[n_{\RR,m}^f,f_{\RR,m}^\dag]\gamma_{\kk,a} \nonumber \\  
=&
\hh{\sum_{\RR, \kk, m,a} UV_{\RR\kk,ma}R_{\RR\kk,ma} \sum_{m',m'\ne m}n_{\RR,m'}^f f_{\RR,m}^\dag \gamma_{\kk, a}}\\
&= \hh{\sum_{\RR, \kk, m,a} UV_{\RR\kk,ma}R_{\RR\kk,ma}  (\sum_{m'} n_{\RR,m}^f-1) f_{\RR,m}^\dag \gamma_{\kk, a}=\sum_{\RR, \kk, m,a} UV_{\RR\kk,ma}R_{\RR\kk,ma}  (n_\RR^f-1) f_{\RR,m}^\dag \gamma_{\kk, a}}\\
[\hat{h}_{E_f},s_1^\dag]& = \sum_{\RR,\kk, m,a}E_{f}V_{\RR\kk,ma}R_{\RR\kk,ma}  [n_{\RR,m}^f, f_{\RR,m}^\dag] \gamma_{\kk, a} = \sum_{\RR,\kk, m,a}E_{f} V_{\RR\kk,ma}R_{\RR\kk,ma}  f_{\RR,m}^\dag \gamma_{\kk, a}
\\
[\hat{h}_{c},s_1^\dag] &= \sum_{\RR,\kk, m,a}V_{\RR\kk,ma}R_{\RR\kk,ma}  \sum_{\kk', a'} \epsilon_{k',a'} (-f_{\RR,m}^\dag \gamma_{\kk, a} \gamma_{\kk',a'}^\dag \gamma_{\kk',a'} +\gamma_{\kk',a'}^\dag \gamma_{\kk',a'} f_{\RR,m}^\dag \gamma_{\kk, a} )  \\
&
= \sum_{\RR,\kk, m,a}V_{\RR\kk,ma}R_{\RR\kk,ma} \sum_{\kk', a'} \epsilon_{k',a'}f_{\RR,m}^\dag (- \gamma_{\kk, a} \gamma_{\kk',a'}^\dag -\gamma_{\kk',a'}^\dag  \gamma_{\kk, a} )
\gamma_{\kk',a'}\\
&=\sum_{\RR,\kk, m,a}V_{\RR\kk,ma}R_{\RR\kk,ma} \sum_{\kk', a'} \epsilon_{k,a} (-f_{\RR,m}^\dag \delta_{\kk,\kk'}\delta_{a,a'}\gamma_{\kk',a'}  )
=-\sum_{\RR,\kk, m,a}V_{\RR\kk,ma}R_{\RR\kk,ma}  \epsilon_{k,a} f_{\RR,m}^\dag \gamma_{\kk,a}  
\ea  
\ba 
[\hat{h}_{V_c}, s_1^\dag] &= \sum_{\RR,\kk,m,a}V_{\RR\kk,ma}R_{\RR\kk,ma} f_{\RR,m}^\dag[V_cN_c^2+E_cN_c,\gamma_{\kk, a}] \nonumber \\
=&
\sum_{\RR,\kk,m,a}V_{\RR\kk,ma}R_{\RR\kk,ma} f_{\RR,m}^\dag[V_c(N_c^2-(N_c+1)^2)\gamma_{\kk,\hh{a}} - E_c \gamma_{\kk,\hh{a}}] \\
&=
\sum_{\RR,\kk,m,a}V_{\RR\kk,ma}R_{\RR\kk,ma} f_{\RR,m}^\dag[V_c(-2N_c-1)\gamma_{\kk,a} - E_c \gamma_{\kk,a} 
]
\\
[\hat{h}_W,s_1^\dag] &=
\frac{1}{N_M}W\sum_{a,\RR,\qq, \kk }\sum_{\RR',\kk',m,n}V_{\RR'\kk',mn} R_{\RR'\kk',mn}e^{-i\qq \RR} [n_{\RR}^f \gamma_{\kk+\qq,a}^\dag \gamma_{\kk,a},f_{\RR',m}^\dag \gamma_{\kk',n}] 
\\
&=
\frac{1}{N_M}W\sum_{a,\RR,\qq, \kk }\sum_{\RR',\kk',m,n}V_{\RR'\kk',mn} R_{\RR'\kk',mn}e^{-i\qq \RR} 
\bigg(
n_{\RR}^f f_{\RR',m}^\dag \gamma_{\kk+\qq,a}^\dag \gamma_{\kk,a}, \gamma_{\kk',n} 
-f_{\RR',m}^\dag n_{\RR}^f \gamma_{\kk',n}\gamma_{\kk+\qq,a}^\dag \gamma_{\kk,a} \bigg) \\
&=
\frac{1}{N_M}W\sum_{a,\RR,\qq, \kk }\sum_{\RR',\kk',m,n}V_{\RR'\kk',mn} R_{\RR'\kk',mn}e^{-i\qq \RR} \delta_{\RR,\RR'} f_{\RR',m}^\dag 
\bigg( \gamma_{\kk+\qq,a}^\dag \gamma_{\kk,a}, \gamma_{\kk',n} 
- \gamma_{\kk',n}\gamma_{\kk+\qq,a}^\dag \gamma_{\kk,a}\bigg) \\
&=
\frac{1}{N_M}W\sum_{a,\RR,\qq, \kk }\sum_{\RR',\kk',m,n}V_{\RR'\kk',mn} R_{\RR'\kk',mn}e^{-i\qq \RR} \delta_{\RR,\RR'} f_{\RR',m}^\dag (-\delta_{\kk+\qq,\kk'}\delta_{n,a} )
\gamma_{\kk, a}\\ 
&=
\frac{1}{N_M}W\sum_{\RR,\qq, \kk }\sum_{m,n}V_{\RR(\kk+\qq),mn} R_{\RR\kk+\qq,mn}e^{-i\qq \RR}  (-1)f_{\RR,m}^\dag
\gamma_{\kk, n}\\ 
[\hat{h}_{const},s_1^\dag]& =0
\ea
We notice that $P_L$ only contains the density operator of $f$ electron, and it commutes with $\hH_0$. Then we have $[\hH_0,S_1^\dag] = [\hH_0, P_Ls_1^\dag + s_1^\dag P_L = P_L[\hH_0,s_1^\dag] + [\hH_0,s_1^\dag] P_L$. 

Combining all terms, we have 
\baa 
[\hat{H}_0,S_1^\dag] =
&\sum_{\RR,\kk,m,a}V_{\RR\kk,ma}R_{\RR\kk,ma} 
\bigg[ 
U(n_{\RR}^f-1)+E_f -\epsilon_{\kk, a} +V_c(-2N_c-1)-E_c
\bigg] \bigg( P_Lf_{\RR,m}^\dag \gamma_{\kk,a}
+f_{\RR,m}^\dag \gamma_{\kk,a}P_L\bigg)  
\nonumber 
\\
& +\frac{1}{N_M}W\sum_{\RR,\qq, \kk }\sum_{m,n}V_{\RR(\kk+\qq),mn} R_{\RR\kk+\qq,mn}e^{-i\qq \RR}  (-1)\bigg( P_Lf_{\RR,m}^\dag 
\gamma_{\kk, n} 
+f_{\RR,m}^\dag 
\gamma_{\kk, n} 
P_L
\bigg) 
\label{eq: commute_h0_s1}
\eaa 
From the definition of $P_L$ (\hh{Eq.~\ref{eq:proj}}), we note that 
\ba 
n_{\RR}^f P_L = (\nu_f +4 )P_L
\ea 
with $\nu_f$ the integer number that characterizes the filling of the low-energy state. We let $R_{\RR\kk,ma}$
\baa  
R_{\RR\kk,ma} =- \frac{1}{U(\nu_f+3)+E_f -\epsilon_{\kk, a} +V_c(-2N_c-1)-E_c-W }
\label{eq:sw_def_R}
\eaa  
\hh{Combining the definition of $R_{\RR\kk,ma}$ (Eq.~\ref{eq:sw_def_R}) and Eq.~\ref{eq: commute_h0_s1}, we have}
\ba 
[\hH_0,S_1^\dag] =&- 
\sum_{\RR,\kk,m,a}V_{\RR\kk,ma}
\frac{ 
U(n_{\RR}^f-1)+E_f -\epsilon_{\kk, a} +V_c(-2N_c-1)-E_c}
{U(\nu_f+3) +E_f-\epsilon_{\kk,a} +V_c(-2N_c-1)-E_c-W}
\bigg( P_Lf_{\RR,m}^\dag \gamma_{\kk,a}
+f_{\RR,m}^\dag \gamma_{\kk,a}P_L\bigg)  
\nonumber \\
&-
\frac{1}{N_M}W\sum_{\qq,\kk}\sum_{m,n}
\frac{1}{U(\nu_f+3) +E_f-\epsilon_{\kk+\qq,a} +V_c(-2N_c-1)-E_c-W}
\ea 
which gives the correct commutation relations
\ba 
&[\hH_0,-S_1^\dag] = \sum_{\RR,\kk,m,a}V_{\RR\kk,ma}f_{\RR,m}^\dag \gamma_{\kk,a} \\
&[\hH_0, S_1] =\bigg([\hH_0, -S_1^\dag] \bigg)^\dag  =  \sum_{\RR,\kk,m,a}V^*_{\RR\kk,ma} \gamma_{\kk,a}^\dag f_{\RR,m} \, ,  `
\ea 
and 
\ba 
[\hH_0,S] = [\hH_0,S_1-S_1^\dag] = \hH_1  
\ea

\subsection{Effective Hamiltonian from SW transformation}
\hh{We now derive the effective Hamiltonian $\hH_{eff} = \hH_0 + \frac{1}{2} [S,\hH_1]$. Combining Eq.~\ref{eq:sw_def_s} and Eq.~\ref{eq:sw_def_s1}, we have }
\ba 
&\frac{1}{2} [S,\hH_1] =\hh{ 
\frac{1}{2}(P_Ls_1+s_1P_L-P_Ls_1^\dag - s_1^\dag P_L)(P_L\hH_{fc} +\hH_{fc}P_L )
-\frac{1}{2} (P_L\hH_{fc} +\hH_{fc}P_L )(P_Ls_1+s_1P_L-P_Ls_1^\dag - s_1^\dag P_L)
}
\ea 
\hh{Since $\hH_{fc}$ and $s_1$ change the filling of $f$ by $\pm 1$, both $\hH_{fc}$ and $s_1$ map a low-energy to high-energy state, which indicates 
\baa 
&P_Ls_1P_L =P_L\hH_{fc}P_L =0 \nonumber \\
&P_L\hH_{fc} = P_L\hH_{fc}P_H\quad,\quad 
\hH_{fc}P_L = P_H\hH_{fc}P_L  
\quad,\quad 
P_Ls_1= P_Ls_1P_H\quad,\quad s_1P_L= P_Hs_1P_L \nonumber
\eaa  
Then we find}
\hh{ 
\baa  
\frac{1}{2} [S,\hH_1]  = -\frac{1}{2}P_L
\bigg[ 
(s_1^\dag-s_1)\hH_{fc} +  \hH_{fc}(-s_1^\dag+s_2) 
\bigg] P_L
-\frac{1}{2}P_H
 \bigg[ 
 (s_1^\dag -s_1)P_L\hH_{fc} + \hH_{fc}P_L(s_1-s_1^\dag)
 \bigg] P_H
 \label{eq:sh1_commute_v0}
\eaa  
}
\hh{We note that the effective Hamiltonian $\hH_{eff}=\hH_0+\frac{1}{2}[S,\hH_1]$ will only map a low-energy state to a low-energy state or a high-energy state to a high-energy state. We focus on the effective Hamiltonian that controls the low-energy physics, which is defined in the low-energy Hilbert space: $\hH_{eff} = P_L\hH_0P_L+\frac{1}{2}P_L[S,\hH_1]P_L$. 
Combining Eq.~\ref{eq:sh1_commute_v0}, Eq.~\ref{eq:sw_def_R} and Eq.~\ref{eq:sw_ham_1}, we find}
\baa  
\frac{1}{2}P_L[S,\hH_1]P_L=&-\frac{1}{2}P_L\bigg\{\sum_{\RR,\kk,m,a,\RR',\kk',m',a'}\bigg[ \nonumber \\
&\bigg(V_{\RR\kk,ma}R_{\RR\kk,ma} f_{\RR,m}^\dag \gamma_{\kk,a} -V_{\RR\kk,ma}^*
 \gamma_{\kk,a}^\dag f_{\RR,m} R^\dag_{\RR\kk,ma} \bigg)\bigg(V_{\RR'\kk',m'a'}f_{\RR',m'}^\dag \gamma_{\kk',a'} 
 +V^*_{\RR'\kk',m'a'} \gamma_{\kk',a'}^\dag f_{\RR',m'}\bigg) \nonumber \\
 &+ \bigg(V_{\RR'\kk',m'a'}f_{\RR',m'}^\dag \gamma_{\kk',a'} 
 +V^*_{\RR'\kk',m'a'} \gamma_{\kk',a'}^\dag f_{\RR',m'}\bigg)\bigg(-
 V_{\RR\kk, ma}R_{\RR\kk,ma}f_{\RR,m}^\dag \gamma_{\kk,a} +
 \gamma_{\kk,a}^\dag f_{\RR,m}V_{\RR\kk,ma}^* R^\dag_{\RR\kk,ma}\bigg)\bigg] \nonumber \\
 &\bigg\}P_L 
 \label{eq:sh1_commute_v1}
\eaa  
Annihilating or creating two electrons would obviously violate the uniform charge distribution of $f$, and will map low-energy space to high energy space. Therefore,
\ba 
&P_Lf_{\RR,m'}f_{\RR',m}P_L=0
\quad, \quad P_Lf^\dag_{\RR,m'}f^\dag_{\RR',m}P_L=0
\\
&P_Lf_{\RR,m'}^\dag f_{\RR',m'}P_L\propto \delta_{\RR,\RR'} 
\quad, \quad 
P_Lf_{\RR,m'}f_{\RR',m'}^\dag P_L\propto \delta_{\RR,\RR'} \, ,
\ea 
and \hh{Eq.~\ref{eq:sh1_commute_v1} can be written as}
\baa  
&\frac{1}{2}P_L[S,\hH_1]P_L \nonumber \\
 =&-\frac{1}{2}\sum_{\RR,\kk,\kk',ma,m'a'} \bigg[ P_L 
  (V^*_{\RR\kk',m'a'}
 V_{\RR\kk,ma}R_{\RR\kk,ma} f_{\RR,m}^\dag f_{\RR,m'}
\gamma_{k,a}\gamma_{k',a'}^\dag  )P_L \nonumber \\ 
&
-P_L (V_{\RR\kk',m'a'}V^*_{\RR\kk,ma}\gamma_{\kk,a}^\dag f_{\RR,m}R^\dag_{\RR\kk,ma} f^\dag_{\RR,m'}\gamma_{\kk',a'} )P_L \nonumber \\
 &
+P_L(V_{\RR\kk',m'a'}V_{\RR \kk,ma}^*f_{\RR,m'}^\dag f_{\RR,m} \gamma_{\kk',a'}\gamma^\dag_{\kk,a}R_{\RR\kk,ma}^\dag 
)P_L -P_L 
(V_{\RR'\kk',m'a'}^*V_{\RR\kk,ma}\gamma^\dag_{\kk',a'}f_{\RR,m'}R_{\RR\kk,ma}f_{\RR,m}^\dag \gamma_{\kk,a}
)P_L
\bigg] 
\label{eq:sw_pl_hs_pl}
\eaa  
We next evaluate each term in the above equation (\hh{Eq.~\ref{eq:sw_pl_hs_pl}}). 
The first term \hb{of Eq.~\ref{eq:sw_pl_hs_pl}} reads
\baa  
&-V^*_{\RR\kk',m'a'}
 V_{\RR\kk,ma}R_{\RR\kk,ma} f_{\RR,m}^\dag f_{\RR,m'}
\gamma_{k,a}\gamma_{k',a'}^\dag \nonumber \\
= &V^*_{\RR\kk',m'a'}V_{\RR\kk,ma} \frac{1}{U(n_{\RR}^f-1)+E_f -\epsilon_{\kk, a} +V_c(-2N_c-1)-E_c+W/N_M(N_c-N_f+1) }f_{\RR,m}^\dag f_{\RR,m'}\gamma_{\kk,a}\gamma_{\kk',a'}^\dag \nonumber\\
=&V^*_{\RR\kk',m'a'}V_{\RR\kk,ma} f_{\RR,m}^\dag f_{\RR,m'}\gamma_{\kk,a}\gamma_{\kk',a'}^\dag \frac{1}{U(n_{\RR}^f-1)+E_f -\epsilon_{\kk, a} +V_c(-2N_c-1)-E_c+W/N_M(N_c-N_f+1) } \nonumber\\
=&V^*_{\RR\kk',m'a'}V_{\RR\kk,ma} f_{\RR,m}^\dag f_{\RR,m'}(\delta_{\kk,\kk'}\delta_{a,a'}-\gamma_{\kk',a'}^\dag \gamma_{\kk,a}) \frac{1}{U(n_{\RR}^f-1)+E_f -\epsilon_{\kk, a} +V_c(-2N_c-1)-E_c+W/N_M(N_c-N_f+1) }\nonumber  \\ 
=&- V^*_{\RR\kk',m'a'}V_{\RR\kk,ma} f_{\RR,m}^\dag f_{\RR,m'}\gamma_{\kk',a'}^\dag \gamma_{\kk,a} \frac{1}{U(n_{\RR}^f-1)+E_f -\epsilon_{\kk, a} +V_c(-2N_c-1)-E_c+W/N_M(N_c-N_f+1) } \nonumber \\
& +\delta_{\kk,\kk'}\delta_{a,a'}V^*_{\RR\kk,m'a}V_{\RR\kk,ma} f_{\RR,m}^\dag f_{\RR,m'} \frac{1}{U(n_{\RR}^f-1)+E_f -\epsilon_{\kk, a} +V_c(-2N_c-1)-E_c+W/N_M(N_c-N_f+1) } 
\label{eq:sw_pl_hs_pl_1} 
\eaa 
The second term \hb{of Eq.~\ref{eq:sw_pl_hs_pl}} reads
\baa 
&V_{\RR\kk',m'a'}V^*_{\RR\kk,ma}\gamma_{\kk,a}^\dag f_{\RR,m}R^\dag_{\RR\kk,ma} f^\dag_{\RR,m'}\gamma_{\kk',a'}  \nonumber \\
= &-V_{\RR\kk',m'a'}V^*_{\RR\kk,ma} \gamma_{\kk,a}^\dag f_{\RR,m}\frac{1}{U(n_{\RR}^f-1)+E_f -\epsilon_{\kk, a} +V_c(-2N_c-1)-E_c+W/N_M(N_c-N_f+1) }f_{\RR,m'}^\dag \gamma_{\kk',a'} \nonumber \\
= &-V_{\RR\kk',m'a'}V^*_{\RR\kk,ma} 
\gamma_{\kk,a}^\dag f_{\RR,m}
f_{\RR,m'}^\dag \gamma_{\kk',a'} \nonumber  \\  
&
\frac{1}{U(n_{\RR}^f+1-1)+E_f -\epsilon_{\kk, a} +V_c(-2(N_c-1)-1)-E_c+W/N_M((N_c-1)-(N_f+1)+1) }
\nonumber \\
= &-V_{\RR\kk',m'a'}V^*_{\RR\kk,ma}\gamma_{\kk,a}^\dag f_{\RR,m}
f_{\RR,m'}^\dag \gamma_{\kk',a'}\frac{1}{U(n_{\RR}^f)+E_f -\epsilon_{\kk, a} +V_c(-2N_c+1)-E_c+W/N_M(N_c-N_f-1) }\nonumber \\
= &-V_{\RR\kk',m'a'}V^*_{\RR\kk,ma} f_{\RR,m}
f_{\RR,m'}^\dag  \gamma_{\kk,a}^\dag \gamma_{\kk',a'}\frac{1}{U(n_{\RR}^f)+E_f -\epsilon_{\kk, a} +V_c(-2N_c+1)-E_c+W/N_M(N_c-N_f-1) }
\nonumber \\
= &-V_{\RR\kk',m'a'}V^*_{\RR\kk,ma} (\delta_{m,m'}-
f_{\RR,m'}^\dag  f_{\RR,m}) \gamma_{\kk,a}^\dag \gamma_{\kk',a'}\frac{1}{U(n_{\RR}^f)+E_f -\epsilon_{\kk, a} +V_c(-2N_c+1)-E_c+W/N_M(N_c-N_f-1) }
\nonumber \\
= &V_{\RR\kk',m'a'}V^*_{\RR\kk,ma} 
f_{\RR,m'}^\dag  f_{\RR,m} \gamma_{\kk,a}^\dag \gamma_{\kk',a'}\frac{1}{U(n_{\RR}^f)+E_f -\epsilon_{\kk, a} +V_c(-2N_c+1)-E_c+W/N_M(N_c-N_f-1) }\nonumber \\
&-\delta_{m,m'}V_{\RR\kk',ma'}V^*_{\RR\kk,ma}  \gamma_{\kk,a}^\dag \gamma_{\kk',a'}\frac{1}{U(n_{\RR}^f)+E_f -\epsilon_{\kk, a} +V_c(-2N_c+1)-E_c+W/N_M(N_c-N_f-1) }
\label{eq:sw_pl_hs_pl_2}
\eaa  
The third term \hb{of Eq.~\ref{eq:sw_pl_hs_pl}} reads
\baa  
&-V_{\RR\kk',m'a'}V_{\RR \kk,ma}^*f_{\RR,m'}^\dag f_{\RR,m} \gamma_{\kk',a'}\gamma^\dag_{\kk,a}R_{\RR\kk,ma}^\dag  \nonumber
\\
=&V_{\RR\kk',m'a'}V^*_{\RR\kk,ma}f_{\RR,m'}^\dag f_{\RR,m} \gamma_{\kk',a'}\gamma^\dag_{\kk,a}\frac{1}{U(n_{\RR}^f-1)+E_f -\epsilon_{\kk, a} +V_c(-2N_c-1)-E_c+W/N_M(N_c-N_f+1) }\nonumber  \\
=&V_{\RR\kk',m'a'}V^*_{\RR\kk,ma}f_{\RR,m'}^\dag f_{\RR,m} (\delta_{\kk,\kk'}\delta_{a,a'}-\gamma^\dag_{\kk,a}\gamma_{\kk',a'})\frac{1}{U(n_{\RR}^f-1)+E_f -\epsilon_{\kk, a} +V_c(-2N_c-1)-E_c+W/N_M(N_c-N_f+1) } \nonumber \\
=&-V_{\RR\kk',m'a'}V^*_{\RR\kk,ma}f_{\RR,m'}^\dag f_{\RR,m} \gamma^\dag_{\kk,a} \gamma_{\kk',a'}\frac{1}{U(n_{\RR}^f-1)+E_f -\epsilon_{\kk, a} +V_c(-2N_c-1)-E_c+W/N_M(N_c-N_f+1) } \nonumber \\
&+\delta_{\kk,\kk'}\delta_{a,a'}V_{\RR\kk,m'a}V^*_{\RR\kk,ma} f_{\RR,m}^\dag f_{\RR,m'}\frac{1}{U(n_{\RR}^f-1)+E_f -\epsilon_{\kk, a} +V_c(-2N_c-1)-E_c+W/N_M(N_c-N_f+1) }
\label{eq:sw_pl_hs_pl_3}
\eaa 
The fourth term \hb{of Eq.~\ref{eq:sw_pl_hs_pl}} reads
\baa  
&V_{\RR'\kk',m'a'}^*V_{\RR\kk,ma}\gamma^\dag_{\kk',a'}f_{\RR,m'}R_{\RR\kk,ma}f_{\RR,m}^\dag \gamma_{\kk,a}\nonumber \\
=&-V_{\RR'\kk',m'a'}^*V_{\RR\kk,ma}\gamma^\dag_{\kk',a'}f_{\RR,m'}\frac{1}{U(n_{\RR}^f-1)+E_f -\epsilon_{\kk, a} +V_c(-2N_c-1)-E_c+W/N_M(N_c-N_f+1) }f_{\RR,m}^\dag \gamma_{\kk,a} \nonumber  \\
=&-V_{\RR'\kk',m'a'}^*V_{\RR\kk,ma}\gamma^\dag_{\kk',a'}f_{\RR,m'}f_{\RR,m}^\dag \gamma_{\kk,a}\\
&\frac{1}{U(n_{\RR}^f+1-1)+E_f -\epsilon_{\kk, a} +V_c(-2(N_c-1)-1)-E_c+W/N_M(N_c-1-(N_f+1)+1) }
\nonumber 
\\
=&-V_{\RR'\kk',m'a'}^*V_{\RR\kk,ma}\gamma^\dag_{\kk',a'}f_{\RR,m'}f_{\RR,m}^\dag \gamma_{\kk,a}\frac{1}{U(n_{\RR}^f)+E_f -\epsilon_{\kk, a} +V_c(-2N_c+1)-E_c+W/N_M(N_c-N_f-1) }
\nonumber 
\\
=&-V_{\RR'\kk',m'a'}^*V_{\RR\kk,ma}f_{\RR,m'}f_{\RR,m}^\dag \gamma^\dag_{\kk',a'}\gamma_{\kk,a}\frac{1}{U(n_{\RR}^f)+E_f -\epsilon_{\kk, a} +V_c(-2N_c+1)-E_c+W/N_M(N_c-N_f-1) }
\nonumber 
\\
=&-V_{\RR'\kk',m'a'}^*V_{\RR\kk,ma}(\delta_{m,m'}-f_{\RR,m}^\dag f_{\RR,m'}) \gamma^\dag_{\kk',a'}\gamma_{\kk,a}\frac{1}{U(n_{\RR}^f)+E_f -\epsilon_{\kk, a} +V_c(-2N_c+1)-E_c+W/N_M(N_c-N_f-1) }
\nonumber 
\\
=&V_{\RR'\kk',m'a'}^*V_{\RR\kk,ma}f_{\RR,m}^\dag f_{\RR,m'} \gamma^\dag_{\kk',a'}\gamma_{\kk,a}\frac{1}{U(n_{\RR}^f)+E_f -\epsilon_{\kk, a} +V_c(-2N_c+1)-E_c+W/N_M(N_c-N_f-1) }
\nonumber \\
&-\delta_{m,m'}V_{\RR'\kk',ma'}^*V_{\RR\kk,ma} \gamma^\dag_{\kk',a'}\gamma_{\kk,a}\frac{1}{U(n_{\RR}^f)+E_f -\epsilon_{\kk, a} +V_c(-2N_c+1)-E_c+W/N_M(N_c-N_f-1) } 
\label{eq:sw_pl_hs_pl_4}
\eaa  

\hh{Summing over all terms (Eqs.~\ref{eq:sw_pl_hs_pl_1}, \ref{eq:sw_pl_hs_pl_2}, \ref{eq:sw_pl_hs_pl_3}, \ref{eq:sw_pl_hs_pl_4})}, we have
\baa  
&\frac{1}{2}P_L(S\hH_1 + \hH_1S^\dag)P_L \nonumber\\
=&\sum_{\RR,\kk \kk',m,a,m',a'}V^*_{\RR\kk',m'a'}V_{\RR\kk,ma}P_Lf_{\RR,m}^\dag f_{\RR,m'}\gamma_{\kk',a'}^\dag \gamma_{\kk,a}\nonumber\\
&\bigg[ -
\frac{1}{U(n_{\RR}^f-1)+E_f -\epsilon_{\kk, a} +V_c(-2N_c-1)-E_c+W/N_M(N_c-N_f+1) } \nonumber\\
&+\frac{1}{U(n_{\RR}^f)+E_f -\epsilon_{\kk, a} +V_c(-2N_c+1)-E_c+W/N_M(N_c-N_f-1) }
\bigg] P_L  \nonumber \\
&+\bigg[ \sum_{\RR,\kk,,m,m',a}V_{\RR\kk,m'a}^*V_{\RR\kk,ma}P_Lf_{\RR,m}^\dag f_{\RR,m'}\frac{1}{U(n_{\RR}^f-1)+E_f -\epsilon_{\kk, a} +V_c(-2N_c-1)-E_c+W/N_M(N_c-N_f+1) } P_L \nonumber\\ 
&
-\sum_{\RR,\kk,\kk',m,a,a'}V_{\RR\kk',ma'}V^*_{\RR\kk,ma}  P_L\gamma_{\kk,a}^\dag \gamma_{\kk',a'} 
\bigg[ \frac{1}{U(n_{\RR}^f)+E_f -\epsilon_{\kk, a} +V_c(-2N_c+1)-E_c+W/N_M(N_c-N_f-1) }P_L
\eaa  
Since we fix the filling of $f$ at each site, we can replace $n_{\RR}^f$ with $\nu_f+4$. We also replace $N_c$ by
$ 
N_c = \sum_{\kk, a }\gamma_{\kk,a}^\dag \gamma_{\kk,a} = \sum_{\kk, a}c_{\kk,a}^\dag c_{\kk,a} =  N_M \hat{\nu}_c +\sum_{\kk} 8 \, .
$
Based on this, the pre-factors that appear in the effective Hamiltonian can be written as
\baa  
\frac{1}{D_{1,\kk,a}[\hat{\nu}_c,\nu_f]}=&\frac{1}{U(n_{\RR}^f-1)+E_f -\epsilon_{\kk, a} +V_c(-2N_c-1)-E_c+W/N_M(N_c-N_f+1) }\nonumber\\
=&\frac{1}{U(\nu_f+3) +E_f -\epsilon_{\kk,a}+V_c(-2N_M\hat{\nu_c} -16\sum_{\kk}1 -1)-E_c +W(-\nu_f-4 +\hat{\nu}_c+8\sum_{\kk}1/N_M+1/N_M) } 
\eaa  
and 
\baa 
\frac{1}{D_{2,\kk,a}[\hat{\nu}_c,\nu_f]}=&\frac{1}{U(n_{\RR}^f)+E_f -\epsilon_{\kk, a} +V_c(-2N_c+1)-E_c+W/N_M(N_c-N_f-1) } \nonumber \\
=&\frac{1}{U(\nu_f+4) +E_f -\epsilon_{\kk,a}+V_c(-2N_M\hat{\nu_c} -16\sum_{\kk}1 +1)-E_c +W(-\nu_f-4 +\hat{\nu}_c+8\sum_{\kk}1/N_M-1/N_M) } P_L 
\eaa 
Using explicit formula of $E_f,V_c,E_c$ \hh{in Eq.~\ref{eq:sw_ham_const}}, the denominators becomes
\hh{
\ba 
D_{1,\kk,a}[\hat{\nu}_c,\nu_f]=&U(\nu_f+3)-\frac{7U}{2}-\frac{8W}{N_M}\sum_{\kk }1 -\epsilon_{\kk, a} +\frac{V_0}{2\Omega_0N_M}(-2N_M\hat{\nu_c}-16\sum_{\kk}1-1)-(-4W -\frac{8V_0}{\Omega_0 N_M} \sum_{\kk} 1 )\\
&-W(\nu_f -\hat{\nu}_c-4+8\sum_{\kk}1+1/N_M) \\
=&(U-W)\nu_f +(\frac{-V_0}{\Omega_0}+W)\hat{\nu}_c 
-\epsilon_{\kk, a} -\frac{U}{2} 
+\frac{1}{2N_M }(-\frac{V_0}{2\Omega_0} -W)
\ea 
and
\ba 
D_{2,\kk,a}[\hat{\nu_c},\nu_f]=&
=U(\nu_f+4) -\frac{7U}{2} -\frac{8W}{N_M}\sum_{\kk} 1  -\epsilon_{\kk,a}+V_c(-2N_M\hat{\nu_c} -16\sum_{\kk}1 +1)
-(-4W-\frac{8V_0}{\Omega_0N_M}\sum_{\kk}1)\\ &-W(\nu_f -\hat{\nu}_c+4-\sum_{\kk}1/N_M-1/N_M)\\
=&(U-W)\nu_f  +(\frac{-V_0}{\Omega_0}+W)\hat{\nu}_c 
-\epsilon_{\kk, a} +\frac{U}{2} 
+\frac{1}{2N_M }(\frac{V_0}{2\Omega_0} - W)
\ea 
}

We comment on the value of denominators:
\begin{itemize}
    \item We drop the term at the order of $1/N_M$ which is negligibly small. 
    \item In the original model, there \bh{is} a damping term, which suppresses the hybridization between $f$ fermions and $c$-fermions with large momentum. Then the dominant hybridization comes from the $c$ electron with momentum near $\Gamma_M$. These conduction electrons have small $|\epsilon_{\kk,a}|$, so we drop this term in the denominators. (A similar approximation has also been taken in the standard Kondo model.).
    \item We also replace the operator $\hat{\nu}_c$ by a real number $\nu_c$ which represents the average filling of conduction electrons.
\end{itemize}
Taking above approximations, we can replace $D_{1/2,\kk,a}[\hat{\nu}_c,\nu_f]$ with real numbers $D_{1/2,\nu_c,\nu_f}$
\baa  
&D_{1,\kk,a}[\nu_c,\nu_f] \approx D_{1,\nu_c,\nu_f}=
(U-W)\nu_f -\frac{U}{2}  +(\frac{-V_0}{\Omega_0}+W){\nu}_c
\nonumber \\
&D_{2,\kk,a}[\nu_c,\nu_f] \approx  D_{2,\nu_c,\nu_f}=(U-W)\nu_f +\frac{U}{2}   +(\frac{-V_0}{\Omega_0}+W){\nu}_c
\label{eq:D12_def}
\eaa 
\hh{ 
and we define $D_{\nu_c,\nu_f}$ as
\baa 
D_{\nu_c,\nu_f} = \bigg[-\frac{1}{D_{1,\nu_c,\nu_f}}
+\frac{1}{D_{2,\nu_c,\nu_f} }\bigg]^{-1}
\label{eq:D_def}
\eaa  
} 
We give the value at $\nu_c=0$, $U=  58$meV, $W =48$meV
\hh{ 
\begin{center}
\begin{tabular}{c|c|c|c|c  }
$\nu_f$ & 0 & -1 & -2 \\
\hline 
 $\frac{1}{D_{\nu_c=0,\nu_f}}$&  0.0690 (meV)$^{-1}$    & 0.0805 (meV)$^{-1}$  & 0.161 (meV)$^{-1}$   
 \\
 \end{tabular}
\end{center}
}
The final formula of \hb{the} effective Hamiltonian is
\baa  
\hH_{eff} 
&=\hH_0+
\sum_{\RR,\kk, \kk',m,a,m',a'}\frac{e^{i(\kk-\kk')\RR}}{N_M}
\hb{
\frac{V_{\RR\kk',m'a'}^*V_{\RR\kk,ma}}{D_{\nu_c,\nu_f} } 
}
f_{\RR,m}^\dag f_{\RR,m'}\bh{\gamma_{\kk',a'}^\dag \gamma_{\kk,a}}
 \nonumber  \\
&+\bigg[ \sum_{\RR,\kk,m',a}\frac{V_{\RR\kk,m'a'}^*V_{\RR\kk,ma}}{N_M}f_{\RR,m}^\dag f_{\RR,m'} \frac{1}{D_{1,\nu_c,\nu_f}}-\sum_{\kk,m,a,a'}V_{\RR\kk',m'a'}^*V_{\RR\kk,ma}\gamma_{\kk,a}^\dag \gamma_{\kk,a'}\frac{1}{D_{2,\nu_c,\nu_f}}\bigg]
\, .
\eaa  
Since, we only consider the state with fixed fillings of $f$ at each site, we drop the projection operator $P_L$. However, we need to remember that our Hilbert space is spanned by the states with the fixed filling of $f$-electrons $\nu_f$.

We now transform everything from band basis $\gamma_{\kk,m}$ to conduction electron basis $c_{\kk,m}$. Using $\gamma_{\kk,a}=\sum_a U_{\kk,ba}^* c_{\kk,b}$ and $V_{\RR\kk,mn} = \frac{1}{\sqrt{N_M}}\sum_{m'}e^{i\kk \RR}H^{fc}_{mm'}(\kk)U_{\kk,m'n}$
\baa  
\hH_{eff} 
&=\hH_0
+
\sum_{\RR,\kk, \kk',m,a,m',a'}\frac{e^{i(\kk-\kk')\RR}}{N_M}
\hh{
\frac{H^{fc,*}_{m'b'}(\kk')H^{fc}_{mb}(\kk)}{D_{\nu_c,\nu_f} } 
}
f_{\RR,m}^\dag f_{\RR,m'}c_{\kk',b'}^\dag c_{\kk,b}
\nonumber  \\
&+\bigg[ \sum_{\RR,\kk,,m,m',b}\frac{H^{fc,*}_{m'b}(\kk)H^{fc}_{mb}(\kk)}{N_M}f_{\RR,m}^\dag f_{\RR,m'} \frac{1}{D_{1,\nu_c,\nu_f}}-\sum_{\kk,m,b,b'}H^{fc}_{ma'}(\kk)H_{ma}^{fc,*}(\kk)c_{\kk,b}^\dag c_{\kk,b'}\frac{1}{D_{2,\nu_c,\nu_f}}\bigg] \, . 
\eaa  
We next go to the original labeling. 
\baa  
\hH_{eff} &= \hat{H}_0 \nonumber \\
&+
\sum_{\RR,\kk, \kk',\eta,\eta',s,s'\alpha,\alpha}\sum_{a,a'\in\{1,2\}}\frac{e^{i(\kk-\kk')\RR}}{N_M}
H^{(fc,\eta),*}_{\alpha'a'}(\kk')
H^{(fc,\eta')}_{\alpha a}(\kk)
\hh{\frac{F(|\kk|)^2}{D_{\nu_c,\nu_f}}} 
\bigg[ 
f_{\RR,\alpha \eta s}^\dag f_{\RR,\alpha'\eta's'}c_{\kk',a'\eta's'}^\dag c_{\kk,a\eta s}
\bigg] \nonumber  \\
&+\bigg[ \sum_{\kk,\alpha,\alpha',\eta,s}\sum_{a\in \{1,2\}}\frac{F(|\kk|)^2H^{(fc,\eta),*}_{\alpha'a}(\kk)H^{(fc,\eta)}_{\alpha a}(\kk)}{N_M D_{1,\nu_c,\nu_f}}\bh{\sum_{\RR}}f_{\RR,\alpha \eta s}^\dag f_{\RR,\alpha' \eta s} \nonumber \\
&-\sum_{\kk,\alpha,\eta,s}\sum_{a,a'\in \{1,2\}}\frac{F(|\kk|)^2H^{(fc,\eta)}_{\alpha a'}(\kk)H_{\alpha a}^{(fc,\eta),*}}{D_{2,\nu_c,\nu_f}}(\kk)c_{\kk,a\eta s}^\dag c_{\kk,a'\eta s}\bigg] \, . 
\eaa 
\bh{where 
\baa  
F(|\kk|)=e^{-\lambda^2 |\kk|^2/2}
\label{eq:def_damp_fact_F}
\eaa 
is the damping factor of $f$-$c$ hybridization (Eq.~\ref{eq:def_hfc}).}
With normal ordering, we have 
\baa 
&\hH_{eff} =\hH_0 \nonumber 
\\
&+
\sum_{\RR,\kk, \kk',\eta,\eta',s,s'\alpha,\alpha}\sum_{a,a'\in\{1,2\}}\frac{e^{i(\kk-\kk')\RR}}{N_M}H^{(fc,\eta),*}_{\alpha'a'}(\kk')H^{(fc,\eta')}_{\alpha a}(\kk)
\hh{\frac{F(|\kk|)^2}{D_{\nu_c,\nu_f}}} 
\bigg[ 
:f_{\RR,\alpha \eta s}^\dag f_{\RR,\alpha'\eta's'}::c_{\kk',a'\eta's'}^\dag c_{\kk,a\eta s}:
\bigg]  \nonumber\\
&+\sum_{\RR,\kk, \kk',\eta,\eta',s,s'\alpha,\alpha}\sum_{a,a'\in\{1,2\}}\frac{e^{i(\kk-\kk')\RR}}{N_M}H^{(fc,\eta),*}_{\alpha'a'}(\kk')H^{(fc,\eta')}_{\alpha a}(\kk)
\hh{\frac{ |F(|\kk|)|^2}{D_{\nu_c,\nu_f}}} 
\bigg[ \frac{ |F(|\kk|)|^2}{2}\delta_{\alpha,\alpha'}\delta_{\eta,\eta'}\delta_{s,s'}:c_{\kk',a'\eta's'}^\dag c_{\kk,a\eta s}:
\bigg] \nonumber  \\
&+\sum_{\RR,\kk, \kk',\eta,\eta',s,s'\alpha,\alpha}\sum_{a,a'\in\{1,2\}}\frac{e^{i(\kk-\kk')\RR}}{N_M}H^{(fc,\eta),*}_{\alpha'a'}(\kk')H^{(fc,\eta')}_{\alpha a}(\kk)
\hh{\frac{ |F(|\kk|)|^2}{D_{\nu_c,\nu_f}}} 
\bigg[:f_{\RR,\alpha \eta s}^\dag f_{\RR,\alpha'\eta's'}:\delta_{\kk,\kk'}\delta_{a,a'}\delta_{\eta,\eta'}\delta_{s,s'}
\bigg]  \nonumber \\
&+\bigg[ \sum_{\RR,\kk,\alpha,\alpha',\eta,s}\sum_{a\in \{1,2\}}\frac{ |F(|\kk|)|^2H^{(fc,\eta),*}_{\alpha'a}(\kk)H^{(fc,\eta)}_{\alpha a}(\kk)}{N_M D_{1,\nu_c,\nu_f}}:f_{\RR,\alpha \eta s}^\dag f_{\RR,\alpha' \eta s}:\nonumber \\
&-\sum_{\kk,\alpha,\eta,s}\sum_{a,a'\in \{1,2\}}\frac{ |F(|\kk|)|^2H^{(fc,\eta)}_{\alpha a'}(\kk)H_{\alpha a}^{(fc,\eta),*}(\kk)}{D_{2,\nu_c,\nu_f}}:c_{\kk,a\eta s}^\dag c_{\kk,a'\eta s}:\bigg] \nonumber \\
&+\text{const} \nonumber  \\
=&\hH_0
+
\sum_{\RR,\kk, \kk',\eta,\eta',s,s'\alpha,\alpha}\sum_{a,a'\in\{1,2\}}\frac{e^{i(\kk-\kk')\RR}}{N_M}H^{(fc,\eta),*}_{\alpha'a'}(\kk')H^{(fc,\eta')}_{\alpha a}(\kk)
\hh{\frac{ |F(|\kk|)|^2}{D_{\nu_c,\nu_f}}} 
\bigg[ 
:f_{\RR,\alpha \eta s}^\dag f_{\RR,\alpha'\eta's'}::c_{\kk',a'\eta's'}^\dag c_{\kk,a\eta s}:
\bigg] \nonumber  \\
&+\bigg[ \sum_{\kk,\alpha,\alpha',\eta,s}\sum_{a\in \{1,2\}}\frac{ |F(|\kk|)|^2H^{(fc,\eta),*}_{\alpha'a}(\kk)H^{(fc,\eta)}_{\alpha a}(\kk)}{2N_M}\bigg( 
\frac{1}{D_{1,\nu_c,\nu_f}} +\frac{1}{D_{2,\nu_c,\nu_f}}
\bigg)\sum_{\RR}:f_{\RR,\alpha \eta s}^\dag f_{\RR,\alpha' \eta s}: \nonumber \\
&-\sum_{\kk,\alpha,\eta,s}\sum_{a,a'\in \{1,2\}}\frac{ |F(|\kk|)|^2H^{(fc,\eta)}_{\alpha a'}(\kk)H_{\alpha a}^{(fc,\eta),*}(\kk)}{2}\bigg( 
\frac{1}{D_{1,\nu_c,\nu_f}} +\frac{1}{D_{2,\nu_c,\nu_f}}
\bigg):c_{\kk,a\eta s}^\dag c_{\kk,a'\eta s}:\bigg] +\text{const}
\label{eq:sw_eff_ham_v1}
\eaa

\subsection{Effect of $\hH_J$ and $P_H\hH_{fc}P_H$ }
 We have ignored the effect of \bh{ferromagnetic exchange} coupling term $\hat{H}_J$ and also part of the hybridization term $P_H\hH_{fc}P_H$. We now add these terms to $\hH_0$
\baa  
&\hH_0^{complete} = \hH_0 +\lambda \hH_{\lambda} \nonumber \\
&\hH_{\lambda} =\hH_J  + P_H\hH_{fc}P_H
\eaa  
with $\lambda=1$ a dimensionless parameter \hh{introduced to keep track of the $\hH_{\lambda}$}. We can perform SW transformation based on $\hat{H}^{complete}_0$ and $\hat{H}_1$. The corresponding $S'$ operator needs to satisfy
\baa 
[S',\hat{H}^{complete}_0] = -\hat{H}_1 \, .
\eaa  
The analytical expression of $S'$ is complicated to find, but we can expand $S'$ in powers of $\lambda$ 
\ba 
S' =  \sum_{n=0}^{\infty}\lambda^n S_n \, .
\ea 
Then we have
\ba 
&[S', \hat{H}_0^{complete}] =-\hat{H}_1 \\
\Leftrightarrow &[\sum_{n=0}^\infty \lambda^n S_n, \hat{H}_0 +\lambda \hat{H}_\lambda] =-\hat{H}_1 \\
\Leftrightarrow&[S_0,\hat{H}_0] +\sum_{n=1}^\infty \hh{\lambda^n} \bigg([S_n,\hat{H}_0] + [S_{n-1},\hat{H}_\lambda]\bigg) = -\hat{H}_1 .
\ea 
We can find $S_n$ iteratively by requiring
\ba 
&[S_0,\hat{H}_0] = -\hat{H}_1 \\
&[S_{n},\hat{H}_0] = -[S_{n-1},\hat{H}_\lambda] \quad,\quad n\ge 1 . 
\ea 
$S_0=S$ is the operator we derived previously \hh{(Eq.~\ref{eq:sw_def_s})}.

The effective Hamiltonian after SW transformation now becomes
\ba 
 &(\hat{H}_0+\hat{H}_\lambda) + \frac{1}{2}([S',\hat{H}_1])
=\hat{H}_0 +\lambda\hat{H}_J
+\frac{1}{2} [S_0,\hat{H}_1] + \frac{1}{2} \sum_{n=1}^\infty \lambda^n [S_n, \hat{H}_1]\, .
\ea  
We only keep the leading-order ($\lambda^0$) contribution and the effective Hamiltonian is
\baa  
\hat{H}'_{eff} = \bigg( \hat{H}_0+\hat{H}_J + \frac{1}{2}[S,\hat{H}_1]  
\bigg)  = \hat{H}_{eff} + \hH_J 
\label{eq:sw_eff_ham_v2}
\eaa 
\hh{where $\hH_{eff}$ is defined in Eq.~\ref{eq:sw_eff_ham_v1}.}
We are only interested in the states in the low-energy space, where the filling of $f$ at each site is $\nu_f$, so we drop $P_H\hH_{fc}P_H$ term.

\subsection{Effective Kondo model}
\hh{We are now in the position to write down the effective Kondo model derived from SW transformation. Combining Eq.~\ref{eq:sw_eff_ham_v2} and Eq.~\ref{eq:sw_eff_ham_v1}, we have the following Kondo Hamiltonian}
\baa  
\hh{ \hH_{Kondo} = \hH_{eff}' } =&\hH_0 + \hH_J  \nonumber  \\
&+
\sum_{\RR,\kk, \kk',\eta,\eta',s,s'\alpha,\alpha}\sum_{a,a'\in\{1,2\}}\frac{e^{i(\kk-\kk')\RR}}{N_M}H^{(fc,\eta),*}_{\alpha'a'}(\kk')H^{(fc,\eta')}_{\alpha a}(\kk)
\frac{F(|\kk|)^2}{D_{\nu_c,\nu_f} }
\bigg[ 
:f_{\RR,\alpha \eta s}^\dag f_{\RR,\alpha'\eta's'}::c_{\kk',a'\eta's'}^\dag c_{\kk,a\eta s}:
\bigg]  \nonumber \\
&+\bigg[ \sum_{\RR,\kk,\alpha,\alpha',\eta,s}\sum_{a\in \{1,2\}}\frac{H^{(fc,\eta),*}_{\alpha'a}(\kk)H^{(fc,\eta)}_{\alpha a}(\kk)F(|\kk|)^2}{2N_M}\bigg( 
\frac{1}{D_{1,\nu_c,\nu_f}} +\frac{1}{D_{2,\nu_c,\nu_f}}
\bigg):f_{\RR,\alpha \eta s}^\dag f_{\RR,\alpha' \eta s}: \nonumber \\
&-\sum_{\kk,\alpha,\eta,s}\sum_{a,a'\in \{1,2\}}\frac{H^{(fc,\eta)}_{\alpha a'}(\kk)H_{\alpha a}^{(fc,\eta),*}(\kk)F(|\kk|)^2}{2}\bigg( 
\frac{1}{D_{1,\nu_c,\nu_f}} +\frac{1}{D_{2,\nu_c,\nu_f}}
\bigg):c_{\kk,a\eta s}^\dag c_{\kk,a'\eta s}:\bigg]
\label{eq:sw_hkondo_v0}
\eaa  
\hh{We expand the hybridization matrix in powers of $\kk$ and only keep the zeroth-order and linear order terms:
\baa 
H^{(fc,\eta)}( \kk) = 
 \begin{bmatrix}
\gamma +O(|\kk|^2)& v_\star^\prime (\eta k_x - ik_y)+O(|\kk|^2) & 0 & 0\\
v_\star^\prime (\eta k_x + ik_y)+O(|\kk|^2)  &\gamma +O(|\kk|^2) &0 &0
\end{bmatrix}
\label{eq:hyb_mat}
\eaa
}
\hh{ 
The hybridization between $f$-fermions and $\Gamma_1 \oplus \Gamma_2$ $c$-fermions is \hb{relatively} weak and has been omitted.
}

Besides the four-fermion interactions term, \hb{the} SW transformation introduces two additional fermion-bilinear term 
\hh{as shown in Eq.~\ref{eq:sw_hkondo_v0}}. The first one is
\hh{ 
\ba 
& \sum_{\RR,\kk,\alpha,\alpha',\eta,s}\sum_{a}\frac{F(|\kk|)^2H^{(fc,\eta),*}_{\alpha'a}(\kk)H^{(fc,\eta)}_{\alpha a}(\kk)}{2N_M}\bigg( 
\frac{1}{D_{1,\nu_c,\nu_f}} +\frac{1}{D_{2,\nu_c,\nu_f}}
\bigg):f_{\RR,\alpha \eta s}^\dag f_{\RR,\alpha' \eta s}:\\
=&\sum_{\RR,\kk,\alpha,\eta,s} F(|\kk|)^2\frac{\gamma^2 +O(|\kk|^2)}{2N_M}\bigg( 
\frac{1}{D_{1,\nu_c,\nu_f}} +\frac{1}{D_{2,\nu_c,\nu_f}}
\bigg):f_{\RR,\alpha \eta s}^\dag f_{\RR,\alpha \eta s}:
\approx \sum_{\kk} F(|\kk|)^2\frac{\gamma^2 }{2}\bigg( 
\frac{1}{D_{1,\nu_c,\nu_f}} +\frac{1}{D_{2,\nu_c,\nu_f}}
\bigg)\nu_f \\
\ea 
}
which is a constant. The second one corresponds to a conduction electron scattering term
\hh{ 
\baa   
\hH_{cc} = &-\sum_{\kk,\eta,s}\sum_{a,a'\in \{1,2\}}\frac{F(|\kk|)^2H^{(fc,\eta)}_{\alpha a'}(\kk)H_{\alpha a}^{(fc,\eta),*}(\kk)}{2}\bigg( 
\frac{1}{D_{1,\nu_c,\nu_f}} +\frac{1}{D_{2,\nu_c,\nu_f}}
\bigg):c_{\kk,a\eta s}^\dag c_{\kk,a'\eta s}:  \nonumber \\
= &-\sum_{\kk,\eta,s}
\bigg( 
\frac{1}{D_{1,\nu_c,\nu_f}} +\frac{1}{D_{2,\nu_c,\nu_f}}
\bigg) F(|\kk|)^2
\sum_{a,a' \in \{1,2\} }\Bigg[ 
\begin{bmatrix}
\gamma^2/2 & \gamma v_\star^\prime (\eta k_x -ik_y) \nonumber  \\
\gamma v_\star^\prime (\eta k_x + ik_y) &\gamma^2/2
\end{bmatrix}_{a,a'} +O(|\kk|^2) \Bigg] 
:c_{\kk,a\eta s}^\dag c_{\kk, a'\eta s}:
\\
=& \sum_{\kk,\eta ,s}\sum_{a,a' \in \{1,2\}} [H_{cc}^\eta(\kk)]_{aa'}:c_{\kk, a \eta s}^\dag c_{\kk,a' \eta s}:
\label{eq:hcc_v0} 
\eaa  
where $H_{cc}^\eta(\kk) = - \bigg( 
\frac{1}{D_{1,\nu_c,\nu_f}} +\frac{1}{D_{2,\nu_c,\nu_f}}
\bigg) F(|\kk|)^2 \bigg[ 
\frac{\gamma^2}{2}\sigma_0 +\gamma v_\star^\prime( \eta k_x\sigma_x +k_y\sigma_y)
\bigg] $
}


\hh{We define the four-fermion interaction induced by SW transformation in Eq.~\ref{eq:sw_hkondo_v0} as Kondo interactions $\hH_{K}$ and rewrite it in a more compact form:}
\hh{ 
\baa  
\hH_K 
=
&\sum_{\RR,\kk,\kk',\alpha,\alpha',a,a',\eta,\eta',s,s'}e^{i(\kk-\kk') \RR}
\frac{H_{\alpha a }^{(fc,\eta)}(\kk)(H^{(fc,\eta')}_{\alpha'a'}(\kk'))^* }{N_M D_{\nu_c,\nu_f} } 
:f_{\RR,\alpha \eta s}^\dag  f_{\RR,\alpha'\eta' s'}: :c_{\kk',a'\eta' s'}^\dag c_{\kk, a\eta s}:
\nonumber \\
=&\sum_{\RR,\kk,\kk',\alpha,\alpha',a,a',\eta,\eta',s,s'}e^{i(\kk-\kk') \RR}
F(|\kk|)F(|\kk'|)
\bigg[ 
\frac{
\gamma ^2 \delta_{\alpha',a'} \delta_{\alpha, a}
}{N_M D_{\nu_c,\nu_f}} +\frac{
\gamma v_\star^\prime \delta_{\alpha, a}[\eta' k'_x\sigma_x - k'_y\sigma_y]_{\alpha'a'}
}{N_M D_{\nu_c,\nu_f}} +
\frac{
\gamma v_\star^\prime \delta_{\alpha', a'}[\eta k_x\sigma_x + k_y\sigma_y]_{\alpha a}
}{N_M D_{\nu_c,\nu_f}}  \nonumber \\
&+O(|\kk|^2,|\kk'|^2,|\kk||\kk'|)
\bigg]  
:f_{\RR,\alpha \eta s}^\dag  f_{\RR,\alpha'\eta' s'}: :c_{\kk',a'\eta' s'}^\dag c_{\kk, a\eta s}:   \nonumber \\
\approx &\sum_{\RR,\kk,\kk',\alpha,\alpha',a,a',\eta,\eta',s,s'}e^{i(\kk-\kk') \RR}F(|\kk|)F(|\kk'|)
\bigg[ 
\frac{
\gamma ^2 \delta_{\alpha',a'} \delta_{\alpha, a}
}{N_MD_{\nu_c,\nu_f}} +\frac{
\gamma v_\star^\prime \delta_{\alpha, a}[\eta' k'_x\sigma_x - k'_y\sigma_y]_{\alpha'a'}
}{N_MD_{\nu_c,\nu_f}} +
\frac{
\gamma v_\star^\prime \delta_{\alpha', a'}[\eta k_x\sigma_x + k_y\sigma_y]_{\alpha a}
}{N_MD_{\nu_c,\nu_f}} 
\bigg] \nonumber  \\
&
:f_{\RR,\alpha \eta s}^\dag  f_{\RR,\alpha'\eta' s'}: :c_{\kk',a'\eta' s'}^\dag c_{\kk, a\eta s}:
\label{eq:kondo_ham_mom}
\eaa  
} 
\hh{Using the U(4) momentum defined in Sec.~\ref{sec:u4mom} and Eq.~\ref{eq:u4_mom_sum_2}}, we rewrite $\hH_K$ as
\baa 
\hH_K =
&\sum_{\RR,\kk,\kk',\alpha,\alpha',a,a',\eta,\eta',s,s'}e^{i(\kk-\kk') \RR}
\hh{F(|\kk|)F(|\kk'|)}
\bigg[ 
\frac{
\gamma ^2 \delta_{\alpha',a'} \delta_{\alpha, a}
}{N_MD_{\nu_c,\nu_f}} +\frac{
\gamma v_\star^\prime \delta_{\alpha, a}[\eta' k'_x\sigma_x - k'_y\sigma_y]_{\alpha'a'}
}{N_MD_{\nu_c,\nu_f}} +
\frac{
\gamma v_\star^\prime \delta_{\alpha', a'}[\eta k_x\sigma_x + k_y\sigma_y]_{\alpha a}
}{N_MD_{\nu_c,\nu_f}} 
\bigg] \nonumber  \\
&
:f_{\RR,\alpha \eta s}^\dag  f_{\RR,\alpha'\eta' s'}: :c_{\kk',a'\eta' s'}^\dag c_{\kk, a\eta s}: \nonumber \\
=&
\sum_{\RR,\kk,\kk'}e^{-i(\kk'-\kk)\cdot \RR}
\frac{\hh{F(|\kk|)F(|\kk'|)}}{D_{\nu_c,\nu_f}N_M} \sum_{\mu\nu, \xi\xi'} \bigg[
\gamma^2 :\UF_{\mu\nu}^{(f,\xi\xi')}(\RR) : :\UF_{\mu\nu}^{(c',\xi'\xi)}(\kk,\kk'-\kk) : 
\nonumber \\
&+ \gamma v_\star^\prime  (k_x' +i\xi' k_y') :\UF_{\mu\nu}^{(f,\xi\xi')}(\RR) ::\UF_{\mu\nu}^{(c',-\xi'\xi)}(\kk,\kk'-\kk):
+\gamma v_\star^\prime( k_x -i\xi k_y) :\UF_{\mu\nu}^{(f,\xi\xi')}(\RR) ::\UF_{\mu\nu}^{(c',\xi',-\xi)}(\kk,\kk'-\kk):
\bigg] 
\label{eq:hk_exp_l}
\eaa  
\hb{where we have neglected the $(v_\star^\prime \gamma)^2$ which is the second order in $\kk$, and outsiede the scope of the Hamiltonian approximation in Eq.~\ref{eq:hyb_mat}. 
}

We next use the real-space representation of $U(8)$ moments \hh{as defined in Eq.~\ref{eq:u4_ft}}, which indicates
\ba 
\UF_{\mu\nu}^{(c',\xi\xi')}(\kk,\kk')
=\frac{1}{N_M} \sum_{\kk,\kk'}e^{-i\kk\cdot \rr' +i\kk'\cdot \rr} \UF_{\mu\nu}^{(c',\xi\xi')}(\rr,\rr') 
\ea 
Then \hh{Eq.~\ref{eq:hk_exp_l}} becomes 
\baa  
\hH_K 
=&
\sum_{\RR,\rr,\rr',\kk,\kk'}
\hh{\frac{F(|\kk|)F(|\kk'|)}{D_{\nu_c,\nu_f}N_M^2} }
e^{-i\kk'\cdot(\RR-\rr) +i\kk\cdot(\RR-\rr')}
\sum_{\mu\nu, \xi\xi'} \bigg[
\gamma^2 :\UF_{\mu\nu}^{(f,\xi\xi')}(\RR) : :\UF_{\mu\nu}^{(c',\xi'\xi)}(\rr,\rr') : 
\nonumber \\
&+ \gamma v_\star^\prime  (-i\partial_{r_x} +\xi' \partial_{r_y}) :\UF_{\mu\nu}^{(f,\xi\xi')}(\RR) ::\UF_{\mu\nu}^{(c',-\xi'\xi)}(\rr,\rr'):
+\gamma v_\star^\prime( i\partial_{r'_x} +\xi \partial_{r_y'}) :\UF_{\mu\nu}^{(f,\xi\xi')}(\RR) ::\UF_{\mu\nu}^{(c',\xi',-\xi)}(\rr,\rr'):
\bigg] 
\label{eq:hk_exp_l_2}
\eaa  
\hh{where the derivative is defined as
\ba 
&\partial_{r_\alpha} :\Sigma_{\mu\nu}^{(c',\xi'\xi)}(\rr,\rr'): = \partial_{r_\alpha}\frac{1}{N_M}\sum_{\kk} e^{i\kk \cdot\rr' -i\kk'\cdot \rr} :\Sigma_{\mu\nu}^{(c',\xi'\xi)}(\kk,\kk'-\kk) 
=\frac{1}{N_M}\sum_{\kk} ( -ik'_\alpha) :\Sigma_{\mu\nu}^{(c',\xi'\xi)}(\kk,\kk'-\kk) e^{i\kk \cdot\rr' -i\kk'\cdot \rr} \\
&\partial_{r'_\alpha} :\Sigma_{\mu\nu}^{(c',\xi'\xi)}(\rr,\rr'): = \partial_{r'_\alpha}\frac{1}{N_M}\sum_{\kk} e^{i\kk \cdot\rr' -i\kk'\cdot \rr} :\Sigma_{\mu\nu}^{(c',\xi'\xi)}(\kk,\kk'-\kk) 
=\frac{1}{N_M}\sum_{\kk} ( ik_{\alpha}) :\Sigma_{\mu\nu}^{(c',\xi'\xi)}(\kk,\kk'-\kk) e^{i\kk \cdot\rr' -i\kk'\cdot \rr} 
\ea 
}
\hh{ 
The $\kk$ summation appearing in Eq.~\ref{eq:hk_exp_l_2} takes form of 
\ba 
\frac{1}{N_M}\sum_{\kk}F(|\kk|)e^{i\kk\cdot(\RR-\rr')} \approx \frac{1}{A_{MBZ}}\int \exp(-\frac{1}{2}|\kk|^2\lambda^2)e^{i\kk\cdot(\RR-\rr')}d^2\kk 
=\frac{2\pi}{A_{MBZ}}\frac{e^{-|\RR-\rr'|^2/\lambda^2/2 } }{\lambda^2 } \, .
\ea 
For an arbitrary function of $\rr$, $F(\rr)$, we notice 
\ba 
\sum_{\rr} F(\rr)\bigg(\frac{1}{N_M}\sum_{\kk}F(|\kk|)e^{i\kk\cdot(\RR-\rr')}\bigg)  =& \sum_{\rr} F(\rr) \frac{2\pi}{A_{MBZ}}\frac{e^{-|\RR-\rr'|^2/\lambda^2/2 } }{\lambda^2 } \approx \frac{1}{\Omega_0} \int_{\rr}F(\rr) \frac{2\pi}{A_{MBZ}}\frac{e^{-|\RR-\rr'|^2/\lambda^2/2 } }{\lambda^2 } 
\ea 
For small $\lambda = 0.3375 a_M$, we can approximate $F(\rr)$ with its value at $\RR$. Inserting above equations into Eq.~\ref{eq:hk_exp_l_2}, we have
\ba 
\sum_{\rr} F(\rr)\bigg(\frac{1}{N_M}\sum_{\kk}F(|\kk|)e^{i\kk\cdot(\RR-\rr')}\bigg)  \approx & \frac{1}{\Omega_0} F(\RR) \int_{\rr} \frac{2\pi}{A_{MBZ}}\frac{e^{-|\RR-\rr'|^2/\lambda^2/2 } }{\lambda^2 }  = F(\RR) = \sum_{\rr} \delta_{\rr,\RR}F(\rr) 
\ea 
}

Finally, we have the following approximated real-space expression of $\hH_K$
\baa  
\hH_K\approx &
\sum_{\RR,\kk,\kk',,\xi,\xi'}e^{-i(\kk'-\kk)\cdot \RR}\frac{1}{D_{\nu_c,\nu_f}N_M}\nonumber  \\
&\sum_{\mu\nu, \xi\xi'} \bigg[
\gamma^2 :\UF_{\mu\nu}^{(f,\xi\xi')}(\RR) :
+ \gamma v_\star^\prime  (k_x' -i\xi' k_y') :\UF_{\mu\nu}^{(f,\xi-\xi')}(\RR) :
+\gamma v_\star^\prime( k_x +i\xi k_y) :\UF_{\mu\nu}^{(f,-\xi \xi')}(\RR) :
\bigg] :\UF_{\mu\nu}^{(c',\xi',\xi)}(\kk,\kk'-\kk): \nonumber \\
=&
\sum_{\RR,\xi,\xi'}\frac{\gamma^2}{D_{\nu_c,\nu_f}}
 :\UF_{\mu\nu}^{(f,\xi\xi')}(\RR) : :\UF_{\mu\nu}^{(c',\xi',\xi)}(\RR,\RR): \nonumber  \\
&
+\sum_{\RR,\rr,\xi,\xi'}\frac{\gamma v_\star^\prime }{D_{\nu_c,\nu_f}}
\delta_{\rr,\RR} \bigg[ :\UF_{\mu\nu}^{(f,\xi-\xi')}(\RR) :
 (i\partial_{r_x}+\xi' \partial_{r_y}) :\UF_{\mu\nu}^{(c',\xi'\xi)}(\RR,\rr):
 +:\UF_{\mu\nu}^{(f,-\xi\xi')}(\RR) :
 (-i\partial_{r_x}+\xi \partial_{r_y}) :\UF_{\mu\nu}^{(c',\xi'\xi)}(\rr,\RR):
\bigg] 
\label{eq:sw_kondo_int}
\eaa 

In summary, we have the following two four-fermion interactions: Kondo coupling $\hH_K$ and \bh{ferromagnetic exchange} $\hH_J$:
\hh{ 
\baa  
\hH_K=&
\sum_{\RR,\rr,\rr',\xi,\xi'}\frac{\delta_{\rr,\RR}\delta_{\rr',\RR} }{D_{\nu_c,\nu_f}} 
\sum_{\mu\nu} \bigg[
\gamma^2 :\UF_{\mu\nu}^{(f,\xi\xi')}(\RR) : \nonumber \\
&
+ \gamma v_\star^\prime   :\UF_{\mu\nu}^{(f,\xi-\xi')}(\RR) :(i\partial_{r_x'}+\xi' \partial_{r_y'})
+\gamma v_\star^\prime :\UF_{\mu\nu}^{(f,-\xi \xi')}(\RR) :( -i\partial_{r_x} +\xi \partial_{r_y})
\bigg]  :\UF_{\mu\nu}^{(c',\xi',\xi)}(\rr,\rr'):\nonumber  \\
\hH_J=&(-J)\sum_{\RR,\mu\nu,\xi} :\UF_{\mu\nu}^{(f,\xi\xi)}(\RR)::\UF_{\mu\nu}^{(c'',\xi\xi)}(\RR,\RR)
\label{eq:sw_kondo_int}
\eaa  
\bh{Here, we also provide the interactions in the momentum space ($\kk$-space of $c$-electrons) }
\baa  
\hH_K 
=&
\sum_{\RR,\kk,\kk'}e^{-i(\kk'-\kk)\cdot \RR}
\frac{\hh{F(|\kk|)F(|\kk'|)}}{D_{\nu_c,\nu_f}N_M} \sum_{\mu\nu, \xi\xi'} \bigg[
\gamma^2 :\UF_{\mu\nu}^{(f,\xi\xi')}(\RR) : :\UF_{\mu\nu}^{(c',\xi'\xi)}(\kk,\kk'-\kk) : 
\nonumber \\
&+ \gamma v_\star^\prime  (k_x' +i\xi' k_y') :\UF_{\mu\nu}^{(f,\xi\xi')}(\RR) ::\UF_{\mu\nu}^{(c',-\xi'\xi)}(\kk,\kk'-\kk):
+\gamma v_\star^\prime( k_x -i\xi k_y) :\UF_{\mu\nu}^{(f,\xi\xi')}(\RR) ::\UF_{\mu\nu}^{(c',\xi',-\xi)}(\kk,\kk'-\kk):
\bigg] \nonumber \\ 
\hH_J 
=&-J
\sum_{\RR,\kk,\kk',\xi}
\frac{e^{-i(\kk'-\kk)\cdot \RR}}{N_M} \sum_{\mu\nu,\xi} :\UF_{\mu\nu}^{(f,\xi\xi)}(\RR): :\UF_{\mu\nu}^{(c'',\xi\xi)}(\kk,\kk'-\kk):
\label{eq:kondo_model_int}
\eaa  
}
\hh{
The one-body scattering term from SW transformation is given in Eq.~\ref{eq:hcc_v0}, and is listed below 
\baa   
\hH_{cc} 
=& \sum_{\kk,\eta ,s}\sum_{a,a' \in \{1,2\}} [H_{cc}^\eta(\kk)]_{aa'}:c_{\kk, a \eta s}^\dag c_{\kk,a' \eta s}:
 \nonumber \\
H_{cc}^\eta(\kk) =& - \bigg( 
\frac{1}{D_{1,\nu_c,\nu_f}} +\frac{1}{D_{2,\nu_c,\nu_f}}
\bigg) F(|\kk|)^2 \bigg[ 
\frac{\gamma^2}{2}\sigma_0 +\gamma v_\star^\prime( \eta k_x\sigma_x +k_y\sigma_y)
\bigg] 
\label{eq:hcc_def_v2}
\eaa  
The effective Kondo model now reads
\baa  
\hH_{Kondo} = \hH_0 +\hH_K +\hH_J +\hH_{cc} \, .
\label{eq:kondo_model}
\eaa  
}
Here, we also want to mention that the model is defined with respect to the Hilbert space with the filling of $f$ electrons being $\nu_f$ at each site. We estimate the coupling strength here 
\hh{ 
\begin{center}
\begin{tabular}{|c|c|c|c|c}
$\nu_f$ & 0 & -1 & -2  \\
\hline 
 $J$ & 16meV & 16meV & 16meV  \\
 \hline 
$ \frac{\gamma^2}{D_{\nu_c=0,\nu_f}} $ & 42meV & 49meV &99meV
\\ 
\hline 
$\frac{\gamma |v_\star^\prime|/ a_M}{D_{\nu_c=0,\nu_f}}$
&
20.6meV& 24meV& 48meV
 \end{tabular}
\end{center}
}
where $\gamma = -24.75meV$, $v_\star^\prime/a_M = 12.08meV$, $a_M$ is the moir\'e lattice constant and $1/D  = 1/D_{\nu_c=0,\nu_f}$.

\section{RKKY interactions at $M=0$}
\label{sec:rkky}
\bh{Based on the Kondo model defined in Eq.~\ref{eq:kondo_model}, we derive the corresponding RKKY interactions by integrating out conduction electrons.}

The action of the effective Hamiltonian $\hH_{eff}$ (Eq.~\ref{eq:kondo_model}) can be separated into 
\hh{ 
\ba 
&S = S_0+S_1 \\
&S_0=S_{f} +S_c\quad,\quad 
S_1= S_K +S_J +S_{cc} \, . 
\ea 
}
\hh{$S_{f}$ and $S_c$ denote the effective action of $f$- and $c$- electrons that have been introduced in Eq.~\ref{eq:action_sf_lag} and Eq.~\ref{eq:action_sc} and are also given below
\bh{  
\baa  
S_f &= \sum_{\RR, \alpha \eta s } \int_0^\beta d\tau f_{\RR,\alpha\eta s}^\dag(\tau)  \partial_\tau  f_{\RR,\alpha \eta s}(\tau)  d\tau 
+i\sum_{\RR}\int_0^\beta \lambda_\RR(\tau)
[ \sum_{ \alpha \eta s } f_{\RR,\alpha \eta s}^\dag(\tau)f_{\RR,\alpha \eta s}(\tau)-4 -\nu_f)] d\tau 
  \nonumber \\
  S_c = &\sum_{\kk,a\eta s} \int_0^{\beta} d\tau c_{\kk, a\eta s}^\dag(\tau)  \partial_\tau c_{\kk , a \eta s}(\tau)d\tau   + \int_0^\beta \sum_{\eta,s,a,a',\kk}\bigg(H^{(c,\eta)}_{a,a'}
+(-\mu + W \nu_f + \frac{V_0}{\Omega_0}\nu_c)\delta_{a,a'}
\bigg) 
c_{\kk,a\eta s}^\dag (\tau)
c_{\kk,a'\eta s}(\tau) d\tau \nonumber \\
 &- \frac{ V_0N_M}{2\Omega_0} \nu_c^2 - (W\nu_f +\frac{V_0}{\Omega_0}\nu_c)\sum_{\kk} 8
\eaa 
where we have introduced Lagrangian multiplier $\lambda_\RR(\tau)$ to fix the filling of $f$-electrons. $\hH_U$ and chemical potential term of $f$-electrons are a constants after fixing the filling of $f$-electrons and have been omitted. We also treat $\hH_V$ via a mean-field approximation. 
$S_K,S_J,S_{cc}$ are defined as
\baa  
S_K=\int_0^\beta \hH_K(\tau) d\tau , \quad S_J=\int_0^\beta \hH_J(\tau)d\tau ,\quad 
 S_{cc}=\int_0^\beta \hH_{cc}(\tau)d\tau \, .
 \label{eq:action_sk_sj_scc}
\eaa 
}
\hb{$\hH_K(\tau), \hH_J(\tau ), \hH_{cc}(\tau)$ take the same for as $\hH_K$ 
 (Eq.~\ref{eq:kondo_model_int}), $\hH_J$ (Eq.~\ref{eq:kondo_model_int}), $\hH_{cc}$ (Eq.~\ref{eq:hcc_def_v2}), but with $f_{\RR,\alpha \eta s}$ replaced  by $f_{\RR,\alpha \eta s}(\tau)$, $c_{\kk,a \eta  s}$ replaced by $c_{\kk,a\eta s}(\tau)$, where $\tau$ denotes imaginary time.}}

We now integrate out \bh{the} $c$-electrons in the partition functions and perform the cumulant expansion: $\langle e^{O}\rangle = \exp\bigg( \langle O \rangle +\frac{1}{2}\langle O^2\rangle - \frac{1}{2}\langle O\rangle ^2 +...\bigg)$. 
\hh{
\ba 
Z=&\int D[f_{\RR,\alpha\eta s}^\dag(\tau), f_{\RR,\alpha \eta s}(\tau),\lambda_{\RR}(\tau)]\int D[c_{\kk,a\eta s}^\dag(\tau) ,c_{\kk,a\eta s}(\tau)] e^{-S_0-S_1  } \\
=& \int D[f_{\RR,\alpha\eta s}^\dag(\tau), f_{\RR,\alpha \eta s}(\tau),\lambda_{\RR}(\tau)]e^{-S_f}\langle e^{-S_1}\rangle_0 \\
\approx &\int D[f_{\RR,\alpha\eta s}^\dag(\tau), f_{\RR,\alpha \eta s}(\tau),\lambda_{\RR}(\tau)] \exp\bigg[-S_f
-\langle S_1\rangle_0 +\frac{1}{2} \langle S_1^2\rangle_0 - \frac{1}{2} (\langle S_1\rangle_0 )^2 +...
\bigg] 
\ea 
}
where 
\ba 
&\langle O \rangle_0 :=\frac{1}{Z_0}  \int D[c_{\kk,a\eta s}^\dag(\tau) ,c_{\kk,a\eta s}(\tau)] O e^{-S_c } \quad,\quad 
Z_0 :=\int  D[c_{\kk,a\eta s}^\dag(\tau) ,c_{\kk,a\eta s}(\tau)]  e^{-S_c} 
\ea 

The effective actions of $f$ fermions then becomes \hh{ 
\baa  
S_{eff} = S_f + \langle S_K +S_J+S_{cc}\rangle_0 
-\bigg[ 
\frac{\langle S_J^2 \rangle_{0,con}}{2} 
+\frac{\langle S_K^2 \rangle_{0,con}}{2}
+\langle S_{J}S_K \rangle_{0,con} 
+\frac{\langle S_{cc}^2 \rangle_{0,con}}{2} 
+\langle (S_K+S_J)S_{cc} \rangle_{0,con}
\bigg] 
\label{eq:eff_action_spin}
\eaa  }
with $\langle A B\rangle_{0,con} = \langle A B\rangle_{0}-\langle A\rangle_0 \langle B\rangle_0$. \hh{We treat $S_1$ as perturbation, since all terms in $S_1$ are either generated from SW transformation that is proportional to $\gamma^2,(v_\star^\prime)^2,\gamma v_\star^\prime$, or coming from the \bh{ferromagnetic exchange coupling} which is proportional to $J$}. We derive the effective action \bh{at $M=0$} and integer filling $\nu=0,-1,-2$ with $\nu_f=\nu,\nu_c=0$.

\hh{We first note that, $S_{cc}$ \bh{(see Eq.~\ref{eq:action_sk_sj_scc} and Eq.~\ref{eq:hcc_def_v2})} does not contain any $f$ electron operators, so $\langle S_{cc}\rangle_{0}$ and $\langle S_{cc}^2\rangle_{0,con}$ only give constant contributions.}
As for the linear term: $\langle S_K+S_J\rangle_0$, $\langle S_K\rangle_0$ gives
\baa 
\langle S_K\rangle_0  
=&\int_0^\beta 
\sum_{\RR,\kk,\kk',,\xi,\xi'}e^{-i(\kk'-\kk)\cdot \RR}\frac{1}{D_{\nu_c,\nu_f}N_M}\langle :\UF_{\mu\nu}^{(c',\xi',\xi)}(\kk,\kk'-\kk,\tau):\rangle_0  \nonumber \\
&\sum_{\mu\nu, \xi\xi'} \bigg[
\gamma^2 :\UF_{\mu\nu}^{(f,\xi\xi')}(\RR,\tau) :
+ \gamma v_\star^\prime  (k_x' -i\xi' k_y') :\UF_{\mu\nu}^{(f,\xi-\xi')}(\RR,\tau) :
+\gamma v_\star^\prime( k_x +i\xi k_y) :\UF_{\mu\nu}^{(f,-\xi \xi')}(\RR,\tau) :
\bigg]  
d\tau 
\eaa 
We need to calculate $\langle :\UF_{\mu\nu}^{(c',\xi',\xi)}(\kk,\kk'-\kk,\tau):\rangle_0 $. 
Due to the flat $U(4)$ symmetry, only $\mu\nu=00$ component can be non-zero \bh{ for the same reason proved around Eq.~\ref{eq:zero_linear_uc}}. The momentum conservation requires $\kk'=\kk$. Due to imaginary-time translational symmetry, we only need to consider $\tau=0$
\baa 
\langle :\UF_{\mu\nu}^{(c',\xi'\xi)}(\kk,0)\rangle  =\delta_{\mu,0}\delta_{\nu,0}\langle\sum_m \psi_{\kk,m}^{c',\xi,\dag} \psi_{\kk,m}^{c',\xi'} \rangle_0 
 =\delta_{\mu,0}\delta_{\nu,0}\delta_{\xi,\xi'}\langle\sum_m \psi_{\kk,m}^{c',\xi,\dag} \psi_{\kk,m}^{c',\xi} \rangle_0 \, 
\eaa 
The above quantity is just the filling of conduction electrons in orbitals $1,2$ with index $\xi$. At $\nu_c=0$, this becomes zero. Then $\langle S_K\rangle_0 =0$. 
Another linear term is
$
\langle S_J\rangle_0 
$. We have 
\baa 
\langle S_J\rangle_0 = \int_0^\beta (-J)\sum_{\RR,\mu\nu,\xi} :\UF_{\mu\nu}^{(f,\xi\xi)}(\RR,\tau):
\langle :\UF_{\mu\nu}^{(c'',\xi\xi)}(\RR,\RR,\tau)\rangle_0d\tau 
\eaa 
which requires the calculation of 
$
\langle :\UF_{\mu\nu}^{(c'',\xi\xi)}(\kk,\kk'-\kk)\rangle_0
$. 
Similarly, the only non-zero component corresponds to $\mu\nu=00$ and $\kk'-\kk$. We have
\baa 
\langle :\UF_{\mu\nu}^{(c'',\xi\xi)}(\kk,0)\rangle  =\langle\sum_m \psi_{\kk,m}^{c'',\xi,\dag} \psi_{\kk,m}^{c'',\xi} \rangle_0 \, 
\eaa 
which is the filling of conduction electrons in orbitals $3,4$ with index $\xi$. At $\nu_c=0$, this becomes zero. Then $\langle S_J\rangle_0 =0$.

The remaining terms give the following spin-spin interaction
\baa  
S_{RKKY} = -
\bigg[ 
\frac{\langle S_J^2 \rangle_{0,con}}{2} 
+\frac{\langle S_K^2 \rangle_{0,con}}{2}
+\langle S_{J}S_K \rangle_{0,con}  
\bigg] 
\label{eq:rkky_action}
\eaa  
The effective action becomes
\baa  
S_{eff} =S_f+S_{RKKY}\, .
\eaa  

\hb{
Before calculating $S_{RKKY}$, here, we first introduce the following single-particle Green's function at $M=0$, as also derived in Sec.~\ref{sec:green},
\baa  
G_{aa',\eta}( \kk,\tau) = -\langle T_\tau c_{\kk,a\eta s}(\tau) c_{\kk,a'\eta s}^\dag(0)\rangle  
\eaa 
where $T_\tau$ denotes time ordering. We further introduce $g_0(\kk,\tau)$ and $g_{2,\eta}(\kk,\tau)$ (Sec.~\ref{sec:green}, Eq.~\ref{eq:green_def_2}) as 
\baa  
&g_0(\kk,\tau) =G_{11,\eta}(\kk,\tau) =G_{22,\eta}(\kk,\tau) =G_{33,\eta}(\kk,\tau) = G_{44,\eta}(\kk,\tau) \nonumber \\
&g_{2,\eta}(\kk,\kk)=G_{13,\eta}(\kk,\tau) =G^*_{24,\eta}(\kk,-\tau)
=G^*_{31,\eta}(\kk,-\tau)=G_{42,\eta}(\kk,\tau) 
\label{eq:g0_g2_def}
\eaa 
}

\subsection{ $-\langle (S_K+S_J)S_{cc}\rangle_{0,con}$ }
\hh{ 
We now prove 
$-\langle (S_K+S_J)S_{cc}\rangle_{0,con}$ only produces a constant contribution. For $\langle S_K S_{cc}\rangle_{0,con}$, we have 
\baa  
\langle S_K S_{cc}\rangle_{0,con} =&\int_0^\beta \int_0^\beta  \sum_{\RR,\kk,\kk',\kk_2,\xi,\xi'}e^{-i(\kk'-\kk)\cdot \RR} \sum_{\eta_2s_2}\sum_{a_2,a_2'\in 1,2}\sum_{\mu\nu} \frac{1}{D_{\nu_c,\nu_f}N_M}
[H_{cc}^{\eta_2}(\kk_2)]_{a_2a_2'}
\nonumber  \\
&\bigg[
\gamma^2 :\UF_{\mu\nu}^{(f,\xi\xi')}(\RR) :
+ \gamma v_\star^\prime  (k_x' -i\xi' k_y') :\UF_{\mu\nu}^{(f,\xi-\xi')}(\RR,\tau_1) :
+\gamma v_\star^\prime( k_x +i\xi k_y) :\UF_{\mu\nu}^{(f,-\xi \xi')}(\RR,\tau_1) :
\bigg] \nonumber \\
&
\langle :\UF_{\mu\nu}^{(c',\xi',\xi)}(\kk,\kk'-\kk,\tau_1):  :c_{\kk_2,a_2\eta_2 s_2}^\dag(\tau_2) c_{\kk_2,a_2'\eta_2s_2}(\tau_2):\rangle_{0,con}
d\tau_1d\tau_2 
\label{eq:skscc0_0}
\eaa 
We need to evaluate $\langle  :\UF_{\mu\nu}^{(c',\xi',\xi)}(\kk,\kk'-\kk,\tau_1):  :c_{\kk_2,a_2\eta_2 s_2}^\dag(\tau_2) c_{\kk_2,a_2'\eta_2s_2}(\tau_2):\rangle_{0,con}$. For $\xi'=\xi$, we have 
\baa  
&\langle  :\UF_{\mu\nu}^{(c',\xi,\xi)}(\kk,\kk'-\kk,\tau_1):  :c_{\kk_2,a_2\eta_2 s_2}^\dag(\tau_2) c_{\kk_2,a_2'\eta_2s_2}(\tau_2):\rangle_{0,con}  \nonumber \\
=&\sum_{\eta ,\eta',s,s'}\sum_{a,a'\in\{1,2\}} 
\frac{1}{2}A_{a\eta s,a'\eta's'}^{\mu\nu} \delta_{\xi,(-1)^{a+1}\eta}
\delta_{\xi,(-1)^{a'+1}\eta'}
\langle : c_{\kk',a\eta s}^\dag(\tau_1) c_{\kk,a'\eta' s'}(\tau_1):  :c_{\kk_2,a_2\eta_2 s_2}^\dag(\tau_2) c_{\kk_2,a_2'\eta_2s_2}(\tau_2):\rangle_{0,con}  \nonumber \\
=& \sum_{\eta ,\eta',s,s'}\sum_{a,a'\in\{1,2\}} 
\frac{1}{2}A_{a\eta s,a'\eta's'}^{\mu\nu} \delta_{\xi,(-1)^{a+1}\eta}
\delta_{\xi,(-1)^{a'+1}\eta'}
\delta_{a,a_2'}\delta_{s,s_2}\delta_{s',s_2}\delta_{\eta,\eta_2}\delta_{\eta',\eta_2}\delta_{\kk,\kk_2}\delta_{\kk',\kk_2}
\nonumber 
\\
&(-1) g_0(\tau_1-\tau_2,\kk_2)g_0(\tau_2-\tau_1,\kk_2)  \nonumber 
\\
=& -2 \delta_{\mu,0}\delta_{\nu,0}g_0(\tau_1-\tau_2,\kk_2)g_0(\tau_2-\tau_1,\kk_2)\delta_{\kk,\kk_2}\delta_{\kk',\kk_2}\delta_{a_2,a_2'}
\label{eq:skscc0_1}
\eaa
where we use Wick's theorem \bh{and the definition of Green's function in Eq.\ref{eq:g0_g2_def}.}
As for $\xi'=-\xi=-1$, we have 
\baa  
&\langle  :\UF_{\mu\nu}^{(c',-1,+1)}(\kk,\kk'-\kk,\tau_1):  :c_{\kk_2,a_2\eta_2 s_2}^\dag(\tau_2) c_{\kk_2,a_2'\eta_2s_2}(\tau_2):\rangle_{0,con} \nonumber \\
=&\sum_{\eta ,\eta',s,s'}\sum_{a,a'\in\{1,2\}} 
\frac{1}{2} A_{a\eta s,a'\eta's'}^{\mu\nu} \delta_{-1,(-1)^{a+1}\eta}
\delta_{+1,(-1)^{a'+1}\eta'}
\langle : \eta' c_{\kk',a\eta s}^\dag(\tau_1) c_{\kk,a'\eta' s'}(\tau_1):  :c_{\kk_2,a_2\eta_2 s_2}^\dag(\tau_2) c_{\kk_2,a_2'\eta_2s_2}(\tau_2):\rangle_{0,con} \nonumber\\
=&\sum_{a,a'\in\{1,2\}} 
\frac{1}{2} A_{a_2'\eta_2 s_2,a_2\eta_2s_2}^{\mu\nu} \delta_{-1,(-1)^{a_2'+1}\eta_2}
\delta_{+1,(-1)^{a_2+1}\eta_2}\eta_2(-1) g_0(\tau_1-\tau_2,\kk_2)g_2(\tau_2-\tau_1,\kk_2)  \nonumber \\ 
=&\bh{ g_0(\tau_1-\tau_2,\kk_2)g_2(\tau_2-\tau_1,\kk_2) \sum_{a,a'\in\{1,2\}} 
\frac{1}{2} A_{a_2'\eta_2 s_2,a_2\eta_2s_2}^{\mu\nu} \delta_{-1,(-1)^{a_2'+1}\eta_2}
\delta_{+1,(-1)^{a_2+1}\eta_2}\eta_2(-1) } \nonumber \\ 
= &\bh{g_0(\tau_1-\tau_2,\kk_2)g_2(\tau_2-\tau_1,\kk_2)\times 0 }= 0
\label{eq:skscc0_2}
\eaa 
where we use Wick's theorem \bh{and the definition of Green's function in Eq.\ref{eq:g0_g2_def}}.
Similarly, for $\xi'=-\xi=+1$
\baa 
&\langle  :\UF_{\mu\nu}^{(c',+1,+-1)}(\kk,\kk'-\kk,\tau_1):  :c_{\kk_2,a_2\eta_2 s_2}^\dag(\tau_2) c_{\kk_2,a_2'\eta_2s_2}(\tau_2):\rangle_{0,con} \nonumber \\
=&\sum_{\eta ,\eta',s,s'}\sum_{a,a'\in\{1,2\}} 
\frac{1}{2} A_{a\eta s,a'\eta's'}^{\mu\nu} \delta_{1,(-1)^{a+1}\eta}
\delta_{-1,(-1)^{a'+1}\eta'}
\langle : \eta c_{\kk',a\eta s}^\dag(\tau_1) c_{\kk,a'\eta' s'}(\tau_1):  :c_{\kk_2,a_2\eta_2 s_2}^\dag(\tau_2) c_{\kk_2,a_2'\eta_2s_2}(\tau_2):\rangle_{0,con} \nonumber \\
=&\sum_{a,a'\in\{1,2\}} 
\frac{1}{2} A_{a_2'\eta_2 s_2,a_2\eta_2s_2}^{\mu\nu} \delta_{1,(-1)^{a_2'+1}\eta_2}
\delta_{-1,(-1)^{a_2+1}\eta_2}\eta(-1) g_0(\tau_1-\tau_2,\kk_2)g_2(\tau_2-\tau_1,\kk_2) \nonumber \\ 
=& \bh{g_0(\tau_1-\tau_2,\kk_2)g_2(\tau_2-\tau_1,\kk_2) 
 \sum_{a,a'\in\{1,2\}} 
\frac{1}{2} A_{a_2'\eta_2 s_2,a_2\eta_2s_2}^{\mu\nu} \delta_{1,(-1)^{a_2'+1}\eta_2}
\delta_{-1,(-1)^{a_2+1}\eta_2}\eta(-1)}\nonumber \\
= &\bh{g_0(\tau_1-\tau_2,\kk_2)g_2(\tau_2-\tau_1,\kk_2) \times 0}=0
\label{eq:skscc0_3}
\eaa 
where we use Wick's theorem \bh{and the definition of Green's function in Eq.\ref{eq:g0_g2_def}}
}

\hhb{Combining Eq.~\ref{eq:skscc0_0}, Eq.~\ref{eq:skscc0_1}, Eq.~\ref{eq:skscc0_2} and Eq.~\ref{eq:skscc0_3}, }
\hh{ 
we have
\baa 
&\langle S_KS_{cc}\rangle_{0,con} \nonumber \\
=&\int_0^\beta \int_0^\beta  \sum_{\RR,\kk,\kk',\kk_2\xi}e^{-i(\kk'-\kk)\cdot \RR} \sum_{\eta_2s_2}\sum_{a_2,a_2'\in 1,2}\sum_{\mu\nu} \frac{1}{D_{\nu_c,\nu_f}N_M}
[H_{cc}^{\eta_2}(\kk_2)]_{a_2a_2'}
\nonumber  \\
&\bigg[
\gamma^2 :\UF_{\mu\nu}^{(f,\xi\xi)}(\RR) :
+ \gamma v_\star^\prime  (k_x' -i\xi k_y') :\UF_{\mu\nu}^{(f,\xi-\xi)}(\RR,\tau_1) :
+\gamma v_\star^\prime( k_x +i\xi k_y) :\UF_{\mu\nu}^{(f,-\xi \xi)}(\RR,\tau_1) :
\bigg] \nonumber \\
&
\langle :\UF_{\mu\nu}^{(c',\xi,\xi)}(\kk,\kk'-\kk,\tau_1):  :c_{\kk_2,a_2\eta_2 s_2}^\dag(\tau_2) c_{\kk_2,a_2'\eta_2s_2}(\tau_2):\rangle_{0,con}
d\tau_1d\tau_2 \nonumber  \\
=&
\int_0^\beta \int_0^\beta  \sum_{\RR,\kk,\xi} \sum_{\eta_2s_2}\sum_{a_2\in 1,2}\sum_{\mu\nu} \frac{1}{D_{\nu_c,\nu_f}N_M}
[H_{cc}^{\eta_2}(\kk_2)]_{a_2a_2}
\nonumber  \nonumber \\
&\bigg[
\gamma^2 :\UF_{00}^{(f,\xi\xi)}(\RR) :
+ \gamma v_\star^\prime  (k_x -i\xi k_y) :\UF_{00}^{(f,\xi-\xi)}(\RR,\tau_1) :
+\gamma v_\star^\prime( k_x +i\xi k_y) :\UF_{00}^{(f,-\xi \xi)}(\RR,\tau_1) :
\bigg]   \nonumber \\
&(-2)g_0(\tau_1-\tau_2,\kk)g_0(\tau_2-\tau_1,\kk) 
d\tau_1d\tau_2    \nonumber \\ 
=& \bigg[ \int_0^\beta \int_0^\beta  \sum_{\RR,\kk} \sum_{\eta_2s_2}\sum_{a_2\in 1,2}\sum_{\mu\nu} \frac{1}{D_{\nu_c,\nu_f}N_M}
[H_{cc}^{\eta_2}(\kk_2)]_{a_2a_2}
\gamma^2
(-1)g_0(\tau_1-\tau_2,\kk)g_0(\tau_2-\tau_1,\kk) \, .
\eaa  
We note that $g_0(\kk,\tau) $ only depends on $|\kk|$ and $\tau$ (Sec.~\ref{sec:green}, Eq.~\ref{eq:g0}). 
\bh{Then the terms proportional to $\gamma v_\star^\prime$ vanish after $\kk$ summation due to the linear-$k$ factor.} 
We find $\langle S_KS_{cc}\rangle_{0,con}$ is a constant.
}

\hh{ 
For $\langle S_J S_{cc}\rangle_{0,con}$, we have 
\baa  
\langle S_J S_{cc}\rangle_{0,con} =&\int_0^\beta \int_0^\beta  \sum_{\RR,\kk,\kk',\kk_2,\xi}e^{-i(\kk'-\kk)\cdot \RR} \sum_{\eta_2s_2}\sum_{a_2,a_2'\in 1,2}\sum_{\mu\nu} \frac{-J}{N_M}
[H_{cc}^{\eta_2}(\kk_2)]_{a_2a_2'}:\UF_{\mu\nu}^{(f,\xi\xi)}(\RR,\tau):
\nonumber \\
&
\langle :\UF_{\mu\nu}^{(c'',\xi\xi)}(\kk,\kk'-\kk,\tau) :c_{\kk_2,a_2\eta_2 s_2}^\dag(\tau_2) c_{\kk_2,a_2'\eta_2s_2}(\tau_2):\rangle_{0,con}
d\tau_1d\tau_2 
\label{eq:sj_scc_1}
\eaa  
We need to evaluate $\langle :\UF_{\mu\nu}^{(c'',\xi\xi)}(\kk,\kk'-\kk,\tau) :c_{\kk_2,a_2\eta_2 s_2}^\dag(\tau_2) c_{\kk_2,a_2'\eta_2s_2}(\tau_2):\rangle_{0,con}$, which gives
\baa
&\langle :\UF_{\mu\nu}^{(c'',\xi\xi)}(\kk,\kk'-\kk,\tau) :c_{\kk_2,a_2\eta_2 s_2}^\dag(\tau_2) c_{\kk_2,a_2'\eta_2s_2}(\tau_2):\rangle_{0,con}
 \nonumber 
\\
=&\frac{1}{2}\sum_{\eta,\eta',s,s'}\sum_{a,a'\in \{3,4\}}B^{\mu\nu}_{a\eta s,a'\eta' s'}\delta_{\xi,(-1)^{a+1}\eta}\delta_{\xi',(-1)^{a'+1}\eta'}\langle 
 :c_{\kk',a\eta s}^\dag(\tau_1) c_{\kk, a'\eta's'} (\tau_1)::
c_{\kk_2,a_2\eta_2s_2}^\dag(\tau_2)c_{\kk_2,a_2'\eta_2s_2}(\tau_2):
\rangle_{0,con} \nonumber  \\
=&-\frac{1}{2}\sum_{\eta,\eta',s,s'}\sum_{a,a'\in \{1,2\}}B^{\mu\nu}_{a\eta s,a'\eta' s'}\delta_{\xi,(-1)^{a+1}\eta}\delta_{\xi',(-1)^{a'+1}\eta'} \delta_{\eta,\eta_2}\delta_{\eta',\eta_2}\delta_{s,s_2}\delta_{s',s_2}\delta_{a+2,a_2'}\delta_{a'+2,a_2}\delta_{a,a'}\nonumber \\
&
(-1)\delta_{\kk_2,\kk}\delta_{\kk_2,\kk'} |g_{2,\eta}(\kk_2,\tau_1-\tau_2)|^2 \nonumber \\
=&\delta_{\mu,0}\delta_{\nu,0} (-\frac{1}{2}) \delta_{\xi,(-1)^{a_2+1}\eta_2}\delta_{a_2,a_2'}
\delta_{\kk_2,\kk'}\delta_{\kk_2,\kk}|g_{2,\eta}(\kk,\tau_1-\tau_2)|^2
\label{eq:sj_scc_2}
\eaa  
where we find only $\mu=0,\nu=0$ components can be non-zero, due to structures of $B^{\mu\nu}$ (Eq.~\ref{eq:flat-U4-maintext}). 
\bh{Then combine Eq.~\ref{eq:sj_scc_1} and Eq.~\ref{eq:sj_scc_2}, we find}
\ba 
\langle S_J S_{cc}\rangle_{0,con} =\bigg[ \int_0^\beta \int_0^\beta \sum_{\RR,\kk}\sum_{a_2,\eta}\frac{-J}{N_M}[H_{cc}^\eta]_{a_2,a_2} |g_{2,\eta}(\kk,\tau_1-\tau_2)|^2(-\frac{1}{2}) \tau_1d\tau_2\bigg] \nu_f 
\ea 
which is a constant.
}

\subsection{ $-\langle S_K S_J\rangle_{0,con}$} 
We then calculate $-\langle S_K S_J\rangle_{0,con}$:
\ba 
&-\langle S_KS_J \rangle_{0,con} \\
=&\int_{-\beta/2}^{\beta/2}\int_{-\beta/2}^{\beta/2} 
\sum_{\RR,\RR_2,\rr,\rr'}\frac{J}{D_{\nu_c,\nu_f}}\delta_{\rr,\RR}\delta_{\rr_2,\RR} :\UF_{\mu_2\nu_2}^{(f,\xi_2\xi_2)}(\RR_2,\tau_2):\\
&\sum_{\mu\nu, \xi\xi'} \sum_{\mu_2\nu_2,\xi_2\xi_2}\bigg[
\gamma^2 :\UF_{\mu\nu}^{(f,\xi\xi')}(\RR,\tau) :
+ \gamma v_\star^\prime  (i\partial_{r_x'}+\xi' \partial_{r_y'}) :\UF_{\mu\nu}^{(f,\xi-\xi')}(\RR,\tau) :
+\gamma v_\star^\prime( -i\partial_{r_x} +\xi \partial_{r_y}) :\UF_{\mu\nu}^{(f,-\xi \xi')}(\RR,\tau) :
\bigg] \\
&
\langle :\UF_{\mu_2\nu_2}^{(c',\xi'\xi)}(\rr,\rr',\tau)::\UF_{\mu\nu}^{(c'',\xi_2\xi_2)}(\RR_2,\tau_2):\rangle_{0,con}  d\tau d\tau_2 
\ea

\hh{As we derived in Sec.~\ref{sec:flat_u4_corre}}, Eq.~\ref{eq:u4u4correlator}, we represent $\langle :\UF_{\mu_2\nu_2}^{(c',\xi'\xi)}(\rr,\rr',\tau)::\UF_{\mu\nu}^{(c'',\xi_2\xi_2)}(\RR_2,\tau_2):\rangle_{0,con}$ with single-particle Green's function and find
\ba 
&-\langle S_K S_J \rangle_{0,con} \\
=&\int_{-\beta/2}^{\beta/2}\int_{-\beta/2}^{\beta/2} 
\sum_{\RR,\RR_2,\rr,\rr'}\frac{J}{D_{\nu_c,\nu_f}}
\sum_{\mu\nu, \xi} \bigg[
\gamma^2 :\UF_{\mu\nu}^{(f,\xi\xi)}(\RR,\tau) :
+( v_\star^\prime)^2 (-i\partial_{r_x} +\xi \partial_{r_y})(i\partial_{r_x'}+ \xi'\partial_{r_y'})
 :\UF_{\mu\nu}^{(f,-\xi-\xi)}(\RR,\tau) : 
\\
&+ \gamma v_\star^\prime  (i\partial_{r_x'}+\xi' \partial_{r_y'}) :\UF_{\mu\nu}^{(f,\xi-\xi)}(\RR,\tau) :
+\gamma v_\star^\prime( -i\partial_{r_x} +\xi \partial_{r_y}) :\UF_{\mu\nu}^{(f,-\xi \xi)}(\RR,\tau) :
\bigg]  :\UF_{\mu \nu}^{(f,\xi\xi)}(\RR_2,\tau_2):\\
&(-1)g^*_{2,\xi}(\rr'-\RR_2,\tau-\tau_2)g_{2,\xi}(\rr-\RR_2,\tau-\tau_2)
d\tau d\tau_2 \\ 
=&\int_{-\beta/2}^{\beta/2}\int_{-\beta/2}^{\beta/2} 
\sum_{\RR,\RR_2,\rr,\rr'}(-J)\bigg( -\frac{1}{D_{1,\nu_c,\nu_f}} +\frac{1}{D_{2,\nu_c,\nu_f}}
\bigg) \sum_{\mu\nu, \xi} \bigg[\\
&
\gamma^2 :\UF_{\mu\nu}^{(f,\xi\xi)}(\RR,\tau): :\UF_{\mu\nu}^{(f,\xi\xi)}(\RR_2,\tau_2): g^*_{2,\xi}(\RR-\RR_2,\tau-\tau_2)g_{2,\xi}(\RR-\RR_2,\tau-\tau_2) \\
& +(v_\star^\prime)\gamma  :\UF_{\mu\nu}^{(f,\xi-\xi)}(\RR,\tau): :\UF_{\mu\nu}^{(f,\xi\xi)}(\RR_2,\tau_2):
[(i\partial_{R_x} + \xi\partial_{R_y})g^*_{2,\xi}(\RR-\RR_2,\tau-\tau_2)][g_{2,\xi}(\RR-\RR_2,\tau-\tau_2)]\\
& +(v_\star^\prime)\gamma    :\UF_{\mu\nu}^{(f,-\xi\xi)}(\RR,\tau): :\UF_{\mu\nu}^{(f,\xi\xi)}(\RR_2,\tau_2):
[g^*_{2,\xi}(\RR-\RR_2,\tau-\tau_2)][(-i\partial_{R_x} + \xi\partial_{R_y})g_{2,\xi}(\RR-\RR_2,\tau-\tau_2)
\bigg] 
d\tau d\tau_2 
\ea 
\hh{where the Green's function is defined in Sec.~\ref{sec:green}, Eq.~\ref{eq:green_def} and Eq.~\ref{eq:green_def_2}. }

Using the analytical expression of single-particle Green's function derived in \hh{Sec.~\ref{sec:green}, Eq.~\ref{eq:g0} and Eq.~\ref{eq:g2}} at zero temperature and infinite momentum cutoff, we have 
\baa  
&-\langle S_KS_J\rangle_{0,con} \nonumber \\
=&\int_{-\infty}^{\infty}\int_{-\infty}^{\infty} 
\sum_{\RR,\RR_2}(-J)\bigg( -\frac{1}{D_{1,\nu_c,\nu_f}} +\frac{1}{D_{2,\nu_c,\nu_f}}
\bigg) \sum_{\mu\nu, \xi} \bigg[ \nonumber \\
&
\gamma^2 :\UF_{\mu\nu}^{(f,\xi\xi)}(\RR,\tau): :\UF_{\mu\nu}^{(f,\xi\xi)}(\RR_2,\tau_2): 
\frac{ \pi^2 |\RR-\RR_2|^2}{A_{MBZ}^2(|v_\star(\tau-\tau_2)|^2 + |\RR-\RR_2|^2)^3} \nonumber 
 \\ 
& +(v_\star^\prime)\gamma  :\UF_{\mu\nu}^{(f,\xi-\xi)}(\RR,\tau): :\UF_{\mu\nu}^{(f,\xi\xi)}(\RR_2,\tau_2):
\frac{-3\pi^2 |\RR-\RR_2|^2}{A_{MBZ}^2
(|v_\star(\tau-\tau_2)|^2 + |\RR-\RR_2|^2)^4
}\bigg(\xi (R_y-R_{2,y})+i(R_{x}-R_{2,x})
\bigg) \nonumber 
\\
& +(v_\star^\prime)\gamma    :\UF_{\mu\nu}^{(f,-\xi\xi)}(\RR,\tau): :\UF_{\mu\nu}^{(f,\xi\xi)}(\RR_2,\tau_2):
\frac{-3\pi^2 |\RR-\RR_2|^2}{A_{MBZ}^2
(|v_\star(\tau-\tau_2)|^2 + |\RR-\RR_2|^2)^4
}\bigg(\xi (R_y-R_{2,y})-i(R_{x}-R_{2,x}\bigg)
\bigg]
d\tau d\tau_2 
\label{eq:sKsJ}
\eaa

\subsection{$-\frac{1}{2}\langle S_K^2\rangle_{0,con}$}
We now calculate $-\frac{1}{2}\langle S_K^2\rangle_{0,con}$:
\ba 
&-\frac{1}{2}\langle  S_K^2 \rangle_{0,con}\\
=& -\int_{-\beta/2}^{\beta/2} \int_{-\beta/2}^{\beta/2} 
\sum_{\RR,\rr,\rr'}
\sum_{\RR_2,\rr_2,\rr'_2}
\frac{1}{2D_{\nu_c,\nu_f}^2}
\delta_{\rr,\RR}\delta_{\rr',\RR}
\delta_{\rr_2,\RR_2}\delta_{\rr'_2,\RR_2}
\\
&\sum_{\mu\nu, \xi\xi'}
\bigg[
\gamma^2 :\UF_{\mu\nu}^{(f,\xi\xi')}(\RR,\tau) :
+ \gamma v_\star^\prime  (i\partial_{r_x'}+\xi' \partial_{r_y'}) :\UF_{\mu\nu}^{(f,\xi-\xi')}(\RR,\tau) :
+\gamma v_\star^\prime( -i\partial_{r_x} +\xi \partial_{r_y}) :\UF_{\mu\nu}^{(f,-\xi \xi')}(\RR,\tau) :
\bigg]  \\
&\sum_{\mu_2\nu_2, \xi_2\xi'_2}
\bigg[
\gamma^2 :\UF_{\mu_2\nu_2}^{(f,\xi_2\xi'_2)}(\RR_2,\tau_2) :
+ \gamma v_\star^\prime  (i\partial_{r_{2,x}'}+\xi_2' \partial_{r_{2,y}'}) :\UF_{\mu_2\nu_2}^{(f,\xi_2-\xi_2')}(\RR_2,\tau_2) :
+\gamma v_\star^\prime( -i\partial_{r_{2,x}} +\xi_2 \partial_{r_{2,y}}) :\UF_{\mu_2\nu_2}^{(f,-\xi_2 \xi'_2)}(\RR_2,\tau_2) :
\bigg]  \\
&\langle :\UF_{\mu\nu}^{(c',\xi',\xi)}(\rr,\rr',\tau):
:\UF_{\mu_2\nu_2}^{(c',\xi'_2,\xi_2)}
(\rr_2,\rr'_2,\tau_2):\rangle_{0,con} 
d\tau d\tau_2 \\
\ea 
We rewrite the above equations with single-particle Green's function as derived in \hh{Sec.~\ref{sec:flat_u4_corre}, Eq.~\ref{eq:u4u4correlator_0}}
\baa 
&-\frac{1}{2}\langle  S_K^2 \rangle_{0,con} \nonumber \\
=& -\int_{-\beta/2}^{\beta/2}\int_{-\beta/2}^{\beta/2}
\sum_{\RR,\rr,\rr'}
\sum_{\RR_2,\rr_2,\rr'_2}\sum_{\xi,\xi'}
\frac{1}{2D_{\nu_c,\nu_f}^2} \delta_{\rr,\RR}\delta_{\rr',\RR}
\delta_{\rr_2,\RR_2}\delta_{\rr'_2,\RR_2}  \nonumber \\
&\sum_{\mu\nu}
\bigg[
\gamma^2 :\UF_{\mu\nu}^{(f,\xi\xi')}(\RR,\tau) :
+ \gamma v_\star^\prime  (i\partial_{r_x'}+\xi' \partial_{r_y'}) :\UF_{\mu\nu}^{(f,\xi-\xi')}(\RR,\tau) :
+\gamma v_\star^\prime( -i\partial_{r_x} +\xi \partial_{r_y}) :\UF_{\mu\nu}^{(f,-\xi \xi')}(\RR,\tau) :
\bigg]   \nonumber \\
&
\bigg[
\gamma^2 :\UF_{\mu \nu }^{(f,\xi' \xi)}(\RR_2,\tau_2) :
+ \gamma v_\star^\prime  (i\partial_{r_{2,x}'}+\xi \partial_{r_{2,y}'}) :\UF_{\mu \nu}^{(f,\xi'-\xi)}(\RR_2,\tau_2) :
+\gamma v_\star^\prime( -i\partial_{r_{2,x}} +\xi' \partial_{r_{2,y}}) :\UF_{\mu_2\nu_2}^{(f,-\xi' \xi)}(\RR_2,\tau_2) :
\bigg]   \nonumber \\
&(-1) g_{0}(\rr_2'-\rr,\tau_2-\tau)g_0(\rr'-\rr_2, \tau-\tau_2)
d\tau d\tau_2  \nonumber \\  
=&\int_{-\beta/2}^{\beta/2} \int_{-\beta/2}^{\beta/2} 
\sum_{\RR,\rr,\rr'}
\sum_{\RR_2,\rr_2,\rr'_2}\sum_{\xi,\xi'}
\frac{1}{2D_{\nu_c,\nu_f}^2}
\sum_{\mu\nu} \nonumber \\
&\Bigg[ :\UF_{\mu\nu}^{(f,\xi\xi')}(\RR,\tau) ::\UF_{\mu\nu}^{(f,\xi'\xi)}(\RR_2,\tau_2): 
\gamma^4 g_0(\RR_2-\RR,\tau_2-\tau) g_0(\RR-\RR_2,\tau-\tau_2) \nonumber \\
&+:\UF_{\mu\nu}^{(f,\xi\xi')}(\RR,\tau) ::\UF_{\mu\nu}^{(f,-\xi'-\xi)}(\RR_2,\tau_2): 
2
\gamma^2(v_\star^\prime)^2 [(i\partial_x+\xi \partial_y)g_0(\RR_2-\RR,\tau_2-\tau)] [(i\partial_x-\xi' \partial_y)g_0(\RR-\RR_2,\tau-\tau_2)]   \nonumber \\
&+:\UF_{\mu\nu}^{(f,\xi\xi')}(\RR,\tau) ::\UF_{\mu\nu}^{(f,\xi'-\xi)}(\RR_2,\tau_2): 2\gamma^3v_\star^\prime (i\partial_{R_x}+\xi \partial_{R_y})g_0(\RR_2-\RR,\tau_2-\tau)]g_0(\RR-\RR_2,\tau-\tau_2) \nonumber  \\
&+:\UF_{\mu\nu}^{(f,\xi\xi')}(\RR,\tau) ::\UF_{\mu\nu}^{(f,-\xi'\xi)}(\RR_2,\tau_2): 2\gamma^3v_\star^\prime g_0(\RR_2-\RR,\tau_2-\tau)(i\partial_{R_x}-\xi \partial_{R_y})g_0(\RR-\RR_2,\tau-\tau_2)] 
\bigg] d\tau d\tau_2 
\label{eq:sksk_exp}
\eaa  
\hh{where the Green's function is defined in Sec.~\ref{sec:green}, Eq.~\ref{eq:green_def} and Eq.~\ref{eq:green_def_2}. }
\hhb{ Here, we mention that we expand Kondo interaction in powers of $\kk$ (note that the hybridization matrix has been expanded in powers of $\kk$ in Eq.~\ref{eq:hk_exp_l}). Thus, $\gamma^4, \gamma^3v_\star^\prime$ correspond to zeroth and linear-order terms and are kept. In addition, we notice that $\gamma^2 (v_\star^\prime)^2$ term in Eq.~\ref{eq:sksk_exp} provide leading-order contributions in the interaction channel $:\UF_{\mu\nu}^{(f,\xi\xi')}(\RR,\tau) ::\UF_{\mu\nu}^{(f,-\xi'-\xi)}(\RR_2,\tau_2):$. In other words, even if we expand to Kondo interaction to higher orders in $\kk$, the current $\gamma^2 (v_\star^\prime)^2$ term still gives the leading order contributions and should be kept. 
}

At zero temperature, we utilize the analytical expression of single-particle Green's function derived in \hh{Sec.~\ref{sec:green}, Eq.~\ref{eq:g0}} and obtain
\baa  
\small 
&-\frac{1}{2}\langle S_K^2\rangle_{0,con} \nonumber \\
=&\int_{-\infty}^{\infty} \int_{-\infty}^{\infty}
\sum_{\RR,\RR_2}
\sum_{\xi,\xi'}
\frac{1}{2D_{\nu_c,\nu_f}^2}
\sum_{\mu\nu} \Bigg[ -:\UF_{\mu\nu}^{(f,\xi\xi')}(\RR,\tau) ::\UF_{\mu\nu}^{(f,\xi'\xi)}(\RR_2,\tau_2): 
\gamma^4
 \frac{\pi^2}{A_{MBZ}^2}\frac{v_\star^2 |\tau-\tau_2|^2}{\bigg( |v_\star|^2\tau^2 + |\RR-\RR_2|^2 \bigg)^{3} }
 \nonumber 
 \\
 &-:\UF_{\mu\nu}^{(f,\xi\xi')}(\RR,\tau) ::\UF_{\mu\nu}^{(f,-\xi'-\xi)}(\RR_2,\tau_2): 
\frac{9\gamma^2 (v_\star^\prime)^2 \pi^2 |v_\star(\tau-\tau_2)|^2
}{A_{MBZ}^2\bigg( 
v_\star^2\tau^2 +|\RR-\RR_2|^2
\bigg)^5} (i(R_x-R_{2,x})+\xi (R_y-R_{2,y} )) \\
&(-i(R_x-R_{2,x})+\xi' (R_y-R_{2,y} )) \nonumber \\
&+:\UF_{\mu\nu}^{(f,\xi\xi')}(\RR,\tau) ::\UF_{\mu\nu}^{(f,\xi'-\xi)}(\RR_2,\tau_2): \frac{ 6\gamma^3v_\star^\prime \pi^2 |v_\star(\tau-\tau)|^2 \bigg( i(R_{2,x}-R_x) +\xi( R_{2,y}-R_y)\bigg) }{A_{MBZ}^2 \bigg(v_\star^2\tau^2 +|\RR-\RR_2|^2 \bigg)^4 }
\nonumber 
\\
&+:\UF_{\mu\nu}^{(f,\xi\xi')}(\RR,\tau) ::\UF_{\mu\nu}^{(f,-\xi'\xi)}(\RR_2,\tau_2): \frac{ 6\gamma^3v_\star^\prime \pi^2 |v_\star(\tau-\tau)|^2 \bigg( -i(R_{2,x}-R_x) +\xi( R_{2,y}-R_y)\bigg) }{A_{MBZ}^2 \bigg(v_\star^2\tau^2 +|\RR-\RR_2|^2 \bigg)^4 }
\bigg] d\tau d\tau_2 
\label{eq:sKsK}
\eaa

\subsection{$-\frac{1}{2}\langle S_J^2\rangle_{0,con} $} 
Finally, we calculate $-\frac{1}{2}\langle S_J^2\rangle_{0,con}$. 
\ba 
&-\frac{1}{2}\langle S_J^2\rangle_{0,con}\\
=&-\frac{J^2}{2}\int_{-\beta/2}^{\beta/2}
\int_{-\beta/2}^{\beta/2}
\sum_{\mu\nu\mu_2\nu_2,\xi\xi_2,\RR,\RR_2} :\UF_{\mu\nu}^{(f,\xi\xi)}(\RR,\tau):
:\UF_{\mu_2\nu_2}^{(f,\xi_2\xi_2)}(\RR_2,\tau_2):
\langle ::\UF_{\mu\nu}^{(c'',\xi\xi)}(\RR,\RR,\tau):
:\UF_{\mu_2\nu_2}^{(c'',\xi_2\xi_2)}(\RR_2,\RR,\tau_2):
d\tau d\tau_2 
\ea 
Expressing in terms of single-particle Green's function as we derived in \hh{Sec.\ref{sec:flat_u4_corre}, Eq.~\ref{eq:u4u4correlator_2}}, we have 
\ba 
&-\frac{1}{2}\langle S_J^2\rangle_{0,con}\\
=&-\frac{J^2}{2}\int_{-\beta/2}^{\beta/2}
\int_{-\beta/2}^{\beta/2}
\sum_{\mu\nu,\xi,\RR,\RR_2} :\UF_{\mu\nu}^{(f,\xi\xi)}(\RR,\tau):
:\UF_{\mu\nu}^{(f,\xi\xi)}(\RR_2,\tau_2):
(-1)g_0(\RR_2-\RR,\tau_2-\tau) g_0(\RR-\RR_2,\tau-\tau_2)
d\tau d\tau_2 
\ea 
\hh{where the Green's function is defined in Sec.~\ref{sec:green}, Eq.~\ref{eq:green_def} and Eq.~\ref{eq:green_def_2}.} Using the analytical expression of single-particle Green's function at zero temperature and infinite momentum cutoff as derived in
\hh{Sec.~\ref{sec:green}, Eq.~\ref{eq:g0}}, we have 
\baa  
&-\frac{1}{2}\langle S_J^2\rangle_{0,con} 
=-\frac{J^2}{2}\int_{-\infty}^{\infty}
\int_{-\beta/2}^{\beta/2}
\sum_{\mu\nu,\xi,\RR,\RR_2} :\UF_{\mu\nu}^{(f,\xi\xi)}(\RR,\tau):
:\UF_{\mu\nu}^{(f,\xi\xi)}(\RR_2,\tau_2):
\frac{\pi^2 |v_\star(\tau-\tau_2)|^2}{A_{MBZ}^2
\bigg[ 
(v_\star(\tau-\tau_2))^2 + |\RR-\RR_2|^2
\bigg] ^{3}
}
d\tau d\tau_2 
\label{eq:sJsJ}
\eaa 

\subsection{RKKY interactions} 
We sum over all three terms in Eq.~\ref{eq:sKsJ}, Eq.~\ref{eq:sKsK}, Eq.~\ref{eq:sJsJ}, and find the analytical formula of $S_{RKKY} = -\langle S_KS_J\rangle_{0,con} -\frac{1}{2}\langle S_{K}^2\rangle_{0,con}
-\frac{1}{2}\langle S_{J}^2\rangle_{0,con}$:
\baa  
&S_{RKKY} 
=\int \int \sum_{\RR,\RR_2,\mu\nu} 
\bigg[ 
\sum_{\xi\xi'}:\UF_{\mu\nu}^{(f,\xi\xi')}(\RR,\tau):
:\UF_{\mu\nu}^{(f,\xi'\xi)}(\RR_2,\tau_2): \chi_0(\RR-\RR_2,\tau-\tau_2) \nonumber \\
&+\sum_\xi 
:\UF_{\mu\nu}^{(f,\xi\xi)}(\RR,\tau):
:\UF_{\mu\nu}^{(f,\xi\xi)}(\RR_2,\tau_2): \chi_1(\RR-\RR_2,\tau-\tau_2) 
+\sum_\xi 
:\UF_{\mu\nu}^{(f,\xi\xi)}(\RR,\tau):
:\UF_{\mu\nu}^{(f,-\xi-\xi)}(\RR_2,\tau_2): \chi_2(\RR-\RR_2,\tau-\tau_2) \nonumber \\
&+\sum_\xi 
:\UF_{\mu\nu}^{(f,\xi \xi)}(\RR,\tau):
:\UF_{\mu\nu}^{(f,-\xi \xi)}(\RR_2,\tau_2): \chi_{3,\xi}(\RR-\RR_2,\tau-\tau_2) 
+:\UF_{\mu\nu}^{(f,\xi \xi)}(\RR,\tau):
:\UF_{\mu\nu}^{(f,\xi -\xi)}(\RR_2,\tau_2): \chi_{4,\xi}(\RR-\RR_2,\tau-\tau_2) \nonumber \\
&+\sum_\xi 
:\UF_{\mu\nu}^{(f,\xi-\xi)}(\RR,\tau):
:\UF_{\mu\nu}^{(f,\xi-\xi)}(\RR_2,\tau_2): \chi_{5,\xi}(\RR-\RR_2,\tau-\tau_2) \bigg] d\tau d\tau_2 
\label{eq:action_rkky_v0}
\eaa 
where \hb{
\baa  
\chi_0(\RR,\tau) = &-\frac{\gamma^4}{2D_{\nu_c,\nu_f}^2}   \frac{ \pi^2|v_\star\tau|^2}{A_{MBZ}^2\bigg( |v_\star\tau|^2+ |\RR|^2 \bigg)^{3} }
\nonumber \\
\chi_1(\RR,\tau)=&
\frac{-J\gamma^2}{D_{\nu_c,\nu_f}}\frac{ \pi^2 |\RR|^2}{A_{MBZ}^2(|v_\star \tau|^2 + |\RR|^2)^3} 
-\frac{J^2}{2}\frac{\pi^2 |v_\star\tau|^2}{A_{MBZ}^2
\bigg[ 
|v_\star\tau|^2 + |\RR|^2
\bigg] ^{3}
} 
\nonumber \\
\chi_2(\RR,\tau) =&- \frac{9\gamma^2 (v_\star^\prime)^2}{D_{\nu_c,\nu_f}^2} \frac{ \pi^2 |v_\star\tau|^2|\RR|^2}{A_{MBZ}^2\bigg( 
|v_\star\tau|^2 +|\RR|^2
\bigg)^5} \nonumber  \\
\chi_{3,\xi}(\RR,\tau) = & \frac{-3J v_\star^\prime \gamma}{D_{\nu_c,\nu_f}} \frac{\pi^2 |\RR|^2}{A_{MBZ}^2
(|v_\star \tau|^2 + |\RR|^2)^4
}\bigg( -\xi R_y+iR_{x}\bigg) 
+\frac{ 3\gamma^3v_\star^\prime }{D_{\nu_c,\nu_f}^2} \frac{\pi^2 |v_\star\tau|^2 \bigg( - \xi R_y+iR_x\bigg) }{A_{MBZ}^2 \bigg(v_\star^2\tau^2 +|\RR|^2 \bigg)^4 }\nonumber \\
\chi_{4,\xi}(\RR,\tau) = &\gamma\frac{3Jv_\star^\prime}{D_{\nu_c,\nu_f}} \frac{\pi^2 |\RR|^2}{A_{MBZ}^2
(|v_\star \tau|^2 + |\RR|^2)^4
}\bigg(\xi R_y+i R_{x}\bigg) 
+\frac{3\gamma^3v_\star^\prime }{D_{\nu_c,\nu_f}^2}  \frac{ \pi^2 |v_\star(\tau-\tau)|^2 \bigg( -\xi R_y -iR_x\bigg) }{A_{MBZ}^2 \bigg( |v_\star \tau|^2 +|\RR|^2 \bigg)^4 }\nonumber \\
\chi_{5,\xi}(\RR,\tau) =&-\frac{9\gamma^2 (v_\star^\prime)^2}{D_{\nu_c,\nu_f}^2A}
 \frac{ \pi^2 |v_\star \tau |^2
}{A_{MBZ}^2\bigg( 
v_\star^2\tau^2 +|\RR|^2
\bigg)^5}\bigg[ 
[R_x^2 - R_{y}^2]
-2i\xi R_x R_{y}]
\bigg] 
\label{eq:chi_v1}
\eaa  
} 

The corresponding RKKY interactions can be derived by taking the zero-frequency contribution of $\chi$ in Eq.~\ref{eq:chi_v1}, which leads to the following RKKY Hamiltonian 
\baa  
&H_{RKKY} \nonumber \\
=&
\sum_{\RR,\RR_2,\mu\nu} 
\sum_{\xi\xi'}:\UF_{\mu\nu}^{(f,\xi\xi')}(\RR):
:\UF_{\mu\nu}^{(f,\xi'\xi)}(\RR_2): J^{RKKY}_0(\RR-\RR_2) \nonumber \\
&+\sum_\xi 
:\UF_{\mu\nu}^{(f,\xi\xi)}(\RR):
:\UF_{\mu\nu}^{(f,\xi\xi)}(\RR_2): J^{RKKY}_1(\RR-\RR_2) 
+\sum_\xi 
:\UF_{\mu\nu}^{(f,\xi\xi)}(\RR):
:\UF_{\mu\nu}^{(f,-\xi-\xi)}(\RR_2): J^{RKKY}_2(\RR-\RR_2) \nonumber \\
&+\sum_\xi 
:\UF_{\mu\nu}^{(f,\xi \xi)}(\RR):
:\UF_{\mu\nu}^{(f,-\xi \xi)}(\RR_2): J^{RKKY}_{3,\xi}(\RR-\RR_2) 
+:\UF_{\mu\nu}^{(f,\xi \xi)}(\RR):
\UF_{\mu\nu}^{(f,\xi -\xi)}(\RR_2): J^{RKKY}_{4,\xi}(\RR-\RR_2)\nonumber \\
&+\sum_\xi 
:\UF_{\mu\nu}^{(f,\xi -\xi)}(\RR):
:\UF_{\mu\nu}^{(f,\xi -\xi)}(\RR_2): J^{RKKY}_{5,\xi}(\RR-\RR_2) 
\label{eq:rkky_ham}
\eaa  
\hhb{
where the RKKY interactions are defined as 
\baa 
&J_0^{RKKY} = \int_{-\infty}^{\infty} \chi_{0}(\RR,\tau)) d\tau ,\quad 
J_1^{RKKY} = \int_{-\infty}^{\infty} \chi_{1}(\RR,\tau)) d\tau
,\quad 
J_2^{RKKY} = \int_{-\infty}^{\infty} \chi_{2}(\RR,\tau)) d\tau \nonumber \\
&J_{3,\xi}^{RKKY} = \int_{-\infty}^{\infty} \chi_{3,\xi}(\RR,\tau)) d\tau ,\quad 
J_{4,\xi}^{RKKY} = \int_{-\infty}^{\infty} \chi_{4,\xi}(\RR,\tau)) d\tau
,\quad 
J_{5,\xi}^{RKKY} = \int_{-\infty}^{\infty} \chi_{5,\xi}(\RR,\tau)) d\tau
\eaa 
}
\hh{ 
The analytical expressions of RKKY interactions are given below 
\baa  
&J^{RKKY}_0(\RR) = \frac{-\pi^3}{ |v_\star| A_{MBZ}^2}
\bigg[ 
\frac{\gamma^4}{16D_{\nu_c,\nu_f}^2}\bigg]\frac{1}{|\RR|^3}
 \nonumber \\
&J^{RKKY}_1(\RR) 
=-\frac{ \pi^3}{8A_{MBZ}^2 |v_\star| }\bigg[\frac{3J \gamma^2}{D_{\nu_c,\nu_f}} +\frac{J^2}{2} \bigg]\frac{1}{|\RR|^3} 
\nonumber \\
&J^{RKKY}_2(\RR) =-\frac{\pi^3}{A_{MBZ}^2|v_\star|} 
 \frac{45\gamma^2 (v_\star^\prime)^2}{128D_{\nu_c,\nu_f}^2}
 \frac{1}{\RR^5} \nonumber \\
&J^{RKKY}_{3,\xi}(\RR) =-\frac{3\pi^3}{16A_{MBZ}^2|v_\star|} 
\bigg[ 
\frac{5 v_\star^\prime \gamma J}{D_{\nu_c,\nu_f}} +\frac{v_\star^\prime \gamma^3}{D_{\nu_c,\nu_f}^2}
\bigg] \frac{(\xi R_y -i R_x)}{|\RR|^5} 
\nonumber \\
&J^{RKKY}_{4,\xi}(\RR) =-\frac{3\pi^3}{16A_{MBZ}^2|v_\star|}
\bigg[ 
\frac{5 v_\star^\prime \gamma J}{D_{\nu_c,\nu_f}} +\frac{v_\star^\prime \gamma^3}{D_{\nu_c,\nu_f}^2}
\bigg] \frac{(\xi R_y +i R_x)}{|\RR|^5} \nonumber  \\
&J^{RKKY}_{5,\xi}(\RR) =-\frac{\pi^3}{A_{MBZ}^2|v_\star|} \frac{45 \gamma^2 (v_\star^\prime)^2}{128D_{\nu_c,\nu_f}^2}
\frac{[R_x^2-R_y^2] - 2i\xi R_x R_y}{|\RR|^7}
\label{eq:j_rkky_1}
\eaa 
\bh{We also comment that RKKY interactions diverge at $R\rightarrow 0$. During the derivation of RKKY interactions, when we perform momentum integral, we have set the cutoff of momentum integral to infinity to obtain an analytical expression. This introduces a short-distance divergence. The divergence can be cured by introducing a short-distance cutoff ($a_M$). In practice, we can replace $|\RR|$ in the denominators of Eq.~\ref{eq:j_rkky_1} by $\sqrt{|\RR|^2+a_M^2}$.}}

\hh{
Finally, we provide the Fourier transformation of $J^{RKKY}_{0/1/2/3,\xi/4,\xi}(\kk) = \sum_{\RR}J^{RKKY}_{0/1/2/3,\xi/4,\xi}(\RR)e^{-i\kk\cdot \RR}$ using the Fourier transformation derived in Sec.~\ref{sec:ft}, Eq.~\ref{eq:ft_power}. During the Fourier transformation, we also introduce a short-distance cutoff $a_M$ (moir\'e lattice constant)}
\hb{ RKKY interactions in the momentum space now take the form of 
\baa  
J^{RKKY}_0(\kk) = &\frac{-\pi^3}{ |v_\star| a_MA_{MBZ}}
\frac{\gamma^4}{64D_{\nu_c,\nu_f}^2}(I_0(ka_M) -L_0(ka_M))
 \nonumber \\
J^{RKKY}_1(\kk) 
=& -\frac{ \pi^3}{32A_{MBZ} a_M |v_\star| }\bigg[\frac{3J \gamma^2}{D_{\nu_c,\nu_f}} +\frac{J^2}{2} \bigg](I_0(ka_M) -L_0(ka_M))
\nonumber \\
J^{RKKY}_2(\kk) =&-\frac{\pi^2}{A_{MBZ}|v_\star| a_M^3} 
 \frac{45\gamma^2 (v_\star^\prime)^2}{3072D_{\nu_c,\nu_f}^2} 
 \frac{1}{ ka_M } 
\bigg[ 
3\pi(-2 +k^2a_M^2)L_1(ka_M) + qa_M 
\bigg( 4qa_M + 3\pi I_0(q a_M) \nonumber \\
&-3 \pi qa_M I_1(qa_M)- 3\pi L_2(qa_M)
\bigg) 
\bigg] \nonumber \\ 
J^{RKKY}_{3,\xi}(\kk) =&-\frac{3\pi^2}{32A_{MBZ}^2|v_\star|a_M^2} 
\bigg[ 
\frac{5 v_\star^\prime \gamma J}{D_{\nu_c,\nu_f}} +\frac{v_\star^\prime \gamma^3}{D_{\nu_c,\nu_f}^2}
\bigg]\frac{ 
1-e^{-k a_M}(1+ka_M)
}{k^2a_M^2} (-i \xi k_y - k_x )
\nonumber \\
J^{RKKY}_{4,\xi}(\kk) =&-\frac{3\pi^2}{32A_{MBZ}^2|v_\star|a_M^2} 
\bigg[ 
\frac{5 v_\star^\prime \gamma J}{D_{\nu_c,\nu_f}} +\frac{v_\star^\prime \gamma^3}{D_{\nu_c,\nu_f}^2}
\bigg]\frac{ 
1-e^{-k a_M}(1+ka_M)
}{k^2a_M^2} (-i \xi k_y + k_x )
\nonumber \\
J^{RKKY}_{5,\xi}(\kk) =&\frac{\pi^2}{A_{MBZ}^2|v_\star|a_M} \frac{45 \gamma^2 (v_\star^\prime)^2}{256D_{\nu_c,\nu_f}^2}
ka_M\bigg[ 
-\frac{1}{30} +\frac{\pi}{4k^3a_M^3} 
\bigg( I_2(ka_M) +ka_MI_3(ka_M)
\bigg)  \nonumber \\
&
-\frac{\pi}{4k^3a_M^3}
\bigg(L_2(ka_M) +ka_M L_3(ka_M) \bigg) 
\bigg] (k_x^2-k_y^2 +2i\xi k_xk_y)
\label{eq:rkky_ft}
\eaa  
where $I_n(x)$ is the modified Bessel function of the second kind~\cite{abramowitz1988handbook}, $L_n(x)$ is the  modified Struve function~\cite{abramowitz1988handbook}.}
\hh{
We also provide the long-wavelength behavior of RKKY by expanding in powers of $k$ to second order
\baa  
&J^{RKKY}_0(\kk) \approx \frac{-\pi^3}{ |v_\star| A_{MBZ}a_M}
\bigg[ 
\frac{\gamma^4}{64D_{\nu_c,\nu_f}^2}
\bigg] 
(1-\frac{2}{\pi}a_Mk + \frac{1}{4}a_M^2k^2)
\nonumber 
\\
&J^{RKKY}_1(\kk) \approx  -\bigg(\frac{ 3\pi^3}{4A_{MBZ} |v_\star| }\frac{J \gamma^2}{2D_{\nu_c,\nu_f}} +\frac{\pi^3}{16|v_\star|A_{MBZ}}J^2
\bigg)
\frac{1}{4a_M}(1-\frac{2}{\pi}a_Mk + \frac{1}{4}a_M^2k^2)
\nonumber \\ 
& J^{RKKY}_2(\kk) \approx -\frac{\pi^3}{8A_{MBZ}|v_\star|a_M^3}
\bigg[ 
\frac{45 \gamma^2 (v_\star^\prime)^2}{128D_{\nu_c,\nu_f}^2} 
\bigg] (1 - \frac{1}{4}(ka_M)^2)\nonumber   \\
&J^{RKKY}_{3,\xi}(\kk) \approx \frac{\pi^3}{A_{MBZ}|v_\star|a_M} 
\bigg[ 
\frac{15 v_\star^\prime \gamma J}{16D_{\nu_c,\nu_f}} +\frac{3v_\star^\prime \gamma^3}{16D_{\nu_c,\nu_f}^2}
\bigg] 
(\frac{1}{2}-\frac{ka_M}{3}) 
(i\xi k_y + k_x) \nonumber \\
&J^{RKKY}_{4,\xi}(\kk)\approx \frac{\pi^3}{A_{MBZ}|v_\star|a_M} 
\bigg[ 
\frac{15 v_\star^\prime \gamma J}{16D_{\nu_c,\nu_f}} +\frac{3v_\star^\prime \gamma^3}{16D^2_{\nu_c,\nu_f}}
\bigg] 
(\frac{1}{2}-\frac{ka_M}{3}) 
(i\xi k_y - k_x) \nonumber \\
&J^{RKKY}_{5,\xi}(\kk) \approx \frac{\pi^3}{A_{MBZ}|v_\star|a_M} \frac{45\gamma^2(v_\star^\prime)^2}{8192D_{\nu_c,\nu_f}^2 }(k_x^2-k_y^2+2i\xi k_xk_y)
\label{eq:rkky_ft}
\eaa  
}

\section{Ground state of $f$-moments} 
\label{sec:gnd_state}
\bh{
In this section, we solve the RKKY Hamiltonian (Eq.~\ref{eq:rkky_ham}) and find the corresponding ground state at $\nu=0,-1,-2$, $M=0$. We will consider both $v_\star^\prime =0$ where the system has a $U(4)\times U(4)$ symmetry, and $v_\star^\prime \ne 0$ where the system only has a flat $U(4)$ symmetry. 
}

\subsection{Ground state at $v_\star^\prime =0,M=0$} 
\hh{
Based on the RKKY Hamiltonian we derived in Eq.~\ref{eq:ham_rkky}, we calculate the ground state. We first consider the case of $v_\star^\prime=0$ where we have $U(4)\times U(4)$ symmetry. \bh{The RKKY interactions at $v_\star^\prime =0$ are
\baa 
&J^{RKKY}_0(\RR) = \frac{-\pi^3}{ |v_\star| A_{MBZ}^2}
\bigg[ 
\frac{\gamma^4}{16D_{\nu_c,\nu_f}^2}\bigg]\frac{1}{|\RR|^3} \le  0 \nonumber \\ 
&J^{RKKY}_1(\RR) 
=-\frac{ \pi^3}{8A_{MBZ}^2 |v_\star| }\bigg[\frac{3J \gamma^2}{D_{\nu_c,\nu_f}} +\frac{J^2}{2} \bigg]\frac{1}{|\RR|^3} \le  0 
\nonumber \\
&J^{RKKY}_2(\RR) =J^{RKKY}_{3,\xi}(\RR) =
J^{RKKY}_{4,\xi}(\RR) =
J^{RKKY}_{5,\xi}(\RR) =0
\label{eq:j_rkky_v_zero}
\eaa  
}
The RKKY Hamiltonian (Eq.~\ref{eq:rkky_ham}) now becomes
\baa  
\hH_{RKKY}^{v_\star^\prime=0,M=0} = &\sum_{\RR,\RR_2, \mu\nu,\xi\xi'}J_{0}^{RKKY}(\RR-\RR_2):\Sigma_{\mu\nu}^{(f,\xi\xi')}(\RR)::\Sigma_{\mu\nu}^{(f,\xi'\xi)}(\RR_2) :\nonumber \\
&
+ \sum_{\RR,\RR_2, \mu\nu,\xi}J^{RKKY}_{1}(\RR-\RR_2):\Sigma_{\mu\nu}^{(f,\xi\xi)}(\RR)::\Sigma_{\mu\nu}^{(f,\xi'\xi')}(\RR_2): 
\label{eq:ham_rkky_v0}
\eaa  
where the first term has $U(8)$ symmetry and the second term only has flat $U(4)$ symmetry.  We then introduce the bond operator 
\baa  
A^{\xi,\xi'}_{\RR,\RR_2} = \sum_{i,\xi} (\psi_{\RR,i}^{f,\xi})^\dag  \psi_{\RR_2,i}^{f,\xi'}
\label{eq:bond_a_v0}
\eaa
}
\bh{
where we use the $\psi$ basis defined in Eq.~\ref{eq:psi_basis_def}. Using bond operators and the definition of $U(8)$ moments (Eq.~\ref{eq:flat_u4_general}), we find the following relation 
\baa  
&\sum_{\mu\nu} :\Sigma_{\mu\nu}^{(f,\xi\xi')}(\RR)::\Sigma_{\mu\nu}^{(f,\xi'\xi)}(\RR_2) : \nonumber \\ 
=& \sum_{i,j,l,m}\frac{T^{\mu\nu}_{ij}T^{\mu\nu}_{lm}}{4} 
:\psi_{\RR,i}^{f,\xi,\dag} \psi_{\RR,j}^{f,\xi'}:
:\psi_{\RR_2,l}^{f,\xi',\dag} \psi_{\RR_2,m}^{f,\xi}:  
=\sum_{i,j}:\psi_{\RR,i}^{f,\xi,\dag} \psi_{\RR,j}^{f,\xi'}:
:\psi_{\RR_2,j}^{f,\xi',\dag} \psi_{\RR_2,i}^{f,\xi}: \nonumber \\
=& \sum_{i,j}(\psi_{\RR,i}^{f,\xi,\dag} \psi_{\RR,j}^{f,\xi'}-\frac{\delta_{\xi,\xi'}\delta_{i,j}}{2})
(\psi_{\RR_2,j}^{f,\xi',\dag} \psi_{\RR_2,i}^{f,\xi}- \frac{\delta_{\xi,\xi'}\delta_{i,j}}{2} )
= \sum_{i,j}\psi_{\RR,i}^{f,\xi,\dag} \psi_{\RR,j}^{f,\xi'}\psi_{\RR_2,j}^{f,\xi',\dag} \psi_{\RR_2,i}^{f,\xi}
-\sum_{i}\frac{\delta_{\xi,\xi'}}{2} 
(\psi_{\RR,i}^{f,\xi,\dag} \psi_{\RR,i}^{f,\xi'} +\psi_{\RR_2,i}^{f,\xi',\dag} \psi_{\RR_2,i}^{f,\xi} )+\delta_{\xi,\xi'}
 \nonumber \\ 
 =&\sum_{i,j}\psi_{\RR,i}^{f,\xi,\dag} 
 ( \delta_{\RR,\RR_2}-\psi_{\RR_2,j}^{f,\xi',\dag} \psi_{\RR,j}^{f,\xi'})\psi_{\RR_2,i}^{f,\xi} 
-\sum_{i}\frac{\delta_{\xi,\xi'}}{2} 
(\psi_{\RR,i}^{f,\xi,\dag} \psi_{\RR,i}^{f,\xi'} +\psi_{\RR_2,i}^{f,\xi',\dag} \psi_{\RR_2,i}^{f,\xi} )+\delta_{\xi,\xi'}
 \nonumber \\ 
 =& \sum_{i,j} \psi_{\RR,i}^{f,\xi,\dag}(\delta_{i,j}\delta_{\xi,\xi'} -\psi_{\RR_2,i}^{f,\xi} \psi_{\RR_2,j}^{f,\xi',\dag}) \psi_{\RR,j}^{f,\xi'} +4\delta_{\RR,\RR_2}\sum_i \psi_{\RR,i}^{f,\xi,\dag}\psi_{\RR,i}^{f,\xi} -\sum_{i}\frac{\delta_{\xi,\xi'}}{2} 
(\psi_{\RR,i}^{f,\xi,\dag} \psi_{\RR,i}^{f,\xi'} +\psi_{\RR_2,i}^{f,\xi',\dag} \psi_{\RR_2,i}^{f,\xi} )+\delta_{\xi,\xi'} \nonumber \\ 
=& -\sum_{i,j} \psi_{\RR,i}^{f,\xi,\dag}\psi_{\RR_2,i}^{f,\xi} \psi_{\RR_2,j}^{f,\xi',\dag} \psi_{\RR,j}^{f,\xi'} +4\delta_{\RR,\RR_2}\sum_i \psi_{\RR,i}^{f,\xi,\dag}\psi_{\RR,i}^{f,\xi} + \sum_{i}\frac{\delta_{\xi,\xi'}}{2} 
(\psi_{\RR,i}^{f,\xi,\dag} \psi_{\RR,i}^{f,\xi'} -\psi_{f,\RR_2,i}^{\xi',\dag} \psi_{\RR_2,i}^{f,\xi} )+\delta_{\xi,\xi'} \nonumber \\ 
=& -A_{\RR,\RR_2}^{\xi,\xi}A_{\RR_2,\RR}^{\xi'\xi'} +4\delta_{\RR,\RR_2}\sum_i \psi_{\RR,i}^{\xi,\dag}\psi_{\RR,i}^{\xi} + \sum_{i}\frac{\delta_{\xi,\xi'}}{2} 
(\psi_{\RR,i}^{\xi,\dag} \psi_{\RR,i}^{\xi'} -\psi_{\RR_2,i}^{\xi',\dag} \psi_{\RR_2,i}^{\xi} )+\delta_{\xi,\xi'} 
\label{eq:bond_op_decomp_1}
\eaa  
}

\bh{Using Eq.~\ref{eq:bond_op_decomp_1}}, the first term in RKKY Hamiltonian (Eq.~\ref{eq:ham_rkky_v_0}) can then be written as 
\hh{ 
\baa  
&\sum_{\RR,\RR_2, \mu\nu}\sum_{\xi\xi'}J_{0}^{RKKY}(\RR-\RR_2):\Sigma_{\mu\nu}^{(f,\xi\xi')}(\RR)::\Sigma_{\mu\nu}^{(f,\xi'\xi)}(\RR_2) : \nonumber \\
=&-\sum_{\RR,\RR_2,\xi\xi'}J_0^{RKKY}(\RR-\RR_2)A_{\RR,\RR_2}^{\xi,\xi} A_{\RR,\RR_2}^{\xi',\xi'} 
+ 4\sum_{\RR}J_0^{RKKY}(0)(\nu_f+4)+ \sum_{\RR,\RR_2}J_0^{RKKY}(\RR-\RR_2) (\nu_f - \nu_f)  \nonumber \\ 
& + \sum_{\RR,\RR_2}2J_0^{RKKY}(\RR-\RR_2) \nonumber \\
=& -\sum_{\RR\ne \RR_2,\xi,\xi'}J_0^{RKKY}(\RR-\RR_2)A_{\RR,\RR_2}^{\xi,\xi,\dag}A_{\RR,\RR_2}^{\xi',\xi'} -\sum_{\RR}J_0^{RKKY}(0)(\nu_f+4)^2
+\sum_{\RR}J_0^{RKKY}(0)4(\nu_f+4)
\nonumber \\
&
+ \sum_{\RR,\RR_2}2J_0^{RKKY}(\RR-\RR_2)
\label{eq:ham_rkky_v_0}
\eaa 
}
\hh{ 
\bh{Using Eq.~\ref{eq:bond_op_decomp_1}}, the second term in Eq.~\ref{eq:ham_rkky_v0} can be written as 
\baa  
&\sum_{\RR,\RR_2,\mu\nu}\sum_\xi J^{RKKY}_{1}(\RR-\RR_2):\Sigma_{\mu\nu}^{(f,\xi\xi)}(\RR)::\Sigma_{\mu\nu}^{(f,\xi\xi)}(\RR_2):\nonumber \\
=&- \sum_{\RR\ne \RR_2,\xi}J_1^{RKKY}(\RR-\RR_2)A_{\RR,\RR_2}^{\xi,\xi,\dag} 
A_{\RR,\RR_2}^{\xi,\xi} - \sum_{\RR,\RR_2,\xi}J_1^{RKKY}(\RR=0) (\nu_f^{\xi}+2)^2+4\sum_{\RR}J_1^{RKKY}(0)(\nu_f+4) \nonumber \\
&
+ \sum_{\RR,\RR_2}2J_1^{RKKY}(\RR-\RR_2)
\, .
\label{eq:ham_rkky_v_0}
\eaa 
} 

\hh{ 
The Hamiltonian reads 
\baa 
\hH_{RKKY}^{v_\star^\prime=0,M=0}= &- \sum_{\RR\ne \RR_2,\xi,\xi'}J_{0}^{RKKY}(\RR-\RR_2)A^{\xi,\xi,\dag}_{\RR,\RR_2}A^{\xi',\xi'}_{\RR,\RR_2}
-\sum_{\RR\ne \RR_2,\xi}J_{1}^{RKKY}(\RR-\RR_2) A^{\xi,\xi,\dag}_{\RR,\RR_2}A^{\xi,\xi}_{\RR,\RR_2} 
\\
&\bh{-}\sum_{\RR }J^{RKKY}_{1}(0)\bigg[(\nu_f^{+1}+2)^2 + (\nu_f^{-1}+2)^2
\bigg] + \bh{E_0}
\label{eq:hrkky_vp_0}
\eaa  
\bh{where $E_0= -\sum_{\RR}J_0^{RKKR}(0)(\nu_f+4)^2 + \sum_{\RR}J_0^{RKKY}(0)4(\nu_f+4) +\sum_{\RR,\RR_2}2J_0^{RKKY}(\RR-\RR_2) $ is a constant at fixed $\nu_f$.}
}

\bh{ 
We now solve the Hamiltonian and find the ground states. 
We first note that, for given state $|\psi \rangle$, its energy is
\baa  
&\langle \psi| \hH_{RKKY}^{v_\star^\prime=0,M=0} |\psi\rangle \nonumber \\
=&- \sum_{\RR\ne \RR_2}J_{0}^{RKKY}(\RR-\RR_2)\langle \psi| \sum_{\xi,\xi'}A^{\xi,\xi,\dag}_{\RR,\RR_2}A^{\xi',\xi'}_{\RR,\RR_2} |\psi\rangle 
-\sum_{\RR\ne \RR_2,\xi}J_{1}^{RKKY}(\RR-\RR_2) \langle \psi| A^{\xi,\xi,\dag}_{\RR,\RR_2}A^{\xi,\xi}_{\RR,\RR_2} |\psi\rangle \nonumber 
\\
&-\sum_{\RR }J^{RKKY}_{1}(0)\bigg[\langle \psi| (\nu_f^{+1}+2)^2 + (\nu_f^{-1}+2)^2|\psi\rangle 
\bigg] + E_0 \nonumber \\ 
\ge &- \sum_{\RR}J_1^{RKKY}(0)\bigg[\langle \psi| (\nu_f^{+1}+2)^2 + (\nu_f^{-1}+2)^2|\psi\rangle 
\bigg]+E_0
\eaa  
where the equality holds when $\langle \psi |A_{\RR,\RR_2}^{\xi,\xi,\dag}A_{\RR,\RR_2}^{\xi,\xi}|\psi\rangle =0$ for all $\RR\ne \RR_2,\xi$, and we use the fact that $J_0^{RKKY}(\RR-\RR_2)\le 0 ,J_1^{RKKY}(\RR-\RR_2)\le 0 $. At fixed $\nu_f$, we then note that 
\baa  
E_{\nu} = - \sum_{\RR}J_1^{RKKY}(0)\bigg[\langle \psi| (\nu_f^{+1}+2)^2 + (\nu_f^{-1}+2)^2|\psi\rangle 
\bigg]
\eaa  
is minimized by (similar calculations have been provided in Eq.~\ref{eq:fill_zero_hhyb_rkky})
\baa  
&\nu_f =-2\quad:\quad (\nu_f^{+1} ,\nu_f^{-1}) = (-1,-1) \nonumber \\
&\nu_f =-1\quad:\quad (\nu_f^{+1} ,\nu_f^{-1}) = (0,-1),(\nu_f^{+1} ,\nu_f^{-1}) = (-1,0)\nonumber \\
&\nu_f =0\quad:\quad (\nu_f^{+1} ,\nu_f^{-1}) = (0,0)
\label{eq:gnd_req_u4u4_2} \, . 
\eaa  
We let $E_{\nu}^{min}$ denote the minimum value of $E_{\nu}$ with $E^{min}_{\nu} =-2N_M J_1^{RKKY}(0),-5N_M J_1^{RKKY}(0), -8N_MJ_1^{RKKY}(0)$ at $\nu_f=-2,-1,0$ respectively. Then 
\baa  
\langle \psi| \hH_{RKKY}^{v_\star^\prime=0,M=0} |\psi\rangle  \ge E_{\nu}^{min} +E_0
\eaa 
Therefore, $|\psi\rangle$ with energy $E_{\nu}^{min}+E_0$ must be the ground state. We now prove the following states have energy $E_{\nu}^{min}+E_0$ and are the ground states:
\baa  
|\psi_0\rangle = \prod_\RR \bigg( 
\prod_{n=1}^{\nu_f^{+1}+2}\psi_{\RR,i_n}^{+,\dag} 
\prod_{n=\nu_f^{+1}+3}^{\nu^f+4} \psi_{\RR,i_n}^{-,\dag} |0\rangle 
\bigg) 
\label{eq:gnd_u4u4_psi}
\eaa  
where $\{i_n\}$ are chosen arbitrarily and $(\nu_{f}^{+1}, \nu_{f}^{-1}=\nu_f-\nu_f^{+1})$ satisfy Eq.~\ref{eq:gnd_req_u4u4_2}. We first prove $A_{\RR,\RR_2}^{\xi,\xi}|\psi_0\rangle =0$ with $\RR\ne \RR_2$. Since $A_{\RR,\RR_2}^{\xi,\xi} = \sum_i \psi_{\RR,i}^{\xi,\dag} \psi_{\RR_2,i}^{\xi}$ will move an electrons at site $\RR$ with flavor $\xi,i$ to another site $\RR_2$ with flavor $\xi,i$, then, for the states in Eq.~\ref{eq:gnd_u4u4_psi}, there is either zero electron at $(\RR,\xi,i)$ or one electrons at both $(\RR,\xi,i),(\RR_2,\xi,i)$. Therefore, $A_{\RR,\RR_2}^{\xi,\xi}$ must annihilate $|\psi_0\rangle$. Thus 
\baa 
 \sum_{\RR\ne \RR_2}J_{0}^{RKKY}(\RR-\RR_2)\langle \psi_0| \sum_{\xi,\xi'}A^{\xi,\xi,\dag}_{\RR,\RR_2}A^{\xi',\xi'}_{\RR,\RR_2} |\psi_0\rangle 
-\sum_{\RR\ne \RR_2,\xi}J_{1}^{RKKY}(\RR-\RR_2) \langle \psi_0| A^{\xi,\xi,\dag}_{\RR,\RR_2}A^{\xi,\xi}_{\RR,\RR_2} |\psi_0\rangle =0 \, . 
\label{eq:u4u4_anni}
\eaa 
In addition, from the constructions of $|\psi_0\rangle$ 
\baa
 &- \sum_{\RR}J_1^{RKKY}(0)\bigg[\langle \psi| (\nu_f^{+1}+2)^2 + (\nu_f^{-1}+2)^2|\psi\rangle 
\bigg] = E_{\nu}^{min}
\label{eq:u4u4_fill_cond}
\eaa 
Combining Eq.~\ref{eq:u4u4_anni} and Eq.~\ref{eq:u4u4_fill_cond}, we have 
\baa  
\langle \psi_0 |\hH_{RKKY}^{v_\star^\prime=0,M=0}|\psi_0\rangle = E_{\nu}^{min} +E_0
\eaa  
and $|\psi_0\rangle$ in Eq.~\ref{eq:gnd_u4u4_psi} are the ground states. 
}

\bh{ 
Equivalently, we could also rewrite Eq.~\ref{eq:gnd_u4u4_psi} with $f$-electron operator (Eq.~\ref{eq:psi_basis_def})}
\baa  
|\psi_0\rangle = \prod_\RR 
\bigg( \prod_{i=1}^{\nu_f^{+1}+2} f_{\RR,\alpha_i\eta_is_i}^\dag  \prod_{i={\nu_f^{+1}+3}}^{\nu^f+4}f_{\RR,\alpha_i\eta_i s_i}^\dag \bigg)|0\rangle
\label{eq:gnd_u4u4}
\eaa  
where $(-1)^{\alpha_i+1} \eta_i =1 $ for $i=1,...,\nu_f^{+1}+2$ and
$(-1)^{\alpha_i+1}\eta_i = -1 $ for $i=\nu_f^{+1}+3,...,\nu_f^{-1}+2$, and $\nu_f^+,\nu_f^-$ satisfy Eq.~\ref{eq:gnd_req_u4u4_2}.

\subsection{Ground state at $v_\star^\prime \ne 0,M=0$}
We next consider the ground state at $v_\star^\prime \ne 0, M=0 $. The Hamiltonian now takes the form of 
\baa 
\hH_{RKKY}^{v_\star^\prime\ne 0,M=0}=\hH_{RKKY}^{v_\star^\prime=0,M=0}+\hH_1^{v_\star^\prime} 
\eaa  
where $\hH_{RKKY}^{v_\star^\prime=0,M=0}$ is defined in Eq.~\ref{eq:ham_rkky_v0} and 
\ba 
\hH_1^{v_\star^\prime} = &\sum_{\xi,\RR,\RR_2,\mu\nu}  J^{RKKY}_2(\RR-\RR_2)
:\UF_{\mu\nu}^{(f,\xi\xi)}(\RR):
:\UF_{\mu\nu}^{(f,-\xi-\xi)}(\RR_2): 
\nonumber \\
&
+\sum_{\xi,\RR,\RR_2,\mu\nu} J^{RKKY}_{3,\xi}(\RR-\RR_2) 
:\UF_{\mu\nu}^{(f,\xi \xi)}(\RR):
:\UF_{\mu\nu}^{(f,-\xi \xi)}(\RR_2): 
\nonumber \\
&
+\sum_{\xi,\RR,\RR_2,\mu\nu}J^{RKKY}_{4,\xi}(\RR-\RR_2):\UF_{\mu\nu}^{(f,\xi \xi)}(\RR):
\UF_{\mu\nu}^{(f,\xi -\xi)}(\RR_2): \\
& + \sum_{\xi,\RR,\RR_2,\mu\nu}J_{5,\xi}^{RKKY}(\RR-\RR_2):
:\UF_{\mu\nu}^{(f,\xi -\xi)}(\RR):
\UF_{\mu\nu}^{(f,\xi -\xi)}(\RR_2): \,.
\ea 
\hb{
We next treat $\hH_1^{v_\star^\prime}$ as perturbations and use degenerate perturbation theory to determine the ground states of $\hH_{RKKY}^{v_\star^\prime\ne 0,M=0}$. We let $\{|\psi_{0,i}\rangle\}$ be the ground states defined in Eq.~\ref{eq:gnd_u4u4}, and construct the following matrix.
Then we can construct the following matrix
\ba 
[H_1]_{ij} = \langle \psi_{0,i} | 
\hH_1^{v_\star^\prime} 
|\psi_{0,j}\rangle  \, .
\ea 
We then diagonalize $H_1$ and the states with the lowest eigenvalues are the grounds states. We first note that $|\psi_{0,i}\rangle$ (given in Eq.~\ref{eq:gnd_u4u4}) can be written as the following product states 
\baa  
|\psi_{0,i}\rangle = \prod_{\RR} |\phi_{0,i}(\RR)\rangle
\eaa 
where $|\phi_{0,i}(\RR) \rangle$ is the corresponding state at $\RR$. From the definition (Eq.~\ref{eq:gnd_u4u4}), we find
\baa  
\langle \phi_{0,i}(\RR) | \phi_{0,j}(\RR)\rangle = \delta_{i,j}
\label{eq:phi_ij_orth}
\eaa  
}

\hb{
We now evaluate the off-diagonal terms $[\hH_1]_{ij}$ with $i\ne j$. 
We first consider the effect of $J_{2,\xi}^{RKKY}$. For $i\ne j$
\baa  
&\langle \psi_{0,i} | \sum_{\xi,\RR,\RR_2,\mu\nu} J^{RKKY}_{2}(\RR-\RR_2) 
:\UF_{\mu\nu}^{(f,\xi \xi)}(\RR):
:\UF_{\mu\nu}^{(f,-\xi -\xi)}(\RR_2):   |\psi_{0,j}\rangle \nonumber \\
=&\sum_{\RR,\xi,\mu\nu} J_{2}^{RKKY}(0)
\langle \phi_{0,i}(\RR)| :\UF_{\mu\nu}^{(f,\xi \xi)}(\RR):
:\UF_{\mu\nu}^{(f,-\xi -\xi)}(\RR): |\phi_{0,j}(\RR)\rangle  
\prod_{\RR_2,\RR_2 \ne \RR} 
\langle \phi_{0,i}(\RR_2)| |\phi_{0,j}(\RR_2)\rangle 
\nonumber \\
&
+ \sum_{\RR,\RR_2,\xi,\mu\nu} J_{2}^{RKKY}(\RR-\RR_2) 
\langle \phi_{0,i}(\RR)| :\UF_{\mu\nu}^{(f,\xi \xi)}(\RR):
|\phi_{0,j}(\RR)\rangle  \nonumber \\ 
&
\langle \phi_{0,i}(\RR_2)| 
:\UF_{\mu\nu}^{(f,-\xi -\xi)}(\RR_2): |\phi_{0,j}(\RR_2)\rangle 
\prod_{\RR_3,\RR_3 \ne \RR,\RR_3\ne \RR_2} 
\langle \phi_{0,i}(\RR_3)| |\phi_{0,j}(\RR_3)\rangle =0
\eaa    
where we use Eq.~\ref{eq:phi_ij_orth}. Similarly, the contributions of $J_{3,\xi}^{RKKY},J_{4,\xi}^{RKKY},J_{5,\xi}^{RKKY}$ also vanishes when $i\ne j$. Then $[\hH_1]_{ij}=0$ when $i\ne j$. 
}
\bh{
We only need to consider the diagonal components $[\hH_1]_{ii}$, and the states that minimize $[\hH_1]_{ii}$ are the ground states. 
} 

\bh{
We first consider the effect of $J_{3,\xi}^{RKKY}$
\baa  
&\langle \psi_{0,i} | \sum_{\xi,\RR,\RR_2,\mu\nu} J^{RKKY}_{3,\xi}(\RR-\RR_2) 
:\UF_{\mu\nu}^{(f,\xi \xi)}(\RR):
:\UF_{\mu\nu}^{(f,-\xi \xi)}(\RR_2):   |\psi_{0,i}\rangle \nonumber \\
=&
\sum_{\RR,\RR_2,\xi,\mu\nu} J_{3,\xi}^{RKKY}(\RR-\RR_2) 
\langle \phi_{0,i}(\RR)| :\UF_{\mu\nu}^{(f,\xi \xi)}(\RR):
|\phi_{0,i}(\RR)\rangle  \nonumber \\ 
&
\langle \phi_{0,i}(\RR_2)| 
:\UF_{\mu\nu}^{(f,-\xi \xi)}(\RR_2): |\phi_{0,i}(\RR_2)\rangle 
\prod_{\RR_3,\RR_3 \ne \RR,\RR_3\ne \RR_2} 
\langle \phi_{0,i}(\RR_3)| |\phi_{0,i}(\RR_3)\rangle 
\eaa   
where we use the fact that $J_{3,\xi}^{RKKY}(0)=0$ (Eq.~\ref{eq:j_rkky_1}). For the ferromagnetic states (in Eq.~\ref{eq:gnd_u4u4}), $\langle \phi_{0,i}(\RR)| 
:\UF_{\mu\nu}^{(f,\xi \xi')}(\RR): |\phi_{0,i}(\RR)\rangle  $ 
 does not depend on $\RR$. We let 
\baa  
\sigma_{\mu\nu}^{\xi\xi'} = 
\langle \phi_{0,i}(\RR)| 
:\UF_{\mu\nu}^{(f,\xi \xi')}(\RR): |\phi_{0,i}(\RR)\rangle
\eaa
Then 
\baa  
&\langle \psi_{0,i} | \sum_{\xi,\RR,\RR_2,\mu\nu} J^{RKKY}_{3,\xi}(\RR-\RR_2) 
:\UF_{\mu\nu}^{(f,\xi \xi)}(\RR):
:\UF_{\mu\nu}^{(f,-\xi \xi)}(\RR_2):   |\psi_{0,i}\rangle 
=
\sum_{\RR,\RR_2,\xi,\mu\nu} \sigma_{\mu\nu}^{\xi\xi'}  \sigma_{\mu\nu}^{\xi\xi'} J^{RKKY}_{3,\xi}(\RR-\RR_2) =0
\eaa 
where we use the fact that $\sum_{\RR}J^{RKKY}_{3,\xi}(\RR-\RR_2)=0 $ (Eq.~\ref{eq:j_rkky_1}). Then we find $J_{3,\xi}^{RKKY}$ will not contribute to $[\hH_1]_{ii}$  For the same reason, we also have 
\baa  
\langle \psi_{0,i} | \sum_{\xi,\RR,\RR_2,\mu\nu} J^{RKKY}_{4,\xi}(\RR-\RR_2) 
:\UF_{\mu\nu}^{(f,\xi \xi)}(\RR):
:\UF_{\mu\nu}^{(f,\xi -\xi)}(\RR_2):   |\psi_{0,i}\rangle =0\nonumber \\
\langle \psi_{0,i} | \sum_{\xi,\RR,\RR_2,\mu\nu} J^{RKKY}_{5,\xi}(\RR-\RR_2) 
:\UF_{\mu\nu}^{(f,\xi -\xi)}(\RR):
:\UF_{\mu\nu}^{(f,\xi -\xi)}(\RR_2):   |\psi_{0,i}\rangle =0
\eaa 
(Note that, from Eq.~\ref{eq:j_rkky_1}, $J^{RKKY}_{4,\xi}(0)=J^{RKKY}_{5,\xi}(0)=0  $ and 
$\sum_\RR J^{RKKY}_{4,\xi}(\RR) = \sum_\RR J^{RKKY}_{5,\xi}(\RR) =0$ ).
}

\bh{
Therefore, the only non-zero contributions come from $J_{2}^{RKKY}$ (since $J_{2}^{RKKY}(0)\ne 0, \sum_{\RR} J_{2}^{RKKY}(\RR)\ne 0$). We find
\baa  
[H_1]_{ii} =&\langle \psi_{0,i} | \sum_{\xi,\RR,\RR_2,\mu\nu} J^{RKKY}_{2}(\RR-\RR_2) 
:\UF_{\mu\nu}^{(f,\xi \xi)}(\RR):
:\UF_{\mu\nu}^{(f,-\xi -\xi)}(\RR_2):   |\psi_{0,i}\rangle\nonumber \\
=&\sum_{\RR,\xi,\mu\nu} J_{2}^{RKKY}(0) 
\langle \phi_{0,i}(\RR)| :\UF_{\mu\nu}^{(f,\xi \xi)}(\RR):
 :\UF_{\mu\nu}^{(f,-\xi -\xi)}(\RR):
|\phi_{0,i}(\RR)\rangle  \nonumber \\ 
&+\sum_{\RR\ne \RR_2,\xi,\mu\nu} J_2^{RKKY}(\RR-\RR_2) \sigma_{\mu\nu}^{\xi\xi} \sigma_{\mu\nu}^{-\xi-\xi} 
\label{eq:h1_n_vp_def}
\eaa 
}
\bh{We emphasize that $J_2^{RKKY}$ plays the rule of picking the true ground state in the nonchiral-flat limit.}

\bh{
We direct evaluate $[\hH_1]_{ii}$ (Eq.~\ref{eq:h1_n_vp_def}) and the ground states are the states $|\psi_{0,i}\rangle$ with lowest $[\hH_1]_{ii}$. We use $\{\alpha \eta s\}$ to characterize the states in Eq.~\ref{eq:gnd_u4u4}, where $\{\alpha \eta s\}$ denote the set of flavors with one $f$-electron per site. 
}

\bh{
At $\nu_f=0$, we find 
the following states have $[\hH_1]_{ii}= \sum_{\RR\RR_2}2J_2^{RKKY}(\RR-\RR_2)$
\baa  
&\{ 1+\dn,2-\dn,2+\dn,1-\dn \}, \{2-\up,2-\dn,1-\up,1-\dn\}, \{1+\dn,2-\up,2+\dn,1-\up\},\{1+\up,2-\dn,2+\up,1-\dn \} ,\nonumber \\
&\{1+\up,1+\dn,2+\up,2+\dn\}
,\{1+\up,2-\up,2+\up,1-\up\}
\label{eq:gnd_state_v0_0}. 
\eaa  
The following states have $[\hH_1]_{ii}= -\sum_{\RR\RR_2}2J_2^{RKKY}(\RR-\RR_2)$
\baa  
&\{1+\up,2-\up,2+\up,1-\up \}, \{1+\up,1+\dn,2+\up,2+\dn \}, \{1+\up,2-\dn,2+\up,1-\dn\},\nonumber \\
&\{2-\up,2-\dn,1-\up,1-\dn\},\{1+\dn,2-\dn,1-\dn,2+\dn \}
\eaa 
All other states (of Eq.~\ref{eq:gnd_u4u4}) at $\nu_f=0$ have $[\hH_1]_{ii}=0$. Then Eq.~\ref{eq:gnd_state_v0_0} gives the ground state at $\nu_f=0$. 
}

\bh{
At $\nu_f=-1$, we find the following states have $[\hH_1]_{ii}=-\sum_{\RR\RR_2}J_2^{RKKY}(\RR-\RR_2)$
\baa  
&\{ 1+\up,1+\dn, 2+\up\}, \{ 1+\up,1+\dn, 2+\dn\},
\{1+\up,2-\up,2+\dn\} ,\{1+\up,2-\up,1-\dn\} ,
\{1+\up,2-\dn, 2+\dn\},\{1+\up,2-\dn, 1-\up\}, \nonumber \\
&\{1+\dn,2-\up, 2+\up \}, \{1+\dn,2-\up, 1-\dn\} ,
\{1+\dn,2-\dn,2+\up \} ,\{ 1+\dn,2-\dn,1-\up\}, 
\{2-\up,2-\dn, 2+\up\}, \{2-\up,2-\dn, 2+\dn\} \nonumber \\
&\{ 1+\up, 2+\up,2+\dn\}, \{ 1+\dn, 2+\up,2+\dn\},
\{1+\dn,2+\up,1-\up\} ,\{2-\dn,2+\up,1-\up\} ,
\{1+\dn,2+\up,1-\dn\},\{2-\up,2+\up,1-\dn\}, \nonumber \\
&\{ 1+\up,2+\dn,1-\up \}, \{2-\dn\, 2+\dn,1-\up\} ,
\{1+\up,2+\dn,1-\dn \} ,\{2-\up, 2+\dn,1-\dn\}, 
\{1+\up,1-\up,1-\dn\}, \{1+\dn,1-\up,1-\dn\} 
\label{eq:gnd_state_v0_1}
\eaa 
and all other states at $\nu_f=-1$ have $[\hH_1]_{ii}=\sum_{\RR\RR_2}J_2^{RKKY}(\RR-\RR_2)$. Then, Eq.~\ref{eq:gnd_state_v0_1} gives the ground state at $\nu_f=-1$.
}

\bh{
At $\nu_f=-2$, we find the following states have $[\hH_1]_{ii}=2\sum_{\RR\RR_2}J_2^{RKKY}(\RR-\RR_2)$
\baa  
&\{ 1+\up,2+\up\}, \{ 1+\dn, 2+\dn\}, \{ 2-\up, 1-\up\}, 
\{ 2-\dn, 1-\dn\} 
\label{eq:gnd_state_v0_2}
\eaa 
and all other states at $\nu_f=-2$ have $[\hH_1]_{ii}=0$. Then, Eq.~\ref{eq:gnd_state_v0_2} gives the ground state at $\nu_f=-2$.
}

\bh{
In summary, Eq.~\ref{eq:gnd_state_v0_0}, Eq.~\ref{eq:gnd_state_v0_1}, Eq.~\ref{eq:gnd_state_v0_2} give ground states at $\nu_f=0,-1,-2$ respectively. }
\hh{ 
In a more compact form, ground states at $\nu_f=0,-2$ are
\baa  
|\psi_0\rangle = \prod_\RR \prod_{i=1}^{(\nu_f+4)/2} f_{\RR,1\eta_is_i}^\dag f_{\RR,2\eta_is_i}|0\rangle
\label{eq:gnd_nuf_even}
\eaa  
and the ground states at $\nu_f=-1$ are 
\baa  
|\psi_0\rangle = \prod_\RR f_{\RR,\xi \eta's' }^\dag \prod_{i=1}^{2} f_{\RR,1\eta_is_i}^\dag f_{\RR,2\eta_is_i}|0\rangle 
\label{eq:gnd_nuf_1}
\eaa 
where $(\eta',s') \ne (\eta_i,s_i)$. \bh{We also provide the ground states in the $\psi$ basis (Eq.~\ref{eq:psi_basis_def}). At $\nu_f=0,-2$, the ground states are 
\baa  
|\psi_0\rangle = \prod_\RR \prod_{n=1}^{(\nu_f+4)/2} \psi_{\RR,i_n}^{f,+,\dag}\psi_{\RR,i_n}^{f,+,\dag} |0\rangle 
\eaa  
where ${i_n}$ are chosen arbitrarily. At $\nu_f=-1$, the ground states are
\baa  
|\psi_0\rangle = \psi_{\RR,j_n}^{f,\xi,\dag} \prod_\RR \psi_{\RR,j}^{f,\xi,\dag} \psi_{\RR,i}^{f,+,\dag} \psi_{\RR,i}^{f,-,\dag} |0\rangle 
\eaa  
where $i,j,\xi$ are chosen arbitrarily. The ground states given in Eq.~\ref{eq:gnd_nuf_even} and Eq.~\ref{eq:gnd_nuf_1} also give rise to the same ground states as derived from projected Coulomb Hamiltonian~\cite{tbgiv} in nonchiral-flat limit. 
}
}

\hh{ 
In terms of the Young tableaux, the ground state at $\nu_f=0$ corresponds to 
\[
\begin{ytableau}
       \tikznode{a1}{~} & \tikznode{a1}{~} & \cdots & \tikznode{a1}{~} &\tikznode{a1}{~} \\
       \tikznode{a2}{~} & \tikznode{a3}{~} & \cdots & \tikznode{a4}{~} &\tikznode{a5}{~} \\
\end{ytableau}
\]
\tikz[overlay,remember picture]{%
\draw[decorate,decoration={brace},thick] ([yshift=-2mm,xshift=2mm]a5.south east) -- 
([yshift=-2mm,xshift=-2mm]a2.south west) node[midway,below]{$2N_M$};}
}
\newline 
\hh{ 
There are $2N_M$ boxes in the first and second rows. $N_M$ columns correspond to the $\xi=+1$ and $N_M$ columns correspond to the $\xi=-1$ fermions (note that we have $N_M$ site). The ferromagnetic nature of the ground state indicates the symmetry under the permutation of the position indices. In addition, the ground state in Eq.~\ref{eq:gnd_nuf_even} is also symmetric under the permutation of $\xi$ indices. This in total leads to $2N_M =2 \times N_M$ boxes for each row. 
}
\hh{ 
Similarly, at $\nu_f = -2$, the corresponding Young tableaxu of the ground states are
\[
\begin{ytableau}
       \tikznode{a1}{~} & \tikznode{a2}{~} & \cdots & \tikznode{a3}{~} &\tikznode{a4}{~} \\
\end{ytableau}
\]
\tikz[overlay,remember picture]{%
\draw[decorate,decoration={brace},thick] ([yshift=-2mm,xshift=2mm]a4.south east) -- 
([yshift=-2mm,xshift=-2mm]a1.south west) node[midway,below]{$2N_M$};}
}
\newline 
\hh{
There are $2N_M$ boxes in the first row. $N_M$ columns correspond to the $\xi=+1$ and $N_M$ columns correspond to the $\xi=-1$ fermions (note that we have $N_M$ site). The ferromagnetic nature of the ground state indicates the symmetry under the permutation of the position indices. In addition, the ground state in Eq.~\ref{eq:gnd_nuf_even} is also symmetric under the permutation of $\xi$ indices. This in total leads to $2N_M =2 \times N_M$ boxes for each row. 
}

\hh{ 
At $\nu_f = -1$, the corresponding Young tableaux of the ground states are
\[
\begin{ytableau}
       \tikznode{a1}{~} & \cdots & \tikznode{a2}{~}  &\tikznode{a3}{~} & \cdots & \tikznode{a4}{~} \\
       \tikznode{a5}{~} & \cdots & \tikznode{a6}{~} \\
\end{ytableau}
\]
\tikz[overlay,remember picture]{%
\draw[decorate,decoration={brace},thick] ([yshift=-2mm,xshift=2mm]a6.south east) -- 
([yshift=-2mm,xshift=-2mm]a5.south west) node[midway,below]{$N_M$};
\draw[decorate,decoration={brace},thick] ([yshift=-2mm,xshift=2mm]a4.south east) -- 
([yshift=-2mm,xshift=-2mm]a3.south west) node[midway,below]{$N_M$};
}
\newline 
}
\hh{ 
There are $2N_M$ boxes in the first row and $N_M$ boxes in the second row. In the case of $\nu_f^{+1}=0,\nu_f^{-1}=-1$. The first $N_M$ columns correspond to the $\xi=+1$ and $N$ columns correspond to the $\xi=-1$ fermions. The ferromagnetic nature of the ground state indicates the symmetry under the permutation of the position indices. In addition, the ground state in Eq.~\ref{eq:gnd_nuf_1} tends to be as symmetric as possible under the permutation of $\xi$ indices. Consequently, there are $2N_M$ boxes in the first row which indicates the symmetric properties between $\xi=+1,\xi=-1$ and also between different sites. The remaining $N_M$ boxes in the second row capture the symmetric feature under the permutation of site indices of the remaining $\xi=+1$ fermions. 
}

\subsection{Comparisons of the ground states at different limits} 
\hh{
We now compare the ground states at different limits. In general, we find three types of ground states depending on the values of $\gamma,v_\star^\prime, J$. 
\begin{itemize}
    \item Type I: Defined in Eq.~\ref{eq:gnd_state_m0} or Eq.~\ref{eq:gnd_u4u4} \bh{for $v_\star^\prime = 0, M=0$}.
    \baa  
    \prod_\RR 
    \bigg( \prod_{i=1}^{\nu_f^{+1}+2} f_{\RR,\alpha_i\eta_is_i}^\dag  \prod_{i={\nu_f^{+1}+3}}^{\nu^f+4}f_{\RR,\alpha_i\eta_i s_i}^\dag \bigg)|0\rangle
     \label{eq:u4u4_state_comp}
    \eaa  
    where $(-1)^{\alpha_i+1} \eta_i =1 $ for $i=1,...,\nu_f^{+1}+2$ and
    $(-1)^{\alpha_i+1}\eta_i = -1 $ for $i=\nu_f^{+1}+3,...,\nu_f^{-1}+2$. and  \baa  
    &\nu_f =-2\quad:\quad (\nu_f^{+1} ,\nu_f^{-1}) = (-1,-1) \nonumber \\
    &\nu_f =-1\quad:\quad (\nu_f^{+1} ,\nu_f^{-1}) = (0,-1),(\nu_f^{+1} ,\nu_f^{-1}) = (-1,0)\nonumber \\
    &\nu_f =0\quad:\quad (\nu_f^{+1} ,\nu_f^{-1}) = (2,2)
     \eaa  
    \item Type II: Defined in Eq.~\ref{eq:gnd_nuf_even} and Eq.~\ref{eq:gnd_nuf_1} \bh{for $v_\star^\prime \ne 0, M = 0$}.  
    \newline 
    At $\nu=0,-2$,
    \baa  
    |\psi\rangle = \prod_\RR \prod_{i=1}^{(\nu_f+4)/2} f_{\RR,1\eta_is_i}^\dag f_{\RR,2\eta_is_i}|0\rangle 
    \eaa  
    At $\nu_f=-1$:
    \baa  
    |\psi\rangle = \prod_\RR f_{\RR,\xi \eta's' }^\dag \prod_{i=1}^{2} f_{\RR,1\eta_is_i}^\dag f_{\RR,2\eta_is_i}|0\rangle 
    \eaa 
    where $(\eta',s') \ne (\eta_i,s_i)$.
    \item Type III: Defined in Eq.~\ref{eq:gnd_state_mn0} \bh{for $v_\star^\prime = 0, M \ne 0$}. Type I ground states (Eq.~\ref{eq:u4u4_state_comp}) with additional requirement $ (\eta_i, s_i) \ne (\eta_j, s_j) \quad \text{ for }\quad  i\ne j$.
\end{itemize}
}

\hh{
The symmetry and the ground states at different limits are 
\bh{
\begin{center}
\begin{tabular}{c|c|c | c | c }
Hybridization  & $\gamma=0,v_\star^\prime = 0$ & $\gamma=0,v_\star^\prime = 0$ & $\gamma \ne 0, v_\star^\prime=0$
    & $\gamma \ne0, v_\star^\prime \ne 0$ \\ 
    \hline 
   $M$ & $M= 0$ & $M \ne 0$ & $M=0$   & $M=0$  
    \\
    \hline 
   Symmetry  & $U(4)\times U(4)$ & Chiral $U(4)$  &  $U(4)\times U(4)$ & Flat $U(4)$ \\ 
   \hline 
   Ground states & Type I & Type III & Type I & Type II
\end{tabular}    
\end{center}
}
}

\hh{
For the type I ground states, we have a degenerate ferromagnetic ground state corresponding to $U(4)\times U(4)$ symmetry. 
}
\hh{ 
For the type II ground states, it is stabilized by a non-zero $v_\star^\prime$. Due to the ferromagnetic interactions between two flat $U(4)$ moments with opposite $\xi$ indices, the ground state tends to align two flat $U(4)$ moments.
}
\hh{ 
For the type III ground states, it is stabilized by a non-zero $M$ in the zero-hybridization limit. Due to the antiferromagnetic interactions between two chiral $U(4)$ moments with opposite $\xi$ indices, the ground state tends to align two chiral $U(4)$ moments. 
}

\section{Fluctuation spectrum of $f$-moments based on RKKY interactions} 
\bh{
In this section, we derive the fluctuation spectrum of $f$-moments at $M=0$ and $\nu_f=0,-1,-2$ from the RKKY Hamiltonian in Eq.~\ref{eq:rkky_ham}. Here, we point out that in this approach, the polarization effect of $c$-electrons has not been included. In Sec.~\ref{sec:eff_thy}, we will obtain a more accurate description of the fluctuations of $f$-moments by including the polarization effect of $c$-electrons. 
}

\bh{ To find the fluctuation spectrum we first rewrite the RKKY Hamiltonian (Eq.~\ref{eq:rkky_ham}) with bond operators $A_{\RR,\RR_2}^{\xi,\xi_2}$ (Eq.~\ref{eq:bond_a_v0}). Then we take the ground states (Eq.~\ref{eq:gnd_nuf_even} and Eq.~\ref{eq:gnd_nuf_1}) and obtain the excitation spectrum of $f$-moments on top of the ground state by calculating the commutator between $\psi_{\RR,i}^{f,\xi,\dag}\psi_{\RR,j}^{f,\xi',\dag}$ and RKKY Hamiltonian (Eq.~\ref{eq:rkky_ham})~\cite{tbgv}. 
}

\subsection{RKKY Hamiltonian}
\hh{ 
We first rewrite the RKKY Hamiltonian in Eq.~\ref{eq:rkky_ham} in a more general formula
\baa 
&\hH_{RKKY} = \sum_{\RR,\RR_2,\mu\nu}  \sum_{\xi_1,\xi_2,\xi_3,\xi_4} :\UF_{\mu\nu}^{(f,\xi_1\xi_2)}(\RR): :\UF_{\mu\nu}^{(f,\xi_3\xi_4)}(\RR_2): J_{RKKY}(\RR-\RR_2,\xi_1\xi_2\xi_3\xi_4) 
\label{eq:rkky_ham_general}
\eaa  
\bh{ 
The new RKKY interactions are connected with RKKY interactions in Eq.~\ref{eq:rkky_ham} via the following relations
\baa  
&J_{RKKY}(\RR,\xi\xi\xi\xi) =J_{0}^{RKKY}(\RR) +J_{1}^{RKKY}(\RR) ,\quad
J_{RKKY}(\RR,\xi\xi-\xi-\xi) =J_{2}^{RKKY}(\RR) ,\quad \nonumber\\
&J_{RKKY}(\RR,\xi-\xi-\xi \xi) =J_{0}^{RKKY}(\RR) ,\quad
J_{RKKY}(\RR,\xi-\xi \xi-\xi) =J_{5}^{RKKY}(\RR) \nonumber \\
&J_{RKKY}(\RR,\xi \xi \xi-\xi) =J_{4\xi}^{RKKY}(\RR) ,\quad 
J_{RKKY}(\RR,\xi \xi -\xi \xi) =J_{3\xi}^{RKKY}(\RR) 
\label{eq:def_new_rkky}
\eaa
with all other components of $J_{RKKY}(\RR,\xi_1 \xi_2\xi_3 \xi_4)$ vanish.
}
}

\hh{
\bh{In order to obtain the fluctuation spectrum, we first rewrite all the terms in $\hH_{RKKY}$ (Eq.~\ref{eq:rkky_ham_general}) via the bond operator $A_{\RR,\RR_2}^{\xi,\xi_2}$ defined in Eq.~\ref{eq:bond_a_v0} which is also given below (We note that in the previous calculation in Sec~\ref{sec:gnd_state}, we only rewrite $J_0^{RKKY}, J_1^{RKKY}$ term with bond operators)}
\baa  
A_{\RR,\RR_2}^{\xi ,\xi_2} = \sum_i \psi_{\RR,i}^{f,\xi,\dag} \psi_{\RR_2,i}^{f,\xi_2} \, .
\label{eq:bond_a_def}
\eaa  
} 

\hh{
We introduce the following relations
\baa 
:\UF_{\mu\nu}^{(f,\xi\xi_2)}(\RR) : :\UF_{\mu\nu}^{(f,\xi_3\xi_4)}(\RR_2): 
=& \sum_{i,j} :\psi_{\RR,i}^{f,\xi,\dag} \psi_{\RR,j}^{f,\xi_2} : : \psi_{\RR_2,j}^{f,\xi_3,\dag} \psi_{\RR_2,i}^{f,\xi_4}: \nonumber \\
=& \delta_{\xi,\xi_2}\delta_{\xi_3,\xi_4} -\frac{1}{2}\delta_{\xi,\xi_2}A_{\RR_2,\RR_2}^{\xi_3,\xi_4} -\frac{1}{2}\delta_{\xi_3,\xi_4}A_{\RR,\RR}^{\xi,\xi_2} 
+\sum_{i,j} \psi_{\RR,i}^{f,\xi,\dag} \psi_{\RR,j}^{f,\xi_2}  \psi_{\RR_2,j}^{f,\xi_3,\dag} \psi_{\RR_2,i}^{f,\xi_4}  \nonumber \\
=& \delta_{\xi,\xi_2}\delta_{\xi_3,\xi_4} -\frac{1}{2}\delta_{\xi,\xi_2}A_{\RR_2,\RR_2}^{\xi_3,\xi_4} +\frac{1}{2}\delta_{\xi_3,\xi_4}A_{\RR,\RR}^{\xi,\xi_2} 
+4\delta_{\RR,\RR_2}\delta_{\xi_2,\xi_3}A_{\RR,\RR}^{\xi,\xi_4} 
- A_{\RR,\RR_2}^{\xi,\xi_4}A_{\RR_2,\RR}^{\xi_3,\xi_2}
\label{eq:ss_aa}
\eaa 
Using Eq.~\ref{eq:ss_aa}, we rewrite Hamiltonian in Eq.~\ref{eq:rkky_ham_general} as 
\baa 
\hH_{RKKY} = &\sum_{\RR,\RR_2}\bigg[ -\frac{1}{2}\sum_{\xi_1 \xi_3\xi_4}J_{RKKY}(\RR-\RR_2,\xi_1\xi_1\xi_3\xi_4)A_{\RR_2,\RR_2}^{\xi_3,\xi_4} +\frac{1}{2}\sum_{\xi_1 \xi_2 \xi_3}J_{RKKY}(\RR-\RR_2,\xi_1\xi_2\xi_3\xi_3)A_{\RR,\RR}^{\xi_1,\xi_2} 
\bigg] \nonumber \\
&+\sum_{\RR,\xi_1\xi_2\xi_4}4J_{RKKY}(0,\xi_1\xi_2\xi_2\xi_4)A_{\RR,\RR}^{\xi_1,\xi_4} - \sum_{\RR,\RR_2,\xi_1\xi_2\xi_4\xi_4}J_{RKKY}(\RR-\RR_2,\xi_1\xi_2\xi_3\xi_4)A_{\RR,\RR_2}^{\xi_1,\xi_4}A_{\RR_2,\RR}^{\xi_3,\xi_2}
\label{eq:hrkky_intem}
\eaa
where we drop the constant contribution and use the fact that $A_{\RR,\RR}^{\xi,\xi} = \hat{\nu}^\xi_f(\RR)+2 
$.}

\hh{
\bh{We next simplify the RKKY Hamiltonian in Eq.~\ref{eq:hrkky_intem}.} \bh{From Eq.~\ref{eq:def_new_rkky} and Eq.~\ref{eq:j_rkky_1}}, we have $J_{RKKY}(\RR=0,\xi_1\xi_2\xi_3\xi_1)\propto \delta_{\xi_2,\xi_3}$ and $J_{RKKY}(\RR=0,\xi_1 \xi_2\xi_2\xi_4)\propto\delta_{\xi_1,\xi_4}$. We separate then Hamiltonian \bh{in Eq.~\ref{eq:hrkky_intem}} into two parts and have
\baa 
\hH_{RKKY} =& \hH_{RKKY,0} + \hH_{RKKY,1} \nonumber \\
\hH_{RKKY,0}=&\sum_{\RR,\xi} f_1(\xi) \hat{\nu}^\xi_f(\RR) + \sum_{\RR,\xi _1\ne \xi_2}f_2(\xi_1 \xi_2) A^{\xi_1,\xi_2}_{\RR,\RR} + \sum_{\RR,\xi_1,\xi_2} f_3(\xi_1\xi_2) \hat{\nu}_f^{\xi_1} (\RR) \hat{\nu}_f^{\xi_2}(\RR)  \nonumber \\
 \hH_{RKKY,1}=&\sum_{
\substack{ 
      \RR,\RR_2,\xi_1,\xi_2,\xi_3,\xi_4 \\
      (\RR,\xi_1)\ne (\RR_2,\xi_2)\\
       (\RR_2,\xi_3)\ne (\RR,\xi_4)
}
}J(\RR-\RR_2,\xi_1 \xi_2\xi_3\xi_4)A_{\RR,\RR_2}^{\xi_1,\xi_2}A_{\RR_2,\RR}^{\xi_3,\xi_4}
\label{eq:ham_rkky_bond}
\eaa  
where we have dropped the constant contribution and define
\baa  
&f_1(\xi) = -\frac{1}{2}\sum_{\RR_2,\xi_1}J_{RKKY}(\RR_2-\RR,\xi_1\xi_1\xi\xi) + \frac{1}{2} \sum_{\RR_2,\xi_3}J_{RKKY}(\RR-\RR_2,\xi \xi \xi_3\xi_3) + 2\sum_{\xi_2}(J_{RKKY}(0,\xi \xi_2\xi_2\xi)-J_{RKKY}(0,\xi_2 \xi\xi\xi_2)) \nonumber \\
&f_2(\xi_1\xi_2)=-\frac{1}{2}\sum_{\RR_2,\xi}J_{RKKY}(\RR_2-\RR,\xi\xi\xi_1\xi_2) +\frac{1}{2}\sum_{\RR_2,\xi}J_{RKKY}(\RR-\RR_2,\xi_1\xi_2\xi\xi) + 4 \sum_{\xi}J_{RKKY}(0,\xi_1\xi\xi\xi_2) \nonumber \\
&f_3(\xi_1\xi_2)  = -J_{RKKY}(0,\xi_1\xi_2\xi_2\xi_1)\nonumber  \\
&J(\RR-\RR_2,\xi_1\xi_2\xi_3\xi_4) = -J_{RKKY}(\RR-\RR_2,\xi_1\xi_4\xi_3\xi_2) 
\label{eq:ham_rkky_bond_para}
\eaa  
}

\hh{
\bh{Combining Eq.~\ref{eq:ham_rkky_bond_para} and the definition of RKKY interaction (Eq.~\ref{eq:def_new_rkky}), we find}
\ba 
&f_1(\xi) = 0
\\
&f_2(\xi -\xi) = 0 \quad,\quad \text{($f_2(\xi_1\xi_2)$ is only defined for $\xi_1\ne \xi_2$)}
\quad,\quad \\
&f_3(\xi \xi) = -J_0^{RKKY}(\RR=0)-J_1^{RKKY}(\RR=0)
\quad,\quad 
f_3(\xi-\xi) =-J_0^{RKKY}(\RR=0)
\ea 
Then, $\hH_{RKKY,0}$ (Eq.~\ref{eq:ham_rkky_bond}) can be written as 
\baa  
\hH_{RKKY,0} = -J_0^{RKKY}(\RR_2=0) \sum_{\RR}\hat{\nu}_f(\RR)\hat{\nu}_f(\RR)  
-J_1^{RKKY}(\RR_2=0) \sum_{\RR,\xi}\hat{\nu}^{\xi}_f(\RR)
\hat{\nu}^{\xi}_f(\RR)
\eaa  
Since we fix the filling of $f$ at each site, we can replace $\hat{\nu}_f(\RR)$ with an integer number $\nu_f$. The first term is only a constant and it is sufficient to only keep the second term:
\baa  
\hH_{RKKY,0} = 
-J_1^{RKKY}(\RR_2=0) \sum_{\RR,\xi}\hat{\nu}^{\xi}_f(\RR)
\hat{\nu}^{\xi}_f(\RR)
\label{eq:ham_rkky_0_new}
\eaa  
} 

\hh{
\bh{We next go to momentum space}. We consider the Fourier transformation of RKKY interaction $J(\qq,\xi_1\xi_2\xi_3\xi_4) = \sum_{\RR}J(\RR,\xi_1\xi_2\xi_3\xi_4)e^{-i\qq \cdot\RR }$. \bh{Using Eq.~\ref{eq:def_new_rkky} and Eq.~\ref{eq:ham_rkky_bond_para}}, we find 
\baa  
&J(\qq,\xi\xi\xi\xi) =- (J^{RKKY}_0(\qq) -J_0^{RKKY}(\RR=0) +J^{RKKY}_1(\qq) - J_1^{RKKY}(\RR=0) ) \nonumber  \\
&J(\qq,\xi\xi,-\xi-\xi) = -(J^{RKKY}_0(\qq) -J_0^{RKKY}(\RR=0) )  \nonumber \\
&J(\qq,\xi-\xi-\xi\xi) =-(J^{RKKY}_2(\qq) )  \nonumber \\
&J(\qq,\xi \xi-\xi\xi) =- J_{3,\xi}^{RKKY}(\qq) \nonumber \\
&J(\qq,\xi -\xi \xi\xi) =- J_{4,\xi}^{RKKY}(\qq) \nonumber  \\
&J(\qq,\xi-\xi\xi-\xi)=-J_{5,\xi}^{RKKY}(\qq)
\label{eq:def_j_spin_spec}
\eaa  
}

\bh{In summary, combining Eq.~\ref{eq:ham_rkky_0_new} and Eq.~\ref{eq:ham_rkky_bond}, the final RKKY Hamiltonian in terms of the bond operators can be written as 
\baa 
\hH_{RKKY} =& \hH_{RKKY,0} + \hH_{RKKY,1} \nonumber \\
\hH_{RKKY,0}=&-J_1^{RKKY}(\RR_2=0) \sum_{\RR,\xi}\hat{\nu}^{\xi}_f(\RR)
\hat{\nu}^{\xi}_f(\RR)  \nonumber \\
 \hH_{RKKY,1}=&\sum_{
\substack{ 
      \RR,\RR_2,\xi_1,\xi_2,\xi_3,\xi_4 \\
      (\RR,\xi_1)\ne (\RR_2,\xi_2)\\
       (\RR_2,\xi_3)\ne (\RR,\xi_4)
}
}J(\RR-\RR_2,\xi_1 \xi_2\xi_3\xi_4)A_{\RR,\RR_2}^{\xi_1,\xi_2}A_{\RR_2,\RR}^{\xi_3,\xi_4}
\label{eq:ham_rkky_bond_v2}
\eaa 
where the constant term in $\hH_{RKKY,0}$ has been dropped and the RKKY interactions are given in Eq.~\ref{eq:def_j_spin_spec}. We next use Eq.~\ref{eq:ham_rkky_bond_v2} to calculate the fluctuation of $f$-moments on top of the ground states at $M=0,\nu=\nu_f=0,-1,-2$ given in Eq.~\ref{eq:gnd_nuf_even} and Eq.~\ref{eq:gnd_nuf_1}.
}

\subsection{ {Excitation spectrum from RKKY interaction at $\nu_f=0,-2$}}
\hh{ 
We take the ground states $|\psi_0\rangle$ at $\nu_f=0,-2$ \bh{ that are given in Eq.~\ref{eq:gnd_nuf_even} and are also shown below
\baa  
|\psi_0\rangle = \prod_\RR \prod_{n=1}^{(\nu_f+4)/2} \psi^{f,+,\dag}_{\RR,i_n} \psi_{\RR,i_n}^{f,-,\dag}|0\rangle
\label{eq:gnd_nuf_even_v2}
\eaa 
}
where $i_n$ are chosen arbitrarily. 
We then have
\baa 
&A_{\RR,\RR_2}^{\xi,\xi_2}|\psi_0\rangle =0 \quad,\quad (\RR,\xi)\ne (\RR_2,\xi_2)
\label{eq:gnd_cond}
\eaa  
\bh{
We now prove Eq.~\ref{eq:gnd_cond}.
$A_{\RR,\RR_2}^{\xi,\xi_2} = \sum_{i}\psi_{\RR,i}^{f,\xi,\dag} \psi_{\RR_2,i}^{f,\xi_2,\dag}$ describes the procedure of moving one $f$-electron from $(R_2,\xi_2,i)$ to $(\RR,\xi,i)$ with $(\RR_2,\xi_2)\ne (\RR,\xi)$. However, for the ground states in Eq.~\ref{eq:gnd_nuf_even_v2},  $(\RR,\xi,i)$ and $(\RR_2,\xi_2,i)$ are both filled with either one $f$-electron or zero $f$-electron, because they have same valley-spin flavor $i$. Thus the procedure described by $A_{\RR,\RR_2}^{\xi,\xi_2} $ is forbidden and as a consequence $A_{\RR,\RR_2}^{\xi,\xi_2}|\psi_0\rangle =0$ when $(\RR,\xi)\ne (\RR_2,\xi_2)$.
This feature allows us to derive the excitation of RKKY Hamiltonian exactly by calculating the commutator between Hamiltonian and fermion bilinear operators~\cite{tbgv}. 
}
}

\hh{
We first introduce the following commutation relations
\baa  
&[A_{\RR,\RR_2}^{\xi ,\xi_2}, \psi_{\RR',i}^{ f,\xi'} ] =- \delta_{\RR,\RR'}\delta_{\xi,\xi'} \psi_{\RR_2, i}^{ f,\xi_2} \quad ,\quad
[A_{\RR,\RR_2}^{\xi ,\xi_2} , \psi_{\RR',i}^{f, \xi' ,\dag} ] =\delta_{\RR_2,\RR'}\delta_{\xi_2,\xi'} \psi_{\RR,i}^{ f,\xi,\dag}
\label{eq:commutation_rel_1}
\eaa 
We then calculate the commutator 
\baa  
[A_{\RR,\RR_2}^{\xi_1,\xi_2} A_{\RR_2,\RR}^{\xi_3,\xi_4},\psi_{\RR_3,i}^{f,\xi_1',\dag} \psi_{f,\RR_3,j}^{\xi_2'}] =&\delta_{\RR_2,\RR_3}\delta_{\xi_2,\xi_1'}\psi_{\RR,i}^{f,\xi_1,\dag} \psi_{\RR_3,j}^{\xi_2'}A_{\RR_2,\RR}^{\xi_3,\xi_4} 
- \delta_{\RR,\RR_3}\delta_{\xi_1,\xi_2'}
\psi_{\RR_3,i}^{f,\xi_1',\dag}\psi_{\RR_2,j}^{f,\xi_2}A_{\RR_2,\RR}^{\xi_3,\xi_4}  \nonumber \\
&
+\delta_{\RR,\RR_3}\delta_{\xi_1',\xi_4}A_{\RR,\RR_2}^{\xi_1,\xi_2}\psi_{\RR_2,i}^{f,\xi_3,\dag} \psi_{\RR_3,j}^{f,\xi_2'} 
-\delta_{\RR_2,\RR_3}\delta_{\xi_2',\xi_3}A_{\RR,\RR_2}^{\xi_1,\xi_2}\psi_{\RR_3,i}^{f,\xi_1',\dag} \psi_{\RR,j}^{f,\xi_4}
\label{eq:aa_comute}
\eaa  
Using Eq.~\ref{eq:gnd_cond}, Eq.~\ref{eq:commutation_rel_1} and Eq.~\ref{eq:aa_comute}, for $(\RR,\xi_1) \ne (\RR_2,\xi_2)$ and $(\RR_2,\xi_3)\ne (\RR,\xi_4)$, we find
\baa 
&[A_{\RR,\RR_2}^{\xi_1,\xi_2} A_{\RR_2,\RR}^{\xi_3,\xi_4}, \psi_{\RR',i}^{ f,\xi',\dag} \psi_{\RR',i_2}^{ f,\xi_2'}] |\psi_0\rangle =  
[A_{\RR,\RR_2}^{\xi_1,\xi_2} A_{\RR_2,\RR}^{\xi_3,\xi_4}, \psi_{\RR',i}^{f, \xi',\dag} ]\psi_{\RR',i_2}^{f, \xi_2'} |\psi_0\rangle  + \psi_{\RR',i}^{f, \xi',\dag}[A_{\RR,\RR_2}^{\xi_1,\xi_2} A_{\RR_2,\RR}^{\xi_3,\xi_4},  \psi_{\RR',i_2}^{ f,\xi_2'}] |\psi_0\rangle  \nonumber \\
=&\delta_{\RR,\RR'}\delta_{\xi_2,\xi_3}\delta_{\xi_4,\xi'}\psi_{\RR,i}^{f,\xi_1,\dag} \psi_{\RR,i_2}^{f,\xi_2'}  |\psi_0\rangle 
-\delta_{\RR,\RR'}\delta_{\xi_4,\xi'}\delta_{\xi_1,\xi_2'} 
\psi_{\RR_2,i}^{f,\xi_3,\dag} \psi_{\RR_2,i_2}^{f,\xi_2}|\psi_0\rangle   \nonumber \\
&
-\delta_{\RR_2,\RR'}\delta_{\xi_3,\xi_2'} \delta_{\xi_2,\xi'}\psi_{\RR,i}^{f,\xi_1,\dag} \psi_{\RR,i_2}^{f,\xi_4} |\psi_0\rangle 
+\delta_{\RR_2,\RR'}\delta_{\xi_3,\xi_2'}\delta_{\xi_1,\xi_4} \psi_{\RR',i}^{f,\xi',\dag}\psi_{f,\RR',i_2}^{\xi_2} |\psi_0\rangle 
\label{eq:commutation_rel_2}
\eaa
}

\hh{ 
We next consider the charge-0 excitation created by the following operators~\cite{tbgv}:
\bh{
\baa  
O^I_{\qq,ii_2}= &\frac{1}{N_M}\sum_{\RR,\xi'\xi_2'} \psi_{\RR,i}^{f, \xi',\dag} \psi_{\RR,i_2}^{ f,\xi'_2} u^I_{ii_2,\xi'\xi_2'}(\qq)e^{i\qq\cdot \RR} 
 \label{eq:boson_mode}
\eaa  
}
with $u_{ii_2,\xi'\xi_2'}^I(\qq)$ a complex number.
We comment that, we could also introduce bosonic mode created by $\psi_{\RR,i}^{\xi',\dag} \psi_{\RR_2,i_2}^{\xi_2'}$ with $\RR_2\ne \RR$. However, the uniform charge distribution of $f$ will be violated after acting $\psi_{\RR,i}^{\xi',\dag} \psi_{\RR_2,i_2}^{\xi_2'}$ on the ground state.  
\bh{To find the excitation spectrum, we calculate $[\hH_{RKKY},O^I_{\qq,ii_2}]|\psi_0\rangle = [\hH_{RKKY,0},O^I_{\qq,ii_2}]|\psi_0\rangle+[\hH_{RKKY,1},O^I_{\qq,ii_2}]|\psi_0\rangle$;  ($\hH_{RKKY}$ and $|\psi_0\rangle$ given in Eq.~\ref{eq:ham_rkky_bond_v2} and Eq.~\ref{eq:gnd_cond} respectively).
}
} 

\hh{
\bh{We first consider $[\hH_{RKKY,1},O_{\qq,ii_2}^I]|\psi_0\rangle$. Using Eq.~\ref{eq:ham_rkky_bond_v2}, Eq.~\ref{eq:commutation_rel_2} and Eq.~\ref{eq:boson_mode}}, we have
\baa 
&[\hH_{RKKY,1},O^I_{\qq,ii_2}]|\psi_0\rangle  \nonumber \\ 
=& 
\sum_{\RR,\RR_2}\sum_{\xi_1,\xi_2,\xi_3,\xi_4}J(\RR-\RR_2,\xi_1,\xi_2,\xi_3,\xi_4) 
 \frac{1}{N_M}
 \sum_{\RR',\xi',\xi_2'} 
u^I_{ii_2,\xi'\xi_2'}(\qq) 
e^{ i \qq\cdot \RR'} \nonumber \\
&\bigg[ 
\delta_{\RR,\RR'}\delta_{\xi_2,\xi_3}\delta_{\xi_4,\xi'}\psi_{\RR,i}^{f,\xi_1,\dag} \psi_{\RR,i_2}^{f,\xi_2'}  
-\delta_{\RR,\RR'}\delta_{\xi_4,\xi'}\delta_{\xi_1,\xi_2'} 
\psi_{\RR_2,i}^{f,\xi_3,\dag} \psi_{\RR_2,i_2}^{f,\xi_2} \nonumber \\
&
-\delta_{\RR_2,\RR'}\delta_{\xi_3,\xi_2'} \delta_{\xi_2,\xi'}\psi_{\RR,i}^{f,\xi_1,\dag} \psi_{\RR,i_2}^{f,\xi_4} 
+\delta_{\RR_2,\RR'}\delta_{\xi_3,\xi_2'}\delta_{\xi_1,\xi_4} \psi_{\RR',i}^{f,\xi',\dag}\psi_{\RR',i_2}^{f,\xi_2} 
\bigg]|\psi_0\rangle \nonumber 
\\
=&\frac{1}{N_M}\sum_{\qq} 
\bigg[ 
J(\qq_2=0,\xi_1,\xi',\xi',\xi_3)u_{ii_2,\xi_3\xi_2}^I(\qq) \psi_{\RR,i}^{f,\xi_1,\dag} \psi_{\RR,i_2}^{f,\xi_2} 
+J(\qq_2=0,\xi',\xi_2,\xi_4,\xi')u_{ii_2,\xi_1\xi_4}^I(\qq) \psi_{\RR,i}^{f,\xi_1,\dag}\psi_{\RR,i_2}^{f,\xi_2}\nonumber  \\
&-J(-\qq,\xi_4,\xi_2,\xi_1,\xi_3)
u_{ii_2,\xi_3\xi_4}^I(\qq)
\psi_{\RR,i}^{f,\xi_1,\dag}\psi_{\RR,i_2}^{f,\xi_2}
-J(\qq,\xi_1,\xi_3,\xi_4,\xi_2) u_{ii_2,\xi_3\xi_4}(\qq) 
\psi_{\RR,i}^{f,\xi_1,\dag} \psi_{\RR,i_2}^{f,\xi_2}
\bigg] e^{i\qq \cdot \RR } 
\label{eq:commutation_rel_3}
\eaa 
}

\hh{
As for the $\hH_{RKKY,0}$(Eq.~\ref{eq:ham_rkky_bond_v2}), we first introduce the following commutation relation
\baa 
[\sum_{\xi_1}\sum_{\RR}\hat{\nu_f}^{\xi_1}(\RR)\hat{\nu_f}^{\xi_1}(\RR), O_{\qq,ii_2}^I]|\psi_0\rangle  =& \frac{1}{N_M} \sum_{\RR',\xi',\xi'_2,\xi_1}
u^I_{ii_2,\xi'\xi_2'}(\qq) e^{i\qq\cdot \RR'}
 [\sum_{\RR}\hat{\nu_f}^{\xi_1}(\RR)\hat{\nu_f}^{\xi_1}(\RR),\psi_{\RR',i}^{ f,\xi',\dag} \psi_{\RR',i_2}^{f, \xi'_2} ] |\psi_0\rangle \nonumber \\ 
 =& \frac{1}{N_M} \sum_{\RR,i,i_2,\xi',\xi'_2}
u^I_{ii_2,\xi'\xi_2'}(\qq) e^{i\qq\cdot \RR}
 [\hat{\nu_f}^{\xi_1}(\RR)\hat{\nu_f}^{\xi_1}(\RR),\psi_{\RR,i}^{ f,\xi',\dag} \psi_{\RR,i_2}^{ f,\xi'_2} ] |\psi_0\rangle \nonumber \\ 
=&\frac{1}{N_M} \sum_{\RR,i,i_2,\xi',\xi'_2}
u^I_{ii_2,\xi'\xi_2'}(\qq) e^{i\qq\cdot \RR}
\psi_{\RR,i}^{ f,\xi',\dag} \psi_{\RR,i_2}^{ f,\xi'_2}
2(1-\delta_{\xi',\xi_2'})
|\psi_0\rangle 
\label{eq:commutation_rel_4}
 \eaa 
where we use the fact that 
\ba 
\hat{\nu}_f^{\xi}(\RR)|\psi_0\rangle = \nu_f/2 |\psi_0\rangle 
\ea 
with $\nu_f = 0,-2$. 
}

\hh{ 
Combining Eq.~\ref{eq:ham_rkky_bond_v2}, Eq.~\ref{eq:commutation_rel_3} and Eq.~\ref{eq:commutation_rel_4}, we have
we have 
\baa 
[\hH_{RKKY},O_{\qq,ii_2}^I ]|\psi_0\rangle  = \sum_{\kk,\xi_1,\xi_2}\sum_{\kk',\xi_1',\xi_2'}M(\qq)_{\xi_1\xi_2,\xi_1'\xi_2'}u_{ii_2,\xi_1'\xi_2'}(\qq) \psi_{\RR,i}^{f,\xi_1,\dag}\psi_{\RR,i_2}^{f,\xi_2}e^{i\qq\cdot \RR} 
|\psi_0\rangle 
\label{eq:spin_exct_eigen_eq}
\eaa 
with
\baa 
M(\qq)_{\xi_1\xi_2,\xi_1'\xi_2'}=&
\sum_{\xi}\bigg[ 
J(\qq_2=0,\xi_1,\xi,\xi,\xi_1')\delta_{\xi_2',\xi_2} 
+J(\qq_2=0,\xi,\xi_2,\xi_2',\xi)\delta_{\xi_1,\xi_1'} \bigg] \nonumber \\
&-\bigg[J(\qq,\xi_1,\xi_1',\xi_2',\xi_2) 
+J(-\qq,\xi_2',\xi_2,\xi_1,\xi_1' )
\bigg] 
-2J_1^{RKKY}(\RR_2=0)\delta_{\xi_1,-\xi_2}\delta_{\xi_1,\xi_1'}\delta_{\xi_2,\xi_2'} 
\label{eq:Mmat}
\eaa 
\hb{Here we comment that
 $M(\qq)$ matrix takes the same form at $\nu_f=0,-2$. However, the filling $\nu_f$ will affect the choice of $ii_2$ (because not all choice of $ii_2$ in $u_{ii_2,\xi\xi_2}$ produces valid fluctuations) and the values of RKKY interactions $J(\qq,\xi_1,\xi_2,\xi_3,\xi_4)$}
Using Eq.~\ref{eq:spin_exct_eigen_eq}, dispersion of \hb{$f$-moment fluctuation} ($E_{\qq}^I$) and the corresponding wavefunctions ($u^I_{ii_2,\xi_1\xi_2}(\kk,\qq)$) can be derived by solving the following eigen-equations
\baa  
\sum_{\xi_1',\xi_2'}M(\qq)_{\xi_1\xi_2,\xi_1'\xi_2'}u^I_{ii_2,\xi_1'\xi_2'}(\qq) 
=E_{\qq}^I u^I_{ii_2,\xi_1\xi_2}(\qq)
\label{eq:spin_exct_eigen_eq_2}
\eaa  
\bh{where $M(\qq)_{\xi_1\xi_2,\xi_1'\xi_2}$ can be treated as $4\times 4$ matrix with row and column indices $++,--,+-,-+$.}
More explicitly, $M(\qq)$ can be written as
\baa 
M(\qq)_{++,++} = &-2J_0^{RKKY}(\qq_2=0) + J_0^{RKKY}(\qq) +J_0^{RKKY}(-\qq)
\nonumber \\
&
-2J^{RKKY}_1(\qq_2=0) +J_1^{RKKY}(\qq) + J_1^{RKKY}(-\qq) - 2J_2^{RKKY}(\qq=0)
\nonumber 
\\
M(\qq)_{--,--} = &-2J_0^{RKKY}(\qq_2=0) + J_0^{RKKY}(\qq) +J_0^{RKKY}(-\qq)
\nonumber \\
&
-2J^{RKKY}_1(\qq_2=0) +J_1^{RKKY}(\qq) + J_1^{RKKY}(-\qq)- 2J_2^{RKKY}(\qq=0)
\nonumber 
\\
M(\qq)_{+-,+-} =& J_0^{RKKY}(\qq) +J_0^{RKKY}(-\qq)- 2J_0^{RKKY}(\qq_2=0) -2J_1^{RKKY}(\RR=0)- 2J_2^{RKKY}(\qq=0)
\nonumber 
\\
M(\qq)_{-+,-+} = & J_0^{RKKY}(\qq) +J_0^{RKKY}(-\qq)- 2J_0^{RKKY}(\qq_2=0) -2J_1^{RKKY}(\RR=0)- 2J_2^{RKKY}(\qq=0) 
\nonumber \\
M(\qq)_{+-,-+} =&J_{5,+}^{RKKY}(\qq) +J_{5,+}^{RKKY}(-\qq) 
\quad,\quad 
M(\qq)_{-+,+-} =J_{5,-}^{RKKY}(\qq) +J_{5,-}^{RKKY}(-\qq) 
\nonumber 
\\
M(\qq)_{++,--} =& J_2^{RKKY}(\qq)   +  J_2^{RKKY}(-\qq) \quad,\quad 
M(\qq)_{--,++} = J_2^{RKKY}(\qq)   +  J_2^{RKKY}(-\qq) 
\nonumber \\
M(\qq)_{++,+-} =& -J_{3,+}^{RKKY}(\qq) \quad,\quad M(\qq)_{+-,++} = -J_{4,+}^{RKKY}(-\qq) 
\nonumber \\
M(\qq)_{++,-+} =& -J_{4,+}^{RKKY}(\qq) \quad,\quad 
M(\qq)_{-+,++} = -J_{3,+}^{RKKY}(-\qq) 
\nonumber \\
M(\qq)_{--,+-} =& -J_{4,-}^{RKKY}(\qq) \quad,\quad M(\qq)_{+-,--} = -J_{3,-}^{RKKY}(-\qq) \nonumber \\
M(\qq)_{--,-+} =& -J_{3,-}^{RKKY}(\qq) \quad,\quad 
M(\qq)_{-+,--} = -J_{4,-}^{RKKY}(-\qq) 
\eaa 
We define the following functions 
\baa
h_0(\qq) =& -2J_0^{RKKY}(\qq_2=0) + J_0^{RKKY}(\qq) +J_0^{RKKY}(-\qq)
-2J^{RKKY}_1(\qq_2=0) \nonumber \\ 
&-J_1^{RKKY}(\qq) - J_1^{RKKY}(-\qq) - 2J_2^{RKKY}(\qq=0)\nonumber 
\\
h_1(\qq) =& J_0^{RKKY}(\qq) +J_0^{RKKY}(-\qq)- 2J_0^{RKKY}(\qq_2=0) -2J_1^{RKKY}(\RR=0)- 2J_2^{RKKY}(\qq=0) \nonumber \\
h_2(\qq) = & J_2^{RKKY}(\qq)   +  J_2^{RKKY}(-\qq) \nonumber \\
h_3(\qq) = & -J_{3,+}^{RKKY}(\qq)  \nonumber \\ 
h_4(\qq) = &2J_{5,+}(\qq) +2J_{5,+}(-\qq) 
\label{eq:m_h_def}
\eaa  
where $h_0(\qq),h_1(\qq),h_2(\qq)$ are real numbers $h_3(\qq)$ is a complex number 
Then we express
$M^I(\qq)$ in a matrix form 
\baa 
M^I(\qq) = 
\begin{bmatrix}
h_0(\qq) & h_2(\qq) &h_3(\qq) & -h_3^*(\qq) \\
h_2(\qq) & h_0(\qq) & -h_3(\qq) & h_3^*(\qq)  \\
h_3^*(\qq) & -h_3^*(\qq) & h_1(\qq) & h_4(\qq)\\
-h_3(\qq) & h_3(\qq) & h_4^*(\qq) & h_1(\qq)
\end{bmatrix}
\label{eq:m_def}
\eaa
}

\bh{
We now discuss the choice of $ii_2$ of $u_{ii_2,\xi\xi_2}(\qq)$.
For the purpose of discussion, we pick the following ground state given in Eq.~\ref{eq:gnd_nuf_even_v2} (at $\nu_f=0,-2$)
\baa  
&\nu_f=0\quad:\quad|\psi_0\rangle = \prod_{\RR}
\psi_{\RR,1}^{f,+,\dag}\psi_{\RR,1}^{f,-,\dag}
\psi_{\RR,2}^{f,+,\dag}\psi_{\RR,2}^{f,-,\dag} |0\rangle \label{eq:gnd_nuf_nu_0_v2} \\
&\nu_f=-2\quad:\quad|\psi_0\rangle = \prod_{\RR}
\psi_{\RR,1}^{f,+,\dag}\psi_{\RR,1}^{f,-,\dag} |0\rangle 
\label{eq:gnd_nuf_nu_2_v2}
\eaa 
} 

\bh{We first consider the diagonal component.
We first consider the diagonal components $u_{ii,\xi\xi_2}^I(\qq)$. From Eq.~\ref{eq:boson_mode} and Eq.~\ref{eq:gnd_nuf_nu_0_v2}, we find the corresponding "excitation" state is
\baa  
O_{\qq,ii}^I|\psi_0\rangle =\frac{1}{N_M}\sum_{\xi,\xi_2,\RR}
u_{ii,\xi\xi_2}(\RR)\psi_{\RR,i}^{f,\xi,\dag}\psi_{\RR,i}^{f,\xi_2}e^{i\qq\cdot \RR} |\psi_0\rangle 
= \frac{1}{N_M}\sum_{\xi,\RR}
u_{ii,\xi\xi} (\RR) n(\xi,i) |\psi_0\rangle
\label{eq:diag_oq}
\eaa  
where $n(\xi,i)$ is the filling of $\xi,i$ flavors defined as $\psi_{\RR,i}^{\xi,\dag} \psi_{\RR,i}^{\xi}|\psi_0\rangle =n(\xi,i)|\psi_0\rangle $, and we have used the fact that $\psi^{f,+,\dag}_{\RR,i}\psi_{\RR,i}^{f,-}|\psi_0\rangle =0$ (as proved near  Eq.~\ref{eq:gnd_cond}). From Eq.~\ref{eq:diag_oq}, $O_{\qq,ii_2}^I|\psi_0\rangle =A |\psi_0\rangle $, with $A =  \frac{1}{N_M}\sum_{\xi,\RR}
u_{ii,\xi\xi} (\RR) n(\xi,i)$ which is a complex number. Therefore $O_{\qq,ii}^I|\psi_0\rangle$ is the same state (up to a phase and normalization factor) as $|\psi_0\rangle$ when $A\ne 0$. If $A=0$, $O_{\qq,ii}^I|\psi_0\rangle$ vanish. For both $A\ne 0$
 and $A=0$, $O_{\qq,ii}^I$ does not create an excitation state and will not be considered. We next consider the off-diagonal term
\baa  
O_{\qq,ii_2}^I|\psi_0\rangle =\frac{1}{N_M}\sum_{\xi,\xi_2,\RR}
u_{ii,\xi\xi_2}(\RR)\psi_{\RR,i}^{f,\xi,\dag}\psi_{\RR,i_2}^{f,\xi_2}e^{i\qq\cdot \RR} |\psi_0\rangle 
\label{eq:off_diag_o}
\eaa  
A valid mode should have $O_{\qq,ii_2}^I|\psi_0\rangle \ne 0$. Since 
$\psi_{\RR,i}^{\xi,\dag}\psi_{\RR,i_2}^{\xi_2}$ in Eq.~\ref{eq:off_diag_o} describes the procedure of moving one electron from $(\RR,i_2,\xi_2)$ to $(\RR,i_1,\xi_1)$. This procedure can only be valid when there are zero electrons at $(\RR,i,\xi)$ and one electron at $(\RR,i_2,\xi_2)$. 
We next find the valid procedure at $\nu_f=0,-2$ corresponding to the ground states given in Eq.~\ref{eq:gnd_nuf_nu_0_v2}, Eq.~\ref{eq:gnd_nuf_nu_2_v2}. 
From Eq.~\ref{eq:gnd_nuf_nu_0_v2} and Eq.~ Eq.~\ref{eq:gnd_nuf_nu_2_v2}, the valid choices are
\baa 
&\nu_f=0 \quad: \quad (i_1,i_2) \in \{(1,3),(1,4),(2,3),(2,4)\}\nonumber \\ 
&\nu_f=-2 \quad: \quad (i_1,i_2) \in \{(1,2),(1,3),(1,4)\}
\label{eq:ii2}
\eaa 
}

\bh{
Combining Eq.~\ref{eq:spin_exct_eigen_eq_2} and Eq.~\ref{eq:ii2}, the excitation modes are derived by solving the following equations
\baa 
&\sum_{\xi_1',\xi_2'}M(\qq)_{(\xi_1\xi_2),(\xi_1'\xi_2')}u^I_{ii_2,\xi_1'\xi_2'}(\qq) 
=E_{\qq}^I u^I_{ii_2,\xi_1\xi_2}(\qq) \nonumber \\ 
&(i_1,i_2) \in \{(1,3),(1,4),(2,3),(2,4)\},\quad \text{for } \nu_f=0 \nonumber \\ 
&(i_1,i_2) \in \{(1,2),(1,3),(1,4)\},\quad \text{for } \nu_f=-2
\eaa  
By diagonalizing $M(\qq)$ (Eq.~\ref{eq:m_def}), we plot the spin spectrum at $\nu=0,-2$ in Fig.~\ref{fig:spin_fluc}.
}
\begin{hhc}

We next discuss the Goldstone modes.
At $\qq=0$, we have $h_3(\qq=0)=h_4(\qq=0)=0$ (Eq.~\ref{eq:m_h_def}). The eigenvalues of $M(\qq)$ (Eq.~\ref{eq:m_def}) are 
\baa  
&E_{\qq=0}^1 =  -J_1^{RKKY}(\RR=0) -2J_2^{RKKY}(\qq=0) \nonumber 
\\
&
E_{\qq=0}^2 =0  \nonumber \\
&E_{\qq=0}^3= -4 J_2^{RKKY}(\qq=0)
 \nonumber 
\\
&E_{\qq=0}^4= -J_1^{RKKY}(\RR=0) -2J_2^{RKKY}(\qq=0) 
\eaa 
where $E_{\qq}^2 $ branch gives a Goldstone mode. However, we have four choices of $i,i_2$ at $\nu_f=0$ and three choices of $i,i_2$ at $\nu_f=-2$. Then we find 4 Goldstone modes at $\nu_f=0 $ and three Goldstone modes at $\nu_f=-2$. 
We plot the spin fluctuations at $\nu_f=0,\nu_f=-2$ as shown in  Fig.~\ref{fig:spin_fluc}.

\begin{figure}
    \centering
    \includegraphics[width=1.0\textwidth]{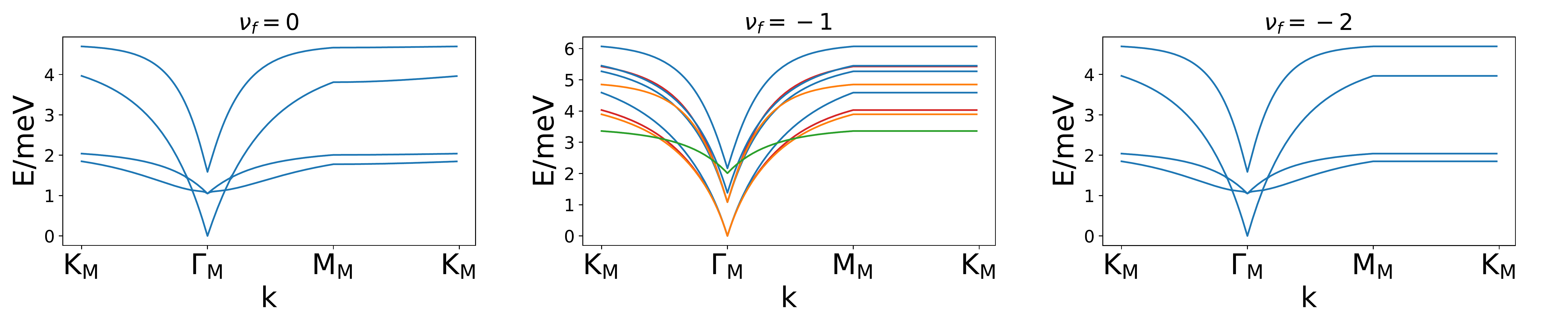}
    \caption{Charge 0 excitations at $\nu_f=0,-2,-3$.}
    \label{fig:spin_fluc}
\end{figure}

\subsection{{Excitation spectrum from RKKY interaction  at $\nu_f=-1$}} 
\bh{
We next calculate the excitation spectrum on top of the ground state at $\nu_f=-1$ (Eq.~\ref{eq:gnd_nuf_1}). We first point out the differences between $\nu_f=0,-2$ and $\nu_f=-1$. 
At $\nu_f=0,-2$, all the valley-spin flavors are filled with either zero or two $f$-electrons. However, at $\nu_f=-1$, there is one valley-spin flavor filled with two $f$-electron, one valley-spin flavor filled with one $f$-electron and two valley-spin flavors filled with zero $f$-electrons (Eq.~\ref{eq:gnd_nuf_1}). Thus the condition in Eq.~\ref{eq:gnd_cond} is not satisfied by the ground state at $\nu_f=-1$. In other words, not all bond operators $A_{\RR,\RR}^{\xi,\xi_2}$ (Eq.~\ref{eq:bond_a_def}) annihilate the ground state at $\nu_f=-1$.
} 

\begin{hhc}
To obtain the excitation spectrum at $\nu_f=-1$, we first \bh{rewrite    $[A_{\RR,\RR_2}^{\xi_1,\xi_2} A_{\RR_2,\RR}^{\xi_3,\xi_4},\psi_{\RR_3,i}^{\xi_1',\dag} \psi_{\RR_3,j}^{\xi_2'}]$ (Eq.~\ref{eq:aa_comute}) with normal order.}  
From Eq.~\ref{eq:aa_comute} and Wick's theorem, we have (consider the case of $(\RR,\xi_1)\ne (\RR_2,\xi_2)$)
\baa  
[A_{\RR,\RR_2}^{\xi_1,\xi_2} A_{\RR_2,\RR}^{\xi_3,\xi_4},\psi_{\RR_3,i}^{f,\xi_1',\dag} \psi_{\RR_3,j}^{f,\xi_2'}] =&
\delta_{\RR_2,\RR_3}\delta_{\xi_2,\xi_1'}:
\psi_{\RR,i}^{f,\xi_1,\dag} \psi_{\RR_3,j}^{f,\xi_2'}A_{\RR_2,\RR}^{\xi_3,\xi_4} :
- \delta_{\RR,\RR_3}\delta_{\xi_1,\xi_2'}:
\psi_{\RR_3,i}^{f,\xi_1',\dag}\psi_{\RR_2,j}^{f,\xi_2}A_{\RR_2,\RR}^{\xi_3,\xi_4} : \nonumber \\
&
+\delta_{\RR,\RR_3}\delta_{\xi_1',\xi_4}:A_{\RR,\RR_2}^{f,\xi_1,\xi_2}\psi_{\RR_2,i}^{f,\xi_3,\dag} \psi_{\RR_3,j}^{f,\xi_2'} :
-\delta_{\RR_2,\RR_3}\delta_{\xi_2',\xi_3}:A_{\RR,\RR_2}^{\xi_1,\xi_2}\psi_{\RR_3,i}^{f,\xi_1',\dag} \psi_{\RR,j}^{f,\xi_4}:
\nonumber 
\\
&+\delta_{\RR_2,\RR_3}\delta_{\xi_2,\xi_1'}
\delta_{\xi_2',\xi_3}(1-n(\xi_2',j)) \psi_{\RR,i}^{f,\xi_1,\dag} \psi_{\RR,j}^{f,\xi_4}
+\delta_{\RR_2,\RR_3}\delta_{\xi_2,\xi_1'}
\delta_{\xi_1,\xi_4}n(\xi_1,i)\psi_{\RR_3,j}^{f,\xi_2'}\psi_{\RR_3,i}^{f,\xi_3,\dag} 
\nonumber 
\\
&
-\delta_{\RR,\RR_3}\delta_{\xi_1,\xi_2'} \delta_{\xi_2,\xi_3}(1-n(\xi_3,j))\psi_{\RR,i}^{f,\xi_1',\dag}\psi_{\RR ,j}^{f,\xi_4} 
-\delta_{\RR,\RR_3}\delta_{\xi_1,\xi_2'}\delta_{\xi_1',\xi_4}n(\xi_1',i)\psi_{f,\RR_2,j}^{\xi_2}\psi_{\RR_2,i}^{f,\xi_3,\dag} 
\nonumber 
\\
& +\delta_{\RR,\RR_3}\delta_{\xi_1',\xi_4}
\delta_{\xi_2,\xi_3}(1-n(\xi_2,i) )\psi_{\RR,i}^{f.\xi_1,\dag}\psi_{\RR_3,j}^{f.\xi_2'}
+\delta_{\RR,\RR_3}\delta_{\xi_1',\xi_4}\delta_{\xi_1,\xi_2'}n(\xi_1,j) \psi_{\RR_2,j}^{f,\xi_2}\psi_{\RR_2,i}^{f,\xi_3,\dag} 
\nonumber 
\\
&-\delta_{\RR_2,\RR_3}\delta_{\xi_3,\xi_2'}\delta_{\xi_2,\xi_1'}(1-n(\xi_2,i) )\psi_{\RR,i}^{f,\xi_1,\dag}\psi_{\RR,j}^{f,\xi_4} 
-\delta_{\RR_2,\RR_3}\delta_{\xi_3,\xi_2'}\delta_{\xi_1,\xi_4}n(\xi_1,j) \psi_{\RR_2,j}^{f,\xi_2}\psi_{\RR_3,i}^{f,\xi_1',\dag} 
\nonumber \\
&+\delta_{i,j}\delta_{\RR,\RR_2}\delta_{\RR,\RR_3}(n(\xi_2',i) -n(\xi_1',i)) 
\bigg[ 
\delta_{\xi_1',\xi_2}\delta_{\xi_1,\xi_2'}
A_{\RR_2,\RR}^{\xi_3,\xi_4}  
+\delta_{\xi_1',\xi_4}\delta_{\xi_3,\xi_2'}A_{\RR,\RR_2}^{\xi_1,\xi_2} \bigg]\nonumber  \\
&-\delta_{i,j}C_0(\RR,\RR_2,\RR_3,\xi_1,\xi_2,\xi_3,\xi_4,\xi_1',\xi_2') 
\label{eq:aa_psipsi_commutator}
\eaa  
where $:O:$ represents the normal ordered form of the operator $O$ with respect to the ground state $|\psi_0\rangle$ in Eq.~\ref{eq:gnd_nuf_1}, and
\baa  
n(\xi,i) = \langle \psi_0|\psi_{\RR,i}^{v\xi,\dag} \psi_{\RR,i}^{f,\xi}|\psi_0\rangle \quad,\quad 
1-n(\xi,i) = \langle \psi_0| \psi_{\RR,i}^{f,\xi}\psi_{\RR,i}^{f,\xi,\dag}|\psi_0\rangle \, . 
\eaa  
The constant $C_0(\RR,\RR_2,\RR_3,\xi_1,\xi_2,\xi_3,\xi_4,\xi_1',\xi_2')$ is defined as
\ba 
&C_0(\RR,\RR_2,\RR_3,\xi_1,\xi_2,\xi_3,\xi_4,\xi_1',\xi_2')\\ =&-\delta_{\RR_2,\RR_3}\delta_{\xi_2,\xi_1'}\delta_{\xi_3,\xi_2'}\delta_{\xi_1,\xi_4} n(\xi_1,i)[-n(\xi_3,i) +n(\xi_2,i)] \\
&
-\delta_{\RR,\RR_3}\delta_{\xi_1,\xi_2'}\delta_{\xi_4,\xi_1'}\delta_{\xi_2,\xi_3}[-n(\xi_4,i) +n(\xi_1,i) + n(\xi_2,i)n(\xi_4,i) -n(\xi_2,i)n(\xi_1,i)] \\
&
-\delta_{\RR,\RR_2}\delta_{\RR,\RR_3}
[n(\xi_2',i)-n(\xi_1',i)]\bigg[\delta_{\xi',\xi_2}\delta_{\xi_1,\xi_2'}(\sum_m \delta_{\xi_3,\xi_4}n(\xi_3,m) +\delta_{\xi_1',\xi_4}\delta_{\xi_2',\xi_3}\delta_{\xi_1,\xi_2}\sum_m n(\xi_1,m) \bigg] 
\ea 

\bh{
We also mention the difference between $\nu_f=-1$ and $\nu_f=0,-2$ again. At $\nu_f=0,-2$, when acting the commutator $[A_{\RR,\RR_2}^{\xi_1,\xi_2} A_{\RR_2,\RR}^{\xi_3,\xi_4},\psi_{\RR_3,i}^{\xi_1',\dag} \psi_{\RR_3,j}^{\xi_2'}]$ (Eq.~\ref{eq:aa_psipsi_commutator}) on the ground state (Eq.~\ref{eq:gnd_nuf_even_v2}), the four-fermion normal ordered term vanishes because of Eq.~\ref{eq:gnd_cond}. However, at $\nu_f=-1$, when acting the commutator $[A_{\RR,\RR_2}^{\xi_1,\xi_2} A_{\RR_2,\RR}^{\xi_3,\xi_4},\psi_{\RR_3,i}^{\xi_1',\dag} \psi_{\RR_3,j}^{\xi_2'}]$ on the ground state (Eq.~\ref{eq:gnd_nuf_1}), the four-fermion normal ordered term will not vanish, because Eq.~\ref{eq:gnd_cond} no longer holds at $\nu_f=-1$.
}

At $\nu_f=-1$, we approximate the commutator by dropping the four-fermion normal-ordering term
\baa 
[A_{\RR,\RR_2}^{\xi_1,\xi_2} A_{\RR_2,\RR}^{\xi_3,\xi_4},\psi_{\RR_3,i}^{f,\xi_1',\dag} \psi_{\RR_3,j}^{f,\xi_2'}] 
\approx &\delta_{\RR_2,\RR_3}\delta_{\xi_2,\xi_1'}
\delta_{\xi_2',\xi_3}(1-n(\xi_2',j)) \psi_{\RR,i}^{f,\xi_1,\dag} \psi_{\RR,j}^{f,\xi_4}
+\delta_{\RR_2,\RR_3}\delta_{\xi_2,\xi_1'}
\delta_{\xi_1,\xi_4}n(\xi_1,i)\psi_{\RR_3,j}^{f,\xi_2'}\psi_{\RR_3,i}^{f,\xi_3,\dag} 
\nonumber 
\\
&
-\delta_{\RR,\RR_3}\delta_{\xi_1,\xi_2'} \delta_{\xi_2,\xi_3}(1-n(\xi_3,j))\psi_{\RR,i}^{f,\xi_1',\dag}\psi_{\RR ,j}^{f,\xi_4} 
-\delta_{\RR,\RR_3}\delta_{\xi_1,\xi_2'}\delta_{\xi_1',\xi_4}n(\xi_1',i)\psi_{\RR_2,j}^{f,\xi_2}\psi_{\RR_2,i}^{f,\xi_3,\dag} 
\nonumber 
\\
& +\delta_{\RR,\RR_3}\delta_{\xi_1',\xi_4}
\delta_{\xi_2,\xi_3}(1-n(\xi_2,i) )\psi_{\RR,i}^{f,\xi_1,\dag}\psi_{\RR_3,j}^{f,\xi_2'}
+\delta_{\RR,\RR_3}\delta_{\xi_1',\xi_4}\delta_{\xi_1,\xi_2'}n(\xi_1,j) \psi_{\RR_2,j}^{f,\xi_2}\psi_{\RR_2,i}^{f,\xi_3,\dag} 
\nonumber 
\\
&-\delta_{\RR_2,\RR_3}\delta_{\xi_3,\xi_2'}\delta_{\xi_2,\xi_1'}(1-n(\xi_2,i) )\psi_{\RR,i}^{f,\xi_1,\dag}\psi_{\RR,j}^{f,\xi_4} 
\nonumber \\
&+\delta_{i,j}\delta_{\RR,\RR_2}\delta_{\RR,\RR_3}(n(\xi_2',i) -n(\xi_1',i)) 
\bigg[ 
\delta_{\xi_1',\xi_2}\delta_{\xi_1,\xi_2'}
A_{\RR_2,\RR}^{\xi_3,\xi_4}  
+\delta_{\xi_1',\xi_4}\delta_{\xi_3,\xi_2'}A_{\RR,\RR_2}^{\xi_1,\xi_2} \bigg] \nonumber \\
& +\delta_{i,j}C_0(\RR,\RR_2,\RR_3,\xi_1,\xi_2,\xi_3,\xi_4,\xi_1',\xi_2') 
\label{eq:wick_commu}
\eaa

\bh{
We now calculate the commutator between $\hH_{RKKY,1}+\hH_{RKKY,0}$ (Eq.~\ref{eq:ham_rkky_bond_v2}) and $O_{q,ij}^I$ (Eq.~\ref{eq:boson_mode}). Using Eq.~\ref{eq:wick_commu}, we find
}
\baa  
&[\hH_{RKKY,1},O_{\qq,ij}^I]|\psi_0\rangle  \nonumber \\
\approx & 
\sum_{\RR,\RR_2}\sum_{\xi_1,\xi_2,\xi_3,\xi_4}J(\RR-\RR_2,\xi_1,\xi_2,\xi_3,\xi_4) 
 \frac{1}{N_M}
 \sum_{\RR_3,\xi',\xi_2'} 
u^I_{ij,\xi_1'\xi_2'}(\qq) 
e^{ i \qq\cdot \RR_3} \nonumber \\
&\bigg[ \delta_{\RR_2,\RR_3}\delta_{\xi_2,\xi_1'}
\delta_{\xi_2',\xi_3}(1-n(\xi_2',j)) \psi_{\RR,i}^{f,\xi_1,\dag} \psi_{\RR,j}^{f,\xi_4}
+\delta_{\RR_2,\RR_3}\delta_{\xi_2,\xi_1'}
\delta_{\xi_1,\xi_4}n(\xi_1,i)\psi_{\RR_3,j}^{f,\xi_2'}\psi_{\RR_3,i}^{f,\xi_3,\dag} 
\nonumber 
\\
&
-\delta_{\RR,\RR_3}\delta_{\xi_1,\xi_2'} \delta_{\xi_2,\xi_3}(1-n(\xi_3,j))\psi_{\RR,i}^{f,\xi_1',\dag}\psi_{\RR ,j}^{f,\xi_4} 
-\delta_{\RR,\RR_3}\delta_{\xi_1,\xi_2'}\delta_{\xi_1',\xi_4}n(\xi_1',i)\psi_{\RR_2,j}^{f,\xi_2}\psi_{\RR_2,i}^{f,\xi_3,\dag} 
\nonumber 
\\
& +\delta_{\RR,\RR_3}\delta_{\xi_1',\xi_4}
\delta_{\xi_2,\xi_3}(1-n(\xi_2,i) )\psi_{\RR,i}^{f,\xi_1,\dag}\psi_{\RR_3,j}^{f,\xi_2'}
+\delta_{\RR,\RR_3}\delta_{\xi_1',\xi_4}\delta_{\xi_1,\xi_2'}n(\xi_1,j) \psi_{\RR_2,j}^{f,\xi_2}\psi_{\RR_2,i}^{f,\xi_3,\dag} 
\nonumber 
\\
&-\delta_{\RR_2,\RR_3}\delta_{\xi_3,\xi_2'}\delta_{\xi_2,\xi_1'}(1-n(\xi_2,i) )\psi_{\RR,i}^{f,\xi_1,\dag}\psi_{\RR,j}^{f,\xi_4} 
-\delta_{\RR_2,\RR_3}\delta_{\xi_3,\xi_2'}\delta_{\xi_1,\xi_4}n(\xi_1,j) \psi_{\RR_2,j}^{f,\xi_2}\psi_{\RR_3,i}^{f,\xi_1',\dag} \nonumber \\
&+\delta_{i,j}\delta_{\RR,\RR_2}\delta_{\RR,\RR_3}(n(\xi_2',i) -n(\xi_1',i))
\bigg(
\delta_{\xi_1',\xi_2}\delta_{\xi_1,\xi_2'}
A_{\RR_2,\RR}^{\xi_3,\xi_4}  
+\delta_{\xi_1',\xi_4}\delta_{\xi_3,\xi_2'}A_{\RR,\RR_2}^{\xi_1,\xi_2} \bigg) \nonumber \\
&-\delta_{i,j}
C_0(\RR,\RR_2,\RR_3,\xi_1,\xi_2,\xi_3,\xi_4,\xi_1',\xi_2')
\bigg] |\psi_0\rangle \nonumber \\
=&
\sum_{\RR,\qq}\sum_{\xi_1,\xi_2,\xi_3,\xi_4}
\frac{e^{i\qq\cdot \RR }}{N_M}
\bigg[
 \nonumber \\
&
J(\qq,\xi_1,\xi_2,\xi_3,\xi_4)(1-n(\xi_3,j) )
u_{ij, \xi_2\xi_3}^I(\qq) 
\psi_{\RR,i}^{f,\xi_1,\dag}\psi_{f,\RR,j}^{\xi_4} 
+\sum_{\xi_2'}J(\qq_2=0,\xi_1,\xi_2,\xi_3,\xi_1)n(\xi_1,i) \psi_{\RR,j}^{\xi_2'}
u_{ij,\xi_2\xi_2'}^I(\qq)
\psi_{\RR,i}^{\xi_3,\dag}\nonumber  \\
&-\sum_{\xi_1'}J(\qq_2=0,\xi_1,\xi_2,\xi_2,\xi_4)(1-n(\xi_2,j))u_{ij,\xi_1'\xi_1}\psi_{\RR,i}^{f,\xi_1',\dag} \psi_{\RR,j}^{f,\xi_4} 
-J(-\qq,\xi_1,\xi_2,\xi_3,\xi_4)n(\xi_4,i)u_{ij,\xi_4\xi_1}^I(\qq) \psi_{\RR,j}^{f,\xi_2}\psi_{\RR,i}^{f,\xi_3,\dag} 
\nonumber \\
& +\sum_{\xi_2'}J(\qq_2=0,\xi_1,\xi_2,\xi_2,\xi_4)(1-n(\xi_2,i))u_{ij,\xi_4\xi_2'}^I(\qq) \psi_{\RR,i}^{f,\xi_1,\dag}\psi_{\RR,j}^{f,\xi_2'} \nonumber  \\
&
+ J(-\qq,\xi_1,\xi_2,\xi_3,\xi_4)n(\xi_1,j)u_{ij,\xi_4\xi_1}^I(\qq)\psi_{\RR,j}^{\xi_2}\psi_{\RR,i}^{\xi_3,\dag} -\sum_{\xi_1'}J(\qq,\xi_1,\xi_2,\xi_3,\xi_4)(1-n(\xi_2,i)) u_{ij,\xi_2\xi_3}^I(\qq) \psi_{\RR,i}^{f,\xi_1,\dag}\psi_{\RR,j}^{f,\xi_4} \nonumber  \\
&
- J(\qq_2=0,\xi_1,\xi_2,\xi_3,\xi_1)n(\xi_1,j)u_{ij,\xi_1' \xi_3}^I(\qq) \psi_{\RR,j}^{f,\xi_2}\psi_{\RR,i}^{f,\xi_1',\dag} 
\bigg] \nonumber \\
&+\sum_{\xi_1,\xi_2,\xi_3,\xi_4,\qq,i}\frac{e^{i\qq\cdot \RR} J(\RR_2=0,\xi_1,\xi_2,\xi_3,\xi_4)}{N_M} 
\delta_{i,j} \bigg[(n(\xi_1,i)-n(\xi_2,i)) u_{ii,\xi_2 \xi_1}^I(\qq) \sum_{m} \psi_{\RR,m}^{f,\xi_3,\dag} \psi_{\RR,m }^{f,\xi_4} \nonumber \\
&+ (n(\xi_4,i) -n(\xi_3,i) )u_{ii,\xi_4\xi_3}^I(\qq)\sum_m \psi_{\RR,m}^{f,\xi_1,\dag} \psi_{\RR,m}^{f,\xi_2}
\bigg] \nonumber \\
&-\sum_{\RR,\RR_2}\sum_{\xi_1,\xi_2,\xi_3,\xi_4}J(\RR-\RR_2,\xi_1,\xi_2,\xi_3,\xi_4)
 \frac{1}{N_M}
 \sum_{\RR_3,\xi',\xi_2',i,j} 
u^I_{ij,\xi_1'\xi_2'}(\qq) 
e^{ i \qq\cdot \RR_3}\delta_{i,j}\nonumber \\
&
C_0(\RR,\RR_2,\RR_3,\xi_1,\xi_2,\xi_3,\xi_4,\xi_1',\xi_2')|\psi_0\rangle 
\label{eq:commute_h1_o}
\eaa  

Using Eq.~\ref{eq:wick_commu}, we find
\baa 
[\hH_{RKKY,0}, O_{\qq,ij}^I]|\psi_0\rangle  =&
-J_1^{RKKY}(\RR_2=0)[\sum_{\xi_1}\sum_{\RR}\hat{\nu_f}^{\xi_1}(\RR)\hat{\nu_f}^{\xi_1}(\RR), O_{\qq,ij}^I]|\psi_0\rangle  \nonumber \\
=&\frac{-J_1^{RKKY}(\RR_2=0)}{N_M} \sum_{\RR',\xi',\xi'_2,\xi_1}
u^I_{ii_2,\xi'\xi_2'}(\qq) e^{i\qq\cdot \RR'}
 [\sum_{\RR}\hat{\nu_f}^{\xi_1}(\RR)\hat{\nu_f}^{\xi_1}(\RR),\psi_{\RR',i}^{f, \xi',\dag} \psi_{\RR',i_2}^{f, \xi'_2} ] |\psi_0\rangle \nonumber \\ 
 =& \frac{-J_1^{RKKY}(\RR_2=0)}{N_M} \sum_{\RR,\xi',\xi'_2,\xi_1}
u^I_{ij,\xi'\xi_2'}(\qq) e^{i\qq\cdot \RR}
 [\hat{\nu_f}^{\xi_1}(\RR)\hat{\nu_f}^{\xi_1}(\RR),\psi_{\RR,i}^{f,\xi',\dag} \psi_{\RR,j}^{ f,\xi'_2} ] |\psi_0\rangle
 \nonumber \\ 
 =& \frac{-J_1^{RKKY}(\RR_2=0)}{N_M} \sum_{\RR,\xi',\xi'_2}
u^I_{ij,\xi'\xi_2'}(\qq) e^{i\qq\cdot \RR}2(\nu_f^{\xi'} - \nu_f^{\xi_2'} +1 )(1-\delta_{\xi',\xi_2'})
 \psi_{\RR,i}^{f, \xi',\dag} \psi_{\RR,j}^{ f,\xi'_2} ] |\psi_0\rangle
 \nonumber \\ 
\label{eq:commute_nn_o}
\eaa

Before calculating the spectrum with the commutator, we first discuss the choice of $ij$ in $O_{\qq,ij}^I$. \bh{  We take the following ground state (from Eq.~\ref{eq:gnd_nuf_1}) as an example}
\baa  
|\psi_0\rangle = \prod_{\RR}\psi_{\RR,1}^{f,+,\dag} \psi_{\RR,2}^{f,+,\dag} \psi_{\RR,1}^{f,-,\dag} |0\rangle  \, .
\label{eq:gnd_spin_nu_1}
\eaa 
\bh{ 
For the same reason given below Eq.~\ref{eq:off_diag_o}, the diagonal components $O_{\qq,ii}^I$ will not contribute to the excitation spectrum. For the off-diagonal term $O_{\qq,ij}^I$, $O_{\qq,ij}^I$ (when acting on the ground state) will move one electron from valley-spin flavor $j$ to valley-spin flavor $i$. Then a valid procedure requires valley-spin flavor $j$ to have at least one filled $\xi$-orbital (remember we have $\xi=\pm 1$ orbitals at each valley-spin flavor) and valley-spin flavor $i$ has one empty $\xi$-orbital. Then only the following choices of $ij$ (in $O_{\qq,ij}$) are valid for the ground state in Eq.~\ref{eq:gnd_spin_nu_1}
\baa  
(i,j) \in \{(3,1), (4,1), (3,2),(4,2), (2,1),(2,2) \} \\
\eaa  
We further classify them into four sectors according to whether flavor $i,j$ are fully-filled, empty or half-filled 
\baa  
&\text{Full-empty sector}: (i,j) \in \{ (3,1),(4,1)\} \nonumber \\ 
&\text{Full-half sector}: (i,j)  = (2,1) \nonumber \\ 
&\text{Half-empty sector}: (i,j) \in \{ (3,2),(4,2)\} \nonumber \\ 
&\text{Half-half sector}: (i,j) =(2,2) 
\label{eq:ij_cond}
\eaa  
}

\bh{
We next give the explicit form of $O_{\qq,ij}^I|\psi_0\rangle$ (from Eq.~\ref{eq:boson_mode})
\baa 
O^I_{\qq,ij}|\psi_0\rangle = &\frac{1}{N_M}\sum_{\RR,\xi'\xi_2'} \psi_{\RR,i}^{f, \xi',\dag} \psi_{\RR,i_2}^{ f,\xi'_2} u^I_{ij,\xi'\xi_2'}(\qq)e^{i\qq\cdot \RR} |\psi_0\rangle 
\label{eq:boson_mode_2}
\eaa  
For full-half sector with $(i,j)=(2,1)$. $\xi=+,i=2$ flavor has been filled with one electron (Eq.~\ref{eq:gnd_spin_nu_1}). The term associated with $u^I_{ij,+\xi_2'}$ annihilate the ground state (Eq.~\ref{eq:boson_mode_2}). In other words,
\baa  
\psi_{\RR,i=2}^{f,+,\dag}\psi_{\RR,j=1}^{f,\xi_2'}|\psi_0\rangle = 0 
\eaa 
Then we can simply set the corresponding $u_{ij,+\xi_2'}^I(\qq)$ to zero
\baa  
u_{ij,+\xi_2'}^I(\qq) = 0 \quad,\quad (i,j) = (2,1)
\label{eq:ii_cond_nu_1_fh}
\eaa  
since $u_{ij,+\xi_2'}^I(\qq)$ does not describe a valid procedure. 
} 

\bh{
For half-empty sector with $(i,j)=(3,2)$ or $(i,j)=(4,2)$. $\xi_2'=-, j=2$ is empty. Then $\psi_{\RR,i}^{+,\dag}\psi_{\RR,j=2}^{\xi_2'=-}|\psi_0\rangle =0$ (Eq.~\ref{eq:boson_mode_2}) and we let 
\baa  
u_{ij,-\xi'}^I(\qq) = 0 \quad,\quad (i,j) \in \{(3,2) , (4,2)\}
\label{eq:ii_cond_nu_1_he}
\eaa  
For half-half sector, with $(i,j)=(2,2)$. $\xi_2'=-,j=2$ flavor is empty and $\xi'=+,i=2$ flavor is filled with one $f$-electron, we then set
\baa  
u_{ii, \xi\xi}^I(\qq)  = 0 \quad,\quad (i,j) = (2,2), \xi \in \{\pm\}
\label{eq:ii_cond_nu_1_fh}
\eaa  
As for full-empty sector, with $(i,j) \in \{(3,1), (4,1)\}$, all components of $u_{ij,\xi\xi'}^I(\qq)$ can be non-zero.
} 

\subsubsection{{Full-empty sector}}
We now consider the full-empty sector with $(i,j)\in\{(3,1),(4,1)\}$.  
Combining Eq.~\ref{eq:commute_h1_o} and Eq.~\ref{eq:commute_nn_o}, we have
\baa 
[\hH_{RKKY},O_{ij,\qq}^I ]|\psi_0\rangle  = \sum_{\RR}\sum_{\xi_1,\xi_2}\sum_{\xi_1',\xi_2'}u_{ij,\xi_1'\xi_2'}(\qq)
M(\qq)_{(\xi_1\xi_2),(\xi_1'\xi_2')} \psi_{\RR,i}^{f,\xi_1,\dag}\psi_{\RR,j}^{f,\xi_2}\frac{e^{i\qq\cdot \RR}}{N_M}
|\psi_0\rangle 
\label{eq:spin_exct_eigen_eq}
\eaa 
where 
\baa  
M(\qq)_{++,++} =& 2[J_0^{RKKY}(\qq) -J_0^{RKKY}(\qq=0) + J_1^{RKKY}(\qq) -J_1^{RKKY}(\qq=0)] 
-2J_2^{RKKY}(\qq_2=0)  \nonumber \\
M(\qq)_{--,--} = &2[J_0^{RKKY}(\qq) -J_0^{RKKY}(\qq_2=0) + J_1^{RKKY}(\qq) -J_1^{RKKY}(\qq_2=0)] 
-2J_2^{RKKY}(\qq_2=0) \nonumber \\
M(\qq)_{++,--} = &-2J_2^{RKKY}(\qq) (-1) \quad,\quad 
M(\qq)_{--,++} = -2J_2^{RKKY}(\qq) (-1) \nonumber \\
M(\qq)_{+-,+-} =& 2[J_0^{RKKY}(\qq)-J_0^{RKKY}(\qq_2=0) +J_1^{RKKY}(\qq)-J_1^{RKKY}(\qq_2=0) -J_2^{RKKY}(\qq_2=0)] \nonumber \\
&-2J_1^{RKKY}(\RR_2=0)
\nonumber \\
M(\qq)_{-+,-+} =& 2[J_0^{RKKY}(\qq)-J_0^{RKKY}(\qq_2=0) +J_1^{RKKY}(\qq)-J_1^{RKKY}(\qq_2=0) -J_2^{RKKY}(\qq_2=0)] 
\nonumber \\ 
M(\qq)_{++,+-} =& J_{3+}^{RKKY}(\qq)
\quad,\quad 
M(\qq)_{+-,++} = J_{4+}^{RKKY}(-\qq) \nonumber 
\\
M(\qq)_{++,-+} =& J_{4+}^{RKKY}(\qq)
\quad,\quad 
M(\qq)_{-+,++} = J_{3+}^{RKKY}(-\qq) \nonumber 
\\
M(\qq)_{--,+-} =& J_{4-}^{RKKY}(\qq)
\quad,\quad 
M(\qq)_{+-,--} = J_{3-}^{RKKY}(-\qq) \nonumber \\
M(\qq)_{--,-+} =& J_{3-}^{RKKY}(\qq)
\quad,\quad  
M(\qq)_{-+,--} = J_{4-}^{RKKY}(-\qq) \nonumber \\
M(\qq)_{+-,-+} = &J_{5,+}(\qq)+J_{5,+}(-\qq)
\quad,\quad 
M(ij,\qq)_{-+,+-} =J_{5,+}^*(\qq) +J_{5,+}^*(\qq) 
\eaa 
There are four modes for each choice of $(i,j)$, and eight modes in total.

At $\qq=0$, we can derive the analytical expressions of four eigenvalues of $M(\qq=0)$:
\baa  
&E^{1}_{\qq=0}=0\quad,\quad E^{2}_{\qq=0}=-4J_2^{RKKY}(\qq_2=0) \nonumber \\ &E^{3}_{\qq=0}=-2J_2^{RKKY}(\qq_2=0)-2J_1^{RKKY}(\RR_2=0)
\quad,\quad E^{4}_{\qq=0}=-2J_2^{RKKY}(\qq_2=0)
 \label{eq:disp_fe}
\eaa 
\bh{ 
where $E^1(\qq)$ corresponds to Goldstone modes.}

\subsubsection{{Full-half sector}}
\bh{We consider the full-half sector with $(i,j)=(2,1)$. Combining Eq.~\ref{eq:commute_h1_o}, Eq.~\ref{eq:commute_nn_o} and also the constraints~\ref{eq:ii_cond_nu_1_fh}, we construct the following eigenequations with eigenvalue $E_{\qq}^I$}
\baa 
&[\hH_{RKKY},O_{21,\qq}^I ]|\psi_0\rangle  = \sum_{\RR}\sum_{\xi_2,\xi_2'}u^I_{21,-\xi_2'}(\qq)
M(\qq)_{\xi_2,\xi_2'} \psi_{\RR,2}^{f,-,\dag}\psi_{\RR,j}^{f,\xi_2}\frac{e^{i\qq\cdot \RR}}{N_M}
|\psi_0\rangle  \nonumber \\ 
=&E_\qq^I\sum_{\RR}u^I_{21,-\xi_2}(\qq)
\psi_{\RR,2}^{f,-,\dag}\psi_{\RR,j}^{f,\xi_2}\frac{e^{i\qq\cdot \RR}}{N_M}
|\psi_0\rangle
\label{eq:spin_exct_eigen_eq_nu1_hf}
\eaa
\bh{where the $2\times 2$ matrix $M(\qq)_{\xi_2,\xi_2'}$ is defined as }
\baa  
M(\qq)_{-,-} 
=&2[J_0^{RKKY}(\qq) -J_0^{RKKY}(\qq_2=0) +J_1^{RKKY}(\qq) -J_1^{RKKY}(\qq_2=0)] \nonumber \\
M(\qq)_{+,+} =&
2[J_0^{RKKY}(\qq) - J_0^{RKKY}(\qq=0) +J_1^{RKKY}(\qq) - J_1^{RKKY}(\qq=0) -J_2^{RKKY}(\qq=0)] 
\nonumber \\ 
M(\qq)_{-,+} =& J_{3,-}^{RKKY}(\qq) \nonumber \\
M(\qq)_{+,-} =& J_{4,-}^{RKKY}(-\qq)
\eaa  
\bh{
Equivalently, Eq.~\ref{eq:spin_exct_eigen_eq_nu1_hf} can be written as 
\baa 
\sum_{\xi_2'}M(\qq)_{\xi_2,\xi_2'}u^I_{21,-\xi_2'}(\qq) =
E_\qq^I u^I_{21,-\xi_2}(\qq)
\eaa
} 
There are two excitation modes with dispersions
\baa 
E^{1,2}_\qq =& 2[J_0^{RKKY}(\qq) -J_0^{RKKY}(\qq_2=0) +J_1^{RKKY}(\qq) -J_1^{RKKY}(\qq_2=0)] -J^{RKKY}_2(\qq=0) \nonumber \\
&
\pm \sqrt{(J^{RKKY}_2(\qq=0))^2  + |J_{3,+}^{RKKY}(\qq)|^2}
\eaa 
At $\qq=0$, two excitation modes are 
\baa 
E_{\qq=0}^{1} =-2J_2^{RKKY}(\qq=0) \quad,\quad E_{\qq=0}^{2}= 0 
\label{eq:disp_fh}
\eaa  
with $E_{\qq=0}^2$ gives the Goldstone modes.

\subsubsection{ {Half-empty sector}}
\bh{ We next discuss the half-empty sector with $(i,j) \in \{(3,2),(4,2)\}$.
Combining Eq.~\ref{eq:commute_h1_o}, Eq.~\ref{eq:commute_nn_o} \bh{and also the constraints~\ref{eq:ii_cond_nu_1_fh}}, we construct following eigenequations (where $i=3,4$) with eigenvalue $E_{\qq}^I$}
\baa 
&[\hH_{RKKY},O_{i2,\qq}^I ]|\psi_0\rangle  = \sum_{\RR}\sum_{\xi_1,\xi_1'}u^I_{i2,\xi_1'+}(\qq)
M(\qq)_{\xi_1,\xi_1'} \psi_{\RR,i}^{f,\xi_1,\dag}\psi_{\RR,2}^{f, +}\frac{e^{i\qq\cdot \RR}}{N_M}
|\psi_0\rangle  \nonumber \\ 
=&E_\qq^I\sum_{\RR}u^I_{i2,\xi_1 +}(\qq)
\psi_{\RR,i}^{f,\xi_2,\dag}\psi_{\RR,2}^{f,+}\frac{e^{i\qq\cdot \RR}}{N_M}
|\psi_0\rangle
\label{eq:spin_exct_eigen_eq_nu2_hf}
\eaa 
\bh{where the $2\times 2$ matrix $M(\qq)_{\xi_1,\xi_1'}$ is defined as}
\baa  
M(\qq)_{+,+} =& 2[J_0^{RKKY}(\qq) -J_0^{RKKY}(\qq_2=0) + J_1^{RKKY}(\qq) -J_1^{RKKY}(\qq_2=0)] \nonumber \\
M(\qq)_{-,-} =& 2[J_0^{RKKY}(\qq)-J^{RKKY}_0(\qq_2=0)+J_1^{RKKY}(\qq)-J_1^{RKKY}(\qq_2=0) -J_2^{RKKY}(\qq_2=0)]
\nonumber \\ 
M(\qq)_{+,-} =& J_{4+}^{RKKY}(\qq)\quad,\quad
M(\qq)_{-,+} =J_{3+}^{RKKY}(-\qq)
\eaa  
\bh{
Equivalently, Eq.~\ref{eq:spin_exct_eigen_eq_nu2_hf} can be written as 
\baa 
\sum_{\xi_1'}M(\qq)_{\xi_1,\xi_1'}u^I_{I2,\xi_1'+}(\qq) =
E_\qq^I u^I_{i2,\xi_1+}(\qq)
\eaa
} 
The two eigenvalues of \bh{$M(\qq)_{\xi_1,\xi_1'}$} are 
\baa 
E^{1}_{\qq} =& 2[J_0^{RKKY}(\qq) -J_0^{RKKY}(\qq_2=0) + J_1^{RKKY}(\qq) -J_1^{RKKY}(\qq_2=0)]-J_2^{RKKY}(\qq_2=0)  \nonumber \\
&+ \sqrt{ J_2^{RKKY}(\qq_2=0)^2 +|J_{3+}^{RKKY}(-\qq)|^2 } \nonumber \\
E^{2}_{\qq} =& 2[J_0^{RKKY}(\qq) -J_0^{RKKY}(\qq_2=0) + J_1^{RKKY}(\qq) -J_1^{RKKY}(\qq_2=0)]-J_2^{RKKY}(\qq_2=0) \nonumber \\
&-\sqrt{ J_2^{RKKY}(\qq_2=0)^2 +|J_{3+}^{RKKY}(-\qq)|^2 }
\eaa 
At $\qq=0$, we have 
\baa 
E_{\qq=0}^{1} = -2J_2^{RKKY}(\qq_2=0) \quad,\quad 
E_{\qq=0}^{2} =0 \label{eq:disp_he}
\eaa 
\bh{where $E_{\qq}^2$ corresponds to the Goldstone mode}

\subsubsection{ {Half-half sector}}
\bh{ We next discuss the half-empty sector with $(i,i) =(2,2)$.
Combining Eq.~\ref{eq:commute_h1_o}, Eq.~\ref{eq:commute_nn_o} \bh{and also the constraints~\ref{eq:ii_cond_nu_1_fh}}, we have}
\ba 
&[\hH_{RKKY},O_{\qq,22}^I]|\psi_0\rangle \nonumber \\
\approx 
&
\sum_{\RR}
\frac{e^{i\qq\cdot \RR }}{N_M}
\bigg[
 -J(\qq_2=0,+,-,-,+) u_{22,-+}^I(\qq)\psi_{\RR,2}^{f,-,\dag}\psi_{f,\RR,2}^{+} 
-J(\qq_2=0,+,-,-,+)u_{22,-+}\psi_{\RR,2}^{f,-,\dag} \psi_{\RR,2}^{f,+} 
\nonumber \\
&
 +J(\qq_2=0,-,-,-,-)u_{22,-+}^I(\qq) \psi_{\RR,2}^{f,-,\dag}\psi_{\RR,2}^{f,+} 
+ J(-\qq,+,+,-,-)u_{22,-+}^I(\qq)(-1)\psi_{\RR,2}^{f,-,\dag}\psi_{\RR,2}^{f,+}
\nonumber \\
&
-J(\qq,-,-,+,+) u_{22,-+}^I(\qq) \psi_{\RR,2}^{f,-,\dag}\psi_{\RR,2}^{f,+} 
- J(\qq_2=0,+,+,+,+)u_{22,- +}^I(\qq)(-1) \psi_{\RR,2}^{f,-,\dag} \psi_{\RR,2}^{f,+}
\nonumber \\ 
&+
J(\RR_2=0,+,-,-,+)u_{22,- +}^I(\qq)  \psi_{\RR,2}^{f,-,\dag} \psi_{\RR,2 }^{f,+} +J(\RR_2=0,-,+,+,-)u_{22,-+}^I(\qq) \psi_{f,\RR,2}^{f,-,\dag} \psi_{f,\RR,2}^{f,+}
\bigg] |\psi_0\rangle \\
=& 2\bigg[ J_2^{RKKY}(\qq_2=0) -J_2^{RKKY}(\RR=0)-2J_1^{RKKY}(\qq_2=0) +J_1^{RKKY}(\RR=0) 
+J_0^{RKKY}(\qq) -J_0^{RKKY}(\qq=0) \bigg]O_{\qq,22}^I|\psi_0\rangle 
\nonumber \\ 
=&E_{\qq}^IO_{\qq,22}^I|\psi_0\rangle 
\ea 
where the dispersion is
\baa 
E_{\qq}^I=2\bigg[ J_2^{RKKY}(\qq_2=0)-J_2^{RKKY}(\RR=0) -J_1^{RKKY}(\qq_2=0) +J_1^{RKKY}(\RR=0) 
+J_0^{RKKY}(\qq) -J_0^{RKKY}(\qq=0) \bigg] 
\eaa 

\subsubsection{Number of Goldstone modes} 
\bh{
We now count the number of Goldstone modes of ground states in Eq.~\ref{eq:gnd_spin_nu_1}. According to Eq.~\ref{eq:disp_fe}, there are one Goldstone modes for each $(i,j) \in \{(3,1),(4,1)\}$ choice in half-empty sector. Then, there are in total two Goldstone modes in the full-empty sector. According to Eq.~\ref{eq:disp_fh}, there are one Goldstone modes for each $(i,j) \in \{(2,1)\}$ choices in the full-half sector. Then, there is one Goldstone mode in the full-half sector. 
According to Eq.~\ref{eq:disp_he}, there are one Goldstone modes for each $(i,j) \in \{(3,2),(4,2)\}$ choices in the half-empty sector. Then, there are in total two Goldstone modes in the half-empty sector. Therefore, we have 5 Goldstone modes in total, which is consistent with Ref.~\cite{tbgv}. 
}

\subsection{Discussion} 
\bh{
First, we mention that our spectrum is calculated from RKKY Hamiltonian (Eq.~\ref{eq:ham_rkky}). Thus the polarization of conduction electrons is ignored and the RKKY interactions are long-range (power-law decay). Consequently, we observe a linear dispersion of the Goldstone modes. Second, we derive the RKKY interaction in the momentum space by performing Fourier transformation integral and introducing a short distance cutoff (as described below Eq.~\ref{eq:rkky_ft} and also in Sec.~\ref{sec:ft}). Therefore, the short-distance information (large momentum) is lost in the spectrum. In the next section (Sec.~\ref{sec:eff_thy}, we will include the polarization of conduction electrons and produce the excitation spectrum that will more accurately describe the fluctuations of $f$-moments.
}

\end{hhc}

\section{Effective theory of $f$-moments} 
\label{sec:eff_thy}
After finding the ground state from RKKY interactions, we now derive the effective theory of $f$-moments \bh{in the nonchiral-flat limit ($v_\star^\prime =0,M\ne 0)$ and at integer filling $\nu=\nu_f=0,-1,-2$}.

We first define the coherent state $|u\rangle$ of $f$-moments as
\baa  
&|u\rangle = \prod_\RR | u(\RR)\rangle_\RR  
\label{eq:u_latt}   \\
&|u(\RR)\rangle_{\RR}  = \hat{R}[u(\RR)]|\psi_0\rangle_{\RR} 
\label{eq:u_state} \,.
\eaa  
$|u(\RR)\rangle_\RR$ is the coherent state of the $f$-moment at site $\RR$ 
where $|\psi_0\rangle_{\RR}$ is a basis state that corresponds to the ground state given in Eq.~\ref{eq:gnd_nuf_even} or Eq.~\ref{eq:gnd_nuf_1}. Without loss of generality, at each filling, we let
\baa  
\nu_f=0 \quad:\quad &  |\psi_0\rangle_{\RR} = \psi_{\RR,1}^{f,+,\dag}
 \psi_{\RR,1}^{f,-,\dag} \psi_{\RR,2}^{f,+,\dag} \psi_{\RR,2}^{f,-,\dag} |0\rangle  \nonumber \\
   \nu_f=-1 \quad:\quad &  |\psi_0\rangle_{\RR} = \psi_{\RR,2}^{f,+,\dag}
\psi_{\RR,1}^{f,+,\dag} \psi_{\RR,1}^{f,-,\dag} |0\rangle 
   \nonumber \\
      \nu_f=-2 \quad:\quad &  |\psi_0\rangle_{\RR} =
\psi_{\RR,1}^{f,+,\dag} \psi_{\RR,1}^{f,-,\dag} |0\rangle 
\label{eq:def_psi_0_r}
\eaa 
The corresponding ground states are then 
\baa  
|\psi_0\rangle = \prod_\RR |\psi_0\rangle_\RR
\label{eq:gnd_order}
\eaa 
$\hat{R}[u(\RR)]$ is a $SU(8)$ rotation defined as 
\baa  
\hat{R}[u(\RR)] = \prod_\RR\exp\bigg( 
-i \sum_{ij,\xi\xi'}u_{i\xi,j\xi'}(\RR)\psi_{\RR,i}^{f,\xi,\dag} \psi_{\RR,j}^{f,\xi'} \bigg) \, . 
\label{eq:ru_def}
\eaa 
(Here we do not include the $U(1)$ charge rotation. Because we fix the filling of $f$ \bh{to be integer}, and the $U(1)$ charge is a good quantum number). If we treat $i\xi$ as row indices and $j\xi'$ as column indices, then $u(\RR)$ can be understood as a $8\times 8$ matric. Since $\hat{R}_\RR[u(\RR)]$ generates a $SU(8)$ rotation of $f$-fermions, $u(\RR)$ is a traceless Hermitian matrix. To observe the nature of the transformation, we act the transformation operator on the $f$ electrons which gives
\baa 
\bigg(\hat{R}^\dag_{\RR}[u(\RR)] \bigg)\psi_{\RR,i}^{f,\xi} \bigg(\hat{R}_{\RR}[u(\RR)]\bigg) = [e^{-iu(\RR)}]_{i\xi,j\xi'} \psi_{\RR,j}^{f,\xi'}
\eaa  
where $\exp(-iu(\RR))$ is a matrix exponential. We let 
\baa 
R_{i\xi,j\xi'}(\RR) = [e^{-iu(\RR)}]_{i\xi,j\xi'}
\label{eq:def_u8_r}
\eaa 
then 
\baa  
\bigg(\hat{R}^\dag_{\RR}[u(\RR)] \bigg)\psi_{\RR,i}^{f,\xi} \bigg(\hat{R}_{\RR}[u(\RR)]\bigg) = R_{i\xi,j\xi'}(\RR) \psi_{\RR,j}^{f,\xi'}
\label{eq:u8_transf_f}
\eaa  

We now derive the path integral of the following Hamiltonian
\baa  
\hH= \hH_{c}' +\hH_{int}  \, .
\label{eq: ham_u_c}
\eaa  
$\hH_{c}'$ contains the fermion bilinear terms and $\hH_{int}$ denotes the interactions between $c$-electrons and $f$-moments. 
$\hH_{c'}$ takes the form of 
\ba 
\hH_{c}' =& \sum_{\kk,i}\bigg[v_\star(k_x+i\xi k_y) \psi_{\kk,i}^{\xi,c',\dag} \psi_{\kk,i}^{\xi,c''} +\text{h.c.} \bigg]
+
(-\mu+W\nu_f+\frac{V_0}{\Omega_0}\nu_c)\sum_{\kk,i} [\psi_{\kk,i}^{\xi,c',\dag}\psi_{\kk,i}^{\xi,c'} +\psi_{\kk,i}^{\xi,c'',\dag}\psi_{\kk,i}^{\xi,c''} ] \nonumber \\
&- \sum_{\kk,\xi, i}
\bigg(\frac{1}{D_{1,\nu_c,\nu_f}} +\frac{1}{D_{2,\nu_c,\nu_f}} \bigg) e^{-\lambda^2|\kk|^2}\bigg[\frac{\gamma^2}{2}\psi_{\kk,i}^{\xi,c',\dag}\psi_{\kk,i}^{\xi,c'} + v_\star^\prime \gamma (k_x-i\xi k_y)
\psi_{\kk,i}^{c',\xi,\dag}
\psi_{\kk,i}^{c',-\xi} \bigg] 
\ea 
where the contributions from $\hH_W$ (Eq.~\ref{eq:hw_def}), $\hH_V$ (Eq.~\ref{eq:hv_def}), \bh{as well as the one-body scattering term from SW transformation $\hH_{cc}$(Eq.~\ref{eq:hcc_def_v2})}, have been included, and $\hH_V$ has been treated by mean-field approximation. In addition, we have dropped the constant terms.  
The interaction term (Eq.~\ref{eq:kondo_model_int}) is
\baa  
\hH_{int} =&
\sum_{\RR,\kk,\kk',\xi,\xi'} 
\sum_{\mu\nu}\frac{e^{-i(\kk'-\kk)\cdot \RR)}F(|\kk)|F(|\kk')}{N_M{D_{\nu_c,\nu_f}}} \bigg[
\gamma^2 :\UF_{\mu\nu}^{(f,\xi\xi')}(\RR) : 
\nonumber \\ 
&+ \gamma v_\star^\prime  : \UF_{\mu\nu}^{(f,\xi-\xi')}(\RR) :(k_x'-i\xi'k_y')
+\gamma v_\star^\prime :\UF_{\mu\nu}^{(f,-\xi\xi')}(\RR):( k_x +i\xi k_y)
\bigg] :\UF_{\mu\nu}^{(c',\xi',\xi)}(\kk,\kk'-\kk):
\nonumber \\
&-J\sum_{\RR,\kk,\kk',\mu\nu,\xi}e^{-i(\kk'-\kk)\cdot \RR} :\UF_{\mu\nu}^{(f,\xi\xi)}(\RR)::\UF_{\mu\nu}^{(c'',\xi\xi)}(\kk,\kk'-\kk):
\label{eq:int_u_c}
\eaa

We now derive the path integral formula of the partition function corresponding to the Hamiltonian $\hH=\hH_c'+\hH_{int}$. The partition function can be written as 
\bh{
\baa  
Z= \lim_{N\rightarrow \infty}\prod_{n=1}^{N} \text{Tr}[e^{-\Delta \tau \hH} ],\quad \text{where }\Delta \tau = \beta /N
\eaa 
}

For each time slice $\tau_n = n \Delta \tau$, we insert the identity operator (with Haar measure~\cite{altland2010condensed})  and have
\baa
Z= &\lim_{N\rightarrow \infty}\int D[u(\RR,\tau_n),c_{\kk,\alpha \eta s}^\dag(\tau_n),c_{\kk,\alpha \eta s}^\dag(\tau_n) ]
\text{Tr}\bigg[  \prod_{n=1}^N
e^{-\Delta \tau H}\frac{ |c(\tau_n)\rangle |u(\tau_n)\rangle \langle u(\tau_n) | \langle c(\tau_n)| }{\bigg|
\langle u(\tau_n) |\langle c(\tau_n) |
\bigg|}
\bigg ] \nonumber \\ 
=&\lim_{N\rightarrow \infty}\int D[u(\RR,\tau_n),c_{\kk,\alpha \eta s}^\dag(\tau_n),c_{\kk,\alpha \eta s}^\dag(\tau_n) ] 
\text{Tr}\bigg[  \prod_{n=1}^N \langle \frac{ \langle u(\tau_n)\langle c(\tau_{n+1} )| 
e^{-\Delta \tau H }|c(\tau_{n+1})\rangle |u(\tau_{n+1})\rangle}{
{\sqrt{\bigg|  \langle u(\tau_n+\Delta \tau) |\langle c(\tau_n+\Delta \tau ) \bigg| 
\bigg|
\langle u(\tau_n) |\langle c(\tau_n) 
\bigg| }} 
}
\bigg ]
\eaa 
\bh{where $|u(\tau_n)\rangle = \prod_\RR |u(\RR,\tau_n)\rangle, |c(\tau_n)\rangle$ are coherent states of $c$ and $u$ fields, respectively, at time slice $\tau_n = n\beta /N$}. For each time slice \bh{and large $N$ (small $\Delta \tau$)}, we have \baa 
&\frac{ \langle u(\tau_n+\Delta \tau) |\langle c(\tau_n+\Delta \tau ) |  
e^{-\Delta \tau \hH} 
| c(\tau_n) \rangle 
| u(\tau_n) \rangle }
{\sqrt{\bigg|  \langle u(\tau_n+\Delta \tau) |\langle c(\tau_n+\Delta \tau ) \bigg| 
\bigg|
\langle u(\tau_n) |\langle c(\tau_n) 
\bigg| }} \nonumber \\
\approx &  
\frac{ \langle c(\tau_n+\Delta \tau ) |  
| c(\tau_n) \rangle  }
{\sqrt{\bigg|   |\langle c(\tau_n+\Delta \tau ) \bigg| 
\bigg|
 |\langle c(\tau_n) 
\bigg| }} 
\frac{ \langle u(\tau_n+\Delta \tau)  |  
| u(\tau_n) \rangle }
{\sqrt{\bigg|  \langle u(\tau_n+\Delta \tau)\bigg| 
\bigg|
\langle u(\tau_n) 
\bigg| }}
-\Delta \tau \frac{ \langle u(\tau_n+\Delta \tau) |\langle c(\tau_n+\Delta \tau ) |  
\hH 
| c(\tau_n) \rangle 
| u(\tau_n) \rangle }
{\sqrt{\bigg|  \langle u(\tau_n+\Delta \tau) |\langle c(\tau_n+\Delta \tau ) \bigg| 
\bigg|
\langle u(\tau_n) |\langle c(\tau_n) 
\bigg| }}\nonumber \\
\approx& \bh{\bigg( 1 -\sum_{\kk,a\eta s}
[c_{\kk,a\eta s}^\dag(\tau_n+\Delta \tau) -c_{\kk,a\eta s}^\dag(\tau_n) ]c_{\kk,a \eta s}(\tau) 
\bigg) 
\bigg\{
1 - 
\bigg[\frac{ \langle u(\tau_n+\Delta \tau)   
| u(\tau_n) \rangle }
{\sqrt{\bigg|  \langle u(\tau_n+\Delta \tau)\bigg| 
\bigg|
\langle u(\tau_n) 
\bigg| }}-1\bigg] \bigg\}
-H(\tau_n) \Delta \tau }\nonumber\\
\approx& 1 -\sum_{\kk,a\eta s}c_{\kk,a\eta s}^\dag(\tau_n) \partial_\tau c_{\kk,a\eta s}(\tau_n) \Delta \tau 
+ 
\bigg[\bh{\frac{\langle u(\tau_n+\Delta \tau)   
| u(\tau_n) \rangle}{ 
\sqrt{\bigg|  \langle u(\tau_n+\Delta \tau)\bigg| 
\bigg|
\langle u(\tau_n) 
\bigg| }
} }-1\bigg] 
-H(\tau_n) \Delta \tau \nonumber\\
\approx &
\exp\bigg[ 
-\sum_{\kk,a\eta s}c_{\kk,a\eta s}^\dag(\tau_n) \partial_\tau c_{\kk,a\eta s}(\tau_n) \Delta \tau 
+ 
\bigg[\bh{
\frac{\langle u(\tau_n+\Delta \tau)   
| u(\tau_n) \rangle}{ 
\sqrt{\bigg|  \langle u(\tau_n+\Delta \tau)\bigg| 
\bigg|
\langle u(\tau_n) 
\bigg| }
} 
}-1\bigg] 
-H(\tau_n) \Delta \tau
\bigg] 
\label{eq:time_slice_evl}
\eaa 
where $H(\tau_n)$ is the original Hamiltonian $\hH$ where each $c_{\kk,a\eta s}$ is replaced by $c_{\kk,a\eta s}(\tau_n)$, and each $\UF_{\mu\nu}^{(f,\xi\xi')} (\RR)
$ is replaced by $
\Sigma_{\mu\nu}^{(f,\xi\xi')} (\RR,\tau_n)=
\langle u(\tau_n) |
:\UF_{\mu\nu}^{(f,\xi\xi')} (\RR): | u(\tau) \rangle 
$. \bh{And we use the fact that $|u(\tau_n)\rangle$ is normalized with $| |u(\tau_n)\rangle |=1$.} 

We then define the action as 
\baa  
S =  \int_0^\beta \sum_{\kk,a\eta s}c_{\kk,a\eta s}^\dag(\tau) \partial_\tau c_{\kk,a\eta s}(\tau) d\tau 
-\lim_{\Delta \tau \rightarrow 0}
\sum_n \bigg[\bh{
\frac{\langle u(\tau_n+\Delta \tau)   
| u(\tau_n) \rangle}{ 
\sqrt{\bigg|  \langle u(\tau_n+\Delta \tau)\bigg| 
\bigg|
\langle u(\tau_n) 
\bigg| }
} 
}-1\bigg]
+\int_0^\beta  H(\tau) d\tau 
\label{eq:action_c_u_v0} \, .
\eaa  

We next evaluate each term in the action. The first term is the dynamical term of $c$ electrons. The second one corresponds to the Berry phase term of $f$-moments. We note that 
\baa  
\langle u | \tilde{u}\rangle \approx& \prod_{\RR}\langle \psi_0| \bigg\{1+ i \sum_{ij,\xi\xi'}u_{i\xi,j\xi'}(\RR)\psi_{\RR,i}^{\xi,\dag} \psi_{\RR,j}^{\xi'}
-\frac{1}{2}\bigg[ 
\sum_{ij,\xi\xi'}u_{i\xi,j\xi'}(\RR)\psi_{\RR,i}^{\xi,\dag} \psi_{\RR,j}^{\xi'}
\bigg] ^2 
\bigg\} \nonumber \\
&
\bigg\{1- i \sum_{i_2j_2,\xi_2\xi'_2}\tilde{u}_{i_2\xi_2,j_2\xi_2'}(\RR)
\psi_{\RR,i_2}^{\xi_2,\dag} \psi_{\RR,j_2}^{\xi'_2}
-\frac{1}{2} \bigg[ 
\sum_{i_2j_2,\xi_2\xi'_2}\tilde{u}_{i_2\xi_2,j_2\xi_2'}(\RR)
\psi_{\RR,i_2}^{\xi_2,\dag} \psi_{\RR,j_2}^{\xi'_2}\bigg]^2
\bigg\} |\psi_0\rangle  \nonumber \\
\approx & \prod_{\RR} \bigg[ 1+\sum_{i\xi} i u_{i\xi,i\xi}n(\xi,i)(\RR)
-i\tilde{u}_{i\xi,i\xi}(\RR) n(\xi,i) \nonumber \\
&
+\sum_{i,\xi,j,\xi',i_2,\xi_2,j_2,\xi_2'}[
\delta_{i,j}\delta_{\xi,\xi'} \delta_{i_2,j_2}\delta_{\xi_2,\xi_2'} n(\xi,i) n(\xi_2,i_2) 
+\delta_{i,j_2}\delta_{\xi,\xi_2'}\delta_{j,i_2}\delta_{\xi',\xi_2}\delta_{j,i_2}n(\xi,i) (1-n(\xi',j) )] \nonumber \\
&[ 
u_{i\xi,j\xi'}(\RR)\tilde{u}_{i_2\xi_2,j_2\xi_2'} (\RR)
-\frac{1}{2}u_{i\xi,j\xi'}(\RR){u}_{i_2\xi_2,j_2\xi_2'} (\RR)
-\frac{1}{2}\tilde{u}_{i\xi,j\xi'}(\RR)\tilde{u}_{i_2\xi_2,j_2\xi_2'} (\RR)
] 
\bigg]
\eaa  
where we use the fact that 
$\langle \psi_0|\psi_{\RR,i}^{\xi,\dag} \psi_{\RR,j}^{\xi'}|\psi_0\rangle \propto \delta_{i,j}\delta_{\xi,\xi'}$, with $|\psi_0\rangle$ the ground state given in Eq.~\ref{eq:gnd_nuf_even} or Eq.~\ref{eq:gnd_nuf_1}. In addition, we let $n(\xi,i) = \langle \psi_0|\psi_{\RR,i}^{\xi,\dag} \psi_{\RR,i}^{\xi}|\psi_0\rangle$. 
We aim to rewrite the above equation in a more compact form. We introduce the following matrix
\baa 
\Lambda_{i\xi,j\xi'} = \frac{1}{2}\langle \psi_0 |: \psi_{\RR,i}^{f,\xi,\dag} \psi_{\RR,j}^{f,\xi'}: |\psi_0\rangle =\delta_{i,j}\delta_{\xi,\xi'}\frac{1}{2} (n(\xi,i)-\frac{1}{2})
\label{eq:def_exp_gnd}
\eaa 
For $|\psi_0\rangle$ defined in Eq.~\ref{eq:def_psi_0_r} \bh{(which are ground states at $v_\star^\prime \ne 0,M= 0$)}, the values of $\Lambda_{i\xi,i\xi}$ are
\baa 
\nu_f=0 \quad:\quad & \Lambda_{1+,1+}=\Lambda_{2+,2+}=\Lambda_{1-,1-}=\Lambda_{2-,2-}=\frac{1}{4} \quad, \quad \Lambda_{3+,3+}=\Lambda_{4+,4+}=\Lambda_{3-,3-}=\Lambda_{4-,4-}=-\frac{1}{4} \nonumber \\
\nu_f=-1 \quad:\quad & \Lambda_{1+,1+}=\Lambda_{2+,2+}=\Lambda_{1-,1-}=\frac{1}{4} \quad, \quad \Lambda_{2-,2-}= \Lambda_{3+,3+}=\Lambda_{4+,4+}=\Lambda_{3-,3-}=\Lambda_{4-,4-}=\frac{1}{4} \nonumber \\
\nu_f=-2 \quad:\quad & \Lambda_{1+,1+}=\Lambda_{1-,1-}=\frac{1}{4} \quad, \quad \Lambda_{2+,2+}=\Lambda_{2-,2-}= \Lambda_{3+,3+}=\Lambda_{4+,4+}=\Lambda_{3-,3-}=\Lambda_{4-,4-}=\frac{1}{4} 
\label{eq:val_lam}
\eaa 
\bh{Here we comment that no matter what types of ferromagnetic order/ground state we considered, we can always make a basis transformation to make $\Lambda$ a diagonal matrix. To observe this, we assume $\Lambda$ is an arbitrary Hermitian matrix. We perform an eigendecomposition $\Lambda = V\tilde{\Lambda}V^\dag$ where $\tilde{\Lambda}$ is a diagonal matrix that characterizes the eigenvalues of $\Lambda$ and $V$ is the matrix formed by eigenvectors of $\Lambda$. Clearly, $V$ can be understood as a $SU(8)$ rotation and we can make a basis change accordingly $\psi^{f} \rightarrow \tilde{\psi}^f$, where $\tilde{\psi}^{f,\xi}_{\RR,i} = \sum_{j,\xi'} \psi^{f,\xi'}_{\RR,j}V_{j\xi',i\xi}$, such that $\frac{1}{2}\langle \psi_0 |:\tilde{\psi}_{\RR,i}^{f,\xi,\dag} \tilde{\psi}_{\RR,j}^{f,\xi'} :|\psi_0\rangle = \tilde{\Lambda}$ which is a diagonal matrix. Then in the new basis, we have a new model with a diagonal $\Lambda$ matrix. However, there are two situations. In the first case, the $SU(8)$ rotation characterized by $V$ is also a symmetry transformation of the flat $U(4)$ group. Since the Hamiltonian is invariant under a flat $U(4)$ group, so the Hamiltonian is invariant under the basis change which is just a flat $U(4)$ transformation (Note that we also need to perform the same $V$-transformation on $c$-electron). Then our effective theory remains the same compared to the effective theory we built for the ground state in Eq.~\ref{eq:def_psi_0_r}. This is also a consequence of the flat $U(4)$ symmetry of the system, namely all the ground states (characterized by $\Lambda$) that are connected by flat $U(4)$ transformation giving rise to the same effective theory. In the second case, the $SU(8)$ rotation characterized by $V$  does not belong to the flat $U(4)$ group. Then we have different types of ground state compared to the ground state in Eq.~\ref{eq:def_psi_0_r}, and we have a different effective theory.
} 
Then we have 
\baa  
\langle u|\tilde{u}\rangle = &\prod_{\RR} 
1+2i\text{Tr}[(u-\tilde{u})\Lambda] -2\bigg[ 
\text{Tr}[(u-\tilde{u})\Lambda ]
\bigg]^2 +2
\text{Tr}\bigg[ [(u-\tilde{u})\Lambda]^2 \bigg] 
-\frac{1}{8}\text{Tr}\bigg[ 
(u-\tilde{u})^2 
\bigg] -\text{Tr}[u\Lambda\tilde{u}] +\text{Tr}[\tilde{u}\Lambda u]
\label{eq:uu_overlap}
\eaa  
Using Eq.~\ref{eq:uu_overlap}, the Berry phase term of $f$-moments becomes
\baa  
\frac{\langle u(\tau_n+\Delta \tau)   
| u(\tau_n) \rangle}{ 
\sqrt{\bigg|  \langle u(\tau_n+\Delta \tau)\bigg| 
\bigg|
\langle u(\tau_n) 
\bigg| }
} \approx  \prod_{\RR} 
\bigg[ 
1+2i\text{Tr}[ \Lambda  \partial_\tau  u(\RR,\tau) ]\Delta \tau -\text{Tr}
[(\Lambda u(\RR,\tau_n)- u(\RR,\tau_n)\Lambda) \partial_\tau u(\RR,\tau_n) ]\Delta \tau 
\bigg] \label{eq:u4_berry} 
\eaa  
Finally, \hh{to the formula of $H(\tau)$(Eq.~\ref{eq:time_slice_evl}), we represent the $U(8)$ moments in the path integral with $u$},
\ba 
\Sigma_{\mu\nu}^{(f,\xi\xi')} (\RR,\tau_n)=
\langle u(\tau_n) |:
\UF_{\mu\nu}^{(f,\xi\xi')} (\RR) :| u(\tau_n) \rangle  
=\frac{1}{2}\sum_{i,j} 
T_{ij}^{\mu\nu}  
\langle \psi_0 | \hat{R}^\dag[u(\RR\bh{,\tau})] :\psi_{\RR,i}^{\xi,\dag} \psi_{\RR,j}^{\xi'} :\hat{R}[u(\RR\bh{,\tau})]  |\psi_0\rangle  
\ea 
Using Eq.~\ref{eq:def_u8_r}, Eq.~\ref{eq:u8_transf_f} and Eq.~\ref{eq:def_exp_gnd}, \bh{we can now express $f$-moments with $u$ fields}
\baa 
\Sigma_{\mu\nu}^{(f,\xi\xi')} (\RR,\tau) = &\frac{1}{2}
\sum_{i,j} 
T_{ij}^{\mu\nu}  R_{j\xi',j_2\xi_2'}(\RR,\tau_n)R^*_{i\xi,i_2\xi_2}(\RR,\tau_n)
\langle \psi_0 |:\psi_{\RR,i_2}^{f,\xi_2,\dag} \psi_{\RR,j_2}^{f,\xi'_2} : |\psi_0\rangle  \nonumber \\ 
=& 
\sum_{i,j,i_2,j_2,\xi_2,\xi_2'} 
T_{ij}^{\mu\nu}  R_{j\xi',j_2\xi_2'}(\RR,\tau)R^*_{i\xi,i_2\xi_2}(\RR,\tau)
\Lambda_{i_2\xi_2,j_2\xi'_2}  \nonumber \\
=&\sum_{ij}T_{ij}^{\mu\nu} [R^\dag(\RR,\tau) \Lambda R(\RR,\tau)]_{i\xi,j\xi'}   = \sum_{ij}T_{ij}^{\mu\nu}[e^{iu(\RR,\tau)} \Lambda e^{-iu(\RR,\tau)}]_{i\xi,j\xi'}
\label{eq:coherent_sate_u_sig}
\eaa 
where we use $R(\RR,\tau) = \exp[-iu(\RR,\tau)]$ (Eq.~\ref{eq:def_u8_r}). We next expand in powers of $u$, which gives
\hb{
\baa  
\Sigma_{\mu\nu}^{(f,\xi\xi')} (\RR,\tau)=&
\sum_{ij}T_{ij}^{\mu\nu} [R^\dag(\RR,\tau) \Lambda R(\RR,\tau)]_{i\xi,j\xi'}   \nonumber \\
\approx &
\sum_{ij}T_{ij}^{\mu\nu} \bigg[ \Lambda + i u(\RR,\tau)\Lambda - i\Lambda u(\RR,\tau) 
-\frac{u(\RR,\tau)u(\RR,\tau)\Lambda +\Lambda u(\RR,\tau)u(\RR,\tau)-2u(\RR,\tau)\Lambda u(\RR,\tau)}{2} \bigg]_{i\xi,j\xi'}
\label{eq:u4_mom_u}
\eaa  
}
We can introduce the following two matrices 
\hhb{ 
\baa 
A(\RR,\tau)_{i\xi,j\xi'} &= [iu(\RR,\tau)\Lambda -i\Lambda u(\RR,\tau)]_{i\xi,j\xi'} 
=i[u(\RR,\tau),\Lambda]_{i\xi,j\xi'} \nonumber \\
B(\RR,\tau)_{i\xi,j\xi'} &= 
\frac{i}{2}\bigg[ u(\RR,\tau) A(\RR,\tau) - A(\RR,\tau)u(\RR,\tau) \bigg]_{i\xi,j\xi'} = \frac{i}{2} [ u(\RR,\tau),A(\RR,\tau)]_{i\xi,j\xi'}
\label{eq:A_mat}\,.
\eaa  
}
We also let 
\baa  
[\delta \Sigma(\RR,\tau)]_{i\xi,j\xi'} = A(\RR,\tau)_{i\xi,j\xi'} +B(\RR,\tau)_{i\xi,j\xi'}
\eaa 
such that 
\baa 
\Sigma^{(f,\xi\xi')}_{\mu\nu}(\RR,\tau) \approx \sum_{ij}T^{\mu\nu}_{ij}\bigg( \Lambda_{i\xi,j\xi'} + [\delta\Sigma(\RR,\tau)]_{i\xi,j\xi'} \bigg)
\eaa 

Using Eq.~\ref{eq:u4_mom_u}, we can rewrite the interaction term (Eq.~\ref{eq:int_u_c}) as 
\baa 
&\hH_{int} \nonumber \\
\approx &
\sum_{\RR,\xi,\xi',\kk,\kk'}
\sum_{\mu\nu} 
\frac{\gamma^2 e^{-i(\kk'-\kk)\cdot\RR}F(|\kk|)F(|\kk'|)}{N_M D_{\nu_c,\nu_f}} [
\sum_{i,j}\Lambda_{i\xi,j\xi'}T^{\mu\nu}_{ij}] :\UF_{\mu\nu}^{(c',\xi'\xi)}(\kk,\kk'-\kk): \nonumber \\ 
&
-J\sum_{\RR,\mu\nu,\xi,\kk,\kk'} \frac{e^{-i(\kk'-\kk)\cdot \RR}}{N_M } [\sum_{ij}
\Lambda_{i\xi,j\xi}T^{\mu\nu}_{ij}]:\UF_{\mu\nu}^{(c'',\xi\xi)}(\kk,\kk'-\kk):  \nonumber \\
& +
\sum_{\RR,\kk,\kk',\xi,\xi'}\frac{e^{-i(\kk'-\kk)\cdot\RR}F(|\kk|)F(|\kk'|)}{N_M }
\bigg[ \gamma v_\star^\prime  
T^{\mu\nu}_{ij} \Lambda_{\xi i,-\xi' j} (k_x'-i\xi'k_y')
+\gamma v_\star^\prime T^{\mu\nu}_{ij} \Lambda_{-\xi i,\xi'j}( k_x +i\xi k_y)
\bigg] :\Sigma_{\mu\nu}^{(c',\xi'\xi)}(\kk,\kk'-\kk): \nonumber \\
&+ \sum_{\RR,\kk,\kk',\xi,\xi',\mu\nu,i,j}\frac{e^{-i(\kk'-\kk)\cdot\RR}F(|\kk|)F(|\kk'|) T^{\mu\nu}_{ij}}{N_M D_{\nu_c,\nu_f}} 
  \bigg[
\gamma^2 
[\delta \Sigma(\RR)]_{i\xi,j\xi'}  
+ \gamma v_\star^\prime  :
[\delta \Sigma(\RR)]_{i\xi,j-\xi'}  
:(k_x'-i\xi'k_y') \nonumber \\ 
&
+\gamma v_\star^\prime
[\delta \Sigma(\RR)]_{i-\xi,j\xi'}  
(k_x+i\xi k_y )
\bigg] \nonumber \\
&:\UF_{\mu\nu}^{(c',\xi',\xi)}(\kk,\kk'-\kk): \nonumber -J\sum_{\RR,\kk,\kk',\mu\nu,\xi,i,j}
\frac{ e^{-i(\kk'-\kk)\cdot \RR} }{N_M }
T^{\mu\nu}_{ij} [\delta \Sigma(\RR)]_{i\xi,j\xi} :\UF_{\mu\nu}^{(c'',\xi\xi)}(\kk,\kk'-\kk):
\label{eq:ham_int_c_u}
\eaa 
where the first two lines describe the polarization of conduction electrons due to the \bh{ferromagnetic} ordering of $f$-moments and the last three lines describe the interactions between conduction electrons and the fluctuations of $f$-moments. \bh{$F(|\kk|)$ (Eq.~\ref{eq:def_damp_fact_F}) is the damping factor.}

Now we are able to write down the full action. Combining Eq.~\ref{eq: ham_u_c}, Eq.~\ref{eq:action_c_u_v0}, Eq.~\ref{eq:u4_berry} and Eq.~\ref{eq:ham_int_c_u}, we have 
\baa 
S = &S_{u} +S_{c,order} +S_{int} \nonumber \\
S_u = &-\int_0^\beta  \sum_{\RR} \bigg[
2i\text{Tr}[\Lambda \partial_\tau u(\RR,\tau)] - \text{Tr}
[\bh{[\Lambda, u(\RR,\tau) ] }\partial_\tau u(\RR,\tau)]
\bigg] 
d\tau \nonumber  \\
S_{c,order} = &\int_0^\beta \bigg[
\sum_{\kk,a\eta s} c_{\kk,a\eta s}^\dag(\tau) \partial_\tau c_{\kk,a\eta s}  +H_{c,order}(\tau)\bigg] d\tau  \nonumber \\
S_{int} =& \int_0^\beta  \sum_{\RR,\kk,\kk',\xi,\xi'}\frac{e^{-i(\kk'-\kk)\cdot \RR}F(|\kk|)F(|\kk'|) }{N_M D_{\nu_c,\nu_f}} 
\sum_{\mu\nu,i,j}T^{\mu\nu}_{ij}  \bigg[
\gamma^2 
A(\RR,\tau)_{i\xi,j\xi'}
+ \gamma v_\star^\prime  
A(\RR,\tau)_{i\xi,j-\xi'}  
(k_x'-i\xi' k_y')\nonumber \\
&
+\gamma v_\star^\prime
A(\RR,\tau)_{i-\xi,j\xi'}  
( k_x+i\xi k_y)
\bigg] :\UF_{\mu\nu}^{(c',\xi',\xi)}(\kk,\kk'-\kk,\tau): \nonumber \\
&-J\sum_{\RR,\kk,\kk',\mu\nu,\xi,i,j}\frac{e^{-i(\kk'-\kk)\cdot \RR}}{N_M } T^{\mu\nu}_{ij} A(\RR,\tau)_{i\xi,j\xi} :\UF_{\mu\nu}^{(c'',\xi\xi)}(\kk,\kk'-\kk,\tau): d\tau \nonumber \\
S_{int,2} =& \int_0^\beta  \sum_{\RR,\kk,\kk',\xi,\xi'}\frac{e^{-i(\kk'-\kk)\cdot \RR} F(|\kk|)F(|\kk'|)}{N_M D_{\nu_c,\nu_f}} 
\sum_{\mu\nu,i,j}T^{\mu\nu}_{ij}  \bigg[
\gamma^2 
B(\RR,\tau)_{i\xi,j\xi'}
+ \gamma v_\star^\prime  
B(\RR,\tau)_{i\xi,j-\xi'}  
(k_x'-i\xi'k_y])\nonumber \\
&
+\gamma v_\star^\prime
B(\RR,\tau)_{i-\xi,j\xi'}  
( k_x+i\xi k_y)
\bigg] :\UF_{\mu\nu}^{(c',\xi',\xi)}(\kk,\kk',\tau): \nonumber \\
&-J\sum_{\RR,\kk,\kk',\mu\nu,\xi,i,j}\frac{e^{-i(\kk'-\kk)\cdot \RR} }{N_M }T^{\mu\nu}_{ij} B(\RR,\tau)_{i\xi,j\xi} :\UF_{\mu\nu}^{(c'',\xi\xi)}(\kk,\kk'-\kk,\tau): d\tau
\eaa  
where $S_u$ describes the Berry phase of the $f$ moments, $S_{c,order}$ contains all the fermion bilinear and is characterized by the Hamiltonian $\hH_{c,order}$. \hb{$\hH_{c,order}$ takes the form of
\baa  
\hH_{c,order} =& \sum_{\kk,i}\bigg[v_\star(k_x+i\xi k_y) \psi_{\kk,i}^{\xi,c',\dag} \psi_{\kk,i}^{\xi,c''} +\text{h.c.} \bigg]
+
(-\mu+W+\frac{V_0}{\Omega_0}\nu_c)\nu_f\sum_{\kk,i} [\psi_{\kk,i}^{\xi,c',\dag}\psi_{\kk,i}^{\xi,c'} +\psi_{\kk,i}^{\xi,c'',\dag}\psi_{\kk,i}^{\xi,c''} ] \nonumber \\
&- \sum_{\kk,\xi, i}
\bigg(\frac{1}{D_{1,\nu_c,\nu_f}} +\frac{1}{D_{2,\nu_c,\nu_f}} \bigg) |F(|\kk|)|^2\bigg[\frac{\gamma^2}{2}\psi_{\kk,i}^{\xi,c',\dag}\psi_{\kk,i}^{\xi,c'} + v_\star^\prime \gamma (k_x-i\xi k_y)
\psi_{\kk,i}^{c',\xi,\dag}
\psi_{\kk,i}^{c',-\xi} \bigg] \nonumber \\
& + \sum_{\kk,\xi,i} \frac{2\gamma^2}{D_{\nu_c,\nu_f}}\Lambda_{i\xi,i\xi}\psi_{\kk,i}^{c',\xi,\dag} \psi_{\kk,i}^{c',\xi} 
- \sum_{\kk, \xi,i} 2J\Lambda_{i\xi,i\xi}\psi_{\kk,i}^{c'',\xi,\dag} \psi_{\kk,i}^{c'',\xi}  
+\sum_{\kk,\xi,i}\frac{2\gamma v_\star^\prime}{D_{\nu_c,\nu_f}} (\Lambda_{i-\xi,i-\xi} \nonumber \\
&+\Lambda_{i\xi,i\xi}) (k_x-ik_y\xi) \psi_{\kk,i}^{\xi,c',\dag} \psi_{\kk,i}^{-\xi,c'}
\label{eq:hcorder} 
\eaa  
where we rewrite the Hamiltonian with $\psi_{\kk,i}^{c'},\psi_{\kk,i}^{c''}$ basis and drop the constant term.}

In Eq.~\ref{eq:hcorder}, The first line contains the contribution from $\hH_c$, $\hH_W$ and $\hH_V$ (mean-field level). The second line comes from $\hH_{cc}$ (Eq.~\ref{eq:hcc_def_v2}). The third line describes the polarization of conduction electrons due to the long-range order of the $f$-moments. In addition, we utilize the fact that $\Lambda_{i\xi,j\xi'}$ is a diagonal matrix. 
\bh{
We also rewrite the $\hH_{c,order}$ in a more compact form 
 \baa  
 &\hH_{c,order} \nonumber \\
 =&\sum_{\kk,i} 
 \begin{bmatrix}
 \psi_{\kk,i}^{\xi=+,c',\dag} & \psi_{\kk,i}^{\xi=+,c'',\dag}
 & \psi_{\kk,i}^{\xi=-,c',\dag} & \psi_{\kk,i}^{\xi=-,c'',\dag} 
 \end{bmatrix}
\begin{bmatrix}
E_{0,\kk}^{+,i} +E_{3,\kk}^{+,i} & v_\star(k_x +i k_y)
& v_\kk^{i} (k_x-i k_y) &0 
\\
v_\star(k_x-i k_y) & E_{0,\kk}^{+,i}-E_{3,\kk}^{+,i} 
& 0 & 0 
\\
 v_\kk^{i} (k_x+i k_y) & 0& E_{0,\kk}^{-,i} +E_{3,\kk}^{-,i} & v_\star(k_x -i k_y)\\
0&0& v_\star(k_x+i k_y) & E_{0,\kk}^{-,i}-E_{3,\kk}^{-,i} 
\end{bmatrix}
  \begin{bmatrix}
 \psi_{\kk,i}^{\xi=+,c'} \\ \psi_{\kk,i}^{\xi=+,c''}
 \\
 \psi_{\kk,i}^{\xi=-,c'} \\ \psi_{\kk,i}^{\xi=-,c''}
 \end{bmatrix}
 \label{eq:formula_hcorder}
 \eaa  
 where 
 \baa  
 E_{0,\kk}^{\xi,i} &= -\mu +W+\frac{V_0}{\Omega_0}\nu_c 
 -
\bigg(\frac{1}{D_{1,\nu_c,\nu_f}} +\frac{1}{D_{2,\nu_c,\nu_f}} \bigg) \frac{e^{-\lambda^2|\kk|^2} \gamma^2}{4} + \frac{\gamma^2\Lambda_{i\xi,i\xi}}{D_{\nu_c,\nu_f}} - J\Lambda_{i\xi,i\xi} \nonumber \\
E_{3,\kk}^{\xi,i} &= 
-\bigg(\frac{1}{D_{1,\nu_c,\nu_f}} +\frac{1}{D_{2,\nu_c,\nu_f}} \bigg) \frac{e^{-\lambda^2|\kk|^2} \gamma^2}{4} + \frac{\gamma^2\Lambda_{i\xi,i\xi}}{D_{\nu_c,\nu_f}} + J\Lambda_{i\xi,i\xi}  \nonumber \\
v_{\kk}^{i} &= - \sum_{\kk,\xi, i}
\bigg(\frac{1}{D_{1,\nu_c,\nu_f}} +\frac{1}{D_{2,\nu_c,\nu_f}} \bigg) e^{-\lambda^2|\kk|^2}v_\star^\prime + \frac{ 2\gamma v_\star^\prime}{D_{\nu_c,\nu_f}} (\Lambda_{i-,i-}+\Lambda_{i+,i+})
\label{eq:formula_e0e3_hc}
 \eaa  
 }

$S_{int}$ and $S_{int,2}$ contain the interaction between $u$ and $c$. We separate $S_{int}$ into two parts $S_K',S_J'$ 
\baa 
S_{int}=& S_K' +S_J'  \nonumber \\
S_K'=&\int_0^\beta  \sum_{\RR,\kk,\kk',\xi,\xi'}\frac{e^{-i(\kk'-\kk)\cdot \RR} F(|\kk|)F(|\kk'|)}{N_M D_{\nu_c,\nu_f}} 
\sum_{\mu\nu,i,j}T^{\mu\nu}_{ij}  \bigg[
\gamma^2 
A(\RR,\tau)_{i\xi,j\xi'}
+ \gamma v_\star^\prime  
A(\RR,\tau)_{i\xi,j-\xi'}  
(k_x' -i\xi' k_y])\nonumber \\
&
+\gamma v_\star^\prime
A(\RR,\tau)_{i-\xi,j\xi'}  
( k_x+i\xi k_y )
\bigg] :\UF_{\mu\nu}^{(c',\xi',\xi)}(\kk,\kk'-\kk,\tau): \nonumber \\
S_J'=&-J\sum_{\RR,\kk,\kk',\mu\nu,\xi,i,j}
\frac{ e^{-i(\kk'-\kk)\cdot \RR}}{N_M }
T^{\mu\nu}_{ij} A(\RR,\tau)_{i\xi,j\xi} :\UF_{\mu\nu}^{(c'',\xi\xi)}(\kk,\kk'-\kk,\tau): d\tau 
\eaa 

We now integrate out conduction electrons to derive the effective action of the $f$-moments described by $u(\RR,\tau)$. The partition function can be written as 
\baa  
Z =& \int D[c_{\kk,a\eta s}^\dag(\tau) ,c_{\kk,a\eta s}(\tau),u(\RR,\tau)]\exp 
\bigg[ 
-S_u -S_{c,order}-S_{int}-S_{int,2}
\bigg] \nonumber  \\  
\propto &  
 \int D[u(\RR,\tau)]
 e^{-S_u} \langle e^{-S_{int}-S_{int,2} }\rangle_0 \nonumber \\  
 \approx & 
  \int D[u(\RR,\tau) ]
 e^{-S_u} \exp\bigg[- \langle S_{int}\rangle_0 -\langle S_{int,2}\rangle_0 +\frac{1}{2}[\langle S_{int}^2\rangle_0 -(\langle S_{int}\rangle_0)^2]\bigg]  \nonumber \\
 \approx & \int D[u(\RR,\tau) ]
  \exp\bigg[-\bh{ \bigg[ S_u+\langle S_{int}\rangle_0 +\langle S_{int,2}\rangle_0 -\frac{1}{2}[\langle S_{int}^2\rangle_0 -(\langle S_{int}\rangle_0)^2]\bigg]\bigg]  }\label{eq:eff_s}
\eaa  
where we integrate out $c$-electrons in the second line and introduce
\baa  
\langle O\rangle_0 = \frac{1}{Z_0}\int D[c_{\kk,a\eta s}^\dag(\tau) ,c_{\kk,a\eta s}(\tau) ]O e^{-S_{c,order}} 
\quad,\quad 
Z_0 = \int D[c_{\kk,a\eta s}^\dag(\tau) ,c_{\kk,a\eta s}(\tau) ] e^{-S_{c,order}} \, .
\eaa  
$S_{int}$\bh{(Eq.~\ref{eq:ham_int_c_u})} is linear in $u(\RR,\tau)$, so we keep its first-order and second-order contributions. $S_{int,2}$\bh{(Eq.~\ref{eq:ham_int_c_u})} contains bilinear term of $u(\RR,\tau)$, so we only keep its first-order contribution. \hhb{Overall, we keep the zeroth order, first order and second-order terms of $u(\RR,\tau)$ in the effective action.}
\bh{From Eq.~\ref{eq:eff_s}. }
We then define the effective action of $f$-moments as
\baa  
S_{eff} =& S_{u} +\langle  S_{int}\rangle_0 \bh{+} \langle S_{int,2}\rangle_0 -\frac{1}{2}\bigg[\langle   S_{int}S_{int}\rangle_{0,con}\bigg]   \label{eq:eff_action_order}
\eaa 
where
\baa  
-\frac{1}{2}\langle T_\tau  S_{int}S_{int}\rangle_{0,con} = 
-\frac{1}{2} \langle T_\tau S_K' S_K' \rangle_{0,con} -\frac{1}{2}\langle T_\tau  S_J' S_J'\rangle_{0,con} -\langle S_J'S_K\rangle_{0,con} 
\label{eq:ss_decomp}
\eaa  

\subsection{Single-particle Green's function of conduction electrons} 
In order to find the effective theory of $f$-moments, it is useful to first consider the following single-particle Green's function of conduction electrons 
\bh{
\baa 
&g^{\xi\xi',i}_{c'c'}(\kk,\tau) = - \langle T_\tau \psi_{\kk,i}^{c',\xi}(\tau)  \psi_{\kk,i}^{c',\xi',\dag}(0)\rangle_0  \quad,\quad g^{\xi\xi',i}_{c''c''}(\RR,\tau) = - \langle T_\tau \psi_{\kk,i}^{c'',\xi}(\tau)  \psi_{\kk,i}^{c'',\xi',\dag}(0)\rangle_0  \nonumber \\
&g^{\xi\xi',i}_{c'c''}(\kk,\tau)  = - \langle T_\tau \psi_{\kk,i}^{c',\xi}(\tau)  \psi_{\kk,i}^{c'',\xi',\dag}(0)\rangle_0 \quad,\quad 
g^{\xi\xi',i}_{c''c'}(\kk,\tau)  = - \langle T_\tau \psi_{\kk,i}^{c'',\xi}(\tau)  \psi_{\kk,i}^{c',\xi',\dag}(0)\rangle_0
\label{eq:single_green_order}
\eaa   
}
where the expectation value is calculated with respect to the Hamiltonian $\hH_{c,order}$ (Eq.~\ref{eq:formula_hcorder}).
Using Wick's theorem we have
\baa 
&\langle T_\tau :\Sigma_{\mu\nu}^{(c',\xi'\xi)}(\kk,\kk'-\kk,\tau)::\Sigma_{\mu_2\nu_2}^{(c',\xi_2\xi_2')}(\kk_2',\kk_2-\kk_2',0):\rangle   =- \sum_{ij}\delta_{\kk,\kk_2}\delta_{\kk,\kk_2'}T^{\mu\nu}_{ij}T^{\mu_2\nu_2}_{ji}g_{c'c'}^{\xi_2'\xi',i}(\kk',-\tau)g_{c'c'}^{\xi\xi_2,j}(\kk,\tau)/4 \nonumber \\
&\langle T_\tau :\Sigma_{\mu\nu}^{(c'',\xi'\xi)}(\kk,\kk'-\kk,\tau)::\Sigma_{\mu_2\nu_2}^{(c'',\xi_2\xi_2')}(\kk_2',\kk_2-\kk_2',0):\rangle   =- \sum_{ij}\delta_{\kk,\kk_2}\delta_{\kk,\kk_2'}T^{\mu\nu}_{ij}T^{\mu_2\nu_2}_{ji}g_{c''c''}^{\xi_2'\xi',i}(\kk',-\tau)g_{c''c''}^{\xi\xi_2,j}(\kk,\tau)/4 \nonumber \\
&\langle T_\tau :\Sigma_{\mu\nu}^{(c',\xi'\xi)}(\kk,\kk'-\kk,\tau)::\Sigma_{\mu_2\nu_2}^{(c'',\xi_2\xi_2')}(\kk_2',\kk_2-\kk_2',0):\rangle   =- \sum_{ij}\delta_{\kk,\kk_2}\delta_{\kk,\kk_2'}T^{\mu\nu}_{ij}T^{\mu_2\nu_2}_{ji}g_{c''c'}^{\xi_2'\xi',i}(\kk',-\tau)g_{c'c''}^{\xi\xi_2,j}(\kk,\tau)/4 
\label{eq:u4u4_order}
\eaa

Now we are in the position to calculate each term in the effective action. 
\subsection{ $   \langle S_{int}\rangle_0$} 
We have 
\baa 
\langle T_\tau S_{int}\rangle  =& \int_0^\beta  \sum_{\RR,\kk,\kk',\xi,\xi'}\frac{e^{-i(\kk'-\kk)\cdot \RR} F(|\kk|)F(|\kk'|)}{N_M D_{\nu_c,\nu_f}} 
\sum_{\mu\nu,i,j}T^{\mu\nu}_{ij}  \bigg[
\gamma^2 
A(\RR,\tau)_{i\xi,j\xi'}
+ \gamma v_\star^\prime  
A(\RR,\tau)_{i\xi,j-\xi'}  
(k_x'-i\xi' k_y')\nonumber \\
&
+\gamma v_\star^\prime
A(\RR,\tau)_{i-\xi,j\xi'}  
( k_x+i\xi k_y)
\bigg] \langle T_\tau :\UF_{\mu\nu}^{(c',\xi',\xi)}(\kk,\kk'-\kk,\tau):\rangle_0 \nonumber \\
&-J\sum_{\RR,\kk,\kk',\mu\nu,\xi,i,j}\frac{ e^{-i(\kk'-\kk)\cdot \RR} }{N_M }T^{\mu\nu}_{ij} A(\RR,\tau)_{i\xi,j\xi} \langle T_\tau :\UF_{\mu\nu}^{(c'',\xi\xi)}(\kk,\kk'-\kk,\tau):\rangle_0 d\tau 
\label{eq:s1_order}
\eaa 
\bh{where $F(|\kk|)$ (Eq.~\ref{eq:def_damp_fact_F}) is the damping factor.}
For the given Hamiltonian of conduction electrons (Eq.~\ref{eq:formula_hcorder}), we have 
\bh{
\baa  
&\langle :\UF_{\mu\nu}^{(c',\xi',\xi)}(\kk,\kk'-\kk,\tau) : \rangle  =\frac{1}{2}\sum_{s,t}  T^{\mu\nu}_{st}\langle :\psi_{\kk,s}^{c',\xi',\dag}(\tau) 
\psi_{\kk',t}^{c',\xi}(\tau) : \rangle_{0}
= \frac{\delta_{\kk,\kk'} }{2}\sum_s T^{\mu\nu}_{ss} \langle :\psi_{\kk,s}^{c',\xi',\dag}(\tau) 
\psi_{\kk,s}^{c',\xi}(\tau): \rangle_{0}  \nonumber  \\
&\langle :\UF_{\mu\nu}^{(c'',\xi',\xi)}(\kk,\kk',\tau) : \rangle  =\frac{1}{2}\sum_{s,t} T^{\mu\nu}_{st}\langle :\psi_{\kk,s}^{c'',\xi',\dag}(\tau)  
\psi_{\kk',s}^{c'',\xi}(\tau) : \rangle_{0} 
= \frac{\delta_{\kk,\kk'} }{2}\sum_{s} T^{\mu\nu}_{ss}\langle :\psi_{\kk,s}^{c'',\xi',\dag}(\tau)  
\psi_{\kk,s}^{c'',\xi}(\tau) :\rangle_{0} 
\eaa 
}
Combining above equation with Eq.~\ref{eq:s1_order}, we find
\baa 
&\langle T_\tau  S_{int}\rangle_{0} \nonumber  \\
=& \int_0^\beta   \sum_{\RR,\kk,\xi,\xi'}\frac{|F(|\kk|)|^2}{D_{\nu_c,\nu_f}N_M} 
\sum_{i}2  \bigg[ 
\gamma^2 
A(\RR,\tau)_{i\xi,i\xi'}
+ \gamma v_\star^\prime  
A(\RR,\tau)_{i\xi,i-\xi'}  
(k_x-i\xi' k_y)
+\gamma v_\star^\prime
A(\RR,\tau)_{i-\xi,i\xi'}  
( k_x+i\xi k_y)
\bigg]  \nonumber \\
&
\langle :\psi_{\kk,i}^{c',\xi',\dag}(\tau) 
\psi_{\kk,i}^{c',\xi}(\tau): \rangle_{0} -J\sum_{\RR,\kk,\xi,i}\frac{2}{N_M} A(\RR,\tau)_{i\xi,i\xi}\langle :\psi_{\kk,i}^{c'',\xi,\dag}(\tau) 
\psi_{\kk,i}^{c'',\xi}(\tau): \rangle_{0} d\tau  
\label{eq:exp_sint_0}
\eaa  
According to the definition of $A(\RR,\tau)_{i\xi,j\xi'}$ (Eq.~\ref{eq:A_mat}), we have 
\baa  
&A(\RR,\tau)_{i\xi,i\xi} =iu(\RR,\tau)(\Lambda_{i\xi,i\xi}-\Lambda_{i\xi,i\xi}) = 0 
\eaa 
Then only $A(\RR,\tau)_{i\xi,i-\xi}(\RR,\tau)$ can give non-zero contribution
\baa  
&\langle T_\tau  S_{int}\rangle_{0} \nonumber  \\
=& \int_0^\beta   \sum_{\RR,\kk,\xi}\frac{|F(|\kk|)|^2}{D_{\nu_c,\nu_f}N_M} 
\sum_{i}2  \bigg[
\gamma^2 
A(\RR,\tau)_{i\xi,i-\xi} 
\langle :\psi_{\kk,i}^{c',-\xi,\dag}(\tau) 
\psi_{\kk,i}^{c',\xi}(\tau) : \rangle_{0}
\nonumber \\
&
+ \bigg( \gamma v_\star^\prime  
A(\RR,\tau)_{i\xi,i-\xi}  
(k_x-i\xi k_y)
+\gamma v_\star^\prime
A(\RR,\tau)_{i-\xi,i\xi}  
( k_x+i\xi k_y)\bigg) 
\langle :\psi_{\kk,i}^{c',\xi,\dag}(\tau) 
\psi_{\kk,i}^{c',\xi}(\tau): \rangle_{0}
\bigg]  
\label{eq:exp_sint_0}
\eaa  
\bh{
In terms of Green's function 
\baa 
&\langle :\psi_{\kk,i}^{c',-\xi,\dag} (\tau)
\psi_{\kk,i}^{c',\xi}(\tau) :\rangle_0 = -g_{c'c',i}^{\xi,-\xi}(\kk,-0^+) \nonumber \\
&\langle :\psi_{\kk,i}^{c',\xi,\dag} (\tau)
\psi_{\kk,i}^{c',\xi}(\tau) :\rangle_0 = -g_{c'c',i}^{\xi,\xi}(\kk,-0^+) -1/2 \nonumber \\
\eaa 
where the $-1/2$ comes from the normal ordering and minus sign comes from the fermion anti-commutation relation. 
By direct solving the $\hH_{c,order}$, we find $g_{c'c'}^{\xi,-\xi,i}(\kk,\tau) = (k_x-i\xi k_y) f_4(|\kk|,\tau,i)$ (see Sec.~\ref{sec:green_order}, Eq.~\ref{eq:green_order_fromula}) where $f_4(|\kk|,\tau,i)$ is a function that only depends on $|\kk|$. Then the following $\kk$ summation vanishes due to the $(k_x \pm i \xi k_y) $ contribution
\baa 
\sum_{\kk} \langle :\psi_{\kk,i}^{c',-\xi,\dag}(\tau) 
\psi_{\kk,i}^{c',\xi}(\tau): \rangle_{0} = 
\sum_{\kk}- (k_x - i\xi k_y) f_4(|\kk|,0^+,i) = 0
\label{eq:Tsint_k_sum_1}
\eaa 
As for $g_{c'c'}^{\xi,\xi,i}(\kk,\tau)$, by direct solving $\hH_{c,order}$, we find it only depends on $|\kk|$ (see Sec.~\ref{sec:green_order}, Eq.~\ref{eq:green_order_fromula}). Then the following $\kk$ summation vanishes due to the $(k_x\pm i \xi k_y)$ term
\baa  
\sum_\kk (k_x \pm i\xi k_y)\langle : \psi_{\kk,i}^{c',\xi,\dag} (\tau)
\psi_{\kk,i}^{c',\xi}(\tau) :\rangle_0 
= \sum_\kk (k_x \pm i\xi k_y)  [-g_{c'c'}^{\xi,\xi}(\kk,-0^+) -1/2]=0
\label{eq:Tsint_k_sum_2}
\eaa  
}
From Eq.~\ref{eq:Tsint_k_sum_1} and Eq.~\ref{eq:Tsint_k_sum_2}, all $\kk$ summation in Eq.~\ref{eq:exp_sint_0} goes to zero and then 
\baa 
\langle T_\tau S_{int} \rangle_0 = 0
\eaa

\subsection{ $\langle S_{int,2}\rangle_0$} 
Expanding $\Sigma_{\mu\nu}^{c'/c'',\xi,\xi')}(\kk,\kk',\tau)$ in $S_{int,2}$, we have 
\baa  
&\langle T_\tau S_{int,2}\rangle_0 \nonumber  \\
= &\int
 \sum_{\RR,\kk,\kk'\xi,\xi'}
\sum_{\mu\nu,i,j}\frac{e^{-i(\kk-\kk')\cdot \RR} 
F(|\kk|)F(|\kk'|)
}{N_M} \sum_{lm}T^{\mu\nu}_{lm}\frac{T_{ij}^{\mu\nu}}{2}\bigg[
\frac{\gamma^2 }{D_{\nu_c,\nu_f}} 
B(\RR,\tau)_{l\xi,m\xi'}
\langle :\psi_{\kk,i}^{c',\xi',\dag}(\tau) \psi_{\kk',j}^{c',\xi}(\tau):\rangle_0 \nonumber \\ 
&-J \delta_{\xi,\xi'}B(\RR,\tau)_{i\xi,j\xi}\langle :\psi_{\kk,i}^{c'',\xi,\dag}(\tau) \psi_{\kk',j}^{c'',\xi}(\tau)\rangle_0  +\gamma v_\star^\prime  
B(\RR,\tau)_{l\xi,m\xi'}  
(k_x'+i\xi'k_y)\langle: \psi_{\kk,i}^{c',-\xi',\dag}(\tau) \psi_{\kk',j}^{c',\xi}(\tau) :\rangle_0  \nonumber \\
&+\gamma v_\star^\prime
B(\RR,\tau)_{l\xi,m\xi'}  
( k_x-i\xi k_y)
\langle: \psi_{\kk,i}^{c',\xi',\dag}(\tau) \psi_{\kk',j}^{c'',-\xi}(\tau):\rangle_0 \bigg] 
d\tau \nonumber \\
=&\int \sum_{\RR,\kk, \xi,\xi',i}B(\RR,\tau)_{i\xi,i\xi'} \frac{|F(|\kk|)|^2}{N_M}\sum_{\kk}\bigg[ \frac{2\gamma^2}{D_{\nu_c,\nu_f}}\langle  :\psi_{\kk,i}^{c',\xi',\dag}(\tau) \psi_{\kk,i}^{c',\xi}(\tau):\rangle_0 
-2J^2 \delta_{\xi,\xi'}\langle  :\psi_{\kk,i}^{c'',\xi,\dag}(\tau) \psi_{\kk,i}^{c'',\xi}(\tau):\rangle_0 \nonumber \\
& + \frac{ 2\gamma v_\star^\prime}{D_{\nu_c,\nu_f}}(k_x+i\xi' k_y) \langle :\psi_{\kk,i}^{c',-\xi',\dag}(\tau) \psi_{\kk,i}^{c',\xi}(\tau):\rangle_0
+ \frac{ 2\gamma v_\star^\prime}{D_{\nu_c,\nu_f}} (k_x-i\xi k_y) \langle :\psi_{\kk,i}^{c',\xi',\dag}(\tau) \psi_{\kk,i}^{c',-\xi}(\tau):\rangle_0
\bigg] d\tau 
\eaa  
\bh{where $F(|\kk|)$ (Eq.~\ref{eq:def_damp_fact_F}) is the damping factor.}
\hb{
Using Eq.~\ref{eq:Tsint_k_sum_1} and Eq.~\ref{eq:Tsint_k_sum_2}, we find 
\baa 
\langle T_\tau S_{int,2}\rangle_0
=&\int \sum_{\RR,\kk, \xi,i}B(\RR,\tau)_{i\xi,i\xi} \frac{|F(|\kk|)|^2}{N_M}\sum_{\kk}\bigg[ \frac{2\gamma^2}{D_{\nu_c,\nu_f}}\langle  :\psi_{\kk,i}^{c',\xi,\dag}(\tau) \psi_{\kk,i}^{c',\xi}(\tau):\rangle_0 
-2J^2 \langle :\psi_{\kk,i}^{c'',\xi,\dag}(\tau) \psi_{\kk,i}^{c'',\xi}(\tau):\rangle_0 \nonumber \\
& + \frac{ 2\gamma v_\star^\prime}{D_{\nu_c,\nu_f}} (k_x+i\xi k_y) \langle :\psi_{\kk,i}^{c',-\xi,\dag}(\tau) \psi_{\kk,i}^{c',\xi}(\tau):\rangle_0
+ \frac{ 2\gamma v_\star^\prime}{D_{\nu_c,\nu_f}} (k_x-i\xi k_y) \langle :\psi_{\kk,i}^{c',\xi,\dag}(\tau) \psi_{\kk,i}^{c',-\xi}(\tau):\rangle_0
\bigg] d\tau 
\eaa 
We then have 
\baa 
&\langle T_\tau S_{int,2}\rangle_0 =\int \sum_{\RR, \xi,i}B(\RR,\tau)_{i\xi,i\xi} N_{i\xi} d\tau 
\label{eq:sint2}
\eaa  
where we have defined $N_{i\xi}$ as 
\baa 
N_{i\xi }=&\frac{1}{N_M}\sum_{\kk}\bigg[ \frac{2|F(|\kk|)|^2\gamma^2}{D_{\nu_c,\nu_f}}\langle  :\psi_{\kk,i}^{c',\xi,\dag}(\tau) \psi_{\kk,i}^{c',\xi}(\tau):\rangle_0 
-2J \langle :\psi_{\kk,i}^{c'',\xi,\dag}(\tau) \psi_{\kk,i}^{c'',\xi}(\tau):\rangle_0 \nonumber \\
& +\frac{ 2\gamma v_\star^\prime |F(|\kk|)|^2}{D_{\nu_c,\nu_f}} (k_x+i\xi k_y) \langle :\psi_{\kk,i}^{c',-\xi,\dag}(\tau) \psi_{\kk,i}^{c',\xi}(\tau):\rangle_0
+ \frac{ 2\gamma v_\star^\prime |F(|\kk|)|^2}{D_{\nu_c,\nu_f}} (k_x-i\xi k_y) \langle :\psi_{\kk,i}^{c',\xi,\dag}(\tau) \psi_{\kk,i}^{c',-\xi}(\tau):\rangle_0
\bigg]
\eaa 
}

\subsection{$-\frac{1}{2}\langle S_K'S_K'\rangle_{0,con}$}
\bh{From Eq.~\ref{eq:eff_s}, we obtain}
\baa  
&-\frac{1}{2}\langle S_K'S_K'\rangle_0 \nonumber  \\
=& 
-\frac{1}{2}\int_0^\beta \int_0^\beta 
\sum_{\RR,\kk,\kk',\xi,\xi'}\sum_{\RR_2,\kk_2,\kk'_2,\xi_2,\xi'_2}\frac{e^{-i(\kk'-\kk)\cdot \RR}F(|\kk|)F(|\kk'|) }{N_MD_{\nu_c,\nu_f}} 
\frac{e^{-i(\kk_2-\kk_2')\cdot \RR_2} F(|\kk|)F(|\kk'|)}{N_MD_{\nu_c,\nu_f}} 
\sum_{\mu\nu,i,j}T^{\mu\nu}_{ij}
\sum_{\mu_2\nu_2,i_2,j_2}T^{\mu_2\nu_2}_{i_2j_2} \nonumber \\
&\bigg[
\gamma^2 
A(\RR,\tau)_{i\xi,j\xi'}
+ \gamma v_\star^\prime  
A(\RR,\tau)_{i\xi,j-\xi'}  
(k_x'-i\xi' k_y')
+\gamma v_\star^\prime
A(\RR,\tau)_{i-\xi,j\xi'}  
( k_x+i\xi k_y)
\bigg] \nonumber \\
&\bigg[
\gamma^2 
A(\RR_2,\tau_2)_{i_2\xi_2,j_2\xi_2'}
+ \gamma v_\star^\prime  
A(\RR_2,\tau_2)_{i_2\xi_2,j_2-\xi'_2}  
(k_{2,x}-i\xi_2'k_{2,y})
+\gamma v_\star^\prime
A(\RR_2,\tau_2)_{i_2-\xi_2,j_2\xi'_2}  
( k_{2,x}'+i\xi_2k_{2,y}')
\bigg] \nonumber \\
&\langle :\UF_{\mu\nu}^{(c',\xi',\xi)}(\kk,\kk'-\kk,\tau):
:\UF_{\mu_2\nu_2}^{(c',\xi'_2,\xi_2)}(\kk_2',\kk_2'-\kk_2,\tau_2):\rangle_{0,con}
d\tau d\tau_2  \nonumber \\
=&- \frac{1}{2}\int_0^\beta \int_0^\beta 
\sum_{\RR,\kk,\kk',\xi,\xi'}\sum_{\RR_2,\xi_2,\xi'_2}\frac{e^{-i(\kk'-\kk)\cdot (\RR-\RR_2) } |F(|\kk|)F(|\kk'|)|^2 }{N_MD_{\nu_c,\nu_f}} 
\frac{1 }{N_MD_{\nu_c,\nu_f}} 
\sum_{\mu\nu,i,j}T^{\mu\nu}_{ij}
\sum_{\mu_2\nu_2,i_2,j_2}T^{\mu_2\nu_2}_{i_2j_2} 
\sum_{s,t}
T^{\mu\nu}_{st} T^{\mu_2\nu_2}_{ts}
\nonumber \\
&\bigg[
\gamma^2 
A(\RR,\tau)_{i\xi,j\xi'}
+ \gamma v_\star^\prime  
A(\RR,\tau)_{i\xi,j-\xi'}  
(k_x'-i\xi' k_y')
+\gamma v_\star^\prime
A(\RR,\tau)_{i-\xi,j\xi'}  
( k_x+i\xi k_y)
\bigg] \nonumber \\
&\bigg[
\gamma^2 
A(\RR_2,\tau_2)_{i_2\xi_2,j_2\xi_2'}
+ \gamma v_\star^\prime  
A(\RR_2,\tau_2)_{i_2\xi_2,j_2-\xi'_2}  
(k_{x}-i\xi_2'k_{y})
+\gamma v_\star^\prime
A(\RR_2,\tau_2)_{i_2-\xi_2,j_2\xi'_2}  
( k_{x}'+i\xi_2k_{y}')
\bigg] \nonumber \\
&(-1)g_{c'c'}^{\xi_2\xi',s}(\kk',\tau_2-\tau)g_{c'c'}^{\xi \xi_2',t}(\kk,\tau-\tau_2 )/4
d\tau d\tau_2  
\eaa  
where we use Eq.~\ref{eq:u4u4_order} in the final step
\bh{and the factor $1/4$ comes from the two $1/2$ factors in the definition of $\Sigma_{\mu\nu}^{(c',\xi'\xi)}$ and  $\Sigma_{\mu_2\nu_2}^{(c',\xi_2',\xi_2)}$ (Eq.~\ref{eq:flat_u4_general})}. 
Using $\sum_{\mu\nu}T^{\mu\nu}_{ij}T^{\mu\nu}_{st} =4 \delta_{i,t}\delta_{i,s}$, we find 
\baa 
&-\frac{1}{2}\langle S_K'S_K'\rangle_0 \nonumber \\
=&2\int_0^\beta \int_0^\beta 
\sum_{\RR,\RR_2,\kk,\kk',\xi,\xi',\xi_2,\xi_2',i,j}\frac{e^{-i(\kk'-\kk)\cdot (\RR-\RR_2) }|F(|\kk|)F(|\kk|)|^2 }{(N_MD_{\nu_c,\nu_f})^2}
A(\RR,\tau)_{i\xi,j\xi'} A_{j\xi_2,i\xi_2'}(\RR_2,\tau_2) 
\nonumber \\
&\bigg\{ 
\gamma^4 g_{c'c'}^{\xi_2\xi',j}(\kk',\tau_2-\tau)g_{c'c'}^{\xi \xi_2',i}(\kk,\tau-\tau_2) \nonumber \\
&+\gamma^3v_\star^\prime \bigg[ 
(k_{x}+i\xi_2'k_{y}) 
 g_{c'c'}^{\xi_2\xi',j}(\kk',\tau_2-\tau)g_{c'c'}^{\xi -\xi_2',i}(\kk,\tau-\tau_2)
 +
 (k_{x}'-i\xi_2k_{y}’) 
 g_{c'c'}^{-\xi_2\xi',j}(\kk',\tau_2-\tau)g_{c'c'}^{\xi \xi_2',i}(\kk,\tau-\tau_2) \nonumber \\
 &
 +
 (k_x'+i\xi' k_y') g_{c'c'}^{\xi_2 -\xi',j}(\kk',\tau_2-\tau)
 g_{c'c'}^{\xi \xi_2',i}(\kk,\tau-\tau_2)
 +
 (k_x-i\xi k_y) g_{c'c'}^{\xi_2 \xi',j}(\kk',\tau_2-\tau)
 g_{c'c'}^{-\xi \xi_2',i}(\kk,\tau-\tau_2) 
\bigg] \nonumber \\
& + \gamma^2 (v_\star^\prime)^2
\bigg[ 
(k_x'+i\xi' k_y')(k_x + i\xi_2' k_y)
g_{c'c'}^{\xi_2 -\xi',j}(\kk',\tau_2-\tau)
 g_{c'c'}^{\xi -\xi_2',i}(\kk,\tau-\tau_2) \nonumber \\
 &
 +
 (k_x-i\xi k_y)(k_x'-i\xi_2 k_y')
 g_{c'c'}^{-\xi_2 \xi',j}(\kk',\tau_2-\tau)
 g_{c'c'}^{-\xi \xi_2',i}(\kk,\tau-\tau_2)  \nonumber \\ 
 &
 +(k_x'+i\xi' k_y) (k_x'-i\xi_2 k_y')g_{c'c'}^{-\xi_2 -\xi',j}(\kk',\tau_2-\tau)
 g_{c'c'}^{\xi \xi_2',i}(\kk,\tau-\tau_2) \nonumber \\
 &+(k_x-i\xi k_y)(k_x+i\xi_2' k_y)g_{c'c'}^{\xi_2 \xi',j}(\kk',\tau_2-\tau)
 g_{c'c'}^{-\xi -\xi_2',i}(\kk,\tau-\tau_2) 
\bigg] 
\bigg\} d\tau d\tau_2 
\label{eq:skdotskdot}
\eaa 
\bh{
Here, we comment that, all terms that are at the second order of $S_K'$ ($\gamma^4,\gamma^2(v_\star^\prime)^2,\gamma^3v_\star^\prime$ terms) need to be kept in order to recover the Goldstone mode in the ordered ground state. This is because all terms we kept in the $\hH_K$, ($\gamma^2,\gamma v_\star^\prime$ terms) introduce a polarization effect to the conduction electrons. This can be observed from Eq.~\ref{eq:formula_hcorder} and Eq.~\ref{eq:formula_e0e3_hc} where both $\gamma^2$ and $\gamma v_\star^\prime$ terms affect the single-particle Hamiltonian of conduction electrons. Therefore, when we derive the effective action, 
all the second-order terms induced by $\gamma^2,\gamma v_\star^\prime$ terms in $\hH_K$, that are $\gamma^4,\gamma^2(v_\star^\prime)^2,\gamma^3v_\star^\prime$ terms, need to be kept, in order to be consistent with the single-particle Hamiltonian of conduction electrons. 
}

\subsection{$-\frac{1}{2}\langle S_J'S_J'\rangle_{0,con}$}
Using Eq.~\ref{eq:u4u4_order}, we find
\baa  
&-\frac{1}{2}\langle S_J'S_J'\rangle_{0,con} \nonumber \\
=&-\int_0^\beta \int_0^\beta \frac{J^2}{2N_M^2} \sum_{\RR,\RR_2,\kk,\kk',\kk_2,\kk_2'}\sum_{\mu\nu,\mu_2\nu_2}\sum_{\xi,\xi_2}\sum_{ij,i_2j_2}\frac{e^{-i(\kk'-\kk)\cdot\RR-i(\kk_2-\kk_2')\cdot\RR_2}}{N_M^2} T^{\mu\nu}_{ij}T^{\mu_2\nu_2}_{i_2j_2}
A(\RR,\tau)_{i\xi,j\xi}A(\RR_2,\tau_2)_{i_2\xi_2,j_2\xi_2} \nonumber \\
&\langle :\UF_{\mu\nu}^{(c'',\xi\xi)}(\kk,\kk'-\kk,\tau)::\UF_{\mu_2\nu_2}^{(c'',\xi_2\xi_2)}(\kk_2',\kk_2-\kk_2',\tau_2):\rangle_{0,con} d\tau d\tau_2\nonumber \\
=&-\int_0^\beta \int_0^\beta \frac{J^2}{2} \sum_{\RR,\RR_2,\kk,\kk'}\sum_{\mu\nu,\mu_2\nu_2}\sum_{\xi,\xi_2}\sum_{ij,i_2j_2}e^{-i(\kk'-\kk)\cdot(\RR-\RR_2)}T^{\mu\nu}_{ij}T^{\mu_2\nu_2}_{i_2j_2} 
A(\RR,\tau)_{i\xi,j\xi}A(\RR_2,\tau_2)_{i_2\xi_2,j_2\xi_2}\nonumber \\
&(-1)
\sum_{i'j'}T^{\mu\nu}_{i'j'}T^{\mu_2\nu_2}_{j'i'}
g_{c''c''}^{\xi_2\xi,i'}(\kk',-\tau+\tau_2)g_{c''c''}^{\xi\xi_2,j'}(\kk,\tau-\tau_2)/4  d\tau d\tau_2
\eaa  
Using $\sum_{\mu\nu}T^{\mu\nu}_{ij}T^{\mu\nu}_{st} =4 \delta_{i,t}\delta_{i,s}$, we find
\baa  
&-\frac{1}{2}\langle S_J'S_J'\rangle_{0,con}
=\int_0^\beta \int_0^\beta \frac{2J^2e^{-i(\kk'-\kk)\cdot(\RR-\RR_2)}}{N_M^2}\sum_{\RR,\RR_2}\sum_{\xi,\xi_2,i,j} 
A(\RR,\tau)_{i\xi,j\xi}A(\RR_2,\tau_2)_{j\xi_2,i\xi_2} g_{c''c''}^{\xi_2\xi,j}(\kk',-\tau+\tau_2)g_{c''c''}^{\xi\xi_2,i}(\kk,\tau-\tau_2) d\tau d\tau_2 
\label{eq:sjdotsjdot}
\eaa  
\subsection{$-\langle S_J'S_K'\rangle_{0,con}$ }
Using Eq.~\ref{eq:u4u4_order}, we find 
\baa 
&-\langle S_J'S_K'\rangle_{0,con} \nonumber \\
=&\int_0^\beta \int_0^\beta J\sum_{\RR,\kk,\kk',\kk_2,\kk_2',\xi,\xi'}\frac{e^{-i(\kk'-\kk)\cdot\RR-i(\kk_2-\kk_2')\cdot\RR_2}  F(|\kk|)F(|\kk'|)}{N_M^2D_{\nu_c,\nu_f}}
\sum_{\mu\nu,i,j}T^{\mu\nu}_{ij} \sum_{\RR_2,\mu_2\nu_2,i,j,\xi_2} T_{i_2j_2}^{\mu_2\nu_2}
\sum_{i'j'}T_{i'j'}^{\mu\nu} T_{j'i'}^{\mu_2\nu_2}\nonumber \\
&\bigg[
\gamma^2 
A(\RR,\tau)_{i\xi,j\xi'} 
+ \gamma v_\star^\prime  
A(\RR,\tau)_{i\xi,j-\xi'}  
(k_x'-i\xi k_y')
+\gamma v_\star^\prime
A(\RR,\tau)_{i-\xi,j\xi'}  
( k_x +i\xi k_y)
\bigg]A(\RR_2,\tau_2)_{i_2\xi_2,j_2\xi_2} \nonumber \\
&
\langle :\UF_{\mu\nu}^{(c',\xi',\xi)}(\kk,\kk'-\kk,\tau): :\UF_{\mu_2\nu_2}^{(c'',\xi_2,\xi_2)}(\kk_2',\kk_2-\kk_2',\tau_2): \rangle_{0,con} d\tau d\tau_2  \nonumber \\
=&\int_0^\beta \int_0^\beta J\sum_{\RR,\kk,\kk',\xi,\xi'}\frac{
e^{-i(\kk'-\kk)\cdot(\RR-\RR_2)}F(|\kk|)F(|\kk'|)
}{N_M^2 D_{\nu_c,\nu_f}} 
\sum_{\mu\nu,i,j}T^{\mu\nu}_{ij}  \sum_{\RR_2,\mu_2\nu_2,i,j,\xi_2} T_{i_2j_2}^{\mu_2\nu_2}
\sum_{i'j'}T_{i'j'}^{\mu\nu} T_{j'i'}^{\mu_2\nu_2}
\nonumber \\
&\bigg[
\gamma^2 
A(\RR,\tau)_{i\xi,j\xi'}
+ \gamma v_\star^\prime  
A(\RR,\tau)_{i\xi,j-\xi'}  
(k_x'-i\xi'k_y')
+\gamma v_\star^\prime
A(\RR,\tau)_{i-\xi,j\xi'}  
( k_x+i\xi k_y) 
\bigg] \nonumber \\
&A(\RR_2,\tau_2)_{i_2\xi_2,j_2\xi_2} 
(-1)g_{c''c'}^{\xi_2\xi',i'}(\kk',-\tau+\tau_2) g_{c'c''}^{\xi\xi_2,j'}(\kk,\tau-\tau_2)/4
d\tau d\tau_2
\eaa 
Using $\sum_{\mu\nu}T^{\mu\nu}_{ij}T^{\mu\nu}_{st} =4 \delta_{i,t}\delta_{i,s}$, we find 
\baa  
&-\langle S_J'S_K'\rangle_{0,con} \nonumber \\
=&-4\int_0^\beta \int_0^\beta J\sum_{\RR,\RR_2,\kk,\kk',\xi,\xi',\xi_2}\frac{ 
e^{-i(\kk'-\kk)\cdot(\RR-\RR_2)}F(|\kk|)F(|\kk|)
}{N_M^2 D_{\nu_c,\nu_f}} 
A(\RR,\tau)_{i\xi,j\xi'} A(\RR_2,\tau_2)_{j\xi_2,i\xi_2}
\bigg[ \nonumber \\
&
\gamma^2 g_{c''c'}^{\xi_2\xi',j}(\kk',-\tau+\tau_2)g^{\xi\xi_2,i}_{c'c''}(\kk,\tau-\tau_2) 
+\gamma v_\star^\prime (k_x'+i\xi' k_y') 
g_{c''c'}^{\xi_2-\xi',j}(\kk',-\tau+\tau_2) g_{c'c''}^{\xi\xi_2,i}(\kk,\tau-\tau_2)\nonumber \\ 
&
+ \gamma v_\star^\prime (k_x -i\xi k_y) 
g_{c''c'}^{\xi_2\xi',j}(\kk',-\tau+\tau_2) g_{c'c''}^{-\xi\xi_2,i}(\kk,\tau-\tau_2)
\bigg] d\tau d\tau_2 
\label{eq:sjdotskdot}
\eaa

\end{hhc} 

\begin{hbb}
\subsection{Effective action}
Combining Eq.~\ref{eq:eff_action_order}, Eq.~\ref{eq:sint2}, Eq.~\ref{eq:skdotskdot}, Eq.~\ref{eq:sjdotsjdot} and Eq.~\ref{eq:sjdotskdot}, we find the following effective action of the $f$-moments. 
\baa 
S_{eff} =& S_u 
+\int \sum_{\xi,i,\RR}B(\RR,\tau)_{i\xi,i\xi}N_{i\xi} d\tau \nonumber  \\
&+ \frac{1}{2\pi }\int \int \sum_{\RR,\RR_2,\qq, \xi,\xi'} A(\RR,\tau)_{i\xi,j\xi'}A(\RR_2,\tau_2)_{j \xi_2,i\xi_2'} \frac{ e^{i\qq \cdot (\RR_2-\RR)-i\omega(\tau_2-\tau) }}{N_M} \chi_{\xi\xi',\xi_2'\xi_2}^A(\qq,i\omega,i,j)  d\tau  d\tau_2 
\label{eq:seff_2nd}
\eaa 
where we consider zero temperature and define 
\baa 
&\chi^A_{\xi\xi',\xi_2' \xi_2}(\qq,i\omega,i,j)\nonumber \\
=&  \int\sum_{\kk,\kk'} \frac{\delta_{\kk'-\kk,\qq} e^{i\omega \tau}  }{N_M}
\bigg\{
\frac{2\gamma^4[F(|\kk|)F(|\kk'|)]^2}{D_{\nu_c,\nu_f}^2}  g_{c'c'}^{\xi_2\xi',j}(\kk',\tau)g_{c'c'}^{\xi \xi_2',i}(\kk,-\tau) \nonumber \\
&+\frac{2\gamma^3v_\star^\prime  [F(|\kk|)F(|\kk'|)]^2 }{D_{\nu_c,\nu_f}^2} \bigg[ 
(k_{x}+i\xi_2'k_{y}) 
 g_{c'c'}^{\xi_2\xi',j}(\kk',\tau)g_{c'c'}^{\xi -\xi_2',i}(\kk,-\tau)
 +
 (k_{x}'-i\xi_2k_{y}’) 
 g_{c'c'}^{-\xi_2\xi',j}(\kk',\tau)g_{c'c'}^{\xi \xi_2',i}(\kk,-\tau) \nonumber \\
 &
 +
 (k_x'+i\xi' k_y') g_{c'c'}^{\xi_2 -\xi',j}(\kk',\tau)
 g_{c'c'}^{\xi \xi_2',i}(\kk,-\tau)
 +
 (k_x-i\xi k_y) g_{c'c'}^{\xi_2 \xi',j}(\kk',\tau)
 g_{c'c'}^{-\xi \xi_2',i}(\kk,-\tau) 
\bigg] \nonumber \\
& +\frac{2\gamma^2(v_\star^\prime)^2[F(|\kk|)F(|\kk'|)]^2 }{D_{\nu_c,\nu_f}^2}
\bigg[ 
(k_x'+i\xi' k_y')(k_x + i\xi_2' k_y)
g_{c'c'}^{\xi_2 -\xi',j}(\kk',\tau)
 g_{c'c'}^{\xi -\xi_2',i}(\kk,-\tau) \nonumber \\
 &
 +
 (k_x-i\xi k_y)(k_x'-i\xi_2 k_y')
 g_{c'c'}^{-\xi_2 \xi',j}(\kk',\tau)
 g_{c'c'}^{-\xi \xi_2',i}(\kk,-\tau) 
 +(k_x'+i\xi' k_y) (k_x'-i\xi_2 k_y')g_{c'c'}^{-\xi_2 -\xi',j}(\kk',\tau)
 g_{c'c'}^{\xi \xi_2',i}(\kk,-\tau) \nonumber \\
 &+(k_x-i\xi k_y)(k_x+i\xi_2' k_y)g_{c'c'}^{\xi_2 \xi',j}(\kk',\tau)
 g_{c'c'}^{-\xi -\xi_2',i}(\kk,-\tau) 
\bigg] 
+2J^2 \delta_{\xi,\xi'}\delta_{\xi_2',\xi_2} 
g_{c''c''}^{\xi_2\xi,j}(\kk',\tau)g_{c''c''}^{\xi\xi_2,i}(\kk,-\tau) \nonumber \\
&-\frac{ 4\delta_{\xi_2,\xi_2'} J F(|\kk|)F(|\kk'|)}{D_{\nu_c,\nu_f} }
\bigg[ 
\gamma^2 g_{c''c'}^{\xi_2\xi',j}(\kk',\tau)g^{\xi\xi_2',i}_{c'c''}(\kk,-\tau)   \nonumber \\
&
+\gamma v_\star^\prime (k_x'+i\xi' k_y') 
g_{c''c'}^{\xi_2-\xi',j}(\kk',\tau) g_{c'c''}^{\xi\xi_2,i}(\kk,-\tau)
+ \gamma v_\star^\prime (k_x -i\xi k_y) 
g_{c''c'}^{\xi_2\xi',j}(\kk',\tau) g_{c'c''}^{-\xi\xi_2,i}(\kk,-\tau )
\bigg] 
\bigg\} d\tau 
\label{eq:chi_A_formula}
\eaa 
where $\omega$ denotes the Matsubara frequency ($w\in 2\pi/\beta\mathbb{Z}$ at finite temperature and can be treated as a continuous variable at zero temperature). 

We next rewrite $A(\RR,\tau)_{i\xi,j\xi'},B(\RR,\tau)_{i\xi,i\xi}$ with $u(\RR,\tau)_{i\xi,j\xi'}$ 
\baa 
A(\RR,\tau)_{i\xi,j\xi'} = i(\Lambda_{j\xi',j\xi'}-\Lambda_{i\xi,i\xi})u(\RR,\tau)_{i\xi,j\xi'} \quad,\quad 
B(\RR,\tau)_{i\xi,i\xi} = u(\RR,\tau)_{i\xi,j\xi'}u(\RR,\tau)_{j\xi',i\xi} \bigg(
 \Lambda_{j\xi',j\xi'}- \Lambda_{i\xi,i\xi}  
\bigg) 
\eaa 
and and introduce the Fourier transformation of $u(\RR,\tau)_{i\xi,j\xi'}$ 
\ba 
u(\qq,i\omega)_{i\xi,j\xi'} = \int \sum_\RR e^{-i\qq \cdot \RR +i\omega \tau }u(\RR,\tau)_{i\xi,j\xi'} d\tau 
\ea 
Then 
\baa 
S_{eff} = &-\int \sum_\RR 2i\sum_{i\xi} \Lambda_{i\xi,i\xi} \partial_\tau u(\RR,\tau)_{i\xi,i\xi }+ \frac{1}{2\pi N_M}\int \sum_\qq  \sum_{i\xi,j\xi' }i\omega 
(\Lambda_{i\xi,i\xi} -\Lambda_{j\xi',j\xi'})u(\qq,i\omega)_{i\xi,j\xi'} u(-\qq,-i\omega)_{j\xi',i\xi}d\omega \nonumber \\
&+ \frac{1}{2\pi N_M}\int \sum_\qq 
\sum_{i\xi,j\xi' }N_{i\xi} \bigg(\Lambda_{j\xi',j \xi'}-\Lambda_{i\xi,i\xi}\bigg)u(\qq,i\omega)_{i\xi,j\xi'} u(-\qq,-i\omega)_{j\xi',i\xi}
d\omega \nonumber \\
& 
 +\frac{1}{2\pi N_M} \int \sum_\qq 
\sum_{i,j,\xi,\xi',\xi_2,\xi_2'}\bigg( \Lambda_{j\xi',j\xi'}-\Lambda_{i\xi,i\xi}\bigg)
\bigg( \Lambda_{j\xi_2,j\xi_2}- 
\Lambda_{i\xi_2',i\xi_2'}
\bigg)  \chi^A_{\xi\xi',\xi_2'\xi_2}(\qq,i\omega,i,j)  \nonumber \\
&u(\qq,i\omega)_{i\xi,j\xi'} u(-\qq,-i\omega)_{j\xi_2,i\xi_2'} 
d\omega 
\label{eq:seff_2nd}
\eaa 
We next to separate $u(\RR,\tau)$ (or $u(\qq,i\omega)$) into two-parts, the diagonal components $u(\RR,\tau)_{i\xi,i\xi}$ and the off-diagonal components $u(\RR,\tau)_{i\xi,j\xi'}$ with $i\xi\ne j\xi'$. Correspondingly, we can separate the effective action into two parts $S_{eff} = S_{eff,diag} +S_{eff,off}$, where $S_{eff,diag}$($S_{eff,off}$) describes the behaviors of diagonal (off-diagonal) components of $u(\RR,\tau)$. $S_{eff,diag}$ can be written as 
\baa 
S_{eff,diag} = -\int \sum_\RR 2i \sum_{i\xi} \Lambda_{i\xi,i\xi}\partial_\tau u(\RR,\tau)_{i\xi,i\xi}  d\tau
\label{eq:seff_diag}
\eaa 
where we write the action in the position and imaginary-time spaces to illustrate its behaviors. 
Clearly, $S_{eff,diag}$ is a total derivative and only matters when $u(\RR,\tau)_{i\xi,i\xi}$ develops topologically non-trivial configurations which give a non-zero winding number $Q_{i\xi}(\RR) \ne 0$ where $Q_{i\xi}(\RR) = \int \partial_\tau u(\RR,\tau)_{i\xi,i\xi} d\tau$. Then, $S_{eff,diag} = -2i\sum_{\RR,i\xi} \Lambda_{i\xi,i\xi} Q_{i\xi}(\RR)$ is just a phase factor. Therefore, we will focus on the off-diagonal components of $u$.

The off-diagonal components give 
\baa 
S_{eff,off} = & \frac{1}{2\pi N_M}\int \sum_\qq  \sum_{i,j,\xi,\xi',\xi_2,\xi_2' }u(\qq,i\omega)_{i\xi,j\xi'} u(-\qq,-i\omega)_{j\xi_2,i\xi_2'} \nonumber \\
& \bigg\{ -
i\omega 
(\Lambda_{i\xi,i\xi} -\Lambda_{j\xi',j\xi'})\delta_{\xi_2,\xi'}\delta_{\xi_2',\xi} 
+N_{i\xi} (\Lambda_{j\xi',j\xi'} -\Lambda_{i\xi,i\xi} ) \delta_{\xi_2,\xi'}\delta_{\xi_2',\xi} \nonumber \\
&
+\bigg( \Lambda_{j\xi',j\xi'}-\Lambda_{i\xi,i\xi}\bigg)
\bigg( \Lambda_{j\xi_2,j\xi_2}- 
\Lambda_{i\xi_2',i\xi_2'}
\bigg) \chi^A_{\xi\xi',\xi_2'\xi_2}(\qq,i\omega,i,j) 
\bigg\} d\omega 
\label{eq:seffoff_v0}
\eaa 
It is worth mentioning that not all the off-diagonal terms appear in the effective action $S_{eff,off}$. To observe that, we first introduce the following two sets of indices:
\baa  
S_{fill}= \{i\xi | \Lambda_{i\xi,i\xi} = 1/4 \} 
\quad,\quad 
S_{emp}= \{i\xi | \Lambda_{i\xi,i\xi} =- 1/4 \} 
\eaa  
where $S_{fill} $ ($S_{emp}$) denotes the set of flavors, that are filled with one (zero) $f$ electron. For the ground states in Eq.~\ref{eq:def_psi_0_r}, we have 
\baa  
\nu_f = 0 & : \quad S_{fill} = \{1+,1-,2+,2-\},\quad S_{emp} = \{3+,3-,4+,4-\} \nonumber \\
\nu_f = -1 & : \quad S_{fill} = \{1+,1-,2+\},\quad S_{emp} = \{2-,3+,3-,4+,4-\} \nonumber \\
\nu_f = -2 & : \quad S_{fill} = \{1+,1-\},\quad S_{emp} = \{2+,2-,3+,3-,4+,4-\} \nonumber \\
\eaa 
and 
\baa  
&\Lambda_{i\xi,i\xi} -\Lambda_{j\xi',j\xi'} = 0 \quad,\quad i\xi,j\xi' \in S_{fill}  \nonumber \\
&\Lambda_{i\xi,i\xi} -\Lambda_{j\xi',j\xi'} = 0 \quad,\quad i\xi,j\xi' \in S_{emp}  \nonumber \\
&\Lambda_{i\xi,i\xi} -\Lambda_{j\xi',j\xi'} =1/2 \quad,\quad i\xi \in S_{fill}\quad,\quad j\xi' \in S_{emp}  \nonumber \\
&\Lambda_{i\xi,i\xi} -\Lambda_{j\xi',j\xi'} = -1/2 \quad,\quad i\xi \in S_{emp}\quad,\quad j\xi' \in S_{fill} \, .
\eaa  
All terms in Eq.~\ref{eq:seffoff_v0} that contain $u(\qq,i\omega)_{i\xi,j\xi'}$ will have a prefactor $\Lambda_{i\xi,i\xi}-\Lambda_{j\xi',j\xi'}$. Then, only $u(\qq,i\omega)_{i\xi,j\xi'}$ with $i\xi \in S_{fill}, j\xi'\in S_{emp}$ or $i\xi \in S_{emp}, j\xi'\in S_{fill}$ produces non-zero contribution. This allows us to write the effective action as 
\baa 
S_{eff,off} = & \frac{1}{2\pi N_M}\int \sum_\qq 
\sum_{i\xi,i\xi_2' \in S_{fill}, j\xi',j\xi_2 \in S_{emp}}u(\qq,i\omega)_{i\xi,j\xi'} u(-\qq,-i\omega)_{j\xi_2,i\xi_2'} \nonumber \\
& \bigg\{ 
i\omega \delta_{\xi_2,\xi'}\delta_{\xi_2',\xi}
-\frac{N_{i\xi}-N_{j\xi'}}{2} \delta_{\xi_2,\xi'}\delta_{\xi_2',\xi} 
+\frac{1}{4} \bigg( 
\chi^A_{\xi\xi',\xi_2'\xi_2}(\qq,i\omega=0,i,j)  + \chi^A_{\xi_2\xi_2',\xi'\xi}(-\qq,i\omega=0,j,i)
\bigg) 
\bigg\} d\omega 
\label{eq:seff_off}
\eaa 
where we also take the low-frequency approximation of the $\chi^A$ by letting $\chi^A_{\xi\xi',\xi_2'\xi_2}(\qq,i\omega,i,j) \approx \chi^A_{\xi\xi',\xi_2'\xi_2}(\qq,i\omega=0,i,j) $. For what follows, we will focus on the $S_{eff,off}$ which describes the spin fluctuations of the system. Here, we also provide the formula of $N_{i\xi}, \chi_{\xi\xi',\xi_2'\xi_2}^{A}(\qq,i\omega,i,j)$
\baa  
&N_{i\xi }\nonumber \\
=&\frac{1}{N_M}\sum_{\kk}\bigg[ \frac{2\gamma^2 [F(|\kk|)]^2}{D_{\nu_c,\nu_f}}\langle  :\psi_{\kk,i}^{c',\xi,\dag}(\tau) \psi_{\kk,i}^{c',\xi}(\tau):\rangle_0 
-2J \langle :\psi_{\kk,i}^{c'',\xi,\dag}(\tau) \psi_{\kk,i}^{c'',\xi}(\tau):\rangle_0 \nonumber \\
& +\frac{ 2\gamma v_\star^\prime [F(|\kk|)]^2}{D_{\nu_c,\nu_f}} (k_x+i\xi k_y) \langle :\psi_{\kk,i}^{c',-\xi,\dag}(\tau) \psi_{\kk,i}^{c',\xi}(\tau):\rangle_0
+ \frac{ 2\gamma v_\star^\prime [F(|\kk|)]^2}{D_{\nu_c,\nu_f}} (k_x-i\xi k_y) \langle :\psi_{\kk,i}^{c',\xi,\dag}(\tau) \psi_{\kk,i}^{c',-\xi}(\tau):\rangle_0
\bigg] \nonumber \\ 
&\chi^A_{\xi\xi',\xi_2' \xi_2}(\qq,i\omega=0,i,j)\nonumber \\
=&  \int\sum_{\kk,\kk'} \frac{\delta_{\kk'-\kk,\qq} }{N_M}
\bigg\{
\frac{2\gamma^4 [F(|\kk|)F(|\kk'|)]^2}{D_{\nu_c,\nu_f}^2}  g_{c'c'}^{\xi_2\xi',j}(\kk',\tau)g_{c'c'}^{\xi \xi_2',i}(\kk,-\tau) \nonumber \\
&+\frac{2\gamma^3v_\star^\prime [F(|\kk|)F(|\kk'|)]^2 }{D_{\nu_c,\nu_f}^2} \bigg[ 
(k_{x}+i\xi_2'k_{y}) 
 g_{c'c'}^{\xi_2\xi',j}(\kk',\tau)g_{c'c'}^{\xi -\xi_2',i}(\kk,-\tau)
 +
 (k_{x}'-i\xi_2k_{y}’) 
 g_{c'c'}^{-\xi_2\xi',j}(\kk',\tau)g_{c'c'}^{\xi \xi_2',i}(\kk,-\tau) \nonumber \\
 &
 +
 (k_x'+i\xi' k_y') g_{c'c'}^{\xi_2 -\xi',j}(\kk',\tau)
 g_{c'c'}^{\xi \xi_2',i}(\kk,-\tau)
 +
 (k_x-i\xi k_y) g_{c'c'}^{\xi_2 \xi',j}(\kk',\tau)
 g_{c'c'}^{-\xi \xi_2',i}(\kk,-\tau) 
\bigg] \nonumber \\
& +\frac{2\gamma^2(v_\star^\prime)^2 [F(|\kk|)F(|\kk'|)]^2 }{D_{\nu_c,\nu_f}^2}
\bigg[ 
(k_x'+i\xi' k_y')(k_x + i\xi_2' k_y)
g_{c'c'}^{\xi_2 -\xi',j}(\kk',\tau)
 g_{c'c'}^{\xi -\xi_2',i}(\kk,-\tau) \nonumber \\
 &
 +
 (k_x-i\xi k_y)(k_x'-i\xi_2 k_y')
 g_{c'c'}^{-\xi_2 \xi',j}(\kk',\tau)
 g_{c'c'}^{-\xi \xi_2',i}(\kk,-\tau) 
 +(k_x'+i\xi' k_y) (k_x'-i\xi_2 k_y')g_{c'c'}^{-\xi_2 -\xi',j}(\kk',\tau)
 g_{c'c'}^{\xi \xi_2',i}(\kk,-\tau) \nonumber \\
 &+(k_x-i\xi k_y)(k_x+i\xi_2' k_y)g_{c'c'}^{\xi_2 \xi',j}(\kk',\tau)
 g_{c'c'}^{-\xi -\xi_2',i}(\kk,-\tau) 
\bigg] 
+2J^2 \delta_{\xi,\xi'}\delta_{\xi_2',\xi_2} 
g_{c''c''}^{\xi_2\xi,j}(\kk',\tau)g_{c''c''}^{\xi\xi_2,i}(\kk,-\tau) \nonumber \\
&-\frac{ 4\delta_{\xi_2,\xi_2'} J F(|\kk|)F(|\kk'|)}{D_{\nu_c,\nu_f} }
\bigg[ 
\gamma^2 g_{c''c'}^{\xi_2\xi',j}(\kk',\tau)g^{\xi\xi_2',i}_{c'c''}(\kk,-\tau)   \nonumber \\
&
+\gamma v_\star^\prime (k_x'+i\xi' k_y') 
g_{c''c'}^{\xi_2-\xi',j}(\kk',\tau) g_{c'c''}^{\xi\xi_2,i}(\kk,-\tau)
+ \gamma v_\star^\prime (k_x -i\xi k_y) 
g_{c''c'}^{\xi_2\xi',j}(\kk',\tau) g_{c'c''}^{-\xi\xi_2,i}(\kk,-\tau )
\bigg] 
\bigg\} d\tau 
\label{eq:chi_n}
\eaa

We next introduce 
\baa 
F^{ij}_{\xi\xi',\xi_2'\xi_2}(\qq) = \frac{N_{i\xi}-N_{j\xi'}}{2} \delta_{\xi_2,\xi'}\delta_{\xi_2',\xi}
-\frac{1}{4} \bigg( 
\chi^A_{\xi\xi',\xi_2'\xi_2}(\qq,i\omega=0,i,j)  + \chi^A_{\xi_2\xi_2',\xi'\xi}(-\qq,i\omega=0,j,i)
\bigg)
\label{eq:def_F}
\eaa 
Then Eq.~\ref{eq:seff_off} can be written as 
\baa 
S_{eff,off} = & \frac{1}{2\pi N_M}\int \sum_\qq 
\sum_{i\xi,i\xi_2' \in S_{fill}, j\xi',j\xi_2 \in S_{emp}}u(\qq,i\omega)_{i\xi,j\xi'} u(-\qq,-i\omega)_{j\xi_2,i\xi_2'} \bigg\{ 
i\omega \delta_{\xi_2,\xi'}\delta_{\xi_2',\xi}
-F_{\xi\xi',\xi_2'\xi_2}^{ij}(\qq)
\bigg) 
\bigg\} d\omega 
\label{eq:seff_f}
\eaa 
We mention that the excitation spectrum is obtained by numerically evaluating Eq.~\ref{eq:chi_n} and Eq.~\ref{eq:def_F}, and then fining the eigenvalues of $F_{\xi\xi',\xi_2'\xi_2}^{ij}(\qq)$.
We now prove $F_{\xi\xi',\xi_2'\xi_2}^{ij}(\qq)$, as a matrix with row and column indices $\xi\xi',\xi_2'\xi_2$, is a Hermitian matrix, in other words
\baa 
F_{\xi\xi',\xi_2'\xi_2}^{ij}(\qq) = [F_{\xi_2'\xi_2,\xi\xi'}(\qq)]^*
\label{eq:f_hert}
\eaa

To prove Eq.~\ref{eq:f_hert}, we first consider $N_{i\xi}$ (Eq.~\ref{eq:chi_n})
\baa 
N_{i\xi}^* =& \frac{1}{N_M}\sum_{\kk}\bigg[ \frac{2\gamma^2 [F(|\kk|)]^2}{D_{\nu_c,\nu_f}}\langle  :\psi_{\kk,i}^{c',\xi,\dag}(\tau) \psi_{\kk,i}^{c',\xi}(\tau):\rangle_0^* 
-2J \langle :\psi_{\kk,i}^{c'',\xi,\dag}(\tau) \psi_{\kk,i}^{c'',\xi}(\tau):\rangle_0^* \nonumber \\
& +\frac{ 2\gamma v_\star^\prime [F(|\kk|)]^2}{D_{\nu_c,\nu_f}} (k_x-i\xi k_y) \langle :\psi_{\kk,i}^{c',-\xi,\dag}(\tau) \psi_{\kk,i}^{c',\xi}(\tau):\rangle_0^*
+ \frac{ 2\gamma v_\star^\prime [F(|\kk|)|^2}{D_{\nu_c,\nu_f}} (k_x+i\xi k_y) \langle :\psi_{\kk,i}^{c',\xi,\dag}(\tau) \psi_{\kk,i}^{c',-\xi}(\tau):\rangle_0^*
\bigg] \nonumber \\ 
= &\frac{1}{N_M}\sum_{\kk}\bigg[ \frac{2\gamma^2 [F(|\kk|)]^2}{D_{\nu_c,\nu_f}}\langle  :\psi_{\kk,i}^{c',\xi,\dag}(\tau) \psi_{\kk,i}^{c',\xi}(\tau):\rangle_0 
-2J \langle :\psi_{\kk,i}^{c'',\xi,\dag}(\tau) \psi_{\kk,i}^{c'',\xi}(\tau):\rangle_0 \nonumber \\
& +\frac{ 2\gamma v_\star^\prime [F(|\kk|)]^2}{D_{\nu_c,\nu_f}} (k_x-i\xi k_y) \langle :\psi_{\kk,i}^{c',\xi,\dag}(\tau) \psi_{\kk,i}^{c',-\xi}(\tau):\rangle_0
+ \frac{ 2\gamma v_\star^\prime [F(|\kk|)|^2}{D_{\nu_c,\nu_f}} (k_x+i\xi k_y) \langle :\psi_{\kk,i}^{c',-\xi,\dag}(\tau) \psi_{\kk,i}^{c',\xi}(\tau):\rangle_0
\bigg] \nonumber \\ 
=&N_{i\xi} 
\label{eq:real_N}
\eaa 
As for $\chi^A$ in Eq.~\ref{eq:chi_n}, we find
\baa  
&[\chi^A_{\xi\xi',\xi_2' \xi_2}(\qq,i\omega=0,i,j)]^*\nonumber \\
=&  \int\sum_{\kk,\kk'} \frac{\delta_{\kk'-\kk,\qq} }{N_M}
\bigg\{
\frac{2\gamma^4 [F(|\kk|)F(|\kk'|)]^2}{D_{\nu_c,\nu_f}^2}  g_{c'c'}^{\xi'\xi_2,j}(\kk',\tau)g_{c'c'}^{ \xi_2'\xi,i}(\kk,-\tau) \nonumber \\
&+\frac{2\gamma^3v_\star^\prime [F(|\kk|)F(|\kk'|)]^2 }{D_{\nu_c,\nu_f}^2} \bigg[ 
(k_{x}-i\xi_2'k_{y}) 
 g_{c'c'}^{\xi'\xi_2,j}(\kk',\tau)g_{c'c'}^{-\xi_2'\xi ,i}(\kk,-\tau)
 +
 (k_{x}'+i\xi_2k_{y}’) 
 g_{c'c'}^{\xi'-\xi_2,j}(\kk',\tau)g_{c'c'}^{\xi_2'\xi ,i}(\kk,-\tau) \nonumber \\
 &
 +
 (k_x'-i\xi' k_y') g_{c'c'}^{ -\xi'\xi_2,j}(\kk',\tau)
 g_{c'c'}^{\xi_2'\xi ,i}(\kk,-\tau)
 +
 (k_x+i\xi k_y) g_{c'c'}^{ \xi'\xi_2,j}(\kk',\tau)
 g_{c'c'}^{ \xi_2'-\xi,i}(\kk,-\tau) 
\bigg] \nonumber \\
& +\frac{2\gamma^2(v_\star^\prime)^2 [F(|\kk|)F(|\kk'|)]^2 }{D_{\nu_c,\nu_f}^2}
\bigg[ 
(k_x'-i\xi' k_y')(k_x - i\xi_2' k_y)
g_{c'c'}^{ -\xi'\xi_2,j}(\kk',\tau)
 g_{c'c'}^{ -\xi_2'\xi,i}(\kk,-\tau) \nonumber \\
 &
 +
 (k_x+i\xi k_y)(k_x'+i\xi_2 k_y')
 g_{c'c'}^{ \xi'-\xi_2,j}(\kk',\tau)
 g_{c'c'}^{-\xi \xi_2'-\xi,i}(\kk,-\tau) 
 +(k_x'-i\xi' k_y) (k_x'+i\xi_2 k_y')g_{c'c'}^{-\xi'-\xi_2 ,j}(\kk',\tau)
 g_{c'c'}^{\xi_2'\xi ,i}(\kk,-\tau) \nonumber \\
 &+(k_x+i\xi k_y)(k_x-i\xi_2' k_y)g_{c'c'}^{\xi'\xi_2 ,j}(\kk',\tau)
 g_{c'c'}^{ -\xi_2'-\xi,i}(\kk,-\tau) 
\bigg] 
+2J^2 \delta_{\xi,\xi'}\delta_{\xi_2',\xi_2} 
g_{c''c''}^{\xi\xi_2,j}(\kk',\tau)g_{c''c''}^{\xi_2\xi,i}(\kk,-\tau) \nonumber \\
&-\frac{ 4\delta_{\xi_2,\xi_2'} J F(|\kk|)F(|\kk'|)}{D_{\nu_c,\nu_f} }
\bigg[ 
\gamma^2 g_{c'c''}^{\xi'\xi_2,j}(\kk',\tau)g^{\xi_2'\xi,i}_{c''c'}(\kk,-\tau)   \nonumber \\
&
+\gamma v_\star^\prime (k_x'-i\xi' k_y') 
g_{c'c''}^{-\xi'\xi_2,j}(\kk',\tau) g_{c''c'}^{\xi_2\xi,i}(\kk,-\tau)
+ \gamma v_\star^\prime (k_x +i\xi k_y) 
g_{c'c''}^{\xi'\xi_2,j}(\kk',\tau) g_{c''c'}^{\xi_2-\xi,i}(\kk,-\tau )
\bigg] 
\bigg\} d\tau  \nonumber \\ 
=& \chi_{\xi_2'\xi_2,\xi\xi'}^A(-\qq,i\omega=0,j,i)
\label{eq:conj_chi}
\eaa 
Combining Eq.~\ref{eq:def_F}, Eq.~\ref{eq:real_N} and Eq.~\ref{eq:conj_chi}, we have 
\baa  
&[F_{\xi_2'\xi_2,\xi\xi'}^{ij}(\qq)]^* = \frac{(N_{i\xi}-N_{j\xi'})^*}{2} \delta_{\xi_2,\xi'}\delta_{\xi_2',\xi}
-\frac{1}{4} \bigg( 
[\chi^A_{\xi_2'\xi_2,\xi\xi'}(\qq,i\omega=0,i,j)]^*  + [\chi^A_{\xi'\xi,\xi_2\xi_2'}(-\qq,i\omega=0,j,i)]^*
\bigg) \nonumber \\  = &
\frac{N_{i\xi}-N_{j\xi'}}{2} \delta_{\xi_2,\xi'}\delta_{\xi_2',\xi}
-\frac{1}{4} \bigg( 
[\chi^A_{\xi\xi',\xi_2'\xi_2}(-\qq,i\omega=0,j,i)]  + [\chi^A_{\xi_2\xi_2',\xi'\xi}(\qq,i\omega=0,i,j)]
\bigg) =F_{\xi_2'\xi_2,\xi\xi'}^{ij}(\qq)
\eaa 
Thus we proved Eq.~\ref{eq:f_hert}.
\subsection{Flat $U(4)$ symmetry and Noether's theorem} 
We consider the model at $M=0$, which has a flat $U(4)$ symmetry. We note that $U(4) = SU(4)\times U(1)_c$ with $U(1)_c$ a $U(1)$ charge symmetry. We focus on the $SU(4)$ part, which will produce the Goldstone modes (note that the ground states (Eq.~\ref{eq:def_psi_0_r}) preserve $U(1)_c$ symmetry). And we will call it flat $SU(4)$ symmetry.

 We first introduce flat $SU(4)$ transformation. We consider a flat $SU(4)$ transformation characterize by real number $v_{\mu\nu}$ with $\mu,\nu \in \{0,x,y,z\}$ and $\mu\nu \ne 00 $. The $SU(4)$ transformation on the $f$ electron is then defined as 
 \baa 
 g = e^{-i \sum_{\RR,\xi,i,j} \phi_{ij} \psi_{\RR,i}^{f,\xi,\dag} \psi_{\RR,j}^{f,\xi} }
 \label{eq:flatu4_trnasf_v0}
 \eaa 
where $\phi = \sum_{\mu\nu \ne 00 } v_{\mu\nu} T^{\mu\nu}$ is a $4\times 4$ matrix and $T^{\mu\nu}$ are the generators of flat $U(4)$ group (Eq.~\ref{eq:flat_u4_general}). In addition, $\phi$ is a $4\times 4 $ traceless Hermitian matrix. We next discuss how $u_{i\xi,j\xi'}(\RR)$ transform under flat $SU(4)$ transformation. We act $SU(4)$ transformation on the $|u \rangle$ state (using Eq.~\ref{eq:u_latt},Eq.~\ref{eq:u_state} and Eq.~\ref{eq:ru_def}) 
\baa  
g|u\rangle   = e^{-i \sum_{\RR,\xi,i,j} \phi_{ij} \psi_{\RR,i}^{f,\xi,\dag} \psi_{\RR,j}^{f,\xi} } |u\rangle  = \prod_\RR  g_\RR\hat{R}[u(\RR)]  |\psi_0\rangle
\eaa 
where we define
\baa 
g_\RR =  e^{-i \sum_{\xi,i,j} \phi_{ij} \psi_{\RR,i}^{f,\xi,\dag} \psi_{\RR,j}^{f,\xi} } \, .
\eaa 
such that $g = \prod_\RR g_\RR$ (note that $\phi_{ij}$ is site $R$-independent). 
Since $g_\RR$ generate a $SU(4)$ transformation, $\hat{R}[u(\RR)]$ (Eq.~\ref{eq:ru_def}) generate a $SU(8)$ transformation (for each site). Then $g_\RR\hat{R}[u(\RR)]$ also generate a $SU(8)$ transformation. Thus there exists $\tilde{u}(\RR)$, such that 
\baa 
g_\RR \hat{R}[u(\RR)] = \hat{R}[\tilde{u}(\RR)]  
\label{eq:def_u_tilde}
\eaa 
To find $\tilde{u}$, we can act the transformation on the $\psi^f$ fermion. From Eq.~\ref{eq:def_u_tilde}
\baa 
\bigg(g_\RR \hat{R}_{\RR}[u(\RR)] \bigg)\psi_{\RR,i}^{f,\xi} \bigg(g_\RR\hat{R}_{\RR}[u(\RR)]\bigg)^{-1} =  \bigg(\hat{R}[\tilde{u}(\RR)]  \bigg)\psi_{\RR,i}^{f,\xi}  \bigg(\hat{R}[\tilde{u}(\RR)] \bigg)^{-1}
\label{eq:find_u_tilde_0}
\eaa 
The left-hand-side (LHS) of Eq.~\ref{eq:find_u_tilde_0} gives 
\baa 
LHS = \sum_{j,\xi'} [ e^{i\Phi}e^{iu(\RR)}]_{i\xi,j\xi'}\psi_{\RR,j}^{f,\xi'}
\eaa 
where $\Phi$ is a $8\times 8$ matrix with column and row indices $i\xi,j\xi'$ and $[\Phi]_{i\xi,j\xi'} = \phi_{ij}\delta_{\xi,\xi'}$. 
The right-hand-side (LHS) of Eq.~\ref{eq:find_u_tilde_0} gives 
\baa 
RHS = \sum_{j,\xi'} [ e^{i\tilde{u}(\RR)}]_{i\xi,j\xi'}\psi_{\RR,j}^{f,\xi'}
\eaa 
Then 
\baa  
LHS = RHS \Rightarrow & \sum_{j,\xi'} [ e^{i\Phi}e^{iu(\RR)}]_{i\xi,j\xi'} \psi_{\RR,j}^{f,\xi'}= \sum_{j,\xi'} [ e^{i\tilde{u}(\RR)}]_{i\xi,j\xi'}\psi_{\RR,j}^{f,\xi'} \Rightarrow \sum_{j\xi'} [ e^{-i\tilde{u}(\RR)}e^{ i\Phi} e^{iu(\RR)}]_{i\xi,j\xi'}\psi_{\RR,j}^{f,\xi'}  = \psi_{\RR,i}^{f,\xi} \nonumber \\
\Rightarrow & e^{-i\tilde{u}(\RR)}e^{ i\Phi} e^{iu(\RR)} = \mathbb{I} \Rightarrow e^{i\tilde{u}(\RR)} = e^{ i\Phi} e^{iu(\RR)} 
\eaa 
where $\mathbb{I}$ is an $8\times 8$ identity matrix. We can now give the expression of $\tilde{u}(\RR)$
\baa 
\tilde{u}(\RR) = -i\log \bigg[e^{ i\Phi} e^{iu(\RR)}  \bigg]
\label{eq:def_tilde_u}
\eaa 

We then take an infinitesimal transformation $g$ with $|\phi_{ij}| <<1$. From Eq.~\ref{eq:def_tilde_u} we have 
\baa 
\tilde{u}(\RR) \approx& i\log \bigg[ (1-i\Phi )e^{-iu(\RR)}\bigg] 
= i\log \bigg[ e^{-iu(\RR)} - i\Phi e^{-iu(\RR) } \bigg] \nonumber \\ 
=& u(\RR) + \int_0^1 \frac{1}{1-t(1-e^{-iu(\RR)}) } [\Phi e^{-iu(\RR)}]  \frac{1}{1-t(1-e^{-iu(\RR)}) } dt 
\label{eq:tilde_u_inf}
\eaa 
where we use the derivative of a matrix logarithm. 

From Eq.~\ref{eq:tilde_u_inf}, we introduce a tensor $Du$ which is a function of $u(\RR)$
\baa 
Du[u(\RR)]_{i\xi,j\xi'}^{mn} = \sum_{\xi_2,j_2\xi_2'}\int_0^1 \bigg[\frac{1}{1-t(1-e^{-iu(\RR)}) }\bigg]_{i\xi,m\xi_2} [e^{-iu(\RR)}]_{n\xi_2,j_2\xi'_2}  \bigg[\frac{1}{1-t(1-e^{-iu(\RR)}) }\bigg]_{j_2\xi'_2,j\xi'} dt \label{Du}
\eaa 
such that Eq.~\ref{eq:tilde_u_inf} can be written as 
\baa  
\tilde{u}(\RR) = u(\RR) + \sum_{mn} Du[u(\RR)]^{mn}\phi_{mn}
\eaa 
Therefore $Du[u(\RR)]$ characterize the transformation properties of $u(\RR)$ under flat $U(4)$ transformation.

We next discuss the consequence of infinitesimal symmetry transformation.  We note that we have integrated out conduction $c$-electron fields and derived an effective action of $u$ fields by performing expansion to second order in $u$. Here, we let $S[u]$ denote the effective action obtained by integrating out conduction $c$ electrons and performing expansion to the infinity order. Thus, $S[u]$ is an exact theory and has flat $U(4)$ symmetry (Here we comment that the effective action we derived in Eq.~\ref{eq:seff_2nd} describes the fluctuations around one symmetry breaking state and thus does not have the flat $U(4)$ symmetry. However, by performing expansion to infinity order, we effectively sum over the contributions from all the possible symmetry-breaking states and the symmetry is recovered.) Then it must be invariant under a symmetry transformation. In other words
\baa  
S[\tilde{u}] = S[u]  
\eaa  
where $u=\{{u}(\RR,\tau)\}$ denotes the configuration of the original $u$ fields 
where $\tilde{u} = \{\tilde{u}(\RR,\tau)\}$ denotes the configuration of fields after acting $SU(4)$ transformation $g$ on each time slice. $\tau$ is the imaginary time. Written explicitly,  for an infinitesimal flat $SU(4)$ transformation, we have
\baa 
S[u]= &S[\tilde{u}] = S\bigg[  
\{ {u}(\RR,\tau) + \sum_{mn} Du[u(\RR,\tau)]^{mn}\phi_{mn} \} 
\bigg] \nonumber \\  
= &S[u] + \int_\tau \sum_\RR\sum_{i\xi,j\xi',m,n}\frac{\delta S[u]}{\delta u_{i\xi,j\xi'}(\RR,\tau)} Du[u(\RR,\tau)]_{i\xi,j\xi'}^{mn} \phi_{mn}  
\eaa 
Therefore we have 
\baa 
\int_\tau \sum_\RR\sum_{i\xi,j\xi',m,n}\frac{\delta S[u]}{\delta u_{i\xi,j\xi'}(\RR,\tau)} Du[u(\RR,\tau)]_{i\xi,j\xi'}^{mn} \phi_{mn}  =0
\eaa 
$\phi_{mn}$ is an infinitesimal arbitrary $4\times 4$ traceless Hermitian matrix. Here we focus on the off-diagonal components and have 
\baa 
\int_\tau \sum_\RR\sum_{i\xi,j\xi',m,n}\frac{\delta S[u]}{\delta u_{i\xi,j\xi'}(\RR,\tau)} Du[u(\RR,\tau)]_{i\xi,j\xi'}^{mn} = 0 \quad, m\ne n
\label{eq:cons_u_1}
\eaa 
We next define the functional $C[u]_{mn}$
\baa 
C[u]_{mn}= \int_\tau \sum_\RR\sum_{i\xi,j\xi',m,n}\frac{\delta S[u]}{\delta  u_{i\xi,j\xi'}(\RR,\tau)} Du[u(\RR,\tau)]_{i\xi,j\xi'}^{mn}  
\label{eq:c_u}
\eaa 
such that Eq.~\ref{eq:cons_u_1} can be written as 
\baa 
C[u]_{mn} = 0 ,\quad m \ne n 
\label{eq:cons_u_2}
\eaa 
Since Eq.~\ref{eq:cons_u_1} and Eq.~\ref{eq:cons_u_2} , holds for any given configuration of $\{u(\RR,\tau)\}$, we must have
\baa  
\frac{\delta C[u]_{mn}}{\delta u(\RR,\tau)_{i\xi,j\xi'}} = 0 ,\quad \text{for all } \RR,\tau,i\xi,j\xi'
\eaa 
where $m \ne n$. And consequently
\baa 
\frac{\delta C[u]_{mn}}{\delta u(\RR,\tau)_{i\xi,j\xi'}} \bigg|_{u=0}= 0, \quad \text{for all } \RR,\tau,i\xi,j\xi'
\label{eq:dcdu} 
\eaa 

To evaluate Eq.~\ref{eq:dcdu}, we perform an expansion in $u$. We first consider $Du[u(\RR,\tau)]$. From Eq.~\ref{Du}, we have 
\baa 
Du[u(\RR,\tau)]_{i\xi,j\xi'}^{mn} \approx  Du[0]^{mn}_{i\xi,j\xi'} + O(u) = \delta_{i,m}\delta_{j,n} \delta_{\xi,\xi'} +o(u) 
\label{eq:du_exp}
\eaa 
We next perform expansion of $S$. We have already derived the effective action up to the second order $S_{eff}$ (Eq.~\ref{eq:seff_2nd})
\baa  
S[u] = S_{eff}[u] + O(u^3)
\eaa  
(for the purpose of discussion, $S_{eff}$ is treated as a as a functional of $u$ and is written as $S_{eff}[u]$). Then we have 
\baa 
\frac{\delta S[u]}{\delta  u_{i\xi,j\xi'}(\RR,\tau)} \approx \frac{\delta  S_{eff}[u]}{\delta u_{i\xi,j\xi'}(\RR,\tau)} +O(u^2)
\eaa 
Since we have separated $S_{eff}$ into diagonal (Eq.~\ref{eq:seff_diag}) and off-diagonal part (Eq.~\ref{eq:seff_off}): $S_{eff} = S_{eff,diag} +S_{eff,off}$, then 
\baa 
\frac{\delta  S[u]}{\delta  u_{i\xi,j\xi'}(\RR,\tau)} \approx \frac{\delta S_{eff,diag}[u]}{\partial u_{i\xi,j\xi'}(\RR,\tau)}
+\frac{\delta  S_{eff,off}[u]}{ \delta u_{i\xi,j\xi'}(\RR,\tau)} +O(u^2) 
\label{eq:s_eff_exp}
\eaa 
Combining Eq.~\ref{eq:c_u}, Eq.~\ref{eq:du_exp} and Eq.~\ref{eq:s_eff_exp}, we find 
\baa 
C[u]_{mn}=&\int_\tau \sum_\RR\sum_{i\xi,j\xi'}\bigg[\frac{\delta S_{eff,diag}[u]}{ \delta u_{i\xi,j\xi'}(\RR,\tau)}
+\frac{\delta S_{eff,off}[u]}{\delta u_{i\xi,j\xi'}(\RR,\tau)}+o(u^2)\bigg]\bigg[\delta_{i,m}\delta_{j,n} \delta_{\xi,\xi'} +o(u) \bigg] \nonumber \\ 
=&\int_\tau \sum_\RR\sum_{i\xi,j\xi'}\bigg[\frac{ \delta S_{eff,diag}[u]}{\delta u_{i\xi,j\xi'}(\RR,\tau)}
+\frac{\delta S_{eff,off}[u]}{\delta  u_{i\xi,j\xi'}(\RR,\tau)}+o(u^2)\bigg]\bigg[\delta_{i,m}\delta_{j,n} \delta_{\xi,\xi'} +o(u) \bigg]
\label{eq:c_2}
\eaa 
We next consider Eq.~\ref{eq:dcdu} with $i_2\xi_2 \ne j_2\xi_2'$
\baa 
0= \frac{\delta C[u]_{mn}}{\delta u(\RR_2,\tau_2)_{i_2\xi_2,j_2\xi_2'}} \bigg|_{u=0},\quad i_2 \xi_2 \ne j_2 \xi_2'
\label{eq:dcdu_2}
\eaa 
Combine Eq.~\ref{eq:c_2} and Eq.~\ref{eq:dcdu_2}
\baa  
0= \frac{\delta C[u]_{mn}}{\delta u(\RR_2,\tau_2)_{i_2\xi_2,j_2\xi_2}}\bigg|_{u=0}  = \bigg\{
\int_\tau\sum_\RR \sum_{i\xi,j\xi'}\bigg[ \frac{\delta \frac{\delta S_{eff,off}[u]}{\delta u_{i\xi,j\xi'}(\RR,\tau)}}{\delta u(\RR_2,\tau_2)_{i_2\xi_2,j_2\xi_2'}}+o(u^2)\bigg]\bigg[\delta_{i,m}\delta_{j,n} \delta_{\xi,\xi'} +o(u) \bigg]  \bigg\}_{u=0}
\eaa 
since $S_{eff,diag}$ (Eq.~\ref{eq:seff_diag} only contains the diagonal components of $u$, its contribution vanishes at leading order. Then 
\baa 
0 = \int_\tau \sum_\RR \sum_\xi \frac{\delta ^2 S_{eff,off}[u]}{ \delta u(\RR_2,\tau_2)_{i\xi_2,j\xi_2'} \delta u_{m\xi,n\xi}(\RR,\tau)} \bigg|_{u=0},\quad \text{for all $\RR_2,\tau_2,i\xi_2,j\xi_2',m,n$ with $i\xi_2\ne j\xi_2'$ and $m\ne n$}
\label{eq:const_v2}
\eaa 
To evaluate Eq.~\ref{eq:const_v2}, we rewrite transform Eq.~\ref{eq:seff_f} into imaginary-time and real space 
\baa 
S_{eff,off}[u] = & \int_\tau \sum_{\RR_1,\RR_2} 
\sum_{i\xi,i\xi_2' \in S_{fill}, j\xi',j\xi_2 \in S_{emp}} u(\RR_1,\tau)_{i\xi,j\xi'} \bigg\{ 
\partial_\tau \delta_{\xi_2,\xi'}\delta_{\xi_2',\xi}\delta_{\RR_1,\RR_2}
-F_{\xi\xi',\xi_2'\xi_2}^{ij}(\RR_2-\RR_1) \bigg\}
u(\RR_2,\tau)_{j\xi_2,i\xi_2'} 
\label{eq:seffoff_real}
\eaa 
where $F_{\xi\xi',\xi_2'\xi_2}^{ij}(\RR) = \frac{1}{N_M}\sum_\qq F_{\xi\xi',\xi_2'\xi_2}^{ij}(\qq)e^{i\qq \cdot \RR}$. We take 
\baa 
&i\xi_2 \in S_{fill}, j\xi_2' \in S_{emp}, i\ne j\nonumber \\
&m=j, n=i
 \label{eq:ind_const}
\eaa 
and then 
\baa 
\frac{\delta S_{eff,off}[u]}{ \delta u(\RR_2,\tau_2)_{i\xi_2,j\xi_2'}} =  \partial_{\tau_2} u(\RR_2,\tau_2)_{j\xi_2,i\xi_2} \delta_{\xi_2',\xi_2}- \sum_{\RR_3}
\sum_{  \substack{ \xi_3 \\ \text{ $j\xi_3 \in S_{emp}$}} } \sum_{ \substack{\xi_3'\\ \text{ $i\xi_3' \in S_{fill}$}}} F^{ij}_{\xi_2\xi_2',\xi_3'\xi_3}(\RR_3-\RR_2)u(\RR_3,\tau)_{j\xi_3,i\xi_3'} \, .
\eaa 
(Note that $i,j$ are not summed over)
We next take $m=j, n=i$ and find 
\baa  
\frac{\delta ^2 S_{eff,off}[u]}{ \delta u(\RR_2,\tau_2)_{i\xi_2,j\xi_2'} \delta u_{m\xi,n\xi}(\RR,\tau)}  = \partial_{\tau_2} \delta(\RR_2,\RR)\delta(\tau_2-\tau) - 
\sum_{  \substack{ \xi \\ \text{ $j\xi \in S_{emp}, i\xi \in S_{fill}$}} }  F^{ij}_{\xi_2\xi_2',\xi\xi}(\RR-\RR_2)
\label{eq:2nd_der}
\eaa 
Combining Eq.~\ref{eq:const_v2} and Eq.~\ref{eq:2nd_der}, we find 
\baa  
&0 = \int_\tau \sum_\RR \sum_\xi \sum_\xi \bigg[\partial_{\tau_2} \delta(\RR_2,\RR)\delta(\tau_2-\tau) - \sum_{j\xi \in S_{emp}, i\xi \in S_{fill}} F^{ij}_{\xi_2\xi_2',\xi\xi}(\RR-\RR_2)]\bigg]
  \nonumber \\
 &=-\int_\tau \sum_{\RR,\xi} \sum_{  \substack{ \xi \\ \text{ $j\xi \in S_{emp}, i\xi \in S_{fill}$}} }  F^{ij}_{\xi_2\xi_2',\xi\xi}(\RR-\RR_2) 
\eaa 
which indicates
\baa 
0= \sum_\RR  \sum_{  \substack{ \xi \\ \text{ $j\xi \in S_{emp}, i\xi \in S_{fill}$}} }   F^{ij}_{\xi_2\xi_2',\xi\xi}(\RR) 
\eaa  
(Note that $i,j$ are not summed over)
Transforming to the momentum space, we reach the following symmetry constraint on $F^{ij}_{\xi_2\xi_2,\xi\xi}$
\baa  
 \sum_{  \substack{ \xi \\ \text{ $j\xi \in S_{emp}, i\xi \in S_{fill}$}} }   F^{ij}_{\xi_2\xi_2',\xi\xi}(\qq=0) = 0
 \label{eq:symmetry_const}
\eaa 
with $\xi_2,\xi_2'$ satisfies $i\xi_2 \in S_{fill}, j\xi_2' \in S_{emp}$ and $i\ne j$. (Eq.~\ref{eq:ind_const}). This constraint will lead to exact Goldstone modes.




\subsection{Symmetry} 
We next discuss the symmetry of the ground state given in Eq.~\ref{eq:def_psi_0_r} and Eq.~\ref{eq:gnd_order}.
The effective theory (Eq.~\ref{eq:seff_off}) has the same symmetry properties as the ground state since the action describes the fluctuation on top of the ground state. 
At $\nu_f=0$, we have $U(2)\times U(2)$. The first $U(2)$ symmetry  corresponds to the flavors $i=1,2$, where we have filled two $f$-electrons $(\xi=\pm 1)$ for each flavor. The second $U(2)$ symmetry corresponds to the flavors $i=3,4$, where both flavors are filled with zero $f$ electrons. At $\nu_f=-2$, we have a $U(1) \times U(3)$ symmetry. The $U(1)$ symmetry corresponds to the flavor $i=1$, where we have filled two $f$-electrons $(\xi=\pm 1)$. The $U(3)$ symmetry comes from the flavors $i=2,3,4$, where all three flavors are filled with zero $f$ electrons. At $\nu_f=-1$, we have $U(1)\times U(1) \times  U(2)$. The first $U(1)$ corresponds to the flavor $i=1$, where we have filled two $f$-electrons. The second $U(1)$ corresponds to the flavor $i=2$, where we have filled one $f$ electron. The $U(2)$ corresponds to the flavor $i=3,4$, where we have filled two $f$ electrons. 

At $\nu_f=0,2$, the ground states (defined in Eq.~\ref{eq:def_psi_0_r} and Eq.~\ref{eq:gnd_order}) also have $C_{3z}$, $C_{2x}$ and $C_{2z}T$ symmetry. As for $\nu_f=-1$, the ground state (defined in Eq.~\ref{eq:def_psi_0_r} and Eq.~\ref{eq:gnd_order}) only has $C_{3z}$ symmetry. This is because, at $\nu_f=-1$, $i=2$ spin-valley only has one $f$-electrons.

We next define the symmetry transformation of the $f$-electrons and conduction $c$-electrons. To begin with, we introduce the representation matrix $D^f(g),D^{c'}(g),D^{c''}(g)$ for a given symmetry operator $g$ as following
\baa 
&g\psi_{\RR, i}^{f,\xi,\dag } g^{-1} = \sum_{j,\xi'} \psi^{f,\xi',\dag}_{g\RR, j}D^f(g)_{j\xi',i\xi}  \nonumber \\
&g\psi_{\RR, i}^{c',\xi,\dag } g^{-1} = \sum_{j,\xi'} \psi^{c',\xi',\dag}_{g\RR, j}D^{c'}(g)_{j\xi',i\xi} \quad,\quad g\psi_{\RR, i}^{c'',\xi,\dag } g^{-1} = \sum_{j,\xi'} \psi^{c'',\xi',\dag}_{g\RR, j}D^{c''}(g)_{j\xi',i\xi} 
\eaa 
For the symmetries we considered, the corresponding representation matrices~\cite{HF_MATBLG} are

\baa 
&
D^f(C_{3z})_{j\xi',i\xi} = \delta_{i,j}\delta_{\xi',\xi} e^{i\frac{2\pi}{3} \xi } 
\quad,\quad 
D^f(C_{2x})_{j\xi',i\xi} = [\varsigma'_0 \rho_z]_{ji}[\zeta_x]_{\xi'\xi} \quad,\quad 
D^f(C_{2z}T)_{j\xi',i\xi} = [\varsigma'_0 \rho_z]_{ji} [\zeta_x]_{\xi'\xi}\nonumber \\
&D^{c'}(C_{3z})_{j\xi',i\xi} = \delta_{i,j}\delta_{\xi',\xi} e^{i\frac{2\pi}{3} \xi} 
\quad,\quad 
D^{c'}(C_{2x})_{j\xi',i\xi} = [\varsigma'_0 \rho_z]_{ji}[\zeta_x]_{\xi'\xi}
\quad,\quad 
D^{c'}(C_{2z}T)_{j\xi',i\xi} = [\varsigma'_0 \rho_z]_{ji} [\zeta_x]_{\xi'\xi}
\nonumber \\ 
&
D^{c'}(C_{3z})_{j\xi',i\xi} = \delta_{i,j}\delta_{\xi',\xi}
\quad,\quad 
D^{c''}(C_{2x})_{j\xi',i\xi} = [\varsigma'_0 \rho_z]_{ji} [\zeta_x]_{\xi'\xi}
\quad,\quad 
D^{c''}(C_{2z}T)_{j\xi',i\xi} = [\varsigma'_0 \rho_z]_{ji} [\zeta_x]_{\xi'\xi}
\label{eq:desc_sym}
\eaa 
where $\zeta_{0,x,y,z}$ are identity matrix and Puli matrices for the $\xi$ degrees of freedom, $\rho_{0,x,y,z},\varsigma_{0,x,y,z}$ are introduced in Eq.~\ref{eq:flat_u4_general}. 

In addition, the system at $M=0$ also has a flat-$U(4)$ symmetry. As introduced in Eq.~\ref{eq:flatu4_trnasf_v0}, we characterize the flat-$U(4)$ transformation with a $4\times 4$ traceless Hermitian matrices $\phi_{ij}$. We use $g_\phi$ to represent the corresponding symmetry operator and introduce the representation matrices as $D^f(U(4), \phi),D^{c'}(U(4), \phi),D^{c''}(U(4), \phi)$. The representation matrices are defined below
\baa 
&g_v \psi_{\RR,i}^{f,\xi,\dag} g_v = \sum_j \psi_{\RR,j}^{f,\xi,\dag}v D^f(U(4),\phi)_{ji} \quad,\quad \nonumber \\
&
g_v \psi_{\kk,i}^{c',\xi,\dag} g_v = \sum_j \psi_{\kk,j}^{c',\xi,\dag} D^{c'}(U(4),\phi)_{ji} \quad,\quad
g_v \psi_{\kk,i}^{c'',\xi,\dag} g_v = \sum_j \psi_{\kk,j}^{c'',\xi,\dag} D^{c''}(U(4),\phi)_{ji} \quad,\quad
\nonumber \\
&D^f(U(4),\phi) =D^{c'}(U(4),\phi) =D^{c''}(U(4),\phi) = e^{i \phi}
\label{eq:u4_sym_d}
\eaa 
where $e^{iv}$ denotes the matrix exponential. 

We next discuss the effect of symmetry on the single-particle Green's function. For a given symmetry $g$ of $\hH_{c,order}$ (Eq.~\ref{eq:formula_e0e3_hc})
\baa 
g\psi_{\kk,i}^{c', \xi,\dag}g^{-1} = \sum_{j,\xi'}\psi_{g\kk,j}^{c',\xi',\dag} D^{c'}(g)_{\xi'j', \xi i}
\quad,\quad 
g\psi_{\kk,i}^{c'', \xi,\dag}g^{-1} = \sum_{j,\xi'}\psi_{g\kk,j}^{c'',\xi',\dag} D^{c''}(g)_{\xi'j', \xi i}
\eaa 
The corresponding Green's functions (Eq.~\ref{eq:single_green_order}) satisfy (for the unitary transformation)
\baa 
& \langle T_\tau \psi_{\kk,i}^{a,\xi}(\tau) \psi_{\kk,j}^{a',\xi',\dag}(\kk',0) \rangle_0 = \sum_{i_2,\xi_2,j_2,\xi'_2}D^{a,*}(g)_{\xi_2i_2,\xi i}
D^{a'}(g)_{\xi' j,\xi_2' j_2}
\langle T_\tau \psi_{g\kk,i_2}^{a,\xi,\dag}(\tau) \psi_{\kk,j_2}^{a',\xi'}(g\kk',0) \rangle_0   \nonumber \\
\Rightarrow & g_{aa'}^{\xi\xi',i}(\kk,\tau) = \sum_{i_2,\xi_2,\xi'_2}D^{a}(g)^*_{\xi_2i_2,\xi i}
D^{a'}(g)_{\xi' i,\xi_2' i_2} g_{aa'}^{\xi_2\xi'_2,i_2}(g\kk,\tau) 
\label{eq:sym_green}
\eaa 
where $a,a' \in \{c',c''\}$ and we use the fact that $\langle T_\tau  \psi_{\kk,i}^{a,\xi}(\tau) \psi_{\kk,j}^{a,\xi',\dag}(0) \rangle =0$ when $i\ne j$ (because $\hH_{c,order}$ is blocked diagonal with respect to the flavor indices $i$ as shown in Eq.~\ref{eq:formula_e0e3_hc}).
Then the product of two Green's functions satisfy (for the unitary transformation)
\baa  
g_{aa'}^{\xi\xi',i}(\kk,\tau) g_{a_2a_2'}^{\xi_2\xi_2',j}(\kk',\tau') = \sum_{i_2,\xi_3,\xi_3'}\sum_{j_2,\xi_4,\xi_4'}D^a(g)_{\xi_3i_2,\xi i}^*D^{a'}(g) _{\xi'i,\xi_3'i_2}
D^{a_2}(g)_{\xi_4j_2,\xi_2j}^*D^{a_2'}(g) _{\xi_2'j,\xi_4'j_2} g_{aa'}^{\xi_2\xi_2',i_2}(g\kk,\tau) g_{a_2a_2'}^{\xi_3\xi_3',j_2}(g\kk',\tau') 
\label{eq:sym_order_chi}
\eaa 

For anti-unitary transformation, we have 
\baa  
&  g_{aa'}^{\xi\xi',i}(\kk,\tau) = \sum_{i_2,\xi_2,\xi'_2}D^{a}(g)^*_{\xi_2i_2,\xi i}
D^{a'}(g)_{\xi' i,\xi_2' i_2} \bigg( g_{aa'}^{\xi_2\xi'_2,i_2}(g\kk,\tau) \bigg)^* \nonumber \\
&\ g_{aa'}^{\xi\xi',i}(\kk,\tau) g_{a_2a_2'}^{\xi_2\xi_2',j}(\kk',\tau') = \sum_{i_2,\xi_3,\xi_3'}\sum_{j_2,\xi_4,\xi_4'}D^a(g)_{\xi_3i_2,\xi i}^*D^{a'}(g) _{\xi'i,\xi_3'i_2}
D^{a_2}(g)_{\xi_4j_2,\xi_2j}^*D^{a_2'}(g)_{\xi_2'j,\xi_4'j_2} \nonumber \\
&\bigg( g_{aa'}^{\xi_2\xi_2',i_2}(g\kk,\tau)\bigg)^* \bigg( g_{a_2a_2'}^{\xi_3\xi_3',j_2}(g\kk',\tau') \bigg)^*
\label{eq:anti_unitary}
\eaa  

\subsection{$\nu_f=0$} 
We now analyze the effective theory at $\nu_f=0$ and $\nu_f=-2$. 

At $\nu_f=0$, we consider the $U(2)\times U(2)\in U(4)$ symmetry $g_\phi$ (in Eq.~\ref{eq:u4_sym_d} ,note that $U(2)\times U(2)$ is a subgroup of flat $U(4)$). From Eq.~\ref{eq:sym_green}. 
\baa  
[e^{i\phi}]_{ij}g_{aa'}^{\xi\xi',i}(\kk,\tau)[e^{-i\phi}]_{ji} = g_{aa'}^{\xi\xi',j}(\kk,\tau)
\label{eq:u2u2_g}
\eaa 
For the ground state we considered in Eq.~\ref{eq:def_psi_0_r}, one $U(2)$ corresponding to the rotations of flavor $i=1,2$ and the other $U(2)$ corresponding to the rotations of flavor $i=3,4$. Thus $e^{i\phi}$ is block diagonal
\baa  
[e^{i\phi}]_{ij} = \begin{bmatrix}
    e^{i \phi_1} & 0_{2\times 2} \\
    0_{2\times 2} & e^{i\phi_2}
\end{bmatrix}_{ij}
\eaa  
where $\phi_1,\phi_2$ are two traceless Hermitian matrices that characterize two $U(2)$ symmetries. Therefore, Eq.~\ref{eq:u2u2_g} can only be satisfied when
\baa 
g_{aa'}^{\xi\xi',i}(\kk,\tau) = g_{aa'}^{\xi\xi',j}(\kk,\tau) \quad,\quad a,a' \in \{c',c''\}\quad,\quad i,j\in \{1,2\} \nonumber \\
g_{aa'}^{\xi\xi',i}(\kk,\tau) = g_{aa'}^{\xi\xi',j}(\kk,\tau) \quad,\quad a,a' \in \{c',c''\}\quad,\quad i,j\in \{3,4\} 
\label{eq:gaa_nu0}
\eaa 
Consequently, from Eq.~\ref{eq:chi_n} and Eq.~\ref{eq:gaa_nu0}, we have
\baa 
&\chi_{\xi\xi',\xi_2'\xi_2}^A(\qq,i\omega=0,i,j) = \chi_{\xi\xi',\xi_2'\xi_2}^A(\qq,i\omega=0,i',j') \quad,\quad i,i' \in \{1,2\}\quad,\quad j,j'\in \{3,4\}\nonumber \\
&N_{i\xi}-N_{j\xi'} = N_{i'\xi}-N_{j'\xi'} \quad,\quad i,i' \in \{1,2\}\quad,\quad j,j'\in \{3,4\}
\eaa 
We can thus define (Eq.~\ref{eq:def_F})
\baa 
&H^{\nu_f=0}(\qq)_{\xi\xi',\xi_2'\xi_2} = F_{\xi\xi',\xi_2'\xi_2}^{ij}(\qq) \nonumber \\ 
&i\in\{1,2\},  j\in \{ 3,4 \} 
\label{eq:n_u0}
\eaa  
which does not depend on $i,j$. Then the effective action in Eq.~\ref{eq:seff_f} can be written as 
\baa
S_{eff,off} = & \frac{1}{2\pi N_M}\int \sum_\qq 
\sum_{i\xi,i\xi_2' \in S_{fill}, j\xi',j\xi_2 \in S_{emp}}u(\qq,i\omega)_{i\xi,j\xi'} u(-\qq,-i\omega)_{j\xi_2,i\xi_2'} \bigg\{ 
i\omega \delta_{\xi_2,\xi'}\delta_{\xi_2',\xi}
-H^{\nu_f=0}(\qq)_{\xi\xi',\xi_2'\xi_2}
\bigg\} d\omega 
\label{eq:eff_h_nu_0}
\eaa

 We then discuss the discrete symmetries. We first illustrate the symmetry properties of single-particle Green's function. We consider $C_{2z}T$, $C_{2x}$ and $C_{3z}$. From Eq.~\ref{eq:desc_sym}, Eq.~\ref{eq:sym_green} and Eq.~\ref{eq:anti_unitary}, we have
\baa 
&g_{aa'}^{\xi\xi',i}(\kk,\tau) = (g_{aa'}^{-\xi-\xi',i}(C_{2z}T\kk,\tau) )^* \quad,\quad a,a' \in \{c',c''\}
\nonumber \\
&g_{aa'}^{\xi\xi',i}(\kk,\tau) = g_{aa'}^{-\xi-\xi',i}(C_{2x}\kk,\tau)
\quad,\quad a,a' \in \{c',c''\}
\nonumber \\
&g_{c'c'}^{\xi\xi',i}(\kk,\tau) = g_{c'c'}^{\xi\xi',i}(C_{3z}\kk,\tau) e^{i\frac{2\pi}{3}(\xi'-\xi)} \quad ,\quad 
g_{c''c''}^{\xi\xi',i}(\kk,\tau) = g_{c''c''}^{\xi\xi',i}(C_{3z}\kk,\tau) \nonumber \\
&
g_{c'c''}^{\xi\xi',i}(\kk,\tau) = g_{c'c''}^{\xi\xi',i}(C_{3z}\kk,\tau) e^{i\frac{2\pi}{3}(-\xi)} \quad,\quad 
g_{c''c'}^{\xi\xi',i}(\kk,\tau) = g_{c''c'}^{\xi\xi',i}(C_{3z}\kk,\tau) e^{i\frac{2\pi}{3}(\xi')}  \, .
\label{eq:green_descrete} 
\eaa 
Thus using Eq.~\ref{eq:chi_n}, Eq.~\ref{eq:eff_h_nu_0} and Eq.~\ref{eq:green_descrete}, we find
\baa 
&H^{\nu_f=0}(\qq)_{\xi\xi',\xi_2'\xi_2} = 
(H^{\nu_f}(C_{2z}T\qq)_{-\xi-\xi',-\xi_2'-\xi_2})^* \nonumber \\
&H^{\nu_f=0}(\qq)_{\xi\xi',\xi_2\xi_2'} = 
H^{\nu_f=0}(C_{2x}\qq)_{-\xi-\xi',-\xi_2'-\xi_2}\nonumber \\
&H^{\nu_f=0}(\qq)_{\xi\xi',\xi_2\xi_2'} = H^{\nu_f=0}(C_{3z}\qq)_{\xi\xi',\xi_2'\xi_2} e^{i\frac{2\pi}{3}(\xi'-\xi_2+\xi_2'-\xi)} 
\label{eq:m_nu_eve_sym}
\eaa 
In addition, from Eq.~\ref{eq:f_hert} and Eq.~\ref{eq:n_u0} indicates $\hH^{\nu_f}(\qq)$ is a Hermitian matrix
\baa 
H^{\nu_f}(\qq)_{\xi\xi',\xi_2'\xi_2} =[ H^{\nu_f}(\qq)_{\xi_2\xi_2',\xi'\xi} ]^*
\label{eq:m_nu_eve_def}
\eaa

We next consider the long-wavelength limit (or small $\qq$ expansion) of $H^{\nu_f=0}(\qq)_{\xi\xi',\xi_2\xi_2'}$. We let 
\baa 
M^{\nu_f}(\qq)_{\xi\xi',\xi_2'\xi_2} \approx C_{0,\xi\xi',\xi_2'\xi_2} +\sum_{\mu=\{x,y\}}C_{\mu,\xi\xi',\xi_2'\xi_2}q_\mu + \sum_{\mu,\nu \in \{x,y\} }
C_{\mu\nu ,\xi\xi',\xi_2'\xi_2}q_{\mu}q_{\nu}
\label{eq:m_nu_eve_q_exp} 
\eaa 
where $C_{0,\xi\xi',\xi_2'\xi_2},C_{\mu,\xi\xi',\xi_2'\xi_2},C_{\mu\nu,\xi\xi',\xi_2'\xi_2}$ are the coefficients of the expansion. We next combine Eq.~\ref{eq:m_nu_eve_def}, Eq.~\ref{eq:m_nu_eve_def} and Eq.~\ref{eq:m_nu_eve_q_exp} and derive the symmetry constraints of the coefficients.
For $C_{0,\xi\xi',\xi_2'\xi_2}$, we find 
\baa 
&C_{0,\xi\xi',\xi_2'\xi_2} =(  C_{0,\xi_2'\xi_2,\xi\xi'} )^*\nonumber \\
&C_{0,\xi\xi',\xi_2\xi_2'} = C_{0,-\xi-\xi',-\xi_2'-\xi_2} \nonumber \\ 
&C_{0,\xi\xi',\xi_2'\xi_2} = C_{0,\xi\xi',\xi_2'\xi_2} e^{i\frac{2\pi}{3}(\xi'-\xi_2+\xi_2'-\xi)}
\nonumber \\
&C_{0,\xi\xi',\xi_2'\xi_2} = C_{0,\xi_2\xi_2',\xi'\xi} 
\eaa 
Thus, we can introduce three real numbers $C_{0,0},C_{0,1},C_{0,2}$:
\baa 
&C_{0,0} = C_{0,++,++} =C_{0,--,--}\quad,\quad 
C_{0,1} = C_{0,+-,+-} = C_{0,-+,-+} \nonumber \\
&C_{0,2} = C_{0,++,--} =C_{0,--,++}
\eaa
and all other components of $C_{0,\xi\xi',\xi_2'\xi_2}$ are zero.

As for $C_{\mu,\xi\xi',\xi_2\xi_2'}$, we find (using Eq.~\ref{eq:m_nu_eve_def}, Eq.~\ref{eq:m_nu_eve_def} and Eq.~\ref{eq:m_nu_eve_q_exp})
\baa 
&C_{\mu,\xi\xi',\xi_2'\xi_2} = C_{\mu,\xi_2'\xi_2,\xi\xi'} ^* \nonumber \\
&C_{x,\xi\xi',\xi_2'\xi_2} = C_{x,-\xi-\xi',-\xi_2'-\xi_2}  
\quad,\quad 
C_{y,\xi\xi',\xi_2'\xi_2} = - C_{y,-\xi-\xi',-\xi_2'-\xi_2}  \nonumber \\
&C_{x,\xi\xi',\xi_2'\xi_2} =(-\frac{1}{2} C_{x,\xi\xi',\xi_2' \xi_2}  +\frac{\sqrt{3}}{2} \nonumber \\ 
&C_{y,\xi\xi',\xi_2' \xi_2})e^{i\frac{2\pi}{3}(\xi'-\xi_2+\xi_2'-\xi)}
\quad,\quad 
C_{y,\xi\xi',\xi_2'\xi_2} = ( -\frac{\sqrt{3}}{2}C_{x,\xi\xi',\xi_2' \xi_2}  
-\frac{1}{2}C_{y,\xi\xi',\xi_2' \xi_2})e^{i\frac{2\pi}{3}(\xi'-\xi_2+\xi_2'-\xi)}  
\nonumber \\
&C_{\mu, \xi\xi',\xi_2'\xi_2} = -C_{\mu, \xi_2\xi_2',\xi'\xi} 
\eaa 
Then we can introduce a real number $C_{1}$
\baa 
C_{1,1} =& C_{x,++,-+} =C_{x,--,+-} =C_{x,-+,++}=C_{x,+-,--}
=-C_{x,+-,++} =- C_{x,-+,--} =-C_{x,++,+-}=-C_{x,--,-+} \nonumber \\
=& i C_{y,++,-+} = -iC_{y,--,+-}
=-iC_{y,-+,++} = iC_{y,+-,--} 
=-iC_{y,+-,++} = iC_{y,-+,--} 
= iC_{y,++,+-} = -iC_{y,--,-+} 
\eaa 

As for the $C_{\mu\nu,\xi\xi',\xi_2'\xi_2}$, Eq.~\ref{eq:m_nu_eve_def}, Eq.~\ref{eq:m_nu_eve_def} and Eq.~\ref{eq:m_nu_eve_q_exp} we have 
\baa 
&C_{\mu\nu,\xi\xi',\xi_2'\xi_2} = 
(C_{\mu\nu,\xi_2'\xi_2,\xi \xi'})^*  \nonumber \\
&C_{xx,\xi\xi',\xi_2'\xi_2} = 
C_{xx,-\xi-\xi',-\xi_2'-\xi_2}
\quad,\quad 
C_{yy,\xi\xi',\xi_2'\xi_2} = 
C_{yy,-\xi-\xi',-\xi_2'-\xi_2}
\nonumber \\
&C_{xy,\xi\xi',\xi_2'\xi_2} = 
-C_{xy,-\xi-\xi',-\xi_2'-\xi_2}
\quad,\quad 
C_{yx,\xi\xi',\xi_2'\xi_2} = -
C_{yx,-\xi-\xi',-\xi_2'-\xi_2}
\nonumber \\
&C_{xx,\xi\xi',\xi_2'\xi_2} =\frac{  C_{xx,\xi\xi',\xi_2'\xi_2} +3C_{yy,\xi\xi',\xi_2'\xi_2} -\sqrt{3}C_{xy,\xi\xi',\xi_2'\xi_2} -\sqrt{3}C_{yx,\xi\xi',\xi_2'\xi_2}  }{4} e^{i\frac{2\pi}{3} (\xi'-\xi_2+\xi_2'-\xi)} \nonumber \\
&C_{yy,\xi\xi',\xi_2'\xi_2} =\frac{ 3 C_{xx,\xi\xi',\xi_2'\xi_2} +C_{yy,\xi\xi',\xi_2'\xi_2} +\sqrt{3}C_{xy,\xi\xi',\xi_2'\xi_2} +\sqrt{3}C_{yx,\xi\xi',\xi_2'\xi_2}  }{4} e^{i\frac{2\pi}{3} (\xi'-\xi_2+\xi_2'-\xi)} \nonumber \\
&C_{xy,\xi\xi',\xi_2'\xi_2} =\frac{ \sqrt{3} C_{xx,\xi\xi',\xi_2'\xi_2} -\sqrt{3}C_{yy,\xi\xi',\xi_2'\xi_2} +C_{xy,\xi\xi',\xi_2'\xi_2} -3C_{yx,\xi\xi',\xi_2'\xi_2}  }{4} e^{i\frac{2\pi}{3} (\xi'-\xi_2+\xi_2'-\xi)} \nonumber \\
&C_{yx,\xi\xi',\xi_2'\xi_2} =\frac{ \sqrt{3} C_{xx,\xi\xi',\xi_2'\xi_2} -\sqrt{3}C_{yy,\xi\xi',\xi_2'\xi_2} -3C_{xy,\xi\xi',\xi_2'\xi_2} +C_{yx,\xi\xi',\xi_2'\xi_2}  }{4} e^{i\frac{2\pi}{3} (\xi'-\xi_2+\xi_2'-\xi)}
\eaa 
We can then introduce the following real numbers \baa 
&C_{2,0} = C_{xx,++,++}
=C_{yy,++,++}=
C_{xx,--,--}
=C_{yy,--,--} \nonumber \\
&C_{2,1} = C_{xx,+-,+-}
=C_{yy,+-,+-}=
C_{xx,-+,-+}
=C_{yy,-+,-+} \nonumber \\
&C_{2,2} = C_{xx,++,--}
=C_{yy,++,--}=
C_{xx,--,++}
=C_{yy,--,++} \nonumber \\
&C_{2,3} = C_{xx,+-,-+} =-C_{yy,+-,-+} =-iC_{xy,+-,-+}=-iC_{yx,+-,-+} 
= C_{xx,-+,+-}=-C_{yy,-+,+-} \nonumber \\
&=iC_{xy,+-,-+}=iC_{yx,-+,+-} 
\eaa 
and all other components of $C_{\mu\nu,\xi\xi',\xi_2\xi_2'}$ vanishes.

In summary, we have the following $q$ expansion of $H^{\nu_f}(\qq)$
\baa 
H^{\nu_f}(\qq) = 
\begin{bmatrix}
C_{0,0} + C_{2,0} |\qq|^2
&C_{0,2} +C_{2,2}|\qq|^2 & C_{1,1}(-q_x -iq_y) & C_{1,1}(q_x -iq_y) 
\\
C_{0,2} +C_{2,2}|\qq|^2 & C_{0,0} + C_{2,0} |\qq|^2 & C_{1,1}(q_x+iq_y) & C_{1,1}(-q_x+iq_y) \\ 
C_{1,1}(-q_x+iq_y)& C_{1,1}(q_x-iq_y) & C_{0,1}+C_{2,1}|\qq|^2 &C_{2,3}(q_x^2 -q_y^2 -2i q_x q_y) \\
C_{1,1}(q_x+iq_y)& C_{1,1}(-q_x-iq_y) & C_{2,3}(q_x^2 -q_y^2 +2i q_x q_y)& C_{0,1} +C_{2,1}|\qq|^2 
\end{bmatrix}
\label{eq:mat_struct_nu_even}
\eaa 
Finally, we utilize the symmetry constraints introduced in Eq.~\ref{eq:symmetry_const}. We take $i,j=(1,3)$ (or equivalently $(i,j)=(1,4),(2,4),(2,3)$) and $\xi_2=\xi_2'=++$ in Eq.~\ref{eq:symmetry_const} and find 
\baa  
&H^{\nu_f}(\qq=0)_{++,++} +H^{\nu_f}(\qq=0)_{++,--}  =0\nonumber  \\ \Rightarrow &C_{0,0}+C_{0,2} = 0
\label{eq:h_nu_0_sym_const}
\eaa  
Then we have $C_{0,0}=-C_{0,2}$ and 
\baa 
H^{\nu_f}(\qq) = 
\begin{bmatrix}
C_{0,0} + C_{2,0} |\qq|^2
&-C_{0,0} +C_{2,2}|\qq|^2 & C_{1,1}(-q_x -iq_y) & C_{1,1}(q_x -iq_y) 
\\
-C_{0,0} +C_{2,2}|\qq|^2 & C_{0,0} + C_{2,0} |\qq|^2 & C_{1,1}(q_x+iq_y) & C_{1,1}(-q_x+iq_y) \\ 
C_{1,1}(-q_x+iq_y)& C_{1,1}(q_x-iq_y) & C_{0,1}+C_{2,1}|\qq|^2 &C_{2,3}(q_x^2 -q_y^2 -2i q_x q_y) \\
C_{1,1}(q_x+iq_y)& C_{1,1}(-q_x-iq_y) & C_{2,3}(q_x^2 -q_y^2 +2i q_x q_y)& C_{0,1} +C_{2,1}|\qq|^2 
\end{bmatrix}
\label{eq:mat_struct_nu_even}
\eaa 
where we note that $C_{0,0}=-C_{0,2}$ leads to a Goldstone mode at $\qq=0$.

We also provide the expression of the parameters
\baa 
&C_{0,0} =[ H^{\nu_f}(\qq=0)]_{++,++} \nonumber \\ 
&C_{2,0} = [ (\partial_{q_x})^2H^{\nu_f}(\qq=0)]_{++,++} \nonumber \\
&C_{2,2} = [ (\partial_{q_x})^2H^{\nu_f}(\qq=0)]_{++,--} \nonumber \\
&C_{1,1} = -[ (\partial_{q_x})^2H^{\nu_f}(\qq=0)]_{++,+-} \nonumber \\
&C_{0,1} = [H^{\nu_f}(\qq=0)]_{+-,+-} \nonumber \\
&C_{2,1} = [\partial_{q_x}^2H^{\nu_f}(\qq=0)]_{+-,+-} \nonumber \\
&C_{2,3} = [\partial_{q_x}^2H^{\nu_f}(\qq=0)]_{+-,-+}
\label{eq:old_lag_para} \, .
\eaa 
In practice, we first calculate $N_{i\xi}$, $\chi^A$ and $H^{\nu_f}(\qq)$ (Eq.~\ref{eq:chi_n}, Eq.~\ref{eq:m_nu_eve_q_exp}) and then find the values of $C_{0,0},C_{2,0},C_{2,2},C_{1,1},C_{0,1},C_{2,1},C_{2,3}$ using Eq.~\ref{eq:old_lag_para}. The values of parameters are discussed in Sec.~\ref{sec:lag_long_nu_even}.


We now discuss the number of Goldstone modes. The original model has (flat) $U(4)$ symmetry with rank (number of independent generators) $16$. At $\nu_f=0$, the symmetry group of the ground state (Eq.~\ref{eq:def_psi_0_r}) is $U(2)\times U(2)$ with rank $4+4=8$. Then the number of Goldstone modes is $(16-8)/2 = 4$, where $1/2$ comes from the fact that each Goldstone mode is a complex boson~\cite{tbgv}. We find $M^{\nu_f=0}_{\xi\xi',\xi_2\xi_2'}(\qq)$ has one zero eigenvalues at $\qq=0$, which corresponds to the Goldstone mode. However, since we have four ways to pick $i,j$ indices (of the bosnoic fields $u_{i\xi,j\xi'}$ for the given ground state in Eq.~\ref{eq:def_psi_0_r}), with $(i,j)=(1,3),(2,3),(1,4),(2,4)$, we have one Goldstone mode for each $i,j$ choices and $4$ Goldstone modes in total. 

In the spectrum of spin excitation, we also observe a quasi-degeneracy between gapped modes at $\Gamma_M$ at $\nu=0$. At $\qq=0$, we note (Eq.~\ref{eq:mat_struct_nu_even})
\baa 
H^{\nu_f}(\qq=0) = \begin{bmatrix}
C_{0,0} & -C_{0,0} & 0 & 0 \\
-C_{0,0} & C_{0,0} & 0 & 0 \\
0 & 0 & C_{0,1} & 0 \\
0 & 0& 0 & C_{0,1}
\end{bmatrix}
\eaa 
Four eigenstates are $0, 2C_{0,0}, C_{0,1},C_{0,1}$. Then there is one Goldstone zero mode and three gapped modes. We next estimate the energy difference between the gapped modes
\baa 
\Delta E =& C_{0,1}-2C_{0,0} = \frac{1}{4} \bigg[ 
\chi_{\xi \xi,\xi \xi}(\qq=0,i\omega=0,i,j) - \chi_{\xi\xi,-\xi-\xi}(\qq=0,i\omega=0,i,j) - \chi_{\xi -\xi,\xi -\xi}(\qq=0,i\omega=0,i,j)\bigg] \nonumber  \\
& i\in \{1,2\}, j\in \{3,4\} 
\eaa 
Approximately, we take the non-interacting single-particle Green's functions (given in Sec.~\ref{sec:green}) to calculate $\Delta E$. From Eq.~\ref{eq:chi_n}, we have From Eq.~\ref{eq:chi_A_formula}, we find 
\baa 
\Delta E \approx &\frac{1}{4} \int \sum_{\kk}\frac{1}{N_M}
\bigg\{ 
\frac{2\gamma^4}{D_{\nu_c,\nu_f}^2} \bigg[2 g_{c'c'}^{\xi\xi,j}(\kk,\tau)g_{c'c'}^{\xi\xi,i}(\kk,-\tau)
-g_{c'c'}^{\xi\xi,j}(\kk,\tau)g_{c'c'}^{-\xi-\xi,i}(\kk,-\tau)-g_{c'c'}^{\xi\xi,i}(\kk,\tau)g_{c'c'}^{-\xi-\xi,j}(\kk,-\tau)
\bigg] \nonumber \\
&+2J^2\bigg[ 2g_{c''c''}^{\xi \xi,j}(\kk,\tau)g_{c''c''}^{\xi\xi ,i}(\kk,-\tau)
\bigg] 
-\frac{2J\gamma^2}{D_{\nu_c,\nu_f}} \bigg[2g_{c''c'}^{\xi\xi,j}(\kk,\tau)g^{\xi\xi,i}_{c'c''}(\kk,-\tau)\bigg] \nonumber \\
&+\frac{2\gamma^2 (v_\star^\prime)^2}{D_{\nu_f,\nu_c}^2} 
|\kk|^2 \bigg[ -2g_{c'c'}^{\xi\xi,j}(\kk,\tau)g^{\xi\xi,i}_{c'c'}(\kk,-\tau)
-2g_{c'c'}^{\xi\xi,i}(\kk,\tau)g^{\xi\xi,j}_{c'c'}(\kk,-\tau)
\bigg] 
\bigg\} d\tau 
\eaa 
Using non-interacting single-particle Green's function (Eq.~\ref{eq:single_green_order}, Eq.~\ref
{eq:green_def_2}, Eq.~\ref{eq:g0}, Eq.~\ref{eq:g2}), we find 
\baa 
\Delta E \approx& \frac{1}{4}\sum_{\kk}\frac{1}{N_M} 
\bigg[ 
4J^2  \frac{2}{4}\frac{1}{2 |v_\star| |\kk| } 
+\frac{4J\gamma^2e^{- |\kk|^2 \lambda^2 }}{D_{\nu_c,\nu_f}}\frac{2}{4}\frac{1}{2|v_\star|\kk|}
-\frac{ 8\gamma^2 (v_\star^\prime)^2 e^{-2 |\kk|^2 \lambda^2 }}{D_{\nu_c,\nu_f}^2} e^{- 2|\kk|^2 \lambda^2 }
|\kk|^2 \frac{2}{4} \frac{1}{2|v_\star |\kk|}
\bigg] \\
\approx& 
\frac{1}{4A_{MBZ}} \int_0^{2\pi} \int_0^{\Lambda_c} \bigg( 
\frac{J^2}{|v_\star|} +\frac{J \gamma^2e^{-k^2\lambda^2}}{D_{\nu_c,\nu_f}|v_\star|}
-\frac{2\gamma^2 (v_\star^\prime)^2 e^{- 2k^2 \lambda^2 }  k^2}{D_{\nu_c,\nu_f}^2|v_\star|}\bigg) 
dk d\theta  \nonumber \\
\approx & \frac{\pi}{2A_{MBZ}|v_\star|} 
\bigg[ 
J^2 \Lambda_c + \frac{J\gamma^2 }{D_{\nu_{c,\nu_f}} }\frac{\sqrt{\pi}Erf[\lambda\Lambda_c]}{2\lambda} 
-\frac{2\gamma^2 (v_\star^\prime)^2 }{D_{\nu_c,\nu_f}^2}
\frac{1}{16\lambda^3}
\bigg( 
-4\lambda \Lambda_c e^{-2\lambda^2\Lambda_c^2} +\sqrt{2\pi} Erf[\sqrt{2}\Lambda_c\lambda]
\bigg) 
\bigg] 
\eaa 
where $Erf[x]$ is the error function. 
Approximately, we take the momentum cutoff $\Lambda_c=1/\lambda$ and find the energy difference between two gapped modes is
\baa 
\Delta E = \frac{ \pi}{2A_{MBZ}|v_\star|\lambda} \bigg[ J^2+\frac{0.747J\gamma^2}{D_{\nu_c,\nu_f}} -
\frac{0.231\gamma^2 (v_\star ^\prime)^2 }{D_{\nu_c,\nu_f}^2 \lambda^2 }\bigg] 
\eaa 
where $\lambda = 0.3375a_M$ is the damping factor of the hybridization term. 
The degenerate condition $\Delta E=0$ is
\baa 
\alpha = \frac{\gamma^2}{JD_{\nu_c,\nu_f}}\bigg[ -0.374  + \sqrt{ 0.140 + 0.231 \frac{ (v_\star^\prime) ^2}{\gamma^2 \lambda^2} }\bigg] =1 
\eaa 
Using the realistic values of each parameter, we find $\alpha=1.074\approx 1$.

\subsection{$\nu_f=-2$}
At $\nu_f= -2$, 
we have $U(1) \times U(3)$ symmetry. $U(1)$ acts on the flavor $i=1$ and $U(3)$ acts on the flavor $i=2,3,4$. Thus Green's function in flavor $i=2,3,4$ are equivalent 
\baa 
g_{aa'}^{\xi\xi',i}(\kk,\tau) = g_{aa'}^{\xi\xi',j}(\kk,\tau) \quad,\quad a,a' \in \{c',c''\}\quad,\quad i,j\in \{2,3,4\} 
\label{eq:u3u1_g}
\eaa 
Consequently, from Eq.~\ref{eq:chi_n} and Eq.~\ref{eq:u3u1_g}, we have
\baa 
&\chi_{\xi\xi',\xi_2'\xi_2}^A(\qq,i\omega=0,i=1,j) = \chi_{\xi\xi',\xi_2'\xi_2}^A(\qq,i\omega=0,i=1,j') \quad,\quad  j,j'\in \{2,3,4\}\nonumber \\
&N_{1\xi}-N_{j\xi'} = N_{1\xi}-N_{j'\xi'} \quad,\quad j,j'\in \{2,3,4\}
\label{eq:chi_n_nu_2}
\eaa 
We can thus define (Eq.~\ref{eq:def_F})
\baa 
&H^{\nu_f=-2}(\qq)_{\xi\xi',\xi_2'\xi_2} = F_{\xi\xi',\xi_2'\xi_2}^{ij}(\qq) \nonumber \\
& i=1,  j\in \{2, 3,4 \} 
\eaa 
which utilize the symmetry properties of Eq.~\ref{eq:chi_n_nu_2}. 
Then the effective action in Eq.~\ref{eq:seff_off} can be written as 
\baa
S_{eff,off} = & \frac{1}{2\pi N_M}\int \sum_\qq 
\sum_{i\xi,i\xi_2' \in S_{fill}, j\xi',j\xi_2 \in S_{emp}}u(\qq,i\omega)_{i\xi,j\xi'} u(-\qq,-i\omega)_{j\xi_2,i\xi_2'} \bigg\{ 
i\omega \delta_{\xi_2,\xi'}\delta_{\xi_2',\xi}
-H^{\nu_f=-2}(\qq)_{\xi\xi',\xi_2'\xi_2}
\bigg\} d\omega 
\label{eq:eff_h_nu_2}
\eaa

We next consider the discrete symmetry. We have $C_{3z},C_{2z}T,C_{2x}$. Using Eq.~\ref{eq:green_descrete}, we find
\baa 
&H^{\nu_f=-2}(\qq)_{\xi\xi',\xi_2'\xi_2} = 
(H^{\nu_f}(C_{2z}T\qq)_{-\xi-\xi',-\xi_2'-\xi_2})^* \nonumber \\
&H^{\nu_f=-2}(\qq)_{\xi\xi',\xi_2\xi_2'} = 
H^{\nu_f=-2}(C_{2x}\qq)_{-\xi-\xi',-\xi_2'-\xi_2}\nonumber \\
&H^{\nu_f=-2}(\qq)_{\xi\xi',\xi_2\xi_2'} = H^{\nu_f=-2}(C_{3z}\qq)_{\xi\xi',\xi_2'\xi_2} e^{i\frac{2\pi}{3}(\xi'-\xi_2+\xi_2'-\xi)} 
\label{eq:m_nu_eve_2_sym}
\eaa 
In addition, the definition of $H^{\nu_f=-2}(\qq)_{\xi\xi',\xi_2'\xi_2}$ (Eq.~\ref{eq:n_u0}) and Eq.~\ref{eq:f_hert}, we have
\baa 
H^{\nu_f=-2}(\qq)_{\xi\xi',\xi_2'\xi_2} = [H^{\nu_f=-2}(\qq)_{\xi_2\xi_2',\xi'\xi} ]^*
\label{eq:m_nu_eve_def_2}
\eaa 

Finally, we utilize the symmetry constraints in Eq.~\ref{eq:symmetry_const}. 
We take $i,j=(1,2)$ (or equivalently $(i,j)=(1,2),(1,4)$) and $\xi_2=\xi_2'=++$ in Eq.~\ref{eq:symmetry_const} and find 
\baa 
H^{\nu_f=-2}(\qq=0)_{++,++} +H^{\nu_f=-2}(\qq=0)_{++,--} =0
\label{eq:sym_const_nu_2}
\eaa 

Combining Eq.~\ref{eq:m_nu_eve_sym}, Eq.~\ref{eq:m_nu_eve_2_sym} and Eq.~\ref{eq:sym_const_nu_2}), we conclude $\hH^{\nu_f=-2}(\qq)$ follows the same constraint under symmetry as $\hH^{\nu_f=0}(\qq)$ (Eq.~\ref{eq:m_nu_eve_sym}, Eq.~\ref{eq:m_nu_eve_def}, Eq.~\ref{eq:h_nu_0_sym_const}). 
Therefore, the long-wavelength behavior of $\hH^{\nu_f=-2}(\qq)$ also takes the form of Eq.~\ref{eq:mat_struct_nu_even}.

We now discuss the number of Goldstone modes. At $\nu_f=-2$, the symmetry group of the ground state (Eq.~\ref{eq:def_psi_0_r}) is $U(1)\times U(3)$ with rank $1+9=10$. 
The symmetry of the original Hamiltonian at $M=0$ has flat $U(4)$ symmetry with rank $16$. 
The number of Goldstone modes is $(16-10)/2=3$~\cite{tbgv}. From Eq.~\ref{eq:mat_struct_nu_even}, $H^{\nu_f=-2}_{\xi\xi',\xi_2\xi_2'}(\qq)$ has one zero eigenvalues at $\qq=0$, which corresponds to the Goldstone mode. Since we have three ways to pick $i,j$ indices (of the bosonic fields $u_{i\xi,j\xi'}$), with $(i,j)=(1,2),(1,3),(1,4)$ (for the given ground state in Eq.~\ref{eq:def_psi_0_r}), we have one Goldstone mode for each $i,j$ choices and $3$ Goldstone modes in total.

We also comment that, due to the fact that values of $\gamma^2/D_{\nu_c,\nu_f}$ at $\nu_f=2$ is much larger than its value at $\nu_f=0,-1$. The Kondo interaction is much stronger at $\nu_f=-2$, which leads to a larger bandwidth of the excitation spectrum.

\subsection{Lagrangian at $\nu_f=0,-2$ in the long-wavelength limit}
\label{sec:lag_long_nu_even}
In this section, we provide the Lagrangian of effective theory at $\nu_f=0,-2$ in a more compact form. From Eq.~\ref{eq:eff_h_nu_0}, Eq~\ref{eq:eff_h_nu_2} and Eq.~\ref{eq:mat_struct_nu_even}, we have the following Lagrangian density in the long-wavelength limit at $\nu_f=0,-2$
\baa  
L = 
\sum_{i\xi,i\xi_2' \in S_{fill}, j\xi',j\xi_2 \in S_{emp}}u(\qq,i\omega)_{i\xi,j\xi'} u(-\qq,-i\omega)_{j\xi_2,i\xi_2'} \bigg\{ 
i\omega \delta_{\xi_2,\xi'}\delta_{\xi_2',\xi}
-H^{\nu_f}(\qq)_{\xi\xi',\xi_2'\xi_2} \bigg\} 
\label{eq:lag}
\eaa
where we introduce a conventional parameterization of $H^{\nu_f}(\qq)_{\xi\xi',\xi_2'\xi_2}$ 
\baa  
H^{\nu_f}(\qq)_{\xi\xi',\xi_2'\xi_2} = 
\begin{bmatrix}
    \frac{\Delta_1}{2} + (\frac{1}{4m_0} +\frac{1}{4m_1})|\qq|^2& -\frac{\Delta_1}{2} +(\frac{1}{4m_0} - \frac{1}{4m_1})|\qq|^2& \frac{V}{\sqrt{2}} q_+ & -\frac{V}{\sqrt{2}} q_-  \\
    \frac{\Delta_1}{2} + (\frac{1}{4m_0} - \frac{1}{4m_1})|\qq|^2& \frac{\Delta_1}{2} + (\frac{1}{4m_0} - \frac{1}{4m_1})|\qq|^2 &- \frac{V}{\sqrt{2}} q_+  &  \frac{V}{\sqrt{2}} q_-\\
    \frac{V}{\sqrt{2}} q_- & -\frac{V}{\sqrt{2}}q_- & \Delta_2 + \frac{|\qq|^2}{2m_2} & \frac{q_-^2}{2m_3} \\ 
    -\frac{V}{\sqrt{2}}q_+ & \frac{V}{\sqrt{2}}q_+  & \frac{q_+^2}{2m_3}  &   \Delta_2 + \frac{|\qq|^2}{2m_2} 
\end{bmatrix}_{\xi\xi',\xi_2'\xi_2}
\eaa 
where the row and column indices are $(++,--,+-,-+)$, $q_\pm = q_x \pm i q_y$. The parameters are 
\baa  
&\Delta_1 = 2C_{0,0} ,\quad \Delta_2 = C_{0,1} \nonumber \\
&m_0 = \frac{1}{2(C_{2,0}+C_{2,2})}, \quad  m_1 = \frac{1}{2(C_{2,0}-C_{2,2})}, \quad  m_2 =\frac{1}{2C_{2,1}}, \quad m_3 = \frac{1}{2C_{2,3}}\nonumber \\
& V=-\sqrt{2}C_{1,1}
\label{eq:new_lag_para}
\eaa  

We then take the following new basis
\baa 
&u_{j,i} (\qq,i\omega)= (u_{(j+,i+)} (\qq,\omega)+u_{(j-,i-)} (\qq,i\omega))/\sqrt 2 \nonumber\\ 
&U_{j,i}^T (\qq,i\omega)=((u_{(j+,i+)} (\qq,i\omega)-u_{(j-,i-)}) (\qq,i\omega)/\sqrt 2, u_{(j+,i-)} (\qq,i\omega),u_{(j-,i+)} (\qq,i\omega)) 
\eaa 
The Lagrangian (Eq.~\ref{eq:lag}) can then be written as 
\begin{eqnarray}
&& L = L_{Goldstone} + L_{gapped}, \nonumber \\
&& L_{Goldstone} = \sum_{\substack{j \xi \in {S}_{fill}\\ i \xi\in {S}_{emp}}}u_{j,i}^\dag({\bf q},i\omega)\Big(i\omega  - q^2/2m_0\Big)u_{j,i}({\bf q},i\omega), \nonumber\\
&&L_{gapped} = \sum_{\substack{j \xi \in {S}_{fill}\\ i \xi\in {S}_{emp}}}U_{j,i}^\dag({\bf q},i\omega)[i\omega \hat I -  H(\qq)]U_{j,i}({\bf q},i\omega),
\end{eqnarray}
where 
\begin{eqnarray}
H(\qq)
= \left(
\begin{array}{ccc}
\frac{q_+q_- }{2m_1}+\Delta_1 & Vq_+& - Vq_- \\
 V q_- &\frac{q_+q_-}{2m_2} +\Delta_2 & \frac{q_-^2}{2m_3}\\
- V q_+ & \frac{q_+^2}{2m_3} & \frac{q_+q_-}{2m_2} +\Delta_2 
\end{array}
\right) \nonumber \\
\label{eq:lag_nu_even}
\end{eqnarray}
and $\hat I$ is a $4\times 4$ identity matrix. After transforming from Matsubara frequency to real frequency ($i\omega \rightarrow \omega$), we reach the same formula given in the main text (Eq.[11] and Eq.[12]).

We also provide the dispersion of gapped modes. By diagonalizing $H(\qq)$ in Eq.~\ref{eq:lag_nu_even}, we find the dispersion of three gapped modes are
\baa 
&E^{gap,1}_\qq = \Delta_2 + (\frac{1}{2m_2} +\frac{1}{2m_3})|\qq|^2 \nonumber \\ 
&E^{gap,2}_\qq = \frac{\Delta_1+\Delta_2}{2}+\frac{|\qq|^2}{4}\bigg[\frac{1}{m_1}+\frac{1}{m_2}-\frac{1}{m_3}\bigg] +\sqrt{2V^2|\qq|^2  +\bigg[\frac{\Delta_1-\Delta_2}{2} + \frac{|\qq|^2}{4}
\bigg[\frac{1}{m_1}-\frac{1}{m_2} +\frac{1}{m_3} \bigg]
\bigg]^2  } \nonumber \\ 
&E^{gap,3}_\qq - \frac{\Delta_1+\Delta_2}{2}+\frac{|\qq|^2}{4}\bigg[\frac{1}{m_1}+\frac{1}{m_2}-\frac{1}{m_3}\bigg] -\sqrt{2V^2|\qq|^2  +\bigg[\frac{\Delta_1-\Delta_2}{2} + \frac{|\qq|^2}{4}
\bigg[\frac{1}{m_1}-\frac{1}{m_2} +\frac{1}{m_3} \bigg]
\bigg]^2  }
\eaa 

We provide the numerical values of parameters (derived from Eq.~\ref{eq:old_lag_para} and Eq.~\ref{eq:new_lag_para}) at $\nu_f=0,-2$ in Tab.~\ref{tab:lag_nu_even}. In Fig.~\ref{fig:comp_even}, we also compare the dispersion from the effective model (Eq.~\ref{eq:lag_nu_even}) with the dispersion from directly evaluating Eq.~\ref{eq:seff_f}. We note that the effective model~\ref{eq:lag_nu_even} correctly predicts the long wavelength (small $\kk$) behaviors but failed to capture the large momentum behaviors such as the roton modes. To recover the roton mode, we need to keep high-order contribution ($\qq^3$ term).

\begin{table}[]
    \centering
    \begin{tabular}{c|c|c|c|c|c|c |c }
        Parameter  &  $\Delta_1$ & $\Delta_2$ & $a_M^2/m_0$ & $a_M^2/m_1$ &$a_M^2/m_2$ & $a_M^2/m_3$ & $V$  \\ 
        \hline 
    Value at $\nu_f=0$ (meV) &6.3 & 6.7 & 6.0 & 0.9 & 0.7 & 1.0 & 2.5
    \\ \hline 
    Value at $\nu_f=-2$ (meV) &22.8 & 16.4 & 11.8 & -3.1 & 0.9 & 3.3 & 7.3 \\ \hline 
    \end{tabular}
    \caption{Numeircal values of parameters in Eq.~\ref{eq:lag_nu_even} at $\nu_f=0,-2$.}
    \label{tab:lag_nu_even} 
\end{table}

\begin{figure}
    \centering
    \includegraphics[width=0.6\textwidth]{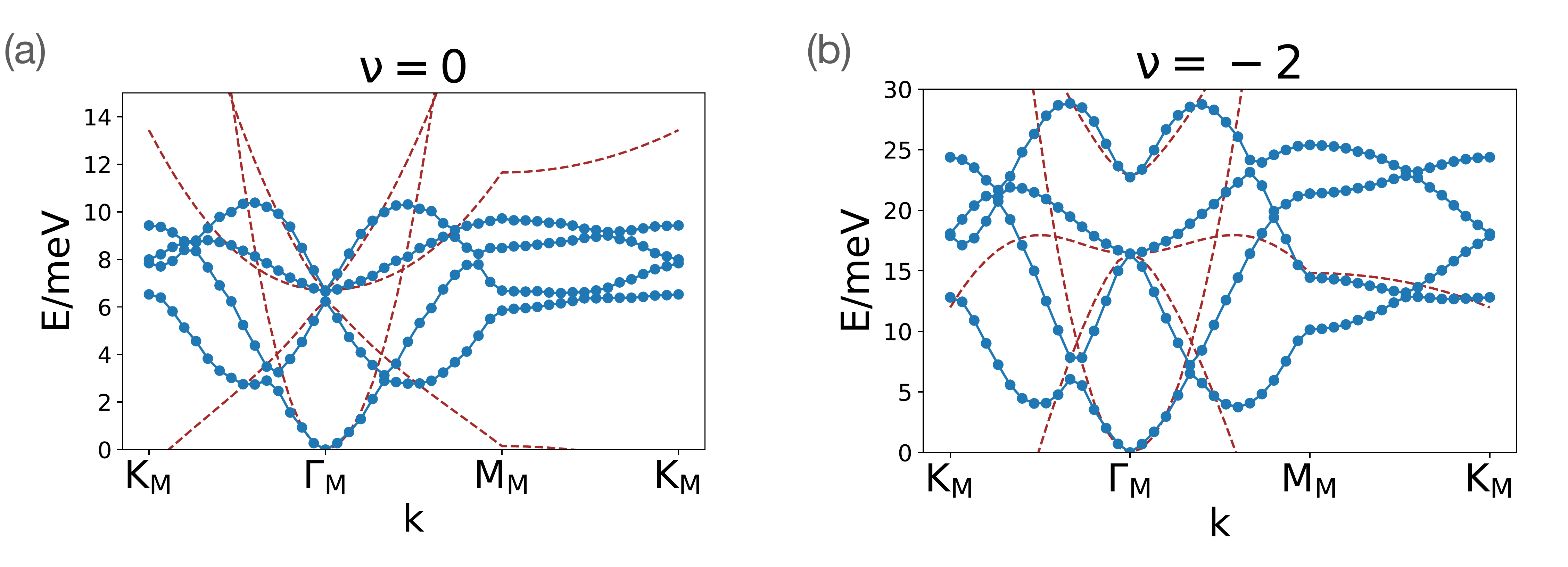}
    \caption{Excitation spectrum from the effective model in Eq.~\ref{eq:lag_nu_even} (brown) and from numerical evaluation of Eq.~\ref{eq:seff_f} (blue). }
    \label{fig:comp_even}
\end{figure}

\subsection{$\nu_f=-1$} 
We now consider $\nu_f=-1$. 
We first discuss the symmetry of $\nu_f=-1$ ground state. 
It turns out, the remaining discrete symmetry of the ground state (Eq.~\ref{eq:def_psi_0_r}) at $\nu_f=-1$ is $C_{3z}$, because one of the valley-spin flavors at $\nu_f=-1$ is filled with only one electron. In addition, the original flat $U(4)$ symmetry is broken to $U(1)\times U(1) \times U(2)$ symmetry. 

We next discuss the fluctuations at $\nu_f=-1$. From Eq.~\ref{eq:u_state} and Eq.~\ref{eq:ru_def}, we find 
\baa  
|u(\RR)\rangle_{\RR}  = \prod_\RR\exp\bigg( 
-i \sum_{ij,\xi\xi'}u_{i\xi,j\xi'}(\RR)\psi_{\RR,i}^{f,\xi,\dag} \psi_{\RR,j}^{f,\xi'} \bigg) |\psi_0\rangle \approx \bigg[ 
1-i \sum_{ij,\xi\xi'}u_{i\xi,j\xi'}(\RR)\psi_{f,\RR,i}^{\xi,\dag} \psi_{\RR,j}^{f,\xi'}
\bigg] |\psi_0\rangle 
\eaa  
Therefore at the leading-order, $u_{i\xi,j\xi'}(\RR)$ ($i\xi \ne j\xi'$) describe the procedure of moving one $f$-electron from $j\xi'$ flavor at site $\RR$ to $i\xi$ flavor at the same site. This allows us to classify the $u$ fields into four groups. Taking the ground state in Eq.~\ref{eq:def_psi_0_r} at $\nu_f=-1$, four groups of $u_{i\xi,j\xi'}(\RR)$ are 
 (1) full-empty sector with $i = 3,4, j=1$ or $i= 1, j=3,4$; (2) half-empty sector with $i = 2, j=3,4$ or $i = 3,4, j=2$ ; (3) full-half sector with $i=1,j=2$ or $j=2,i=1$; (4) half-half sector with $i=2,j=2$. Group (1) describes the fluctuations between fully filled and empty spin-valley flavors.
 Group (2) describes the fluctuations between half-filled and empty spin-valley flavors. Group (3) describes the fluctuations between fully filled and empty spin-valley flavors. Group (4) describes the fluctuations between half-filled and half-filled spin-valley flavors. 
 Correspondingly, we can separate the $S_{eff,off}$ into four parts:
\baa 
S_{eff,off} = S_{eff,off,fe}+ S_{eff,off,he} +S_{eff,off,fh} + 
S_{eff,off,hh} 
\eaa 
where $ S_{eff,off,fe}, S_{eff,off,he} ,S_{eff,off,fh} ,
S_{eff,off,hh} $ describe the effective theory of full-empty sector, half-empty sector, full-half sector, and half-half sector respectively.

\subsubsection{Full-empty sector}
For the full-empty sector, we have action (Eq.~\ref{eq:seff_f})
\baa 
S_{eff,off,fe} =  & \frac{1}{2\pi N_M}\int \sum_\qq \sum_{i \in \{1\}, j\in \{3,4\} }
\sum_{\xi,\xi_2' ,\xi',\xi_2}u(\qq,i\omega)_{i\xi,j\xi'} u(-\qq,-i\omega)_{j\xi_2,i\xi_2'} [ 
i\omega \delta_{\xi_2,\xi'}\delta_{\xi_2',\xi}
- F^{ij}_{\xi\xi',\xi_2'\xi_2}(\qq) ]d\omega 
\eaa 
where 
 $i=1, j \in \{3,4\}$. $i=1$ flavor is filled with two $f$ electrons and $j = 3,4$ are empty (we take the ground sate in Eq.~\ref{eq:def_psi_0_r}).
 Utilizing the $U(1) \times U(1) \times U(2) $, symmetry, we have 
\baa 
&\chi_{\xi\xi',\xi_2'\xi_2}^A(\qq,i\omega=0,i=1,j) = \chi_{\xi\xi',\xi_2'\xi_2}^A(\qq,i\omega=0,i=1,j') \quad,\quad  j,j'\in \{3,4\}\nonumber \\
&N_{1\xi}-N_{j\xi'} = N_{1\xi}-N_{j'\xi'} \quad,\quad j,j'\in \{3,4\}
\eaa 
Therefore $F^{ij}_{\xi\xi',\xi_2'\xi_2' }$ take the same value for $i=1, j\in\{3,4\}$. 
We thus define 
\baa 
&H^{\nu_f=-1,fe}(\qq)_{\xi\xi',\xi_2'\xi_2} =F^{ij}_{\xi\xi',\xi_2'\xi_2' } (\qq)
\quad,\quad i=1,  j\in \{ 3,4 \} 
\label{eq:h1_def}
\eaa  
The effective theory can be written as 
\baa 
S_{eff,off,fe} =  & \frac{1}{2\pi N_M}\int \sum_\qq \sum_{i \in \{1\}, j\in \{3,4\} }
\sum_{\xi,\xi_2' ,\xi',\xi_2}u(\qq,i\omega)_{i\xi,j\xi'} u(-\qq,-i\omega)_{j\xi_2,i\xi_2'} \bigg\{ 
i\omega \delta_{\xi_2,\xi'}\delta_{\xi_2',\xi}
-H^{\nu_f=-1,fe}(\qq)_{\xi\xi',\xi_2'\xi_2}
\bigg\} d\omega 
\label{eq:eff_s_fe}
\eaa 
The excitation spectrum can be derived by finding the eigenvalues of $H^{\nu_f=-1,fe}(\qq)$.

Here we comment that $\hH_{c,order}$ has a larger symmetry group than the symmetry group of the ground state. Even though, the ground state does not have $C_{2z}T$ and $C_{2x}$ symmetries, $\hH_{c,order}$ has a special type of $C_{2z}T$ and $C_{2x}$ symmetry, which we call $(C_{2z}T)'$ and $(C_{2x})'$. For the ground state in Eq.~\ref{eq:def_psi_0_r}, $(C_{2z}T)'$ ($(C_{2x})'$) are defined as $C_{2z}T$($C_{2x}$) transformations that only act on the valley-spin flavors $i=1,3,4$ (Note that, $C_{2z}T$ and $C_{2x}$ will not flip valley and spin indices). 
Then $\hH_{c,order}$ satisfies $(C_{2z}T)'$ and $(C_{2x})'$ symmetries, so is the $H^{\nu_f=-1,fe}(\qq)_{\xi\xi',\xi_2'\xi_2} $. We comment that $(C_{2z}T)',(C_{2x})'$ are not real symmetries of the system.

Consequently, we have 
\baa 
&g_{aa'}^{\xi\xi',i}(\kk,\tau,) = (g_{aa'}^{-\xi-\xi',i}(C_{2z}T\kk,\tau) )^* \quad,\quad a,a' \in \{c',c''\}
\nonumber \\
&g_{aa'}^{\xi\xi',i}(\kk,\tau) = g_{aa'}^{-\xi-\xi',i}(C_{2x}\kk,\tau)
\quad,\quad a,a' \in \{c',c''\}
\nonumber \\
&g_{c'c'}^{\xi\xi',i}(\kk,\tau) = g_{c'c'}^{\xi\xi',i}(C_{3z}\kk,\tau) e^{i\frac{2\pi}{3}(\xi'-\xi)} \quad ,\quad 
g_{c''c''}^{\xi\xi',i}(\kk,\tau) = g_{c''c'',i}^{\xi\xi'}(C_{3z}\kk,\tau) \nonumber \\
&
g_{c'c''}^{\xi\xi',i}(\kk,\tau) = g_{c'c''}^{\xi\xi',i}(C_{3z}\kk,\tau) e^{i\frac{2\pi}{3}(-\xi)} \quad,\quad 
g_{c''c'}^{\xi\xi',i}(\kk,\tau) = g_{c''c'}^{\xi\xi',i}(C_{3z}\kk,\tau) e^{i\frac{2\pi}{3}(\xi')}  \, .
\label{eq:green_descrete_nu_1} 
\eaa 
where $i=1,3,4$. 
Thus, using Eq.~\ref{eq:h1_def}, Eq.~\ref{eq:chi_n} and Eq.~\ref{eq:green_descrete_nu_1}, we find
\baa 
&H^{\nu_f=-1,fe}(\qq)_{\xi\xi',\xi_2'\xi_2} = 
(H^{\nu_f}(C_{2z}T\qq)_{-\xi-\xi',-\xi_2'-\xi_2})^* \nonumber \\
&H^{\nu_f=-1,fe}(\qq)_{\xi\xi',\xi_2\xi_2'} = 
H^{\nu_f=-1,fe}(C_{2x}\qq)_{-\xi-\xi',-\xi_2'-\xi_2}\nonumber \\
&H^{\nu_f=-1,fe}(\qq)_{\xi\xi',\xi_2\xi_2'} = H^{\nu_f=-1,fe}(C_{3z}\qq)_{\xi\xi',\xi_2'\xi_2} e^{i\frac{2\pi}{3}(\xi'-\xi_2+\xi_2'-\xi)} 
\label{eq:m_nu_1_sym}
\eaa 
In addition from Eq.~\ref{eq:symmetry_const} (with $i=1,j=3,\xi=\xi'=+$)
\baa 
H^{\nu_f=-1,fe}(\qq=0)_{++,++} +H^{\nu_f=-1,fe}(\qq=0)_{+-,--} =0
\eaa 

Clearly, $H^{\nu_f=-1,fe}(\qq)$ satisfy the same constrain as $\hH^{\nu_f=0}(\qq)$ (Eq.~\ref{eq:m_nu_eve_sym}). Consequently, $H^{\nu_f=-1,fe}(\qq)$ has the same long-wavelength behavior as shown in Eq.~\ref{eq:mat_struct_nu_even}.

\subsubsection{Full-half sector}
For the full-half sector, we take $i=1,j=2$ where $i=1$ flavor is filled with two $f$ electrons, and $j=2$ flavor is filled with one $f$ electrons (Eq.~\ref{eq:def_psi_0_r}). 
\baa 
S_{eff,off,fh}  =  & \frac{1}{2\pi N_M}\int \sum_\qq 
\sum_{\xi,\xi'}u(\qq,i\omega)_{1\xi,2-} u(-\qq,-i\omega)_{2-,1\xi'} \bigg\{ 
i\omega \delta_{\xi,\xi'}
-F_{\xi-,\xi'-}^{12}(\qq)
\bigg\} d\omega 
\eaa 
We thus introduce the following matrix 
\baa 
H^{\nu_f=-1,fh}(\qq)_{\xi,\xi'} =F_{\xi-,\xi'-}^{12}(\qq)
\label{eq:h_nu_1_fh}
\eaa  
which is a $2\times 2$ matrix. 
The effective action can be written as 
\baa 
S_{eff,off,fh}  =  & \frac{1}{2\pi N_M}\int \sum_\qq 
\sum_{\xi,\xi'}u(\qq,i\omega)_{1\xi,2-} u(-\qq,-i\omega)_{2-,1\xi'}  \bigg\{ 
i\omega \delta_{\xi,\xi'}
-H^{\nu_f=-1,fh}(\qq)_{\xi,\xi'} 
\bigg\} d\omega 
\label{eq:seff_fh_h}
\eaa 

Ground state at $\nu_f=-3$ (Eq.~\ref{eq:def_psi_0_r}) only has $C_{3z}$ symmetry. We use Eq.~\ref{eq:desc_sym}, Eq.~\ref{eq:sym_green} and find
\baa 
&g_{c'c'}^{\xi\xi',i}(\kk,\tau) = g_{c'c'}^{\xi\xi',i}(C_{3z}\kk,\tau) e^{i\frac{2\pi}{3}(\xi'-\xi)} \quad ,\quad 
g_{c''c''}^{\xi\xi',i}(\kk,\tau) = g_{c''c''}^{\xi\xi',i}(C_{3z}\kk,\tau) \nonumber \\
&
g_{c'c''}^{\xi\xi',i}(\kk,\tau) = g_{c'c''}^{\xi\xi',i}(C_{3z}\kk,\tau) e^{i\frac{2\pi}{3}(-\xi)} \quad,\quad 
g_{c''c'}^{\xi\xi',i}(\kk,\tau) = g_{c''c'}^{\xi\xi',i}(C_{3z}\kk,\tau) e^{i\frac{2\pi}{3}(\xi')}  \, .
\label{eq:green_descrete_fh} 
\eaa 
Then combining Eq.~\ref{eq:green_descrete_fh} and Eq.~\ref{eq:chi_n}, we have 
\baa
&\chi^A_{\xi -, \xi'-}(\qq,i\omega=0,2,1) =\chi^A_{\xi -, \xi'-}(C_{3z}\qq,i\omega=0,2,1)e^{i2\pi/3(\xi'-\xi)}
\label{eq:chi_sym_fh}
\eaa
Thus using Eq.~\ref{eq:chi_n}, Eq.~\ref{eq:h_nu_1_fh} and Eq.~\ref{eq:chi_sym_fh} 
\baa  
\hH^{\nu_f=-1,fh}(\qq)_{\xi,\xi'} = \hH^{\nu_f=-1,fh}(C_{3z}\qq)_{\xi,\xi'}e^{i2\pi/3(\xi'-\xi)}
\label{eq:h_fh_sym}
\eaa 


In the long-wavelength limit, we assume $H_1^{\nu_f=1,fh}(\qq)$ takes the form of 
\baa  
\hH_1^{\nu_f=1,fh}(\qq)_{\xi\xi'} = \begin{bmatrix}
C_{0,0}^{fh} +C_{\mu, 1,0}^{fh}q_\mu + C_{\mu\nu,2,0}^{fh}q_\mu q_\nu &  
 C_{0,1}^{fh} +C_{\mu, 1,1}^{fh}q_\mu + C_{\mu\nu,2,1}^{fh}q_\mu q_\nu \\ C_{0,3}^{fh} +C_{\mu, 1,3}^{fh}q_\mu + C_{\mu\nu,2,3}^{fh}q_\mu q_\nu & 
 C_{0,4}^{fh} +C_{\mu, 1,4}^{fh}q_\mu + C_{\mu\nu,2,4}^{fh}q_\mu q_\nu 
\end{bmatrix}_{\xi\xi'}
\label{eq:small_q_fh}
\eaa  
Combining Eq.~\ref{eq:small_q_fh} and Eq.~\ref{eq:h_fh_sym} we find
\baa  
&C_{0,1}^{fh} = e^{-i4\pi/3}C_{0,1}^{fh} ,\quad  C_{0,3}^{fh} = e^{i4\pi/3}C_{0,3}^{fh} \nonumber \\
&C_{x,1,0}^{fh} = -\frac{1}{2}C_{x,1,0}^{fh} +\frac{\sqrt{3}}{2} C_{y,1,0}^{fh}   , \quad C_{y,1,0}^{fh} = -\frac{-\sqrt{3}}{2}C_{x,1,0}^{fh}  -\frac{1}{2} C_{y,1,0}^{fh}  \nonumber \\ 
&C_{x,1,4}^{fh} = -\frac{1}{2}C_{x,1,4}^{fh}  +\frac{\sqrt{3}}{2} C_{y,1,4}^{fh}   , \quad C_{y,1,4}^{fh} = -\frac{-\sqrt{3}}{2}C_{x,1,4}^{fh}  -\frac{1}{2} C_{y,1,4}^{fh}  \nonumber \\ 
&C_{x,1,1}^{fh} = (-\frac{1}{2}C_{x,1,1}^{fh}  +\frac{\sqrt{3}}{2} C_{y,1,1}^{fh}  )e^{-i4\pi/3} , \quad C_{y,1,1}^{fh} = (-\frac{-\sqrt{3}}{2}C_{x,1,1}^{fh}  -\frac{1}{2} C_{y,1,1}^{fh}  )e^{-i4\pi/3}\nonumber \\ 
&C_{x,1,3}^{fh} = (-\frac{1}{2}C_{x,1,3} ^{fh} +\frac{\sqrt{3}}{2} C_{y,1,3}^{fh} )e^{i4\pi/3}  , \quad C_{y,1,3}^{fh} = (-\frac{-\sqrt{3}}{2}C_{x,1,3}^{fh}  -\frac{1}{2} C_{y,1,3}^{fh} )e^{i4\pi/3} \nonumber \\ 
&C_{xx, 2,0/4 }^{fh} = \frac{
C_{xx, 2,0/4 }^{fh} +3 C_{yy, 2,0/4 }^{fh} -\sqrt{3}C_{xy, 2,0/4 }^{fh} -\sqrt{3}C_{yx, 2,0/4 }^{fh}
}{4} 
\nonumber \\ 
& C_{yy,  2,0/4 }^{fh} = \frac{3
C_{xx, 2,0/4 }^{fh} +C_{yyv}^{fh} +\sqrt{3}C_{xy, 2,0/4 }^{fh} +\sqrt{3}C_{yx, 2,0/4 }^{fh}
}{4} 
\nonumber \\ 
&C_{xy,  2,0/4 }^{fh} = \frac{\sqrt{3}
C_{xx, 2,0/4}^{fh} -\sqrt{3}C_{yy, 2,0/4 }^{fh} +C_{xy, 2,0/4 }^{fh} -3C_{yx, 2,0/4 }^{fh}
}{4} 
\nonumber \\ 
&
C_{yx,  2,0/4 }^{fh} = \frac{\sqrt{3}
C_{xx, 2,0/4 }^{fh} -\sqrt{3}C_{yy, 2,0/4 }^{fh} -3C_{xy, 2,0/4 }^{fh} +C_{yx, 2,0/4 }^{fh}
}{4}  \nonumber \\ 
&C_{xx, 2,1}^{fh} = \frac{
C_{xx,2,1}^{fh} +3 C_{yy,2,1}^{fh} -\sqrt{3}C_{xy,2,1}^{fh} -\sqrt{3}C_{yx,2,1}^{fh}
}{4} e^{-i4\pi/3}
\nonumber \\ 
&
C_{yy, 2,1}^{fh} = \frac{3
C_{xx,2,1}^{fh} +C_{yy,n,0}^{fh} +\sqrt{3}C_{xy,2,1}^{fh} +\sqrt{3}C_{yx,2,1}^{fh}
}{4} e^{-i4\pi/3},
\nonumber \\ 
&C_{xy, 2,1}^{fh} = \frac{\sqrt{3}
C_{xx,2,1}^{fh} -\sqrt{3}C_{yy,2,1}^{fh} +C_{xy,2,1}^{fh} -3C_{yx,2,1}^{fh}
}{4} e^{-i4\pi/3}
\nonumber \\ 
&
C_{yx, 2,1}^{fh} = \frac{\sqrt{3}
C_{xx,2,1}^{fh} -\sqrt{3}C_{yy,2,1}^{fh} -3C_{xy,2,1}^{fh} +C_{yx,2,1}^{fh}
}{4} e^{-i4\pi/3}\nonumber \\ 
&C_{xx, 2,3}^{fh} = \frac{
C_{xx,2,3}^{fh} +3 C_{yy,2,1}^{fh} -\sqrt{3}C_{xy,2,1}^{fh} -\sqrt{3}C_{yx,2,3}^{fh}
}{4} e^{i4\pi/3}
\nonumber \\ 
&C_{yy, 2,3}^{fh} = \frac{3
C_{xx,2,1}^{fh} +C_{yy,n,0}^{fh} +\sqrt{3}C_{xy,2,3}^{fh} +\sqrt{3}C_{yx,2,3}^{fh}
}{4} e^{i4\pi/3},
\nonumber \\ 
&C_{xy, 2,3}^{fh} = \frac{\sqrt{3}
C_{xx,2,3}^{fh} -\sqrt{3}C_{yy,2,3}^{fh} +C_{xy,2,3}^{fh} -3C_{yx,2,3}^{fh}
}{4} e^{i4\pi/3},
\nonumber \\ 
&
C_{yx, 2,3}^{fh} = \frac{\sqrt{3}
C_{xx,2,3}^{fh} -\sqrt{3}C_{yy,2,3}^{fh} -3C_{xy,2,3}^{fh} +C_{yx,2,3}^{fh}
}{4} e^{i4\pi/3}
\eaa  
We also utilize the fact that $H_1^{\nu_f=1,fh}(\qq)$ is a Hermitian matrix (from Eq.~\ref{eq:h_nu_1_fh} and Eq.~\ref{eq:f_hert}). Then we find the non-vanishing components are 
\baa  
C_{0,0}^{fh},\quad   C_{xx/yy,2,0}^{fh},\quad  
C_{0,4}^{fh},\quad   C_{xx/yy,2,4}^{fh},\quad  C_{\mu,1,3},\quad   C_{\mu,1,1}^{fh},\quad  C_{\mu,1,3}^{fh}
\eaa  with 
\baa  
&C_{0,0}^{fh} \in \mathbb{R}, \quad C_{0,4}^{fh} \in \mathbb{R} ,\quad 
C_{\mu\mu,2,0}^{fh} \in \mathbb{R},\quad 
C_{\mu\mu,2,4}^{fh} \in \mathbb{R} \nonumber \\ 
&C_{xx,2,0}^{fh}=C_{yy,2,4}^{fh} ,\quad C_{xx,2,0}^{fh}=C_{yy,2,4}^{fh} \nonumber \\ 
&C_{y,1,1}^{fh} = -i C_{x,1,1}^{fh} = -C_{y,1,3}^{fh} = -iC_{x,1,3}^{fh}
\eaa  
Thus we have 
\baa  
\hH^{\nu_f=1,fh}(\qq)_{\xi\xi'} = \begin{bmatrix}
C_{0,0}^{fh} +C_{xx,2,0}^{fh}|\qq|^2 &  
 C_{x, 1,1}^{fh}(q_x-iq_y)\\  
 C_{x, 1,1}^{fh}(q_x+iq_y) & 
 C_{0,4}^{fh} +C_{xx,2,4}^{fh}|\qq|^2
\end{bmatrix}_{\xi\xi'}
\eaa 
We next use Eq.~\ref{eq:symmetry_const} (we take $i=1,j=2,\xi_2=+,\xi_2'=-$) and Eq.~\ref{eq:eff_s_fe}, and find 
\baa  
H^{\nu_f=1,fh}(\qq=0)_{--} =0\Rightarrow C_{0,4}^{fh}=0
\eaa  
Then we have 
\baa  
H^{\nu_f=1,fh}(\qq)_{\xi\xi'} = \begin{bmatrix}
C_{0,0}^{fh} +C_{xx,2,0}^{fh}|\qq|^2 &  
 C_{x, 1,1}^{fh}(q_x-iq_y)\\  
 C_{x, 1,1}^{fh}(q_x+iq_y) & 
 C_{xx,2,4}^{fh}|\qq|^2
\end{bmatrix}_{\xi\xi'}
\label{eq:h1_fh}
\eaa 
where we observe $C^{fh}_{0,4} =0$ produces a Goldstone mode. The eigenvalues of Eq.~\ref{eq:h1_fh} are 
\baa 
E(\qq) = \frac{C_{0,0}^{fh} +C_{xx,2,0}^{fh}|\qq|^2 +C_{xx,2,4}^{fh}|\qq|^2}{2} \pm 
\sqrt{\bigg( \frac{C_{0,0}^{fh} +C_{xx,2,0}^{fh}|\qq|^2 -C_{xx,2,4}^{fh}|\qq|^2}{2}\bigg)^2 + |C_{x,1,1}^{fh}|^2 |\qq|^2}
\eaa 

\subsubsection{Half-empty sector}
For the half-empty sector, we take $i=2, j=3,4$, where $i=2$ flavor is filled with two $f$ electrons, and $j=3,4$ flavors are empty (Eq.~\ref{eq:def_psi_0_r}). Then  
\baa 
S_{eff,off,he}  = &  \frac{1}{2\pi N_M}\int \sum_\qq \sum_{ j\in \{3,4\} }
\sum_{\xi,\xi_2}u(\qq,i\omega)_{2+,j\xi} u(-\qq,-i\omega)_{j\xi_2,2+} \bigg\{ 
i\omega \delta_{\xi_2,\xi} -F^{2j}_{+\xi,+\xi_2}(\qq)
\bigg\}d\omega ,\quad j=3,4
\eaa 
The $U(2)$ symmetry that act on the valley-spin flavors $i=3,4$ indicates 
\baa 
&-\frac{N_{2+}-N_{j\xi}}{2} = -\frac{N_{2+}-N_{j'\xi}}{2} \quad,\quad j,j' \in \{3,4\}  \nonumber \\
&\frac{1}{4} \bigg( 
\chi^A_{+\xi,+\xi_2}(\qq,i\omega=0,i=2,j)  + \chi^A_{\xi_2+,\xi+}(-\qq,i\omega=0,j,i=2)
\bigg) =
\frac{1}{4} \bigg( \nonumber \\
&
\chi^A_{+\xi,+\xi_2}(\qq,i\omega=0,i=2,j')  + \chi^A_{\xi_2+,\xi+}(-\qq,i\omega=0,j',i=2)\bigg) \quad,\quad j,j' \in\{3,4\}
\eaa 
Therefore, $F^{23}_{+\xi,+\xi_2}(\qq)=F^{24}_{+\xi,+\xi_2}(\qq)$. 
We can introduce the following matrix 
\baa 
H^{\nu_f=-1,he}(\qq)_{\xi,\xi_2}
=F^{2j}_{+\xi,+\xi_2}(\qq)\quad,\quad j \in \{3,4\}
\eaa 
which is a $2\times 2$ matrix and does not depend on $j$.   
The effective action can be written as 
\baa 
S_{eff,off,he}  = &  \frac{1}{2\pi N_M}\int \sum_\qq \sum_{ j\in \{3,4\} }
\sum_{\xi,\xi_2}u(\qq,i\omega)_{2+,j\xi} u(-\qq,-i\omega)_{j\xi_2,2+} 
[i\omega \delta_{\xi,\xi_2} - H^{\nu_f=-1,he}(\qq)_{\xi,\xi_2}]
d\omega 
\label{seff_he}
\eaa 

Wround state at $\nu_f=-3$ (Eq.~\ref{eq:def_psi_0_r}) has $C_{3z}$ symmetry.
We combine Eq.~\ref{eq:green_descrete_fh} and Eq.~\ref{eq:chi_n}, and find $(j=3,4)$
\baa
&\chi^A_{+ \xi, +\xi_2}(\qq,i\omega=0,j,2) =\chi^A_{+\xi, +\xi_2}(C_{3z}\qq,i\omega=0,j,2)e^{i2\pi/3(\xi'-\xi)}
\label{eq:chi_sym_fh}
\eaa
Thus using Eq.~\ref{eq:chi_n}, Eq.~\ref{eq:h_nu_1_fh} and Eq.~\ref{eq:chi_sym_fh} 
\baa  
H^{\nu_f=-1,he}(\qq)_{\xi,\xi_2} = H^{\nu_f=-1,he}(C_{3z}\qq)_{\xi,\xi_2}e^{i2\pi/3(\xi-\xi_2)}
\label{eq:h_he_sym}
\eaa 

We next study the long-wavelength limit, we assume $H_1^{\nu_f=1,fh}(\qq)$ takes the form of 
\baa  
H^{\nu_f=1,he}(\qq)_{\xi\xi_2} = \begin{bmatrix}
C_{0,0}^{fe} +C_{\mu, 1,0}^{fe}q_\mu + C_{\mu\nu,2,0}^{fh}q_\mu q_\nu &  
 C_{0,1}^{fe} +C_{\mu, 1,1}^{fe}q_\mu + C_{\mu\nu,2,1}^{fe}q_\mu q_\nu \\ C_{0,3}^{fh} +C_{\mu, 1,3}^{fe}q_\mu + C_{\mu\nu,2,3}^{fe}q_\mu q_\nu & 
 C_{0,4}^{fh} +C_{\mu, 1,4}^{fe}q_\mu + C_{\mu\nu,2,4}^{fe}q_\mu q_\nu 
\end{bmatrix}_{\xi\xi2}
\label{eq:small_q_he}
\eaa  
with row and column indices $\{+,-\}$. 
Combining Eq.~\ref{eq:small_q_he} and Eq.~\ref{eq:h_he_sym} we find
\baa  
&C_{0,1}^{he} = e^{-i4\pi/3}C_{0,1}^{he} ,\quad  C_{0,3}^{he} = e^{i4\pi/3}C_{0,3}^{he} \nonumber \\
&C_{x,1,0}^{he} = -\frac{1}{2}C_{x,1,0}^{he} +\frac{\sqrt{3}}{2} C_{y,1,0}^{he}   , \quad C_{y,1,0}^{he} = -\frac{-\sqrt{3}}{2}C_{x,1,0}^{he}  -\frac{1}{2} C_{y,1,0}^{he}  \nonumber \\ 
&C_{x,1,4}^{he} = -\frac{1}{2}C_{x,1,4}^{he}  +\frac{\sqrt{3}}{2} C_{y,1,4}^{he}   , \quad C_{y,1,4}^{he} = -\frac{-\sqrt{3}}{2}C_{x,1,4}^{he}  -\frac{1}{2} C_{y,1,4}^{he}  \nonumber \\ 
&C_{x,1,1}^{he} = (-\frac{1}{2}C_{x,1,1}^{he}  +\frac{\sqrt{3}}{2} C_{y,1,1}^{he}  )e^{i4\pi/3} , \quad C_{y,1,1}^{he} = (-\frac{-\sqrt{3}}{2}C_{x,1,1}^{he}  -\frac{1}{2} C_{y,1,1}^{he}  )e^{i4\pi/3}\nonumber \\ 
&C_{x,1,3}^{he} = (-\frac{1}{2}C_{x,1,3} ^{he} +\frac{\sqrt{3}}{2} C_{y,1,3}^{he} )e^{-i4\pi/3}  , \quad C_{y,1,3}^{he} = (-\frac{-\sqrt{3}}{2}C_{x,1,3}^{he}  -\frac{1}{2} C_{y,1,3}^{he} )e^{-i4\pi/3} \nonumber \\ 
&C_{xx, 2,0/4 }^{he} = \frac{
C_{xx, 2,0/4 }^{he} +3 C_{yy, 2,0/4 }^{he} -\sqrt{3}C_{xy, 2,0/4 }^{he} -\sqrt{3}C_{yx, 2,0/4 }^{he}
}{4} 
\nonumber \\ 
& C_{yy,  2,0/4 }^{he} = \frac{3
C_{xx, 2,0/4 }^{he} +C_{yyv}^{he} +\sqrt{3}C_{xy, 2,0/4 }^{he} +\sqrt{3}C_{yx, 2,0/4 }^{he}
}{4} 
\nonumber \\ 
&C_{xy,  2,0/4 }^{he} = \frac{\sqrt{3}
C_{xx, 2,0/4}^{he} -\sqrt{3}C_{yy, 2,0/4 }^{he} +C_{xy, 2,0/4 }^{he} -3C_{yx, 2,0/4 }^{he}
}{4} 
\nonumber \\ 
&
C_{yx,  2,0/4 }^{he} = \frac{\sqrt{3}
C_{xx, 2,0/4 }^{he} -\sqrt{3}C_{yy, 2,0/4 }^{he} -3C_{xy, 2,0/4 }^{he} +C_{yx, 2,0/4 }^{he}
}{4}  \nonumber \\ 
&C_{xx, 2,1}^{he} = \frac{
C_{xx,2,1}^{he} +3 C_{yy,2,1}^{he} -\sqrt{3}C_{xy,2,1}^{he} -\sqrt{3}C_{yx,2,1}^{he}
}{4} e^{i4\pi/3}
\nonumber \\ 
&
C_{yy, 2,1}^{he} = \frac{3
C_{xx,2,1}^{he} +C_{yy,n,0}^{he} +\sqrt{3}C_{xy,2,1}^{he} +\sqrt{3}C_{yx,2,1}^{he}
}{4} e^{i4\pi/3},
\nonumber \\ 
&C_{xy, 2,1}^{he} = \frac{\sqrt{3}
C_{xx,2,1}^{he} -\sqrt{3}C_{yy,2,1}^{he} +C_{xy,2,1}^{he} -3C_{yx,2,1}^{he}
}{4} e^{i4\pi/3}
\nonumber \\ 
&
C_{yx, 2,1}^{he} = \frac{\sqrt{3}
C_{xx,2,1}^{he} -\sqrt{3}C_{yy,2,1}^{he} -3C_{xy,2,1}^{he} +C_{yx,2,1}^{he}
}{4} e^{-i4\pi/3}\nonumber \\ 
&C_{xx, 2,3}^{he} = \frac{
C_{xx,2,3}^{he} +3 C_{yy,2,1}^{he} -\sqrt{3}C_{xy,2,1}^{he} -\sqrt{3}C_{yx,2,3}^{he}
}{4} e^{-i4\pi/3}
\nonumber \\ 
&C_{yy, 2,3}^{he} = \frac{3
C_{xx,2,1}^{he} +C_{yy,n,0}^{he} +\sqrt{3}C_{xy,2,3}^{he} +\sqrt{3}C_{yx,2,3}^{he}
}{4} e^{-i4\pi/3},
\nonumber \\ 
&C_{xy, 2,3}^{he} = \frac{\sqrt{3}
C_{xx,2,3}^{he} -\sqrt{3}C_{yy,2,3}^{he} +C_{xy,2,3}^{he} -3C_{yx,2,3}^{he}
}{4} e^{-i4\pi/3},
\nonumber \\ 
&
C_{yx, 2,3}^{he} = \frac{\sqrt{3}
C_{xx,2,3}^{he} -\sqrt{3}C_{yy,2,3}^{he} -3C_{xy,2,3}^{he} +C_{yx,2,3}^{he}
}{4} e^{-i4\pi/3}
\eaa  
We also utilize the fact that $H^{\nu_f=1,he}(\qq)$ is a Hermitian matrix (from Eq.~\ref{eq:h_he_sym} and Eq.~\ref{eq:f_hert}). Then we find the non-vanishing components are 
\baa  
C_{0,0}^{he},\quad   C_{xx/yy,2,0}^{he},\quad  
C_{0,4}^{he},\quad   C_{xx/yy,2,4}^{he},\quad  C_{\mu,1,3},\quad   C_{\mu,1,1}^{he},\quad  C_{\mu,1,3}^{he}
\eaa  with 
\baa  
&C_{0,0}^{he} \in \mathbb{R}, \quad C_{0,4}^{he} \in \mathbb{R} ,\quad 
C_{\mu\mu,2,0}^{he} \in \mathbb{R},\quad 
C_{\mu\mu,2,4}^{he} \in \mathbb{R} \nonumber \\ 
&C_{xx,2,0}^{he}=C_{yy,2,4}^{he} ,\quad C_{xx,2,0}^{he}=C_{yy,2,4}^{he} \nonumber \\ 
&C_{y,1,1}^{he} = i C_{x,1,1}^{he} = -C_{y,1,3}^{he} = iC_{x,1,3}^{he}
\eaa  
Thus we have 
\baa  
\hH^{\nu_f=1,he}(\qq)_{\xi\xi'} = \begin{bmatrix}
C_{0,0}^{he} +C_{xx,2,0}^{he}|\qq|^2 &  
 C_{x, 1,1}^{he}(q_x+iq_y)\\  
 C_{x, 1,1}^{he}(q_x-iq_y) & 
 C_{0,4}^{he} +C_{xx,2,4}^{he}|\qq|^2
\end{bmatrix}_{\xi\xi'}
\eaa 
We next use Eq.~\ref{eq:symmetry_const} (we let $i=2,j=3,\xi_2=+,\xi_2'=+$) and Eq.~\ref{seff_he}, and find 
\baa  
H^{\nu_f=1,fh}(\qq=0)_{++} =0\Rightarrow C_{0,0}^{fe}=0
\eaa  
Then 
\baa  
H^{\nu_f=1,he}(\qq)_{\xi\xi'} = \begin{bmatrix}
C_{xx,2,0}^{he}|\qq|^2 &  
 C_{x, 1,1}^{he}(q_x+iq_y)\\  
 C_{x, 1,1}^{he}(q_x-iq_y) & 
 C_{0,4}^{he} +C_{xx,2,4}^{he}|\qq|^2
\end{bmatrix}_{\xi\xi'}
\label{eq:h1_he}
\eaa 
where we notice $C_{0,0}^{fe} = 0=0$ produces a Goldstone mode for each $\qq$.
The eigenvalues of Eq.~\ref{eq:h1_he} are 
\baa 
E(\qq) = \frac{C_{0,4}^{he} +C_{xx,2,4}^{he}|\qq|^2 +C_{xx,2,0}^{he}|\qq|^2}{2} \pm 
\sqrt{\bigg( \frac{C_{0,4}^{he} +C_{xx,2,4}^{he}|\qq|^2 -C_{xx,2,0}^{he}|\qq|^2}{2}\bigg)^2 + |C_{x,1,1}^{he}|^2 |\qq|^2}
\eaa

\subsubsection{Half-half sector}

For the half-half sector, we take $i=2,j=2$, where $i=j=2$ flavor is filled with one $f$ electron (Eq.~\ref{eq:def_psi_0_r}). We then have (Eq.~\ref{eq:seff_f})
\baa 
S_{eff,off,hh}  = & \frac{1}{2\pi N_M}\int \sum_\qq u(\qq,i\omega)_{2+,2-} u(-\qq,-i\omega)_{2-,2+} [
i\omega
-F_{-+,-+}^{22}(\qq)]d\omega 
\eaa 
We introduce $H^{\nu_f=-1,hh}(\qq)$
\baa  
H^{\nu_f=-1,hh}(\qq) = F_{-+,-+}^{22}(\qq)
\label{eq:h_f_half_half}
\eaa  
and have
\baa 
S_{eff,off,hh}  = & \frac{1}{2\pi N_M}\int \sum_\qq u(\qq,i\omega)_{2+,2-} u(-\qq,-i\omega)_{2-,2+} [
i\omega
-H^{\nu_f=-1,hh}(\qq)]d\omega 
\label{eq:seff_hh_h}
\eaa 
From Eq.~\ref{eq:f_hert}, $F_{-+,-+}^{22}(\qq)$ and also $H^{\nu_f=-1,hh}(\qq)$ is real. 

We now discuss the effect of $C_{3z}$ symmetry. From Eq.~\ref{eq:green_descrete_fh}, Eq.~\ref{eq:chi_n} and Eq.~\ref{eq:def_F}, we have
\baa  
F_{-+,-+}^{22}(\qq) = F_{-+,-+}^{22}(C_{3z}\qq) \nonumber \\ 
H^{\nu_f=-1,hh}(\qq) = H^{\nu_f=-1,hh}(C_{3z}\qq) 
\label{eq:h_sym_hh}
\eaa  

We perform a small $\qq$ expansions of $H^{\nu_f=-1,hh}(\qq)$
\baa  
H^{\nu_f=-1,hh}(\qq) = C_{0}^{hh} +\sum_{\mu}C_{\mu, 1}^{hh}q_\mu + \sum_{\mu\nu}C_{\mu\nu,2}^{hh}q_\mu q_\nu
\eaa  
From Eq.~\ref{eq:h_sym_hh}, we find 
\baa 
&C_{x,1}^{hh} = -\frac{1}{2}C_{x,1}^{hh} +\frac{\sqrt{3}}{2} C_{y,1}^{hh}  , \quad C_{y,1}^{hh} = -\frac{-\sqrt{3}}{2}C_{x,1}^{hh} -\frac{1}{2} C_{y,1}^{hh} \nonumber \\ 
&C_{xx, 2 }^{hh} = \frac{
C_{xx, 2 }^{hh} +3 C_{yy,2 }^{hh} -\sqrt{3}C_{xy, 2 }^{hh} -\sqrt{3}C_{yx, 2 }^{hh}
}{4} ,
\quad C_{yy,  2 }^{hh} = \frac{3
C_{xx, 2 }^{hh}+C_{yy,2 }^{hh} +\sqrt{3}C_{xy,2 }^{hh}+\sqrt{3}C_{yx, 2 }^{hh}
}{4} 
\nonumber \\ 
&C_{xy,  2 }^{hh}= \frac{\sqrt{3}
C_{xx, 2 }^{hh} -\sqrt{3}C_{yy, 2 }^{hh}+C_{xy, 2 }^{hh} -3C_{yx, 2 }^{hh}
}{4} 
,\quad 
C_{yx,  2 }^{hh} = \frac{\sqrt{3}
C_{xx, 2 }^{hh} -\sqrt{3}C_{yy, 2 }^{hh}-3C_{xy, 2 }^{hh} +C_{yx,2 }^{hh}
}{4} \nonumber \\ 
\eaa 
The non-zero components are 
\baa  
 C_{0}^{hh},  C_{xx,2}^{hh},  C_{yy,2 }^{hh} \, , 
\eaa 
and we also have 
\baa  
 &C_{0}^{hh} \in \mathbb{R},\quad  C_{xx,2}^{hh} \in \mathbb{R} \quad,  C_{yy,2}^{hh} \in \mathbb{R} \nonumber \\
 &C_{xx,2}^{hh} = C_{yy,2}^{hh} \, .
\eaa 
Then we have
\baa  
H^{\nu_f=-1,hh}(\qq) = C_{0}^{hh} +C_{xx,2}^{hh}|\qq|^2
\label{eq:h1_hh}
\eaa  

\subsubsection{Effective theory in the long-wavelength limit}
Here, we summarize the effective theory in the long-wavelength limit at $\nu_f=-1$. From Eq.~\ref{eq:eff_s_fe}, Eq.~\ref{eq:seff_fh_h}, Eq.~\ref{seff_he} and Eq.~\ref{eq:seff_hh_h}
\baa  
&S_{eff,off} = S_{eff,off,fe}+ S_{eff,off,he} +S_{eff,off,fh} + 
S_{eff,off,hh}  \nonumber \\ 
&S_{eff,off,fe} =   \frac{1}{2\pi N_M}\int \sum_\qq \sum_{i \in \{1\}, j\in \{3,4\} }
\sum_{\xi,\xi_2' ,\xi',\xi_2}u(\qq,i\omega)_{i\xi,j\xi'} u(-\qq,-i\omega)_{j\xi_2,i\xi_2'} \bigg\{ 
i\omega \delta_{\xi_2,\xi'}\delta_{\xi_2',\xi}
-H^{\nu_f=-1,fe}(\qq)_{\xi\xi',\xi_2'\xi_2}
\bigg\} d\omega  \nonumber \\ 
&S_{eff,off,fh}  =   \frac{1}{2\pi N_M}\int \sum_\qq 
\sum_{\xi,\xi'}u(\qq,i\omega)_{1\xi,2-} u(-\qq,-i\omega)_{2-,1\xi'}  \bigg\{ 
i\omega \delta_{\xi,\xi'}
-H^{\nu_f=-1,fh}(\qq)_{\xi,\xi'} 
\bigg\} d\omega \nonumber \\ 
&S_{eff,off,he}  =   \frac{1}{2\pi N_M}\int \sum_\qq \sum_{ j\in \{3,4\} }
\sum_{\xi,\xi_2}u(\qq,i\omega)_{2+,j\xi} u(-\qq,-i\omega)_{j\xi_2,2+} 
\bigg\{i\omega \delta_{\xi,\xi_2} - H^{\nu_f=-1,he}(\qq)_{\xi,\xi_2}\bigg\}
d\omega \nonumber \\ 
&
S_{eff,off,hh}  = \frac{1}{2\pi N_M}\int \sum_\qq u(\qq,i\omega)_{2+,2-} u(-\qq,-i\omega)_{2-,2+} \bigg\{
i\omega
-H^{\nu_f=-1,hh}(\qq)\bigg\}d\omega 
\label{eq:seff_nu_1_h}
\eaa 
where (from Eq.~\ref{eq:h1_fh}, Eq.~\ref{eq:h1_he} and Eq.~\ref{eq:h1_hh})
\baa 
&H^{\nu_f=-1,fe} =
\begin{bmatrix}
C^{fe}_{0,0} + C^{fe}_{2,0} |\qq|^2
&-C^{fe}_{0,0} +C^{fe}_{2,2}|\qq|^2 & C^{fe}_{1,1}(-q_x -iq_y) & C^{fe}_{1,1}(q_x -iq_y) 
\\
-C^{fe}_{0,0} +C^{fe}_{2,2}|\qq|^2 & C^{fe}_{0,0} + C^{fe}_{2,0} |\qq|^2 & C^{fe}_{1,1}(q_x+iq_y) & C^{fe}_{1,1}(-q_x+iq_y) \\ 
C^{fe}_{1,1}(-q_x+iq_y)& C^{fe}_{1,1}(q_x-iq_y) & C^{fe}_{0,1}+C^{fe}_{2,1}|\qq|^2 &C^{fe}_{2,3}(q_x^2 -q_y^2 -2i q_x q_y) \\
C^{fe}_{1,1}(q_x+iq_y)& C^{fe}_{1,1}(-q_x-iq_y) & C^{fe}_{2,3}(q_x^2 -q_y^2 +2i q_x q_y)& C^{fe}_{0,1} +C^{fe}_{2,1}|\qq|^2 
\end{bmatrix} \nonumber \\ 
&H^{\nu_f=1,fh}(\qq)_{\xi\xi'} = \begin{bmatrix}
C_{0,0}^{fh} +C_{xx,2,0}^{fh}|\qq|^2 &  
 C_{x, 1,1}^{fh}(q_x-iq_y)\\  
 C_{x, 1,1}^{fh}(q_x+iq_y) & 
 C_{xx,2,4}^{fh}|\qq|^2
\end{bmatrix}_{\xi\xi'}\nonumber \\ 
&H^{\nu_f=1,he}(\qq)_{\xi\xi'} = \begin{bmatrix}
C_{xx,2,0}^{he}|\qq|^2 &  
 C_{x, 1,1}^{he}(q_x+iq_y)\\  
 C_{x, 1,1}^{he}(q_x-iq_y) & 
 C_{0,4}^{he} +C_{xx,2,4}^{he}|\qq|^2
\end{bmatrix}_{\xi\xi'} \nonumber \\ 
&H^{\nu_f=-1,hh}(\qq) = C_{0}^{hh} +C_{xx,2}^{hh}|\qq|^2
\label{eq:nu_1_h_def_comp}
\eaa 
(We note that $H^{\nu_f=-1,fe}$ takes the same structure as $H^{\nu_f=0,-2}(\qq)$ in Eq.\ref{eq:mat_struct_nu_even}). We provide the numerical values of the parameters in Tab.~\ref{tab:nu_1_para}. In Fig.~\ref{fig:comp_nu1} (a), we also compare the dispersion from effective model~\ref{eq:seff_nu_1_h} with the dispersion from directly evaluating Eq.~\ref{eq:seff_f}. We note that the effective model (Eq.~\ref{eq:seff_nu_1_h}) correctly predicts the long wavelength (small $\kk$) behaviors. In addition, we observe a soft mode with a tiny gap in the half-half sector, as shown in Fig.~\ref{fig:comp_nu1} (b). Here we also provide the expression of the gap $\Delta_{hh} (=C_0^{hh})$ in the half-half sector. Using Eq.~\ref{eq:h1_hh} and Eq.~\ref{eq:seff_hh_h}
\baa  
\Delta = C_0^{hh}  = H^{\nu_f=-1,hh}(\qq=0)=F_{-+,-+}^{22}(\qq=0)
\eaa 
Using the expression of $F_{-+,-+}^{22}(\qq)$ (Eq.~\ref{eq:def_F}), we have 
\baa 
\Delta = \frac{N_{2-}-N_{2+}}{2}
-\frac{1}{4} \bigg( 
\chi^A_{-+,-+}(\qq=0,i\omega=0,2,2)  + \chi^A_{+-,+-}(\qq=0,i\omega=0,2,2)
\bigg)
\label{eq:gap_half_half}
\eaa 
where $N_{i\xi},\chi_{\xi\xi',\xi_2'\xi_2}^A(\qq,i\omega,i,j)$ are defined in Eq.~\ref{eq:gap_half_half}. A direct numerical evaluation of Eq.~\ref{eq:gap_half_half} gives $\Delta \approx 0.1$meV, which indicates a small gap at $\Gamma_M$.

\begin{figure}
    \centering
    \includegraphics[width=0.8\textwidth]{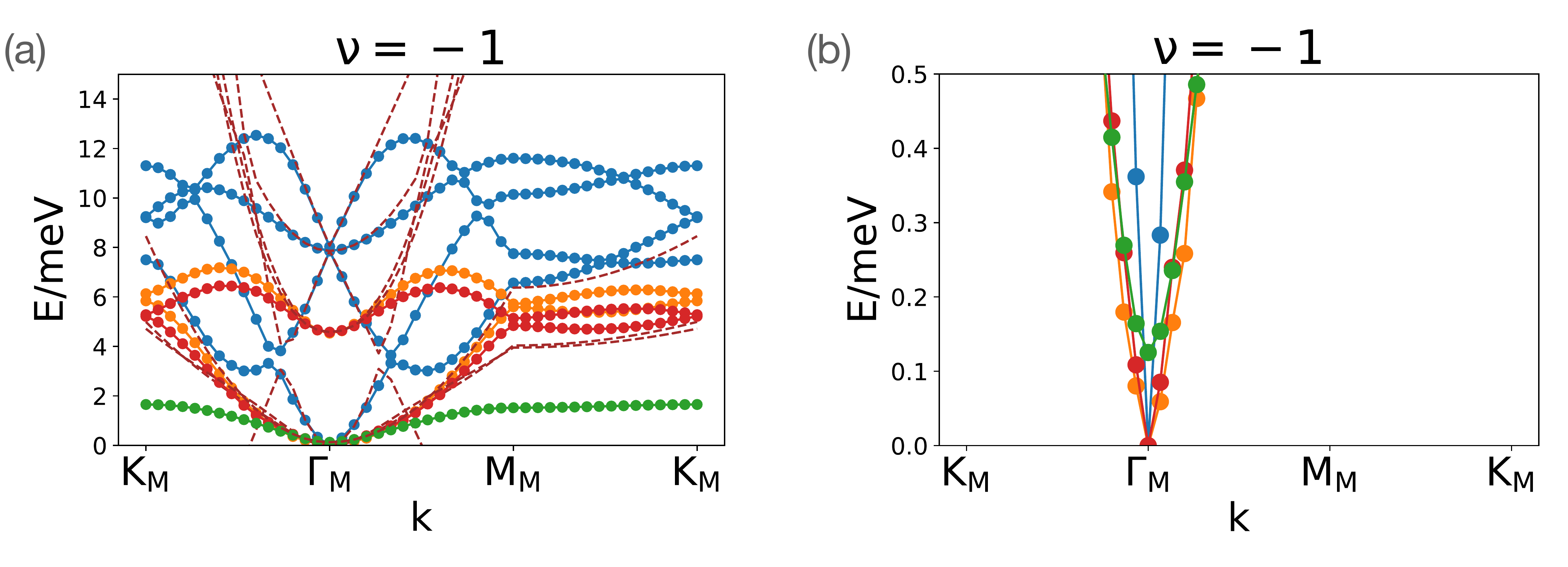}
    \caption{ (a) Excitation spectrum from the effective model in Eq.~\ref{eq:seff_nu_1_h} (brown) and from numerical evaluation of Eq.~\ref{eq:seff_f} (blue, orange, red, green). Blue, orange, red and green denote the full-empty sector, half-empty sector, full-half sector, and half-half sector respectively. (b) Illustration of soft mode in the half-half sector (green). }
    \label{fig:comp_nu1}
\end{figure}

\begin{table}[]
    \centering
    \begin{tabular}{c|c|c|c|c|c|c|c}
    Parameter & $C_{0,0}^{fe}$ & $C_{2,0}^{fe}a_M^2$ & $C_{2,1}^{fe}a_M^2$ & $C_{0,1}^{fe}$ & $C_{2,2}^{fe}a_M^2$ &$C_{2,3}^{fe}a_M^2$ & $C_{1,1}^{fe}a_M$ \\ 
    \hline 
    Value (meV) &  4.0 & 1.8 & 0.4 & 7.9 & 1.5 & 0.6 & -2.2 \\ 
    \hline \hline 
     Parameter &  $C_{0,0}^{fh}$ &  $C_{2,0}^{fh}a_M^2$
     & $C^{fh}_{2,4}a_M^2$ & $C_{1,1}^{fh}a_M$ \\ 
    \hline 
    Values (meV) &4.5 & 0.2 & 2.0 & -2.3 \\ 
    \hline \hline 
     Parameter &  $C_{0,4}^{he}$ &  $C_{2,0}^{he}a_M^2$
     & $C^{he}_{2,4}a_M^2$ & $C_{1,1}^{he}a_M$ \\ 
    \hline 
    Value (meV) & 4.5 & 1.8 & 0.2 & 2.1  \\ 
    \hline \hline 
     Parameter &  $C_{0}^{hh}$ &  $C_{xx,2}^{hh}a_M^2$ \\ 
    \hline 
    Value (meV) & 0.1 & 0.5  \\ 
    \end{tabular}
    \caption{Parameters of the effective model in Eq.~\ref{eq:seff_nu_1_h} (and also Eq.~\ref{eq:nu_1_h_def_comp}). }
    \label{tab:nu_1_para}
\end{table}

\subsubsection{Number of Goldstone modes}
We discuss the number of Goldstone modes at $\nu_f=-1$. The symmetry of the ground state at $\nu_f=-1$ is $U(1)\times U(1) \times U(2)$ (Eq.~\ref{eq:def_psi_0_r}), which has rank $1+1+4=6$. The symmetry of the original Hamiltonian at $M=0$ has flat $U(4)$ symmetry with rank $16$. We thus have $(16-6)/2=5$ Goldstone modes~\cite{tbgv}. We find two of the Goldstone modes are given by $S_{eff,off,fe}$ (full-empty sector), one of the Goldstone modes is given by $S_{eff,off,fh}$ (full-half sector) and the last two Goldstone modes are given by $S_{eff,off,he}$(half-empty sector).

\end{hbb}

\newpage

\section{Single-particle Green's function}
\label{sec:green}
\subsection{ Single-particle Green's function at $M=0$}
\bh{In this section, we evaluate the single-particle propagator of conduction $c$ electrons in the non-ordered state at $\nu_c=0$ and $M=0$, which has been used in Sec.~\ref{sec:sec_zero_hyb_j} and Sec.~\ref{sec:rkky}. 
The Hamiltonian takes the form of 
\baa 
\hH_c'= \hH_c + \hH^{MF}_V + \hH_W -\mu \sum_{\kk,\alpha \eta s}c_{\kk,\alpha \eta s}^\dag c_{\kk,\alpha \eta s}
\eaa 
where $\mu$ is the chemical potential. 
We treat $\hH^{MF}_V$ is the mean-field Hamiltonian of $\hH^V$ (Eq.~\ref{eq:hv_mf_def}) and has the form of 
\baa 
 \hH_V^{MF}=\frac{V(0)}{2\Omega_0}N_M\nu_c^2 +\frac{V(0)}{\Omega_0}\nu_c\sum_{\kk,a\eta s} (c_{\kk,a \eta s}^\dag c_{\kk, a \eta s}-1/2)
\eaa 
The filling of $f$-electron is fixed to be $\nu_f$ at each site. Then $\hH_W$ (Eq.~\ref{eq:hw_def} becomes
\baa  
\hH_W = W \sum_{\kk, a \eta s } \nu_f c_{\kk,a \eta s}^\dag c_{\kk, a \eta s}
\eaa  
Then $\hH_c'$ only contains the one-body term 
\baa  
\hH_c'= \sum_{\kk, a,a',\eta ,s} H_{a,a'}^{(c,\eta)}(\kk)c_{\kk,a \eta s}^\dag c_{\kk,a'\eta s}^\dag + (\frac{V(0)\nu_c}{\Omega_0}+W-\mu) \sum_{\kk}c_{\kk,a \eta s}^\dag c_{\kk,a \eta s} +\text{const}
\eaa 
We then consider $\nu_c=0$ ($\nu_f=\nu_t=0,-1,-2$). This can be realized by setting $\mu =\frac{V(0)\nu_c}{\Omega_0}+W = -W$. Then the single-particle Hamiltonian becomes 
\baa  
&\hH_c' =\sum_{\kk, a,a',\eta ,s} H_{a,a'}^{(c,\eta)}(\kk)c_{\kk,a \eta s}^\dag c_{\kk,a'\eta s}^\dag +\text{const} \nonumber \\
&H^{(c,\eta)}(\kk) = \begin{bmatrix}
0_{2\times 2} &v_{\star}(\eta k_x  \sigma_0 +ik_y \sigma_z)\\
v_{\star}(\eta k_x  \sigma_0 \bh{-}ik_y \sigma_z)& 0_{2\times 2 }
\end{bmatrix} 
 \label{eq:single_particle_ham}
\eaa  
which is equivalent to the non-interacting Hamiltonian of conduction electrons (with an additional constant term). We consider $M=0$ limit, and calculate the single-particle Green's function. 
}
The Green's functions in the imaginary-time $\tau$ are defined as
\baa
G_{aa',\eta}(\kk,\tau) = -\langle T_\tau c_{\kk, a \eta s}(\tau) c_{\kk,a'\eta s}^\dag(0)\rangle 
\label{eq:green_def}
\eaa  
The corresponding Green's function \bh{in the Matsubara frequency domain} is
\ba 
(i\omega_n -H^{(c,\eta)}(\kk))^{-1} 
= \frac{1}{-\omega^2 -|v_\star|^2k^2}
\begin{bmatrix}
i\omega_n & 0& v_\star(\eta k_x +ik_y)&0 \\
0&i\omega_n &0 & v_\star(\eta k_x -ik_y)\\
v_\star (\eta k_x-ik_y) & 0 & i\omega_n &0 \\
0& v_\star(\eta k_x+ik_y) & 0 & i\omega_n
\end{bmatrix}
\ea 
where $k= \sqrt{k_x^2+k_y^2}$, \bh{$\omega_n=(2n+1)\pi/\beta$ and $\beta=1/T$ the inverse temperature}.
where
\baa 
&G_{11,\eta}(\kk,\tau) = -\langle T_\tau c_{\kk,1\eta s}(\tau)c^\dag_{\kk, 1\eta s}(0)\rangle = \frac{1}{\beta}\sum_{i\omega_n} \frac{i\omega_n}{-\omega_n^2 -|v_\star|^2 k^2 }e^{-i\omega_n \tau} \nonumber  \\
&G_{22,\eta}(\kk,\tau) = -\langle T_\tau c_{\kk,2\eta s}(\tau)c^\dag_{\kk, 2\eta s}(0)\rangle=\frac{1}{\beta}\sum_{i\omega_n} \frac{i\omega_n}{-\omega_n^2 -|v_\star|^2 k^2 }e^{-i\omega_n \tau}  \nonumber \\
&G_{33,\eta}(\kk,\tau) = -\langle T_\tau c_{\kk,3\eta s}(\tau)c^\dag_{\kk, 3\eta s}(0)\rangle = \frac{1}{\beta}\sum_{i\omega_n} \frac{i\omega_n}{-\omega_n^2 -|v_\star|^2 k^2 }e^{-i\omega_n \tau}  \nonumber \\
&G_{44,\eta}(\kk,\tau) = -\langle T_\tau c_{\kk,4\eta s}(\tau)c^\dag_{\kk, 4\eta s}(0)\rangle = \frac{1}{\beta}\sum_{i\omega_n} \frac{i\omega_n}{-\omega_n^2 -|v_\star|^2 k^2 }e^{-i\omega_n \tau} \nonumber  \\
&G_{13,\eta}(\kk,\tau)=-\langle T_\tau c_{\kk,1\eta s}(\tau)c^\dag_{\kk 3 \eta s}(0)\rangle = \frac{1}{\beta}\sum_{i\omega_n}\frac{v_\star(\eta k_x+ik_y)}{-\omega_n^2 -|v_\star|^2 k^2}e^{-i\omega_n\tau} \nonumber  \\
&G_{24,\eta}(\kk,\tau)=-\langle T_\tau c_{\kk,2\eta s}(\tau)c^\dag_{\kk 4 \eta s}(0)\rangle =\frac{1}{\beta}\sum_{i\omega_n}\frac{v_\star(\eta k_x-ik_y)}{-\omega_n^2 -|v_\star|^2 k^2}e^{-i\omega_n\tau}  \nonumber \\
&G_{31,\eta}(\kk,\tau)=-\langle T_\tau c_{\kk,3\eta s}(\tau) c^\dag_{\kk 1 \eta s}(0)\rangle = \frac{1}{\beta}\sum_{i\omega_n}\frac{v_\star(\eta k_x-ik_y)}{-\omega_n^2 -|v_\star|^2 k^2}e^{-i\omega_n\tau}  \nonumber \\
&G_{42,\eta}(\kk,\tau)=-\langle T_\tau c_{\kk,4\eta s}(\tau) c^\dag_{\kk 2 \eta s}(0)\rangle =\frac{1}{\beta}\sum_{i\omega_n}\frac{v_\star(\eta k_x+ik_y)}{-\omega_n^2 -|v_\star|^2 k^2}e^{-i\omega_n\tau} 
\eaa

We let 
\ba 
&g_0(\tau,\kk) = \frac{1}{\beta}\sum_{i\omega_n} \frac{i\omega_n}{-\omega_n^2-|v_\star|^2k^2}e^{-i\omega_n\tau}
&g_{2,\eta}(\tau,\kk) =  \frac{1}{\beta}\sum_{i\omega_n}\frac{v_\star(\eta k_x+ik_y)}{-\omega_n^2 -|v_\star|^2 k^2}e^{-i\omega_n\tau}
\, .
\ea 
Then, we have 
\baa  
&G_{11,\eta}(\kk,\tau) =G_{22,\eta}(\kk,\tau) =G_{33,\eta}(\kk,\tau) = G_{44,\eta}(\kk,\tau) =g_0(\kk,\tau) \nonumber \\
&G_{13,\eta}(\tau,\kk) = g_{2,\eta}(\tau,\kk) \quad,\quad G_{24,\eta}(\tau,\kk) =g^*_{2,\eta}(-\tau,\kk)
\quad,\quad 
G_{31,\eta}(\tau,\kk) = g_{2,\eta}^*(-\tau,\kk) \quad,\quad G_{42,\eta}(\tau,\kk) = g_{2,\eta}(\tau,\kk) 
\label{eq:green_def_2}
\eaa  
and the corresponding Fourier transformation
\ba 
g_0(\RR,\tau) = \frac{1}{N} \sum_{\kk}g_1(\kk,\tau)e^{i\kk \cdot \RR} \quad,\quad 
g_2(\RR,\tau) = \frac{1}{N} \sum_{\kk}g_2(\kk,\tau)e^{i\kk \cdot \RR} 
\ea 

We next give the analytical expression of $g_0(\RR,\tau),g_1(\RR,\tau)$.

\subsubsection{$g_0(\RR,\tau)$}
We calculate $g_0(\RR,\tau)$ \bh{in the} thermodynamic limit. 
\ba 
g_0(\RR,\tau) = 
&\frac{1}{A_{MBZ}}\int g_1(\kk,\tau)e^{i\kk \cdot \RR} d^2\kk  \\ 
=& \frac{1}{A_{MBZ}}\int  \frac{1}{\beta}\sum_{i\omega_n}\frac{i\omega_n}{-\omega_n^2 -|v_\star|^2k^2}e^{-i\omega_n\tau} e^{i\kk \cdot \RR} d^2\kk  \\ 
\ea 
where we have replace $\sum_{\kk}$ with $\frac{1}{A_{MBZ}}\int d^2\kk$ where $A_{MBZ}$ is the area of moire Brillouin zone. We introduce the following expansion
\baa  
e^{i\kk \cdot \RR}  = J_0(kr) + 2 \sum_{m=1}i^m J_m(kr) \cos(m(\theta-\theta_\RR))
\label{eq:exp_eikr}
\eaa  
where $k = |\kk|, r = |\RR|$, $\theta = \arctan(k_y/k_x), \theta_{\RR} = \arctan(R_y/R_x)$ and $J_m(x)$ is the Bessel function~\cite{abramowitz1988handbook}. We then have 
\baa  
g_0(\RR,\tau) =& \frac{1}{A_{MBZ}}\int_0^{2\pi} \int_0^{\Lambda_c}  \frac{1}{\beta}\sum_{i\omega_n}\frac{i\omega_n}{-\omega_n^2 -|v_\star|^2k^2}[ J_0(kr) + 2 \sum_{m=1}i^m J_m(kr)]\bh{e^{-i\omega_n\tau}} k dk d\theta  \nonumber  \\ 
=&\frac{2\pi}{A_{MBZ}}\int_0^{\Lambda_c}  \frac{1}{\beta}\sum_{i\omega_n}\frac{i\omega_n}{-\omega_n^2 -|v_\star|^2k^2} J_0(kr) \bh{e^{-i\omega_n\tau}} k dk d\theta  
\label{eq:g0_v0}
\eaa  
where $\Lambda_c$ is the momentum cutoff. 

\hh{ 
We then perform Matsubara summation. We use the following identity 
\baa  
\frac{1}{\beta}\sum_{i\omega_n} \frac{1}{i\omega_n-\epsilon} e^{-i\omega_n\tau} = - (1-f(\epsilon) ) e^{-\epsilon \tau} 
\label{eq:matsu_sum_pos_tau}
\eaa 
where $\beta >\tau >0 $ and $f(x)$ is the Fermi-Driac function. This can be derived by replacing the summation with contour integral
\baa  
\frac{1}{\beta}\sum_{i\omega_n} \frac{1}{i\omega_n - \epsilon}e^{-i\omega_n \tau}  = \int_C \frac{dz}{2\pi i} \frac{e^{-z \tau}}{z-\epsilon}  (1-f(z))
\eaa  
where $C= [0^+ -i\infty, 0^++i\infty ]\cup[0-+i\infty,0--i\infty]$. \bh{We comment that depending on the sign of $\tau$, we could have either $1-f(z)$ prefactor or $f(z)$ prefactor~\cite{coleman2015introduction}. For $\beta>\tau>0$ as we considered here, because $e^{-z\tau}$ goes to zero at $\text{Re}[z]>0$, we need to choose $1-f(z)$ prefactor, such that the integrand also goes to zero at $\text{Re}[z]<0$.}
There is a pole on the real axis at $z=\epsilon$. We use the residue theorem and find 
\baa 
\frac{1}{\beta}\sum_{i\omega_n} \frac{1}{i\omega_n-\epsilon} e^{-i\omega_n\tau}=\bh{2\pi i (-1)\bigg[ \frac{1}{2\pi i}\frac{e^{-z \tau}}{z-\epsilon}(1-f(\epsilon))\bigg]_{z= \epsilon} = 
2\pi i \frac{1}{2\pi i }(-1) e^{-\epsilon \tau} (1-f(\epsilon)) }= - e^{-\epsilon \tau} (1-f(\epsilon))
\label{eq:mats_summ}
\eaa  
}

We then perform summation over Matsubara frequency of Eq.~\ref{eq:g0_v0}. \bh{Combining Eq.~\ref{eq:g0_v0} and Eq.~\ref{eq:mats_summ}}, at $\beta>\tau>0$, we find
\baa 
g_0(\RR,\tau)=& \frac{2\pi}{A_{MBZ}}\int_0^{\Lambda_c}  \frac{1}{\beta}\sum_{i\omega_n}\frac{i\omega_n}{-\omega_n^2 -|v_\star|^2k^2} J_0(kr) \bh{e^{-i\omega_n\tau}} k dk d\theta  \nonumber \\
=& \frac{2\pi}{A_{MBZ}}\int_0^{\Lambda_c} \frac{1}{\beta}
\sum_{i\omega_n} \frac{1}{2} \bigg[ 
\frac{1}{i\omega_n - |v_\star|k} + \frac{1}{i\omega_n +|v_\star| k }
\bigg] e^{-i\omega_n\tau} J_0(kr)kdk \nonumber  \\
=& \frac{2\pi}{A_{MBZ}}\int_0^{\Lambda_c} \frac{1}{2} (1-f(|v_\star|k))\bigg[ -e^{-|v_\star|k\tau} - e^{-|v_\star|k(\beta-\tau)}
\bigg] J_0(kr)kdk  \, .
\eaa 
Finally, we then take the zero-temperature limit and set $\Lambda_c=\infty$ to find the analytical expression:
\ba 
g_1(\RR,\tau)=& \frac{2\pi}{A_{MBZ}}\int_0^\infty \frac{1}{2} (-e^{-|v_\star| k\tau}) J_0(kr)kdk =\frac{-\pi}{A_{MBZ}}\frac{|v_\star|\tau}{((v_\star\tau)^2 +r^2)^{3/2} }
\ea 

For $ -\beta< -\tau <0$, 
\bh{we note that 
\baa  
\frac{1}{\beta}\sum_{i\omega_n} \frac{1}{i\omega_n-\epsilon} e^{-i\omega_n(-\tau)}=  (-1) \frac{1}{\beta}\sum_{i\omega_n} \frac{1}{i\omega_n-\epsilon} e^{-i\omega_n(\beta-\tau)}
\eaa  
Using Eq.~\ref{eq:matsu_sum_pos_tau}, we have 
\baa  
\frac{1}{\beta}\sum_{i\omega_n} \frac{1}{i\omega_n-\epsilon} e^{-i\omega_n(-\tau)}=  (-1) \frac{1}{\beta}\sum_{i\omega_n} \frac{1}{i\omega_n-\epsilon} e^{-i\omega_n(\beta-\tau)} = (1-f(\epsilon))e^{-(\beta-\tau)\epsilon }
\label{eq:matsu_sum_neg_tau}
\eaa  
Using Eq~\ref{eq:matsu_sum_neg_tau}, $g_0(\RR,-\tau)$ at negative imaginary time $-\beta <-\tau<0$ can be evaluated as}
\ba 
g_0(\RR,-\tau) &= \frac{2\pi}{A_{MBZ}}\int_0^{\Lambda_c} \frac{1}{\beta}
\sum_{i\omega_n} \frac{1}{2} \bigg[ 
\frac{1}{i\omega_n - |v_\star|k} + \frac{1}{i\omega_n +|v_\star| k }
\bigg] e^{i\omega_n\tau} J_0(kr)kdk  \\
&=\frac{2\pi}{A_{MBZ}}\int_0^{\Lambda_c} \frac{1}{2} (1-f(|v_\star|k))\bigg[ e^{-|v_\star|k(\beta-\tau)} + e^{-|v_\star|k\tau}
\bigg] J_0(kr)kdk  \,.
\ea 
Similarly, we let $\Lambda_c=\infty$ and $\beta=\infty$. Then 
\ba 
g_0(\RR,-\tau) = \frac{2\pi}{A_{MBZ}}\int_0^\infty \frac{1}{2} (e^{-|v_\star| k\tau}) J_0(kr)kdk =\frac{\pi}{A_{MBZ}}\frac{|v_\star|\tau}{((v_\star\tau)^2 +r^2)^{3/2} }
\ea 
In summary,
\baa  
g_0(\RR,\tau) = -\text{sgn}(\tau) \frac{ \pi}{A_{MBZ}}\frac{|v_\star\tau|}{((v_\star\tau)^2 +r^2)^{3/2} }
\label{eq:g0}
\eaa  
\bh{ where we observe $g_0(\RR,\tau)\propto \text{sgn}(\tau)$. We note that for the particle-hole symmetric system (realized at $\nu_c=0$), we have $g_0(\RR,\beta-\tau) = g_0(\RR,\tau)$. Then $g_0(\RR,-\tau) = -g(\RR,\beta-\tau) = -g_0(\RR,\tau)$. }

\subsubsection{$g_{2,\eta}(\tau,\RR)$}
We calculate $g_{2,\eta}(\tau,\RR)$ in this section. 
For $\beta>\tau>0$
\ba 
g_{2,\eta}(\tau,\RR) =& \frac{1}{A_{MBZ}}\int  g_{2,\eta}(\tau,\kk)e^{i\kk \cdot\RR }d^2\kk \\
=&\frac{1}{A_{MBZ}\beta} \int \sum_{i\omega_n}\frac{v_{\star} (\eta k_x+ik_y)}{-\omega_n^2 -|v_\star|^2 k^2 }e^{-i\omega_n\tau} e^{i\kk \cdot \RR}d^2\kk \\
=&\frac{1}{A_{MBZ}\beta} \int \sum_{i\omega_n}\frac{v_{\star} (\eta \cos(\theta)+i\sin(\theta))}{-\omega_n^2 -|v_\star|^2 k^2 }e^{-i\omega_n\tau}\bigg[ 
J_0(kr) +2\sum_{m=1}i^m J_m(kr) \cos(m(\theta-\theta_\RR))
\bigg] k^2dk d\theta  \\
=&\frac{1}{A_{MBZ}\beta} \int \sum_{i\omega_n}\frac{v_{\star} 2\pi i(\eta \cos(\theta_\RR)+i\sin(\theta_\RR))}{-\omega_n^2 -|v_\star|^2 k^2 }e^{-i\omega_n\tau} J_1(kr) k^2dk   \\
=&\frac{1}{A_{MBZ}\beta} \int \sum_{i\omega_n}\frac{|v_{\star}|2\pi i(\eta \cos(\theta_\RR)+i\sin(\theta_\RR))}{2|v_\star|k }
[\frac{1}{i\omega_n- |v_\star k| } -\frac{1}{i\omega_n+ |v_\star k|}]
e^{-i\omega_n\tau} J_1(kr) k^2dk  \\
=&\frac{1}{A_{MBZ}} \int \frac{|v_{\star}| 2\pi i(\eta \cos(\theta_\RR)+i\sin(\theta_\RR))}{2|v_\star|k }
(1-f(|v_\star|k))(e^{-|v_\star|k\tau} - e^{-|v_\star|k(\beta-\tau)}J_1(kr) k^2dk  \\
\ea 
We let $\beta \rightarrow \infty$ and set the momentum cutoff to infinity.
\ba 
g_{2,\eta}(\tau,\RR) =&\frac{i\pi (\eta \cos(\theta_\RR) +i\sin(\theta_\RR)}{A_{MBZ}} \int_0^\infty  e^{-|v_\star|k\tau} J_1(kr) k dk  
=\frac{i\pi (\eta \cos(\theta_\RR) +i\sin(\theta_\RR)r}{A_{MBZ} ((v_\star \tau)^2 +r^2)^{3/2}}
\ea 

For negative time ($-\beta<-\tau<0$) 
\ba 
&g_{2,\eta}(-\tau, \RR) = -g_{2,\eta}(\beta-\tau,\RR) \\
=&-\frac{1}{A_{MBZ}} \int \frac{|v_{\star}| 2\pi i(\eta \cos(\theta_\RR)+i\sin(\theta_\RR))}{2|v_\star|k }
(1-f(|v_\star|k))(e^{-|v_\star|k(\beta-\tau)} - e^{-|v_\star|k(\tau)}J_1(kr) k^2dk  \\
\ea 
We let $\beta \rightarrow \infty$ and set the momentum cutoff to infinity.
\ba   
&g_{2,\eta}(-\tau,\RR)=- \frac{i\pi (\eta \cos(\theta_\RR) +i\sin(\theta_\RR)}{A_{MBZ}} \int_0^\infty  e^{-|v_\star|k\tau} J_1(kr) k dk  
=-\frac{i\pi (\eta \cos(\theta_\RR) +i\sin(\theta_\RR)r}{A_{MBZ} ((v_\star \tau)^2 +r^2)^{3/2}}
\ea

In summary, 
\baa 
g_{2,\eta}(\tau,\RR) = \text{sgn}(\tau) \frac{i\pi (\eta \cos(\theta_\RR) +i\sin(\theta_\RR)r}{A_{MBZ} ((v_\star \tau)^2 +r^2)^{3/2}}
\label{eq:g2}
\eaa  
and also 
$
g_{2,-}(\tau,\RR) = [g_{2,+}(\tau,\RR)]^*\, .
$

\subsection{Single-particle Green's function at $M\ne 0$}
\bh{In this section, we calculate the single-particle Green's function at $M\ne 0 $, $\nu_c=0$, which has been used in Sec.~\ref{sec:sec_zero_hyb_j}. The Hamiltonian is given in Eq.~\ref{eq:single_particle_ham} (but with a non-zero $M$) and is also listed below. }
\baa  
 &\hH_c'=\sum_{\kk, \eta, a,a',s}H_{aa'}^{(c,\eta)}(\kk) c_{\kk,a\eta s}^\dag c_{\kk,a'\eta s} + \text{const} \nonumber \\
&H^{(c,\eta)}(\kk) = \begin{bmatrix}
0_{2\times 2} &v_{\star}(\eta k_x  \sigma_0 +ik_y \sigma_z)\\
v_{\star}(\eta k_x  \sigma_0 +ik_y \sigma_z)& M\sigma_x
\end{bmatrix}  
\eaa 
The corresponding Green's function in the Matsubara frequency $\omega_n$ is
\baa  
&(i\omega_n -H^{(c,\eta)}(\kk))^{-1} \nonumber \\
= &\frac{1}{-\omega^2 -|v_\star|^2k^2}
\begin{bmatrix}
i\omega_n & 0& v_\star(\eta k_x +ik_y)&0 \\
0&i\omega_n &0 & v_\star(\eta k_x -ik_y)\\
v_\star (\eta k_x-ik_y) & 0 & i\omega_n &0 \\
0& v_\star(\eta k_x+ik_y) & 0 & i\omega_n
\end{bmatrix} \nonumber \\
&+\frac{M}{(-\omega_n^2-|v_\star|^2k^2)^2}
\begin{bmatrix}
0 & |v_\star|^2(\eta k_x+ik_y)^2 & 0 & i\omega v_\star(\eta k_x +ik_y) \\
 |v_\star|^2(\eta k_x-ik_y)^2 & 0 & i\omega v_\star(\eta k_x -ik_y) & 0 \\
 0 & i\omega v_\star(\eta k_x +ik_y) & 0& -\omega^2 \\
 i\omega v_\star(\eta k_x -ik_y) & 0 & -\omega^2 & 0 
\end{bmatrix} +O(M^2)
\eaa  
where $k= \sqrt{k_x^2+k_y^2}$ and we expand in powers of $M$. We consider the following single-particle Green's functions in this case
\ba 
&G_{33,\eta}(\kk,\tau) = -\langle T_\tau c_{\kk,3\eta s}(\tau)c_{\kk, 3\eta s}(0)\rangle = \frac{1}{\beta}\sum_{i\omega_n}  \frac{i\omega_n}{-\omega_n^2 -|v_\star|^2 k^2 }e^{-i\omega_n \tau} +\bh{O(M^2)}\\
&G_{44,\eta}(\kk,\tau) = -\langle T_\tau c_{\kk,4\eta s}(\tau)c_{\kk, 4\eta s}(0)\rangle = \frac{1}{\beta}\sum_{i\omega_n} \frac{i\omega_n}{-\omega_n^2 -|v_\star|^2 k^2 }e^{-i\omega_n \tau}+\bh{O(M^2)} \\
&G_{34,\eta}(\kk,\tau)=-\langle T_\tau c_{\kk,3\eta s}(\tau)^\dag c_{\kk, 4 \eta s}\rangle =\frac{1}{\beta}\sum_{i\omega_n}\frac{-M\omega_n^2}{(-\omega_n^2 -|v_\star|^2 k^2)^2}e^{-i\omega_n\tau} +\bh{O(M^2)} \\
&G_{43,\eta}(\kk,\tau)=-\langle T_\tau c_{\kk,4\eta s}(\tau)^\dag c_{\kk, 3 \eta s}\rangle =\frac{1}{\beta}\sum_{i\omega_n}\frac{-M\omega_n^2}{(-\omega_n^2 -|v_\star|^2 k^2)^2}e^{-i\omega_n\tau}  +\bh{O(M^2)}
\ea 
We let 
\baa  
g_1(\kk,\tau)= \frac{1}{\beta}\sum_{i\omega_n}\frac{-M\omega_n^2}{(-\omega_n^2 -|v_\star|^2 k^2)^2}e^{-i\omega_n\tau}  \, . 
\label{eq:gfun}
\eaa 
Then, we have 
\ba 
&
G_{33,\eta}(\kk,\tau) = G_{44,\eta}(\kk,\tau) =g_0(\kk,\tau)\\
&G_{34,\eta}(\kk, \tau) = G_{43,\eta}(\kk,\tau) = g_1(\kk,\tau) 
\ea 
and the corresponding Fourier transformation
\ba 
g_1(\RR,\tau) = \frac{1}{N_M} \sum_{\kk}g_1(\kk,\tau)e^{i\kk \cdot \RR} 
\ea 

We next give the analytical expression of $g_1(\RR,\tau)$ at zero temperature $\beta\rightarrow \infty$ and thermodynamic limit. In the thermodynamic limit, we can replace the momentum summation with the momentum integral 
\ba 
\frac{1}{N_M}\sum_{\kk} \rightarrow \frac{1}{A_{MBZ}}\int_{|\kk|<\Lambda_c}
\ea 
where $A_{MBZ}$ is the area of first moir\' e Brillouin zone and $\Lambda_c$ is the momentum cutoff. In addition, during the calculation, we take $\Lambda_c =\infty$ to obtain the analytical expression.

\ba 
g_1(\RR,\tau) = 
&\frac{1}{A_{MBZ}}\int g_1(\kk,\tau)e^{i\kk \cdot \RR} d^2\kk  \\ 
=& \frac{M}{A_{MBZ}}\int  \frac{1}{\beta}\sum_{i\omega_n}\frac{-\omega_n^2}{(-\omega_n^2 -|v_\star|^2k^2)^2}e^{-i\omega_n\tau} e^{i\kk \cdot \RR} d^2\kk 
\ea 
Using Eq.~\ref{eq:exp_eikr}, we have 
\baa  
g_1(\RR,\tau) =& \frac{M}{A_{MBZ}}\int_0^{2\pi} \int_0^{\Lambda_c}  \frac{1}{\beta}\sum_{i\omega_n}\frac{-\omega_n^2}{(-\omega_n^2 -|v_\star|^2k^2)^2}[ J_0(kr) + 2 \sum_{m=1}i^m J_m(kr)] e^{-i\omega_n\tau} k dk d\theta  \nonumber  \\ 
=&\frac{2\pi M}{A_{MBZ}}\int_0^{\Lambda_c}  \frac{1}{\beta}\sum_{i\omega_n}\frac{-\omega_n^2}{(-\omega_n^2 -|v_\star|^2k^2)^2} J_0(kr)e^{-i\omega_n\tau}
  k dk
 \label{eq:g1_med}
\eaa  

Now, we evaluate the following Matsubara summation via contour integral:
\ba 
\frac{1}{\beta} \sum_{i\omega_n} \frac{-\omega_n^2}{(-\omega_n^2 - |v_\star|^2k^2)^2} e^{-i\omega_n\tau}
\ea 
We fist consider \bh{$\beta>\tau>0$} case. 
We change Matsubara summation to a contour integral 
\ba 
\frac{1}{\beta} \sum_{i\omega_n} \frac{-\omega_n^2}{(-\omega_n^2 - |v_\star|^2k^2)^2} e^{-i\omega_n\tau}
=&\int_C \frac{dz}{2\pi i}(f(z)-1) 
\frac{z^2}{(z^2- |v_\star|^2k^2)^2} e^{-z\tau}
\ea 
where the contour $C= [0^+ + i\infty,0^+-i\infty] \cup [0^--i\infty,0^-+i\infty] $. Since, the poles only appear on the real axis ($\text{Im}[z]=0$), We distort the contour to $C'=[\infty +i0^+, -\infty+i0^+]\cup [-\infty -i0^+, \infty -i0^+]$. 
This gives 
\ba 
&\frac{1}{\beta} \sum_{i\omega_n} \frac{-\omega_n^2}{(-\omega_n^2 - |v_\star|^2k^2)^2} e^{-i\omega_n\tau}= \int_{C'} \frac{dz}{2\pi i}(f(z)-1) 
\frac{z^2}{(z^2- |v_\star|^2k^2)^2} e^{-z\tau}
\ea 
We aim to evaluate the integral via residual theorem. The residues around two poles at $z = \pm |v_\star k| $ are 
\ba 
&\text{Res}\bigg((f(z)-1) 
\frac{z^2}{(z^2- |v_\star|^2k^2)^2} e^{-z\tau} ,|v_\star k| \bigg)  =
\frac{e^{-|v_\star k|\tau }}{4|v_\star k| }
\bigg[ 
(1-|v_\star k|\tau ) f(-|v_\star k|)^2 
+ [1+|v_\star k|(\beta-\tau)] f(-|v_\star k|) (1-f(-|v_\star k|))
\bigg]  \\
&\text{Res}\bigg((f(z)-1) 
\frac{z^2}{(z^2- |v_\star|^2k^2)^2} e^{-z\tau} ,-|v_\star k| \bigg)  =-
\frac{e^{|v_\star k|\tau }}{4|v_\star k| }
\bigg[ 
(1+|v_\star k|\tau ) f(|v_\star k|)^2 
+ [1-|v_\star k|(\beta-\tau)] f(-|v_\star k|) (1-f(-|v_\star k|))
\bigg]  \\
\ea 
Using the residue theorem, we have 
\baa  
&\frac{1}{\beta} \sum_{i\omega_n} \frac{-\omega_n^2}{(-\omega_n^2 - |v_\star|^2k^2)^2} e^{-i\omega_n\tau}=
\text{Res}\bigg((f(z)-1) 
\frac{z^2}{(z^2- |v_\star|^2k^2)^2} e^{-z\tau} ,|v_\star k| \bigg) 
+ \text{Res}\bigg((f(z)-1) 
\frac{z^2}{(z^2- |v_\star|^2k^2)^2} e^{-z\tau} ,-|v_\star k| \bigg) \nonumber   \\
=&\frac{- |v_\star k| \tau \cosh(|v_\star k|(\beta-\tau)) + |v_\star k|(\beta-\tau)\cosh(|v_\star k|\tau) + \sinh(|v_\star k|(\beta-\tau)) - \sinh(|v_\star k|\tau) }{4 |v_\star k | (1+\cosh(|v_\star k|\beta)} 
\label{eq:matsu_sum}
\eaa  
where $\beta>\tau>0$. 
For the negative time, we can use the fact that 
\baa  
\frac{1}{\beta} \sum_{i\omega_n} \frac{-\omega_n^2}{(-\omega_n^2 - |v_\star|^2k^2)^2} e^{-i\omega_n(-\tau) } 
 = - \frac{1}{\beta} \sum_{i\omega_n} \frac{-\omega_n^2}{(-\omega_n^2 - |v_\star|^2k^2)^2} e^{-i\omega_n(\beta-\tau) } 
\eaa  
Then we can replace $-\tau$ with $\beta -\tau$ and add a minus sign to Eq.~\ref{eq:matsu_sum}. This leads to 
\baa  
&\frac{1}{\beta} \sum_{i\omega_n} \frac{-\omega_n^2}{(-\omega_n^2 - |v_\star|^2k^2)^2} e^{-i\omega_n(-\tau) } \nonumber  \\
=&-\frac{- |v_\star k| \tau \cosh(|v_\star k|\tau) + |v_\star k|\tau\cosh(|v_\star k|(\beta-\tau)) + \sinh(|v_\star k|\tau) - \sinh(|v_\star k|(\beta-\tau)) }{4 |v_\star k | (1+\cosh(|v_\star k|\beta)}  \label{eq:matsu_sum_2}
\eaa  

Combining Eq.~\ref{eq:g1_med} and Eq.~\ref{eq:matsu_sum} we have ($\tau>0$)
\baa  
&g_1(\RR,\tau)  \nonumber \\ 
=&\frac{2\pi M}{A_{MBZ}}\int_0^{\Lambda_c}   J_0(kr)
\frac{- |v_\star k| \tau \cosh(|v_\star k|(\beta-\tau)) + |v_\star k|(\beta-\tau)\cosh(|v_\star k|\tau) + \sinh(|v_\star k|(\beta-\tau)) - \sinh(|v_\star k|\tau) }{4 |v_\star k | (1+\cosh(|v_\star k|\beta)} 
  k dk 
\eaa  
At zero temperature and $\Lambda_c=\infty$, it becomes 
\ba 
g_1(\RR,\tau)  =& \frac{2\pi M}{A_{MBZ}}\int_0^{\infty} 
J_0(kr) \frac{-e^{-|v_\star k|\tau } (-1 + |v_\star k|\tau)}{4|v_\star k| }k dk  \\
&
=-\frac{\pi  M }{2A_{MBZ} |v_\star|}\frac{r^2}{\bigg( |v_\star \tau|^2 + r^2\bigg)^{3/2} } 
\ea 

For the negative time, combining  Eq.~\ref{eq:g1_med} and Eq.~\ref{eq:matsu_sum}, we have  $(\tau>0)$
\ba 
g_1(\RR,-\tau)
=&\frac{2\pi M}{A_{MBZ}}\int_0^{\Lambda_c}   J_0(kr)
(-1)\frac{- |v_\star k| \tau \cosh(|v_\star k|\tau) + |v_\star k|\tau\cosh(|v_\star k|(\beta-\tau)) + \sinh(|v_\star k|\tau) - \sinh(|v_\star k|(\beta-\tau)) }{4 |v_\star k | (1+\cosh(|v_\star k|\beta)} 
  k dk \\
\ea 
At zero temperature and $\Lambda_c=\infty$, it becomes 
\ba 
g_1(\RR,-\tau)  =&\frac{2\pi M}{A_{MBZ}}\int_0^{\infty} 
J_0(kr) \frac{-e^{-|v_\star k|\tau } (-1 + |v_\star k|\tau)}{4|v_\star k| }k dk  \\
&
=-\frac{\pi  M }{2A_{MBZ} |v_\star|}\frac{r^2}{\bigg( |v_\star \tau|^2 + r^2\bigg)^{3/2} } 
\ea 

In summary, we have 
\baa  
g_1(\RR,\tau) = -\frac{\pi  M }{2A_{MBZ} |v_\star|}\frac{r^2}{\bigg( |v_\star \tau|^2 + r^2\bigg)^{3/2} } 
\label{eq:g1}
\eaa  
\hb{Here, unlike $g_0(\RR,\tau)$, $g_1(\RR,\tau)$ is not proportional to $\text{sgn}(\tau)$. This is because $g_1(\RR,\tau) = -\langle T_\tau c_{\kk,a\eta s}(\tau)c_{\kk,a\eta s}(0)\rangle $ which is the correlation functions of two fermionic operators with same $\kk,\alpha \eta s$ indices. However, $g_2(\RR,\tau) = -\langle T_\tau c_{\kk,3\eta s}(\tau)c_{\kk,4\eta s}(0)\rangle $ is the correlation functions of two fermionic operators with different $\kk,\alpha \eta s$ indices. Thus, the particle-hole symmetry will not enforce e $g_2(\RR,\tau) \propto \text{sgn}(\tau)$.}

\section{Four-fermioin correlation function I}
\label{sec:chiral_u4_corre}
We now calculate the following correlation function in Eq.~\ref{eq:chic_real} 
\ba 
 \chi_c(\RR,\tau,\xi,\xi_2) =-&\sum_{\kk,\qq}\sum_{a_1,a_2 =3,4}\sum_{\eta s   } \frac{1}{2}\frac{1}{2N_M^2} \delta_{\xi, (-1)^{a-1}\eta} \delta_{\xi_2,(-1)^{a_2-1}\eta}
     G_{a_2a,\eta }(\kk+\qq,-\tau)G_{aa_2,\eta }(\kk, \tau) ^{i\qq\cdot \RR } 
 \ea 
 
For $\xi=\xi_2$,
\baa 
 \chi_c(\RR,\tau,\xi,\xi) =-&\sum_{\kk,\qq}\sum_{a_1,a_2 =3,4}\sum_{\eta s   } \frac{1}{2}\frac{1}{2N_M^2} \delta_{\xi, (-1)^{a-1}\eta} \delta_{\xi,(-1)^{a_2-1}\eta}
     G_{a_2a,\eta }(\kk+\qq,-\tau)G_{aa_2,\eta }(\kk, \tau) ^{i\qq\cdot \RR } \nonumber \\
=& - \sum_{\kk} \frac{1}{N_M^2}g_0(\kk+\qq,-\tau)g_0(\kk,\tau)e^{i\qq \cdot \RR} +o(M^2) \nonumber \\
=& - g_0(\RR,-\tau) g_0(-\RR,\tau)+o(M^2) = \frac{\pi^2}{A_{MBZ}^2}\frac{|v_\star \tau|^2}{(|v_\star \tau|^2 +r^2)^{3} } +O(M^3) 
\label{eq:chic_ana}
 \eaa  
 where we use the analytical expression of Green's function at $\beta=\infty$, $\Lambda_c=\infty$ as shown in Eq.~\ref{eq:g0}.

For $\xi=-\xi_2$,
\baa 
 \chi_c(\RR,\tau,\xi,-\xi) =-&\sum_{\kk,\qq}\sum_{a_1,a_2 =3,4}\sum_{\eta s   } \frac{1}{2}\frac{1}{2N_M^2} \delta_{\xi, (-1)^{a-1}\eta} \delta_{-\xi,(-1)^{a_2-1}\eta}
     G_{a_2a,\eta }(\kk+\qq,-\tau)G_{aa_2,\eta }(\kk, \tau) ^{i\qq\cdot \RR } \nonumber \\
=& - \sum_{\kk} \frac{1}{N_M^2}g_1(\kk+\qq,-\tau)g_1(\kk,\tau)e^{i\qq \cdot \RR}  +O(M^3)  \nonumber  \\
=& - g_1(\RR,-\tau) g_1(-\RR,\tau) +o(M^2)  =
-\frac{\pi^2M^2}{4A_{MBZ}^2 |v_\star|^2} \frac{r^4}{\bigg( |v_\star \tau|^2 + r^2\bigg)^3}+O(M^3) 
\label{eq:chic_ana_2}
 \eaa  
 where we use the analytical expression of Green's function at $\beta=\infty$, $\Lambda_c=\infty$ as shown in Eq.~\ref{eq:g1}.

\section{Four-fermioin correlation function II}
\label{sec:flat_u4_corre}
In this section, we evaluate the four-fermion correlations at $M=0, \nu=\nu_f=0,-1,-2,\nu_c=0$, that has been used in Sec.~\ref{sec:rkky}. \bh{Unlike Sec.~\ref{sec:chiral_u4_corre}, we consider $M=0$ limit and work in the $\psi$ basis.} 
We consider the following correlators
\baa  
&\langle :\UF_{\mu\nu}^{(c',\xi \xi')}(\rr_1,\rr_1',\tau): :\UF_{\mu_2\nu_2}^{(c',\xi_2'\xi_2)} (\rr_2',\rr_2,0) : \rangle_{0,con}  \nonumber \\
=& -\sum_{n_1,n_2,n_3,n_4} 
[T^{\mu\nu}]_{n_1,n_2}
[T^{\mu_2\nu_2}]_{n_3,n_4}
\langle 
\psi^{c',\xi'}_{\rr_1',n_2}(\tau) \psi^{c',\xi_2',\dag}_{\rr_2',n_3}(0)\rangle_0
\langle 
\psi^{c',\xi_2}_{\rr_2,n_4}(\tau) \psi^{c',\xi,\dag}_{\rr_1n_1}(0)\rangle_0 \nonumber 
\\
&\langle :\UF_{\mu\nu}^{(c'',\xi \xi')}(\rr_1,\rr_1',\tau): :\UF_{\mu_2\nu_2}^{(c'',\xi_2\xi_2')} (\rr_2',\rr_2,0) : \rangle_{0,con} \nonumber \\
=& -\sum_{n_1,n_2,n_3,n_4} 
[T^{\mu\nu}]_{n_1,n_2}
[T^{\mu_2\nu_2}]_{n_3,n_4}
\langle 
\psi^{c'',\xi'}_{\rr_1',n_2}(\tau) \psi^{c'',\xi_2',\dag}_{\rr_2',n_3}(0)\rangle_0
\langle 
\psi^{c'',\xi_2}_{\rr_2,n_4}(\tau) \psi^{c'',\xi,\dag}_{\rr_1n_1}(0)\rangle_0\nonumber 
\\
&\langle :\UF_{\mu\nu}^{(c',\xi \xi')}(\rr_1,\rr_1',\tau): :\UF_{\mu_2\nu_2}^{(c'',\xi_2'\xi_2)} (\rr_2',\rr_2,0) : \rangle_{0,con} \nonumber \\
=& -\sum_{n_1,n_2,n_3,n_4} 
[T^{\mu\nu}]_{n_1,n_2}
[T^{\mu_2\nu_2}]_{n_3,n_4}
\langle 
\psi^{c',\xi'}_{\rr_1',n_2}(\tau) \psi^{c'',\xi_2',\dag}_{\rr_2',n_3}(0)\rangle_0
\langle 
\psi^{c'',\xi_2}_{\rr_2,n_4}(\tau) \psi^{c',\xi,\dag}_{\rr_1n_1}(0)\rangle_0
\label{eq:def_u4u4_corre}
\eaa  
Since $S_0$ has flat $U(4)$ symmetry, only components with $\mu\nu = \mu_2\nu_2$ can be non-zero. In addition, at $M=0$, the single-particle Green's function $-\langle T_\tau c_{\kk,\alpha\eta s}(\tau) c_{\kk',\alpha'\eta's'} (0)\rangle $ equals to zero if $(-1)^{\alpha+1}\eta \ne (-1)^{\alpha'+1}\eta'$. This is because the single-particle Hamiltonian of conduction electrons does not have a hybridization term between two $c$ electrons with opposite $\xi$ indices. 
Therefore, only components with $\xi=\xi_2, \xi'=\xi_2'$ is non-zero, and we only need to calculate
\ba 
&\langle :\UF_{\mu\nu}^{(c',\xi \xi')}(\rr_1,\rr_1',\tau): :\UF_{\mu \nu}^{(c',\xi'\xi)} (\rr_2',\rr_2,0) : \rangle_{0,con} 
\\
&\langle :\UF_{\mu\nu}^{(c'',\xi \xi')}(\rr_1,\rr_1',\tau): :\UF_{\mu \nu}^{(c'',\xi'\xi)}  (\rr_2',\rr_2,0) : \rangle_{0,con} 
\\
&\langle :\UF_{\mu\nu}^{(c',\xi \xi')}(\rr_1,\rr_1',\tau): :\UF_{\mu \nu}^{(c'',\xi'\xi)}  (\rr_2',\rr_2,0) : \rangle_{0,con} 
\ea 
where $\langle A B \rangle_{0,con} = \langle A B \rangle_0- \langle A \rangle_0 \langle B \rangle_0$

We next prove all $\mu\nu $ components are equivalent. \bh{For the components of $\mu\nu \ne 00$, they are connected by flat $U(4)$ rotation and are equivalent. To show it, we consider a $SU(4) \subset U(4)$ rotation $g$ that will rotates $\UF_{\mu\nu}^{(c',\xi\xi')}$ ($\mu\nu \ne 00$).
\baa  
g \UF_{\mu\nu}^{(c',\xi\xi')} g^{-1} = \sum_{\mu_2\nu_2} R_{\mu\nu,\mu_2\nu_2}(g) \UF_{\mu_2\nu_2}^{(c',\xi\xi')}
\eaa  
where $R_{\mu\nu,\mu_2\nu_2}(g)$ is the representation matrix of $SU(4)$ rotation corresponding to the adjoint representation of $SU(4)$ (Note that $\{\UF_{\mu\nu}^{(c',\xi\xi')}\}_{\mu\nu\ne 00}$ form an adjoint representation of $SU(4)$ group (subgroup of $U(4)$). Due to the $SU(4)$ symmetry, we have 
\baa  
&\langle :\UF_{\mu\nu}^{(c',\xi \xi')}(\rr_1,\rr_1',\tau): :\UF_{\mu \nu}^{(c',\xi'\xi)} (\rr_2',\rr_2,0) : \rangle_{0,con}  = \langle g:\UF_{\mu\nu}^{(c',\xi \xi')}(\rr_1,\rr_1',\tau): :\UF_{\mu \nu}^{(c',\xi'\xi)} (\rr_2',\rr_2,0) : g^{-1}\rangle_{0,con} 
\nonumber \\
= &\sum_{\mu_2\nu_2,\mu_3\nu_3}R_{\mu\nu,\mu_2\nu_2}(g) R_{\mu\nu,\mu_3\nu_3}(g) \langle :\UF_{\mu_2\nu_2}^{(c',\xi \xi')}(\rr_1,\rr_1',\tau): :\UF_{\mu_3\nu_3}^{(c',\xi'\xi)} (\rr_2',\rr_2,0) : \rangle_{0,con}  \nonumber \\
= &\sum_{\mu_2\nu_2}R_{\mu\nu,\mu_2\nu_2}(g) R_{\mu\nu,\mu_3\nu_3}(g) \langle :\UF_{\mu_2\nu_2}^{(c',\xi \xi')}(\rr_1,\rr_1',\tau): :\UF_{\mu_2\nu_2}^{(c',\xi'\xi)} (\rr_2',\rr_2,0) : \rangle_{0,con} 
\label{eq:chiralchiral_u4u4}
\eaa  
Eq.~\ref{eq:chiralchiral_u4u4} needs to be satisfied for any $SU(4)$ rotation $g$, which requires  all the $\mu\nu \ne 00$ components are equivalent (here we pick $0z$ components as a representation)
\baa  
\langle :\UF_{\mu\nu}^{(c',\xi \xi')}(\rr_1,\rr_1',\tau): :\UF_{\mu \nu}^{(c',\xi'\xi)} (\rr_2',\rr_2,0) : \rangle_{0,con}  = \langle :\UF_{0z}^{(c',\xi \xi')}(\rr_1,\rr_1',\tau): :\UF_{0z}^{(c',\xi'\xi)} (\rr_2',\rr_2,0) : \rangle_{0,con} ,\quad \mu\nu \ne 00
\eaa
Similarly, for other correlators, we also have
\baa  
\langle :\UF_{\mu\nu}^{(c'',\xi \xi')}(\rr_1,\rr_1',\tau): :\UF_{\mu \nu}^{(c'',\xi'\xi)} (\rr_2',\rr_2,0) : \rangle_{0,con}  = \langle :\UF_{0z}^{(c'',\xi \xi')}(\rr_1,\rr_1',\tau): :\UF_{0z}^{(c'',\xi'\xi)} (\rr_2',\rr_2,0) : \rangle_{0,con} ,\quad \mu\nu \ne 00 \nonumber \\
\langle :\UF_{\mu\nu}^{(c',\xi \xi')}(\rr_1,\rr_1',\tau): :\UF_{\mu \nu}^{(c'',\xi'\xi)} (\rr_2',\rr_2,0) : \rangle_{0,con}  = \langle :\UF_{0z}^{(c',\xi \xi')}(\rr_1,\rr_1',\tau): :\UF_{0z}^{(c'',\xi'\xi)} (\rr_2',\rr_2,0) : \rangle_{0,con} ,\quad \mu\nu \ne 00
\eaa
}

We next prove $\langle :\UF_{0z}^{(c',\xi \xi')}(\rr_1,\rr_1',\tau): :\UF_{0z}^{(c',\xi'\xi)} (\rr_2',\rr_2,0) : \rangle_{0,con} = \langle :\UF_{00}^{(c',\xi \xi')}(\rr_1,\rr_1',\tau): :\UF_{00}^{(c',\xi'\xi)} (\rr_2',\rr_2,0) : \rangle_{0,con} $.
$\langle :\UF_{0z}^{(c',\xi \xi')}(\rr_1,\rr_1',\tau): :\UF_{0z}^{(c',\xi'\xi)} (\rr_2',\rr_2,0) : \rangle_{0,con}$ takes the form of
\ba 
&\langle :\UF_{0z}^{(c',\xi \xi')}(\rr_1,\rr_1',\tau): :\UF_{0z}^{(c',\xi'\xi)} (\rr_2',\rr_2,0) : \rangle_{0,con}  \\
=  &-\sum_{n_1,n_2,n_3,n_4} 
[T^{0z}]_{n_1,n_2}
[T^{0z}]_{n_3,n_4}
\langle 
\psi^{c',\xi'}_{\rr_1',n_2}(\tau) \psi^{c',\xi_2',\dag}_{\rr_2',n_3}(0)\rangle_0
\langle 
\psi^{c',\xi_2}_{\rr_2,n_4}(\tau) \psi^{c',\xi,\dag}_{\rr_1,n_1}(0)\rangle_0
\\
=&-\frac{1}{4}\sum_{n_1,n_2} 
s_{n_1}s_{n_3} 
\langle 
\psi^{c',\xi'}_{\rr_1',n_2}(\tau) \psi^{c',\xi_2',\dag}_{\rr_2',n_2}(0)\rangle_0
\langle 
\psi^{c',\xi_2}_{\rr_2,n_1}(\tau) \psi^{c',\xi,\dag}_{\rr_1n_1}(0)\rangle_0 \delta_{n_1, n_2} \delta_{n_3,n_4} \delta_{n_2,n_3} \delta_{n_1,n_4} \\
=&-\frac{1}{4}\sum_{n_1,n_2} 
\langle 
\psi^{c',\xi'}_{\rr_1',n_2}(\tau) \psi^{c',\xi_2',\dag}_{\rr_2',n_2}(0)\rangle_0
\langle 
\psi^{c',\xi_2}_{\rr_2,n_1}(\tau) \psi^{c',\xi,\dag}_{\rr_1n_1}(0)\rangle_0  
\ea 
where $s_n \in \{+1,-1\}$ is the spin index of $\psi_{\rr,n}^{c',\xi'}$ electrons.  For $\mu\nu=00$ component, we have 
\ba 
&\langle :\UF_{00}^{(c',\xi \xi')}(\rr_1,\rr_1',\tau): :\UF_{00}^{(c',\xi'\xi)} (\rr_2',\rr_2,0) : \rangle_{0,con}  \\
=  &-\frac{1}{4}\sum_{n_1,n_2,n_3,n_4} 
[T^{00}]_{n_1,n_2}
[T^{00}]_{n_3,n_4}
\langle 
\psi^{c',\xi'}_{\rr_1',n_2}(\tau) \psi^{c',\xi_2',\dag}_{\rr_2',n_3}(0)\rangle_0
\langle 
\psi^{c',\xi_2}_{\rr_2,n_4}(\tau) \psi^{c',\xi,\dag}_{\rr_1,n_1}(0)\rangle_0
\\
=&-\frac{1}{4}\sum_{n_1,n_2} 
\langle 
\psi^{c',\xi'}_{\rr_1',n_2}(\tau) \psi^{c',\xi_2',\dag}_{\rr_2',n_2}(0)\rangle_0
\langle 
\psi^{c',\xi_2}_{\rr_2,n_1}(\tau) \psi^{c',\xi,\dag}_{\rr_1n_1}(0)\rangle_0 \delta_{n_1, n_2} \delta_{n_3,n_4} \delta_{n_2,n_3} \delta_{n_1,n_4} \\
=&\langle :\UF_{0z}^{(c',\xi \xi')}(\rr_1,\rr_1',\tau): :\UF_{0z}^{(c',\xi'\xi)} (\rr_2',\rr_2,0) : \rangle_{0,con} \, . 
\ea 
Therefore all $\mu\nu$ components are equivalent. For the same reason, this is true for all three correlators, and we have 
\baa  
&\langle :\UF_{\mu\nu}^{(c',\xi \xi')}(\rr_1,\rr_1',\tau): :\UF_{\mu \nu}^{(c',\xi'\xi)} (\rr_2',\rr_2,0) : \rangle_{0,con} =\langle :\UF_{00}^{(c',\xi \xi')}(\rr_1,\rr_1',\tau): :\UF_{00}^{(c',\xi'\xi)} (\rr_2',\rr_2,0) : \rangle_{0,con}  \nonumber 
\\
&\langle :\UF_{\mu\nu}^{(c'',\xi \xi')}(\rr_1,\rr_1',\tau): :\UF_{\mu \nu}^{(c'',\xi'\xi)}  (\rr_2',\rr_2,0) : \rangle_{0,con} 
=\langle :\UF_{00}^{(c'',\xi \xi')}(\rr_1,\rr_1',\tau): :\UF_{00}^{(c'',\xi'\xi)} (\rr_2',\rr_2,0) : \rangle_{0,con}  \nonumber 
\\
&\langle :\UF_{\mu\nu}^{(c',\xi \xi')}(\rr_1,\rr_1',\tau): :\UF_{\mu \nu}^{(c'',\xi'\xi)}  (\rr_2',\rr_2,0) : \rangle_{0,con} 
=\langle :\UF_{00}^{(c',\xi \xi')}(\rr_1,\rr_1',\tau): :\UF_{00}^{(c'',\xi'\xi)} (\rr_2',\rr_2,0) : \rangle_{0,con} 
\eaa

Using Wick's theorem and single-particle Green's function, the first correlator \bh{in Eq.~\ref{eq:def_u4u4_corre}} becomes
\baa  
&\langle :\UF_{\mu\nu}^{(c',\xi \xi')}(\rr_1,\rr_1',\tau): :\UF_{\mu \nu}^{(c',\xi'\xi)} (\rr_2',\rr_2,0) : \rangle_{0,con}  \nonumber \\
=&\langle :\UF_{00}^{(c',\xi \xi')}(\rr_1,\rr_1',\tau): :\UF_{00}^{(c',\xi'\xi)} (\rr_2',\rr_2,0) : \rangle_{0,con} 
=\frac{1}{4}\sum_{a,b} \langle 
:\psi_{\rr_1,a}^{c',\xi,\dag}(\tau) \psi_{\rr_1',a}^{c',\xi'}(\tau) ::
\psi_{\rr_2',b}^{c',\xi',\dag}(0) \psi_{\rr_2,b}^{c',\xi}(0):
\rangle_{0,con} \nonumber \\
=&-\frac{1}{4}\sum_{a,b} \langle 
\psi_{\rr_2,b}^{c',\xi}(0)\psi_{\rr_1,a}^{c',\xi,\dag}(\tau)\rangle \langle \psi_{\rr_1',a}^{c',\xi'}(\tau) 
\psi_{\rr_2',b}^{c',\xi',\dag}(0)\rangle 
=-g_0(\rr_2-\rr_1,-\tau) g_0(\rr_1'-\rr_2',\tau)
\label{eq:u4u4correlator_0} 
\, .
\eaa  
The second correlator \bh{in Eq.~\ref{eq:def_u4u4_corre}} is
\baa  
&\langle :\UF_{\mu\nu}^{(c'',\xi \xi')}(\rr_1,\rr_1',\tau): :\UF_{\mu \nu}^{(c'',\xi'\xi)} (\rr_2',\rr_2,0) : \rangle_{0,con}   \nonumber  \\
=&\langle :\UF_{00}^{(c'',\xi \xi')}(\rr_1,\rr_1',\tau): :\UF_{00}^{(c'',\xi'\xi)} (\rr_2',\rr_2,0) : \rangle_{0,con} 
=\frac{1}{4}\sum_{a,b} \langle 
:\psi_{\rr_1,a}^{c'',\xi,\dag}(\tau) \psi_{\rr_1',a}^{c'',\xi'}(\tau) ::
\psi_{\rr_2',b}^{c'',\xi',\dag}(0) \psi_{\rr_2,b}^{c'',\xi}(0):
\rangle_{0,con} \nonumber  \\
=&-\frac{1}{4}\sum_{a,b} \langle 
\psi_{\rr_2,b}^{c'',\xi}(0)\psi_{\rr_1,a}^{c'',\xi,\dag}(\tau)\rangle \langle \psi_{\rr_1',a}^{c'',\xi'}(\tau) 
\psi_{\rr_2',b}^{c'',\xi',\dag}(0)\rangle  =-g_0(\rr_2-\rr_1,-\tau) g_0(\rr_1'-\rr_2',\tau)
\label{eq:u4u4correlator_2}
\,.
\eaa  
The third correlator \bh{in Eq.~\ref{eq:def_u4u4_corre}} is 
\baa  
&\langle :\UF_{\mu\nu}^{(c',\xi \xi')}(\rr_1,\rr_1',\tau): :\UF_{\mu \nu}^{(c'',\xi'\xi)} (\rr_2',\rr_2,0) : \rangle_{0,con}  \nonumber \\
=&\langle :\UF_{00}^{(c',\xi \xi')}(\rr_1,\rr_1',\tau): :\UF_{00}^{(c'',\xi'\xi)} (\rr_2',\rr_2,0) : \rangle_{0,con} 
=\frac{1}{4}\sum_{a,b} \langle 
:\psi_{\rr_1,a}^{c',\xi,\dag}(\tau) \psi_{\rr_1',a}^{c',\xi'}(\tau) ::
\psi_{\rr_2',b}^{c'',\xi',\dag}(0) \psi_{\rr_2,b}^{c'',\xi}(0):
\rangle_{0,con} \nonumber \\
=&-\frac{1}{4}\sum_{a,b} \langle 
\psi_{\rr_2,b}^{c'',\xi}(0)\psi_{\rr_1,a}^{c',\xi,\dag}(\tau)\rangle \langle \psi_{\rr_1',a}^{c',\xi'}(\tau) 
\psi_{\rr_2',b}^{c'',\xi',\dag}(0)\rangle
=-\frac{1}{4}\sum_{a} \langle 
\psi_{\rr_2,a}^{c'',\xi}(0)\psi_{\rr_1,a}^{c',\xi,\dag}(\tau)\rangle \langle \psi_{\rr_1',a}^{c',\xi'}(\tau) 
\psi_{\rr_2',a}^{c'',\xi',\dag}(0)\rangle\, .
\label{eq:u4u4correlator_3_v0}
\eaa  
For Eq.~\ref{eq:u4u4correlator_3_v0}, we consider each $\xi,\xi'$ combination. For $\xi=\xi' = +1$: 
\baa  
&\langle :\UF_{\mu\nu}^{(c',+1 +1)}(\rr_1,\rr_1',\tau): :\UF_{\mu \nu}^{(c'',+1 +1)} (\rr_2',\rr_2,0) : \rangle_{0,con} =
-\frac{1}{4}\sum_{a} \langle 
\psi_{\rr_2,a}^{c'',+1}(0)\psi_{\rr_1,a}^{c',+1,\dag}(\tau)\rangle \langle \psi_{\rr_1',a}^{c',+1}(\tau) 
\psi_{\rr_2',a}^{c'',+1,\dag}(0)\rangle \nonumber \\
=&
-\frac{1}{2} \bigg( 
G_{31,+}(\rr_2-\rr_1,-\tau)G_{13,+}(\rr_1'-\rr_2',\tau) +
G_{42,-}(\rr_2-\rr_1,-\tau)G_{24,-}(\rr_1'-\rr_2',\tau)
\bigg) \nonumber  \\
=&-\frac{1}{2}[g_{2,+}^*(\rr_1-\rr_2,\tau)g_{2,+}(\rr_1'-\rr_2',\tau) 
+
g_{2,-}(\rr_2-\rr_1,-\tau)g_{2,-}^*(\rr_2'-\rr_1',-\tau)
] \, .
\label{eq:u4u4_corre_pp}
\eaa  
For $\xi=\xi'=-1$:
\baa  
&\langle :\UF_{\mu\nu}^{(c',-1 -1)}(\rr_1,\rr_1',\tau): :\UF_{\mu \nu}^{(c'',-1 -1)} (\rr_2',\rr_2,0) : \rangle_{0,con} =
-\frac{1}{4}\sum_{a} \langle 
\psi_{\rr_2,a}^{c'',-1}(0)\psi_{\rr_1,a}^{c',-1,\dag}(\tau)\rangle \langle \psi_{\rr_1',a}^{c',-1}(\tau) 
\psi_{\rr_2',a}^{c'',-1,\dag}(0)\rangle \nonumber \\
=&
-\frac{1}{2} \bigg( 
G_{31,-}(\rr_2-\rr_1,-\tau)G_{13,-}(\rr_1'-\rr_2',\tau) +
G_{42,+}(\rr_2-\rr_1,-\tau)G_{24,+}(\rr_1'-\rr_2',\tau)
\bigg) \nonumber  \\
=&-\frac{1}{2}[g_{2,-}^*(\rr_1-\rr_2,\tau)g_{2,-}(\rr_1'-\rr_2',\tau) 
+
g_{2,+}(\rr_2-\rr_1,-\tau)g_{2,+}^*(\rr_2'-\rr_1',-\tau)
] \, .
\label{eq:u4u4_corre_nn}
\eaa  
For $\xi=1,\xi'=-1$,
\baa  
&\langle :\UF_{\mu\nu}^{(c',+1 -1)}(\rr_1,\rr_1',\tau): :\UF_{\mu \nu}^{(c'',-1 +1)} (\rr_2',\rr_2,0) : \rangle_{0,con} =
-\frac{1}{4}\sum_{a} \langle 
\psi_{\rr_2,a}^{c'',+1}(0)\psi_{\rr_1,a}^{c',-1,\dag}(\tau)\rangle \langle \psi_{\rr_1',a}^{c',-1}(\tau) 
\psi_{\rr_2',a}^{c'',+1,\dag}(0)\rangle \nonumber \\
=&
-\frac{1}{2} \bigg( 
G_{31,+}(\rr_2-\rr_1,-\tau)G_{13,-}(\rr_1'-\rr_2',\tau) +
G_{42,-}(\rr_2-\rr_1,-\tau)G_{24,+}(\rr_1'-\rr_2',\tau)
\bigg)  \nonumber \\
=&-\frac{1}{2}[g_{2,+}^*(\rr_1-\rr_2,\tau)g_{2,-}(\rr_1'-\rr_2',\tau) 
+
g_{2,-}(\rr_2-\rr_1,-\tau)g_{2,+}^*(\rr_2'-\rr_1',-\tau)
] \, .
\label{eq:u4u4_corre_pn}
\eaa  
For $\xi=-1,\xi'=1$,
\baa  
&\langle :\UF_{\mu\nu}^{(c',-1 +1)}(\rr_1,\rr_1',\tau): :\UF_{\mu \nu}^{(c'',+1 -1)} (\rr_2',\rr_2,0) : \rangle_{0,con} =
-\frac{1}{4}\sum_{a} \langle 
\psi_{\rr_2,a}^{c'',-1}(0)\psi_{\rr_1,a}^{c',+1,\dag}(\tau)\rangle \langle \psi_{\rr_1',a}^{c',+1}(\tau) 
\psi_{\rr_2',a}^{c'',-1,\dag}(0)\rangle \nonumber \\
=&
-\frac{1}{2} \bigg( 
G_{31,-}(\rr_2-\rr_1,-\tau)G_{13,+}(\rr_1'-\rr_2',\tau) +
G_{42,+}(\rr_2-\rr_1,-\tau)G_{24,-}(\rr_1'-\rr_2',\tau)
\bigg) \nonumber  \\
=&-\frac{1}{2}[g_{2,-}^*(\rr_1-\rr_2,\tau)g_{2,+}(\rr_1'-\rr_2',\tau) 
+
g_{2,+}(\rr_2-\rr_1,-\tau)g_{2,-}^*(\rr_2'-\rr_1',-\tau)
] \, .
\label{eq:u4u4_corre_np}
\eaa  
In summary, \bh{combining Eq.~\ref{eq:u4u4_corre_pp}, Eq.~\ref{eq:u4u4_corre_nn}, Eq.~\ref{eq:u4u4_corre_pn} and Eq.~\ref{eq:u4u4_corre_np}}
\baa  
\langle :\UF_{\mu\nu}^{(c',\xi\xi')}(\rr_1,\rr_1',\tau): :\UF_{\mu \nu}^{(c'',\xi'\xi)} (\rr_2',\rr_2,0) : \rangle_{0,con} =&-\frac{1}{2}[g_{2,\xi}^*(\rr_1-\rr_2,\tau)g_{2,\xi'}(\rr_1'-\rr_2',\tau) 
+
g_{2,-\xi}(\rr_2-\rr_1,-\tau)g_{2,-\xi'}^*(\rr_2'-\rr_1',-\tau)
]\nonumber \\
=& -g_{2,\xi}^*(\rr_1-\rr_2,\tau)g_{2,\xi'}(\rr_1'-\rr_2',\tau) \, . 
\label{eq:u4u4correlator}
\eaa

\begin{hhc}
\section{Fourier transformations}
\label{sec:ft}
\bh{In this section, we provide the Fourier transformation of the following functions \bh{that have been used in Sec.~\ref{sec:rkky}, Eq.~\ref{eq:rkky_ft}.}
\baa  
&\sum_{\RR}\frac{1}{|\RR|^3} e^{-i\kk\cdot\ \RR} ,\quad 
\sum_{\RR}\frac{1}{|\RR|^5} e^{-i\kk\cdot\ \RR}  ,\quad 
&\sum_{\RR}\frac{R_x}{|\RR|^5} e^{-i\kk\cdot\ \RR}  ,\quad 
&\sum_{\RR}\frac{R_y}{|\RR|^5} e^{-i\kk\cdot\ \RR} ,\quad  
&\sum_{\RR}\frac{R_x^2-R_y^2 -2i\xi R_xR_y}{|\RR|^7} e^{-i\kk\cdot\ \RR} 
\eaa 
}

We first consider the Fourier transformation of $|\RR|^{-3}$
\ba 
\sum_{\RR}\frac{1}{|\RR|^3} e^{-i\kk\cdot\ \RR}
\ea 
We replace the summation with the integral, $\sum_{\RR} \rightarrow  \frac{A_{MBZ}}{4\pi^2}\int d^2\RR $ and have
\ba 
&\frac{A_{MBZ}}{4\pi^2} 
\int d\theta_{\RR} r dr 
\frac{1}{r^3 }\bigg[ 
J_0(kr) +2 \sum_{m=1} (-i)^m J_m(kr) \cos(m(\theta_{\kk} -\theta_{\RR} ) )
\bigg]  \\
=&\frac{A_{MBZ}}{2\pi} 
\int \frac{1}{r^{2}} J_0(kr) dr 
\ea 
We then \bh{regularize} the integral by replacing $r$ with $\sqrt{r^2+a_M^2}$ where $a_M$ is the moir\'e lattice constant. Then we find 
\ba 
\sum_{\RR}\frac{1}{|\RR|^3} e^{-i\kk\cdot\ \RR} \approx 
\frac{A_{MBZ}}{2\pi} 
\int \frac{1}{(r^2+a_M^2)} J_0(kr) dr 
= \frac{ A_{MBZ}}{4 a_{M}}(I_0(ka_M) -L_0(ka_M))
\ea 
where $I_n(x)$ is the modified Bessel function of the first kind~\cite{abramowitz1988handbook}, $L_n(x)$ is the modified Struve function~\cite{abramowitz1988handbook}. It is also useful to evaluate its behavior at small $k$
\ba 
\frac{ A_{MBZ}}{4 a_{M}}(I_0(ka_M) -L_0(k a_M)) =  \frac{ A_{MBZ}}{4 a_{M}}(1 - \frac{2  a_M}{\pi}k + \frac{a_M^2k^2}{4}) +o(k^2)
\ea

We next consider $1/|\RR|^5$. Following the same strategy 
\ba 
\sum_{\RR}\frac{1}{|\RR|^5} e^{-i\kk\cdot\ \RR}
\approx&\frac{A_{MBZ}}{4\pi^2} 
\int d\theta_{\RR}  dr 
\frac{1}{(r^2+a_M^2)^{2} }\bigg[ 
J_0(kr) +2 \sum_{m=1} (-i)^m J_m(kr) \cos(m(\theta_{\kk} -\theta_{\RR} ) )
\bigg]  \\
=&\frac{A_{MBZ}}{2\pi} 
\int \frac{1}{(r^{2}+a_M^2)^{2}} J_0(kr) dr \\
=& \frac{A_{MBZ}}{24\pi ka_M^4 } 
\bigg[ 
3\pi(-2 +k^2a_M^2)L_1(ka_M) + qa_M 
\bigg( 4qa_M + 3\pi I_0(q a_M) -3 \pi qa_M I_1(qa_M) - 3\pi L_2(qa_M)
\bigg) 
\bigg]  \\
\approx & \frac{A_{MBZ}}{8a_M^3}(1 - \frac{1}{4}(ka_M)^2) +o(k^2)
\ea 
In the final line, we evaluated the behavior at small $k$.

We next consider the Fourier transformation of $R_x/r^5$. Following the same strategy 
\ba 
\sum_{\RR}\frac{R_x}{|\RR|^5} e^{-i\kk\cdot\ \RR}
\approx&\frac{A_{MBZ}}{4\pi^2} 
\int d\theta_{\RR}  dr 
\frac{\cos(\theta_\RR)}{(r^2+a_M^2)^{3/2} }\bigg[ 
J_0(kr) +2 \sum_{m=1}(-i)^m J_m(kr) \cos(m(\theta_{\kk} -\theta_{\RR} ) )
\bigg]  \\
=&\frac{-i \cos(\theta_\kk)A_{MBZ}}{2\pi} 
\int   dr 
\frac{1}{(r^2+a_M^2)^{3/2} } J_1(kr)  \\
=&\frac{-iA_{MBZ}k_x}{2\pi k^2a_M^3} 
\bigg[ 1-e^{-k a_M}(1+ka_M)\bigg] \\
\approx & \frac{-iA_{MBZ}k_x}{2\pi a_M}(\frac{1}{2}-\frac{ka_M}{3}) +o(k^2)
\ea 
In the final line, we evaluated the behavior at small $k$.

We next consider the Fourier transformation of $R_y/r^5$. Following the same strategy 
\ba 
\sum_{\RR}\frac{R_y}{|\RR|^5} e^{-i\kk\cdot\ \RR}
\approx&\frac{A_{MBZ}}{4\pi^2} 
\int d\theta_{\RR}  dr 
\frac{\sin(\theta_\RR)}{(r^2+a_M^2)^{3/2} }\bigg[ 
J_0(kr) +2 \sum_{m=1} (-i)^m J_m(kr) \cos(m(\theta_{\kk} -\theta_{\RR} ) )
\bigg]  \\
=&\frac{-i \sin(\theta_\kk)A_{MBZ}}{2\pi} 
\int   dr 
\frac{1}{(r^2+a_M^2)^{3/2} } J_1(kr)  \\
=&\frac{-iA_{MBZ}k_y}{2\pi k^2a_M^3} 
\bigg[ 1-e^{-k a_M}(1+ka_M)\bigg] \\
\approx & \frac{-iA_{MBZ}k_y}{2\pi a_M}(\frac{1}{2}-\frac{ka_M}{3}) +o(k^2)
\ea 
In the final line, we evaluated the behavior at small $k$.

We next consider the Fourier transformation of $\frac{R_x^2-R_y^2-2i\xi R_xR_y}{|\RR|^7}$. 
\ba 
&\sum_{\RR}\frac{R_x^2-R_y^2 -2i\xi R_xR_y}{|\RR|^7} e^{-i\kk\cdot\ \RR}\\
\approx&\frac{A_{MBZ}}{4\pi^2} 
\int d\theta_{\RR}  dr 
\frac{\cos(2\theta_{\RR}) -i\xi\sin(2\theta_{\RR})}{(r^2+a_M^2)^{2} }\bigg[ 
J_0(kr) +2 \sum_{m=1} (-i)^m J_m(kr) \cos(m(\theta_{\kk} -\theta_{\RR} ) )
\bigg]  \\
=&\frac{-2\pi A_{MBZ}}{4\pi^2} 
\int   dr 
\frac{\cos(2\theta_{\kk}) +i\xi\sin(2\theta_{\kk})}{(r^2+a_M^2)^{2} }J_2(kr)\\
=&-\frac{A_{MBZ}k}{2\pi}
(k_x^2-k_y^2 +2i\xi k_xk_y) 
\bigg[ 
-\frac{1}{30} +\frac{\pi}{4k^3a_M^3} 
\bigg( I_2(ka_M) +ka_MI_3(ka_M)
\bigg) 
-\frac{\pi}{4k^3a_M^3}\bigg(L_2(ka_M) +ka_M L_3(ka_M) \bigg) 
\bigg] 
\\
\approx & -\frac{A_{MBZ}}{64a_M}(k_x^2-k_y^2 +2i\xi k_xk_y)
\ea 
In the final line, we \bh{evaluated the} behavior at small $k$. \bh{$I_n(x)$ is the modified Bessel function of the first kind~\cite{abramowitz1988handbook}, $L_n(x)$ is the modified Struve function~\cite{abramowitz1988handbook}.}

In summary, 
\baa 
\sum_{\RR}\frac{1}{|\RR|^3} e^{-i\kk\cdot\ \RR} =&\frac{ A_{MBZ}}{4 a_{M}}(I_0(ka_M) -L_0(ka_M)) \approx \frac{ A_{MBZ}}{4 a_{M}}(1 - \frac{2  a_M}{\pi}k + \frac{a_M^2k^2}{4})\nonumber \\
\sum_{\RR}\frac{1}{|\RR|^5} e^{-i\kk\cdot\ \RR}= &\frac{A_{MBZ}}{24\pi ka_M^4 } 
\bigg[ 
3\pi(-2 +k^2a_M^2)L_1(ka_M) + qa_M 
\bigg( 4qa_M + 3\pi I_0(q a_M) -3 \pi qa_M I_1(qa_M) \nonumber \\
& - 3\pi L_2(qa_M)
\bigg) 
\bigg] 
\approx \frac{A_{MBZ}}{8a_M^3}(1 - \frac{1}{4}(ka_M)^2) \nonumber \\
\sum_{\RR}\frac{R_x}{|\RR|^5} e^{-i\kk\cdot\ \RR} =&\frac{-iA_{MBZ}k_x}{2\pi k^2a_M^3} 
\bigg[ 1-e^{-k a_M}(1+ka_M)\bigg]\approx  \frac{-iA_{MBZ}k_x}{2\pi a_M}(\frac{1}{2}-\frac{ka_M}{3}) \nonumber \\
\sum_{\RR}\frac{R_y}{|\RR|^5} e^{-i\kk\cdot\ \RR}=&\frac{-iA_{MBZ}k_y}{2\pi k^2a_M^3} 
\bigg[ 1-e^{-k a_M}(1+ka_M)\bigg] \approx  \frac{-iA_{MBZ}k_y}{2\pi a_M}(\frac{1}{2}-\frac{ka_M}{3}) \nonumber \\
\sum_{\RR}\frac{R_x^2-R_y^2 -2i\xi R_xR_y}{|\RR|^7} e^{-i\kk\cdot\ \RR}=&-\frac{A_{MBZ}k}{2\pi}
(k_x^2-k_y^2 +2i\xi k_xk_y) 
\bigg[ 
-\frac{1}{30} +\frac{\pi}{4k^3a_M^3} 
\bigg( I_2(ka_M) +ka_MI_3(ka_M)
\bigg) 
-\frac{\pi}{4k^3a_M^3}\nonumber \\ 
&\bigg(L_2(ka_M) +ka_M L_3(ka_M) \bigg) 
\bigg] 
\approx-\frac{A_{MBZ}}{64a_M}(k_x^2-k_y^2 -2i\xi k_xk_y) \label{eq:ft_power}
\eaa

\bh{ 
\section{Single particle Green's function in the ordered state}
\label{sec:green_order}
In this section, we discuss the single-particle Green's function of the conduction electrons in the ordered phase that has been used in Sec.~\ref{sec:eff_thy}. The Hamiltonian of $c$ electrons is defined in Eq.~\ref{eq:formula_hcorder} and is also written here 
 \baa  
 &\hH_{c,order} \nonumber \\
 =&\sum_{\kk,i} 
 \begin{bmatrix}
 \psi_{\kk,i}^{\xi=+,c',\dag} & \psi_{\kk,i}^{\xi=+,c'',\dag}
 & \psi_{\kk,i}^{\xi=-,c',\dag} & \psi_{\kk,i}^{\xi=-,c'',\dag} 
 \end{bmatrix}
\begin{bmatrix}
E_{0,\kk}^{+,i} +E_{3,\kk}^{+,i} & v_\star(k_x +i k_y)
& v_\kk^{i} (k_x-i k_y) &0 
\\
v_\star(k_x-i k_y) & E_{0,\kk}^{+,i}-E_{3,\kk}^{+,i} 
& 0 & 0 
\\
 v_\kk^{i} (k_x+i k_y) & 0& E_{0,\kk}^{-,i} +E_{3,\kk}^{-,i} & v_\star(k_x -i k_y)\\
0&0& v_\star(k_x+i k_y) & E_{0,\kk}^{-,i}-E_{3,\kk}^{-,i} 
\end{bmatrix}
  \begin{bmatrix}
 \psi_{\kk,i}^{\xi,c'} \\ \psi_{\kk,i}^{\xi,c''}
 \\
 \psi_{\kk,i}^{\xi=-,c'} \\ \psi_{\kk,i}^{\xi=-,c''}
 \end{bmatrix}
 \eaa  
 The single-particle Green's function can be obtained by
 \baa 
g_{aa'}^{\xi\xi',i}(\kk, i\omega) = \left\{ i\omega \mathbb{I}_4 - 
 \begin{bmatrix}
E_{0,\kk}^{+,i} +E_{3,\kk}^{+,i} & v_\star(k_x +i k_y)
& v_\kk^{i} (k_x-i k_y) &0 
\\
v_\star(k_x-i k_y) & E_{0,\kk}^{+,i}-E_{3,\kk}^{+,i} 
& 0 & 0 
\\
 v_\kk^{i} (k_x+i k_y) & 0& E_{0,\kk}^{-,i} +E_{3,\kk}^{-,i} & v_\star(k_x -i k_y)\\
0&0& v_\star(k_x+i k_y) & E_{0,\kk}^{-,i}-E_{3,\kk}^{-,i} 
\end{bmatrix} \right\}_{\xi a,\xi'a'} 
\eaa 
where $a,a' \in \{c',c''\}$. By diagonalizing the matrix, we have 
\baa 
&g_{c'c'}^{\xi\xi,i}(\kk,i\omega) = f_1(|\kk|,i\omega,\xi,i) \nonumber \\ 
&g_{c''c''}^{\xi\xi,i}(\kk,i\omega) = f_2(|\kk|,i\omega,\xi,i) \nonumber \\
&g_{c'c''}^{\xi\xi,i}(\kk,i\omega) =
v_\star(k_x +i\xi k_y) f_3(|\kk|,i\omega,\xi,i) \nonumber \\
&g_{c''c'}^{\xi\xi,i}(\kk,i\omega) = 
v_\star(k_x - i\xi k_y) f_3(|\kk|,i\omega,\xi,i)\nonumber \\
&g_{c'c'}^{\xi-\xi,i}(\kk,i\omega) = v_\star^\prime( k_x-i\xi k_y)f_4(|\kk|,i\omega,i) \nonumber \\
&g_{c''c''}^{\xi-\xi,i}(\kk,i\omega) =  v_\star^\prime v_\star^2 ( k_x-i\xi k_y)^3  f_5(|\kk|,i\omega,i)\nonumber \\
&g_{c'c''}^{\xi-\xi,i}(\kk,i\omega) =  v_\star^\prime v_\star ( k_x-i\xi k_y)^2 f_6(|\kk|,i\omega,\xi,i) \nonumber \\
&g_{c''c'}^{\xi-\xi,i}(\kk,i\omega) =  v_\star^\prime v_\star ( k_x+i\xi k_y)^2f_6(|\kk|,i\omega,-\xi,i)
\label{eq:green_order_fromula}
\eaa 
where we introduce the following functions 
\baa 
&F_{|\kk|,i} = 
\bigg(
\bigg(i\omega - E_{0,\kk}^{+,i} \bigg)^2 -v_\star^2 |\kk|^2 -\bigg(E_{3,\kk}^{+,i}\bigg)^2 -v_\star^2 |\kk|^2 \bigg) 
\bigg( 
\bigg(i\omega - E_{0,\kk}^{-,i} \bigg)^2 -v_\star^2 |\kk|^2 -\bigg(E_{3,\kk}^{-,i}\bigg)^2 -v_\star^2 |\kk|^2 \bigg) \nonumber \\
&f_1(|\kk|,i\omega,\xi,i) = \frac{1}{F_{|\kk|,i} }(i\omega +E_{3,\kk}^{\xi,i} -E_{0,\kk}^{\xi,i} )\bigg[
(i\omega -E_{0,\kk}^{-\xi,i})^2 -(E_{3,\kk}^{-\xi,i})^2 -v_\star^2 |\kk|^2 
\bigg] \nonumber \\ 
&f_2(|\kk|,i\omega,\xi,i) = \frac{1}{F_{|\kk|,i} }\bigg\{(i\omega -E_{3,\kk}^{\xi,i} -E_{0,\kk}^{\xi,i} )\bigg[
(i\omega -E_{0,\kk}^{-\xi,i})^2 -(E_{3,\kk}^{-\xi,i})^2 -v_\star^2 |\kk|^2 
\bigg]
+(i\omega +E_{3,\kk}^{-\xi,i}-E_{0,\kk}^{-\xi,i}) (v_\star^\prime)^2 |\kk|^2 \bigg\} \nonumber \\
&f_3(|\kk|,i\omega,\xi,i)= \frac{1}{F_{|\kk|,i}}
\bigg[ 
(i\omega - E_{0,\kk}^{-\xi,i})^2 -(E_{3,\kk}^{-\xi,i})^2 -v_\star^2 |\kk|^2 
\bigg] \nonumber \\
&f_4(|\kk|,i\omega,i) = \frac{1}{F_{|\kk|,i}}(i\omega + E_{3,\kk}^{+,i} - E_{0,\kk}^{+, i} )
(i\omega + E_{3,\kk}^{-,i} - E_{0,\kk}^{-, i}) \nonumber \\
&f_5(|\kk|,i\omega,i)=  \frac{1}{F_{|\kk|,i}} \nonumber \\
&f_6(|\kk|,i\omega,\xi,i) =  \frac{1}{F_{|\kk|,i}} (i\omega +E_{3,\kk}^{\xi,i} -E_{0,\kk}^{\xi,i} ) 
\eaa 
where 
\baa 
F_{|\kk|,i} = &
\bigg(
\bigg(i\omega - E_{0,\kk}^{+,i} \bigg)^2 -v_\star^2 |\kk|^2 -\bigg(E_{3,\kk}^{+,i}\bigg)^2 -v_\star^2 |\kk|^2 \bigg) 
\bigg( 
\bigg(i\omega - E_{0,\kk}^{-,i} \bigg)^2 -v_\star^2 |\kk|^2 -\bigg(E_{3,\kk}^{-,i}\bigg)^2 -v_\star^2 |\kk|^2 \bigg) \nonumber \\
&
-(v_\star^\prime)^2 |\kk|^2 (i\omega -E_{0,\kk}^{+,i} +E_{3,\kk}^{+,i}) 
(i\omega -E_{0,\kk}^{-,i}+E_{3,\kk}^{-,i}) \, .
\eaa 
We also mention that  $E_{3,\kk}^{\xi,i},E_{0,\kk}^{\xi,i}, F_{|\kk|,i}, f_1(|\kk|,i\omega,\xi,i),f_2(|\kk|,i\omega,\xi,i),f_3(|\kk|,i\omega,\xi,i),f_4(|\kk|,i\omega,i),f_5(|\kk|,i\omega,i),f_6(|\kk|,i\omega,\xi,i)$ are functions of $|\kk|$ instead of $\kk$. 
}

\end{hhc}

\end{document}